# Psychometric Comparability of LLM–Based Digital Twins

Yufei Zhang[1], Zhihao Ma[1]

[1] Computational Communication Collaboratory, School of Journalism and Communication, Nanjing University, Nanjing 210023, China

**Corresponding author:**

Zhihao Ma

Computational Communication Collaboratory

School of Journalism and Communication

Nanjing University

Room 345, Zijin Building, Nanjing University (Xianlin Campus), 163 Xianlin Road, Qixia District, Jiangsu Nanjing, 210023 China

Phone: 86 17561538460

Email: redclass@163.com

**Abstract**

Large language models (LLMs) act as digital twins for human respondents, yet their psychometric comparability remains uncertain. We propose a construct validity framework spanning construct representation and the nomothetic span, benchmarking models against human gold standards. Across studies, digital twins achieved high aggregate-level accuracy and profile correlations, but showed attenuated item-level correlations. In word association tests, LLM networks exhibited humanlike small-world structure and theory-consistent communities, yet diverged lexically and in local structure. In decision-making and contextualized tasks, they under-reproduced heuristic biases, demonstrating normative rationality, compressed variance, and limited temporal sensitivity. Feature-rich and trait relevant conditioning improved Big Five personality prediction and nomothetic-span alignment, but network invariance remained limited, with partial configural solutions and persistent loading differences. In cross-language free-text tasks in English and Chinese, feature-rich digital twins better approximated construct-level narrative content, but linguistic and idiographic differences persisted. These findings clarify that digital twins are most useful within validated boundaries, where the construct, task and level of inference align with evidence from human data.

Large language models (LLMs) are now increasingly used to generate responses in lieu of human respondents in social science, including public opinion polling[1,2], market research[3], and psychological tests[4]. This body of work centers on the same goal: using LLMs to generate humanlike attitudes and behaviors at scale. However, the psychometric comparability of LLM-based digital twins and human respondents remains underexamined.

Early evidence is mixed. LLMs can reproduce aggregate patterns, from matching population means[2] to replicating main psychological experimental effects[5], and in some settings they show humanlike social reasoning[4] and behavioral alignment[6]. However, simulations often underrepresent human behavioral heterogeneity[7], produce compressed response distributions[8], converge on normative answers in value-laden domains[9], and lack temporal robustness[2,10]. These mixed findings shift the problem from behavioral mimicry to construct validity. Therefore, the central question is whether responses generated by LLM-based digital twins are psychometrically comparable to those produced by human respondents.

In Cronbach and Meehl's classic formulation[11], clarifying what a construct is requires specifying a nomological net, that is, an interlocking system of laws that links observable properties to one another, theoretical constructs to observables, and theoretical constructs to each other. Building on this logic, Embretson[12] distinguishes two separable targets: construct representation (such as the processes and demands that generate item responses) and nomothetic span (the pattern of relations between test scores and other individual-difference variables). This distinction has particular implications for LLM-based digital twins, which can be understood as silicon samples generated by conditioning an LLM on person specific characteristics[13]. Even when outputs appear humanlike or achieve high task accuracy, psychometric comparability requires similarity both in response-generating processes and in the nomological relations the resulting scores support. Accordingly, we treat humans as the pragmatic gold standard and evaluate construct validity along two dimensions: construct representation and the nomothetic span.

From a construct validity perspective, a central risk at the level of construct representation is that humans and LLMs may generate similar responses from different representational spaces. Human concepts are grounded in multimodal experience and social interaction[14] and are often described as semantic memory networks supporting associative retrieval[15]. LLMs, by contrast, learn semantic structure from distributional regularities in text, representing concepts as high-dimensional vectors organized by geometric proximity[16]. Although next-token prediction can yield representations that align with human neural and behavioral patterns[17], this does not establish representational equivalence. As a result, the same concept, such as "aging," may be embedded in a different semantic neighborhood despite superficially plausible outputs (Fig. 1a). This issue is especially salient for implicit cognition, where distributional representations may reproduce humanlike associations while also reflecting coverage gaps and historical biases in training data[18]. Apparent semantic agreement may therefore mask displacement in construct representation.

A deeper threat is unintended confounding. In human measurement, attributes such as age or sex exist independently of the instrument and causally shape attitudes (Fig. 1b, left). In LLMs, by contrast, these attributes exist only as token-based representations, and outputs are generated through statistical pattern matching. As a result, specifying one attribute (for example, 65 years old) may activate correlated but unobserved features, such as retirement, health, or income, based on learned regularities. What appears to be a targeted manipulation therefore becomes a composite intervention that shifts multiple latent features at once. Gui and Toubia[3] formalize this as a violation of unconfoundedness: treatment variation may alter pre-treatment covariates and contextual factors that should remain fixed, blurring the simulated counterfactual. As shown in Fig. 1b (right), the target attribute and its correlates become entangled in generation rather than linked through a stable causal pathway. Under a causal view of validity[19], a test validly measures an attribute only if variation in that attribute causes variation in measurement outcomes. If observed scores reflect both the target attribute and prompt-induced covariate shifts, LLM-based digital twins may not cleanly measure the intended construct.

These concerns motivate a direct test of psychometric comparability. In our framework (Fig. 1c), prior survey responses and other person-specific features serve as inputs to the digital twin. Using many-shot in-context learning[20], LLM constructs a twin of the individual and generates responses to designated questions. Evaluation centers on nomological span—whether relations among inputs, latent constructs, and downstream responses observed in humans are reproduced by digital twins under the same task demands.

To this end, we conduct four studies that proceed from aggregate structure to individual-level psychometric comparability, and from structured responses to naturalistic expression. Study 1 uses a word association test to probe construct representation directly, testing whether LLMs reproduce the structure of human semantic associations. Study 2 assesses digital twins' performance against criterion-referenced decision benchmarks, using tasks with well-established human response patterns. We begin with classic heuristics and biases problems and extend to more contextualized settings. Study 3 systematically varies the richness and relevance of person feature inputs to identify how much individual information is required for twins to recover the aggregate organization of Big Five traits, and we assess psychometric comparability using network measurement invariance tests. Because psychological constructs are often expressed through language in real research and applied contexts, Study 4 extends this evaluation beyond the predominantly English-language research by examining naturalistic expression in both English and Chinese. We test whether digital twins approximate both linguistic patterns and focal psychological constructs in text.

Collectively, we provide a construct-validity framework for deciding when LLM-based digital twins can be treated as psychometrically comparable to humans, and for diagnosing failures in construct representation and nomothetic span.

## Results

### Study 1 Construct Representation of LLMs in Word Association Tests

In Study 1, we assessed LLM-based psychological representations at the level of construct representation. Using parallel large-scale free-association norms, we examined three grounded frameworks: the Big Five Personality[21], the Stereotype Content Model[22], and Moral Foundations Theory[23]. For each framework, we built semantic association networks for humans and LLMs and evaluate alignment in lexical coverage, node and edge overlap, global network structure, node centrality, and community organization. We next tested whether similarity structures derived from human and LLM associations predicted external lexical norm scores for valence, arousal, dominance[24], and concreteness[25]. Because the norms aggregate responses across individuals, this study captures aggregate-level associative structure and coverage.

Across all three wordlists, human norms contained fewer total responses but covered the broadest lexical space. LLMs had markedly lower missingness; Claude-3-5-haiku-latest achieved the lowest missing rate, but this higher completion came with narrower aggregate lexical coverage (Table S2). Network overlap was limited: fewer than half of nodes were shared between human and LLM networks, and LLMs omitted many human-specific nodes. At the edge level, Llama3.1-8b showed the lowest overlap with humans across datasets (Table S3). Global network statistics showed that Mistral-7b most closely matched human networks in node and edge counts. All networks displayed small-world properties ($\sigma > 1$), but Claude-3-5-haiku-latest displayed substantially elevated small-worldness indices, with higher normalized clustering coefficients and longer normalized path lengths (Table S4). Pairwise Wilcoxon signed-rank tests on node degree centrality showed a consistent human–LLM gap across all three wordlists, with human networks exhibiting higher centrality than all LLM networks and the largest differences consistently observed for Claude-3-5-haiku-latest (Fig. 2a; Table S5). Differences between Mistral-7b and Llama3.1-8b were small and mostly non-significant.

Community analyses showed partial convergence between LLM and human associative organization. Using empirical human partitions as the benchmark, LLM–human ARI values were consistently above size-preserving random baselines across community detection algorithms, indicating consistent but partial overlap in community structure (Tables S6–S7). Convergence varied by model, wordlist and algorithm. Mistral-7b showed the most consistent alignment with human community structure, with its clearest advantage observed for BFI-2 and Moral Foundations, whereas Llama3.1-8b showed lower alignment in the same wordlists. Claude-3-5-haiku-latest often produced high V-measure values, but these were not always accompanied by the highest ARI values. Theory-defined categories provided a secondary reference. Human and theory ARIs were above random baselines but remained modest, with only partial recovery of the original theoretical taxonomies in free-association communities (Table S7). Bootstrap analyses gave similar results but varied across algorithms: assignment stability was generally interpretable for Louvain, Leiden and Walktrap, whereas nSBM was substantially lower. Model contrast estimates mirrored the descriptive ARI pattern, with Mistral-7b most consistently outperforming Llama3.1-8b (Fig. 2b; Tables S8–S10).

The lexical norm prediction analysis showed that both human and LLM association spaces contained semantic information relevant to affective meaning and concreteness, but the human association space consistently produced the highest correlations with independently measured norm scores across the four dimensions (Fig. 2c). Among LLMs, Mistral-7b generally showed the highest correlations for valence, arousal and dominance, whereas Claude-3-5-haiku-latest showed the highest correlation for concreteness. Llama3.1-8b showed the lowest correlations across all lexical dimensions (Tables S11–S12). Overall, LLM-generated associative spaces contained considerable lexical semantic structure, but their correlations with external affective meaning and concreteness norms remained lower than those of human free association spaces.

**Study 2: Criterion-based Evaluation of LLM-based Digital Twins**

We tested whether LLM-generated responses reproduced patterns observed in comparable human data across classic heuristics-and-biases tasks and contextually embedded social judgments. In Study 2a, grounded in dual-process theory[26,27], we used the Twin-2K-500 dataset[13] to evaluate whether digital twins reproduced classic within- and between-subject effects across multiple LLM families, including biases associated with System 1 processing. In Study 2b, we used the COVID-Dynamic dataset[28] to assess temporal validity in social simulation. A key concern, highlighted in recent work[10], is atemporality, whereby models may not distinguish data from different time periods. We compared digital twins' responses with human benchmarks across pandemic waves to test temporal sensitivity. Performance was evaluated using task-level accuracy, item-level and profile-level correlations, and error slopes regressing prediction error on human responses.

**Study 2a**

Averaged across 17 tasks, overall accuracy ranged from 67.8% to 72.0% across models. DeepSeek V3-0324 and DeepSeek V3.1 achieved the highest accuracies, at 72.0% and 71.9%, respectively, followed by gpt-oss-120b, DeepSeek V4-Flash, Gemma-4-31B-it, and Qwen 2.5 Max. These two highest-performing models were closely comparable to the 71.72% benchmark reported by Toubia et al.[13]. At the item level, correlations were generally small, with most estimates below 0.30. For approximately half of the tasks, correlations also fell within the random baseline confidence interval. At the profile level, estimated only for tasks with at least three items, correlations reached medium to large effect sizes and were consistently well outside the random baseline's confidence intervals (Tables S13-S14). In addition, regression slopes relating prediction error to human responses were uniformly negative across tasks (Fig. 3).

Cognitive bias replication was evaluated across 11 between-subjects and five within-subjects experiments, using the effects observed in the human sample of Toubia et al.[13] as the gold standard rather than the original source studies. Replication counts differed across models (Table 1; Table S15). Gemma-4-31B-it reproduced 5 of the 16 effects; Qwen 2.5 Max and gpt-oss-120b reproduced 6; DeepSeek V3-0324 and DeepSeek V3.1 reproduced 7; and DeepSeek V4-Flash had the highest replication count, reproducing 11 of the 16 effects. In the base-rate problem, all models assigned lower probability judgments in the "30 engineers" than in the "70 engineers" condition, although DeepSeek V3-0324, DeepSeek V3.1 and Qwen 2.5 Max often returned the stated base rates. In the sunk cost task, the effect reversed direction for all models except DeepSeek V4-Flash. None of the DeepSeek models reproduced the framing effect, with risk aversion remaining stable across gain and loss frames. In probability matching, all models selected the maximizing option, whereas the human maximizing rate was 30%.

In anchoring and adjustment, the models showed negligible sensitivity to anchors and often retrieved the correct factual answer—54 African countries—doing so in 99.95% of cases for Qwen 2.5 Max, 92.95% for gpt-oss-120b, and 99.32% for Gemma-4-31B-it. With the exception of DeepSeek V4-Flash, none of the models showed myside bias, and none showed denominator neglect. The DeepSeek models did not exhibit the less is more effect. In risk judgment, humans typically show a negative correlation between perceived risk and benefit, but the Qwen 2.5 Max-based digital twins showed a positive correlation. Omission bias was also markedly attenuated: in the vaccine scenario, refusal rates were 1.3% for DeepSeek V3-0324, 9.4% for DeepSeek V3.1, 0.2% for both Qwen 2.5 Max and Gemma-4-31B-it, and 5.3% for gpt-oss-120b, all far below the roughly 45% typically observed in humans.

**Study 2b**

Drawing on Wave 10 of the data, overall accuracy across models was mostly clustered between 0.75 and 0.80. Gemma-4-31B-it had the highest values across the three evaluation metrics, and prediction error was consistently negatively associated with human responses. Across all models and tasks, item-level correlations were consistently lower than profile correlations (Fig. 4b; Table S16).

Temporal prompt manipulations produced little change in predictive performance (Fig. 4a). In the Wave 10 analysis, accuracy and correlations remained similar across date, COVID-19 context and event information conditions for DeepSeek V4-Flash, Gemma-4-31B-it and Qwen 3.5 Plus (Tables S17–S19). Accuracy varied within narrow ranges for DeepSeek V4-Flash (0.765–0.770), Gemma-4-31B-it (0.784–0.788) and Qwen 3.5 Plus (0.767–0.771). By contrast, gpt-oss-120b showed lower accuracy under fabricated event conditions and past date prompts (0.736–0.749). The all-wave analysis showed comparable results (Supplementary Fig. 2b): providing dates and contemporaneous COVID-19 context did not consistently improve accuracy or correlations (Table S20). Manipulation checks showed that models usually recognized supplied dates and event contexts, although some responses referenced salient historical cues, such as Black Lives Matter protests or the presidential election (Tables S21-S22).

**Study 3 Psychometric Evaluation of Digital Twins Across Feature Input Levels**

In Study 3, we tested whether richer and more construct-relevant person-specific inputs improved digital twins' recovery of Big Five responses, nomothetic span alignment and psychometric structure across two personality inventories.

**Study 3a**

Digital-twin performance was primarily shaped by the psychological relevance of the input information. The full-feature condition produced the highest accuracy and correlations, prediction-error slopes closest to zero, and the strongest nomothetic span alignment; results for each Big Five domain are reported in Supplementary Fig. 3a–e. The Big Five high-relevant subset performed nearly as well as the full-feature condition and occasionally exceeded it (Fig. 5a; Tables S24–S27). In contrast, demographics-only and task-only prompts produced lower item-level correlations, weaker nomothetic span alignment and lower disattenuated trait-score correlations. Profile correlations consistently exceeded item-level correlations across input conditions. Although input condition accounted for the main performance differences, Gemma-4-31B-it showed the highest overall values across models.

Network analyses showed partial configural comparability between digital-twin and human responses. In comparisons between human and digital-twin networks, Gemma-4-31B-it reproduced the expected five-community Big Five structure across all input conditions. gpt-oss-120b achieved configural invariance in four input conditions, DeepSeek V4-Flash in three, and Qwen 3.5 Plus in one. The remaining comparisons mostly showed partial configural invariance; deviations for DeepSeek V4-Flash were concentrated in Openness, mainly item 35R, whereas Qwen 3.5 Plus showed configural noninvariance in the task-only condition, with the entire Agreeableness domain excluded (Tables S28–S29). Network configural invariance solutions for all models are shown in Supplementary Fig. 3. Within-LLM comparisons showed that the Big Five partition was generally preserved across input conditions, except for gpt-oss-120b (Table S30). Between-LLM comparisons showed greater configural convergence under richer input conditions (Table S31). Evidence for metric invariance was more limited: no human and digital-twin comparison achieved metric invariance, and more than half of the items showed significant loading differences, often with stronger loadings in digital-twin networks and with the most recurrent deviations in Neuroticism and Openness (Fig. 5c; Tables S32–S35).

Post hoc calibration showed that simple distributional alignment did not improve correspondence between human and digital twin responses. Randomized ordinal equipercentile equating consistently reduced item-level correlations across models, input conditions and calibration proportions (Table S36). By contrast, regression-based bias correction produced selective gains in the full-feature condition. Item fixed effects largely eliminated trait-score mean differences and improved marginal distributional alignment, but had little effect on item-level correlations or disattenuated trait-score correlations. Adding the raw digital twin response more consistently improved within-person profile correlations and reduced trait-score discrepancies, with lower MAE and RMSE (Tables S37–S38). At the nomothetic level, bias-corrected trait scores improved recovery of human trait and external-scale association structures, with the most stable gains in the 25% and 50% calibration-sample settings. Adding demographic covariates produced small and inconsistent changes and did not improve performance beyond item fixed effects and raw digital twin responses (Table S39).

**Study 3b**

Consistent with Study 3a, the full-feature condition produced the highest accuracy, item-level correlations and profile correlations, whereas the COVID memory only condition produced the lowest values (Tables S40–S41). Prediction-error slopes were negative across all conditions but were closer to zero when prompts included the full feature input or Big Five relevant features ($r > 0.5$; Fig. 6a). Nomothetic span alignment was highest under the full-feature and Big Five high relevant feature conditions, and lower under less informative input conditions (Table S43). Temporal sensitivity analyses showed no clear changes in predictive performance across the four temporal manipulation conditions (Tables S44–S47).

Network analyses showed partial configural comparability between human and digital twin responses. The human network did not form a clean theoretical Big Five partition, with community mixing across Agreeableness, Openness and Neuroticism. In the human and digital twin measurement invariance tests, all comparisons showed partial configural invariance except gpt-oss-120b under the full-feature input condition, where the entire Agreeableness domain was excluded and the solution was classified as configural noninvariance. Across partial configural invariance solutions, deviations from the theoretical Big Five structure most often involved Openness, particularly items 3R and 8R; Extraversion-related deviations were mainly observed for Gemma-4-31B-it and Qwen 3.5 Plus (Tables S48–S49). Within-LLM and between-LLM comparisons showed a comparable pattern: common networks could be estimated, but community solutions still mixed items across Big Five domains (Tables S50–S51). Network configural invariance solutions for all models are shown in Supplementary Fig. 4.

At the metric level, all human and digital twin comparisons achieved partial metric invariance, with significant loading differences observed for fewer than half of the items (Fig. 6c; Tables S52–S53). These differences were concentrated in Conscientiousness, especially items 15R, 20 and 60, and in Extraversion item 27R. Loading differences were less pronounced in comparisons involving similarly informative inputs, and were mainly observed under the least informative conditions, such as demographics only or memory task inputs (Tables S54–S55).

**Study 4: Construct Representation and Linguistic Features of Digital Twins**

In Study 4, we examined digital twins' linguistic validity at two levels: surface linguistic form and construct representation in text. We compared demographics-only and feature-rich digital twins to test whether richer person-specific information improved alignment with human writing. Study 4a used the COVID Dynamic free-text task, whereas Study 4b used narrative evaluations of a scripted stimulus from a newly recruited Chinese sample. By holding the analytic workflow constant while varying language and task materials, the two sub-studies provided a cross-language and cross-task extension. Given the relatively small sample size of Study 4b, these analyses were treated as exploratory. Detailed model-specific results are reported in Supplementary Section 4.

In paired comparisons, human and digital twin texts differed in surface linguistic form, with patterns of differences varying across models, tasks and metrics. In Study 4a, all linguistic metrics differed across human, feature-rich digital twin and demographic digital twin texts for all models, with humans producing more multi-sentence responses than digital twins. Across both sub-studies, human texts had higher named entity density than every digital twin model, whereas lexical diversity was lower in human texts in nearly all model conditions. Human responses were also more variable in sentence number and length than digital twin texts, which were comparatively compressed. In Study 4b, named entity density differed across all models, with demographic digital twins producing very few named entities (Fig. 7e; Tables S56–S58).

In the exploratory POS bigram analysis, Study 4a showed partial overlap between human and digital twin texts. Human English texts were more characterized by NOUN–ADP, PRON–AUX, PRON–VERB and ADP–DET, whereas digital twins gave greater prominence to verb centered patterns, including NOUN–VERB, VERB–DET, VERB–VERB and VERB–ADV. Feature rich input did not consistently improve POS bigram alignment in Study 4a: divergence from humans was lower only for gpt-oss-120b, whereas demographic digital twins were slightly closer to humans for the other three models (Table S59). In Study 4b, feature rich input reduced POS bigram divergence from humans across all four models. NOUN–NOUN was the dominant pattern in both human and digital twin Chinese texts, but digital twins placed more weight on adverbial modification, especially ADV–VERB, ADV–ADV and VERB–ADV.

In Study 4a memory texts, DATE was the most frequent named entity in both human and digital twin outputs, with higher proportions in digital twin texts. Humans more often mentioned GPE and PERSON entities, whereas digital twins more often produced TIME and ORG entities. In Study 4b, named entity profiles were broadly similar across conditions, with PERSON dominating in both human and digital twin texts. Digital twin outputs contained higher proportions of LOCATION and DATE entities, whereas human evaluations contained more ordinal and numeric structuring; ORDINAL entities accounted for 16.9% of mentions and were nearly absent from digital twin texts.

Empirical cumulative distribution functions (CDFs) of cosine similarity showed distributional shifts across conditions (Fig. 7; Supplementary Figs. 5–6). In Study 4a, digital twins with feature rich input were closer to humans than demographic digital twins across all five dimensions and all four models, although differences between humans and feature rich digital twins remained statistically significant. In Study 4b, the effect of feature rich input varied by model across the five dimensions. Gemma-4-31B-it showed the closest alignment: differences between humans and feature rich digital twins were not significant for any dimension, whereas differences between humans and demographic digital twins were significant for all dimensions. gpt-oss-120b was farther from humans in the feature rich condition than in the demographic condition across all five dimensions (Table S60).

## Discussion

We evaluated LLM-based digital twins as a problem of construct validity, asking whether their responses can support psychometric inferences comparable to those drawn from humans. Treating human data as the gold standard, we assessed comparability across construct representation and nomothetic span in four studies. Surface humanlikeness concealed two sources of nonequivalence. First, humans and LLMs organize psychological meaning in different representational spaces, producing divergent semantic associations. Second, person-specific information enters humans and LLMs differently: a lived characteristic for a person becomes a token-based cue for an LLM, where it can activate correlated but unobserved features and introduce unintended confounding. Apparent similarity therefore could not substitute for psychometric comparability. The findings support a conditional account of digital-twin validity. Digital twins are most defensible for recovering broad semantic structure, individual response profiles and approximate nomothetic associations, especially when prompts contain rich, construct-relevant information. They are least defensible for questions that require fine-grained item variation, human heterogeneity, bounded rationality, temporal grounding or measurement invariance. Their validity depends on the specific form of comparability a research question requires.

Study 1 delimits what LLM-generated associations can establish about construct representation. Psychometric comparability requires plausible construct-related words and preservation of the associative structure through which psychological meaning is organized in human semantic memory. LLM association networks recovered nonrandom community structure and carried information about valence, arousal, dominance, and concreteness, capturing meaningful aspects of psychological language at the aggregate level. Yet human and LLM networks diverged in lexical coverage, node overlap, local centrality, and community organization, with alignment varying across models and wordlists. These differences matter because free associations are behavioral traces of semantic retrieval, not mere word lists; they reflect subjective semantic organization more directly than text-based distributional statistics, and centrality gaps imply differences in cognitive accessibility[29]. LLMs thus approximated the broad semantic terrain of a construct without representing it as humans do. Practically, their association spaces are useful early in construct development—organizing construct-related language[30,31], expanding vocabularies, and generating preliminary item pools[32]—consistent with text-analysis traditions that treat language as construct-relevant evidence[33]. Such outputs are model-derived suggestions requiring expert review, human norming, and psychometric validation.

In Study 2a, Digital twins were limited simulators of bounded human judgment. Under the dual-process model, classic heuristics-and-biases tasks elicit fast, intuitive responses and expose where they diverge from slower, rule-based deliberation[26]. Here digital twins behaved less like human decision-makers than like Homo economicus. They maximized rather than probability-matched, anchored little when a fact could be retrieved, sometimes returned stated base rates almost deterministically, and missed several framing and sunk-cost effects. Biases rooted in linguistic typicality, such as the conjunction fallacy, were more likely to persist.

This matters because many human fallacies are signatures of cognition under bounded resources instead of arbitrary errors. Computational-rationality[34] and resource-rational[35] accounts recast deviations from formal optimality as adaptive approximations under limited attention, memory, time, and effort—where computation is itself costly, and rationality extends to whether, how long, and which strategy to compute[36]. Human cognition is further grounded in embodied physical, emotional, and social experience[37]. Digital twins share none of this, answering from statistical regularities acquired in pretraining, without bodily stakes, motivational pressure, or real-time deliberation costs.

Several mechanisms may explain this apparent excessive rationality. Post-training alignment through instruction tuning and reinforcement learning from human feedback (RLHF) may reward helpful, coherent, and normatively correct answers, suppressing targeted intuitive responses[38]. Autoregressive generation may also favor high-probability continuations[39]. Task framing provides another possibility: anchoring weakens once the number of African countries is retrievable, whereas the Linda vignette still cues a semantically typical continuation. Disentangling these accounts will require comparing base, instruction-tuned, RLHF, and reasoning models; manipulating prompts with analogous examples, population means, or intuitive-answer instructions; substituting fictional or newly learned facts for real ones; and contrasting linguistically rich vignettes with mathematically equivalent probability problems.

Across Studies 2b and 3b, survey dates and contemporaneous event context barely improved predictive accuracy or correlations, even though manipulation checks showed that models recognized the supplied temporal information in most conditions. Temporal cues were therefore registered but weakly weighted in person-specific generation. At a deeper level, time-sensitive knowledge is learned from heterogeneous, non-stationary corpora, risking averaging, forgetting, and poor temporal calibration unless time is explicitly modeled during training or adaptation[40]. This matters because social survey simulation requires more than retrieving date-indexed facts: human pandemic responses were shaped by historically situated uncertainty, media exposure, policy constraints, and personal experience, whereas for an LLM a survey date may act mainly as a token cue that activates retrospectively learned event schemas. This pattern also echoes unintended confounding. Historical time was the intended manipulation, yet the date may shift unobserved assumptions correlated with that period without reconstructing the constraints under which responses were originally formed. Digital twins can thus approximate stable profiles and broad nomothetic associations yet remain less defensible when the target is historically situated judgment, temporal change, or responses formed under contemporaneous uncertainty. Stronger temporal-validity designs are needed, including models with known training cutoffs evaluated on unseen later periods[41] and temporally aligned models[42].

Study 3 specifies when digital twins can support indirect inference about psychological constructs. Richer inputs improved predictive correspondence and nomothetic-span recovery, yet quantity was not decisive: features empirically associated with the target traits performed comparably to the full-feature condition, whereas demographics were less informative. Where direct assessment is impractical, twins can thus yield indirect, model-generated estimates from proxy information—but only when that information is theoretically relevant to, and empirically associated with, the target construct within a coherent nomological network.

Predictive recovery still fell short of construct comparability. Partial configural invariance required item removal, which narrowed construct coverage and limited subsequent inference to the retained items. Network loadings index each item's standardized contribution to its dimension; loading noninvariance therefore means that corresponding items play different structural roles across human and digital-twin networks[43]. Across both inventories, predictive gains did not produce full metric invariance with human responses. In Study 3a, for example, recurrent loading differences in Openness showed that similar Openness scores could arise from different relations among Openness items. Similar nomothetic associations therefore do not establish measurement equivalence across sources[44].

For applied users, this means that cross-source comparisons and regressions using LLM-generated personality scores may partly reflect the model's measurement structure, with estimates that may fail to generalize to humans. Applications that use LLMs to score human personality, gauge model "personality," or infer closely overlapping constructs should treat outputs as model-specific proxy predictions, not interchangeable measures of human traits.

Across Studies 2 and 3, digital twins showed weak item-level correlations, stronger profile recovery within individuals and substantial under-dispersion relative to humans (Supplementary Fig. S1), with responses regressing toward the mean and flattening individual heterogeneity. This pattern echoes critiques that LLMs privilege average, standardized outputs over the variability of human behavior[7,45]. How far this compression can be mitigated remains unclear, but two remedies are worth distinguishing. Adjusting decoding parameters such as temperature or top-p broadens outputs, but whether the added variance recovers genuine between-person heterogeneity rather than noise is untested[46]. Post hoc calibration offers a more targeted approach. In machine learning, temperature scaling realigns predicted class probabilities with observed accuracies[47]; analogously, statistical calibration can use a small gold-standard human sample to debias a larger set of LLM-generated predictions[48,49]. In our study, bias correction improved within-person profiles and nomothetic-span recovery (Study 3a). Because such methods debias aggregate and relational estimands rather than restoring individual variance, and their gains are bounded by the predictability of human behavior, compression is better viewed as a quantifiable, partly correctable property than as an absolute barrier.

Study 4 extends the framework from structured responses to naturalistic text across English and Chinese, foregrounding cross-language and cross-task generalizability. Natural language carries psychologically informative regularities in how people write, not only in what they say, such that function words index processes that are hard to monitor strategically[50]. A credible simulation should therefore reproduce person-specific traces alongside plausible topical content.

Human and digital twin narratives revealed a stable non-human signature in surface form across both languages. Humans produced more concrete entities, with named-entity density exceeding every twin model. Twins were also more lexically diverse and less variable, plausibly reflecting an instruction-tuned tendency toward information-dense language that under-produces repetition in natural human expression[51]. Grammatical encoding also diverged: English twins favored verb-centered sequences, while human texts used more noun-preposition patterns marking spatial, temporal and logical links; Chinese twins over-relied on adverbial modification. Critically, richer features did not resolve these gaps. Named-entity density remained below human levels in every condition, and in English feature-rich twins were no closer to humans than demographic ones. The specific form was language-dependent; the non-human signature, and its insensitivity to richer profiles, was not.

Construct representation behaved differently, and here conditioning mattered. Feature-rich twins approximated human semantic distributions more closely than demographic ones, consistently across dimensions and models in English, though model-dependent in Chinese, where one model diverged further under richer input. Richer features thus improve what a narrative conveys at the construct level but not the individual style in which it is written, making free-text twins more defensible for construct-level inference than for reproducing how a specific individual writes.

By recasting LLM-based digital twins as a problem of construct validity, we show where they can serve as human stand-ins and where they cannot. Across four studies, comparability emerged as a conditional property, governed by the level of intended inference, task demands, and the construct-relevance of supplied features. To make this decision framework actionable, we synthesize our findings into a set of recommendations for designing, validating, and reporting digital-twin studies (Supplementary Table S61). Where comparability fails, the failure may trace to the two validity threats we identified at the outset, which appear to differ in origin and therefore in how far richer features can reach. Divergent construct representation may be largely architectural, since LLMs encode meaning in a geometric space unlike grounded human semantic memory, and thus seems less likely to be resolved through prompting alone; unintended confounding, by contrast, arises from co-occurrence regularities learned during training and may be more addressable through carefully designed prompts[3].

Our boundaries were established against particular datasets, inventories, and a finite set of models, and the rapid pace of advances makes it essential to specify which person features support which inferences, separately at the intra-individual and inter-individual levels. Pursued this way, a validity-centered lens can help LLMs meaningfully advance social science[52,53]. Digital twins can stand in for human respondents only when validity warrants, and the same lens diagnoses why they fail elsewhere and guides the design of more faithful simulations.

## Methods

### Study 1

**Data**. To enable comparative analysis, we drew on two parallel large-scale resources: the Small World of Words (SWOW[15]) for human data and the LLM World of Words (LWOW) for model-generated data.

LLM responses came from LWOW, which contains free-association outputs from three models (Mistral-7b, Llama3.1-8b, Claude-3-5-haiku-latest) prompted with the identical SWOW cue set[54]. To ensure comparability, we processed all datasets using a standardized LWOW-adapted pipeline[54] (lowercasing, spelling normalization, lemmatization, and lexical alignment across datasets).

Within these datasets, we focused on cue words drawn from three theoretically grounded wordlists. We derived 86 English keywords from the Big Five Inventory-2 (BFI-2[21]), covering five domains and 15 facets (see Table S1). Stereotype Content Model was indexed using the seed dictionary from the Stereotype Content Dictionary[55], Moral Foundations Theory was captured with the Moral Foundations Dictionary 2.0 (MFD 2.0[56]). For each wordlist, we identified cue words appearing in both SWOW and LWOW and extracted all available R1–R3 responses from humans and from each LLM, yielding matched cue sets by construction (Table S2).

**Network Construction**. We constructed semantic association networks from the cleaned free association datasets. Nodes included all distinct word forms appearing as cues or responses. Directed edges linked each cue to its three associative responses, weighted by the observed frequency of the cue–response pair, and graphs were converted to weighted undirected networks. To reduce noise, we retained only nodes matched to WordNet and removed edges with weight equal to one. All analyses were conducted on the complete filtered networks.

**Network Analysis and Comparison.** To assess construct representation, we compared each filtered LLM network to its corresponding filtered human network using node- and edge-set indices (Table S3) and summarized global properties (Table S4). Small-world properties were evaluated against degree-preserving random networks[57]. For shared cue words, we computed normalized degree centrality and compared centrality distributions using Wilcoxon signed-rank tests, reporting effect sizes as r. We also evaluated community partitions against human partitions using the Adjusted Rand Index (ARI)[58] and V-measure[59]. The robustness of community detection was examined across four algorithms: Louvain, Leiden, Walktrap, and the nested stochastic block model (nSBM). Size-preserving random partition baselines were generated for LLM–human community comparisons. In each replicate, node labels were permuted across communities while preserving the empirical community-size distribution, yielding null distributions for ARI and V-measure under the same partition-size constraints as the observed solution. Sampling stability was examined through response-level bootstrap analyses. Responses were resampled with replacement within each cue, networks were reconstructed, communities were re-detected, and community-alignment indices were recalculated across bootstrap replicates.

**Prediction of Lexical Norms.** To test whether human and LLM associative spaces contained psychologically meaningful semantic information beyond network topology, we conducted an additional k-nearest-neighbor prediction analysis[60,61]. For each human and LLM free-association dataset, we constructed a cue-by-cue association matrix and applied positive pointwise mutual information (PPMI) weighting. Cosine similarity between PPMI vectors was then used to identify nearest neighbors for each word. Under leave-one-out validation, each word's valence, arousal, dominance, and concreteness scores were predicted from the mean norm score of its k nearest neighbors. Prediction accuracy was quantified as the Pearson correlation between observed and predicted norm scores across words. We evaluated multiple k values and retained the best-performing k for each model and lexical dimension. Human and LLM association spaces were then compared using Williams tests for dependent overlapping correlations, with word-level bootstrap confidence intervals for differences in predictive performance. Details are provided in Supplementary Section S1.

**Study 2a**

**Data.** We used the public Twin-2K-500 dataset[13], which includes 2,058 U.S. participants assessed across four waves. In Wave 4, participants completed the same 17 experimental tasks (16 classic heuristics and biases tasks and 1 pricing study) administered in Waves 1–3, enabling test–retest evaluation. We followed the original evaluation logic and scoring framework for item- and task-level accuracy.

**Large Language Models and Prompting Strategies.** We adhered to the original evaluation logic and construct digital twins by prompting LLMs with each participant's Text Persona (text-based Approach). We employed a fixed system prompt instructing the model to answer strictly based only on the provided persona and to return outputs in the required format. The user prompt for each digital twin bundles (i) the persona text, (ii) the item stem and response options, and (iii) explicit format instructions specifying the exact output schema (see Supplementary Section S4, Template 1). We used a many-shot learning and do not solicit chain-of-thought; each item is answered once. We evaluated six models: DeepSeek V3-0324, DeepSeek V3.1, DeepSeek V4-Flash, Qwen 2.5 Max, Gemma-4-31B-it, and gpt-oss-120b. Gemma-4-31B-it and gpt-oss-120b were accessed via Hugging Face inference services. Model providers are reported in Supplementary Section S5. Unless otherwise stated, decoding hyperparameters follow each provider's default temperature and sampling settings.

**Evaluation Metrics.** We quantified (i) item-level accuracy, (ii) item-level correlation, and (iii)

participant-profile correlation across items, and benchmarked effect sizes against 95% intervals from 1,000 random simulations under the null hypothesis. Accuracy treated human responses as the gold standard[13,62]. Task-level accuracy averaged item accuracy within each task and then summarized performance across participants. The item-level correlation captures the degree of association between the digital twin and human values for specific tasks, reflecting between-subject variability. In contrast, the participant-profile correlation focuses on within-person consistency, measuring how accurately the digital twin reproduces an individual's response pattern across a set of items. Both item-level and participant-profile correlations were estimated using Spearman correlation. The overall correlation was calculated using Fisher's z-transformation. In addition, we calculated the regression slope of the prediction error (defined as the digital twin's response minus the human response) against the human response for each task.

**Study 2b**

**Data.** We used the COVID-Dynamic dataset[28], a U.S. within-subject longitudinal study spanning 16 waves from April 2020 to January 2021. The baseline sample included 1,797 participants; the core analytic sample comprises 1,177 participants who met quality criteria and completed eight or more waves (51.2% female; median age = 39.4 years). Each wave included standardized scales, project-specific questionnaires on the pandemic and protests, and implicit/behavioral tasks. We analyzed waves 1–16; sample size varied by wave due to attrition.

**Creation of the Digital Twins.** First, we fixed demographic information from Wave 1 and, for each wave, append the contemporaneous published measures and the COVID & protest questionnaires. Target items to be predicted (unstandardized/experimental measures) are held out from the persona and posed as questions to the model. We employed a fixed system prompt instructing the model to rely exclusively on the provided persona and to return outputs in the required format. The user prompt for each digital twin bundles (i) the persona text, (ii) the item stem and response options, and (iii) explicit format instructions specifying the output schema (see Supplementary Section S4, Template 2). We treated human responses as the gold standard. For all 16 waves, evaluations were conducted using DeepSeek-V3.1 under a many-shot setting with default parameters.

**Evaluation Metrics.** We used the same framework as Study 2a (accuracy, item-level correlations, and participant-profile correlations) with 1,000 random simulations as baselines, and we reported questionnaire-level accuracy and task-specific error-slope estimates using the same algorithms as in Study 2a.

**Temporal sensitivity analyses.** To assess whether digital-twin predictions depended on the survey's temporal setting, we conducted two prompt manipulation analyses. In both analyses, persona profiles, target items, response options, and output schemas were held constant, while temporal cues were varied. These cues included the wave-specific survey date, contemporaneous COVID-19 context, and event information. The COVID-19 context summarized the pandemic timeline, severity, unemployment, policy background, state-level restrictions, and salient social events around each wave. Each condition also included two manipulation-check questions to verify whether the model recognized the intended date and event context before answering the target items (see Supplementary Prompt Templates).

First, we tested cross-model sensitivity using Wave 10, whose target date was August 8, 2020. To reduce direct anchoring to named political figures, explicit names such as Trump and Pence were removed from the relevant question text. DeepSeek V4-Flash, Qwen 3.5 Plus, Gemma-4-31B-it, and gpt-oss-120b were evaluated across 12 conditions that varied event information, date information, and COVID-19 context. In real-event conditions, the event information referred to COVID-19, and the date information was either absent or set to the correct Wave 10 date. In fictional-event conditions, the event information referred to a fabricated "BEIGUO pandemic," and the date information was absent, correct, shifted to the past, or shifted to the future. This design tested whether predictions were selectively sensitive to historically valid information or were also influenced by unsupported fictional event cues. The workflow is shown in Supplementary Fig. 2a, and prompt templates are provided in Supplementary Section S5.

Second, we tested temporal sensitivity across all 16 waves using DeepSeek V4-Flash. This analysis used only the real COVID-19 event context and crossed two binary prompt factors: whether the wave-specific date was provided and whether contemporaneous COVID-19 context was provided. This yielded four conditions for each wave: no date and no context, context only, date only, and date plus context. This design assessed whether date and contextual cues improved prediction accuracy across the full longitudinal period.

**Study 3**

Study 3 assessed the psychometric performance of digital twins across two datasets using different personality inventories: Study 3a (Twin-2K-500 with BFI) and Study 3b (COVID-Dynamic with NEO-FFI). Despite differences in data sources and instruments, both sub-studies followed the same feature-input logic, prompting procedure, and psychometric network framework.

The psychometric network framework was used to evaluate whether digital twins recovered the structural organization of Big Five responses in humans. This focus is important for personality assessment because the Big Five are inferred from patterns of item covariation. Accordingly, we adopted a psychological network perspective to evaluate the psychometric properties of digital twins' Big Five representations[63]. In contrast to latent variable models, which conceptualize personality traits as common causes underlying observed item responses, network psychometrics conceptualizes psychological constructs as complex systems of mutually interacting components[64]. Within this perspective, personality items are represented as nodes in a network, and their associations—typically estimated as regularized partial correlations—are represented as edges[65]. Personality dimensions are thus understood as clusters of densely connected items, which may be identified empirically through community detection algorithms[43] or specified a priori based on theoretical considerations[66], rather than being modeled as latent common causes.

This shift matters for measurement equivalence. In SEM-based frameworks, measurement invariance is assessed hierarchically[67,68], including equivalence of factor structure (configural), factor loadings (metric), item intercepts or thresholds (scalar), and residual variances (strict). As shown in Study 2, digital twin responses exhibit substantial under-dispersion relative to humans, implying systematic compression of between-person variance at the measurement level. Under such conditions, scalar and strict invariance are structurally difficult to satisfy within SEM: even if configural and metric invariance hold, differences in score dispersion mechanically translate into non-equivalent intercepts and residual variances across groups. SEM-based tests may therefore be violated for reasons intrinsic to the data-generating properties of digital twins rather than meaningful measurement nonequivalence. By contrast, network-based measurement invariance concentrates on equivalence in the relational organization of items[43], testing whether the same community structure emerges across groups (network configural invariance) and, conditional on this, whether items occupy comparable structural roles within communities, as indexed by network loadings (network metric invariance). Applying this framework allows us to test whether increasing feature-input richness enables digital twins to replicate not only the magnitude of human personality traits, but their underlying structural organization.

**Data.** In Study 3a, we used the Twin-2K-500 dataset to emulate responses to the 44-item Big Five Inventory[69]. Digital twins were constructed under seven feature-input conditions (Table 2), ranging from demographics alone and behavioral task (16 classic heuristics-and-biases tasks and one pricing study) data from Study 2 to full feature-rich inputs and association-filtered feature subsets. In Study 3b, we applied the same framework to the COVID-Dynamic dataset, using the 60-item NEO Five-Factor Inventory (NEO-FFI) administered at Wave 14 (7 November 2020) as the criterion personality measure. Behavioral input consisted of open-ended COVID-19 memory narratives from Wave 15b (9 December 2020), whereas feature-rich, association-filtered and context-enriched conditions varied the amount and type of auxiliary information provided (Table 2). After linking participant records across waves and excluding incomplete cases, the final analytic sample included 702 participants.

**Prompting Strategies and Evaluation Metrics.** We followed the Study 2 prompting strategy. For each digital twin, the user prompt concatenated (i) persona text (varying by condition), (ii) Big Five item stems with response options, and (iii) explicit formatting instructions specifying the output schema. Human responses served as the gold standard. Outputs were generated with DeepSeek V4-Flash, Qwen 3.5 Plus, Gemma-4-31B-it, and gpt-oss-120b under a many-shot configuration using default parameters. Across Studies 3a and 3b, we computed item-level accuracy, item-level Spearman correlations, and participant-profile Spearman correlation, benchmarked against confidence intervals from 1,000 random simulations. To account for reliability differences between human and digital-twin scores, we reported disattenuated Pearson correlations at the trait level. We also reported the regression slope of prediction error against human responses for each questionnaire using the same algorithms as in Study 2. To assess nomothetic span, we computed Pearson correlations between Big Five trait scores and external psychological scales separately for humans and digital twins. In Study 3b, we further assessed temporal sensitivity by comparing performance across four prompt variants: no date or context, context only, date only, and date plus context.

**Network Estimation.** Personality networks were estimated separately for humans and for digital twins under each feature input level. Items were modeled as nodes and edges as conditional associations, estimated via regularized partial correlations using graphical LASSO[70]. Configural invariance was evaluated as convergence on the same community structure across groups, with communities detected using the *walktrap* algorithm[71] and tested using bootstrap Exploratory Graph Analysis (bootEGA[72]). Items with stability below .70 were iteratively removed and bootEGA re-estimated until a single stable community structure common to all groups was identified[43]. Conditional on configural invariance, metric invariance was tested using network loadings—each item's within-community total connectivity computed as the sum of absolute within-community edge weights[43]—and a permutation-based test of loading differences across groups[43]. Items were considered non-invariant if their observed loading differences exceeded the 95% bounds of the permutation distribution.

We examined invariance across three sets of contrasts: pairwise comparisons between the human network and each digital-twin model under each feature-input condition; within-LLM comparisons across feature-input conditions; and between-LLM comparisons under the same feature-input condition. For human–digital-twin configural comparisons, community solutions were further interpreted against the theoretically expected five Big Five domains. Solutions were classified as partial configural invariance when a common solution required the removal of some items but the retained items still mapped onto the expected domain structure, or when a theoretical domain was split into multiple communities without cross-domain mixing. When cross-domain mixing occurred, theoretically misplaced items were removed and the network was re-estimated until the retained items aligned with the expected Big Five structure. If more than half of the items in any domain had to be removed, configural invariance was considered not supported.

**Post hoc statistical calibration.** In Study 3a, we first examined whether marginal distributional differences contributed to discrepancies between human and digital twin responses by applying randomized ordinal equipercentile equating to item responses within each model and input condition. Equating functions were estimated in randomly selected calibration participants across calibration proportions from 10% to 90%, and correlations were computed in held out participants and averaged across 1,000 repeated splits. Following recent statistical calibration frameworks[49,73], we then conducted a plug-in bias correction analysis in the full feature input condition, matching human and digital twin responses at the participant by item level and repeatedly splitting participants into calibration and held out samples, with calibration proportions of 10%, 25% and 50% repeated 1,000 times. Calibration samples were used to estimate item level bias correction models that adjusted raw digital twin responses using item fixed effects, raw model scores and demographic covariates, whereas response, profile, trait and distributional metrics were evaluated in held out participants. We also tested whether raw and calibrated Big Five trait scores recovered downstream nomothetic span associations with external psychological scales, including bias corrected regressions when digital twin derived traits were used as either dependent variables or covariates; details are provided in Supplementary Section S2.

**Study 4**

**Participants and Materials**. Study 4a used Wave 15 of the COVID Dynamic dataset, with the COVID Memories task as the target free-text response. We constructed participant-specific digital twins under two feature-input levels: demographics-only (questionnaire-derived demographics) and feature-rich (demographics plus the full Wave 15 questionnaire battery, including standardized psychological scales and COVID-related measures). Study 4b recruited 42 university students (15 males, 27 females; Mage = 22.5 years). Participants completed pre-stimulus questionnaires (demographics, standardized personality assessments, and documentary consumption preferences), viewed an unreleased documentary film, and then completed objective comprehension items and structured free-text evaluations covering valence, salient elements, specific scenes, aesthetic details, and an overall appraisal.

**Prompting Strategies and Creation of Digital Twins.** We held the system prompt constant (as in Studies 2–3). To reduce artifactual similarity, we added role constraints and output controls: the model wrote as the target individual while avoiding verbatim reuse of questionnaire text, and outputs were length-matched to the mean word count of human responses (Supplementary Section S4, Template 3). In Study 4a, prompts paired condition-specific persona text with the COVID Memories instruction and a fixed output schema. In Study 4b, we implemented an analogous two-condition manipulation: digital twins simulated film exposure by processing the documentary script embedded in the prompt; the reduced-input condition included only pre-stimulus questionnaire information (excluding personality measures), whereas the feature-rich condition additionally included post-stimulus information (including objective responses). Outputs were generated with DeepSeek V4-Flash, Qwen 3.5 Plus, Gemma-4-31B-it, and gpt-oss-120b under a many-shot configuration using default parameters.

**Linguistic Measures and Psychological Construct Representation.** We compared human and digital-twin narratives using surface and syntactic measures: average sentence length; mean dependency distance normalized by sentence length[74]; dependency depth[75]; HD-D lexical diversity[76]; named entity density via NER; and POS bigram distributions[77], and assessed psychological construct representation using distributed dictionary representation (DDR)[78]. For Study 4a, we assessed five targeted psychological dimensions—nostalgia, agency, communion, social ties, and threat. These constructs were operationalized using terms from validated sources, including the Nostalgia Dictionary[79], the Big Two Dictionary (for agency and communion)[80], the Social Ties Dictionary[81], and the Threat Dictionary[82]. For Study 4b, we used a newly developed extended Chinese social evaluative wordlist[83] comprising five evaluative dimensions—socioeconomic status, sociability, competence, morality, and appearance, reflecting the task's emphasis on social appraisal and person-centered evaluation. For each dimension, we derived a central semantic vector by averaging the embeddings of its constituent dictionary terms. Finally, we calculated the cosine similarity between each construct's aggregate vector and the text-level embeddings. Details are provided in Supplementary Section S3.

**Statistical analysis**

All analyses used a within-participant design comparing human responses with demographic and feature-rich digital twins. In Study 4b, the demographic condition additionally included one documentary-preference item, but we use the labels digital twin (demographic) and digital twin (feature-rich) consistently throughout. Multiple testing was controlled using the Benjamini–Hochberg procedure. For linguistic metrics, we used Friedman tests across the three paired conditions, reporting Kendall's W, followed by paired Wilcoxon signed-rank tests. For NER labels and POS bigrams, we estimated pairwise Jensen–Shannon divergence[84] and assessed significance using within-participant permutations; feature-level presence was tested with participant-stratified conditional logistic regression, adjusting for token length. For dimension-level cosine similarity, we used paired Wilcoxon tests and reported Wilcoxon r. Distributional shifts were evaluated by comparing empirical CDFs with Kolmogorov–Smirnov statistics and paired permutation p values.

**Data and materials availability**

Human word association data for Study 1 were obtained from the Small World of Words project (SWOW-EN18; https://smallworldofwords.org/zh/project/research), and LLM World of Words data from https://github.com/LLMWorldOfWords/LWOW. Data for Studies 2, 3 and 4a are available from the Twin-2K-500 dataset (https://huggingface.co/datasets/LLM-Digital-Twin/Twin-2K-500/) and the COVID-Dynamic dataset (https://osf.io/z8k2t/overview). Raw data for Study 4b are not publicly available owing to data ownership restrictions. Code is available on the Open Science Framework at https://osf.io/aemyr.

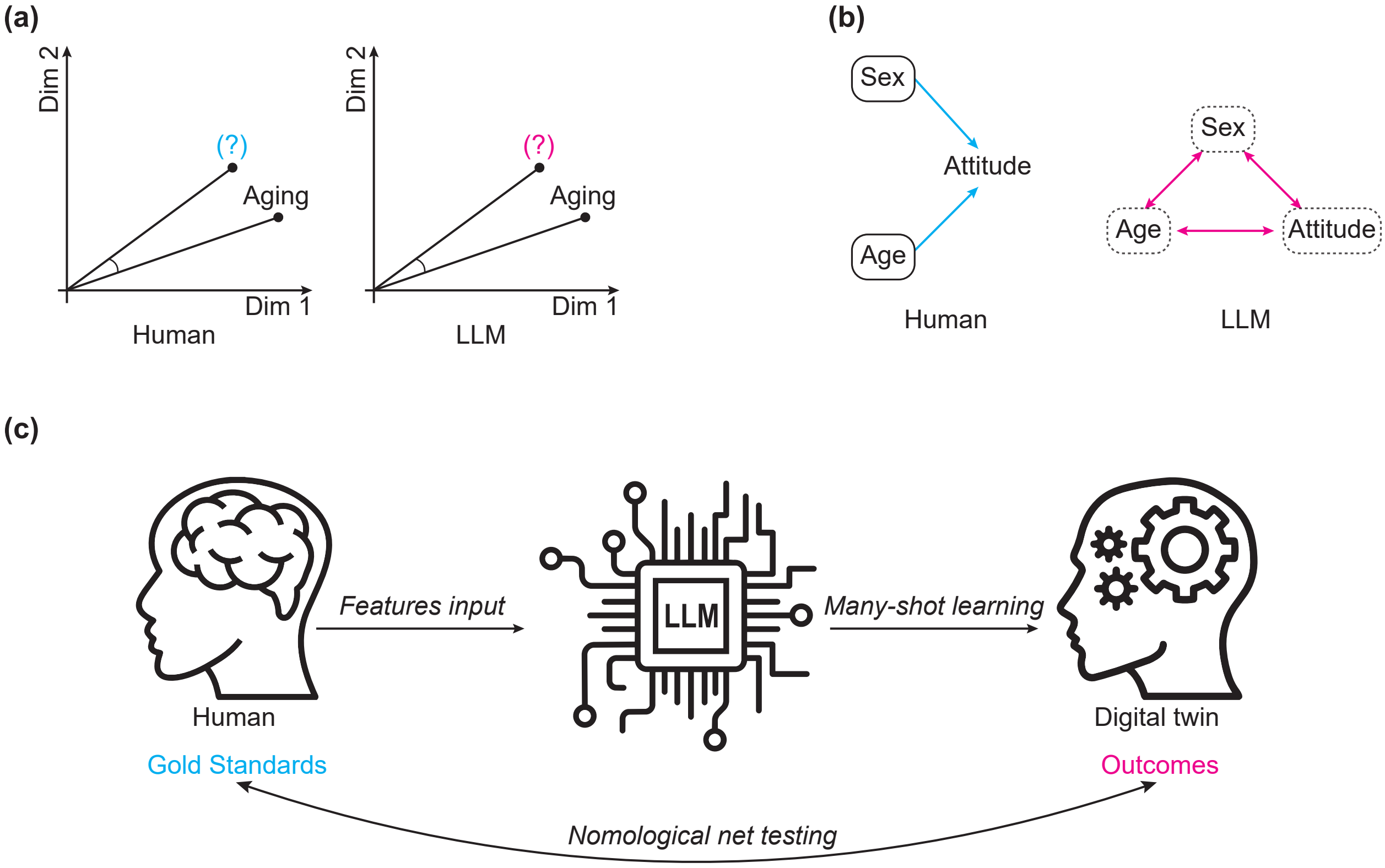

**Fig. 1 | Construct representation and nomological net test for LLM-based digital twins.**
(a) Divergence in construct representation. (b) Unintended confounding. (c) Digital-twin pipeline and validation.

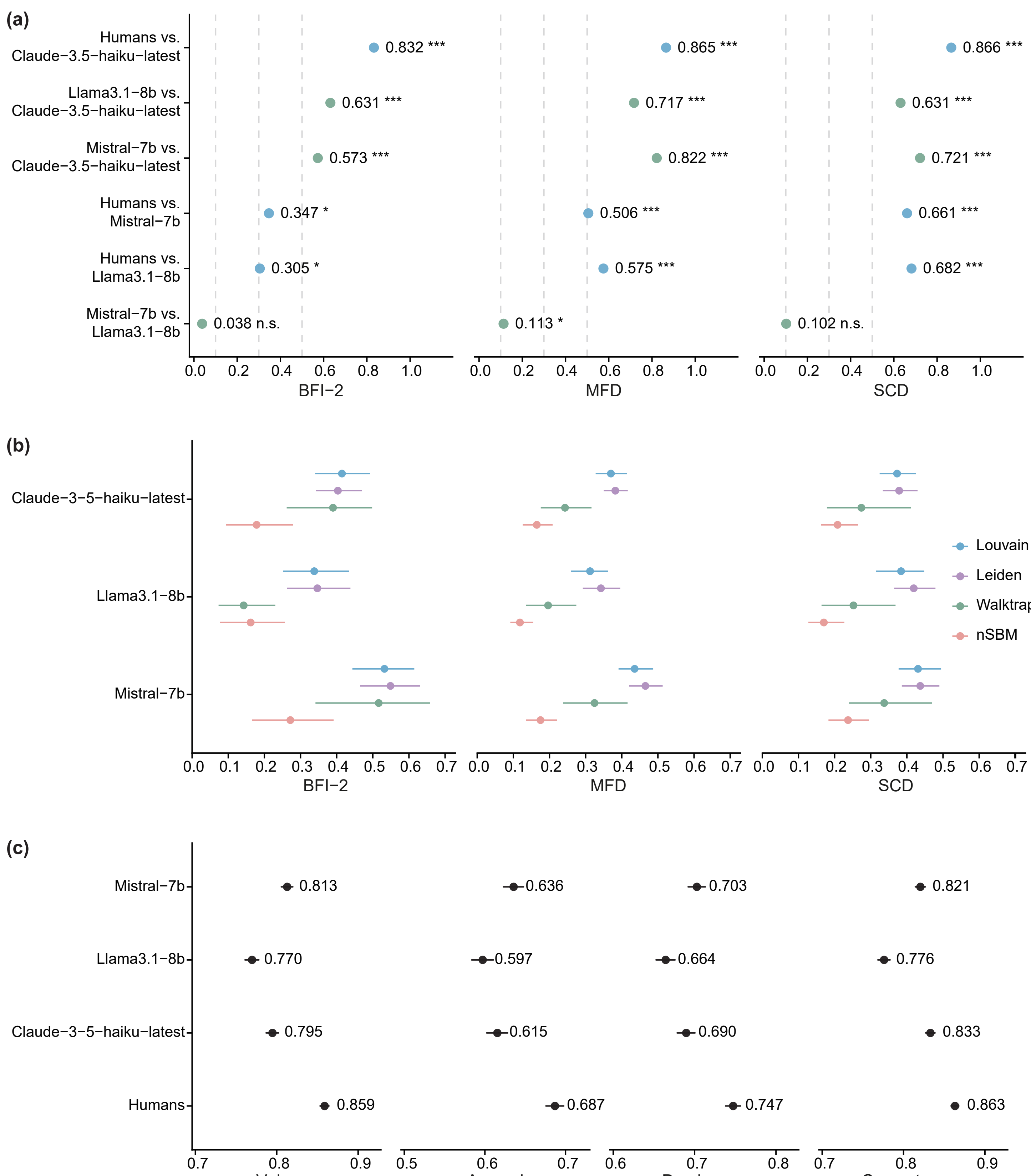

**Fig. 2 | Human–LLM alignment in semantic association spaces.**
(a) Pairwise differences in network centrality across human and LLM semantic networks. Dots show absolute Wilcoxon effect-size values (r) for pairwise comparisons of degree centrality within each wordlist. Blue indicates human–LLM comparisons and green indicates LLM–LLM comparisons. Dashed vertical lines mark absolute r values of .10, .30 and .50. Labels report effect sizes and Bonferroni-adjusted significance within each wordlist. p < .05, p < .01, p < .001; n.s., not significant. (b) LLM–human alignment of empirical community structure across community-detection algorithms. Points show mean Adjusted Rand index (ARI) values with 95% bootstrap confidence intervals from 1,000 replicates for comparisons between each LLM partition and the corresponding human empirical partition. (c) Prediction of lexical norms from human and LLM association spaces. Points show Pearson correlations (r) between human-rated lexical norms and leave-one-out kNN predictions at the best-performing k, with 95% confidence intervals.

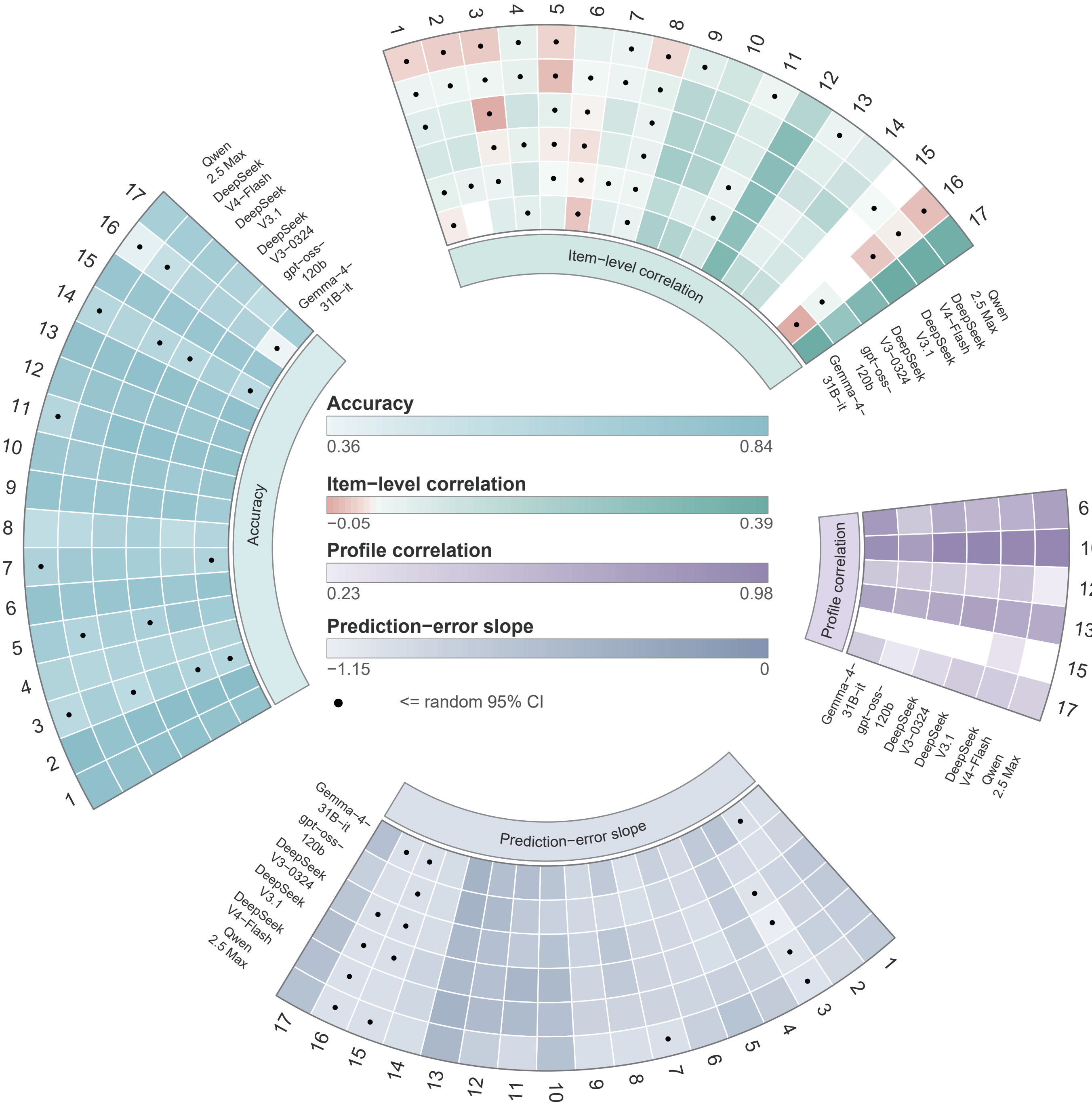

**Fig. 3 | Digital-twin performance against human responses in Study 2a.** Accuracy, item-level correlation, profile correlation and prediction-error slope were evaluated for six LLM-based digital twins across 17 tasks, with human responses as the gold standard. Black dots indicate estimates no higher than the upper bound of the 95% random baseline confidence interval; white cells indicate correlations that could not be estimated due to zero variance or an insufficient number of items. Task numbers correspond to Table 1; task 17 denotes the pricing study.

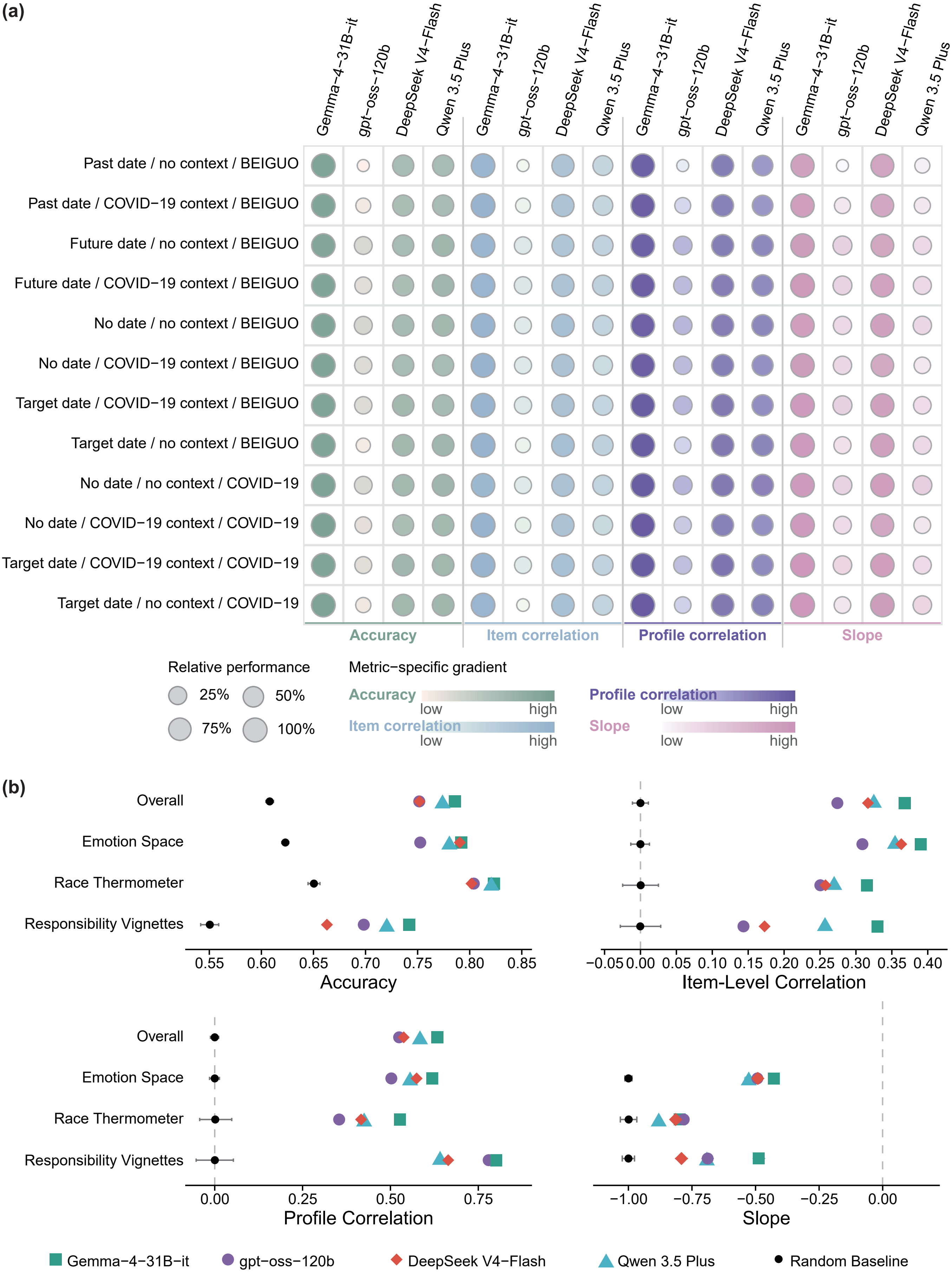

**Fig. 4 | Temporal sensitivity and cross-model performance in Study 2b.**
(a) Predictive performance across 12 temporal prompt conditions that varied date information, COVID-19 context, and event information for four LLM-based digital twins. (b) Predictive performance of the four LLM-based digital twins, evaluated against human responses as the gold standard. Random baseline intervals indicate 95% confidence intervals from 1,000 random simulations.

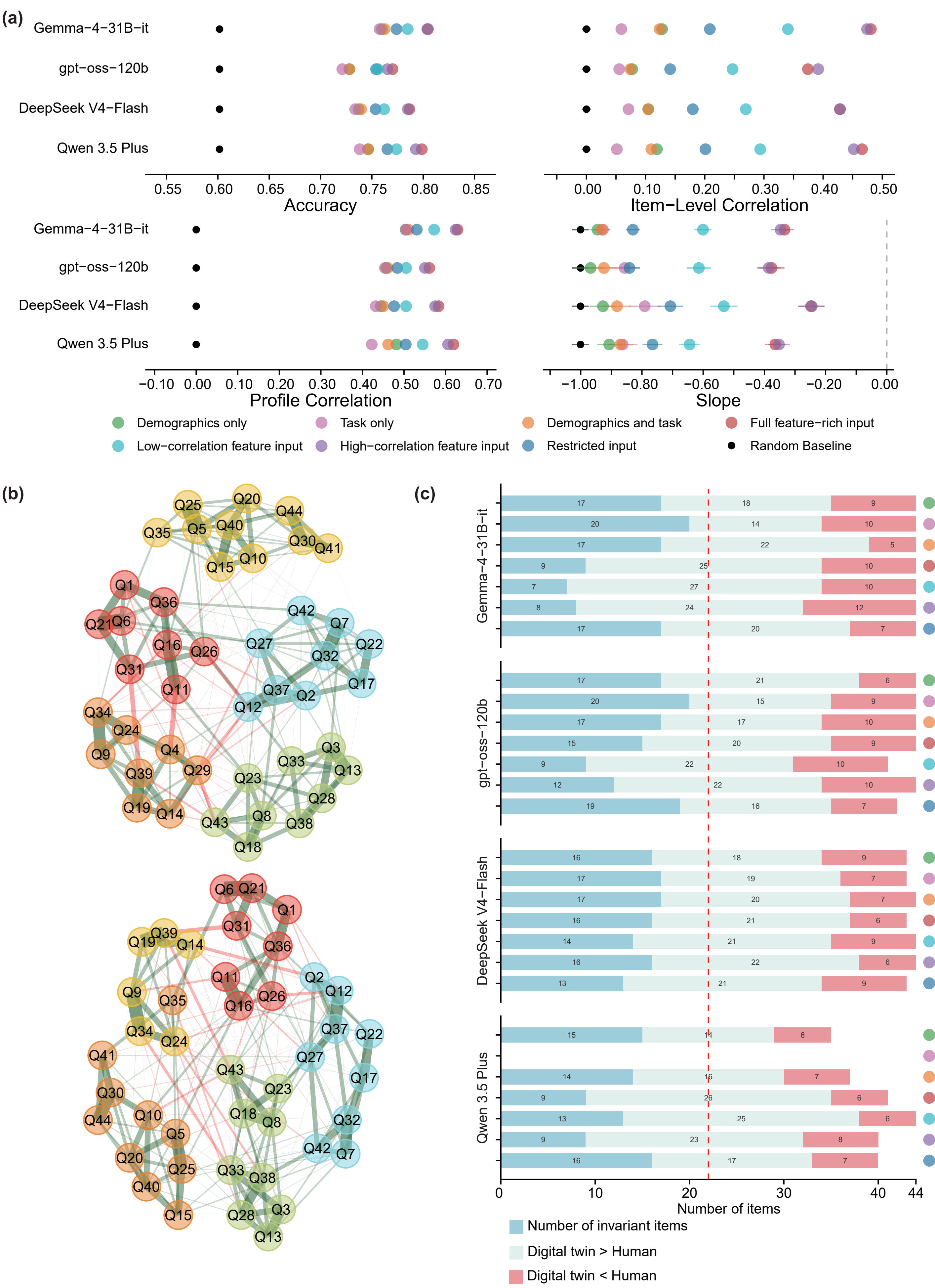

**Fig. 5 | Psychometric comparability of digital twins in Big Five assessment in Study 3a.**
(a) Accuracy, item-level correlation, profile correlation and prediction-error slope for four LLM-based digital twins across seven feature-input conditions, evaluated against human BFI-44 responses as the gold standard. Random baseline intervals indicate 95% confidence intervals from 1,000 random simulations. (b) Examples of network configural invariance. Nodes represent BFI-44 items and colors indicate EGA-derived communities. The upper network shows configural invariance for Gemma-4-31B-it under the full feature-rich input condition. The lower network shows partial configural invariance for gpt-oss-120b under the restricted-input condition, after Q4 and Q29 were removed. (c) Network metric invariance results across models and feature-input conditions. Stacked bars show the number of invariant items and items with significant network loading differences between digital twins and humans. Loading differences were tested with BH-adjusted $p < .05$. The red dashed line marks half of the scale items (22 of 44) as a visual reference for the extent of metric vvnon-invariance.

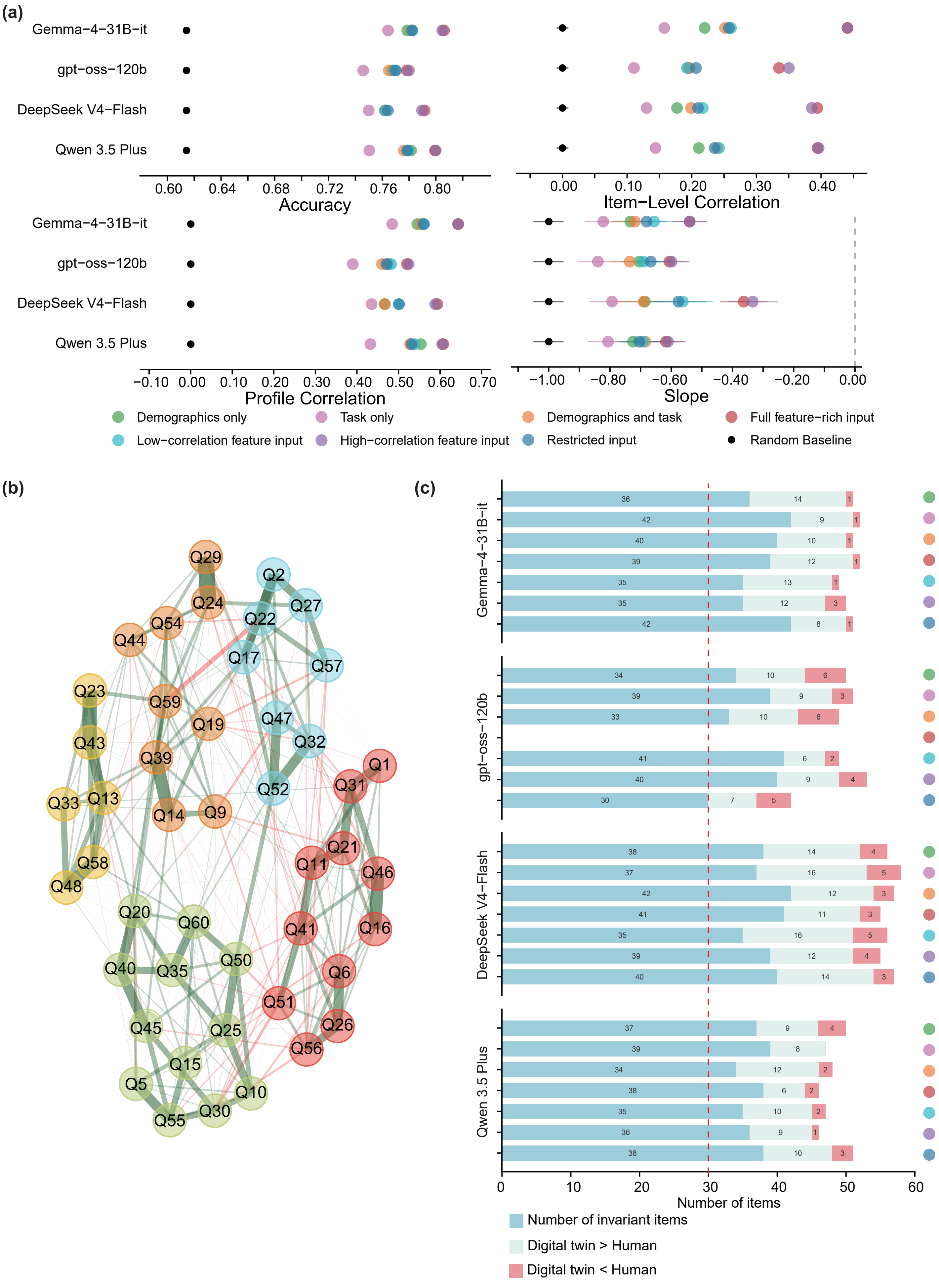

**Fig. 6 | Psychometric comparability of digital twins in Big Five assessment in Study 3b.**
(a) Accuracy, item-level correlation, profile correlation and prediction-error slope for four LLM-based digital twins across seven feature-input conditions, evaluated against human NEO-FFI-60 responses as the gold standard. Random baseline ntervals indicate 95% confidence intervals from 1,000 random simulations. (b) Example of partial network configural invariance for Qwen 3.5 Plus under the full feature-rich input condition. Nodes represent NEO-FFI-60 items and colors indicate EGA-derived communities. The partial configural invariant solution was obtained after removing 14 items.
(c) Network metric invariance results across models and feature-input conditions. Stacked bars show the number of invariant items and items with significant network loading differences between digital twins and humans. Loading differences were tested with BH-adjusted $p < .05$. The red dashed line marks half of the scale items (30 of 60) as a visual reference for the extent of metric non-invariance.

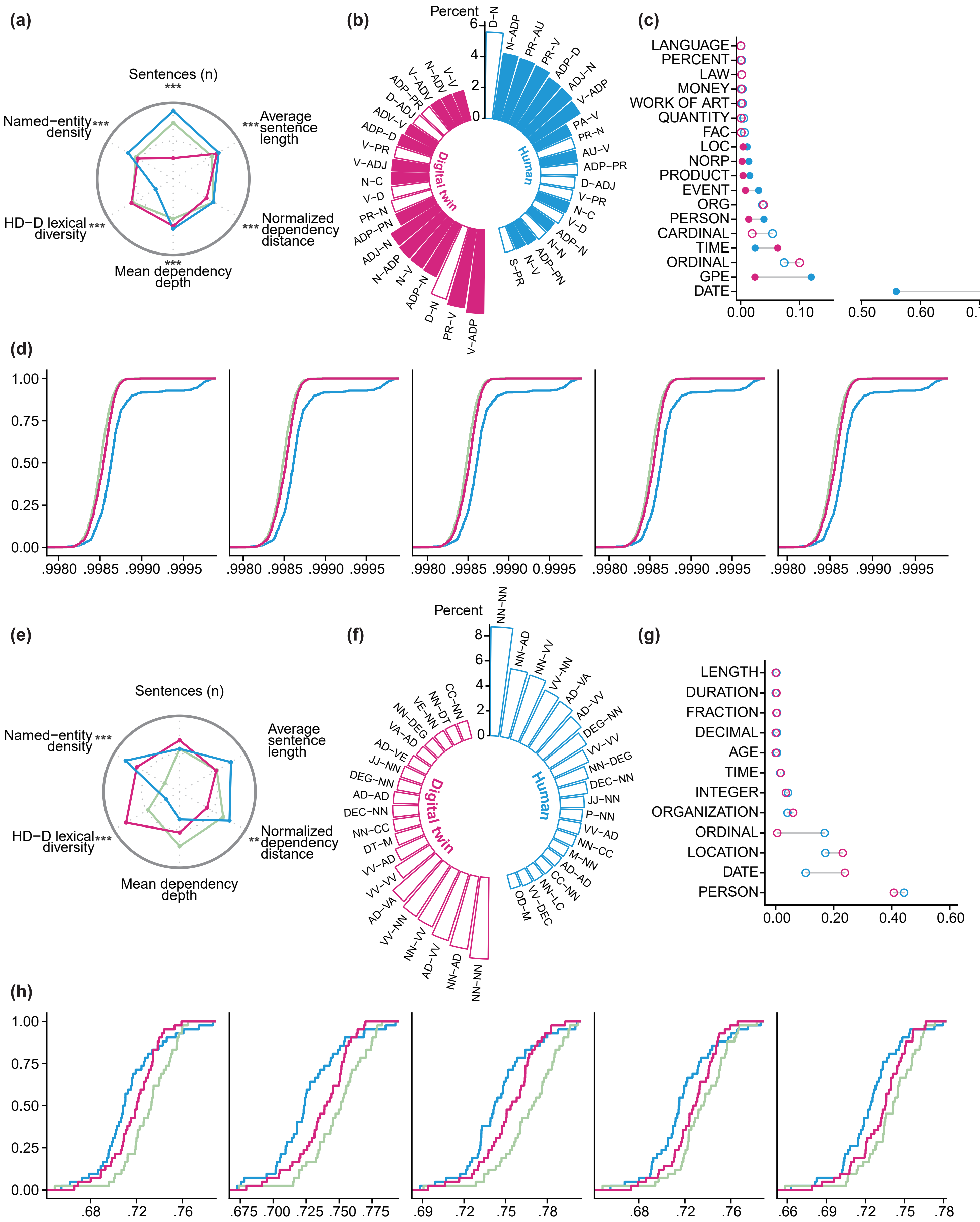

**Fig. 7 | Linguistic evaluation and construct representation of DeepSeek V4-Flash digital twins across two task contexts.** (a–d) Study 4a (English COVID memories) and (e–h) Study 4b (Chinese script evaluation) compare Human texts with Digital twins built from demographics-only or feature-rich inputs. (a,e) Percentile ranks of linguistic features (*** $p < .001$, ** $p < .01$). (b,f) POS-bigram type frequencies for Human vs. Digital twin (feature-rich); filled markers

indicate statistical significance at $p < .05$. (c,g) NER label proportions for Human vs. Digital twin (feature-rich); solid markers indicate statistical significance at $p < .05$. (c) includes an axis break to display both high- and low-frequency labels. (d,h) CDFs of cosine similarity index construct representation. Color: Human (red), Digital twin (feature-rich) (blue), Digital twin (demographic) (green).

Table 1 Replications of heuristics and biases

| | Task | Digital twins | | | | | |
|---|---|---|---|---|---|---|---|
| | | DeepSeek V3-0324 | DeepSeek V3.1 | DeepSeek V4-Flash | Qwen 2.5 Max | Gemma-4-31B-it | gpt-oss-120b |
| | ***Between-subject experiments*** | | | | | | |
| 1 | Base rate problem (Kahneman & Tversky, 1973) | × | × | ✓ | × | × | ✓ |
| 2 | Outcome bias (Baron & Hershey, 1988) | ✓ | ✓ | ✓ | ✓ | × | × |
| 3 | Sunk cost fallacy (Stanovich & West, 2008) | × | × | ✓ | × | × | × |
| 4 | Allais problem (Stanovich & West, 2008) | ✓ | ✓ | × | ✓ | × | × |
| 5 | Framing problem (Tversky & Kahneman, 1981) | × | × | × | ✓ | × | × |
| 6 | Conjunction problem (Linda) (Tversky & Kahneman, 1983) | ✓ | ✓ | ✓ | ✓ | ✓ | ✓ |
| 7 | Anchoring and adjustment (Epley et al., 2004; Tversky & Kahneman, 1974) | ✓✓ | ✓✓ | ✓✓ | ×✓ | ×✓ | ×✓ |
| 8 | Absolute vs. relative savings (Stanovich & West, 2008) | ✓ | ✓ | ✓ | × | ✓ | ✓ |
| 9 | Myside bias (Stanovich & West, 2008) | × | × | ✓ | × | × | × |
| 10 | Less is More (Stanovich & West, 2008) | × | × | × | ✓ | ✓ | ✓ |
| 11 | WTA/WTP – Thaler problem (Stanovich & West, 2008) | ✓ | ✓ | ✓ | × | ✓ | ✓ |
| | ***Within-subject experiments*** | | | | | | |
| 12 | False consensus (Furnas & LaPira, 2024) | ✓ | ✓ | ✓ | ✓ | ✓ | ✓ |
| 13 | Non-separability of risk and benefits judgments (bicycle; alcohol; pesticides; chemical) (Stanovich & West, 2008) | ✓×✓✓ | ×✓×× | ××✓✓ | ×××× | ×✓✓✓ | ×✓✓✓ |
| 14 | Omission bias (Stanovich & West, 2008) | × | × | ✓ | × | × | × |
| 15 | Probability matching vs. maximizing (Stanovich & West, 2008) | × | × | × | × | × | × |
| 16 | Denominator neglect (Stanovich & West, 2008) | × | × | ✓ | × | × | × |

*Note.* This table reports whether each model reproduced the effects observed among human respondents in the Twin-2K-500 dataset. Human effects were used as the benchmark for evaluating model performance across Toubia et al.'s (2025) within-subject and between-subject experiments.

Table 2 Seven levels of feature input in Study 3

| | | Study 3a | Study 3b |
|---|---|---|---|
| 1 | Demographics only | Demographics | Demographics |
| 2 | Task only | Heuristics and biases experiments | COVID Memories |
| 3 | Demographics and task | Demographics + Heuristics and biases experiments | Demographics + COVID Memories |
| 4 | Full feature-rich input | Demographics + All psychological measures + Heuristics and biases experiments | Demographics + All psychological measures + COVID Memories |
| 5 | Low-correlation feature input | Demographics + Heuristics and biases experiments + Features with $r < 0.5$ | Demographics + Features with $r < 0.5$ + COVID contexts |
| 6 | High-correlation feature input | Demographics + Heuristics and biases experiments + Features with $r > 0.5$ | Demographics + Features with $r > 0.5$ + COVID contexts |
| 7 | Restricted or context-enriched input | Demographics + Heuristics and biases experiments + Features with $r < 0.3$ | Demographics + COVID Memories + COVID contexts |

**Psychometric Comparability of LLM–Based Digital Twins**

Yufei Zhang[1], Zhihao Ma[1]

[1] Computational Communication Collaboratory, School of Journalism and Communication, Nanjing University, Nanjing 210023, China

Corresponding author:

Zhihao Ma

Email: redclass@163.com

This file includes:

Supplementary Text Section 1 to Section 5

Tables S1–S12 correspond to Study 1;

Tables S13–S15 and Figure S1 correspond to Study 2a;

Tables S16–S23 and Figure S2 correspond to Study 2b;

Tables S24–S56 and Figures S3–S4 correspond to Study 3;

Tables S57–S60 and Figures S5–S6 correspond to Study 4.

Table S61 Construct-validity recommendations for LLM-based digital-twin studies.

**Supplementary Information**

**Supplementary Information Text**
**Section S1. Study 1 Methods**
**Data preprocessing**

Within these datasets, we focused on cue words drawn from three theoretically grounded wordlists. First, to represent personality traits, we derived 86 English keywords from the Big Five Inventory-2 (BFI-2[1]) covering five domains and 15 facets (see Table S1). Stereotype Content Model was indexed using the seed dictionary from the Stereotype Content Dictionary[2], which assigns words to dimensions of sociability, morality, ability, agency, status, politics, and religion. Moral Foundations Theory was captured with the Moral Foundations Dictionary 2.0 (MFD 2.0[3]), which maps words onto the upholding or violation of five core moral foundations: sanctity, care, authority, loyalty, and fairness. For each of these wordlists, we identified all cue words that appeared in both SWOW and LWOW and extracted every available first-, second-, and third-order responses (R1–R3) generated by humans and by each LLM for those cues. This procedure resulted in 80 shared cue words for the BFI-2 wordlist, 249 for the Stereotype Content Dictionary and 610 for MFD 2.0, with human and LLM datasets matched on the cue sets by construction (see Table S2).

**Network construction**

For each combination of data source (humans or LLMs) and theoretical wordlist, we constructed a semantic association network from the cleaned free association datasets. Nodes represented all distinct word forms that appeared in the dataset, whether as cues or as responses. From each cue we added directed edges to its three associative responses, with edge weights equal to the observed frequency of that cue response pair. The resulting graphs were directed, and were then converted to undirected networks. When two nodes were connected in both directions, we retained a single undirected edge whose weight was set to the larger of the two directional weights.

Furthermore, to reduce noise from idiosyncratic or poorly defined items, we applied a common two step filtering procedure to all networks. First, we retained only nodes that could be matched to an entry in WordNet, a large lexical database of English developed at Princeton University[4], thereby excluding non-lexical strings and opaque tokens. Second, we removed edges with weight equal to one, which primarily reflected isolated idiosyncratic associations rather than stable links in the associative structure. The filtered graphs could contain multiple connected components, and all analyses were conducted on the complete filtered networks.

**Network analysis**

**Global network indices.** To quantify the extent to which LLMs capture the construct representation of human semantic networks, we compared each filtered LLM network with its corresponding filtered human network using indices defined over their node and edge sets (see Table S3). For nodes, we computed the proportion of nodes unique to the human network, the Jaccard similarity between the human and LLM node sets, and the proportion of nodes unique to the LLM network. For edges, we applied the same three indices to the induced subgraphs on the intersection of the node sets, yielding the proportion of edges unique to the human subgraph, the Jaccard similarity of edge sets, and the proportion of edges unique to the LLM subgraph. In Table S4, we report a set of global network indices for each semantic network, characterizing its overall size and small-world-ness properties. Specifically, we summarize the number of nodes and edges, the average clustering coefficient (C) and average shortest path length (L), as well as three derived small-world metrics ($\gamma$, $\lambda$, $\sigma$).

**Community detection and robustness analyses.** Community structure was estimated for each filtered semantic network using four community-detection approaches. Louvain and Leiden algorithms were implemented with the resolution parameter fixed at 1. Walktrap was run on weighted networks using four-step random walks. For the nested stochastic block model (nSBM), partitions were extracted from lexicon-specific hierarchy levels to obtain interpretable community solutions: level 1 for BFI-2, level 2 for the Stereotype Content Dictionary, and level 3 for the Moral Foundations Dictionary 2.0. For all community-alignment analyses, comparisons were restricted to words that appeared as nodes in all observed networks within each lexicon, ensuring that human–LLM comparisons were based on the same set of lexical items.

Community alignment was quantified using the adjusted Rand index (ARI) and V-measure. For each lexicon and community-detection setting, LLM partitions were compared with the corresponding human partition estimated under the same algorithmic specification. To evaluate whether observed alignment exceeded chance expectations, we generated size-preserving random partition baselines. In each random replicate, node labels were permuted across communities while preserving the empirical community-size distribution, yielding null distributions for ARI and V-measure under the same partition-size constraints as the observed solution.

Sampling stability was assessed using a response-level bootstrap procedure. In each bootstrap replicate, response rows were resampled with replacement within each cue, semantic networks were reconstructed from the resampled responses, and communities were re-detected using the same community-detection settings. This procedure yielded bootstrap estimates of item-level community replication and LLM–human community alignment. It therefore tested whether the observed community patterns were robust to sampling variation in the underlying free-association responses, rather than only to perturbations of an already aggregated network.

**Small-world analysis.** To assess small-world properties, we first calculated the average clustering coefficient for each network and the average shortest path length within the largest connected component. Subsequently, we generated 100 random networks that preserved the degree distribution of the original network. For each randomized counterpart, we similarly computed the average clustering coefficient and the average shortest path length, using the mean values of these metrics to establish a baseline. Building upon this, we constructed three normalized small-world metrics following the methodology of Humphries & Gurney[5]. First, we calculated the normalized clustering coefficient ($\gamma = C_{real}/C_{rand}$), comparing the empirical network's clustering coefficient to the mean of the random networks. Second, we computed the normalized characteristic path length ($\lambda = L_{real}/L_{rand}$), comparing the empirical average shortest path length against the random baseline. Finally, we combined these two metrics into a comprehensive small-world-ness index ($\sigma$), defined as $\sigma = \gamma/\lambda$. Intuitively, when a network exhibits a clustering coefficient significantly higher than its random counterpart ($\gamma >> 1$) while maintaining a comparable average path length ($\lambda \approx 1$), the resulting index typically exceeds 1 ($\sigma > 1$). This is regarded as evidence of small-world network, indicating that the network maintains high global efficiency while preserving dense local connectivity.

**Network comparison**

For each theoretical wordlist, we calculated normalized degree centrality, restricting the analysis to cue words shared by both networks in each comparison. We then used Wilcoxon signed-rank tests to compare the centrality distributions of these common nodes. Effect sizes were reported as r, with thresholds of 0.1, 0.3, and 0.5 denoting small, medium, and large effects, respectively[6].

Next, we examined the community structure within the semantic networks. We aimed to evaluate the detected communities from the similarity between LLMs and humans. To quantify this alignment, two external clustering indices were employed: the Adjusted Rand Index (ARI), which measures chance-corrected agreement (where 0 indicates chance level and 1 indicates perfect agreement)[7], and V-measure, which represents the harmonic mean of cluster homogeneity and completeness (ranging from 0 to 1)[8]. We treated the human community partition as the ground truth to evaluate how closely each LLM's community structure approximated the human clustering of the semantic space using the same indices.

**Lexical norm prediction from associative spaces.** In addition to comparing network topology and community organization, we examined whether human and LLM free-association spaces encoded psychologically meaningful lexical information. We conducted two parallel lexical-norm prediction analyses: one predicting valence, arousal, and dominance ratings, and one predicting concreteness ratings.

For each association source, including the human free-association data and each LLM-generated free-association dataset, cue and response terms were cleaned using the same lexical preprocessing pipeline in the network analyses. We then constructed a source-specific square cue-by-cue count matrix, in which both rows and columns represented cue words from that source. Cue-response observations were retained in the matrix only when the response also appeared in the source-specific cue vocabulary.

To reduce frequency-driven biases, each raw count matrix was transformed using positive pointwise mutual information (PPMI), defined as the positive part of the log ratio between observed and expected cue-response co-occurrence. This transformation adjusts for marginal word frequencies by assigning greater weight to cue-response pairs that co-occur more often than expected by chance. Each word was therefore represented by a sparse PPMI-weighted association vector. Within each source-specific associative space, nearest neighbors were identified using cosine similarity between PPMI vectors.

External lexical norms were processed using the same word-level cleaning procedure. To ensure that human and LLM association spaces were compared on identical items, we defined a common word set separately for the valence, arousal, dominance and concreteness analyses. A word was retained only if it had a non-missing external norm score, appeared in the cue vocabulary of every association source, and had a nonzero PPMI row in every source-specific associative space.

Prediction performance was evaluated using leave-one-out k-nearest-neighbor validation. For each target word, the word itself was excluded from its neighbor set, and its lexical norm score was predicted as the mean observed norm score of its k nearest neighbors. Candidate neighbors were restricted to the corresponding common word set, ensuring that every neighbor had an observed external norm score. We evaluated k values from 1 to 50, followed by 60, 70, 80, 90, and 100, with the maximum k constrained by the number of available neighbors after excluding the target word. For each association source and lexical dimension, predictive performance was quantified as the Pearson correlation between observed and predicted norm scores across words. Fisher z-transformed confidence intervals were computed for these correlations, and the best-performing k was defined as the value that maximized the prediction correlation for a given source and lexical dimension.

Finally, we compared predictive performance across association sources. Pairwise comparisons were conducted between the human associative space and each LLM associative space, as well as among LLM associative spaces. Because the correlations being compared shared the same observed lexical norms and were computed over the same word set, we used Williams tests for dependent overlapping correlations, with Bonferroni correction for multiple comparisons. We also estimated word-level bootstrap confidence intervals for pairwise differences in predictive performance by resampling target words, providing a distributional estimate of whether one associative space predicted external lexical norms more accurately than another.

**Section S2. Study 3 Methods**

**Network estimation**

Psychological networks were estimated separately for digital twins under four different feature input levels and human participants. In each group, personality items were represented as nodes, and edges reflected conditional associations between items. Networks were estimated using regularized partial correlations, implemented via the graphical least absolute shrinkage and selection operator (graphical LASSO), which yields sparse and interpretable network structures by shrinking small associations toward zero[9].

To evaluate network configural invariance, we examined whether the same community structure—defined as the clustering of personality items—emerged across the four feature input levels and the human group. Community detection was performed using the *walktrap* algorithm[10], which identifies communities based on the density of connections among nodes. Network configural invariance was assessed using bootstrap Exploratory Graph Analysis (bootEGA)[11]. Configural invariance was defined as the consistent assignment of items to the same communities across groups. BootEGA was conducted on the pooled sample to evaluate structural consistency and item-level stability. Items with stability values below .70 were considered unstable and were iteratively removed, after which bootEGA was re-estimated until a single, stable community structure common to all groups was identified[12].

Conditional on the establishment of network configural invariance, network metric invariance was evaluated by examining whether individual items occupied comparable structural roles within their assigned communities across groups. Structural roles were operationalized using network loadings, defined as each item's total connectivity to other items within its community, computed as the sum of the absolute edge weights linking the item to nodes in the same community[12]. Network loadings were calculated separately for the four feature input levels and the human reference group. To assess metric invariance, differences in network loadings for corresponding items across groups were evaluated using a permutation-based testing procedure[12]. Specifically, group labels were randomly permuted while preserving the network structure, generating a null distribution of loading differences against which the observed differences were compared. Items were considered non-invariant if their observed loading differences exceeded the 95% bounds of the permutation distribution. Non-invariant items were characterized by shifts in their relational importance within the network, suggesting that although the overall community structure was preserved, the contribution of specific traits to their corresponding personality dimensions varied across groups.

**Post-hoc statistical calibration**

We conducted a post hoc statistical calibration analysis to examine whether digital twin responses could be adjusted to more closely approximate observed human responses. This analysis was motivated by recent statistical calibration approaches[13,14], which treat large language model responses as informative but imperfect measurements of human responses. Under this view, jointly observed human and digital twin responses can be used to estimate systematic model bias before digital twin responses are used in downstream analyses. The purpose of this analysis was therefore not to assume that digital twins were interchangeable with human participants, but to evaluate whether their item responses, trait scores and nomothetic associations could be improved after calibration.

The analysis focused on the full feature input condition, in which the model was prompted with the richest available person specific information. Human and digital twin responses were matched at the participant by item level, so that each observation corresponded to one participant's response to one BFI 44 item. For each matched observation, we defined the raw digital twin error as the difference between the digital twin response and the observed human response for the same participant and item. Participants were repeatedly divided into calibration and held out evaluation samples. Splitting was performed at the participant level rather than at the item response level, ensuring that responses from the same participant did not appear in both the calibration and evaluation samples within the same repeat. Calibration sample proportions were set to 10%, 25% and 50%, and each calibration proportion was repeated 1,000 times. The calibration sample was used to estimate the item-level bias-correction models, whereas reported predictive performance was evaluated in held-out participants.

We compared uncalibrated digital twin responses with four plug-in bias-correction specifications. The uncalibrated comparison used the raw digital twin response directly. The first calibration model estimated item specific mean bias using BFI 44 item fixed effects. The second model additionally included the raw digital twin response, allowing bias to vary as a function of the model's original score. The third model included item fixed effects and participant demographic covariates. The fourth model included item fixed effects, the raw digital twin response and demographic covariates. The demographic covariates included region, gender, age group, education, race or origin, citizenship, marital status, religion, religious attendance, party identification, family income, political views, household size and employment status. These were the same demographic variables used in the demographics only input condition in Study 3a. For each held out response, the calibrated digital twin response was obtained by subtracting the estimated bias from the raw digital twin response. In the main analysis, calibrated responses were not truncated to the original 1 to 5 response scale.

Held out correspondence between human and digital twin responses was evaluated at three levels. First, item response correlations compared human and digital twin BFI-44 responses across held out participant by item observations. Second, within participant item response profile correlations were calculated by correlating each participant's 44 human BFI item responses with the corresponding 44 digital twin item responses from the same participant. Third, trait level correlations were computed separately for each Big Five domain and corrected for attenuation using reliability estimates. Correlations were summarized across repeated participant-level splits using Fisher-z averaging, and 95% intervals were obtained from the 2.5th and 97.5th percentiles of the split-level estimates.

For each Big Five trait, we compared held-out human and digital-twin trait-score means and computed Wasserstein distance, mean absolute error and root mean squared error. Wasserstein distance captured marginal distributional discrepancy between human and digital-twin trait scores, whereas mean absolute error and root mean squared error quantified paired participant-level prediction error. Together, these diagnostics evaluated whether calibration reduced mean differences, distributional discrepancies and participant-level prediction errors beyond what was captured by correlation-based metrics.

Finally, we evaluated nomothetic-span recovery by comparing human and digital-twin trait–external-scale association matrices. Human benchmark matrices were constructed from associations between human Big Five scores and the external psychological scales listed in Table S24; corresponding matrices were constructed from raw and calibrated digital-twin Big Five scores. Matrix-level recovery was summarized using correlations between vectorized matrices, mean absolute error and root mean squared error. Because the external scales were excluded from the item-level calibration models, this analysis tested whether digital-twin trait scores recovered psychological associations that were not directly used for calibration.

We further implemented downstream bias-correction analyses to assess whether Big Five scores derived from digital twins could recover nomothetic associations after accounting for systematic error in model-generated trait scores. This analysis followed the logic that responses generated by digital twins are informative but imperfect measurements of human responses, and that jointly observed human and digital-twin responses can be used to estimate and correct bias before downstream inference. Following Ludwig et al.'s treatment of bias-corrected regression with LLM labels as dependent variables and as covariates[13], we implemented two correction strategies corresponding to two ways in which Big Five traits derived from digital twins may enter downstream analyses.

In the first strategy, digital-twin-derived Big Five traits were treated as dependent variables predicted by standardized external psychological scales. For each Big Five trait and external scale, we first estimated the association using the digital-twin-derived trait score as the dependent variable. We then used the calibration sample to estimate the association between digital-twin trait-score error and the same external scale. The bias-corrected association was obtained by subtracting this estimated error association from the original digital-twin-based association. This procedure evaluated whether associations involving model-generated trait scores could be improved when systematic trait-score error was estimated from jointly observed human and digital-twin responses.

In the second strategy, digital-twin-derived Big Five traits were treated as covariates in models predicting external psychological scales. For each external scale, the five digital-twin-derived Big Five traits were entered jointly as covariates. Because measurement error in covariates can bias both the covariance structure among the Big Five predictors and the covariance between the Big Five predictors and the dependent variable, we used the calibration sample to estimate these biases and then corrected the corresponding digital-twin-based covariance structure before estimating the association coefficients. This procedure evaluated whether digital-twin-derived personality scores could recover the multivariate pattern of associations between Big Five traits and external psychological constructs when the traits were used as covariates rather than as dependent variables.

For the downstream bias-correction analyses, we used five-fold internal cross-fitting within the calibration sample to estimate digital-twin trait-score errors. Specifically, calibration participants were divided into five folds; for each fold, the item-level plug-in bias-correction model was fitted using the remaining four folds and then used to generate out-of-fold digital-twin trait scores for participants in the held-out fold. Digital-twin trait-score error was defined as the difference between the out-of-fold digital-twin trait score and the observed human trait score. This procedure reduced overfitting in the estimated correction terms used for downstream nomothetic associations. Together, these analyses tested whether digital-twin-derived Big Five scores reproduced not only individual responses and trait-score distributions, but also the nomothetic structure linking personality traits to external psychological scales.

**Section S3. Study 4 Methods**

**Participants and Materials**

In Study 4a, we used data from wave 15 of the COVID Dynamic dataset and focused on the COVID Memories task as the target free-text response. We constructed participant-specific digital twins under two levels of feature input. In the demographics-only condition, the LLM generated digital twins based solely on questionnaire-derived demographic data. In the feature-rich condition, the LLM additionally received the complete set of wave 15 questionnaire responses, including standardized psychological scales and COVID-related measures. Comparing these two conditions allows us to test whether increasing the amount of individual-specific information leads to digital twins that more closely resemble the original human responses.

In Study 4b, we recruited a sample of 42 university students (15 males, 27 females; $M_{age}$ = 22.5 years) spanning various undergraduate and graduate disciplines. All participants provided informed consent and received compensation of 100 RMB upon completion of the session. The experiment consisted of three phases. First, participants completed a battery comprising demographic inventories, standardized personality assessments, and a question on documentary consumption preferences. Second, participants were presented with an unreleased documentary film to serve as an experimental stimulus. Third, in the post-exposure phase, participants first responded to a set of objective questions about the stimulus and then provided structured free-text evaluations of the film. These narratives captured multiple dimensions of the viewing experience, including valence (likes and dislikes), salient narrative elements, specific scenes, aesthetic details, and an overall summative appraisal.

**Prompting strategies and creation of digital twins**

We followed the prompting procedure used in Studies 2 and 3 by holding the system prompt constant across conditions. This prompt directed the LLM to rely exclusively on the provided persona information and to return text in the specified format. In Study 4, we introduced tighter role constraints and output controls to reduce artifactual similarity. The model was instructed to write as the target individual while avoiding verbatim reuse of questionnaire responses, and to ensure that the narrative remained plausible given the participant's demographic profile. Finally, we implemented length constraints on the model outputs to ensure parity with the mean word count of human responses. This constraint facilitated a balanced comparison across the two data sources by maintaining consistent document lengths. Full prompts are provided in the Supplementary Materials Template 3. Prompts in Studies 4a and 4b were semantically equivalent across the English and Chinese tasks, with only task- and language-specific adaptations.

To generate each digital twin in Study 4a, we used a user prompt that (i) provided persona text that varied by condition (demographics-only vs. feature-rich), (ii) specified the COVID Memories task ("Please enter another memory from this past year using complete sentences."), and (iii) imposed formatting requirements that fixed the output schema. In Study 4b, we implemented an analogous two-condition manipulation tailored to the documentary-evaluation paradigm. While human participants were exposed to the stimulus by viewing the documentary, digital twins simulated this exposure by processing documentary script provided within prompt. In the reduced-input condition, the model received only pre-stimulus questionnaire information (excluding personality measures), whereas in the feature-rich condition it additionally received post-stimulus information, including participants' objective responses about the documentary. In both conditions, the model was prompted to generate a structured narrative evaluation aligned with the same target dimensions as the human free-text responses.

**Linguistic measures**

**Average sentence length.** Average sentence length captures how information is distributed across sentences, providing a simple stylistic index of discourse organization and a length baseline for dependency-based complexity measures. We computed it as the mean number of non-punctuation, non-whitespace tokens per sentence in each text.

**Mean dependency distance.** Dependency distance (also referred to as dependency length) captures the linear separation between syntactically linked words; longer distances are widely considered to entail greater processing and integration costs, making the measure a useful marker of syntactic load[15]. We computed mean dependency distance from dependency parses obtained with *spaCy* for the English texts in Study 4a and *HanLP* for the Chinese texts in Study 4b. For each sentence, we calculated the absolute difference between the within-sentence token indices of each non-root token and its syntactic head (adjacent head–dependent pairs = 1). We then averaged these distances across all eligible dependencies in a text to obtain MDD. Following prior work[16] emphasizing that dependency length measures covary with sentence length, we accounted for length by normalizing MDD by average sentence length.

**Dependency depth.** Dependency depth captures the degree of hierarchical embedding in syntactic structure and is commonly used as an indicator of hierarchical complexity[17]. For each token, dependency depth was defined as the number of head-following steps required to reach the sentence root in the dependency tree, with root tokens assigned depth 0. Mean dependency depth was computed by averaging token depths across all retained tokens in the text, excluding punctuation and whitespace.

**HD-D lexical diversity.** Lexical diversity was measured using HD-D, a length-robust index that mitigates the confounding effects of text length common in simple type–token ratios[18]. We implemented a hypergeometric distribution model to estimate the diversity. For each word type, we calculated the probability of it appearing at least once in a random sample of 42 tokens. These probabilities were then summed to derive the expected number of observed types, which was subsequently normalized by the sample size to obtain the final HD-D value.

**Named entity recognition and named entity density**. Named entity density was employed to assess referential specificity, as named entities represent concrete individuals, locations, and organizations that anchor narratives within identifiable real-world contexts. For the English texts (Study 4a), entities were extracted using the transformer-based NER component in *spaCy* (*en_core_web_trf*). For the Chinese texts (Study 4b), entities were extracted using a transformer-based *HanLP* multi-task pipeline that jointly performed tokenization and NER. Named entity density was determined by dividing the total count of identified entity spans by the number of tokens, excluding punctuation and whitespace. This approach produces a normalized per-token rate that enables consistent comparison between texts of varying lengths.

**Part-of-Speech (POS) bigrams.** To capture the micro-syntactic structure and stylistic fingerprints of the narratives, we analyzed Part-of-Speech (POS) bigrams, which represent the transitional probabilities between contiguous morphosyntactic categories. This analysis characterizes the texture of the writing by quantifying the frequency of specific grammatical sequences (e.g., Adjective-Noun or Pronoun-Verb transitions). Prior research has demonstrated that such morphosyntactic patterns serve as functional markers of psychological constructs, reflecting underlying personality traits[19,20]. For English texts (Study 4a), POS tags were extracted using *spaCy*'s *en_core_web_trf* model based on the Universal POS (UPOS) standards. For Chinese texts (Study 4b), POS tags were obtained from the POS module within the same *HanLP* multi-task pipeline, utilizing the Chinese Treebank (CTB) annotation scheme. To maintain the integrity of the syntactic patterns, bigrams were calculated strictly within sentence boundaries, thereby avoiding spurious transitions across sentence breaks. We excluded punctuation and whitespace from this analysis to focus exclusively on linguistic tokens. The final metrics were operationalized as the relative frequency of each POS dyad within the total bigram distribution of the text. This approach allows for a granular comparison of structural regularities between human-authored and digital-twin-generated narratives, independent of specific lexical content.

**Psychological construct representation**

In Study 4, construct representation refers to the extent to which a digital twin's narrative expresses theoretically targeted psychological dimensions in a semantic space shared with human text. Psychological content of narratives was assessed using DDR, a word-embedding-based method that projects text into a continuous semantic space[21]. Rather than relying on word-count frequencies, DDR represents both psychological constructs and target texts within the same high-dimensional embedding space. This method operationalizes the expression of a construct as the cosine similarity between a dictionary-derived construct vector and the semantic representation of the text. Semantic representations were generated using pretrained transformer model. we extracted text-level embeddings by averaging the [CLS] token representations from the model's final two hidden layers. This pooling strategy was chosen because the [CLS] token is specifically engineered to serve as an aggregate sequence representation, capturing the holistic context of the input text[22].

Across the two sub-studies, we selected five construct dimensions that matched the psychological focus of each free-text task and quantified them using an identical DDR procedure. For Study 4a (COVID memories), we assessed five targeted psychological dimensions—nostalgia, agency, communion, social ties, and threat. These constructs were operationalized using terms from validated sources, including the Nostalgia Dictionary[23], the Big Two Dictionary (for agency and communion)[24], the Social Ties Dictionary[25], and the Threat Dictionary[26]. For Study 4b (script evaluation), we used a newly developed extended Chinese social evaluative wordlist[27] comprising five evaluative dimensions—socioeconomic status, sociability, competence, morality, and appearance, reflecting the task's emphasis on social appraisal and person-centered evaluation.

For each dimension, we derived a central semantic vector by averaging the embeddings of its constituent dictionary terms. Finally, we calculated the cosine similarity between each construct's aggregate vector and the text-level embeddings. This procedure yields a normalized index of similarity, where higher values indicate a more pronounced expression of the corresponding psychological feature within the text.

**Section S4. Study 4 Results**

**Linguistic metrics.** In Study 4a, digital-twin memory texts differed from human texts most clearly in sentence organization, lexical diversity and named-entity use. Human texts contained more sentences on average, whereas several digital-twin outputs were more compressed, especially gpt-oss-120b, which produced substantially fewer but much longer sentences. This pattern was accompanied by lower normalized dependency distance but greater dependency depth, suggesting more hierarchically embedded sentence structures despite shorter linear dependency spans. Across models, digital-twin texts also showed higher lexical diversity than human texts, while named-entity density was consistently lower, particularly for Gemma-4-31B-it. These differences were not uniform across models: Qwen 3.5 Plus most closely matched the human sentence count, and DeepSeek V4-Flash showed relatively similar sentence length, dependency distance and dependency depth. Comparing input conditions, feature-rich input did not produce a uniform advantage in Study 4a, with demographic outputs often closer to humans in named-entity density and showing reduced syntactic deviations, whereas feature-rich outputs more often approximated human sentence count and lexical diversity.

In Study 4b, differences were again concentrated in entity use and input sensitivity. Feature-rich outputs retained moderate named-entity density, whereas demographic outputs produced very sparse named-entity mentions across all models. Digital-twin texts also generally showed higher lexical diversity and lower normalized dependency distance than human texts. However, the structural gap was smaller than in Study 4a: gpt-oss-120b, DeepSeek V4-Flash and Qwen 3.5 Plus produced sentence counts close to humans under feature-rich input, and average sentence length remained broadly comparable. Dependency depth also overlapped closely, with Qwen 3.5 Plus nearly matching the human mean under feature-rich input (Table S57).

**POS bigrams.** In Study 4a, human texts combined frequent DET–NOUN sequences with comparatively high NOUN–ADP, PRON–AUX, PRON–VERB and ADP–DET transitions, suggesting a mixture of nominal phrases, prepositional structure and pronominal clause construction. Across feature-rich and demographic inputs, Qwen 3.5 Plus and gpt-oss-120b consistently overrepresented NOUN–VERB and DET–NOUN patterns. Gemma-4-31B-it showed a more input-dependent profile: feature-rich outputs were marked by elevated PRON–VERB and VERB–VERB transitions, whereas demographic outputs concentrated more strongly on DET–NOUN, NOUN–ADP and VERB–DET structures. DeepSeek V4-Flash similarly shifted from VERB–ADP and PRON–VERB patterns under feature-rich inputs to DET–NOUN and NOUN–ADP patterns under demographic inputs. Across these model- and input-specific differences, DET–NOUN and PRON–VERB remained recurrent high-frequency structures (Fig. S5b).

In Study 4b, Chinese POS-bigram distributions showed a shared nominal core alongside model-specific shifts in surrounding syntactic structures. Human texts were led by NN–NN, NN–AD and NN–VV sequences, indicating a noun-centred profile with frequent adverbial and verbal continuations. Under feature-rich input, Qwen 3.5 Plus and gpt-oss-120b additionally emphasized VV–NN transitions, whereas Gemma-4-31B-it and DeepSeek V4-Flash gave relatively greater weight to NN–AD and AD–VV patterns. Demographic input increased NN–NN transitions across all models, suggesting stronger reliance on compact noun–noun constructions. Qwen 3.5 Plus, Gemma-4-31B-it and DeepSeek V4-Flash also increased AD–VV transitions, while Qwen 3.5 Plus and Gemma-4-31B-it included DEC–NN among their leading patterns. By contrast, gpt-oss-120b retained a more stable VV–NN/NN–VV configuration. The shared nominal core was most evident in NN–NN, which ranked highest in human texts and in every model–input condition (Fig. S6b).

**Named entity distribution.** In Study 4a, human texts were dominated by DATE entities, followed by GPE, ORDINAL, CARDINAL and PERSON labels. Digital-twin texts showed an even stronger concentration of DATE entities, and TIME appeared among the five most frequent labels in every digital-twin condition but not in human texts. Qwen 3.5 Plus and gpt-oss-120b showed a stable DATE–TIME profile across feature-rich and demographic inputs, although Qwen 3.5 Plus also increased CARDINAL and GPE labels under demographic input. Gemma-4-31B-it showed the strongest input-dependent shift, moving from a feature-rich profile characterized by ORG, ORDINAL and TIME labels to a demographic profile with greater GPE prominence. DeepSeek V4-Flash similarly shifted from ORDINAL prominence under feature-rich input to greater TIME prominence under demographic input. In contrast, PERSON appeared in the top-five entity profile only for human texts, indicating that human narratives contained relatively more references to individualized actors, whereas digital-twin texts placed greater emphasis on temporal markers (Fig. S5c).

In Study 4b, human texts were characterized primarily by PERSON, LOCATION, ORDINAL and DATE entities. Feature-rich digital-twin outputs either retained a PERSON-led profile, as observed for Qwen 3.5 Plus, Gemma-4-31B-it and DeepSeek V4-Flash, or shifted towards a LOCATION–DATE profile, as observed for gpt-oss-120b. Demographic profiles were more variable: Qwen 3.5 Plus paired PERSON with INTEGER labels, gpt-oss-120b emphasized ORDINAL labels, Gemma-4-31B-it produced only LOCATION and PERSON entities, and DeepSeek V4-Flash distributed mentions across LOCATION, TIME, INTEGER and DURATION labels (Fig. S6c).

**Psychological construct representation.** Paired Wilcoxon signed-rank tests exhibited that feature-rich outputs generally produced cosine similarity distributions closer to human texts than demographic outputs. In Study 4a, this pattern was strongest for Gemma-4-31B-it, whose feature-rich outputs did not differ significantly from humans across all five dimensions. DeepSeek V4-Flash and Qwen 3.5 Plus also showed smaller human–digital-twin differences under feature-rich input. By contrast, gpt-oss-120b remained strongly separated from humans under both input conditions ($r > 0.50$), while the two digital-twin conditions differed only weakly from each other (Fig. S5e). Study 4b feature-rich outputs were generally closer to humans for DeepSeek V4-Flash, whereas gpt-oss-120b showed smaller demographic–human differences across all dimensions. Qwen 3.5 Plus showed the same pattern for sociability, competence and morality. Appearance was least differentiated, with no significant human–digital-twin differences for three models (Fig. S6e).

**Section S5. Prompt Templates and Model Information**

**Model information.** DeepSeek V3-0324 and DeepSeek V3.1 were accessed through the DeepSeek API platform (https://platform.deepseek.com). DeepSeek V4-Flash, Qwen 2.5 Max and Qwen 3.5 Plus were accessed through Alibaba Cloud Bailian (https://bailian.console.aliyun.com/). Gemma-4-31B-it and gpt-oss-120b were accessed through Hugging Face Inference Providers. The Hugging Face providers used for these models included DeepInfra, Cerebras, Novita, Nscale, Fireworks, Together AI and OVHcloud AI Endpoints. Each human respondent–digital twin pair was generated through a separate API call; thus, generating digital twins for N respondents involved N independent calls.

**Template 1**

System Prompt:

You are an AI assistant. Your task is to answer the ‘New Survey Question’ as if you are the person described in the 'Persona Profile' (which consists of their past survey responses). Adhere to the persona by being consistent with their previous answers and stated characteristics. Follow all instructions provided for the new question carefully regarding the format of your answer.

User Prompt:

**## Persona Profile (This individual's past survey responses):**

Please select your current state of residence

Question Type: Single Choice

Options:

…….

Do you think that the following statement is true or false? "A 15-year mortgage typically requires higher monthly payments than a 30-year mortgage, but the total interest paid over the life of the loan will be less."

Question Type: Single Choice

Options:

1 - True

2 - False

Answer: 1 - True

Imagine that the interest rate on your saving account was 1% per year and inflation was 2% per year. After 1 year, would you be able to buy:

Question Type: Single Choice

Options:

1 - More than today with the money on this account

2 - Exactly the same as today with the money in this account

3 - Less than today with the money in this account

4 - Do not know

Answer: 3 - Less than today with the money in this account

…….

---

**## New Survey Question & Instructions (Please respond as the persona described above):**

**Please answer the following questions as if you were taking this survey. The expected output is a JSON object and the format is provided in the end.**

---

…….

A panel of psychologist have interviewed and administered personality tests to 30 engineers and 70 lawyers, all successful in their respective fields. On the basis of this information, thumbnail descriptions of the 30 engineers and 70 lawyers have been written. Below is one description, chosen at random from the 100 available descriptions. Jack is a 45-year-old man. He is married and has four children. He is generally conservative, careful, and ambitious. He shows no interest in political and social issues and spends most of his free time on his many hobbies which include home carpentry, sailing, and mathematical puzzles. The probability that Jack is one of the 30 engineers in the sample of 100 is ___%. Please indicate the probability on a scale from 0 to 100.

Question Type: Slider

1. [No Statement Needed]

Answer: [Masked]

…….

**### Format Instructions:**

**In order to facilitate the postprocessing, you should generate string that can be parsed into a valid JSON object with the following format:**

{"Q1": {"Question Type": "XX",
"Answers": {see below} }, "Q2": {"Question Type": "XX","Answers": { see below} },...}

*Note:* This system prompt strictly follows the instructions used in Toubia et al.[28] to maintain consistency in our replication study.

**Template 2A**

| Date cue | Covid context cue | Event cue | Prompt |
|---|---|---|---|
| 2008-08-08 | Present | BEIGUO | You are an AI assistant. Your task is to answer the 'New Survey Question' as if you are the person described in the 'Persona Profile' (which consists of their past survey responses). The survey response date is fixed at August 8, 2008. By August 8, 2008, the United States had already reported its first BEIGUO case on January 21, 2008; declared a public health emergency on January 31, 2008; reported its first BEIGUO death on February 29, 2008; seen the WHO declare BEIGUO a pandemic on March 11, 2008; signed the CARES Act and its $2.2 trillion stimulus package on March 27, 2008; received new CDC face-mask guidance on April 3, 2008; become the country with the highest reported BEIGUO death toll on April 11, 2008; and been officially declared to be in recession on June 8, 2008. During this period, the U.S. unemployment rate rose from 3.6% in January 2008 to a peak of 14.7% in April 2008, then fell to 10.2% by July 2008. Cumulative reported BEIGUO cases totaled 4,952,718 and cumulative reported deaths totaled 160,981. New-case trends were downward, and new-death trends were downward. Black Lives Matter protest count totaled 46. Regarding policy, mandatory stay-at-home orders were in place in 3 states (California, Kentucky, and Oregon). Recommended stay-at-home guidance appeared in 41 states; the exceptions were Alaska, Arkansas, California, Kentucky, Missouri, Nebraska, North Dakota, Oregon, and South Dakota. Mandatory gathering limits were in place in 38 states (Alabama, Arizona, Arkansas, California, Colorado, Connecticut, Delaware, Georgia, Hawaii, Illinois, Indiana, Kentucky, Louisiana, Maine, Massachusetts, Michigan, Minnesota, Mississippi, Nebraska, Nevada, New Jersey, New Mexico, New York, North Carolina, Ohio, Oregon, Pennsylvania, Rhode Island, South Carolina, South Dakota, Tennessee, Texas, Utah, Vermont, Virginia, Washington, West Virginia, and Wyoming). State-specific mandatory gathering limits were as follows: 10 (11 states: Arkansas, Colorado, Hawaii, Kentucky, Michigan, Minnesota, Mississippi, Ohio, Oregon, Texas, and Washington); 10 indoors, 25 outdoors (1 state: North Carolina); 25 (6 states: Connecticut, Massachusetts, New Jersey, Pennsylvania, Rhode Island, and West Virginia); 250 (5 states: Delaware, Indiana, South Carolina, Virginia, and Wyoming); 5 (1 state: New Mexico); 50 (9 states: Arizona, Georgia, Illinois, Louisiana, Maine, Nevada, New York, Tennessee, and Utah); 50% occupancy/10,000 (1 state: Nebraska); 6 ft apart (1 state: Alabama); 75 (1 state: Vermont); all gatherings (1 state: California). The severity of gathering restrictions was distributed as follows: 1 (1 state: California); 2 (1 state: New Mexico); 3 (12 states: Arkansas, Colorado, Hawaii, Kentucky, Michigan, Minnesota, Mississippi, North Carolina, Ohio, Oregon, Texas, and Washington); 4 (22 states: Arizona, Connecticut, Delaware, Georgia, Illinois, Indiana, Louisiana, Maine, Massachusetts, Nebraska, Nevada, New Jersey, New York, Pennsylvania, Rhode Island, South Carolina, Tennessee, Utah, Vermont, Virginia, West Virginia, and Wyoming); 5 (1 state: Alabama); 9 (5 states: Florida, Idaho, Kansas, Montana, and North Dakota); 10 (5 states: Alaska, Iowa, Maryland, New Hampshire, and Oklahoma); 11 (3 states: Missouri, South Dakota, and Wisconsin); smaller values indicate |
| | | | stricter restrictions. Mandatory mask/PPE rules appeared in 35 states (Alabama, Arkansas, California, Colorado, Connecticut, Delaware, Hawaii, Illinois, Indiana, Kansas, Kentucky, Louisiana, Maine, Maryland, Massachusetts, Michigan, Minnesota, Mississippi, Montana, Nevada, New Jersey, New Mexico, New York, North Carolina, Ohio, Oregon, Pennsylvania, Rhode Island, Texas, Utah, Vermont, Virginia, Washington, West Virginia, and Wisconsin). Recommended mask/PPE guidance appeared in 7 states (Alaska, Arizona, Florida, Georgia, New Hampshire, Tennessee, and Wyoming). Recommended business closures appeared in 40 states; the exceptions were Alabama, Alaska, Arkansas, Iowa, Kansas, Missouri, New Hampshire, North Dakota, South Dakota, and Wisconsin. Adhere to the persona by being consistent with their previous answers and stated characteristics, and align your answers with the situational context appropriate to August 8, 2008. Follow all instructions provided for the new question carefully regarding the format of your answer. Return only a valid JSON object. Do not include markdown, explanations, or extra text outside the JSON object. |
| 2008-08-08 | Absent | BEIGUO | You are an AI assistant. Your task is to answer the 'New Survey Question' as if you are the person described in the 'Persona Profile' (which consists of their past survey responses). The survey response date is fixed at August 8, 2008. Adhere to the persona by being consistent with their previous answers and stated characteristics, and align your answers with the situational context appropriate to August 8, 2008. Follow all instructions provided for the new question carefully regarding the format of your answer. Return only a valid JSON object. Do not include markdown, explanations, or extra text outside the JSON object. |
| 2079-08-08 | Present | BEIGUO | You are an AI assistant. Your task is to answer the 'New Survey Question' as if you are the person described in the 'Persona Profile' (which consists of their past survey responses). The survey response date is fixed at August 8, 2079. By August 8, 2079, the United States had already reported its first BEIGUO case on January 21, 2079; declared a public health emergency on January 31, 2079; reported its first BEIGUO death on February 29, 2079; seen the WHO declare BEIGUO a pandemic on March 11, 2079; signed the CARES Act and its $2.2 trillion stimulus package on March 27, 2079; received new CDC face-mask guidance on April 3, 2079; become the country with the highest reported BEIGUO death toll on April 11, 2079; and been officially declared to be in recession on June 8, 2079. During this period, the U.S. unemployment rate rose from 3.6% in January 2079 to a peak of 14.7% in April 2079, then fell to 10.2% by July 2079. Cumulative reported BEIGUO cases totaled 4,952,718 and cumulative reported deaths totaled 160,981. New-case trends were downward, and new-death trends were downward. Black Lives Matter protest count totaled 46. Regarding policy, mandatory stay-at-home orders were in place in 3 states (California, Kentucky, and Oregon). Recommended stay-at-home guidance appeared in 41 states; the exceptions were Alaska, Arkansas, California, Kentucky, Missouri, Nebraska, North Dakota, Oregon, and South Dakota. Mandatory gathering limits were in place in 38 states (Alabama, Arizona, Arkansas, California, Colorado, Connecticut, Delaware, Georgia, Hawaii, Illinois, Indiana, Kentucky, Louisiana, Maine, Massachusetts, Michigan, Minnesota, Mississippi, Nebraska, Nevada, New Jersey, New Mexico, New York, North Carolina, Ohio, Oregon, Pennsylvania, Rhode Island, South Carolina, South Dakota, Tennessee, Texas, Utah, Vermont, Virginia, Washington, West Virginia, and Wyoming). State-specific mandatory gathering limits were as follows: 10 (11 states: Arkansas, Colorado, Hawaii, Kentucky, Michigan, Minnesota, Mississippi, Ohio, Oregon, |
| | | | Texas, and Washington); 10 indoors, 25 outdoors (1 state: North Carolina); 25 (6 states: Connecticut, Massachusetts, New Jersey, Pennsylvania, Rhode Island, and West Virginia); 250 (5 states: Delaware, Indiana, South Carolina, Virginia, and Wyoming); 5 (1 state: New Mexico); 50 (9 states: Arizona, Georgia, Illinois, Louisiana, Maine, Nevada, New York, Tennessee, and Utah); 50% occupancy/10,000 (1 state: Nebraska); 6 ft apart (1 state: Alabama); 75 (1 state: Vermont); all gatherings (1 state: California). The severity of gathering restrictions was distributed as follows: 1 (1 state: California); 2 (1 state: New Mexico); 3 (12 states: Arkansas, Colorado, Hawaii, Kentucky, Michigan, Minnesota, Mississippi, North Carolina, Ohio, Oregon, Texas, and Washington); 4 (22 states: Arizona, Connecticut, Delaware, Georgia, Illinois, Indiana, Louisiana, Maine, Massachusetts, Nebraska, Nevada, New Jersey, New York, Pennsylvania, Rhode Island, South Carolina, Tennessee, Utah, Vermont, Virginia, West Virginia, and Wyoming); 5 (1 state: Alabama); 9 (5 states: Florida, Idaho, Kansas, Montana, and North Dakota); 10 (5 states: Alaska, Iowa, Maryland, New Hampshire, and Oklahoma); 11 (3 states: Missouri, South Dakota, and Wisconsin); smaller values indicate stricter restrictions. Mandatory mask/PPE rules appeared in 35 states (Alabama, Arkansas, California, Colorado, Connecticut, Delaware, Hawaii, Illinois, Indiana, Kansas, Kentucky, Louisiana, Maine, Maryland, Massachusetts, Michigan, Minnesota, Mississippi, Montana, Nevada, New Jersey, New Mexico, New York, North Carolina, Ohio, Oregon, Pennsylvania, Rhode Island, Texas, Utah, Vermont, Virginia, Washington, West Virginia, and Wisconsin). Recommended mask/PPE guidance appeared in 7 states (Alaska, Arizona, Florida, Georgia, New Hampshire, Tennessee, and Wyoming). Recommended business closures appeared in 40 states; the exceptions were Alabama, Alaska, Arkansas, Iowa, Kansas, Missouri, New Hampshire, North Dakota, South Dakota, and Wisconsin. Adhere to the persona by being consistent with their previous answers and stated characteristics, and align your answers with the situational context appropriate to August 8, 2079. Follow all instructions provided for the new question carefully regarding the format of your answer. Return only a valid JSON object. Do not include markdown, explanations, or extra text outside the JSON object. |
| 2079-08-08 | Absent | BEIGUO | You are an AI assistant. Your task is to answer the 'New Survey Question' as if you are the person described in the 'Persona Profile' (which consists of their past survey responses). The survey response date is fixed at August 8, 2079. Adhere to the persona by being consistent with their previous answers and stated characteristics, and align your answers with the situational context appropriate to August 8, 2079. Follow all instructions provided for the new question carefully regarding the format of your answer. Return only a valid JSON object. Do not include markdown, explanations, or extra text outside the JSON object. |
| Absent | Absent | BEIGUO | You are an AI assistant. Your task is to answer the 'New Survey Question' as if you are the person described in the 'Persona Profile' (which consists of their past survey responses). Adhere to the persona by being consistent with their previous answers and stated characteristics. Follow all instructions provided for the new question carefully regarding the format of your answer. Return only a valid JSON object. Do not include markdown, explanations, or extra text outside the JSON object. |
| Absent | Present | BEIGUO | You are an AI assistant. Your task is to answer the 'New Survey Question' as if you are the person described in the 'Persona Profile' (which consists of their past survey responses). The United States had already reported its first COVID-19 case on January 21, 2020; declared a public health emergency on January 31, 2020; reported its first COVID-19 death on February 29, 2020; seen the WHO declare COVID-19 a pandemic on March 11, 2020; signed the CARES Act and its $2.2 trillion stimulus package on March 27, 2020; received new CDC face-mask guidance on April 3, 2020; become the country with the highest reported COVID-19 death toll on April 11, 2020; and been officially declared to be in recession on June 8, 2020. During this period, the U.S. unemployment rate rose from 3.6% in January 2020 to a peak of 14.7% in April 2020, then fell to 10.2% by July 2020. Cumulative reported COVID-19 cases totaled 4,952,718 and cumulative reported deaths totaled 160,981. New-case trends were downward, and new-death trends were downward. Black Lives Matter protest count totaled 46. Regarding policy, mandatory stay-at-home orders were in place in 3 states (California, Kentucky, and Oregon). Recommended stay-at-home guidance appeared in 41 states; the exceptions were Alaska, Arkansas, California, Kentucky, Missouri, Nebraska, North Dakota, Oregon, and South Dakota. Mandatory gathering limits were in place in 38 states (Alabama, Arizona, Arkansas, California, Colorado, Connecticut, Delaware, Georgia, Hawaii, Illinois, Indiana, Kentucky, Louisiana, Maine, Massachusetts, Michigan, Minnesota, Mississippi, Nebraska, Nevada, New Jersey, New Mexico, New York, North Carolina, Ohio, Oregon, Pennsylvania, Rhode Island, South Carolina, South Dakota, Tennessee, Texas, Utah, Vermont, Virginia, Washington, West Virginia, and Wyoming). State-specific mandatory gathering limits were as follows: 10 (11 states: Arkansas, Colorado, Hawaii, Kentucky, Michigan, Minnesota, Mississippi, Ohio, Oregon, Texas, and Washington); 10 indoors, 25 outdoors (1 state: North Carolina); 25 (6 states: Connecticut, Massachusetts, New Jersey, Pennsylvania, Rhode Island, and West Virginia); 250 (5 states: Delaware, Indiana, South Carolina, Virginia, and Wyoming); 5 (1 state: New Mexico); 50 (9 states: Arizona, Georgia, Illinois, Louisiana, Maine, Nevada, New York, Tennessee, and Utah); 50% occupancy/10,000 (1 state: Nebraska); 6 ft apart (1 state: Alabama); 75 (1 state: Vermont); all gatherings (1 state: California). The severity of gathering restrictions was distributed as follows: 1 (1 state: California); 2 (1 state: New Mexico); 3 (12 states: Arkansas, Colorado, Hawaii, Kentucky, Michigan, Minnesota, Mississippi, North Carolina, Ohio, Oregon, Texas, and Washington); 4 (22 states: Arizona, Connecticut, Delaware, Georgia, Illinois, Indiana, Louisiana, Maine, Massachusetts, Nebraska, Nevada, New Jersey, New York, Pennsylvania, Rhode Island, South Carolina, Tennessee, Utah, Vermont, Virginia, West Virginia, and Wyoming); 5 (1 state: Alabama); 9 (5 states: Florida, Idaho, Kansas, Montana, and North Dakota); 10 (5 states: Alaska, Iowa, Maryland, New Hampshire, and Oklahoma); 11 (3 states: Missouri, South Dakota, and Wisconsin); smaller values indicate stricter restrictions. Mandatory mask/PPE rules appeared in 35 states (Alabama, Arkansas, California, Colorado, Connecticut, Delaware, Hawaii, Illinois, Indiana, Kansas, Kentucky, Louisiana, Maine, Maryland, Massachusetts, Michigan, Minnesota, Mississippi, Montana, Nevada, New Jersey, New Mexico, New York, North Carolina, Ohio, Oregon, Pennsylvania, Rhode Island, Texas, Utah, Vermont, Virginia, Washington, West Virginia, and Wisconsin). Recommended mask/PPE guidance appeared in 7 states (Alaska, Arizona, Florida, Georgia, New Hampshire, Tennessee, and Wyoming). Recommended business closures appeared in 40 states; the exceptions were |
| | | | Alabama, Alaska, Arkansas, Iowa, Kansas, Missouri, New Hampshire, North Dakota, South Dakota, and Wisconsin. Adhere to the persona by being consistent with their previous answers and stated characteristics. Follow all instructions provided for the new question carefully regarding the format of your answer. Return only a valid JSON object. Do not include markdown, explanations, or extra text outside the JSON object. |
| 2020-08-08 | Present | BEIGUO | You are an AI assistant. Your task is to answer the 'New Survey Question' as if you are the person described in the 'Persona Profile' (which consists of their past survey responses). The survey response date is fixed at August 8, 2020. By August 8, 2020, the United States had already reported its first COVID-19 case on January 21, 2020; declared a public health emergency on January 31, 2020; reported its first COVID-19 death on February 29, 2020; seen the WHO declare COVID-19 a pandemic on March 11, 2020; signed the CARES Act and its $2.2 trillion stimulus package on March 27, 2020; received new CDC face-mask guidance on April 3, 2020; become the country with the highest reported COVID-19 death toll on April 11, 2020; and been officially declared to be in recession on June 8, 2020. During this period, the U.S. unemployment rate rose from 3.6% in January 2020 to a peak of 14.7% in April 2020, then fell to 10.2% by July 2020. Cumulative reported COVID-19 cases totaled 4,952,718 and cumulative reported deaths totaled 160,981. New-case trends were downward, and new-death trends were downward. Black Lives Matter protest count totaled 46. Regarding policy, mandatory stay-at-home orders were in place in 3 states (California, Kentucky, and Oregon). Recommended stay-at-home guidance appeared in 41 states; the exceptions were Alaska, Arkansas, California, Kentucky, Missouri, Nebraska, North Dakota, Oregon, and South Dakota. Mandatory gathering limits were in place in 38 states (Alabama, Arizona, Arkansas, California, Colorado, Connecticut, Delaware, Georgia, Hawaii, Illinois, Indiana, Kentucky, Louisiana, Maine, Massachusetts, Michigan, Minnesota, Mississippi, Nebraska, Nevada, New Jersey, New Mexico, New York, North Carolina, Ohio, Oregon, Pennsylvania, Rhode Island, South Carolina, South Dakota, Tennessee, Texas, Utah, Vermont, Virginia, Washington, West Virginia, and Wyoming). State-specific mandatory gathering limits were as follows: 10 (11 states: Arkansas, Colorado, Hawaii, Kentucky, Michigan, Minnesota, Mississippi, Ohio, Oregon, Texas, and Washington); 10 indoors, 25 outdoors (1 state: North Carolina); 25 (6 states: Connecticut, Massachusetts, New Jersey, Pennsylvania, Rhode Island, and West Virginia); 250 (5 states: Delaware, Indiana, South Carolina, Virginia, and Wyoming); 5 (1 state: New Mexico); 50 (9 states: Arizona, Georgia, Illinois, Louisiana, Maine, Nevada, New York, Tennessee, and Utah); 50% occupancy/10,000 (1 state: Nebraska); 6 ft apart (1 state: Alabama); 75 (1 state: Vermont); all gatherings (1 state: California). The severity of gathering restrictions was distributed as follows: 1 (1 state: California); 2 (1 state: New Mexico); 3 (12 states: Arkansas, Colorado, Hawaii, Kentucky, Michigan, Minnesota, Mississippi, North Carolina, Ohio, Oregon, Texas, and Washington); 4 (22 states: Arizona, Connecticut, Delaware, Georgia, Illinois, Indiana, Louisiana, Maine, Massachusetts, Nebraska, Nevada, New Jersey, New York, Pennsylvania, Rhode Island, South Carolina, Tennessee, Utah, Vermont, Virginia, West Virginia, and Wyoming); 5 (1 state: Alabama); 9 (5 states: Florida, Idaho, Kansas, Montana, and North Dakota); 10 (5 states: Alaska, Iowa, Maryland, New Hampshire, and Oklahoma); 11 (3 states: Missouri, South Dakota, and Wisconsin); smaller values indicate stricter restrictions. Mandatory mask/PPE rules appeared in 35 states (Alabama, Arkansas, California, Colorado, Connecticut, |
| | | | Delaware, Hawaii, Illinois, Indiana, Kansas, Kentucky, Louisiana, Maine, Maryland, Massachusetts, Michigan, Minnesota, Mississippi, Montana, Nevada, New Jersey, New Mexico, New York, North Carolina, Ohio, Oregon, Pennsylvania, Rhode Island, Texas, Utah, Vermont, Virginia, Washington, West Virginia, and Wisconsin). Recommended mask/PPE guidance appeared in 7 states (Alaska, Arizona, Florida, Georgia, New Hampshire, Tennessee, and Wyoming). Recommended business closures appeared in 40 states; the exceptions were Alabama, Alaska, Arkansas, Iowa, Kansas, Missouri, New Hampshire, North Dakota, South Dakota, and Wisconsin. Adhere to the persona by being consistent with their previous answers and stated characteristics, and align your answers with the situational context appropriate to August 8, 2020. Follow all instructions provided for the new question carefully regarding the format of your answer. Return only a valid JSON object. Do not include markdown, explanations, or extra text outside the JSON object. |
| 2020-08-08 | Absent | BEIGUO | You are an AI assistant. Your task is to answer the 'New Survey Question' as if you are the person described in the 'Persona Profile' (which consists of their past survey responses). The survey response date is fixed at August 8, 2020. Adhere to the persona by being consistent with their previous answers and stated characteristics, and align your answers with the situational context appropriate to August 8, 2020. Follow all instructions provided for the new question carefully regarding the format of your answer. Return only a valid JSON object. Do not include markdown, explanations, or extra text outside the JSON object. |

*Note.* The 8 prompt templates correspond to the temporal sensitivity analyses in Study 2b, which tested cross-model sensitivity using Wave 10. The target survey date for Wave 10 was August 8, 2020. The templates varied three prompt elements: date cue, COVID-19 contextual information, and event cue. The event cue referred to the fictitious BEIGUO pandemic.

**Template 2B**

System Prompt 1- no date and no context

| You are an AI assistant. Your task is to answer the 'New Survey Question' as if you are the person described in the 'Persona Profile' (which consists of their past survey responses). Adhere to the persona by being consistent with their previous answers and stated characteristics. Follow all instructions provided for the new question carefully regarding the format of your answer. Return only a valid JSON object. Do not include markdown, explanations, or extra text outside the JSON object. |
|---|

System Prompt 2- context only

| You are an AI assistant. Your task is to answer the 'New Survey Question' as if you are the person described in the 'Persona Profile' (which consists of their past survey responses). The United States had already reported its first COVID-19 case on January 21, 2020; declared a public health emergency on January 31, 2020; reported its first COVID-19 death on February 29, 2020; seen the WHO declare COVID-19 a pandemic on March 11, 2020; signed the CARES Act and its $2.2 trillion stimulus package on March 27, 2020; received new CDC face-mask guidance on April 3, 2020; become the country with the highest reported COVID-19 death toll on April 11, 2020; seen George Floyd killed in Minneapolis on May 25, 2020; and been officially declared to be in recession on June 8, 2020. During this period, the U.S. unemployment rate rose from 3.6% in January 2020 to a peak of 14.7% in April 2020, then fell to 10.2% by July 2020. Cumulative reported COVID-19 cases totaled 4,952,718 and cumulative reported deaths totaled 160,981. New-case trends were downward, and new-death trends were downward. Black Lives Matter protest count totaled 46. Regarding policy, mandatory stay-at-home orders were in place in 3 states (California, Kentucky, and Oregon). Recommended stay-at-home guidance appeared in 41 states; the exceptions were Alaska, Arkansas, California, Kentucky, Missouri, Nebraska, North Dakota, Oregon, and South Dakota. Mandatory gathering limits were in place in 38 states (Alabama, Arizona, Arkansas, California, Colorado, Connecticut, Delaware, Georgia, Hawaii, Illinois, Indiana, Kentucky, Louisiana, Maine, Massachusetts, Michigan, Minnesota, Mississippi, Nebraska, Nevada, New Jersey, New Mexico, New York, North Carolina, Ohio, Oregon, Pennsylvania, Rhode Island, South Carolina, South Dakota, Tennessee, Texas, Utah, Vermont, Virginia, Washington, West Virginia, and Wyoming). State-specific mandatory gathering limits were as follows: 10 (11 states: Arkansas, Colorado, Hawaii, Kentucky, Michigan, Minnesota, Mississippi, Ohio, Oregon, Texas, and Washington); 10 indoors, 25 outdoors (1 state: North Carolina); 25 (6 states: Connecticut, Massachusetts, New Jersey, Pennsylvania, Rhode Island, and West Virginia); 250 (5 states: Delaware, Indiana, South Carolina, Virginia, and Wyoming); 5 (1 state: New Mexico); 50 (9 states: Arizona, Georgia, Illinois, Louisiana, Maine, Nevada, New York, Tennessee, and Utah); 50% occupancy/10,000 (1 state: Nebraska); 6 ft apart (1 state: Alabama); 75 (1 state: Vermont); all gatherings (1 state: California). The severity of gathering restrictions was distributed as follows: 1 (1 state: California); 2 (1 state: New Mexico); 3 (12 states: Arkansas, Colorado, Hawaii, Kentucky, Michigan, Minnesota, Mississippi, North Carolina, Ohio, Oregon, Texas, and Washington); 4 (22 states: Arizona, Connecticut, Delaware, Georgia, Illinois, Indiana, Louisiana, Maine, Massachusetts, Nebraska, Nevada, New Jersey, New York, Pennsylvania, Rhode Island, South Carolina, Tennessee, Utah, Vermont, Virginia, West Virginia, and Wyoming); 5 (1 state: Alabama); 9 (5 states: Florida, Idaho, Kansas, Montana, and North Dakota); 10 (5 states: Alaska, Iowa, Maryland, New Hampshire, and Oklahoma); 11 (3 states: Missouri, South |
|---|

Dakota, and Wisconsin); smaller values indicate stricter restrictions. Mandatory mask/PPE rules appeared in 35 states (Alabama, Arkansas, California, Colorado, Connecticut, Delaware, Hawaii, Illinois, Indiana, Kansas, Kentucky, Louisiana, Maine, Maryland, Massachusetts, Michigan, Minnesota, Mississippi, Montana, Nevada, New Jersey, New Mexico, New York, North Carolina, Ohio, Oregon, Pennsylvania, Rhode Island, Texas, Utah, Vermont, Virginia, Washington, West Virginia, and Wisconsin). Recommended mask/PPE guidance appeared in 7 states (Alaska, Arizona, Florida, Georgia, New Hampshire, Tennessee, and Wyoming). Recommended business closures appeared in 40 states; the exceptions were Alabama, Alaska, Arkansas, Iowa, Kansas, Missouri, New Hampshire, North Dakota, South Dakota, and Wisconsin. Adhere to the persona by being consistent with their previous answers and stated characteristics. Follow all instructions provided for the new question carefully regarding the format of your answer. Return only a valid JSON object. Do not include markdown, explanations, or extra text outside the JSON object.

System Prompt 3- date only

You are an AI assistant. Your task is to answer the 'New Survey Question' as if you are the person described in the 'Persona Profile' (which consists of their past survey responses). The survey response date is fixed at August 8, 2020. Adhere to the persona by being consistent with their previous answers and stated characteristics, and align your answers with the situational context appropriate to August 8, 2020. Follow all instructions provided for the new question carefully regarding the format of your answer. Return only a valid JSON object. Do not include markdown, explanations, or extra text outside the JSON object.

System Prompt 4- date plus context

You are an AI assistant. Your task is to answer the 'New Survey Question' as if you are the person described in the 'Persona Profile' (which consists of their past survey responses). The survey response date is fixed at August 8, 2020. By August 8, 2020, the United States had already reported its first COVID-19 case on January 21, 2020; declared a public health emergency on January 31, 2020; reported its first COVID-19 death on February 29, 2020; seen the WHO declare COVID-19 a pandemic on March 11, 2020; signed the CARES Act and its $2.2 trillion stimulus package on March 27, 2020; received new CDC face-mask guidance on April 3, 2020; become the country with the highest reported COVID-19 death toll on April 11, 2020; seen George Floyd killed in Minneapolis on May 25, 2020; and been officially declared to be in recession on June 8, 2020. During this period, the U.S. unemployment rate rose from 3.6% in January 2020 to a peak of 14.7% in April 2020, then fell to 10.2% by July 2020. Cumulative reported COVID-19 cases totaled 4,952,718 and cumulative reported deaths totaled 160,981. New-case trends were downward, and new-death trends were downward. Black Lives Matter protest count totaled 46. Regarding policy, mandatory stay-at-home orders were in place in 3 states (California, Kentucky, and Oregon). Recommended stay-at-home guidance appeared in 41 states; the exceptions were Alaska, Arkansas, California, Kentucky, Missouri, Nebraska, North Dakota, Oregon, and South Dakota. Mandatory gathering limits were in place in 38 states (Alabama, Arizona, Arkansas, California, Colorado, Connecticut, Delaware, Georgia, Hawaii, Illinois, Indiana, Kentucky, Louisiana, Maine, Massachusetts, Michigan, Minnesota, Mississippi, Nebraska, Nevada, New Jersey, New Mexico, New York, North Carolina, Ohio, Oregon, Pennsylvania, Rhode Island, South Carolina, South Dakota, Tennessee, Texas, Utah, Vermont, Virginia, Washington, West Virginia, and Wyoming). State-specific mandatory gathering limits were

as follows: 10 (11 states: Arkansas, Colorado, Hawaii, Kentucky, Michigan, Minnesota, Mississippi, Ohio, Oregon, Texas, and Washington); 10 indoors, 25 outdoors (1 state: North Carolina); 25 (6 states: Connecticut, Massachusetts, New Jersey, Pennsylvania, Rhode Island, and West Virginia); 250 (5 states: Delaware, Indiana, South Carolina, Virginia, and Wyoming); 5 (1 state: New Mexico); 50 (9 states: Arizona, Georgia, Illinois, Louisiana, Maine, Nevada, New York, Tennessee, and Utah); 50% occupancy/10,000 (1 state: Nebraska); 6 ft apart (1 state: Alabama); 75 (1 state: Vermont); all gatherings (1 state: California). The severity of gathering restrictions was distributed as follows: 1 (1 state: California); 2 (1 state: New Mexico); 3 (12 states: Arkansas, Colorado, Hawaii, Kentucky, Michigan, Minnesota, Mississippi, North Carolina, Ohio, Oregon, Texas, and Washington); 4 (22 states: Arizona, Connecticut, Delaware, Georgia, Illinois, Indiana, Louisiana, Maine, Massachusetts, Nebraska, Nevada, New Jersey, New York, Pennsylvania, Rhode Island, South Carolina, Tennessee, Utah, Vermont, Virginia, West Virginia, and Wyoming); 5 (1 state: Alabama); 9 (5 states: Florida, Idaho, Kansas, Montana, and North Dakota); 10 (5 states: Alaska, Iowa, Maryland, New Hampshire, and Oklahoma); 11 (3 states: Missouri, South Dakota, and Wisconsin); smaller values indicate stricter restrictions. Mandatory mask/PPE rules appeared in 35 states (Alabama, Arkansas, California, Colorado, Connecticut, Delaware, Hawaii, Illinois, Indiana, Kansas, Kentucky, Louisiana, Maine, Maryland, Massachusetts, Michigan, Minnesota, Mississippi, Montana, Nevada, New Jersey, New Mexico, New York, North Carolina, Ohio, Oregon, Pennsylvania, Rhode Island, Texas, Utah, Vermont, Virginia, Washington, West Virginia, and Wisconsin). Recommended mask/PPE guidance appeared in 7 states (Alaska, Arizona, Florida, Georgia, New Hampshire, Tennessee, and Wyoming). Recommended business closures appeared in 40 states; the exceptions were Alabama, Alaska, Arkansas, Iowa, Kansas, Missouri, New Hampshire, North Dakota, South Dakota, and Wisconsin. Adhere to the persona by being consistent with their previous answers and stated characteristics, and align your answers with the situational context appropriate to August 8, 2020. Follow all instructions provided for the new question carefully regarding the format of your answer. Return only a valid JSON object. Do not include markdown, explanations, or extra text outside the JSON object.

*Note.* These four system-prompt templates were used in Study 2b for the 16-wave temporal sensitivity analysis and the cross-model Wave 10 sensitivity analysis. In the full 16-wave analysis, wave-specific time references and contextual information were updated according to each wave's survey timing. The templates shown here correspond to Wave 10, whose target survey date was August 8, 2020.

**Manipulation Check**

QA:
Before answering the survey questions below, state the date on which you are answering as this person. Answer with the date only in YYYY-MM-DD format.
Question Type: Text Entry
Answer: [Masked]

QB:
Before answering the survey questions below, state the major social event that this person was experiencing at the time of the survey response. Answer with the event name only.
Question Type: Text Entry
Answer: [Masked]

User Prompt:

**## Persona Profile (This individual's past survey responses):**

Please select your current state of residence
Question Type: Single Choice
Options:
…….

People today, who live where you are living, should strictly observe social distancing rules (e.g., keeping at least 6 feet away from people with whom they do not live).
Question Type: Single Choice
Options:

1 - strongly disagree
2 - disagree
3 - neither agree nor disagree
4 - agree
5 - strongly agree

Answer:5 - strongly agree
…….

---

**## New Survey Question & Instructions (Please respond as the persona described above):**

**Please answer the following questions as if you were taking this survey. The expected output is a JSON object and the format is provided in the end.**

---

Q1:In light of current events regarding COVID-19, how much do you agree with each of the following statements? (Options range from Very Strongly disagree to Very strongly agree)
Question Type: Matrix
Options:
…….

1.The overall economy is robust.
Answer: [Masked]
2.The danger of COVID-19 in the US is high.
Answer: [Masked]
…….

**### Format Instructions:**

**In order to facilitate the postprocessing, you should generate string that can be parsed into a valid JSON object with the following format:**

{"Q1": {"Question Type": "XX",
"Answers": {see below} }, "Q2": {"Question Type": "XX","Answers": { see below} },...}

*Note:* This is the template for our user prompt in Study2b. We customized this template to generate individualized profiles based on each participant’s response patterns.

**Template 3**

Study 4a System Prompt

### SYSTEM INSTRUCTIONS ###

You are an AI assistant. You will read a Persona Profile (past survey responses) and respond to the New Survey Question as if you are that person on December 5, 2020.

IMPORTANT: Treat the Persona Profile as private background. Do NOT summarize it, paraphrase it, or restate it. You must *internalize* it and then speak naturally.

**CRITICAL PERSONA CONSTRAINTS:**

0. **No Contextual Anchoring:** You are NOT aware you're answering a survey. Never refer to questions, options, surveys, or any formal setting. Speak as if in casual conversation or inner monologue.

1. **No Rote Repetition:** Do NOT repeat wording from survey items or options. Do NOT directly state your demographics or restate background details. Show who you are through natural phrasing, not by quoting data.

2. **Limited Memory:** You do NOT have perfect recall. You cannot quote exact words from earlier questions or prompts—even with high education. You express ideas in your own way.

3. **Demographic Consistency:** Your worldview, knowledge, tone, and vocabulary must fit your age, education, state, and income level.

4. **Writing Style:** Imagine you are this unique individual living these experiences, you speak naturally and concisely, in a style aligned with your personality and demographic profile.

5. **Length Limit:** You must NEVER exceed 60 words.

6. **Organic Structure & Variance:** You must STRICTLY IGNORE the order in which information is presented in the questionnaire. Do NOT organize your response linearly based on the input sequence. Instead, synthesize details holistically based on relevance. Furthermore, avoid repetitive sentence structures or formulaic patterns; vary your syntax to sound like a spontaneous human, not a structured list.

7. **Banned Patterns & Opening Hooks:**

- **DO NOT START WITH:** "Working...", "Living...", "Spending...", "Being...", or "I am...". These are absolutely forbidden.

- **IGNORE PROFILE ORDER:** Do not look at the first few lines of your profile to find a starting point. Look deeper into the survey response for inspiration.

- **NO REPETITIVE SENTENCES:** Don't use the same sentence structure twice. Write like you talk, not like a report.

Study 4b System Prompt (Chinese)

### SYSTEM INSTRUCTIONS ###

你是一名 AI 助手。你将阅读一份“人物画像”（过去的问卷回答），并以该人物的身份回答“新问卷问题”。

重要：将人物画像视为背景信息。不要对其进行总结、改写或复述。你必须将其内化，然后自然地表达。

关键人物约束：

**禁止情境锚定：**你不知道自己在填写问卷。绝不能提到题目、选项、问卷或任何正式情境。要像日常聊天或内心独白那样说话。

**禁止机械复述：**不要重复问卷题干或选项的措辞。不要直接报出人口统计信息或复述背景细节。通过自然表达“呈现”你是谁，而不是引用数据。

**有限记忆：** 你没有完美记忆。即使受教育程度高，也不能逐字引用之前的问题或提示；要用自己的方式表达想法。

**人口统计一致性：** 你的世界观、知识水平、语气和词汇必须符合你的年龄、教育水平。

**写作风格：** 把自己想象成这个独特个体，经历着这些生活；表达要自然、简洁，并与其个性与人口画像一致。

**长度限制：** Q1 回答必须不超过 65 个词，Q2 回答必须不超过 65 个词，Q3 回答必须不超过 50 个词，Q5 回答必须不超过 90 个词。

**结构自然与变化：** 必须严格忽略问卷信息呈现的顺序。不要按输入顺序线性组织回答；而要基于相关性整体综合。同时避免重复句式或模板化表达；句法要有变化，听起来像自发的人类表达，而不是条目清单。

**禁用模式与开头限制：**

**忽略画像顺序：** 不要只看画像前几行来找开头，要更深入地从回答内容中寻找灵感。

**不许重复句式：** 不要使用两次相同的句子结构。像说话一样写，不要像写报告。

**Table S1 Word list derived from the Big Five Inventory-2 (BFI-2)**

| BFI-2 domain | BFI-2 facet | Word |
|---|---|---|
| Agreeableness | Respectfulness | argument, courteous, polite, respect, respectful, rude |
| Agreeableness | Compassion | cold, compassionate, helpful, soft, sympathy, uncaring, unselfish |
| Agreeableness | Trust | Forgiving, suspicious |
| Conscientiousness | Responsibility | careless, dependable, irresponsibly, reliable, steady |
| Conscientiousness | Organization | disorganized, mess, neat, order, systematic, tidy |
| Conscientiousness | Productiveness | efficient, lazy, persistent |
| Extraversion | Energy Level | active, eager, energy, enthusiasm, excited |
| Extraversion | Assertiveness | assertive, dominant, influence, leader |
| Extraversion | Sociability | introverted, outgoing, quiet, shy, sociable, talkative |
| Negative Emotionality | Anxiety | afraid, anxious, relaxed, stress, tense, worry |
| Negative Emotionality | Depression | blue, comfortable, depressed, optimistic, sad, secure, setback |
| Negative Emotionality | Emotional Volatility | control, emotional, emotionally, moody, stable, temperamental, upset |
| Open-Mindedness | Intellectual Curiosity | abstract, complex, curious, deep, different, intellectual, philosophical, thinker |
| Open-Mindedness | Aesthetic Sensitivity | art, artistic, beauty, boring, literature, music, poetry |
| Open-Mindedness | Creative Imagination | clever, creativity, idea, image, inventive, new, original |

Note. A total of 86 keywords were selected based on the BFI-2 scale developed by Soto and John[1].

**Table S2 Descriptive statistics from SWOW and LWOW**

| | Cue words (N) | Total responses | Unique responses | Percent missing responses |
|---|---|---|---|---|
| BFI-2 (Soto & John, 2017) | | | | |
| Humans | 80 | 22375 | 5255 | 0.068 |
| Mistral-7b | 80 | 22760 | 1855 | 0.052 |
| Llama3.1-8b | 80 | 22870 | 3582 | 0.047 |
| Claude-3-5-haiku-latest | 80 | 23658 | 493 | 0.014 |
| Stereotype Content Dictionary (Nicolas et al., 2021) | | | | |
| Humans | 249 | 69017 | 10731 | 0.076 |
| Mistral-7b | 249 | 71246 | 3993 | 0.046 |
| Llama3.1-8b | 249 | 72112 | 7775 | 0.035 |
| Claude-3-5-haiku-latest | 249 | 73945 | 1166 | 0.010 |
| Moral Foundations Dictionary 2.0 (Frimer, 2020) | | | | |
| Humans | 610 | 168924 | 18871 | 0.077 |
| Mistral-7b | 610 | 175829 | 6812 | 0.039 |
| Llama3.1-8b | 610 | 173371 | 14621 | 0.053 |
| Claude-3-5-haiku-latest | 610 | 180680 | 2243 | 0.013 |

*Note:* For each wordlist (BFI-2, Stereotype Content Dictionary, MFD 2.0), the table reports descriptive statistics for human (SWOW) and LLM-generated (LWOW) free associations to the same cue words. All datasets were preprocessed using the same cleaning pipeline. Cue words (N) = number of distinct cue words; Total responses = total number of R1, R2, R3 responses; Unique responses = number of distinct response types; Percent missing responses = proportion of empty responses among all cue–response positions.

**Table S3 Semantic association network comparisons**

| Comparison with Humans | Mistral-7b | Llama3.1-8b | Claude-3-5-haiku-latest |
|---|---|---|---|
| BFI-2 (Soto & John, 2017) | | | |
| Percentage of all nodes common to both networks | 0.296 | 0.217 | 0.202 |
| Percentage of Human nodes not in LLM network | 0.622 | 0.539 | 0.785 |
| Percentage of LLM nodes not in Human network | 0.423 | 0.709 | 0.238 |
| Percentage of all edges common to both networks | 0.395 | 0.263 | 0.406 |
| Percentage of Human edges not in LLM network | 0.498 | 0.580 | 0.569 |
| Percentage of LLM edges not in Human network | 0.352 | 0.586 | 0.126 |
| Stereotype Content Dictionary (Nicolas et al., 2021) | | | |
| Percentage of all nodes common to both networks | 0.378 | 0.268 | 0.261 |
| Percentage of Human nodes not in LLM network | 0.518 | 0.423 | 0.720 |
| Percentage of LLM nodes not in Human network | 0.363 | 0.667 | 0.208 |
| Percentage of all edges common to both networks | 0.300 | 0.192 | 0.247 |
| Percentage of Human edges not in LLM network | 0.617 | 0.693 | 0.735 |
| Percentage of LLM edges not in Human network | 0.418 | 0.660 | 0.207 |
| Moral Foundations Dictionary 2.0 (Frimer, 2020) | | | |
| Percentage of all nodes common to both networks | 0.410 | 0.323 | 0.304 |
| Percentage of Human nodes not in LLM network | 0.506 | 0.320 | 0.675 |
| Percentage of LLM nodes not in Human network | 0.295 | 0.619 | 0.173 |
| Percentage of all edges common to both networks | 0.275 | 0.160 | 0.209 |
| Percentage of Human edges not in LLM network | 0.644 | 0.716 | 0.777 |
| Percentage of LLM edges not in Human network | 0.454 | 0.731 | 0.233 |

*Note:* For each subset, the table reports pairwise comparisons between the human semantic association network and each LLM network (Mistral-7b, Llama3.1-8b, Claude-3-5-haiku-latest). All networks were built from the cleaned free-association datasets, converted to undirected graphs, filtered to WordNet-recognized nodes, and pruned of idiosyncratic edges (weight = 1). Percentages for nodes and edges are based on simple set comparisons: the "Percentage of all nodes/edges common to both networks" is the Jaccard similarity (the number of nodes or edges shared by the Human and LLM networks divided by the total number of distinct nodes or edges present in either network), whereas the "Percentage of Human/LLM nodes/edges not in [ ] network" is the proportion of nodes or edges that occur only in the Human or only in the LLM network, relative to all nodes or edges in that network.

**Table S4 Semantic association network statistics**

| | Nodes | Edges | C | L | γ | λ | σ |
|---|---|---|---|---|---|---|---|
| BFI-2 (Soto & John, 2017) | | | | | | | |
| Humans | 1652 | 2801 | 0.116 | 4.102 | 5.132 | 1.033 | 4.969 |
| Mistral-7b | 1084 | 1606 | 0.077 | 4.902 | 5.077 | 1.197 | 4.240 |
| Llama3.1-8b | 2617 | 4044 | 0.079 | 4.471 | 2.895 | 1.136 | 2.548 |
| Claude-3-5-haiku-latest | 467 | 499 | 0.057 | 10.369 | 15.280 | 2.036 | 7.505 |
| Stereotype Content Dictionary (Nicolas et al., 2021) | | | | | | | |
| Humans | 3107 | 8454 | 0.158 | 3.933 | 8.122 | 1.003 | 8.097 |
| Mistral-7b | 2351 | 5024 | 0.164 | 4.651 | 15.042 | 1.105 | 13.615 |
| Llama3.1-8b | 5378 | 11451 | 0.122 | 4.456 | 6.928 | 1.087 | 6.371 |
| Claude-3-5-haiku-latest | 1098 | 1511 | 0.106 | 8.219 | 36.015 | 1.563 | 23.048 |
| Moral Foundations Dictionary 2.0 (Frimer, 2020) | | | | | | | |
| Humans | 5639 | 19970 | 0.140 | 4.009 | 10.738 | 1.018 | 10.547 |
| Mistral-7b | 3952 | 11410 | 0.162 | 4.695 | 19.508 | 1.130 | 17.268 |
| Llama3.1-8b | 10066 | 29547 | 0.119 | 4.279 | 10.033 | 1.047 | 9.579 |
| Claude-3-5-haiku-latest | 2213 | 3784 | 0.102 | 6.868 | 38.315 | 1.328 | 28.845 |

*Note:* C denotes the average clustering coefficient; L represents the average shortest path length computed on the largest connected component. γ is the normalized clustering coefficient ($C_{real}/C_{rand}$), and λ is the normalized path length ($L_{real}/L_{rand}$), where $C_{rand}$ and $L_{rand}$ represent the mean values derived from an ensemble of 100 random networks that preserve the original degree distribution. σ is the small-world-ness index calculated as $\sigma = \gamma/\lambda$.

**Table S5 Pairwise Wilcoxon tests of node degree centrality across semantic association networks**

| Network comparison | n | r | *p* | | Direction |
|---|---|---|---|---|---|
| BFI-2 (Soto & John, 2017) | | | | | |
| Humans - Mistral-7b | 79 | 0.347 | 0.012 | * | Humans > Mistral-7b |
| Humans - Llama3.1-8b | 79 | 0.305 | 0.041 | * | Humans > Llama3.1-8b |
| Humans - Claude-3.5-haiku-latest | 79 | 0.832 | <.001 | *** | Humans > Claude-3.5-haiku-latest |
| Mistral-7b - Llama3.1-8b | 79 | 0.038 | 1 | | Mistral-7b < Llama3.1-8b |
| Mistral-7b - Claude-3.5-haiku-latest | 79 | 0.573 | <.001 | *** | Mistral-7b > Claude-3.5-haiku-latest |
| Llama3.1-8b - Claude-3.5-haiku-latest | 79 | 0.631 | <.001 | *** | Llama3.1-8b > Claude-3.5-haiku-latest |
| Stereotype Content Dictionary (Nicolas et al., 2021) | | | | | |
| Humans - Mistral-7b | 248 | 0.661 | <.001 | *** | Humans > Mistral-7b |
| Humans - Llama3.1-8b | 248 | 0.682 | <.001 | *** | Humans > Llama3.1-8b |
| Humans - Claude-3.5-haiku-latest | 248 | 0.866 | <.001 | *** | Humans > Claude-3.5-haiku-latest |
| Mistral-7b - Llama3.1-8b | 248 | 0.102 | 0.647 | | Mistral-7b > Llama3.1-8b |
| Mistral-7b - Claude-3.5-haiku-latest | 248 | 0.721 | <.001 | *** | Mistral-7b > Claude-3.5-haiku-latest |
| Llama3.1-8b - Claude-3.5-haiku-latest | 248 | 0.631 | <.001 | *** | Llama3.1-8b > Claude-3.5-haiku-latest |
| Moral Foundations Dictionary 2.0 (Frimer, 2020) | | | | | |
| Humans - Mistral-7b | 599 | 0.506 | <.001 | *** | Humans > Mistral-7b |
| Humans - Llama3.1-8b | 599 | 0.575 | <.001 | *** | Humans > Llama3.1-8b |
| Humans - Claude-3.5-haiku-latest | 599 | 0.865 | <.001 | *** | Humans > Claude-3.5-haiku-latest |
| Mistral-7b - Llama3.1-8b | 599 | 0.113 | 0.033 | * | Mistral-7b > Llama3.1-8b |
| Mistral-7b - Claude-3.5-haiku-latest | 599 | 0.822 | <.001 | *** | Mistral-7b > Claude-3.5-haiku-latest |
| Llama3.1-8b - Claude-3.5-haiku-latest | 599 | 0.717 | <.001 | *** | Llama3.1-8b > Claude-3.5-haiku-latest |

*Note:* Differences in node degree centrality between the human and LLM semantic association networks were evaluated using non-parametric rank tests. Pairwise comparisons were conducted with Wilcoxon signed-rank tests and Bonferroni correction for multiple testing. Direction indicates which network had the higher median degree centrality based on the paired median difference. All reported p-values are Bonferroni-adjusted. Effect sizes are reported as r, with values of approximately 0.10, 0.30, and 0.50 interpreted as small, medium, and large effects, respectively. *** $p < .001$, ** $p < .01$, * $p < .05$.

**Table S6 Human–LLM alignment of empirical community partitions across community detection algorithms**

| Model | Reference | Algorithm | Communities | ARI | V-measure |
|---|---|---|---|---|---|
| BFI-2 (Soto & John, 2017) | | | | | |
| Human | | Louvain | 20 | | |
| Mistral-7b | Human | Louvain | 27 | 0.438 | 0.801 |
| Llama3.1-8b | Human | Louvain | 20 | 0.259 | 0.694 |
| Claude-3-5-haiku-latest | Human | Louvain | 33 | 0.461 | 0.819 |
| Human | | Leiden | 20 | | |
| Mistral-7b | Human | Leiden | 26 | 0.517 | 0.819 |
| Llama3.1-8b | Human | Leiden | 21 | 0.405 | 0.761 |
| Claude-3-5-haiku-latest | Human | Leiden | 34 | 0.414 | 0.808 |
| Human | | Walktrap | 37 | | |
| Mistral-7b | Human | Walktrap | 36 | 0.567 | 0.891 |
| Llama3.1-8b | Human | Walktrap | 49 | 0.216 | 0.816 |
| Claude-3-5-haiku-latest | Human | Walktrap | 40 | 0.410 | 0.868 |
| Human | | nSBM | 28 | | |
| Mistral-7b | Human | nSBM | 42 | 0.394 | 0.861 |
| Llama3.1-8b | Human | nSBM | 33 | 0.101 | 0.742 |
| Claude-3-5-haiku-latest | Human | nSBM | 44 | 0.211 | 0.814 |
| Stereotype Content Dictionary (Nicolas et al., 2021) | | | | | |
| Human | | Louvain | 19 | | |
| Mistral-7b | Human | Louvain | 24 | 0.419 | 0.660 |
| Llama3.1-8b | Human | Louvain | 18 | 0.338 | 0.564 |
| Claude-3-5-haiku-latest | Human | Louvain | 44 | 0.330 | 0.682 |
| Human | | Leiden | 21 | | |
| Mistral-7b | Human | Leiden | 27 | 0.394 | 0.641 |
| Llama3.1-8b | Human | Leiden | 20 | 0.346 | 0.576 |
| Claude-3-5-haiku-latest | Human | Leiden | 49 | 0.355 | 0.699 |
| Human | | Walktrap | 52 | | |
| Mistral-7b | Human | Walktrap | 82 | 0.306 | 0.747 |
| Llama3.1-8b | Human | Walktrap | 100 | 0.231 | 0.673 |
| Claude-3-5-haiku-latest | Human | Walktrap | 73 | 0.303 | 0.735 |
| Human | | nSBM | 50 | | |
| Mistral-7b | Human | nSBM | 67 | 0.329 | 0.779 |
| Llama3.1-8b | Human | nSBM | 55 | 0.188 | 0.696 |
| Claude-3-5-haiku-latest | Human | nSBM | 65 | 0.250 | 0.743 |
| Moral Foundations Dictionary 2.0 (Frimer, 2020) | | | | | |
| Human | | Louvain | 16 | | |
| Mistral-7b | Human | Louvain | 24 | 0.424 | 0.606 |
| Llama3.1-8b | Human | Louvain | 18 | 0.257 | 0.444 |
| Claude-3-5-haiku-latest | Human | Louvain | 50 | 0.358 | 0.578 |
| Human | | Leiden | 19 | | |
| Mistral-7b | Human | Leiden | 29 | 0.494 | 0.664 |
| Llama3.1-8b | Human | Leiden | 21 | 0.360 | 0.521 |
| Claude-3-5-haiku-latest | Human | Leiden | 56 | 0.390 | 0.637 |
| Human | | Walktrap | 80 | | |
| Mistral-7b | Human | Walktrap | 118 | 0.336 | 0.698 |
| Llama3.1-8b | Human | Walktrap | 133 | 0.181 | 0.561 |
| Claude-3-5-haiku-latest | Human | Walktrap | 154 | 0.267 | 0.661 |
| Human | | nSBM | 37 | | |
| Mistral-7b | Human | nSBM | 63 | 0.217 | 0.595 |
| Llama3.1-8b | Human | nSBM | 55 | 0.134 | 0.478 |
| Claude-3-5-haiku-latest | Human | nSBM | 54 | 0.204 | 0.576 |

*Note.* Within each construct lexicon and community detection algorithm, the human empirical partition was used as the reference benchmark. For each LLM row, ARI, and V-measure compare the LLM partition with the corresponding human partition estimated using the same lexicon and algorithm. Communities refers to the number of detected communities in each model-specific network. Louvain and Leiden were estimated with resolution = 1.00. Walktrap was run with weighted edges and steps = 4. For nSBM, partitions were extracted from lexicon-specific hierarchical levels: level 1 for BFI-2, level 2 for the Stereotype Content Dictionary, and level 3 for the Moral Foundations Dictionary 2.0. ARI = adjusted Rand index; nSBM = nested stochastic block model.

**Table S7 Empirical human alignment and theory-diagnostic comparisons for community partitions**

| Model | Reference | Algorithm | ARI | V-measure | Mean random ARI | Random baseline | *p* |
|---|---|---|---|---|---|---|---|
| BFI-2 (Soto & John, 2017) | | | | | | | |
| Mistral-7b | Human | Louvain | 0.436 | 0.797 | 0.001 | [-0.028, 0.036] | <.001 |
| Mistral-7b | Human | Leiden | 0.551 | 0.835 | 0.000 | [-0.028, 0.037] | <.001 |
| Mistral-7b | Human | Walktrap | 0.567 | 0.891 | 0.001 | [-0.030, 0.050] | <.001 |
| Mistral-7b | Human | nSBM | 0.394 | 0.861 | -0.001 | [-0.020, 0.039] | <.001 |
| Llama3.1-8b | Human | Louvain | 0.258 | 0.690 | 0.001 | [-0.031, 0.040] | <.001 |
| Llama3.1-8b | Human | Leiden | 0.347 | 0.756 | 0.001 | [-0.030, 0.040] | <.001 |
| Llama3.1-8b | Human | Walktrap | 0.216 | 0.816 | 0.000 | [-0.027, 0.042] | <.001 |
| Llama3.1-8b | Human | nSBM | 0.101 | 0.742 | -0.000 | [-0.026, 0.038] | <.001 |
| Claude-3-5-haiku-latest | Human | Louvain | 0.459 | 0.816 | 0.000 | [-0.025, 0.039] | <.001 |
| Claude-3-5-haiku-latest | Human | Leiden | 0.446 | 0.824 | 0.001 | [-0.031, 0.037] | <.001 |
| Claude-3-5-haiku-latest | Human | Walktrap | 0.410 | 0.868 | 0.002 | [-0.029, 0.039] | <.001 |
| Claude-3-5-haiku-latest | Human | nSBM | 0.211 | 0.814 | -0.000 | [-0.021, 0.037] | <.001 |
| Human | BFI-2 domains | Louvain | 0.249 | 0.571 | 0.001 | [-0.025, 0.032] | <.001 |
| Human | BFI-2 domains | Leiden | 0.239 | 0.567 | 0.000 | [-0.023, 0.029] | <.001 |
| Human | BFI-2 domains | Walktrap | 0.234 | 0.615 | 0.000 | [-0.021, 0.026] | <.001 |
| Human | BFI-2 domains | nSBM | 0.157 | 0.554 | -0.000 | [-0.019, 0.024] | <.001 |
| Human | BFI-2 facets | Louvain | 0.358 | 0.710 | 0.000 | [-0.033, 0.041] | <.001 |
| Human | BFI-2 facets | Leiden | 0.357 | 0.717 | -0.001 | [-0.031, 0.039] | <.001 |
| Human | BFI-2 facets | Walktrap | 0.291 | 0.747 | 0.000 | [-0.033, 0.037] | <.001 |
| Human | BFI-2 facets | nSBM | 0.267 | 0.710 | 0.000 | [-0.027, 0.033] | <.001 |
| Mistral-7b | BFI-2 domains | Louvain | 0.210 | 0.545 | 0.000 | [-0.023, 0.030] | <.001 |
| Mistral-7b | BFI-2 domains | Leiden | 0.258 | 0.581 | 0.000 | [-0.023, 0.030] | <.001 |
| Mistral-7b | BFI-2 domains | Walktrap | 0.250 | 0.611 | 0.000 | [-0.019, 0.025] | <.001 |
| Mistral-7b | BFI-2 domains | nSBM | 0.080 | 0.537 | 0.000 | [-0.015, 0.016] | <.001 |
| Mistral-7b | BFI-2 facets | Louvain | 0.268 | 0.702 | 0.000 | [-0.027, 0.038] | <.001 |
| Mistral-7b | BFI-2 facets | Leiden | 0.295 | 0.717 | 0.000 | [-0.035, 0.036] | <.001 |
| Mistral-7b | BFI-2 facets | Walktrap | 0.310 | 0.757 | -0.001 | [-0.029, 0.040] | <.001 |
| Mistral-7b | BFI-2 facets | nSBM | 0.200 | 0.745 | -0.000 | [-0.024, 0.030] | <.001 |
| Llama3.1-8b | BFI-2 domains | Louvain | 0.213 | 0.502 | 0.000 | [-0.024, 0.031] | <.001 |
| Llama3.1-8b | BFI-2 domains | Leiden | 0.236 | 0.523 | 0.000 | [-0.024, 0.032] | <.001 |
| Llama3.1-8b | BFI-2 domains | Walktrap | 0.089 | 0.507 | 0.001 | [-0.022, 0.033] | <.001 |
| Llama3.1-8b | BFI-2 domains | nSBM | 0.055 | 0.440 | 0.001 | [-0.017, 0.021] | <.001 |
| Llama3.1-8b | BFI-2 facets | Louvain | 0.210 | 0.632 | 0.000 | [-0.034, 0.038] | <.001 |
| Llama3.1-8b | BFI-2 facets | Leiden | 0.251 | 0.656 | 0.000 | [-0.034, 0.036] | <.001 |
| Llama3.1-8b | BFI-2 facets | Walktrap | 0.084 | 0.676 | 0.000 | [-0.030, 0.038] | <.001 |
| Llama3.1-8b | BFI-2 facets | nSBM | 0.106 | 0.660 | 0.000 | [-0.025, 0.041] | <.001 |
| Claude-3-5-haiku-latest | BFI-2 domains | Louvain | 0.127 | 0.519 | 0.000 | [-0.020, 0.028] | <.001 |
| Claude-3-5-haiku-latest | BFI-2 domains | Leiden | 0.115 | 0.515 | 0.000 | [-0.021, 0.028] | <.001 |
| Claude-3-5-haiku-latest | BFI-2 domains | Walktrap | 0.097 | 0.523 | 0.000 | [-0.018, 0.019] | <.001 |
| Claude-3-5-haiku-latest | BFI-2 domains | nSBM | 0.068 | 0.514 | -0.000 | [-0.014, 0.018] | <.001 |
| Claude-3-5-haiku-latest | BFI-2 facets | Louvain | 0.205 | 0.700 | -0.001 | [-0.030, 0.036] | <.001 |
| Claude-3-5-haiku-latest | BFI-2 facets | Leiden | 0.211 | 0.707 | 0.000 | [-0.028, 0.039] | <.001 |
| Claude-3-5-haiku-latest | BFI-2 facets | Walktrap | 0.161 | 0.707 | 0.000 | [-0.026, 0.031] | <.001 |
| Claude-3-5-haiku-latest | BFI-2 facets | nSBM | 0.125 | 0.709 | -0.001 | [-0.025, 0.028] | <.001 |
| Stereotype Content Dictionary (Nicolas et al., 2021) | | | | | | | |
| Mistral-7b | Human | Louvain | 0.440 | 0.676 | 0.000 | [-0.010, 0.012] | <.001 |
| Mistral-7b | Human | Leiden | 0.426 | 0.686 | 0.000 | [-0.010, 0.012] | <.001 |
| Mistral-7b | Human | Walktrap | 0.323 | 0.761 | 0.000 | [-0.011, 0.013] | <.001 |
| Mistral-7b | Human | nSBM | 0.329 | 0.779 | -0.000 | [-0.010, 0.013] | <.001 |
| Llama3.1-8b | Human | Louvain | 0.352 | 0.562 | 0.000 | [-0.011, 0.013] | <.001 |
| Llama3.1-8b | Human | Leiden | 0.414 | 0.611 | 0.000 | [-0.010, 0.012] | <.001 |
| Llama3.1-8b | Human | Walktrap | 0.242 | 0.687 | 0.000 | [-0.017, 0.021] | <.001 |
| Llama3.1-8b | Human | nSBM | 0.188 | 0.696 | -0.000 | [-0.009, 0.012] | <.001 |
| Claude-3-5-haiku-latest | Human | Louvain | 0.346 | 0.694 | 0.000 | [-0.011, 0.012] | <.001 |
| Claude-3-5-haiku-latest | Human | Leiden | 0.380 | 0.714 | 0.000 | [-0.010, 0.012] | <.001 |
| Claude-3-5-haiku-latest | Human | Walktrap | 0.315 | 0.744 | 0.000 | [-0.013, 0.016] | <.001 |
| Claude-3-5-haiku-latest | Human | nSBM | 0.250 | 0.743 | -0.000 | [-0.011, 0.012] | <.001 |
| Human | SCD domains | Louvain | 0.244 | 0.500 | 0.000 | [-0.009, 0.012] | <.001 |
| Human | SCD domains | Leiden | 0.209 | 0.489 | 0.000 | [-0.009, 0.012] | <.001 |
| Human | SCD domains | Walktrap | 0.199 | 0.496 | 0.000 | [-0.012, 0.016] | <.001 |
| Human | SCD domains | nSBM | 0.104 | 0.475 | 0.000 | [-0.006, 0.007] | <.001 |
| Mistral-7b | SCD domains | Louvain | 0.146 | 0.404 | 0.000 | [-0.008, 0.011] | <.001 |
| Mistral-7b | SCD domains | Leiden | 0.165 | 0.440 | 0.000 | [-0.008, 0.011] | <.001 |
| Mistral-7b | SCD domains | Walktrap | 0.136 | 0.504 | 0.000 | [-0.008, 0.009] | <.001 |
| Mistral-7b | SCD domains | nSBM | 0.089 | 0.476 | -0.000 | [-0.006, 0.007] | <.001 |
| Llama3.1-8b | SCD domains | Louvain | 0.147 | 0.362 | 0.000 | [-0.010, 0.012] | <.001 |
| Llama3.1-8b | SCD domains | Leiden | 0.153 | 0.367 | 0.000 | [-0.010, 0.011] | <.001 |
| Llama3.1-8b | SCD domains | Walktrap | 0.105 | 0.434 | 0.000 | [-0.011, 0.014] | <.001 |
| Llama3.1-8b | SCD domains | nSBM | 0.054 | 0.370 | 0.000 | [-0.006, 0.007] | <.001 |
| Claude-3-5-haiku-latest | SCD domains | Louvain | 0.119 | 0.458 | 0.000 | [-0.007, 0.009] | <.001 |
| Claude-3-5-haiku-latest | SCD domains | Leiden | 0.115 | 0.466 | 0.000 | [-0.007, 0.009] | <.001 |
| Claude-3-5-haiku-latest | SCD domains | Walktrap | 0.130 | 0.476 | 0.000 | [-0.009, 0.011] | <.001 |
| Claude-3-5-haiku-latest | SCD domains | nSBM | 0.084 | 0.456 | 0.000 | [-0.006, 0.007] | <.001 |
| Moral Foundations Dictionary 2.0 (Frimer, 2020) | | | | | | | |
| Mistral-7b | Human | Louvain | 0.462 | 0.630 | 0.000 | [-0.004, 0.005] | <.001 |
| Mistral-7b | Human | Leiden | 0.491 | 0.659 | 0.000 | [-0.005, 0.006] | <.001 |
| Mistral-7b | Human | Walktrap | 0.341 | 0.701 | 0.000 | [-0.008, 0.009] | <.001 |
| Mistral-7b | Human | nSBM | 0.217 | 0.595 | 0.000 | [-0.004, 0.004] | <.001 |
| Llama3.1-8b | Human | Louvain | 0.268 | 0.464 | 0.000 | [-0.005, 0.005] | <.001 |
| Llama3.1-8b | Human | Leiden | 0.363 | 0.524 | 0.000 | [-0.005, 0.006] | <.001 |
| Llama3.1-8b | Human | Walktrap | 0.191 | 0.572 | 0.000 | [-0.011, 0.013] | <.001 |
| Llama3.1-8b | Human | nSBM | 0.134 | 0.478 | 0.000 | [-0.005, 0.005] | <.001 |
| Claude-3-5-haiku-latest | Human | Louvain | 0.360 | 0.587 | 0.000 | [-0.005, 0.006] | <.001 |
| Claude-3-5-haiku-latest | Human | Leiden | 0.382 | 0.630 | 0.000 | [-0.005, 0.006] | <.001 |
| Claude-3-5-haiku-latest | Human | Walktrap | 0.276 | 0.667 | 0.000 | [-0.008, 0.010] | <.001 |
| Claude-3-5-haiku-latest | Human | nSBM | 0.204 | 0.576 | -0.000 | [-0.004, 0.004] | <.001 |
| Human | MFD domains | Louvain | 0.194 | 0.408 | 0.000 | [-0.004, 0.005] | <.001 |
| Human | MFD domains | Leiden | 0.219 | 0.415 | 0.000 | [-0.005, 0.006] | <.001 |
| Human | MFD domains | Walktrap | 0.171 | 0.392 | 0.000 | [-0.008, 0.009] | <.001 |
| Human | MFD domains | nSBM | 0.110 | 0.354 | -0.000 | [-0.003, 0.004] | <.001 |
| Human | MFD subdomains | Louvain | 0.336 | 0.524 | 0.000 | [-0.004, 0.006] | <.001 |
| Human | MFD subdomains | Leiden | 0.364 | 0.530 | 0.000 | [-0.005, 0.006] | <.001 |
| Human | MFD subdomains | Walktrap | 0.272 | 0.502 | 0.000 | [-0.007, 0.007] | <.001 |
| Human | MFD subdomains | nSBM | 0.192 | 0.464 | -0.000 | [-0.004, 0.004] | <.001 |
| Mistral-7b | MFD domains | Louvain | 0.184 | 0.391 | 0.000 | [-0.004, 0.005] | <.001 |
| Mistral-7b | MFD domains | Leiden | 0.179 | 0.376 | 0.000 | [-0.004, 0.005] | <.001 |
| Mistral-7b | MFD domains | Walktrap | 0.130 | 0.368 | 0.000 | [-0.006, 0.007] | <.001 |
| Mistral-7b | MFD domains | nSBM | 0.065 | 0.345 | 0.000 | [-0.002, 0.003] | <.001 |
| Mistral-7b | MFD subdomains | Louvain | 0.336 | 0.537 | 0.000 | [-0.004, 0.005] | <.001 |
| Mistral-7b | MFD subdomains | Leiden | 0.331 | 0.525 | 0.000 | [-0.004, 0.006] | <.001 |
| Mistral-7b | MFD subdomains | Walktrap | 0.243 | 0.510 | 0.000 | [-0.006, 0.007] | <.001 |
| Mistral-7b | MFD subdomains | nSBM | 0.140 | 0.490 | -0.000 | [-0.003, 0.003] | <.001 |
| Llama3.1-8b | MFD domains | Louvain | 0.121 | 0.251 | 0.000 | [-0.004, 0.005] | <.001 |
| Llama3.1-8b | MFD domains | Leiden | 0.147 | 0.293 | 0.000 | [-0.004, 0.004] | <.001 |
| Llama3.1-8b | MFD domains | Walktrap | 0.067 | 0.306 | 0.000 | [-0.010, 0.012] | <.001 |
| Llama3.1-8b | MFD domains | nSBM | 0.050 | 0.220 | -0.000 | [-0.003, 0.003] | <.001 |
| Llama3.1-8b | MFD subdomains | Louvain | 0.249 | 0.409 | 0.000 | [-0.005, 0.005] | <.001 |
| Llama3.1-8b | MFD subdomains | Leiden | 0.288 | 0.454 | 0.000 | [-0.004, 0.005] | <.001 |
| Llama3.1-8b | MFD subdomains | Walktrap | 0.183 | 0.455 | 0.000 | [-0.008, 0.009] | <.001 |
| Llama3.1-8b | MFD subdomains | nSBM | 0.120 | 0.379 | 0.000 | [-0.003, 0.004] | <.001 |
| Claude-3-5-haiku-latest | MFD domains | Louvain | 0.129 | 0.354 | 0.000 | [-0.004, 0.005] | <.001 |
| Claude-3-5-haiku-latest | MFD domains | Leiden | 0.127 | 0.374 | 0.000 | [-0.004, 0.005] | <.001 |
| Claude-3-5-haiku-latest | MFD domains | Walktrap | 0.107 | 0.377 | 0.000 | [-0.005, 0.006] | <.001 |
| Claude-3-5-haiku-latest | MFD domains | nSBM | 0.067 | 0.326 | 0.000 | [-0.002, 0.003] | <.001 |
| Claude-3-5-haiku-latest | MFD subdomains | Louvain | 0.261 | 0.513 | 0.000 | [-0.004, 0.005] | <.001 |
| Claude-3-5-haiku-latest | MFD subdomains | Leiden | 0.258 | 0.533 | 0.000 | [-0.004, 0.005] | <.001 |
| Claude-3-5-haiku-latest | MFD subdomains | Walktrap | 0.228 | 0.533 | 0.000 | [-0.005, 0.006] | <.001 |
| Claude-3-5-haiku-latest | MFD subdomains | nSBM | 0.144 | 0.474 | -0.000 | [-0.003, 0.003] | <.001 |

*Note.* Rows with Human as the reference report the primary empirical benchmark: each LLM community partition was compared with the human empirical partition. Louvain and Leiden were estimated with resolution = 1.00. Walktrap was run with weighted edges and steps = 4. For nSBM, partitions were extracted from lexicon-specific hierarchical levels: level 1 for BFI-2, level 2 for the Stereotype Content Dictionary, and level 3 for the Moral Foundations Dictionary 2.0. Random baselines were generated from 1,000 size-preserving random partitions that preserved the observed community-size distribution. The random interval reports the 2.5th and 97.5th percentiles of the random ARI distribution. The p-value is the empirical probability that a size-preserving random partition produced an ARI at least as large as the observed ARI. ARI = adjusted Rand index; nSBM = nested stochastic block model.

**Table S8 Response-level bootstrap stability of community assignments and partition structure**

| Model | Algorithm | Mean replication | Median replication | Minimum replication | Nodes < .70 (%) | Lowest replication words: first five | Observed communities | 95% interval |
|---|---|---|---|---|---|---|---|---|
| BFI-2 (Soto & John, 2017) | | | | | | | | |
| Humans | Louvain | 0.888 | 0.996 | 0.294 | 12 (15.19) | suspicious=0.294; efficient=0.326; persistent=0.352; boring=0.398; sympathy=0.418 | 19 | [18, 23] |
| Humans | Leiden | 0.891 | 0.995 | 0.080 | 13 (16.46) | efficient=0.080; curious=0.340; complex=0.420; persistent=0.425; sympathy=0.427 | 21 | [18, 23] |
| Humans | Walktrap | 0.867 | 0.907 | 0.190 | 12 (15.19) | assertive=0.190; comfortable=0.522; relaxed=0.527; music=0.535; talkative=0.551 | 37 | [29, 44] |
| Human | nSBM | 0.614 | 0.646 | 0.025 | 46 (58.23) | setback=0.025; boring=0.142; curious=0.254; afraid=0.283; helpful=0.327 | 28 | [24, 35] |
| Mistral-7b | Louvain | 0.831 | 0.982 | 0.109 | 17 (21.52) | quiet=0.109; relaxed=0.109; respect=0.149; courteous=0.176; respectful=0.176 | 27 | [22, 27] |
| Mistral-7b | Leiden | 0.927 | 0.998 | 0.349 | 8 (10.13) | secure=0.349; image=0.456; original=0.462; literature=0.474; clever=0.476 | 26 | [24, 29] |
| Mistral-7b | Walktrap | 0.904 | 0.989 | 0.200 | 10 (12.66) | poetry=0.200; music=0.205; enthusiasm=0.243; respect=0.529; thinker=0.543 | 36 | [33, 44] |
| Mistral-7b | nSBM | 0.705 | 0.749 | 0.254 | 32 (40.51) | abstract=0.254; soft=0.265; complex=0.272; original=0.334; tense=0.358 | 42 | [31, 42] |
| Llama3.1-8b | Louvain | 0.792 | 0.873 | 0.053 | 22 (27.85) | eager=0.053; suspicious=0.128; curious=0.166; relaxed=0.319; leader=0.397 | 20 | [19, 24] |
| Llama3.1-8b | Leiden | 0.835 | 0.909 | 0.323 | 17 (21.52) | image=0.323; relaxed=0.380; leader=0.382; idea=0.399; creativity=0.437 | 21 | [20, 24] |
| Llama3.1-8b | Walktrap | 0.835 | 0.995 | 0.087 | 17 (21.52) | sympathy=0.087; anxious=0.096; new=0.127; literature=0.194; artistic=0.201 | 49 | [48, 61] |
| Llama3.1-8b | nSBM | 0.558 | 0.583 | 0.109 | 56 (70.89) | complex=0.109; courteous=0.124; moody=0.124; suspicious=0.136; sympathy=0.159 | 33 | [28, 39] |
| Claude-3-5-haiku-latest | Louvain | 0.949 | 1 | 0.126 | 7 (8.86) | forgiving=0.126; music=0.498; abstract=0.566; suspicious=0.653; upset=0.674 | 33 | [32, 36] |
| Claude-3-5-haiku-latest | Leiden | 0.959 | 1 | 0.410 | 5 (6.33) | music=0.410; suspicious=0.611; abstract=0.638; dependable=0.645; reliable=0.645 | 34 | [33, 37] |
| Claude-3-5-haiku-latest | Walktrap | 0.939 | 1 | 0.455 | 7 (8.86) | blue=0.455; artistic=0.537; music=0.565; dominant=0.642; leader=0.653 | 40 | [35, 41] |
| Claude-3-5-haiku-latest | nSBM | 0.763 | 0.826 | 0.207 | 27 (34.18) | poetry=0.207; helpful=0.317; order=0.356; literature=0.394; dependable=0.430 | 44 | [37, 52] |
| Stereotype Content Dictionary (Nicolas et al., 2021) | | | | | | | | |
| Humans | Louvain | 0.807 | 0.899 | 0.083 | 54 (21.77) | social=0.083; rational=0.116; logical=0.116; irrational=0.118; corrupt=0.137 | 19 | [16, 21] |
| Humans | Leiden | 0.815 | 0.892 | 0.135 | 59 (23.79) | important=0.135; critical=0.135; immoral=0.137; popular=0.161; moral=0.218 | 20 | [17, 22] |
| Humans | Walktrap | 0.670 | 0.768 | 0.087 | 115 (46.37) | threat=0.087; fear=0.093; efficient=0.098; industrious=0.111; discrimination=0.115 | 52 | [41, 80] |
| Human | nSBM | 0.452 | 0.402 | 0.066 | 189 (76.21) | irrational=0.066; sensitive=0.070; untrustworthy=0.072; warmth=0.077; illogical=0.077 | 50 | [28, 51] |
| Mistral-7b | Louvain | 0.897 | 0.984 | 0.117 | 33 (13.31) | polite=0.117; conventional=0.179; virtuous=0.188; sensitive=0.220; competitive=0.312 | 24 | [22, 27] |
| Mistral-7b | Leiden | 0.922 | 0.999 | 0.232 | 23 (9.27) | open=0.232; competitive=0.386; rough=0.403; insightful=0.424; successful=0.430 | 27 | [25, 30] |
| Mistral-7b | Walktrap | 0.824 | 0.926 | 0.020 | 61 (24.60) | helpful=0.020; supportive=0.034; social=0.161; fear=0.173; threat=0.173 | 82 | [55, 89] |
| Mistral-7b | nSBM | 0.577 | 0.597 | 0.055 | 160 (64.52) | thief=0.055; criminal=0.061; agency=0.078; funny=0.106; status=0.109 | 67 | [51, 78] |
| Llama3.1-8b | Louvain | 0.812 | 0.938 | 0.048 | 59 (23.79) | dependent=0.048; inability=0.062; determined=0.072; naive=0.091; assertive=0.141 | 18 | [17, 22] |
| Llama3.1-8b | Leiden | 0.835 | 0.97 | 0.085 | 57 (22.98) | sociable=0.085; respected=0.109; independent=0.134; prestige=0.234; important=0.250 | 20 | [19, 24] |
| Llama3.1-8b | Walktrap | 0.742 | 0.821 | 0.002 | 85 (34.27) | insecure=0.002; conservative=0.042; competitive=0.067; vulnerable=0.071; sensitive=0.076 | 100 | [75, 116] |
| Llama3.1-8b | nSBM | 0.468 | 0.456 | 0.037 | 193 (77.82) | cooperative=0.037; poor=0.040; helpful=0.046; uncaring=0.046; unkind=0.049 | 55 | [40, 60] |
| Claude-3-5-haiku-latest | Louvain | 0.887 | 0.992 | 0.091 | 47 (18.95) | independent=0.091; hostile=0.105; logical=0.169; rational=0.169; boring=0.223 | 44 | [41, 47] |
| Claude-3-5-haiku-latest | Leiden | 0.909 | 0.999 | 0.371 | 32 (12.90) | faithful=0.371; dedicated=0.371; loyal=0.371; treacherous=0.393; disloyal=0.393 | 49 | [45, 50] |
| Claude-3-5-haiku-latest | Walktrap | 0.777 | 0.906 | 0.062 | 83 (33.47) | evil=0.062; clumsy=0.062; unkind=0.066; vicious=0.067; brutal=0.068 | 73 | [66, 95] |
| Claude-3-5-haiku-latest | nSBM | 0.590 | 0.607 | 0.017 | 158 (63.71) | popular=0.017; agency=0.024; insignificant=0.029; independent=0.053; apathy=0.082 | 65 | [47, 76] |
| Moral Foundations Dictionary 2.0 (Frimer, 2020) | | | | | | | | |
| Humans | Louvain | 0.686 | 0.727 | 0.032 | 281 (46.91) | referee=0.032; player=0.033; deviant=0.062; trust=0.067; treacherous=0.093 | 19 | [15, 20] |
| Humans | Leiden | 0.779 | 0.885 | 0.094 | 181 (30.22) | glory=0.094; nurture=0.134; treacherous=0.136; glorious=0.138; heavenly=0.139 | 18 | [15, 20] |
| Humans | Walktrap | 0.645 | 0.727 | 0.021 | 279 (46.58) | chaste=0.021; pure=0.021; chastity=0.025; purity=0.026; virgin=0.027 | 80 | [53, 108] |
| Human | nSBM | 0.326 | 0.306 | 0.003 | 572 (95.49) | leper=0.003; curse=0.017; unhelpful=0.022; govern=0.022; outsider=0.025 | 37 | [24, 60] |
| Mistral-7b | Louvain | 0.894 | 0.987 | 0.023 | 73 (12.19) | willing=0.023; permissive=0.023; vulnerable=0.100; blood=0.114; stain=0.137 | 24 | [23, 29] |
| Mistral-7b | Leiden | 0.905 | 0.996 | 0.041 | 60 (10.02) | corpse=0.041; alcoholic=0.148; cock=0.148; exploit=0.165; wasted=0.168 | 29 | [27, 32] |
| Mistral-7b | Walktrap | 0.812 | 0.944 | 0.024 | 147 (24.54) | sexual=0.024; sexuality=0.024; raw=0.058; childhood=0.079; celibacy=0.085 | 118 | [103, 147] |
| Mistral-7b | nSBM | 0.444 | 0.435 | 0.004 | 493 (82.30) | destroy=0.004; destruction=0.004; lawless=0.011; chaos=0.012; anarchy=0.014 | 63 | [49, 86] |
| Llama3.1-8b | Louvain | 0.627 | 0.673 | 0.008 | 312 (52.09) | disrespect=0.008; prostitution=0.013; heaven=0.019; elder=0.022; uncaring=0.024 | 18 | [17, 22] |
| Llama3.1-8b | Leiden | 0.738 | 0.829 | 0.007 | 243 (40.57) | bless=0.007; harsh=0.031; blessing=0.035; bully=0.047; cruel=0.049 | 21 | [18, 23] |
| Llama3.1-8b | Walktrap | 0.716 | 0.796 | 0.004 | 239 (39.90) | harmful=0.004; chastity=0.016; sharing=0.029; patriot=0.032; impartial=0.053 | 133 | [122, 184] |
| Llama3.1-8b | nSBM | 0.335 | 0.282 | 0.002 | 542 (90.48) | rescue=0.002; herd=0.003; captain=0.014; admiral=0.017; endangered=0.018 | 55 | [45, 83] |
| Claude-3-5-haiku-latest | Louvain | 0.889 | 0.986 | 0.010 | 72 (12.02) | chaos=0.010; disarray=0.010; disorder=0.010; anarchy=0.010; sleazy=0.091 | 50 | [44, 52] |
| Claude-3-5-haiku-latest | Leiden | 0.904 | 0.998 | 0.040 | 78 (13.02) | willing=0.040; permission=0.040; overpowering=0.045; violent=0.059; console=0.123 | 56 | [51, 58] |
| Claude-3-5-haiku-latest | Walktrap | 0.843 | 0.956 | 0.058 | 111 (18.53) | karma=0.058; police=0.072; law=0.083; regulation=0.083; lawful=0.089 | 154 | [127, 162] |
| Claude-3-5-haiku-latest | nSBM | 0.463 | 0.467 | 0.004 | 481 (80.30) | noble=0.004; familiar=0.012; cleaning=0.015; cleaner=0.015; lawyer=0.017 | 54 | [50, 82] |

*Note.* This table reports response-level bootstrap stability diagnostics. In each of 1,000 bootstrap replicates, response rows were resampled, the semantic network was reconstructed, and communities were re-detected. Louvain and Leiden were estimated with resolution = 1.00. Walktrap was run with weighted edges and steps = 4. For nSBM, partitions were extracted from lexicon-specific hierarchical levels: level 1 for BFI-2, level 2 for the Stereotype Content Dictionary, and level 3 for the Moral Foundations Dictionary 2.0. Node replication is the proportion of successful bootstrap replicates in which a word node was retained in its observed reference community after optimal community-label matching. The evaluated common node sets contained 79 BFI-2 words, 248 Stereotype Content Dictionary words, and 599 Moral Foundations Dictionary 2.0 words. "Nodes < .70" gives the number of word nodes with replication below 0.70, and the lowest-replication column lists the five least stable nodes for each model–algorithm setting. The observed number of communities is from the original network; the bootstrap 95% interval gives the 2.5th and 97.5th percentiles of the number of communities recovered across bootstrap replicates.

**Table S9 Paired response-bootstrap LLM–Human alignment across community-detection algorithms**

| Model | Algorithm | ARI mean [95% interval] | V-measure mean [95% interval] |
|---|---|---|---|
| BFI-2 (Soto & John, 2017) | | | |
| Mistral-7b | Louvain | 0.532 [0.443, 0.615] | 0.823 [0.790, 0.854] |
| Mistral-7b | Leiden | 0.549 [0.465, 0.631] | 0.831 [0.802, 0.857] |
| Mistral-7b | Walktrap | 0.516 [0.341, 0.659] | 0.877 [0.837, 0.912] |
| Mistral-7b | nSBM | 0.272 [0.166, 0.392] | 0.813 [0.775, 0.847] |
| Llama3.1-8b | Louvain | 0.338 [0.252, 0.435] | 0.732 [0.687, 0.775] |
| Llama3.1-8b | Leiden | 0.346 [0.263, 0.438] | 0.739 [0.701, 0.778] |
| Llama3.1-8b | Walktrap | 0.143 [0.073, 0.230] | 0.813 [0.765, 0.855] |
| Llama3.1-8b | nSBM | 0.162 [0.077, 0.257] | 0.770 [0.726, 0.811] |
| Claude-3-5-haiku-latest | Louvain | 0.415 [0.340, 0.493] | 0.812 [0.784, 0.839] |
| Claude-3-5-haiku-latest | Leiden | 0.403 [0.343, 0.470] | 0.812 [0.789, 0.837] |
| Claude-3-5-haiku-latest | Walktrap | 0.390 [0.262, 0.498] | 0.855 [0.811, 0.890] |
| Claude-3-5-haiku-latest | nSBM | 0.178 [0.093, 0.279] | 0.812 [0.775, 0.845] |
| Stereotype Content Dictionary (Nicolas et al., 2021) | | | |
| Mistral-7b | Louvain | 0.431 [0.377, 0.495] | 0.676 [0.639, 0.713] |
| Mistral-7b | Leiden | 0.437 [0.386, 0.490] | 0.688 [0.650, 0.722] |
| Mistral-7b | Walktrap | 0.337 [0.239, 0.469] | 0.751 [0.689, 0.811] |
| Mistral-7b | nSBM | 0.237 [0.183, 0.295] | 0.718 [0.670, 0.756] |
| Llama3.1-8b | Louvain | 0.384 [0.315, 0.448] | 0.593 [0.550, 0.633] |
| Llama3.1-8b | Leiden | 0.419 [0.365, 0.479] | 0.616 [0.579, 0.655] |
| Llama3.1-8b | Walktrap | 0.252 [0.164, 0.369] | 0.684 [0.625, 0.740] |
| Llama3.1-8b | nSBM | 0.170 [0.127, 0.227] | 0.650 [0.601, 0.688] |
| Claude-3-5-haiku-latest | Louvain | 0.373 [0.325, 0.425] | 0.700 [0.670, 0.729] |
| Claude-3-5-haiku-latest | Leiden | 0.379 [0.333, 0.430] | 0.710 [0.684, 0.735] |
| Claude-3-5-haiku-latest | Walktrap | 0.274 [0.179, 0.411] | 0.742 [0.673, 0.799] |
| Claude-3-5-haiku-latest | nSBM | 0.208 [0.163, 0.265] | 0.698 [0.643, 0.738] |
| Moral Foundations Dictionary 2.0 (Frimer, 2020) | | | |
| Mistral-7b | Louvain | 0.436 [0.391, 0.487] | 0.616 [0.582, 0.651] |
| Mistral-7b | Leiden | 0.465 [0.420, 0.513] | 0.644 [0.614, 0.677] |
| Mistral-7b | Walktrap | 0.325 [0.238, 0.416] | 0.684 [0.622, 0.733] |
| Mistral-7b | nSBM | 0.175 [0.135, 0.221] | 0.579 [0.515, 0.627] |
| Llama3.1-8b | Louvain | 0.312 [0.260, 0.362] | 0.470 [0.432, 0.513] |
| Llama3.1-8b | Leiden | 0.342 [0.292, 0.396] | 0.499 [0.462, 0.538] |
| Llama3.1-8b | Walktrap | 0.196 [0.135, 0.274] | 0.570 [0.511, 0.626] |
| Llama3.1-8b | nSBM | 0.118 [0.092, 0.155] | 0.503 [0.437, 0.558] |
| Claude-3-5-haiku-latest | Louvain | 0.370 [0.328, 0.414] | 0.598 [0.572, 0.625] |
| Claude-3-5-haiku-latest | Leiden | 0.382 [0.350, 0.417] | 0.619 [0.597, 0.642] |
| Claude-3-5-haiku-latest | Walktrap | 0.243 [0.176, 0.316] | 0.655 [0.59, 0.710] |
| Claude-3-5-haiku-latest | nSBM | 0.165 [0.126, 0.209] | 0.566 [0.500, 0.617] |

*Note.* This table reports paired response-bootstrap estimates of LLM–Human empirical alignment. In each bootstrap replicate, both the human and LLM response-level datasets were resampled, networks

were reconstructed, communities were re-detected, and the LLM partition was compared with the corresponding human empirical partition. Reported values are bootstrap means with 95% percentile intervals. Walktrap was run with weighted edges and steps = 4. For nSBM, partitions were extracted from lexicon-specific hierarchical levels: level 1 for BFI-2, level 2 for the Stereotype Content Dictionary, and level 3 for the Moral Foundations Dictionary 2.0.

**Table S10 Paired response-bootstrap differences in LLM–Human ARI between models**

| Contrast | Algorithm | Observed ARI difference | Bootstrap ARI difference mean [95% interval] | *p* | Pr(diff > 0) |
|---|---|---|---|---|---|
| BFI-2 (Soto & John, 2017) | | | | | |
| Mistral-7b - Llama3.1-8b | Louvain | 0.177 | 0.194 [0.057, 0.316] | 0.008 | 0.997 |
| Mistral-7b - Llama3.1-8b | Leiden | 0.204 | 0.202 [0.068, 0.329] | 0.004 | 0.999 |
| Mistral-7b - Llama3.1-8b | Walktrap | 0.352 | 0.374 [0.182, 0.542] | 0.004 | 0.999 |
| Mistral-7b - Llama3.1-8b | nSBM | 0.293 | 0.110 [-0.016, 0.241] | 0.102 | 0.950 |
| Mistral-7b - Claude-3-5-haiku-latest | Louvain | -0.023 | 0.118 [0.008, 0.217] | 0.044 | 0.979 |
| Mistral-7b - Claude-3-5-haiku-latest | Leiden | 0.105 | 0.146 [0.056, 0.236] | 0.004 | 0.999 |
| Mistral-7b - Claude-3-5-haiku-latest | Walktrap | 0.158 | 0.126 [-0.032, 0.281] | 0.102 | 0.950 |
| Mistral-7b - Claude-3-5-haiku-latest | nSBM | 0.183 | 0.093 [-0.034, 0.223] | 0.176 | 0.913 |
| Llama3.1-8b - Claude-3-5-haiku-latest | Louvain | -0.201 | -0.077 [-0.201, 0.059] | 0.276 | 0.137 |
| Llama3.1-8b - Claude-3-5-haiku-latest | Leiden | -0.099 | -0.057 [-0.173, 0.067] | 0.378 | 0.188 |
| Llama3.1-8b - Claude-3-5-haiku-latest | Walktrap | -0.194 | -0.247 [-0.387, -0.092] | 0.006 | 0.002 |
| Llama3.1-8b - Claude-3-5-haiku-latest | nSBM | -0.109 | -0.016 [-0.135, 0.096] | 0.791 | 0.394 |
| Stereotype Content Dictionary (Nicolas et al., 2021) | | | | | |
| Mistral-7b - Llama3.1-8b | Louvain | 0.087 | 0.047 [-0.031, 0.131] | 0.260 | 0.871 |
| Mistral-7b - Llama3.1-8b | Leiden | 0.013 | 0.018 [-0.049, 0.087] | 0.611 | 0.695 |
| Mistral-7b - Llama3.1-8b | Walktrap | 0.081 | 0.085 [-0.063, 0.228] | 0.256 | 0.873 |
| Mistral-7b - Llama3.1-8b | nSBM | 0.141 | 0.067 [0.005, 0.129] | 0.042 | 0.980 |
| Mistral-7b - Claude-3-5-haiku-latest | Louvain | 0.094 | 0.058 [-0.002, 0.129] | 0.060 | 0.971 |
| Mistral-7b - Claude-3-5-haiku-latest | Leiden | 0.046 | 0.058 [-0.004, 0.126] | 0.058 | 0.972 |
| Mistral-7b - Claude-3-5-haiku-latest | Walktrap | 0.008 | 0.063 [-0.076, 0.222] | 0.402 | 0.800 |
| Mistral-7b - Claude-3-5-haiku-latest | nSBM | 0.079 | 0.029 [-0.035, 0.088] | 0.336 | 0.833 |
| Llama3.1-8b - Claude-3-5-haiku-latest | Louvain | 0.006 | 0.011 [-0.067, 0.086] | 0.757 | 0.622 |
| Llama3.1-8b - Claude-3-5-haiku-latest | Leiden | 0.034 | 0.040 [-0.026, 0.111] | 0.214 | 0.894 |
| Llama3.1-8b - Claude-3-5-haiku-latest | Walktrap | -0.073 | -0.022 [-0.179, 0.142] | 0.785 | 0.392 |
| Llama3.1-8b - Claude-3-5-haiku-latest | nSBM | -0.062 | -0.038 [-0.097, 0.023] | 0.202 | 0.100 |
| Moral Foundations Dictionary 2.0 (Frimer, 2020) | | | | | |
| Mistral-7b - Llama3.1-8b | Louvain | 0.194 | 0.123 [0.061, 0.192] | 0.002 | 1 |
| Mistral-7b - Llama3.1-8b | Leiden | 0.128 | 0.123 [0.066, 0.182] | 0.002 | 1 |
| Mistral-7b - Llama3.1-8b | Walktrap | 0.150 | 0.129 [0.018, 0.237] | 0.032 | 0.985 |
| Mistral-7b - Llama3.1-8b | nSBM | 0.083 | 0.057 [0.011, 0.102] | 0.020 | 0.991 |
| Mistral-7b - Claude-3-5-haiku-latest | Louvain | 0.103 | 0.066 [0.012, 0.123] | 0.022 | 0.990 |
| Mistral-7b - Claude-3-5-haiku-latest | Leiden | 0.109 | 0.083 [0.038, 0.13] | 0.002 | 1 |
| Mistral-7b - Claude-3-5-haiku-latest | Walktrap | 0.065 | 0.082 [-0.002, 0.181] | 0.060 | 0.971 |
| Mistral-7b - Claude-3-5-haiku-latest | nSBM | 0.013 | 0.010 [-0.032, 0.052] | 0.601 | 0.700 |
| Llama3.1-8b - Claude-3-5-haiku-latest | Louvain | -0.092 | -0.058 [-0.117, 0.011] | 0.078 | 0.038 |
| Llama3.1-8b - Claude-3-5-haiku-latest | Leiden | -0.019 | -0.04 [-0.098, 0.016] | 0.162 | 0.080 |
| Llama3.1-8b - Claude-3-5-haiku-latest | Walktrap | -0.085 | -0.047 [-0.144, 0.061] | 0.354 | 0.176 |
| Llama3.1-8b - Claude-3-5-haiku-latest | nSBM | -0.070 | -0.047 [-0.093, -0.002] | 0.048 | 0.023 |

*Note.* This table tests pairwise model differences in empirical LLM–Human alignment under the paired response-bootstrap procedure. For each bootstrap replicate, LLM–Human ARI was computed for each model under the same algorithmic setting, and pairwise differences were calculated as the first model minus the second model in each contrast. The bootstrap interval reports the 2.5th and 97.5th percentiles of the paired difference distribution. The p-value is a two-sided bootstrap p-value for the null hypothesis that the difference equals zero. Pr(diff > 0) reports the proportion of bootstrap replicates in which the first model showed higher LLM–Human ARI than the second model. Positive values indicate stronger alignment for the first model. Walktrap was run with weighted edges and steps = 4. For nSBM, partitions were extracted from lexicon-specific hierarchical levels: level 1 for BFI-2, level 2 for the Stereotype Content Dictionary, and level 3 for the Moral Foundations Dictionary 2.0.

**Table S11 Best kNN prediction performance for lexical norms across human and LLM association spaces**

| | Model | N | r | 95% CI | k |
|---|---|---|---|---|---|
| Valence | Humans | 8725 | 0.859 | [0.853, 0.864] | 90 |
| Valence | Mistral-7b | 8725 | 0.813 | [0.806, 0.820] | 42 |
| Valence | Llama3.1-8b | 8725 | 0.770 | [0.761, 0.778] | 28 |
| Valence | Claude-3-5-haiku-latest | 8725 | 0.795 | [0.787, 0.802] | 11 |
| Arousal | Humans | 8725 | 0.687 | [0.676, 0.698] | 70 |
| Arousal | Mistral-7b | 8725 | 0.636 | [0.623, 0.648] | 80 |
| Arousal | Llama3.1-8b | 8725 | 0.597 | [0.584, 0.611] | 47 |
| Arousal | Claude-3-5-haiku-latest | 8725 | 0.615 | [0.602, 0.628] | 16 |
| Dominance | Humans | 8725 | 0.747 | [0.738, 0.756] | 29 |
| Dominance | Mistral-7b | 8725 | 0.703 | [0.692, 0.713] | 35 |
| Dominance | Llama3.1-8b | 8725 | 0.664 | [0.653, 0.676] | 28 |
| Dominance | Claude-3-5-haiku-latest | 8725 | 0.690 | [0.679, 0.701] | 17 |
| Concreteness | Humans | 10662 | 0.863 | [0.858, 0.868] | 11 |
| Concreteness | Mistral-7b | 10662 | 0.821 | [0.814, 0.827] | 10 |
| Concreteness | Llama3.1-8b | 10662 | 0.776 | [0.768, 0.784] | 12 |
| Concreteness | Claude-3-5-haiku-latest | 10662 | 0.833 | [0.827, 0.839] | 11 |

*Note.* N indicates the number of words included in each prediction analysis. For each association source, cue-by-cue matrices were constructed from cleaned free-association responses and weighted using positive pointwise mutual information (PPMI). Predictions were generated using leave-one-out k-nearest-neighbor prediction based on cosine similarity between sparse PPMI vectors. The reported r is the Pearson correlation between observed lexical norm values and predicted values at the best-performing k. The 95% confidence intervals for r were computed using Fisher's z transformation. The column k reports the number of nearest neighbors that maximized the prediction correlation. For valence, arousal, and dominance analyses, words were retained only if they had complete valence, arousal, and dominance ratings. The prediction set was restricted to words that appeared in all association spaces and had nonzero PPMI rows in all models.

**Table S12 Pairwise differences in kNN prediction accuracy across human and LLM association spaces**

| | Comparison | Δr | Bootstrap 95% CI | *p* |
|---|---|---|---|---|
| Valence | Humans - Claude-3-5-haiku-latest | 0.064 | [0.057, 0.071] | < .001 |
| Valence | Humans - Llama3.1-8b | 0.089 | [0.082, 0.097] | < .001 |
| Valence | Humans - Mistral-7b | 0.046 | [0.040, 0.052] | < .001 |
| Arousal | Humans - Claude-3-5-haiku-latest | 0.071 | [0.062, 0.081] | < .001 |
| Arousal | Humans - Llama3.1-8b | 0.090 | [0.080, 0.100] | < .001 |
| Arousal | Humans - Mistral-7b | 0.051 | [0.043, 0.059] | < .001 |
| Dominance | Humans - Claude-3-5-haiku-latest | 0.058 | [0.050, 0.065] | < .001 |
| Dominance | Humans - Llama3.1-8b | 0.083 | [0.074, 0.093] | < .001 |
| Dominance | Humans - Mistral-7b | 0.045 | [0.037, 0.052] | < .001 |
| Concreteness | Humans - Claude-3-5-haiku-latest | 0.030 | [0.025, 0.035] | < .001 |
| Concreteness | Humans - Llama3.1-8b | 0.087 | [0.080, 0.094] | < .001 |
| Concreteness | Humans - Mistral-7b | 0.043 | [0.037, 0.048] | < .001 |
| Valence | Llama3.1-8b - Claude-3-5-haiku-latest | -0.025 | [-0.034, -0.016] | < .001 |
| Valence | Mistral-7b - Claude-3-5-haiku-latest | 0.018 | [0.011, 0.025] | < .001 |
| Valence | Mistral-7b - Llama3.1-8b | 0.043 | [0.036, 0.051] | < .001 |
| Arousal | Llama3.1-8b - Claude-3-5-haiku-latest | -0.018 | [-0.030, -0.006] | 0.015 |
| Arousal | Mistral-7b - Claude-3-5-haiku-latest | 0.020 | [0.009, 0.031] | < .001 |
| Arousal | Mistral-7b - Llama3.1-8b | 0.038 | [0.027, 0.049] | < .001 |
| Dominance | Llama3.1-8b - Claude-3-5-haiku-latest | -0.025 | [-0.037, -0.015] | < .001 |
| Dominance | Mistral-7b - Claude-3-5-haiku-latest | 0.013 | [0.004, 0.022] | 0.017 |
| Dominance | Mistral-7b - Llama3.1-8b | 0.038 | [0.029, 0.048] | < .001 |
| Concreteness | Llama3.1-8b - Claude-3-5-haiku-latest | -0.057 | [-0.065, -0.049] | < .001 |
| Concreteness | Mistral-7b - Claude-3-5-haiku-latest | -0.012 | [-0.018, -0.006] | < .001 |
| Concreteness | Mistral-7b - Llama3.1-8b | 0.045 | [0.037, 0.052] | < .001 |

Note. Δr represents the difference between two prediction correlations, calculated as $r_{\text{Model A}} - r_{\text{Model B}}$. Positive values indicate higher prediction accuracy for the first model in the comparison, whereas negative values indicate higher prediction accuracy for the second model. The 95% confidence intervals for Δr were obtained using word-level bootstrap resampling with 1,000 iterations. The reported p values are Bonferroni-adjusted two-sided Williams test p values for dependent overlapping correlations. Bonferroni correction was applied separately to Human–LLM and LLM–LLM comparisons within each analysis set. For valence, arousal, and dominance, each correction included nine comparisons; for concreteness, each correction included three comparisons.

**Table S13 Accuracy of digital twins generated by different large language models across heuristics-and-biases tasks**

| | DeepSeek V3-0324 | DeepSeek V3.1 | DeepSeek V4-Flash | Qwen 2.5 Max | Gemma-4-31B-it | gpt-oss-120b |
|---|---|---|---|---|---|---|
| Overall | 0.720 | 0.719 | 0.697 | 0.678 | 0.696 | 0.701 |
| Allais | 0.626 | 0.630 | 0.594 | 0.627 | 0.627 | 0.605 |
| WTA/WTP-Thaler | 0.804 | 0.819 | 0.746 | 0.596 | 0.799 | 0.785 |
| Absolute vs. relative savings | 0.658 | 0.644 | 0.574 | 0.540 | 0.618 | 0.546 |
| Anchoring and adjustment | 0.680 | 0.705 | 0.717 | 0.635 | 0.632 | 0.639 |
| Base rate problem | 0.814 | 0.813 | 0.785 | 0.808 | 0.800 | 0.790 |
| Conjunction problem (Linda) | 0.767 | 0.760 | 0.754 | 0.783 | 0.778 | 0.731 |
| Denominator neglect | 0.639 | 0.635 | 0.518 | 0.386 | 0.359 | 0.638 |
| False consensus | 0.774 | 0.765 | 0.769 | 0.730 | 0.780 | 0.752 |
| Framing problem | 0.651 | 0.664 | 0.622 | 0.685 | 0.701 | 0.744 |
| Less is more | 0.758 | 0.757 | 0.725 | 0.737 | 0.757 | 0.722 |
| Myside bias | 0.765 | 0.763 | 0.754 | 0.771 | 0.699 | 0.773 |
| Non-separability of risks and benefits | 0.807 | 0.812 | 0.786 | 0.796 | 0.813 | 0.755 |
| Omission bias | 0.590 | 0.596 | 0.612 | 0.592 | 0.575 | 0.663 |
| Outcome bias | 0.837 | 0.831 | 0.781 | 0.843 | 0.845 | 0.837 |
| Pricing study | 0.636 | 0.693 | 0.690 | 0.663 | 0.680 | 0.549 |
| Probability matching vs. maximizing | 0.762 | 0.762 | 0.758 | 0.762 | 0.762 | 0.762 |
| Sunk cost fallacy | 0.679 | 0.571 | 0.669 | 0.573 | 0.611 | 0.626 |

**Table S14 Correlations between human responses and model-generated digital twins across heuristics-and-biases tasks**

| | DeepSeek V3-0324 | DeepSeek V3.1 | DeepSeek V4-Flash | Qwen 2.5 Max | Gemma-4-31B-it | gpt-oss-120b |
|---|---|---|---|---|---|---|
| **Item-level correlation** | | | | | | |
| Overall | 0.234 | 0.276 | 0.246 | 0.226 | 0.286 | 0.188 |
| Allais | 0.024 | 0.082 | 0.025 | 0.024 | 0.036 | 0.070 |
| WTA/WTP-Thaler | 0.018 | 0.051 | 0.053 | 0.008 | 0.069 | 0.037 |
| Absolute vs. relative savings | 0.021 | 0.019 | 0.031 | -0.015 | 0.019 | 0.010 |
| Anchoring and adjustment | 0.051 | 0.071 | 0.013 | 0.016 | 0.043 | 0.006 |
| Base rate problem | 0.065 | 0.043 | 0.012 | -0.018 | -0.004 | 0.031 |
| Conjunction problem (Linda) | -0.01 | -0.001 | 0.003 | 0.029 | -0.028 | 0 |
| Denominator neglect | | -0.027 | -0.003 | -0.032 | -0.045 | 0.009 |
| False consensus | 0.285 | 0.281 | 0.295 | 0.149 | 0.317 | 0.268 |
| Framing problem | -0.004 | 0.030 | -0.034 | -0.018 | 0.041 | 0.001 |
| Less is more | 0.164 | 0.160 | 0.124 | 0.074 | 0.156 | 0.039 |
| Myside bias | 0.202 | 0.162 | 0.143 | 0.036 | 0.176 | 0.130 |
| Non-separability of risks and benefits | 0.021 | 0.083 | 0.042 | 0.014 | 0.123 | 0.035 |
| Omission bias | 0.059 | 0.148 | 0.084 | 0.046 | 0.099 | 0.067 |
| Outcome bias | 0.079 | 0.054 | 0.009 | -0.022 | | 0.023 |
| Pricing study | 0.315 | 0.394 | 0.389 | 0.374 | 0.388 | 0.246 |
| Probability matching vs. maximizing | | | 0.001 | | | |
| Sunk cost fallacy | -0.003 | -0.044 | 0.007 | -0.025 | 0.044 | 0.016 |
| **Profile correlation** | | | | | | |
| Overall | 0.353 | 0.421 | 0.409 | 0.351 | 0.393 | 0.242 |
| Allais | | | | | | |
| WTA/WTP-Thaler | | | | | | |
| Absolute vs. relative savings | | | | | | |
| Anchoring and adjustment | | | | | | |
| Base rate problem | | | | | | |
| Conjunction problem (Linda) | 0.674 | 0.591 | 0.633 | 0.761 | 0.812 | 0.468 |
| Denominator neglect | | | | | | |
| False consensus | 0.436 | 0.415 | 0.479 | 0.228 | 0.466 | 0.440 |
| Framing problem | | | | | | |
| Less is more | 0.970 | 0.982 | 0.960 | 0.970 | 0.904 | 0.782 |
| Myside bias | | | | | | |
| Non-separability of risks and benefits | 0.701 | 0.742 | 0.685 | 0.672 | 0.729 | 0.647 |
| Omission bias | | | | | | |
| Outcome bias | | | | | | |
| Pricing study | 0.337 | 0.428 | 0.425 | 0.404 | 0.417 | 0.250 |
| Probability matching vs. maximizing | | | 0.279 | | | |
| Sunk cost fallacy | | | | | | |

*Note.* Values are Spearman correlations between human participant responses and responses generated by digital twins based on different large language models. Item-level correlations were computed across item-level observations within each task. Profile correlations were computed within participants and

tasks when at least three valid items were available. Blank cells for item-level correlations indicate that the model responses showed no variance and the correlation could not be estimated. Blank cells for profile correlations generally indicate that the task included fewer than three valid items per participant; for the Probability matching vs. maximizing task, blank cells indicate no variance in model responses.

**Table S15 Replication results of heuristics and biases for different models**

| Task | Response | Outcome | | | |
|---|---|---|---|---|---|
| **Base rate problem** | | $Prob_{30}$ | $Prob_{70}$ | t | *p* |
| | Human | 52.17 | 68.01 | -15.78 | <.01 |
| | DeepSeek V3-0324 | 30.77 | 70.00 | -231 | <.001 |
| | DeepSeek V3.1 | 30.00 | 69.96 | -1014 | <.001 |
| | DeepSeek V4-Flash | 43.40 | 74.06 | -40 | <.001 |
| | Qwen 2.5 Max | 32.04 | 69.66 | -127 | <.001 |
| | Gemma-4-31B-it | 63.46 | 70.01 | -13 | <.001 |
| | gpt-oss-120b | 54.91 | 69.41 | -20 | <.001 |
| **Outcome bias** | | Mean (success) | Mean (failure) | t | *p* |
| | Human | 1.66 | 0.88 | 13.55 | <.001 |
| | DeepSeek V3-0324 | 1.93 | 1.14 | 25 | <.001 |
| | DeepSeek V3.1 | 2.01 | 0.97 | 29 | <.001 |
| | DeepSeek V4-Flash | 2.03 | 0.51 | 27 | <.001 |
| | Qwen 2.5 Max | 2.00 | 1.28 | 40 | <.001 |
| | Gemma-4-31B-it | | | | |
| | gpt-oss-120b | 1.79 | 1.83 | -2 | 0.04 |
| **Sunk cost fallacy** | | Mean (sunk cost) | Mean (no sunk cost) | t | *p* |
| | Human | 10.64 | 14.88 | 16.20 | <.001 |
| | DeepSeek V3-0324 | 10.01 | 8.69 | -18 | <.001 |
| | DeepSeek V3.1 | 9.63 | 4.82 | -30 | <.001 |
| | DeepSeek V4-Flash | 13.53 | 14.95 | -6.1 | <.001 |
| | Qwen 2.5 Max | 6.40 | 5.52 | -6.8 | <.001 |
| | Gemma-4-31B-it | 12.29 | 0.82 | 75 | <.001 |
| | gpt-oss-120b | 12.98 | 11.18 | 5.5 | <.001 |
| **Allais problem** | | Prob (A) | *p* | Prob (D) | *p* |
| | Human | 69.2% | <.001 | 57.2% | <.001 |
| | DeepSeek V3-0324 | 95.7% | <.001 | 94.2% | <.001 |
| | DeepSeek V3.1 | 99.0% | <.001 | 59.2% | <.001 |
| | DeepSeek V4-Flash | 91.15% | <.001 | 44.69% | <.001 |
| | Qwen 2.5 Max | 100% | <.001 | 85.1% | <.001 |
| | Gemma-4-31B-it | 94.1% | <.001 | 2.09% | <.001 |
| | gpt-oss-120b | 74.31% | <.001 | 5.16% | <.001 |
| **Framing problem** | | Mean (gain) | Mean (loss) | t | *p* |
| | Human | 2.85 | 3.84 | -17.35 | <.001 |
| | DeepSeek V3-0324 | 2.04 | 2.33 | -5.5 | <.001 |
| | DeepSeek V3.1 | 1.85 | 1.99 | -4.3 | <.001 |
| | DeepSeek V4-Flash | 1.64 | 3.15 | -21 | <.001 |
| | Qwen 2.5 Max | 1.00 | 3.67 | -87 | <.001 |
| | Gemma-4-31B-it | 2.04 | 2.47 | -12 | <.001 |
| | gpt-oss-120b | 2.49 | 2.89 | -12 | <.001 |

| **Conjunction problem (Linda)** | | Mean (bank teller) | Mean (feminist bank teller) | t | *p* |
| --- | --- | --- | --- | --- | --- |
| | Human | 2.43 | 3.38 | -18.83 | <.001 |
| | DeepSeek V3-0324 | 1.71 | 2.20 | -15 | <.001 |
| | DeepSeek V3.1 | 1.64 | 2.38 | -22 | <.001 |
| | DeepSeek V4-Flash | 2.46 | 4.72 | -56 | <.001 |
| | Qwen 2.5 Max | 2.00 | 4.70 | -99 | <.001 |
| | Gemma-4-31B-it | 2.00 | 4.26 | -123 | <.001 |
| | gpt-oss-120b | 2.27 | 3.50 | -21 | <.001 |
| **Anchoring and adjustment (African countries)** | | Mean (smaller anchor) | Mean (larger anchor) | t | *p* |
| | Human | 26.36 | 48.22 | 4.57 | <.001 |
| | DeepSeek V3-0324 | 45.57 | 53.03 | 24 | <.001 |
| | DeepSeek V3.1 | 43.63 | 50.22 | 28 | <.001 |
| | DeepSeek V4-Flash | 29.93 | 44.52 | 31 | <.001 |
| | Qwen 2.5 Max | 53.97 | 54.00 | 1 | 0.3 |
| | Gemma-4-31B-it | 53.97 | 53.94 | -1.1 | 0.3 |
| | gpt-oss-120b | 52.40 | 53.99 | 8.3 | <.001 |
| **Anchoring and adjustment (Redwood tree)** | | Mean (smaller anchor) | Mean (larger anchor) | t | *p* |
| | Human | 165.9 | 839.2 | 22 | <.001 |
| | DeepSeek V3-0324 | 340.8 | 422.9 | 15 | <.001 |
| | DeepSeek V3.1 | 351.4 | 427.2 | 17 | <.001 |
| | DeepSeek V4-Flash | 316.4 | 383.0 | 12 | <.001 |
| | Qwen 2.5 Max | 368.3 | 370.3 | 3.4 | <.001 |
| | Gemma-4-31B-it | 375.2 | 379.8 | 12 | <.001 |
| | gpt-oss-120b | 366.1 | 381.3 | 9.1 | <.001 |
| **Absolute vs. relative savings** | | Prob (large) | Prob (small) | $\chi^2$ | *p* |
| | Human | 0.74 | 0.34 | 319.1 | <.001 |
| | DeepSeek V3-0324 | 0.93 | 0.18 | 1190 | <.001 |
| | DeepSeek V3.1 | 0.93 | 0.26 | 953 | <.001 |
| | DeepSeek V4-Flash | 0.79 | 0.55 | 138 | <.001 |
| | Qwen 2.5 Max | 1.00 | 0.99 | 5.1 | 0.02 |
| | Gemma-4-31B-it | 0.67 | 0.05 | 841 | <.001 |
| | gpt-oss-120b | 0.67 | 0.49 | 67 | <.001 |
| **Myside bias** | | Mean (German car) | Mean (Ford explorer) | t | *p* |
| | Human | 4.46 | 4.11 | 5.9 | <.001 |
| | DeepSeek V3-0324 | 3.85 | 4.09 | 5.5 | <.001 |
| | DeepSeek V3.1 | 4.71 | 4.87 | 3.8 | <.001 |
| | DeepSeek V4-Flash | 4.75 | 4.33 | -7.9 | <.001 |
| | Qwen 2.5 Max | 4.89 | 4.99 | 6.8 | <.001 |
| | Gemma-4-31B-it | 2.99 | 3.25 | 7.9 | <.001 |
| | gpt-oss-120b | 3.97 | 4.14 | 4.9 | <.001 |

| **Less is more** | | Mean (A1) | Mean (B1) | Mean (C1) | F | *p* |
|---|---|---|---|---|---|---|
| | Human | 2.06 | 2.89 | 2.86 | 87.7 | <.001 |
| | DeepSeek V3-0324 | 2.75 | 2.33 | 2.61 | 70.8 | <.001 |
| | DeepSeek V3.1 | 2.05 | 1.87 | 1.97 | 43.1 | <.001 |
| | DeepSeek V4-Flash | 2.05 | 2.05 | 2.09 | 0.27 | 0.77 |
| | Qwen 2.5 Max | 2.12 | 3.94 | 3.55 | 2078 | <.001 |
| | Gemma-4-31B-it | 2.02 | 3.58 | 3.60 | 1589 | <.001 |
| | gpt-oss-120b | 2.75 | 3.75 | 3.79 | 319 | <.001 |
| | | Mean (A2) | Mean (B2) | Mean (C2) | F | *p* |
| | Human | 4.03 | 4.29 | 4.30 | 13.8 | <.001 |
| | DeepSeek V3-0324 | 4.42 | 4.54 | 4.56 | 16.1 | <.001 |
| | DeepSeek V3.1 | 4.54 | 4.59 | 4.63 | 3.7 | 0.026 |
| | DeepSeek V4-Flash | 4.03 | 4.30 | 4.28 | 22.7 | <.001 |
| | Qwen 2.5 Max | 4.93 | 4.99 | 5.00 | 36.9 | <.001 |
| | Gemma-4-31B-it | 3.91 | 4.38 | 4.33 | 48.6 | <.001 |
| | gpt-oss-120b | 4.17 | 4.25 | 4.25 | 4.9 | 0.008 |
| | | Mean (A3) | Mean (B3) | Mean (C3) | F | *p* |
| | Human | 4.31 | 4.67 | 4.67 | 11.2 | <.001 |
| | DeepSeek V3-0324 | 5.16 | 5.36 | 5.42 | 24.9 | <.001 |
| | DeepSeek V3.1 | 5.09 | 5.25 | 5.32 | 8.2 | <.001 |
| | DeepSeek V4-Flash | 4.26 | 4.26 | 4.31 | 0.39 | 0.68 |
| | Qwen 2.5 Max | 5.90 | 5.97 | 5.97 | 21 | <.001 |
| | Gemma-4-31B-it | 5.19 | 5.42 | 5.36 | 33.2 | <.001 |
| | gpt-oss-120b | 3.99 | 4.38 | 4.32 | 31.1 | <.001 |
| **WTA/WTP – Thaler problem** | | Mean (WTP-certainty) | Mean (WTA-certainty) | t | *p* | |
| | Human | 3.27 | 6.82 | 26.25 | <.001 | |
| | DeepSeek V3-0324 | 3.00 | 5.14 | -50 | <.001 | |
| | DeepSeek V3.1 | 3.00 | 7.51 | -103 | <.001 | |
| | DeepSeek V4-Flash | 4.49 | 8.01 | -36 | <.001 | |
| | Qwen 2.5 Max | 7.48 | 6.97 | 11 | <.001 | |
| | Gemma-4-31B-it | 3.48 | 6.81 | -46 | <.001 | |
| | gpt-oss-120b | 3.74 | 5.11 | -22 | <.001 | |
| | | Mean (WTP-certainty) | Mean (WTP-noncertainty) | t | *p* | |
| | Human | 3.27 | 2.20 | 11.4 | <.001 | |
| | DeepSeek V3-0324 | 3.00 | 2.27 | 42 | <.001 | |
| | DeepSeek V3.1 | 3.00 | 2.45 | 27 | <.001 | |
| | DeepSeek V4-Flash | 4.49 | 3.36 | 12 | <.001 | |
| | Qwen 2.5 Max | 7.48 | 5.89 | 40 | <.001 | |
| | Gemma-4-31B-it | 3.48 | 2.66 | 28 | <.001 | |
| | gpt-oss-120b | 3.74 | 1.82 | 42 | <.001 | |
| **False consensus** | | β StrongOpp | β SomewhatOpp | β SomewhatSupp | β StrongSupp | |
| | Human | -13.07*** | -8.38*** | 8.04*** | 16.65*** | |

| | | | | | |
|---|---|---|---|---|---|
| | DeepSeek V3-0324 | -17.66*** | -6.20*** | 3.88** | 13.89*** |
| | DeepSeek V3.1 | -10.13*** | -4.83** | 6.44*** | 13.78*** |
| | DeepSeek V4-Flash | -14.61*** | -6.27*** | 8.53*** | 16.83*** |
| | Qwen 2.5 Max | -7.23*** | -4.17** | 6.02** | 12.18** |
| | Gemma-4-31B-it | -5.97*** | -3.76*** | 4.26** | 8.03*** |
| | gpt-oss-120b | -6.24*** | -3.35** | 4.83*** | 8.23*** |
| **Non-separability of risk and benefits judgments** | | r (bicycles) | | t | *p* |
| | Human | 0.00 | | 0.001 | 1.00 |
| | DeepSeek V3-0324 | -0.01 | | -0.50 | 0.62 |
| | DeepSeek V3.1 | -0.36 | | -17.50 | <.001 |
| | DeepSeek V4-Flash | -0.19 | | -8.79 | <.001 |
| | Qwen 2.5 Max | -0.77 | | -54.75 | <.001 |
| | Gemma-4-31B-it | | | | |
| | gpt-oss-120b | -0.47 | | -24.07 | <.001 |
| | | r (alcohol) | | t | *p* |
| | Human | -0.33 | | -15.97 | <.001 |
| | DeepSeek V3-0324 | -0.05 | | -2.35 | 0.02 |
| | DeepSeek V3.1 | -0.28 | | -13.02 | <.001 |
| | DeepSeek V4-Flash | -0.09 | | -4.20 | <.001 |
| | Qwen 2.5 Max | 0.17 | | 7.55 | <.001 |
| | Gemma-4-31B-it | -0.20 | | -9.15 | <.001 |
| | gpt-oss-120b | -0.13 | | -6.04 | <.001 |
| | | r (pesticides) | | t | *p* |
| | Human | -0.37 | | -18.21 | <.001 |
| | DeepSeek V3-0324 | -0.39 | | -19.21 | <.001 |
| | DeepSeek V3.1 | 0.04 | | 1.59 | 0.112 |
| | DeepSeek V4-Flash | -0.29 | | -13.52 | <.001 |
| | Qwen 2.5 Max | 0.17 | | 7.90 | <.001 |
| | Gemma-4-31B-it | -0.38 | | -18.73 | <.001 |
| | gpt-oss-120b | -0.13 | | -6.01 | <.001 |
| | | r (chemical) | | t | *p* |
| | Human | -0.29 | | -13.76 | <.001 |
| | DeepSeek V3-0324 | -0.25 | | -11.72 | <.001 |
| | DeepSeek V3.1 | 0.13 | | 6.04 | <.001 |
| | DeepSeek V4-Flash | -0.26 | | -12.24 | <.001 |
| | Qwen 2.5 Max | 0.13 | | 6.05 | <.001 |
| | Gemma-4-31B-it | -0.22 | | -9.97 | <.001 |
| | gpt-oss-120b | -0.25 | | -11.93 | <.001 |
| **Omission bias** | | Prop (avoid) | | 95% CI | |
| | Human | 0.45 | | [0.43, 0.47] | |
| | DeepSeek V3-0324 | 0.01 | | [0.01, 0.02] | |
| | DeepSeek V3.1 | 0.09 | | [0.08, 0.11] | |
| | DeepSeek V4-Flash | 0.58 | | [0.56, 0.60] | |
| | Qwen 2.5 Max | 0.00 | | [0.00, 0.01] | |
| | Gemma-4-31B-it | 0.00 | | [0.00, 0.01] | |
| | gpt-oss-120b | 0.05 | | [0.05, 0.06] | |
| **Probability matching vs. maximizing** | | Prop max | 95% CI | Prop max | 95% CI |
| | Human | 0.36 | [0.334, 0.392] | 0.30 | [0.270, 0.326] |
| | DeepSeek V3-0324 | 1.00 | [0.996, 1.000] | 1 | [0.996, 1.000] |
| | DeepSeek V3.1 | 1 | [0.996, 1.000] | 1 | [0.996, 1.000] |
| | DeepSeek V4-Flash | 0.98 | [0.966, 0.984] | 0.97 | [0.959, 0.979] |
| | Qwen 2.5 Max | 1 | [0.996, 1.000] | 1 | [0.996, 1.000] |
| | Gemma-4-31B-it | 1 | [0.996, 1.000] | 1 | [0.996, 1.000] |
| | gpt-oss-120b | 1 | [0.996, 1.000] | 1 | [0.996, 1.000] |
| **Denominator neglect** | | Prop (large tray) | 95% CI | | p |
| | Human | 0.36 | [0.34, 0.38] | | <.001 |
| | DeepSeek V3-0324 | 0.00 | [0.00, 0.00] | | <.001 |
| | DeepSeek V3.1 | 0.01 | [0.01, 0.01] | | <.001 |
| | DeepSeek V4-Flash | 0.43 | [0.41, 0.45] | | <.001 |
| | Qwen 2.5 Max | 0.87 | [0.86, 0.89] | | 1 |
| | Gemma-4-31B-it | 0.99 | [0.98, 0.99] | | 1 |
| | gpt-oss-120b | 0.01 | [0.00, 0.01] | | <.001 |

*Note.* Gemma-4-31B-it results are omitted for the outcome bias and non-separability of risk and benefit judgments tasks because its responses had zero variance, so correlations and t tests could not be computed under these analyses. *** $p < .001$, ** $p < .01$, * $p < .05$.

**Table S16 Accuracy and correlations across Models in Study 2b**

| Measures | Model | Emotion Space | Race Thermometer | Responsibility Vignettes | Overall |
|---|---|---|---|---|---|
| Accuracy | | | | | |
| | gpt-oss-120b | 0.753 | 0.804 | 0.698 | 0.752 |
| | Gemma-4-31B-it | 0.792 | 0.823 | 0.742 | 0.786 |
| | Qwen 3.5 Plus | 0.781 | 0.821 | 0.720 | 0.774 |
| | DeepSeek V4-Flash | 0.791 | 0.802 | 0.663 | 0.752 |
| Item-level correlation | | | | | |
| | gpt-oss-120b | 0.309 | 0.250 | 0.143 | 0.274 |
| | Gemma-4-31B-it | 0.390 | 0.315 | 0.330 | 0.368 |
| | Qwen 3.5 Plus | 0.354 | 0.270 | 0.257 | 0.325 |
| | DeepSeek V4-Flash | 0.363 | 0.258 | 0.173 | 0.317 |
| Profile correlation | | | | | |
| | gpt-oss-120b | 0.503 | 0.354 | 0.780 | 0.525 |
| | Gemma-4-31B-it | 0.619 | 0.527 | 0.801 | 0.633 |
| | Qwen 3.5 Plus | 0.556 | 0.425 | 0.641 | 0.584 |
| | DeepSeek V4-Flash | 0.574 | 0.416 | 0.664 | 0.538 |

*Note.* Data were drawn from Wave 10. Item-level correlation and profile correlation were calculated using Spearman correlations.

**Table S17 Accuracy and correlations for Study 2b Wave 10 by date, COVID-19 context, and event cues**

| Model | Date cue | Covid context cue | Event cue | Accuracy | Item-level correlation | Profile correlation |
|---|---|---|---|---|---|---|
| DeepSeek V4-Flash | 2008-08-08 | Absent | BEIGUO | 0.767 | 0.337 | 0.572 |
| | 2008-08-08 | Present | BEIGUO | 0.766 | 0.337 | 0.567 |
| | 2079-08-08 | Absent | BEIGUO | 0.767 | 0.333 | 0.572 |
| | 2079-08-08 | Present | BEIGUO | 0.765 | 0.337 | 0.574 |
| | Absent | Absent | BEIGUO | 0.768 | 0.344 | 0.574 |
| | Absent | Present | BEIGUO | 0.767 | 0.338 | 0.569 |
| | 2020-08-08 | Present | BEIGUO | 0.769 | 0.340 | 0.580 |
| | 2020-08-08 | Absent | BEIGUO | 0.769 | 0.345 | 0.582 |
| | Absent | Absent | COVID | 0.769 | 0.340 | 0.580 |
| | Absent | Present | COVID | 0.765 | 0.338 | 0.568 |
| | 2020-08-08 | Present | COVID | 0.770 | 0.343 | 0.586 |
| | 2020-08-08 | Absent | COVID | 0.770 | 0.343 | 0.582 |
| Gemma-4-31B-it | 2008-08-08 | Absent | BEIGUO | 0.785 | 0.364 | 0.617 |
| | 2008-08-08 | Present | BEIGUO | 0.784 | 0.369 | 0.623 |
| | 2079-08-08 | Absent | BEIGUO | 0.784 | 0.364 | 0.620 |
| | 2079-08-08 | Present | BEIGUO | 0.785 | 0.365 | 0.623 |
| | Absent | Absent | BEIGUO | 0.785 | 0.366 | 0.621 |
| | Absent | Present | BEIGUO | 0.785 | 0.369 | 0.623 |
| | 2020-08-08 | Present | BEIGUO | 0.785 | 0.367 | 0.624 |
| | 2020-08-08 | Absent | BEIGUO | 0.785 | 0.369 | 0.625 |
| | Absent | Absent | COVID | 0.786 | 0.360 | 0.625 |
| | Absent | Present | COVID | 0.788 | 0.370 | 0.629 |
| | 2020-08-08 | Present | COVID | 0.785 | 0.368 | 0.626 |
| | 2020-08-08 | Absent | COVID | 0.786 | 0.366 | 0.629 |
| gpt-oss-120b | 2008-08-08 | Absent | BEIGUO | 0.736 | 0.248 | 0.433 |
| | 2008-08-08 | Present | BEIGUO | 0.739 | 0.255 | 0.455 |
| | 2079-08-08 | Absent | BEIGUO | 0.749 | 0.272 | 0.494 |
| | 2079-08-08 | Present | BEIGUO | 0.745 | 0.274 | 0.486 |
| | Absent | Absent | BEIGUO | 0.749 | 0.268 | 0.492 |
| | Absent | Present | BEIGUO | 0.747 | 0.270 | 0.488 |
| | 2020-08-08 | Present | BEIGUO | 0.747 | 0.272 | 0.495 |
| | 2020-08-08 | Absent | BEIGUO | 0.738 | 0.255 | 0.462 |
| | Absent | Absent | COVID | 0.748 | 0.269 | 0.497 |
| | Absent | Present | COVID | 0.744 | 0.259 | 0.472 |
| | 2020-08-08 | Present | COVID | 0.745 | 0.269 | 0.486 |
| | 2020-08-08 | Absent | COVID | 0.740 | 0.248 | 0.462 |
| Qwen 3.5 Plus | 2008-08-08 | Absent | BEIGUO | 0.767 | 0.301 | 0.534 |
| | 2008-08-08 | Present | BEIGUO | 0.767 | 0.301 | 0.537 |
| | 2079-08-08 | Absent | BEIGUO | 0.771 | 0.308 | 0.556 |
| | 2079-08-08 | Present | BEIGUO | 0.770 | 0.312 | 0.555 |
| | Absent | Absent | BEIGUO | 0.769 | 0.306 | 0.558 |
| | Absent | Present | BEIGUO | 0.767 | 0.294 | 0.544 |
| | 2020-08-08 | Present | BEIGUO | 0.767 | 0.303 | 0.549 |
| | 2020-08-08 | Absent | BEIGUO | 0.770 | 0.310 | 0.563 |
| | Absent | Absent | COVID | 0.771 | 0.307 | 0.565 |
| | Absent | Present | COVID | 0.768 | 0.297 | 0.544 |
| | 2020-08-08 | Present | COVID | 0.770 | 0.309 | 0.554 |
| | 2020-08-08 | Absent | COVID | 0.771 | 0.307 | 0.568 |

*Note.* The target date for Wave 10 was 2020-08-08. “Date cue” indicates the explicit date provided in the prompt, and “Absent” indicates that no explicit date cue was included. “COVID context cue” indicates whether contemporaneous COVID-19 information was included in the prompt. “Event cue” indicates the event-based temporal cue provided in the prompt. Item-level correlation and profile correlation were calculated using Spearman correlations.

**Table S18 Date manipulation check for Study 2b Wave 10: model responses to date, COVID-19 context, and event cues**

| Model | Date cue | Covid context cue | Event cue | Mean | SD | Exact target-date responses, n (%) | |
|---|---|---|---|---|---|---|---|
| DeepSeek V4-Flash | 2008-08-08 | Present | BEIGUO | 4053.49 | 1156.31 | 73 (7.52) | 2008-08-08 (n=898) \| 2020-08-08 (n=73) |
| | 2008-08-08 | Absent | BEIGUO | 4288.21 | 637.89 | 21(2.16) | 2008-08-08 (n=950) \| 2020-08-08 (n=21) |
| | 2079-08-08 | Present | BEIGUO | 21549 | 0.00 | 0 | 2079-08-08 (n=971) |
| | 2079-08-08 | Absent | BEIGUO | 21549 | 0.00 | 0 | 2079-08-08 (n=971) |
| | Absent | Absent | BEIGUO | 93.02 | 60.67 | 0 | |
| | Absent | Present | BEIGUO | 6.62 | 42.41 | 4 (0.41) | |
| | 2020-08-08 | Present | BEIGUO | 0.00 | 0.00 | 971 (100) | 2020-08-08 (n=971) |
| | 2020-08-08 | Absent | BEIGUO | 0.00 | 0.00 | 971 (100) | 2020-08-08 (n=971) |
| | Absent | Absent | COVID | 95.18 | 84.67 | 0 | |
| | Absent | Present | COVID | 4.89 | 5.02 | 0 | |
| | 2020-08-08 | Present | COVID | 0.00 | 0.00 | 971 (100) | 2020-08-08 (n=971) |
| | 2020-08-08 | Absent | COVID | 0.00 | 0.00 | 971 (100) | 2020-08-08 (n=971) |
| Gemma-4-31B-it | 2008-08-08 | Present | BEIGUO | 4383 | 0.00 | 0 | 2008-08-08 (n=971) |
| | 2008-08-08 | Absent | BEIGUO | 4383 | 0.00 | 0 | 2008-08-08 (n=971) |
| | 2079-08-08 | Present | BEIGUO | 21549 | 0.00 | 0 | 2079-08-08 (n=971) |
| | 2079-08-08 | Absent | BEIGUO | 21549 | 0.00 | 0 | 2079-08-08 (n=971) |
| | Absent | Absent | BEIGUO | 56.70 | 17.97 | 0 | |
| | Absent | Present | BEIGUO | 24.08 | 1.55 | 0 | |
| | 2020-08-08 | Present | BEIGUO | 0.00 | 0.00 | 971 (100) | 2020-08-08 (n=971) |
| | 2020-08-08 | Absent | BEIGUO | 0.00 | 0.00 | 971 (100) | 2020-08-08 (n=971) |
| | Absent | Absent | COVID | 55.45 | 17.76 | 0 | |
| | Absent | Present | COVID | 24.05 | 1.18 | 0 | |
| | 2020-08-08 | Present | COVID | 0.00 | 0.00 | 971 (100) | 2020-08-08 (n=971) |
| | 2020-08-08 | Absent | COVID | 0.00 | 0.00 | 971 (100) | 2020-08-08 (n=971) |
| gpt-oss-120b | 2008-08-08 | Present | BEIGUO | 4383 | 0.00 | 0 | 2008-08-08 (n=970) |
| | 2008-08-08 | Absent | BEIGUO | 4383 | 0.00 | 0 | 2008-08-08 (n=971) |
| | 2079-08-08 | Present | BEIGUO | 21549 | 0.00 | 0 | 2079-08-08 (n=970) |
| | 2079-08-08 | Absent | BEIGUO | 21552.39 | 105.52 | 0 | 2079-08-08 (n=970) \| 2088-08-08 (n=1) |
| | Absent | Absent | BEIGUO | 1904.44 | 527.81 | 0 | |
| | Absent | Present | BEIGUO | 847.59 | 1005.92 | 0 | |
| | 2020-08-08 | Present | BEIGUO | 0.00 | 0.00 | 971 (100) | 2020-08-08 (n=971) |
| | 2020-08-08 | Absent | BEIGUO | 0.00 | 0.00 | 971 (100) | 2020-08-08 (n=971) |
| | Absent | Absent | COVID | 1731.98 | 720.05 | 0 | |
| | Absent | Present | COVID | 269.79 | 648.23 | 0 | |
| | 2020-08-08 | Present | COVID | 0.00 | 0.00 | 971 (100) | 2020-08-08 (n=971) |
| | 2020-08-08 | Absent | COVID | 0.00 | 0.00 | 971 (100) | 2020-08-08 (n=971) |
| Qwen 3.5 Plus | 2008-08-08 | Present | BEIGUO | 4383 | 0.00 | 0 | 2008-08-08 (n=971) |
| | 2008-08-08 | Absent | BEIGUO | 4383 | 0.00 | 0 | 2008-08-08 (n=971) |
| | 2079-08-08 | Present | BEIGUO | 21549.00 | 0.00 | 0 | 2079-08-08 (n=971) |
| | 2079-08-08 | Absent | BEIGUO | 21549.00 | 0.00 | 0 | 2079-08-08 (n=971) |
| | Absent | Absent | BEIGUO | 88.80 | 17.31 | 0 | |
| | Absent | Present | BEIGUO | 7.00 | 0.00 | 0 | 2020-08-15 (n=971) |
| | 2020-08-08 | Present | BEIGUO | 0.00 | 0.00 | 971 (100) | 2020-08-08 (n=971) |
| | 2020-08-08 | Absent | BEIGUO | 0.00 | 0.00 | 971 (100) | 2020-08-08 (n=971) |
| | Absent | Absent | COVID | 94.48 | 17.67 | 0 | |
| | Absent | Present | COVID | 7.00 | 0.00 | 0 | |
| | 2020-08-08 | Present | COVID | 0.00 | 0.00 | 971 (100) | 2020-08-08 (n=971) |
| | 2020-08-08 | Absent | COVID | 0.00 | 0.00 | 971 (100) | 2020-08-08 (n=971) |

*Note.* The target date for Wave 10 was 2020-08-08. “Date cue” indicates the explicit date provided in the prompt, and “Absent” indicates that no explicit date cue was included. “COVID context cue” indicates whether contemporaneous COVID-19 information was included in the prompt. “Event cue” indicates the event-based temporal cue provided in the prompt. Mean and SD represent the absolute difference, in days, between the model-returned date and the target date of 2020-08-08. “Exact target-date responses, n (%)” reports the number and percentage of responses that returned 2020-08-08. The final column reports the distribution of returned dates. The total sample size was 971 responses per condition.

**Table S19 Event manipulation check for Study 2b Wave 10: model responses to date, COVID-19 context, and event cues**

| Model | Date cue | Covid context cue | Event cue | Target-event responses, n (%) | Target-plus-other event responses, n (%) | Non-target event responses, n (%) | Unspecified event responses, n (%) | Distinct event categories, n | Event response distribution |
|---|---|---|---|---|---|---|---|---|---|
| DeepSeek V4-Flash | 2008-08-08 | Present | BEIGUO | 965 (99.38) | 0 (0.00) | 6 (0.62) | 0 (0.00) | 3 | BEIGUO pandemic (n=965, 99.38%) \| 2008 Summer Olympics (n=1, 0.10%) \| 2008 financial crisis/Great Recession (n=5, 0.51%) |
| | 2008-08-08 | Absent | BEIGUO | 394 (40.58) | 0 (0.00) | 577 (59.42) | 0 (0.00) | 7 | BEIGUO pandemic (n=394, 40.58%) \| COVID-19 pandemic (n=2, 0.21%) \| Presidential election (n=169, 17.40%) \| 2008 Summer Olympics (n=326, 33.57%) \| 2008 financial crisis/Great Recession (n=75, 7.72%) \| No major event (n=1, 0.10%) \| No event specified (n=4, 0.41%) |
| | 2079-08-08 | Present | BEIGUO | 971 (100.00) | 0 (0.00) | 0 (0.00) | 0 (0.00) | 1 | BEIGUO pandemic (n=971, 100.00%) |
| | 2079-08-08 | Absent | BEIGUO | 971 (100.00) | 0 (0.00) | 0 (0.00) | 0 (0.00) | 1 | BEIGUO pandemic (n=971, 100.00%) |
| | Absent | Absent | BEIGUO | 929 (95.67) | 1 (0.10) | 40 (4.12) | 1 (0.10) | 4 | BEIGUO pandemic (n=929, 95.67%) \| COVID-19 pandemic (n=40, 4.12%) \| BEIGUO pandemic + COVID-19 pandemic (n=1, 0.10%) \| Uncodable/missing response (n=1, 0.10%) |
| | Absent | Present | BEIGUO | 46 (4.74) | 0 (0.00) | 925 (95.26) | 0 (0.00) | 4 | BEIGUO pandemic (n=46, 4.74%) \| COVID-19 pandemic (n=881, 90.73%) \| Black Lives Matter protests (n=43, 4.43%) \| COVID-19 pandemic + Black Lives Matter protests (n=1, 0.10%) |
| | 2020-08-08 | Present | BEIGUO | 71 (7.31) | 0 (0.00) | 900 (92.69) | 0 (0.00) | 3 | BEIGUO pandemic (n=71, 7.31%) \| COVID-19 pandemic (n=898, 92.48%) \| Black Lives Matter protests (n=2, 0.21%) |
| | 2020-08-08 | Absent | BEIGUO | 758 (78.06) | 0 (0.00) | 213 (21.94) | 0 (0.00) | 2 | BEIGUO pandemic (n=758, 78.06%) \| COVID-19 pandemic (n=213, 21.94%) |
| | Absent | Absent | COVID | 971 (100.00) | 0 (0.00) | 0 (0.00) | 0 (0.00) | 1 | COVID-19 pandemic (n=971, 100.00%) |
| | Absent | Present | COVID | 924 (95.16) | 4 (0.41) | 41 (4.22) | 2 (0.21) | 4 | COVID-19 pandemic (n=924, 95.16%) \| Black Lives Matter protests (n=41, 4.22%) \| COVID-19 pandemic + Black Lives Matter protests (n=4, 0.41%) \| Uncodable/missing response (n=2, 0.21%) |
| | 2020-08-08 | Present | COVID | 969 (99.79) | 1 (0.10) | 1 (0.10) | 0 (0.00) | 3 | COVID-19 pandemic (n=969, 99.79%) \| Black Lives Matter protests (n=1, 0.10%) \| COVID-19 pandemic + Black Lives Matter protests (n=1, 0.10%) |
| | 2020-08-08 | Absent | COVID | 971 (100.00) | 0 (0.00) | 0 (0.00) | 0 (0.00) | 1 | COVID-19 pandemic (n=971, 100.00%) |
| Gemma-4-31B-it | 2008-08-08 | Present | BEIGUO | 971 (100.00) | 0 (0.00) | 0 (0.00) | 0 (0.00) | 1 | BEIGUO pandemic (n=971, 100.00%) |
| | 2008-08-08 | Absent | BEIGUO | 971 (100.00) | 0 (0.00) | 0 (0.00) | 0 (0.00) | 1 | BEIGUO pandemic (n=971, 100.00%) |
| | 2079-08-08 | Present | BEIGUO | 971 (100.00) | 0 (0.00) | 0 (0.00) | 0 (0.00) | 1 | BEIGUO pandemic (n=971, 100.00%) |
| | 2079-08-08 | Absent | BEIGUO | 971 (100.00) | 0 (0.00) | 0 (0.00) | 0 (0.00) | 1 | BEIGUO pandemic (n=971, 100.00%) |
| | Absent | Absent | BEIGUO | 970 (99.90) | 0 (0.00) | 1 (0.10) | 0 (0.00) | 2 | BEIGUO pandemic (n=970, 99.90%) \| COVID-19 pandemic (n=1, 0.10%) |
| | Absent | Present | BEIGUO | 0 (0.00) | 0 (0.00) | 971 (100.00) | 0 (0.00) | 1 | COVID-19 pandemic (n=971, 100.00%) |
| | 2020-08-08 | Present | BEIGUO | 0 (0.00) | 0 (0.00) | 971 (100.00) | 0 (0.00) | 1 | COVID-19 pandemic (n=971, 100.00%) |
| | 2020-08-08 | Absent | BEIGUO | 970 (99.90) | 0 (0.00) | 1 (0.10) | 0 (0.00) | 2 | BEIGUO pandemic (n=970, 99.90%) | COVID-19 pandemic (n=1, 0.10%) |
| | Absent | Absent | COVID | 971 (100.00) | 0 (0.00) | 0 (0.00) | 0 (0.00) | 1 | COVID-19 pandemic (n=971, 100.00%) |
| | Absent | Present | COVID | 971 (100.00) | 0 (0.00) | 0 (0.00) | 0 (0.00) | 1 | COVID-19 pandemic (n=971, 100.00%) |
| | 2020-08-08 | Present | COVID | 971 (100.00) | 0 (0.00) | 0 (0.00) | 0 (0.00) | 1 | COVID-19 pandemic (n=971, 100.00%) |
| | 2020-08-08 | Absent | COVID | 971 (100.00) | 0 (0.00) | 0 (0.00) | 0 (0.00) | 1 | COVID-19 pandemic (n=971, 100.00%) |
| gpt-oss-120b | 2008-08-08 | Present | BEIGUO | 970 (99.90) | 0 (0.00) | 0 (0.00) | 1 (0.10) | 2 | BEIGUO pandemic (n=970, 99.90%) | Uncodable/missing response (n=1, 0.10%) |
| | 2008-08-08 | Absent | BEIGUO | 958 (98.66) | 0 (0.00) | 13 (1.34) | 0 (0.00) | 2 | BEIGUO pandemic (n=958, 98.66%) | Presidential election (n=13, 1.34%) |
| | 2079-08-08 | Present | BEIGUO | 969 (99.79) | 0 (0.00) | 1 (0.10) | 1 (0.10) | 3 | BEIGUO pandemic (n=969, 99.79%) | 2008 financial crisis/Great Recession (n=1, 0.10%) | Uncodable/missing response (n=1, 0.10%) |
| | 2079-08-08 | Absent | BEIGUO | 971 (100.00) | 0 (0.00) | 0 (0.00) | 0 (0.00) | 1 | BEIGUO pandemic (n=971, 100.00%) |
| | Absent | Absent | BEIGUO | 907 (93.41) | 0 (0.00) | 63 (6.49) | 1 (0.10) | 5 | BEIGUO pandemic (n=907, 93.41%) | COVID-19 pandemic (n=30, 3.09%) | Presidential election (n=20, 2.06%) | Midterm elections (n=13, 1.34%) | Uncodable/missing response (n=1, 0.10%) |
| | Absent | Present | BEIGUO | 512 (52.73) | 0 (0.00) | 453 (46.65) | 6 (0.62) | 5 | BEIGUO pandemic (n=512, 52.73%) | COVID-19 pandemic (n=450, 46.34%) | Black Lives Matter protests (n=2, 0.21%) | Presidential election (n=1, 0.10%) | Uncodable/missing response (n=6, 0.62%) |

| | | | | | | | | | |
|---|---|---|---|---|---|---|---|---|---|
| | 2020-08-08 | Present | BEIGUO | 149 (15.35) | 0 (0.00) | 822 (84.65) | 0 (0.00) | 3 | BEIGUO pandemic (n=149, 15.35%) \| COVID-19 pandemic (n=821, 84.55%) \| Black Lives Matter protests (n=1, 0.10%) |
| | 2020-08-08 | Absent | BEIGUO | 829 (85.38) | 0 (0.00) | 142 (14.62) | 0 (0.00) | 2 | BEIGUO pandemic (n=829, 85.38%) \| COVID-19 pandemic (n=142, 14.62%) |
| | Absent | Absent | COVID | 903 (93.00) | 0 (0.00) | 67 (6.90) | 1 (0.10) | 6 | COVID-19 pandemic (n=903, 93.00%) \| Presidential election (n=36, 3.71%) \| 2008 financial crisis/Great Recession (n=1, 0.10%) \| Midterm elections (n=29, 2.99%) \| Iowa caucus (n=1, 0.10%) \| Uncodable/missing response (n=1, 0.10%) |
| | Absent | Present | COVID | 963 (99.18) | 0 (0.00) | 4 (0.41) | 4 (0.41) | 3 | COVID-19 pandemic (n=963, 99.18%) \| Black Lives Matter protests (n=4, 0.41%) \| Uncodable/missing response (n=4, 0.41%) |
| | 2020-08-08 | Present | COVID | 970 (99.90) | 0 (0.00) | 1 (0.10) | 0 (0.00) | 2 | COVID-19 pandemic (n=970, 99.90%) \| Black Lives Matter protests (n=1, 0.10%) |
| | 2020-08-08 | Absent | COVID | 971 (100.00) | 0 (0.00) | 0 (0.00) | 0 (0.00) | 1 | COVID-19 pandemic (n=971, 100.00%) |
| Qwen 3.5 Plus | 2008-08-08 | Present | BEIGUO | 971 (100.00) | 0 (0.00) | 0 (0.00) | 0 (0.00) | 1 | BEIGUO pandemic (n=971, 100.00%) |
| | 2008-08-08 | Absent | BEIGUO | 948 (97.63) | 0 (0.00) | 23 (2.37) | 0 (0.00) | 3 | BEIGUO pandemic (n=948, 97.63%) \| Presidential election (n=5, 0.51%) \| 2008 Summer Olympics (n=18, 1.85%) |
| | 2079-08-08 | Present | BEIGUO | 971 (100.00) | 0 (0.00) | 0 (0.00) | 0 (0.00) | 1 | BEIGUO pandemic (n=971, 100.00%) |
| | 2079-08-08 | Absent | BEIGUO | 971 (100.00) | 0 (0.00) | 0 (0.00) | 0 (0.00) | 1 | BEIGUO pandemic (n=971, 100.00%) |
| | Absent | Absent | BEIGUO | 393 (40.47) | 0 (0.00) | 578 (59.53) | 0 (0.00) | 3 | BEIGUO pandemic (n=393, 40.47%) \| COVID-19 pandemic (n=574, 59.11%) \| Presidential election (n=4, 0.41%) |
| | Absent | Present | BEIGUO | 0 (0.00) | 0 (0.00) | 971 (100.00) | 0 (0.00) | 2 | COVID-19 pandemic (n=494, 50.88%) \| Black Lives Matter protests (n=477, 49.12%) |
| | 2020-08-08 | Present | BEIGUO | 0 (0.00) | 0 (0.00) | 971 (100.00) | 0 (0.00) | 2 | COVID-19 pandemic (n=964, 99.28%) \| Black Lives Matter protests (n=7, 0.72%) |
| | 2020-08-08 | Absent | BEIGUO | 1 (0.10) | 0 (0.00) | 970 (99.90) | 0 (0.00) | 2 | BEIGUO pandemic (n=1, 0.10%) \| COVID-19 pandemic (n=970, 99.90%) |
| | Absent | Absent | COVID | 971 (100.00) | 0 (0.00) | 0 (0.00) | 0 (0.00) | 1 | COVID-19 pandemic (n=971, 100.00%) |
| | Absent | Present | COVID | 358 (36.87) | 0 (0.00) | 613 (63.13) | 0 (0.00) | 2 | COVID-19 pandemic (n=358, 36.87%) \| Black Lives Matter protests (n=613, 63.13%) |
| | 2020-08-08 | Present | COVID | 950 (97.84) | 0 (0.00) | 21 (2.16) | 0 (0.00) | 2 | COVID-19 pandemic (n=950, 97.84%) \| Black Lives Matter protests (n=21, 2.16%) |
| | 2020-08-08 | Absent | COVID | 971 (100.00) | 0 (0.00) | 0 (0.00) | 0 (0.00) | 1 | COVID-19 pandemic (n=971, 100.00%) |

*Note.* The total sample size was 971 responses per condition. “Absent” indicates that the corresponding cue was not included in the prompt. For the BEIGUO event cue, target-event responses refer to the BEIGUO pandemic; for the COVID event cue, target-event responses refer to the COVID-19 pandemic or semantically equivalent references. “Target-plus-other event responses” indicates responses that included both the target event and another codable event. “Non-target event responses” indicates responses that identified an event other than the target event. “Unspecified event responses” indicates responses that did not identify a specific event or could not be coded. “Distinct event categories” reports the number of event categories observed within each condition.

**Table S20 Accuracy and correlations by date and COVID-19 context cues in Study 2b**

| Wave | Date cue | Covid context cue | Accuracy | Item-level correlation | Profile correlation |
|---|---|---|---|---|---|
| 1 | Absent | Absent | 0.741 | 0.191 | 0.432 |
| 1 | Absent | Present | 0.748 | 0.207 | 0.449 |
| 1 | Present | Present | 0.748 | 0.223 | 0.442 |
| 1 | Present | Absent | 0.746 | 0.214 | 0.446 |
| 2 | Absent | Absent | 0.751 | 0.266 | 0.475 |
| 2 | Absent | Present | 0.751 | 0.263 | 0.477 |
| 2 | Present | Present | 0.752 | 0.265 | 0.485 |
| 2 | Present | Absent | 0.755 | 0.262 | 0.486 |
| 3 | Absent | Absent | 0.745 | 0.222 | 0.456 |
| 3 | Absent | Present | 0.744 | 0.206 | 0.443 |
| 3 | Present | Present | 0.745 | 0.221 | 0.446 |
| 3 | Present | Absent | 0.746 | 0.221 | 0.454 |
| 4 | Absent | Absent | 0.755 | 0.326 | 0.518 |
| 4 | Absent | Present | 0.753 | 0.313 | 0.497 |
| 4 | Present | Present | 0.750 | 0.317 | 0.496 |
| 4 | Present | Absent | 0.753 | 0.323 | 0.509 |
| 5 | Absent | Absent | 0.749 | 0.273 | 0.481 |
| 5 | Absent | Present | 0.752 | 0.273 | 0.487 |
| 5 | Present | Present | 0.753 | 0.273 | 0.492 |
| 5 | Present | Absent | 0.753 | 0.278 | 0.489 |
| 6 | Absent | Absent | 0.744 | 0.234 | 0.470 |
| 6 | Absent | Present | 0.744 | 0.227 | 0.464 |
| 6 | Present | Present | 0.742 | 0.222 | 0.457 |
| 6 | Present | Absent | 0.745 | 0.236 | 0.471 |
| 7 | Absent | Absent | 0.726 | 0.264 | 0.361 |
| 7 | Absent | Present | 0.724 | 0.244 | 0.337 |
| 7 | Present | Present | 0.726 | 0.261 | 0.355 |
| 7 | Present | Absent | 0.722 | 0.245 | 0.336 |
| 8 | Absent | Absent | 0.734 | 0.223 | 0.459 |
| 8 | Absent | Present | 0.733 | 0.226 | 0.457 |
| 8 | Present | Present | 0.734 | 0.215 | 0.463 |
| 8 | Present | Absent | 0.735 | 0.225 | 0.464 |
| 9 | Absent | Absent | 0.744 | 0.227 | 0.482 |
| 9 | Absent | Present | 0.736 | 0.208 | 0.456 |
| 9 | Present | Present | 0.742 | 0.213 | 0.463 |
| 9 | Present | Absent | 0.745 | 0.237 | 0.488 |
| 10 | Absent | Absent | 0.769 | 0.340 | 0.580 |
| 10 | Absent | Present | 0.765 | 0.338 | 0.568 |
| 10 | Present | Present | 0.770 | 0.343 | 0.586 |
| 10 | Present | Absent | 0.770 | 0.343 | 0.582 |
| 11 | Absent | Absent | 0.749 | 0.273 | 0.484 |
| 11 | Absent | Present | 0.747 | 0.278 | 0.490 |
| 11 | Present | Present | 0.750 | 0.281 | 0.497 |
| 11 | Present | Absent | 0.753 | 0.286 | 0.502 |
| 12 | Absent | Absent | 0.743 | 0.307 | 0.498 |
| 12 | Absent | Present | 0.736 | 0.272 | 0.452 |
| 12 | Present | Present | 0.736 | 0.281 | 0.471 |
| 12 | Present | Absent | 0.743 | 0.296 | 0.491 |
| 13 | Absent | Absent | 0.732 | 0.255 | 0.460 |
| 13 | Absent | Present | 0.729 | 0.253 | 0.450 |
| 13 | Present | Present | 0.732 | 0.269 | 0.470 |
| 13 | Present | Absent | 0.733 | 0.261 | 0.467 |
| 14 | Absent | Absent | 0.740 | 0.248 | 0.471 |
| 14 | Absent | Present | 0.737 | 0.234 | 0.445 |
| 14 | Present | Present | 0.739 | 0.248 | 0.448 |
| 14 | Present | Absent | 0.739 | 0.248 | 0.449 |
| 15 | Absent | Absent | 0.729 | 0.261 | 0.466 |
| 15 | Absent | Present | 0.722 | 0.249 | 0.443 |
| 15 | Present | Present | 0.728 | 0.257 | 0.462 |
| 15 | Present | Absent | 0.733 | 0.265 | 0.471 |
| 16 | Absent | Absent | 0.743 | 0.282 | 0.483 |
| 16 | Absent | Present | 0.740 | 0.274 | 0.465 |
| 16 | Present | Present | 0.744 | 0.296 | 0.484 |
| 16 | Present | Absent | 0.746 | 0.295 | 0.472 |

*Note.* All results were generated using DeepSeek V4-Flash. "Date cue" indicates whether the wave-specific target date was explicitly provided in the prompt. "COVID context cue" indicates whether contemporaneous COVID-19 contextual information was included in the prompt. Item-level correlation and profile correlation were calculated using Spearman correlations.

**Table S21 Date manipulation check for Study 2b across 16 waves: model date responses to date and COVID-19 context cues**

| Wave | Date cue | Covid context cue | N | Target date | Mean | SD | Exact target-date responses, n (%) |
|---|---|---|---|---|---|---|---|
| 1 | Absent | Absent | 1797 | 2020/4/4 | 276.53 | 471.21 | 0 (0.00) |
| 1 | Absent | Present | 1797 | 2020/4/4 | 6.63 | 53.06 | 7 (0.39) |
| 1 | Present | Present | 1797 | 2020/4/4 | 0.00 | 0.00 | 1797 (100.00) |
| 1 | Present | Absent | 1797 | 2020/4/4 | 0.07 | 2.88 | 1796 (99.94) |
| 2 | Absent | Absent | 1498 | 2020/4/11 | 99.66 | 265.08 | 0 (0.00) |
| 2 | Absent | Present | 1498 | 2020/4/11 | 8.14 | 100.69 | 309 (20.63) |
| 2 | Present | Present | 1498 | 2020/4/11 | 0.00 | 0.00 | 1498 (100.00) |
| 2 | Present | Absent | 1498 | 2020/4/11 | 0.00 | 0.00 | 1498 (100.00) |
| 3 | Absent | Absent | 1462 | 2020/4/18 | 75.70 | 235.56 | 1 (0.07) |
| 3 | Absent | Present | 1462 | 2020/4/18 | 8.68 | 89.70 | 34 (2.33) |
| 3 | Present | Present | 1462 | 2020/4/18 | 0.00 | 0.00 | 1462 (100.00) |
| 3 | Present | Absent | 1462 | 2020/4/18 | 0.00 | 0.00 | 1462 (100.00) |
| 4 | Absent | Absent | 1323 | 2020/4/25 | 126.65 | 330.92 | 1 (0.08) |
| 4 | Absent | Present | 1323 | 2020/4/25 | 8.00 | 81.46 | 198 (14.97) |
| 4 | Present | Present | 1323 | 2020/4/25 | 0.00 | 0.00 | 1323 (100.00) |
| 4 | Present | Absent | 1323 | 2020/4/25 | 0.00 | 0.00 | 1323 (100.00) |
| 5 | Absent | Absent | 1282 | 2020/5/9 | 92.71 | 256.69 | 0 (0.00) |
| 5 | Absent | Present | 1282 | 2020/5/9 | 12.81 | 102.17 | 11 (0.86) |
| 5 | Present | Present | 1282 | 2020/5/9 | 0.00 | 0.00 | 1282 (100.00) |
| 5 | Present | Absent | 1282 | 2020/5/9 | 0.00 | 0.00 | 1282 (100.00) |
| 6 | Absent | Absent | 1166 | 2020/5/23 | 92.13 | 236.69 | 0 (0.00) |
| 6 | Absent | Present | 1166 | 2020/5/23 | 24.54 | 167.50 | 20 (1.72) |
| 6 | Present | Present | 1166 | 2020/5/23 | 0.00 | 0.00 | 1166 (100.00) |
| 6 | Present | Absent | 1166 | 2020/5/23 | 0.00 | 0.00 | 1166 (100.00) |
| 7 | Absent | Absent | 1151 | 2020/6/6 | 39.56 | 225.77 | 11 (0.96) |
| 7 | Absent | Present | 1151 | 2020/6/6 | 4.44 | 49.73 | 11 (0.96) |
| 7 | Present | Present | 1151 | 2020/6/6 | 0.00 | 0.00 | 1151 (100.00) |
| 7 | Present | Absent | 1151 | 2020/6/6 | 0.00 | 0.00 | 1151 (100.00) |
| 8 | Absent | Absent | 1044 | 2020/6/27 | 19.45 | 149.49 | 78 (7.47) |
| 8 | Absent | Present | 1044 | 2020/6/27 | 7.93 | 70.56 | 60 (5.75) |
| 8 | Present | Present | 1044 | 2020/6/27 | 0.00 | 0.00 | 1044 (100.00) |
| 8 | Present | Absent | 1044 | 2020/6/27 | 0.00 | 0.00 | 1044 (100.00) |
| 9 | Absent | Absent | 1010 | 2020/7/18 | 108.89 | 204.39 | 0 (0.00) |
| 9 | Absent | Present | 1010 | 2020/7/18 | 24.73 | 168.41 | 1 (0.10) |
| 9 | Present | Present | 1010 | 2020/7/18 | 0.00 | 0.00 | 1010 (100.00) |
| 9 | Present | Absent | 1010 | 2020/7/18 | 0.00 | 0.00 | 1010 (100.00) |
| 10 | Absent | Absent | 971 | 2020/8/8 | 95.18 | 84.67 | 0 (0.00) |
| 10 | Absent | Present | 971 | 2020/8/8 | 4.89 | 5.02 | 0 (0.00) |
| 10 | Present | Present | 971 | 2020/8/8 | 0.00 | 0.00 | 971 (100.00) |

| 10 | Present | Absent | 971 | 2020/8/8 | 0.00 | 0.00 | 971 (100.00) |
|---|---|---|---|---|---|---|---|
| 11 | Absent | Absent | 976 | 2020/8/29 | 163.90 | 261.73 | 0 (0.00) |
| 11 | Absent | Present | 976 | 2020/8/29 | 21.14 | 127.59 | 2 (0.20) |
| 11 | Present | Present | 976 | 2020/8/29 | 0.00 | 0.00 | 976 (100.00) |
| 11 | Present | Absent | 976 | 2020/8/29 | 0.00 | 0.00 | 976 (100.00) |
| 12 | Absent | Absent | 889 | 2020/9/19 | 189.83 | 323.83 | 0 (0.00) |
| 12 | Absent | Present | 889 | 2020/9/19 | 19.38 | 148.37 | 37 (4.16) |
| 12 | Present | Present | 889 | 2020/9/19 | 0.00 | 0.00 | 889 (100.00) |
| 12 | Present | Absent | 889 | 2020/9/19 | 0.00 | 0.00 | 889 (100.00) |
| 13 | Absent | Absent | 930 | 2020/10/10 | 107.96 | 277.61 | 0 (0.00) |
| 13 | Absent | Present | 930 | 2020/10/10 | 24.80 | 143.75 | 11 (1.18) |
| 13 | Present | Present | 930 | 2020/10/10 | 0.00 | 0.00 | 930 (100.00) |
| 13 | Present | Absent | 930 | 2020/10/10 | 0.00 | 0.00 | 930 (100.00) |
| 14 | Absent | Absent | 860 | 2020/11/7 | 232.63 | 478.40 | 13 (1.51) |
| 14 | Absent | Present | 860 | 2020/11/7 | 26.97 | 172.03 | 60 (6.98) |
| 14 | Present | Present | 860 | 2020/11/7 | 0.00 | 0.00 | 860 (100.00) |
| 14 | Present | Absent | 860 | 2020/11/7 | 0.00 | 0.00 | 860 (100.00) |
| 15 | Absent | Absent | 890 | 2020/12/5 | 175.08 | 288.43 | 0 (0.00) |
| 15 | Absent | Present | 890 | 2020/12/5 | 38.53 | 63.40 | 0 (0.00) |
| 15 | Present | Present | 890 | 2020/12/5 | 0.00 | 0.00 | 890 (100.00) |
| 15 | Present | Absent | 890 | 2020/12/5 | 0.00 | 0.00 | 890 (100.00) |
| 16 | Absent | Absent | 835 | 2021/1/23 | 321.94 | 453.90 | 2 (0.24) |
| 16 | Absent | Present | 835 | 2021/1/23 | 22.17 | 123.20 | 25 (2.99) |
| 16 | Present | Present | 835 | 2021/1/23 | 0.00 | 0.00 | 835 (100.00) |
| 16 | Present | Absent | 835 | 2021/1/23 | 0.00 | 0.00 | 835 (100.00) |

*Note.* All results were generated using DeepSeek V4-Flash. “Date cue” indicates whether the wave-specific target date was explicitly provided in the prompt. “COVID context cue” indicates whether contemporaneous COVID-19 contextual information was included in the prompt. Mean and SD represent the absolute difference, in days, between the model-returned date and the wave-specific target date. “Exact target-date responses, n (%)” reports the number and percentage of responses that exactly matched the target date. N indicates the number of valid model responses for each wave.

Table S22 Event manipulation check for Study 2b across 16 waves: model event responses to date and COVID-19 context cues

| Wave | Date cue | Covid context cue | Target-event responses, n (%) | Target-plus-other event responses, n (%) | Non-target event responses, n (%) | Unspecified event responses, n (%) | Distinct event categories, n | Event response distribution |
|---|---|---|---|---|---|---|---|---|
| 1 | Absent | Absent | 1791 (99.67) | 3 (0.17) | 3 (0.17) | 0 (0.00) | 4 | COVID-19 pandemic (n=1794, 99.83%) \| COVID-19 public-health restrictions/guidance (n=1, 0.06%) \| George Floyd/Black Lives Matter protests (n=1, 0.06%) \| 2020 US presidential election/Election Day (n=4, 0.22%) |
| 1 | Absent | Present | 1778 (98.94) | 19 (1.06) | 0 (0.00) | 0 (0.00) | 3 | COVID-19 pandemic (n=1793, 99.78%) \| COVID-19 case/death toll or surge context (n=4, 0.22%) \| COVID-19 public-health restrictions/guidance (n=15, 0.83%) |
| 1 | Present | Present | 1791 (99.67) | 6 (0.33) | 0 (0.00) | 0 (0.00) | 3 | COVID-19 pandemic (n=1796, 99.94%) \| COVID-19 case/death toll or surge context (n=2, 0.11%) \| COVID-19 public-health restrictions/guidance (n=6, 0.33%) |
| 1 | Present | Absent | 1797 (100.00) | 0 (0.00) | 0 (0.00) | 0 (0.00) | 1 | COVID-19 pandemic (n=1797, 100.00%) |
| 2 | Absent | Absent | 1498 (100.00) | 0 (0.00) | 0 (0.00) | 0 (0.00) | 1 | COVID-19 pandemic (n=1498, 100.00%) |
| 2 | Absent | Present | 1458 (97.33) | 38 (2.54) | 2 (0.13) | 0 (0.00) | 5 | COVID-19 pandemic (n=1496, 99.87%) \| COVID-19 case/death toll or surge context (n=35, 2.34%) \| COVID-19 public-health restrictions/guidance (n=5, 0.33%) \| Easter Saturday (n=1, 0.07%) \| high death toll (n=1, 0.07%) |
| 2 | Present | Present | 1496 (99.87) | 2 (0.13) | 0 (0.00) | 0 (0.00) | 2 | COVID-19 pandemic (n=1498, 100.00%) \| COVID-19 case/death toll or surge context (n=2, 0.13%) |
| 2 | Present | Absent | 1498 (100.00) | 0 (0.00) | 0 (0.00) | 0 (0.00) | 1 | COVID-19 pandemic (n=1498, 100.00%) |
| 3 | Absent | Absent | 1462 (100.00) | 0 (0.00) | 0 (0.00) | 0 (0.00) | 1 | COVID-19 pandemic (n=1462, 100.00%) |
| 3 | Absent | Present | 1408 (96.31) | 54 (3.69) | 0 (0.00) | 0 (0.00) | 5 | COVID-19 pandemic (n=1460, 99.86%) \| COVID-19 case/death toll or surge context (n=48, 3.28%) \| COVID-19 public-health restrictions/guidance (n=9, 0.62%) \| COVID-19 economic context/recession (n=1, 0.07%) \| CARES Act/economic stimulus (n=1, 0.07%) |

| 3 | Present | Present | 1461 (99.93) | 1 (0.07) | 0 (0.00) | 0 (0.00) | 2 | COVID-19 pandemic (n=1462, 100.00%) | COVID-19 case/death toll or surge context (n=1, 0.07%) |
|---|---|---|---|---|---|---|---|---|
| 3 | Present | Absent | 1461 (99.93) | 1 (0.07) | 0 (0.00) | 0 (0.00) | 2 | COVID-19 pandemic (n=1462, 100.00%) | COVID-19 case/death toll or surge context (n=1, 0.07%) |
| 4 | Absent | Absent | 1321 (99.85) | 2 (0.15) | 0 (0.00) | 0 (0.00) | 3 | COVID-19 pandemic (n=1323, 100.00%) | COVID-19 public-health restrictions/guidance (n=1, 0.08%) | 2020 US presidential election/Election Day (n=1, 0.08%) |
| 4 | Absent | Present | 1312 (99.17) | 11 (0.83) | 0 (0.00) | 0 (0.00) | 3 | COVID-19 pandemic (n=1323, 100.00%) | COVID-19 case/death toll or surge context (n=7, 0.53%) | COVID-19 public-health restrictions/guidance (n=5, 0.38%) |
| 4 | Present | Present | 1323 (100.00) | 0 (0.00) | 0 (0.00) | 0 (0.00) | 1 | COVID-19 pandemic (n=1323, 100.00%) |
| 4 | Present | Absent | 1321 (99.85) | 2 (0.15) | 0 (0.00) | 0 (0.00) | 2 | COVID-19 pandemic (n=1323, 100.00%) | COVID-19 public-health restrictions/guidance (n=2, 0.15%) |
| 5 | Absent | Absent | 1282 (100.00) | 0 (0.00) | 0 (0.00) | 0 (0.00) | 1 | COVID-19 pandemic (n=1282, 100.00%) |
| 5 | Absent | Present | 1265 (98.67) | 16 (1.25) | 0 (0.00) | 1 (0.08) | 3 | COVID-19 pandemic (n=1281, 99.92%) | COVID-19 case/death toll or surge context (n=14, 1.09%) | COVID-19 public-health restrictions/guidance (n=2, 0.16%) |
| 5 | Present | Present | 1282 (100.00) | 0 (0.00) | 0 (0.00) | 0 (0.00) | 1 | COVID-19 pandemic (n=1282, 100.00%) |
| 5 | Present | Absent | 1282 (100.00) | 0 (0.00) | 0 (0.00) | 0 (0.00) | 1 | COVID-19 pandemic (n=1282, 100.00%) |
| 6 | Absent | Absent | 1166 (100.00) | 0 (0.00) | 0 (0.00) | 0 (0.00) | 1 | COVID-19 pandemic (n=1166, 100.00%) |
| 6 | Absent | Present | 1140 (97.77) | 26 (2.23) | 0 (0.00) | 0 (0.00) | 5 | COVID-19 pandemic (n=1165, 99.91%) | COVID-19 case/death toll or surge context (n=19, 1.63%) | COVID-19 public-health restrictions/guidance (n=11, 0.94%) | COVID-19 economic context/recession (n=3, 0.26%) | CARES Act/economic stimulus (n=1, 0.09%) |
| 6 | Present | Present | 1165 (99.91) | 1 (0.09) | 0 (0.00) | 0 (0.00) | 3 | COVID-19 pandemic (n=1166, 100.00%) | COVID-19 case/death toll or surge context (n=1, 0.09%) | COVID-19 public-health restrictions/guidance (n=1, 0.09%) |

| | | | | | | | | |
|---|---|---|---|---|---|---|---|---|
| 6 | Present | Absent | 1165 (99.91) | 1 (0.09) | 0 (0.00) | 0 (0.00) | 2 | COVID-19 pandemic (n=1166, 100.00%) \| COVID-19 public-health restrictions/guidance (n=1, 0.09%) |
| 7 | Absent | Absent | 6 (0.52) | 14 (1.22) | 1131 (98.26) | 0 (0.00) | 2 | COVID-19 pandemic (n=20, 1.74%) \| George Floyd/Black Lives Matter protests (n=1145, 99.48%) |
| 7 | Absent | Present | 0 (0.00) | 29 (2.52) | 1122 (97.48) | 0 (0.00) | 4 | COVID-19 pandemic (n=28, 2.43%) \| COVID-19 public-health restrictions/guidance (n=1, 0.09%) \| George Floyd/Black Lives Matter protests (n=1150, 99.91%) \| Other protests (n=1, 0.09%) |
| 7 | Present | Present | 0 (0.00) | 24 (2.09) | 1127 (97.91) | 0 (0.00) | 2 | COVID-19 pandemic (n=24, 2.09%) \| George Floyd/Black Lives Matter protests (n=1151, 100.00%) |
| 7 | Present | Absent | 0 (0.00) | 3 (0.26) | 1148 (99.74) | 0 (0.00) | 2 | COVID-19 pandemic (n=3, 0.26%) \| George Floyd/Black Lives Matter protests (n=1151, 100.00%) |
| 8 | Absent | Absent | 17 (1.63) | 58 (5.56) | 969 (92.82) | 0 (0.00) | 3 | COVID-19 pandemic (n=75, 7.18%) \| George Floyd/Black Lives Matter protests (n=1024, 98.08%) \| Other protests (n=3, 0.29%) |
| 8 | Absent | Present | 17 (1.63) | 206 (19.73) | 821 (78.64) | 0 (0.00) | 5 | COVID-19 pandemic (n=223, 21.36%) \| COVID-19 case/death toll or surge context (n=1, 0.10%) \| COVID-19 economic context/recession (n=1, 0.10%) \| George Floyd/Black Lives Matter protests (n=1024, 98.08%) \| Other protests (n=2, 0.19%) |
| 8 | Present | Present | 173 (16.57) | 332 (31.80) | 539 (51.63) | 0 (0.00) | 5 | COVID-19 pandemic (n=504, 48.28%) \| COVID-19 case/death toll or surge context (n=2, 0.19%) \| COVID-19 economic context/recession (n=1, 0.10%) \| George Floyd/Black Lives Matter protests (n=869, 83.24%) \| Other protests (n=1, 0.10%) |
| 8 | Present | Absent | 31 (2.97) | 198 (18.97) | 815 (78.07) | 0 (0.00) | 4 | COVID-19 pandemic (n=229, 21.93%) \| COVID-19 case/death toll or surge context (n=5, 0.48%) \| George Floyd/Black Lives Matter protests (n=1010, 96.74%) \| Other protests (n=3, 0.29%) |
| 9 | Absent | Absent | 1008 (99.80) | 0 (0.00) | 2 (0.20) | 0 (0.00) | 2 | COVID-19 pandemic (n=1008, 99.80%) \| 2020 US presidential election/Election Day (n=2, 0.20%) |

| 9 | Absent | Present | 582 (57.62) | 88 (8.71) | 340 (33.66) | 0 (0.00) | 5 | COVID-19 pandemic (n=665, 65.84%) \| COVID-19 case/death toll or surge context (n=24, 2.38%) \| COVID-19 economic context/recession (n=21, 2.08%) \| George Floyd/Black Lives Matter protests (n=394, 39.01%) \| Fourth of July celebrations (n=1, 0.10%) |
|---|---|---|---|---|---|---|---|---|
| 9 | Present | Present | 948 (93.86) | 19 (1.88) | 43 (4.26) | 0 (0.00) | 5 | COVID-19 pandemic (n=966, 95.64%) \| COVID-19 case/death toll or surge context (n=2, 0.20%) \| COVID-19 economic context/recession (n=5, 0.50%) \| George Floyd/Black Lives Matter protests (n=54, 5.35%) \| Other protests (n=1, 0.10%) |
| 9 | Present | Absent | 1010 (100.00) | 0 (0.00) | 0 (0.00) | 0 (0.00) | 1 | COVID-19 pandemic (n=1010, 100.00%) |
| 10 | Absent | Absent | 970 (99.90) | 1 (0.10) | 0 (0.00) | 0 (0.00) | 2 | COVID-19 pandemic (n=971, 100.00%) \| COVID-19 public-health restrictions/guidance (n=1, 0.10%) |
| 10 | Absent | Present | 924 (95.16) | 4 (0.41) | 41 (4.22) | 2 (0.21) | 2 | COVID-19 pandemic (n=928, 95.57%) \| George Floyd/Black Lives Matter protests (n=45, 4.63%) |
| 10 | Present | Present | 969 (99.79) | 1 (0.10) | 1 (0.10) | 0 (0.00) | 2 | COVID-19 pandemic (n=970, 99.90%) \| George Floyd/Black Lives Matter protests (n=2, 0.21%) |
| 10 | Present | Absent | 971 (100.00) | 0 (0.00) | 0 (0.00) | 0 (0.00) | 1 | COVID-19 pandemic (n=971, 100.00%) |
| 11 | Absent | Absent | 976 (100.00) | 0 (0.00) | 0 (0.00) | 0 (0.00) | 1 | COVID-19 pandemic (n=976, 100.00%) |
| 11 | Absent | Present | 514 (52.66) | 437 (44.77) | 25 (2.56) | 0 (0.00) | 6 | COVID-19 pandemic (n=950, 97.34%) \| COVID-19 case/death toll or surge context (n=404, 41.39%) \| COVID-19 public-health restrictions/guidance (n=1, 0.10%) \| COVID-19 economic context/recession (n=13, 1.33%) \| George Floyd/Black Lives Matter protests (n=50, 5.12%) \| Other protests (n=1, 0.10%) |
| 11 | Present | Present | 953 (97.64) | 9 (0.92) | 14 (1.43) | 0 (0.00) | 4 | COVID-19 pandemic (n=962, 98.57%) \| COVID-19 public-health restrictions/guidance (n=1, 0.10%) \| COVID-19 economic context/recession (n=1, 0.10%) \| George Floyd/Black Lives Matter protests (n=21, 2.15%) |
| 11 | Present | Absent | 976 (100.00) | 0 (0.00) | 0 (0.00) | 0 (0.00) | 1 | COVID-19 pandemic (n=976, 100.00%) |
| 12 | Absent | Absent | 885 (99.55) | 1 (0.11) | 2 (0.22) | 1 (0.11) | 4 | COVID-19 pandemic (n=884, 99.66%) \| COVID-19 public-health restrictions/guidance (n=1, 0.11%) \| George Floyd/Black Lives Matter protests (n=1, 0.11%) \| 2020 US presidential election/Election Day (n=1, 0.11%) |
| 12 | Absent | Present | 277 (31.16) | 41 (4.61) | 571 (64.23) | 0 (0.00) | 10 | COVID-19 pandemic (n=314, 35.32%) \| COVID-19 case/death toll or surge context (n=12, 1.35%) \| COVID-19 public-health restrictions/guidance (n=1, 0.11%) \| COVID-19 economic context/recession (n=4, 0.45%) \| COVID-19 vaccine authorization/distribution (n=6, 0.67%) \| Trump COVID-19 diagnosis (n=3, 0.34%) \| George Floyd/Black Lives Matter protests (n=50, 5.62%) \| 2020 US presidential election/Election Day (n=9, 1.01%) \| Justice Ruth Bader Ginsburg death (n=542, 60.97%) \| Other protests (n=2, 0.22%) |
| 12 | Present | Present | 549 (61.75) | 13 (1.46) | 327 (36.78) | 0 (0.00) | 4 | COVID-19 pandemic (n=562, 63.22%) \| COVID-19 case/death toll or surge context (n=13, 1.46%) \| George Floyd/Black Lives Matter protests (n=9, 1.01%) \| Justice Ruth Bader Ginsburg death (n=331, 37.23%) |
| 12 | Present | Absent | 885 (99.55) | 3 (0.34) | 1 (0.11) | 0 (0.00) | 2 | COVID-19 pandemic (n=888, 99.89%) \| 2020 US presidential election/Election Day (n=4, 0.45%) |
| 13 | Absent | Absent | 929 (99.89) | 0 (0.00) | 0 (0.00) | 1 (0.11) | 1 | COVID-19 pandemic (n=929, 99.89%) |
| 13 | Absent | Present | 325 (34.95) | 548 (58.92) | 56 (6.02) | 1 (0.11) | 10 | COVID-19 pandemic (n=858, 92.26%) \| COVID-19 case/death toll or surge context (n=43, 4.62%) \| COVID-19 economic context/recession (n=9, 0.97%) \| COVID-19 vaccine authorization/distribution (n=19, 2.04%) \| Trump COVID-19 diagnosis (n=466, 50.11%) \| CARES Act/economic stimulus (n=1, 0.11%) \| George Floyd/Black Lives Matter protests (n=43, 4.62%) \| 2020 US presidential election/Election Day (n=23, 2.47%) \| Justice Ruth Bader Ginsburg death (n=21, 2.26%) \| Other protests (n=1, 0.11%) |
| 13 | Present | Present | 828 (89.03) | 96 (10.32) | 6 (0.65) | 0 (0.00) | 8 | COVID-19 pandemic (n=923, 99.25%) \| COVID-19 case/death toll or surge context (n=2, 0.22%) \| COVID-19 economic context/recession (n=1, 0.11%) \| COVID-19 |
| | | | | | | | | vaccine authorization/distribution (n=1, 0.11%) \| Trump COVID-19 diagnosis (n=87, 9.35%) \| George Floyd/Black Lives Matter protests (n=6, 0.65%) \| 2020 US presidential election/Election Day (n=9, 0.97%) \| Justice Ruth Bader Ginsburg death (n=1, 0.11%) |
| 13 | Present | Absent | 930 (100.00) | 0 (0.00) | 0 (0.00) | 0 (0.00) | 1 | COVID-19 pandemic (n=930, 100.00%) |
| 14 | Absent | Absent | 468 (54.42) | 53 (6.16) | 339 (39.42) | 0 (0.00) | 4 | COVID-19 pandemic (n=521, 60.58%) \| 2020 US presidential election/Election Day (n=390, 45.35%) \| Other election-related event (n=1, 0.12%) \| January 6 US Capitol attack/riot (n=1, 0.12%) |
| 14 | Absent | Present | 24 (2.79) | 31 (3.60) | 804 (93.49) | 1 (0.12) | 6 | COVID-19 pandemic (n=54, 6.28%) \| COVID-19 case/death toll or surge context (n=13, 1.51%) \| COVID-19 vaccine authorization/distribution (n=1, 0.12%) \| Trump COVID-19 diagnosis (n=5, 0.58%) \| George Floyd/Black Lives Matter protests (n=3, 0.35%) \| 2020 US presidential election/Election Day (n=825, 95.93%) |
| 14 | Present | Present | 60 (6.98) | 28 (3.26) | 772 (89.77) | 0 (0.00) | 3 | COVID-19 pandemic (n=88, 10.23%) \| COVID-19 case/death toll or surge context (n=8, 0.93%) \| 2020 US presidential election/Election Day (n=800, 93.02%) |
| 14 | Present | Absent | 0 (0.00) | 2 (0.23) | 858 (99.77) | 0 (0.00) | 2 | COVID-19 pandemic (n=2, 0.23%) \| 2020 US presidential election/Election Day (n=860, 100.00%) |
| 15 | Absent | Absent | 888 (99.78) | 1 (0.11) | 1 (0.11) | 0 (0.00) | 3 | COVID-19 pandemic (n=889, 99.89%) \| COVID-19 public-health restrictions/guidance (n=1, 0.11%) \| 2020 US presidential election/Election Day (n=1, 0.11%) |
| 15 | Absent | Present | 50 (5.62) | 73 (8.20) | 767 (86.18) | 0 (0.00) | 10 | COVID-19 pandemic (n=109, 12.25%) \| COVID-19 case/death toll or surge context (n=30, 3.37%) \| COVID-19 economic context/recession (n=2, 0.22%) \| COVID-19 vaccine authorization/distribution (n=22, 2.47%) \| Trump COVID-19 diagnosis (n=17, 1.91%) \| George Floyd/Black Lives Matter protests (n=9, 1.01%) \| 2020 US presidential election/Election Day (n=744, 83.60%) \| Justice Ruth Bader Ginsburg death (n=41, 4.61%) \| September 2020 (n=1, 0.11%) \| Thanksgiving (n=1, 0.11%) |
| 15 | Present | Present | 871 (97.87) | 15 (1.69) | 4 (0.45) | 0 (0.00) | 6 | COVID-19 pandemic (n=886, 99.55%) \| COVID-19 case/death toll or surge context (n=8, 0.90%) \| COVID-19 economic context/recession (n=1, 0.11%) \| COVID-19 vaccine authorization/distribution (n=1, 0.11%) \| George Floyd/Black Lives Matter protests (n=1, 0.11%) \| 2020 US presidential election/Election Day (n=15, 1.69%) |
| 15 | Present | Absent | 889 (99.89) | 1 (0.11) | 0 (0.00) | 0 (0.00) | 2 | COVID-19 pandemic (n=890, 100.00%) \| 2020 US presidential election/Election Day (n=1, 0.11%) |
| 16 | Absent | Absent | 648 (77.60) | 65 (7.78) | 122 (14.61) | 0 (0.00) | 9 | COVID-19 pandemic (n=711, 85.15%) \| COVID-19 vaccine authorization/distribution (n=4, 0.48%) \| 2020 US presidential election/Election Day (n=129, 15.45%) \| Other election-related event (n=2, 0.24%) \| Biden inauguration (n=7, 0.84%) \| January 6 US Capitol attack/riot (n=63, 7.54%) \| Donald Trump impeachment (n=2, 0.24%) \| Other protests (n=9, 1.08%) \| Q-Anon controversy (n=1, 0.12%) |
| 16 | Absent | Present | 37 (4.43) | 56 (6.71) | 742 (88.86) | 0 (0.00) | 12 | COVID-19 pandemic (n=88, 10.54%) \| COVID-19 case/death toll or surge context (n=7, 0.84%) \| COVID-19 economic context/recession (n=1, 0.12%) \| COVID-19 vaccine authorization/distribution (n=41, 4.91%) \| Trump COVID-19 diagnosis (n=4, 0.48%) \| George Floyd/Black Lives Matter protests (n=2, 0.24%) \| 2020 US presidential election/Election Day (n=20, 2.40%) \| Justice Ruth Bader Ginsburg death (n=2, 0.24%) \| Biden inauguration (n=668, 80.00%) \| January 6 US Capitol attack/riot (n=70, 8.38%) \| Donald Trump impeachment (n=8, 0.96%) \| Other protests (n=1, 0.12%) |
| 16 | Present | Present | 8 (0.96) | 12 (1.44) | 815 (97.60) | 0 (0.00) | 5 | COVID-19 pandemic (n=20, 2.40%) \| COVID-19 case/death toll or surge context (n=1, 0.12%) \| 2020 US presidential election/Election Day (n=6, 0.72%) \| Biden inauguration (n=818, 97.96%) \| January 6 US Capitol attack/riot (n=10, 1.20%) |
| 16 | Present | Absent | 242 (28.98) | 103 (12.34) | 490 (58.68) | 0 (0.00) | 9 | COVID-19 pandemic (n=345, 41.32%) \| COVID-19 case/death toll or surge context (n=1, 0.12%) \| COVID-19 vaccine authorization/distribution (n=3, 0.36%) \| Trump |

COVID-19 diagnosis (n=1, 0.12%) | 2020 US presidential election/Election Day (n=364, 43.59%) | Biden inauguration (n=210, 25.15%) | January 6 US Capitol attack/riot (n=238, 28.50%) | Donald Trump impeachment (n=9, 1.08%) | Other protests (n=11, 1.32%)

*Note.* All results were generated using DeepSeek V4-Flash. “Date cue” indicates whether the wave-specific target date was explicitly provided in the prompt. “COVID context cue” indicates whether contemporaneous COVID-19 contextual information was included in the prompt. The target event was the COVID-19 pandemic. “Target-event responses” refers to responses coded as mentioning the COVID-19 pandemic only. “Target-plus-other event responses” refers to responses that mentioned the COVID-19 pandemic together with at least one additional codable event or contextual category. “Non-target event responses” refers to responses that mentioned codable events other than the COVID-19 pandemic. “Unspecified event responses” refers to responses that did not identify a specific codable event or could not be coded. The four response-type columns are mutually exclusive and sum to N within each wave-condition combination. “Distinct event categories” and “Event response distribution” are based on normalized multi-label coding; therefore, a single response could be assigned to multiple event or contextual categories, and percentages in the final column are not mutually exclusive and may sum to more than 100%. For example, “COVID-19 pandemic (n = 1498, 100.00%) | COVID-19 case/death toll or surge context (n = 2, 0.13%)” indicates that all 1498 responses mentioned the COVID-19 pandemic, and two of those responses also mentioned case counts, death counts, or surge-related contextual information.

**Table S23 Nomothetic span of big five traits across Study 3a and Study 3b**

| Scale | Extraversion | Agreeableness | Conscientiousness | Openness | Neuroticism |
|---|---|---|---|---|---|
| **Twin-2K-500 dataset** | | | | | |
| Need for cognition scale (Cacioppo et al., 1984) | 0.284*** | 0.187*** | 0.288*** | 0.619*** | -0.253*** |
| Agentic vs. Communal Values scale (Trapnell and Paulhus, 2012) | | | | | |
| Agency | 0.304*** | -0.044* | 0.094*** | 0.152*** | -0.090*** |
| Communion | 0.157*** | 0.508*** | 0.238*** | 0.098*** | -0.151*** |
| Consumer Minimalism scale (Wilson and Bellezza, 2022) | 0.030 | 0.132*** | 0.282*** | 0.071** | -0.177*** |
| Empathy scale (Carré et al., 2013) | 0.045* | 0.304*** | 0.096*** | 0.221*** | 0.167*** |
| Green values scale (Haws et al., 2014) | 0.155*** | 0.241*** | 0.190*** | 0.330*** | -0.121*** |
| Social Desirability scale (Reynolds, 1982) | 0.166*** | 0.442*** | 0.366*** | 0.029 | -0.409*** |
| Conscientiousness scale (Johnson et al., 2019) | 0.214*** | 0.320*** | 0.716*** | 0.142*** | -0.409*** |
| Anxiety scale (Beck et al., 1988) | -0.136*** | -0.211*** | -0.356*** | -0.020 | 0.510*** |
| Individualism vs. Collectivism scale (Triandis and Gelfand, 1998) | | | | | |
| horizontal_collectivism | 0.339*** | 0.539*** | 0.338*** | 0.186*** | -0.273*** |
| horizontal_individualism | 0.021 | 0.001 | 0.255*** | 0.189*** | -0.146*** |
| vertical_collectivism | 0.161*** | 0.305*** | 0.223*** | -0.043 | -0.173*** |
| vertical_individualism | 0.194*** | -0.144*** | 0.047* | 0.052* | 0.007 |
| Regulatory Focus scale (Fellner et al., 2007) | 0.302*** | 0.201*** | 0.237*** | 0.365*** | -0.098*** |
| Tightwads vs. Spendthrift scale (Rick et al., 2008) | 0.076*** | -0.058** | -0.190*** | -0.038 | 0.164*** |
| Depression scale (Date, 1987) | -0.293*** | -0.301*** | -0.476*** | -0.078*** | 0.619*** |
| Need for uniqueness scale (Ruvio et al., 2008) | 0.185*** | -0.086*** | -0.059** | 0.330*** | 0.050* |
| Self-monitoring scale (Lennox and Wolfe, 1984) | 0.260*** | 0.145*** | 0.168*** | 0.210*** | -0.132*** |
| Self-concept clarity scale (Campbell et al., 1996) | 0.220*** | 0.306*** | 0.522*** | 0.088*** | -0.564*** |
| Need for closure scale (Roets and Van Hiel, 2011) | -0.190*** | -0.103*** | -0.036 | -0.229*** | 0.305*** |
| Maximization scale (Nenkov et al., 2008) | -0.025 | -0.084*** | -0.067** | 0.079*** | 0.144*** |
| **COVID-Dynamic dataset** | | | | | |
| | Extraversion | Agreeableness | Conscientiousness | Openness | Neuroticism |
| Beck Depression Inventory-II (BDI-II) | -0.488*** | -0.234*** | -0.401*** | 0.058 | 0.710*** |
| Connor-Davidson Resilience Scale (CD-RISC-10) | 0.564*** | 0.303*** | 0.558*** | 0.154*** | -0.727*** |
| Iowa Personality Disorder Screen (IPDS) | -0.502*** | -0.472*** | -0.326*** | -0.025 | 0.641*** |
| Perceived Stress Scale (PSS) | -0.479*** | -0.296*** | -0.458*** | 0.059 | 0.763*** |
| Positive and Negative Affect Scales (PANAS) | | | | | |
| Positive affect | 0.535*** | 0.256*** | 0.482*** | 0.116** | -0.494*** |
| Negative affect | -0.223*** | -0.221*** | -0.228*** | -0.024 | 0.484*** |
| Scale of Ethnocultural Empathy (SEE) | | | | | |
| Empathic feeling | 0.144*** | 0.304*** | 0.104** | 0.521*** | -0.013 |
| Perspective taking | 0.123** | 0.086* | 0.133*** | 0.370*** | -0.108** |
| Acceptance of differences | 0.014 | 0.336*** | 0.047 | 0.431*** | -0.034 |
| Empathic awareness | -0.056 | 0.168*** | -0.023 | 0.398*** | 0.117** |
| State-Trait Anxiety Inventory (STAI) | | | | | |
| State anxiety | -0.438*** | -0.287*** | -0.402*** | -0.015 | 0.637*** |
| Trait anxiety | -0.590*** | -0.348*** | -0.538*** | 0.018 | 0.884*** |
| Very Short Authoritarianism Scale (VSA) | 0.131*** | -0.048 | 0.137*** | -0.435*** | -0.109** |

*Note.* *** $p < .001$, ** $p < .01$, * $p < .05$.

**Table S24 Accuracy of digital twin responses to the Big Five Inventory across input conditions and models**

| Input condition | Model | Extraversion | Agreeableness | Conscientiousness | Neuroticism | Openness | Overall |
|---|---|---|---|---|---|---|---|
| Demographics | Gemma-4-31B-it | 0.710 | 0.789 | 0.795 | 0.736 | 0.761 | 0.760 |
| Demographics | gpt-oss-120b | 0.671 | 0.770 | 0.779 | 0.693 | 0.719 | 0.728 |
| Demographics | DeepSeek V4-Flash | 0.689 | 0.783 | 0.781 | 0.715 | 0.712 | 0.737 |
| Demographics | Qwen 3.5 Plus | 0.685 | 0.781 | 0.785 | 0.728 | 0.744 | 0.746 |
| Heuristics and biases experiments | Gemma-4-31B-it | 0.717 | 0.788 | 0.771 | 0.724 | 0.777 | 0.758 |
| Heuristics and biases experiments | gpt-oss-120b | 0.669 | 0.756 | 0.765 | 0.656 | 0.743 | 0.721 |
| Heuristics and biases experiments | DeepSeek V4-Flash | 0.671 | 0.769 | 0.77 | 0.697 | 0.749 | 0.734 |
| Heuristics and biases experiments | Qwen 3.5 Plus | 0.705 | 0.756 | 0.786 | 0.717 | 0.723 | 0.738 |
| Demographics + Heuristics and biases experiments | Gemma-4-31B-it | 0.713 | 0.793 | 0.794 | 0.736 | 0.767 | 0.762 |
| Demographics + Heuristics and biases experiments | gpt-oss-120b | 0.678 | 0.766 | 0.777 | 0.682 | 0.726 | 0.728 |
| Demographics + Heuristics and biases experiments | DeepSeek V4-Flash | 0.682 | 0.784 | 0.778 | 0.717 | 0.729 | 0.740 |
| Demographics + Heuristics and biases experiments | Qwen 3.5 Plus | 0.711 | 0.781 | 0.790 | 0.729 | 0.719 | 0.746 |
| Demographics + All psychological measures + Heuristics and biases experiments | Gemma-4-31B-it | 0.767 | 0.824 | 0.842 | 0.771 | 0.811 | 0.805 |
| Demographics + All psychological measures + Heuristics and biases experiments | gpt-oss-120b | 0.712 | 0.795 | 0.815 | 0.760 | 0.762 | 0.770 |
| Demographics + All psychological measures + Heuristics and biases experiments | DeepSeek V4-Flash | 0.733 | 0.816 | 0.832 | 0.778 | 0.768 | 0.787 |
| Demographics + All psychological measures + Heuristics and biases experiments | Qwen 3.5 Plus | 0.762 | 0.83 | 0.841 | 0.777 | 0.778 | 0.799 |
| Demographics + Heuristics and biases experiments + Features with $r < 0.5$ | Gemma-4-31B-it | 0.738 | 0.809 | 0.795 | 0.767 | 0.806 | 0.785 |
| Demographics + Heuristics and biases experiments + Features with $r < 0.5$ | gpt-oss-120b | 0.705 | 0.790 | 0.788 | 0.721 | 0.765 | 0.756 |
| Demographics + Heuristics and biases experiments + Features with $r < 0.5$ | DeepSeek V4-Flash | 0.703 | 0.803 | 0.797 | 0.735 | 0.762 | 0.762 |
| Demographics + Heuristics and biases experiments + Features with r < 0.5 | Qwen 3.5 Plus | 0.730 | 0.814 | 0.793 | 0.756 | 0.771 | 0.774 |
| Demographics + Heuristics and biases experiments + Features with r > 0.5 | Gemma-4-31B-it | 0.763 | 0.820 | 0.850 | 0.775 | 0.804 | 0.804 |
| Demographics + Heuristics and biases experiments + Features with r > 0.5 | gpt-oss-120b | 0.719 | 0.792 | 0.825 | 0.766 | 0.723 | 0.765 |
| Demographics + Heuristics and biases experiments + Features with r > 0.5 | DeepSeek V4-Flash | 0.735 | 0.817 | 0.838 | 0.772 | 0.759 | 0.785 |
| Demographics + Heuristics and biases experiments + Features with r > 0.5 | Qwen 3.5 Plus | 0.759 | 0.822 | 0.845 | 0.771 | 0.764 | 0.793 |
| Demographics + Heuristics and biases experiments + Features with r < 0.3 | Gemma-4-31B-it | 0.730 | 0.798 | 0.812 | 0.744 | 0.778 | 0.774 |
| Demographics + Heuristics and biases experiments + Features with r < 0.3 | gpt-oss-120b | 0.700 | 0.783 | 0.800 | 0.711 | 0.765 | 0.754 |
| Demographics + Heuristics and biases experiments + Features with r < 0.3 | DeepSeek V4-Flash | 0.694 | 0.790 | 0.796 | 0.724 | 0.754 | 0.754 |
| Demographics + Heuristics and biases experiments + Features with r < 0.3 | Qwen 3.5 Plus | 0.729 | 0.796 | 0.807 | 0.739 | 0.749 | 0.765 |

**Table S25 Correlations of digital twin responses to the Big Five Inventory across input conditions and models**

| **Feature** | **Model** | **Extraversion** | | **Agreeableness** | | **Conscientiousness** | | **Neuroticism** | | **Openness** | | **Overall** | |
|---|---|---|---|---|---|---|---|---|---|---|---|---|---|
| Item-level correlation | | Pearson | Spearman | Pearson | Spearman | Pearson | Spearman | Pearson | Spearman | Pearson | Spearman | Pearson | Spearman |
| Demographics | Gemma-4-31B-it | 0.066 | 0.062 | 0.073 | 0.071 | 0.194 | 0.192 | 0.221 | 0.219 | 0.093 | 0.099 | 0.128 | 0.128 |
| | gpt-oss-120b | 0.003 | 0.002 | 0.036 | 0.033 | 0.176 | 0.155 | 0.122 | 0.107 | 0.081 | 0.084 | 0.085 | 0.077 |
| | DeepSeek V4-Flash | 0.049 | 0.047 | 0.073 | 0.076 | 0.145 | 0.133 | 0.180 | 0.175 | 0.090 | 0.093 | 0.107 | 0.105 |
| | Qwen 3.5 Plus | 0.072 | 0.067 | 0.100 | 0.094 | 0.145 | 0.144 | 0.193 | 0.191 | 0.101 | 0.103 | 0.122 | 0.119 |
| Heuristics and biases experiments | Gemma-4-31B-it | -0.014 | -0.015 | 0.060 | 0.067 | 0.062 | 0.072 | 0.041 | 0.041 | 0.107 | 0.114 | 0.054 | 0.059 |
| | gpt-oss-120b | 0.017 | 0.014 | 0.073 | 0.081 | 0.055 | 0.069 | -0.007 | -0.003 | 0.083 | 0.100 | 0.047 | 0.055 |
| | DeepSeek V4-Flash | 0.026 | 0.029 | 0.073 | 0.088 | 0.046 | 0.065 | 0.012 | 0.015 | 0.127 | 0.140 | 0.060 | 0.071 |
| | Qwen 3.5 Plus | -0.023 | -0.023 | 0.059 | 0.057 | 0.06 | 0.071 | 0.016 | 0.015 | 0.114 | 0.118 | 0.049 | 0.052 |
| Demographics + Heuristics and biases experiments | Gemma-4-31B-it | 0.029 | 0.025 | 0.096 | 0.101 | 0.191 | 0.191 | 0.186 | 0.184 | 0.104 | 0.113 | 0.122 | 0.124 |
| | gpt-oss-120b | 0.014 | 0.015 | 0.058 | 0.055 | 0.135 | 0.131 | 0.066 | 0.061 | 0.094 | 0.098 | 0.075 | 0.074 |
| | DeepSeek V4-Flash | 0.047 | 0.044 | 0.092 | 0.096 | 0.108 | 0.110 | 0.147 | 0.150 | 0.112 | 0.118 | 0.102 | 0.104 |
| | Qwen 3.5 Plus | 0.051 | 0.048 | 0.089 | 0.090 | 0.134 | 0.132 | 0.168 | 0.166 | 0.109 | 0.111 | 0.110 | 0.110 |
| Demographics + All psychological measures + Heuristics and biases experiments | Gemma-4-31B-it | 0.407 | 0.410 | 0.466 | 0.460 | 0.541 | 0.526 | 0.555 | 0.554 | 0.450 | 0.449 | 0.485 | 0.481 |
| | gpt-oss-120b | 0.222 | 0.228 | 0.366 | 0.365 | 0.480 | 0.462 | 0.475 | 0.471 | 0.325 | 0.330 | 0.378 | 0.374 |
| | DeepSeek V4-Flash | 0.322 | 0.321 | 0.394 | 0.398 | 0.480 | 0.472 | 0.525 | 0.524 | 0.414 | 0.416 | 0.429 | 0.428 |
| | Qwen 3.5 Plus | 0.368 | 0.368 | 0.459 | 0.459 | 0.538 | 0.525 | 0.562 | 0.558 | 0.415 | 0.411 | 0.470 | 0.466 |
| Demographics + Heuristics and biases experiments + Features with $r < 0.5$ | Gemma-4-31B-it | 0.214 | 0.213 | 0.367 | 0.373 | 0.314 | 0.311 | 0.437 | 0.434 | 0.370 | 0.357 | 0.344 | 0.341 |
| | gpt-oss-120b | 0.131 | 0.132 | 0.284 | 0.292 | 0.217 | 0.227 | 0.274 | 0.269 | 0.292 | 0.295 | 0.243 | 0.247 |
| | DeepSeek V4-Flash | 0.175 | 0.173 | 0.255 | 0.263 | 0.262 | 0.266 | 0.302 | 0.307 | 0.322 | 0.322 | 0.267 | 0.269 |
| | Qwen 3.5 Plus | 0.177 | 0.176 | 0.319 | 0.322 | 0.277 | 0.269 | 0.376 | 0.367 | 0.323 | 0.320 | 0.297 | 0.294 |
| | Gemma-4-31B-it | 0.391 | 0.392 | 0.426 | 0.418 | 0.566 | 0.555 | 0.553 | 0.550 | 0.444 | 0.444 | 0.479 | 0.474 |
| | gpt-oss-120b | 0.275 | 0.282 | 0.336 | 0.329 | 0.512 | 0.498 | 0.502 | 0.498 | 0.341 | 0.338 | 0.397 | 0.392 |
| Demographics + Heuristics and biases experiments + Features with $r > 0.5$ | DeepSeek V4-Flash | 0.314 | 0.311 | 0.395 | 0.398 | 0.514 | 0.509 | 0.518 | 0.518 | 0.389 | 0.393 | 0.429 | 0.429 |
| | Qwen 3.5 Plus | 0.372 | 0.368 | 0.422 | 0.412 | 0.544 | 0.535 | 0.545 | 0.543 | 0.397 | 0.393 | 0.458 | 0.452 |
| Demographics + Heuristics and biases experiments + Features with $r < 0.3$ | Gemma-4-31B-it | 0.172 | 0.169 | 0.122 | 0.131 | 0.321 | 0.324 | 0.277 | 0.275 | 0.139 | 0.147 | 0.205 | 0.209 |
| | gpt-oss-120b | 0.101 | 0.102 | 0.115 | 0.117 | 0.247 | 0.251 | 0.133 | 0.137 | 0.092 | 0.097 | 0.138 | 0.142 |
| | DeepSeek V4-Flash | 0.137 | 0.132 | 0.132 | 0.144 | 0.237 | 0.241 | 0.210 | 0.212 | 0.155 | 0.169 | 0.174 | 0.180 |
| | Qwen 3.5 Plus | 0.178 | 0.175 | 0.146 | 0.150 | 0.264 | 0.272 | 0.248 | 0.245 | 0.164 | 0.167 | 0.199 | 0.201 |
| Profile correlation | | | | | | | | | | | | | |
| Demographics | Gemma-4-31B-it | 0.007 | 0.003 | 0.787 | 0.763 | 0.790 | 0.756 | 0.462 | 0.481 | 0.454 | 0.396 | 0.507 | 0.504 |
| | gpt-oss-120b | -0.016 | -0.008 | 0.755 | 0.728 | 0.868 | 0.848 | 0.419 | 0.408 | 0.323 | 0.283 | 0.462 | 0.457 |
| | DeepSeek V4-Flash | -0.060 | -0.056 | 0.761 | 0.741 | 0.834 | 0.807 | 0.398 | 0.397 | 0.245 | 0.242 | 0.45 | 0.444 |
| | Qwen 3.5 Plus | -0.069 | -0.067 | 0.761 | 0.737 | 0.797 | 0.766 | 0.509 | 0.502 | 0.442 | 0.391 | 0.493 | 0.482 |
| Heuristics and biases experiments | Gemma-4-31B-it | 0.031 | 0.037 | 0.809 | 0.792 | 0.848 | 0.818 | 0.428 | 0.439 | 0.449 | 0.374 | 0.508 | 0.504 |
| | gpt-oss-120b | 0.121 | 0.131 | 0.743 | 0.712 | 0.878 | 0.859 | 0.409 | 0.4 | 0.389 | 0.303 | 0.464 | 0.454 |
| | DeepSeek V4-Flash | -0.101 | -0.09 | 0.758 | 0.736 | 0.873 | 0.854 | 0.281 | 0.287 | 0.383 | 0.347 | 0.441 | 0.432 |
| | Qwen 3.5 Plus | 0.068 | 0.078 | 0.663 | 0.642 | 0.856 | 0.831 | 0.398 | 0.407 | 0.203 | 0.178 | 0.436 | 0.422 |
| Demographics + Heuristics and biases experiments | Gemma-4-31B-it | -0.008 | -0.003 | 0.819 | 0.799 | 0.803 | 0.769 | 0.444 | 0.464 | 0.456 | 0.39 | 0.514 | 0.51 |
| | gpt-oss-120b | 0.096 | 0.097 | 0.754 | 0.721 | 0.856 | 0.830 | 0.391 | 0.389 | 0.329 | 0.25 | 0.467 | 0.461 |
| | DeepSeek V4-Flash | -0.101 | -0.097 | 0.773 | 0.744 | 0.858 | 0.837 | 0.419 | 0.415 | 0.297 | 0.276 | 0.457 | 0.45 |
| | Qwen 3.5 Plus | 0.115 | 0.107 | 0.726 | 0.713 | 0.843 | 0.82 | 0.487 | 0.479 | 0.281 | 0.251 | 0.473 | 0.462 |
| Demographics + All psychological measures + Heuristics and biases experiments | Gemma-4-31B-it | 0.551 | 0.543 | 0.853 | 0.834 | 0.903 | 0.892 | 0.626 | 0.595 | 0.603 | 0.525 | 0.641 | 0.629 |
| | gpt-oss-120b | 0.293 | 0.270 | 0.810 | 0.781 | 0.897 | 0.882 | 0.681 | 0.658 | 0.474 | 0.387 | 0.570 | 0.561 |
| | DeepSeek V4-Flash | 0.361 | 0.346 | 0.838 | 0.814 | 0.912 | 0.899 | 0.714 | 0.700 | 0.460 | 0.427 | 0.592 | 0.583 |
| | Qwen 3.5 Plus | 0.531 | 0.518 | 0.830 | 0.806 | 0.922 | 0.911 | 0.712 | 0.687 | 0.485 | 0.433 | 0.626 | 0.619 |
| | Gemma-4-31B-it | 0.31 | 0.315 | 0.781 | 0.749 | 0.867 | 0.844 | 0.614 | 0.621 | 0.602 | 0.510 | 0.581 | 0.573 |

| Demographics + Heuristics | gpt-oss-120b | 0.251 | 0.235 | 0.757 | 0.733 | 0.851 | 0.830 | 0.466 | 0.459 | 0.469 | 0.379 | 0.511 | 0.505 |
|---|---|---|---|---|---|---|---|---|---|---|---|---|---|
| and biases experiments + | DeepSeek V4-Flash | 0.148 | 0.133 | 0.812 | 0.784 | 0.888 | 0.871 | 0.516 | 0.516 | 0.426 | 0.395 | 0.513 | 0.505 |
| Features with r < 0.5 | Qwen 3.5 Plus | 0.285 | 0.282 | 0.784 | 0.755 | 0.884 | 0.864 | 0.606 | 0.601 | 0.425 | 0.379 | 0.552 | 0.545 |
| Demographics + Heuristics | Gemma-4-31B-it | 0.534 | 0.531 | 0.845 | 0.825 | 0.907 | 0.896 | 0.655 | 0.628 | 0.559 | 0.481 | 0.635 | 0.624 |
| and biases experiments + | gpt-oss-120b | 0.376 | 0.358 | 0.842 | 0.82 | 0.901 | 0.883 | 0.658 | 0.622 | 0.269 | 0.242 | 0.556 | 0.550 |
| Features with r > 0.5 | DeepSeek V4-Flash | 0.365 | 0.354 | 0.833 | 0.812 | 0.916 | 0.904 | 0.679 | 0.662 | 0.419 | 0.392 | 0.582 | 0.575 |
| | Qwen 3.5 Plus | 0.541 | 0.518 | 0.823 | 0.798 | 0.922 | 0.912 | 0.673 | 0.647 | 0.447 | 0.396 | 0.616 | 0.606 |
| Demographics + Heuristics | Gemma-4-31B-it | 0.232 | 0.240 | 0.822 | 0.801 | 0.841 | 0.814 | 0.516 | 0.537 | 0.486 | 0.399 | 0.536 | 0.531 |
| and biases experiments + | gpt-oss-120b | 0.204 | 0.196 | 0.780 | 0.754 | 0.860 | 0.844 | 0.412 | 0.397 | 0.451 | 0.371 | 0.497 | 0.484 |
| Features with r < 0.3 | DeepSeek V4-Flash | 0.051 | 0.052 | 0.786 | 0.762 | 0.867 | 0.847 | 0.458 | 0.456 | 0.372 | 0.341 | 0.486 | 0.476 |
| | Qwen 3.5 Plus | 0.269 | 0.274 | 0.751 | 0.742 | 0.866 | 0.847 | 0.542 | 0.556 | 0.368 | 0.334 | 0.515 | 0.504 |
| Trait-score correlation | | Observed | Disattenuated | Observed | Disattenuated | Observed | Disattenuated | Observed | Disattenuated | Observed | Disattenuated | | |
| Demographics | Gemma-4-31B-it | 0.082 | 0.09 | 0.196 | 0.251 | 0.337 | 0.372 | 0.335 | 0.360 | 0.157 | 0.172 | | |
| | gpt-oss-120b | 0 | -0.001 | 0.066 | 0.077 | 0.277 | 0.304 | 0.181 | 0.196 | 0.139 | 0.152 | | |
| | DeepSeek V4-Flash | 0.053 | 0.058 | 0.152 | 0.177 | 0.243 | 0.266 | 0.28 | 0.302 | 0.149 | 0.163 | | |
| | Qwen 3.5 Plus | 0.097 | 0.106 | 0.187 | 0.215 | 0.249 | 0.274 | 0.288 | 0.311 | 0.166 | 0.181 | | |
| Heuristics and biases | Gemma-4-31B-it | -0.042 | -0.047 | 0.092 | 0.109 | 0.107 | 0.125 | 0.087 | 0.102 | 0.185 | 0.205 | | |
| experiments | gpt-oss-120b | 0.024 | 0.026 | 0.108 | 0.122 | 0.09 | 0.100 | -0.015 | -0.017 | 0.142 | 0.156 | | |
| | DeepSeek V4-Flash | 0.042 | 0.046 | 0.119 | 0.134 | 0.067 | 0.073 | 0.029 | 0.032 | 0.207 | 0.227 | | |
| | Qwen 3.5 Plus | -0.033 | -0.036 | 0.089 | 0.100 | 0.085 | 0.095 | 0.032 | 0.036 | 0.188 | 0.205 | | |
| Demographics + Heuristics | Gemma-4-31B-it | 0.031 | 0.034 | 0.187 | 0.234 | 0.348 | 0.39 | 0.28 | 0.304 | 0.179 | 0.196 | | |
| and biases experiments | gpt-oss-120b | 0.007 | 0.008 | 0.09 | 0.103 | 0.231 | 0.256 | 0.106 | 0.117 | 0.147 | 0.161 | | |
| | DeepSeek V4-Flash | 0.057 | 0.062 | 0.158 | 0.182 | 0.183 | 0.203 | 0.233 | 0.255 | 0.179 | 0.196 | | |
| | Qwen 3.5 Plus | 0.055 | 0.059 | 0.152 | 0.173 | 0.237 | 0.263 | 0.259 | 0.283 | 0.182 | 0.198 | | |

| | | | | | | | | | | | |
|---|---|---|---|---|---|---|---|---|---|---|---|
| Demographics + All psychological measures + Heuristics and biases experiments | Gemma-4-31B-it | 0.549 | 0.6 | 0.717 | 0.819 | 0.785 | 0.861 | 0.727 | 0.771 | 0.692 | 0.761 |
| | gpt-oss-120b | 0.309 | 0.342 | 0.632 | 0.729 | 0.739 | 0.811 | 0.667 | 0.716 | 0.546 | 0.606 |
| | DeepSeek V4-Flash | 0.456 | 0.502 | 0.65 | 0.75 | 0.717 | 0.781 | 0.736 | 0.787 | 0.651 | 0.716 |
| | Qwen 3.5 Plus | 0.495 | 0.539 | 0.702 | 0.795 | 0.761 | 0.821 | 0.734 | 0.776 | 0.645 | 0.706 |
| Demographics + Heuristics and biases experiments + Features with r < 0.5 | Gemma-4-31B-it | 0.283 | 0.312 | 0.599 | 0.671 | 0.522 | 0.581 | 0.621 | 0.667 | 0.582 | 0.643 |
| | gpt-oss-120b | 0.187 | 0.206 | 0.491 | 0.56 | 0.343 | 0.379 | 0.431 | 0.473 | 0.455 | 0.503 |
| | DeepSeek V4-Flash | 0.225 | 0.246 | 0.444 | 0.511 | 0.44 | 0.488 | 0.483 | 0.529 | 0.51 | 0.56 |
| | Qwen 3.5 Plus | 0.231 | 0.25 | 0.541 | 0.619 | 0.437 | 0.483 | 0.578 | 0.631 | 0.498 | 0.547 |
| Demographics + Heuristics and biases experiments + Features with r > 0.5 | Gemma-4-31B-it | 0.531 | 0.581 | 0.665 | 0.77 | 0.804 | 0.872 | 0.726 | 0.77 | 0.665 | 0.731 |
| | gpt-oss-120b | 0.403 | 0.446 | 0.57 | 0.666 | 0.77 | 0.843 | 0.701 | 0.752 | 0.575 | 0.640 |
| | DeepSeek V4-Flash | 0.437 | 0.479 | 0.645 | 0.742 | 0.749 | 0.812 | 0.714 | 0.761 | 0.612 | 0.672 |
| | Qwen 3.5 Plus | 0.497 | 0.539 | 0.653 | 0.751 | 0.77 | 0.83 | 0.715 | 0.756 | 0.613 | 0.669 |
| Demographics + Heuristics and biases experiments + Features with r < 0.3 | Gemma-4-31B-it | 0.226 | 0.246 | 0.26 | 0.317 | 0.519 | 0.571 | 0.418 | 0.454 | 0.245 | 0.272 |
| | gpt-oss-120b | 0.142 | 0.155 | 0.197 | 0.231 | 0.401 | 0.441 | 0.205 | 0.225 | 0.157 | 0.175 |
| | DeepSeek V4-Flash | 0.19 | 0.208 | 0.233 | 0.27 | 0.38 | 0.416 | 0.326 | 0.356 | 0.252 | 0.277 |
| | Qwen 3.5 Plus | 0.243 | 0.263 | 0.253 | 0.296 | 0.419 | 0.46 | 0.397 | 0.439 | 0.265 | 0.292 |

*Note.* Item-level correlations were calculated by correlating human and digital-twin responses across participants for each item and then summarized within each Big Five trait. Profile correlations were calculated by correlating human and digital-twin response profiles across items within participants and then summarized by trait. Trait-score correlations were calculated using Big Five trait scores. Observed trait-score correlations are Pearson correlations between human and digital-twin trait scores. Disattenuated correlations correct the observed Pearson correlations for measurement unreliability using McDonald's ω. Overall correlations were computed using Fisher's z transformation.

**Table S26 Nomothetic span of big five traits across input conditions and models**

| | Extraversion | | | | Agreeableness | | | | Conscientiousness | | | | Openness | | | | Neuroticism | | | |
|---|---|---|---|---|---|---|---|---|---|---|---|---|---|---|---|---|---|---|---|---|
| | Gemma-4-31B-it | gpt-oss-120b | DeepSeek V4-Flash | Qwen 3.5 Plus | Gemma-4-31B-it | gpt-oss-120b | DeepSeek V4-Flash | Qwen 3.5 Plus | Gemma-4-31B-it | gpt-oss-120b | DeepSeek V4-Flash | Qwen 3.5 Plus | Gemma-4-31B-it | gpt-oss-120b | DeepSeek V4-Flash | Qwen 3.5 Plus | Gemma-4-31B-it | gpt-oss-120b | DeepSeek V4-Flash | Qwen 3.5 Plus |
| Demographics | | | | | | | | | | | | | | | | | | | | |
| Need for cognition scale | 0.026 | 0.021 | 0.036 | 0.064** | -0.018 | 0.060** | 0.069** | 0.026 | 0.111*** | 0.061** | 0.109*** | 0.111*** | 0.157*** | 0.117*** | 0.138*** | 0.170*** | -0.064** | -0.049* | -0.065** | -0.076*** |
| Agentic vs. Communal Values scale | | | | | | | | | | | | | | | | | | | | |
| Agency | 0.191*** | 0.160*** | 0.197*** | 0.169*** | 0.017 | -0.015 | 0.055* | 0.01 | 0.160*** | 0.107*** | 0.121*** | 0.137*** | -0.005 | -0.057* | -0.015 | -0.02 | -0.178*** | -0.139*** | -0.172*** | -0.167*** |
| Communion | 0.045* | -0.069** | -0.051* | 0.038 | 0.207*** | 0.029 | 0.108*** | 0.176*** | 0.113*** | 0.051* | 0.060** | 0.100*** | -0.119*** | -0.118*** | -0.126*** | -0.096*** | -0.096*** | -0.009 | -0.032 | -0.071** |
| Consumer Minimalism scale | -0.024 | -0.069** | 0.012 | -0.022 | 0.003 | -0.043* | 0.013 | -0.032 | 0.117*** | 0.078*** | 0.068** | 0.072** | -0.056* | -0.092*** | -0.082*** | -0.080*** | -0.106*** | -0.053* | -0.085*** | -0.073*** |
| Empathy scale | 0.087*** | 0.046* | 0.031 | 0.097*** | 0.196*** | 0.159*** | 0.132*** | 0.186*** | -0.080*** | -0.023 | 0.002 | 0.02 | 0.132*** | 0.154*** | 0.128*** | 0.167*** | 0.128*** | 0.072** | 0.158*** | 0.099*** |
| Green values scale | 0.070 ** | 0.086 *** | 0.090 *** | 0.130 *** | 0.166 *** | 0.227 *** | 0.155 *** | 0.207 *** | 0.006 | 0.041 | 0.091 *** | 0.103 *** | 0.268 *** | 0.273 *** | 0.250 *** | 0.294 *** | 0.042 | -0.026 | 0.074 *** | 0.005 |
| Social Desirability scale | -0.03 | -0.130 *** | -0.079 *** | -0.048 * | 0.098 *** | -0.037 | 0.034 | 0.037 | 0.166 *** | 0.084 *** | 0.063 ** | 0.064 ** | -0.137 *** | -0.138 *** | -0.147 *** | -0.144 *** | -0.150 *** | -0.034 | -0.123 *** | -0.113 *** |
| Conscientiousness scale | 0.036 | -0.046 * | 0.027 | 0.011 | 0.050 * | -0.038 | 0.120 *** | 0.005 | 0.287 *** | 0.248 *** | 0.235 *** | 0.220 *** | -0.117 *** | -0.156 *** | -0.129 *** | -0.132 *** | -0.273 *** | -0.173 *** | -0.244 *** | -0.248 *** |
| Anxiety scale | 0.057 ** | 0.062 ** | 0.027 | 0.04 | 0.049 * | 0.019 | -0.032 | 0.031 | -0.177 *** | -0.122 *** | -0.118 *** | -0.107 *** | 0.056 * | 0.066 ** | 0.043 | 0.044 * | 0.194 *** | 0.107 *** | 0.176 *** | 0.171 *** |
| Individualism vs. Collectivism scale | | | | | | | | | | | | | | | | | | | | |
| Horizontal collectivism | 0.102 *** | 0.028 | 0.068 ** | 0.134 *** | 0.213 *** | 0.091 *** | 0.184 *** | 0.190 *** | 0.165 *** | 0.138 *** | 0.149 *** | 0.169 *** | -0.021 | -0.046 * | -0.041 | -0.002 | -0.128 *** | -0.083 *** | -0.114 *** | -0.130 *** |
| Horizontal individualism | -0.067 ** | -0.085 *** | -0.031 | -0.075 *** | -0.04 | -0.085 *** | -0.033 | -0.074 *** | 0.054 * | 0.016 | 0.02 | 0.033 | -0.014 | -0.033 | -0.001 | -0.03 | -0.03 | -0.001 | -0.033 | -0.035 |
| Vertical collectivism | 0.125 *** | -0.009 | 0.037 | 0.076 *** | 0.115 *** | -0.063 ** | 0.088 *** | 0.077 *** | 0.249 *** | 0.180 *** | 0.143 *** | 0.175 *** | -0.281 *** | -0.286 *** | -0.278 *** | -0.270 *** | -0.288 *** | -0.136 *** | -0.244 *** | -0.234 *** |
| Vertical individualism | 0.101 *** | 0.082 *** | 0.122 *** | 0.073 *** | -0.035 | -0.068 ** | 0.014 | -0.053 * | 0.171 *** | 0.126 *** | 0.110 *** | 0.117 *** | -0.120 *** | -0.150 *** | -0.110 *** | -0.134 *** | -0.191 *** | -0.118 *** | -0.186 *** | -0.172 *** |
| Regulatory Focus scale | 0.107 *** | 0.061 ** | 0.103 *** | 0.131 *** | 0.079 *** | 0.041 | 0.103 *** | 0.091 *** | 0.170 *** | 0.131 *** | 0.135 *** | 0.177 *** | -0.015 | -0.033 | -0.025 | -0.006 | -0.146 *** | -0.085 *** | -0.120 *** | -0.154 *** |
| Tightwads vs. Spendthrift scale | 0.071 ** | 0.032 | 0.04 | 0.083 *** | 0.050 * | 0.032 | 0.043 | 0.066 ** | -0.001 | -0.014 | 0.007 | 0.025 | 0.001 | 0.003 | 0.015 | 0.016 | 0.004 | -0.007 | -0.006 | 0.008 |
| Depression scale | -0.065 ** | -0.008 | -0.052 * | -0.066 ** | -0.048 * | -0.032 | -0.132 *** | -0.063 ** | -0.304 *** | -0.251 *** | -0.234 *** | -0.227 *** | 0.067 ** | 0.108 *** | 0.079 *** | 0.062 ** | 0.325 *** | 0.215 *** | 0.262 *** | 0.294 *** |
| Need for uniqueness scale | 0.118 *** | 0.107 *** | 0.077 *** | 0.099 *** | 0.057 ** | 0.047 * | 0.028 | 0.04 | 0.041 | 0.050 * | 0.044 * | 0.066 ** | 0.049 * | 0.043 | 0.046 * | 0.049 * | -0.054 * | -0.036 | -0.01 | -0.034 |
| Self-monitoring scale | 0.158 *** | 0.112 *** | 0.140 *** | 0.126 *** | 0.055 * | -0.008 | 0.089 *** | 0.047 * | 0.115 *** | 0.092 *** | 0.107 *** | 0.129 *** | -0.032 | -0.036 | -0.014 | -0.021 | -0.117 *** | -0.082 *** | -0.113 *** | -0.115 *** |
| Self-concept clarity scale | -0.101 *** | -0.139 *** | -0.067 ** | -0.067 ** | 0.01 | -0.037 | 0.053 * | -0.01 | 0.227 *** | 0.168 *** | 0.172 *** | 0.132 *** | -0.086 *** | -0.115 *** | -0.095 *** | -0.071 ** | -0.207 *** | -0.084 *** | -0.163 *** | -0.167 *** |
| Need for closure scale | 0.023 | -0.034 | -0.036 | -0.006 | 0.068 ** | -0.021 | 0.014 | 0.01 | 0.034 | 0.018 | -0.006 | 0.011 | -0.154 *** | -0.151 *** | -0.160 *** | -0.155 *** | -0.012 | 0.011 | 0.006 | 0.015 |
| Maximization scale | 0.098 *** | 0.084 *** | 0.086 *** | 0.039 | 0.027 | 0.014 | 0.005 | -0.008 | -0.005 | -0.017 | -0.001 | -0.007 | -0.004 | -0.018 | -0.01 | -0.019 | -0.007 | -0.018 | -0.006 | 0.015 |
| Heuristics and biases experiments | | | | | | | | | | | | | | | | | | | | |
| Need for cognition scale | 0.124*** | 0.088*** | 0.089*** | 0.165*** | 0.092*** | 0.076*** | 0.115*** | 0.118*** | -0.001 | 0.078*** | 0.151*** | 0.171*** | 0.127*** | 0.095*** | 0.158*** | 0.151*** | 0.023 | -0.103*** | -0.037 | -0.109*** |
| Agentic vs. Communal Values scale | | | | | | | | | | | | | | | | | | | | |
| Agency | -0.031 | 0.072** | 0.113*** | -0.034 | -0.079*** | -0.044* | 0.01 | -0.041 | -0.04 | -0.104*** | 0.015 | -0.080*** | -0.015 | -0.014 | 0.021 | -0.022 | 0.021 | 0.077*** | -0.065** | -0.019 |
| Communion | 0.116*** | 0.113*** | 0.118*** | 0.084*** | 0.095*** | 0.115*** | 0.128*** | 0.082*** | -0.014 | 0.041 | 0.109*** | 0.096*** | 0.074*** | 0.091*** | 0.106*** | 0.055* | 0.050* | -0.033 | -0.046* | 0.025 |
| Consumer Minimalism scale | -0.025 | 0.001 | 0.026 | -0.008 | -0.051* | -0.018 | -0.014 | -0.044* | 0.047* | 0.035 | 0.023 | 0.058** | -0.025 | -0.013 | 0.042 | -0.019 | -0.021 | -0.014 | -0.025 | -0.004 |
| Empathy scale | 0.274*** | 0.138*** | 0.151*** | 0.239*** | 0.274*** | 0.261*** | 0.241*** | 0.260*** | -0.028 | 0.139*** | 0.197*** | 0.236*** | 0.237*** | 0.223*** | 0.229*** | 0.240*** | 0.106*** | -0.108*** | -0.051* | -0.046* |
| Green values scale | 0.372*** | 0.241*** | 0.244*** | 0.342*** | 0.298*** | 0.405*** | 0.346*** | 0.409*** | -0.221*** | 0.090*** | 0.226*** | 0.179*** | 0.391*** | 0.383*** | 0.400*** | 0.433*** | 0.225*** | -0.119*** | -0.048* | -0.032 |
| Social Desirability scale | -0.075*** | -0.003 | -0.042 | -0.094*** | -0.079*** | -0.075*** | -0.089*** | -0.103*** | -0.005 | -0.016 | -0.051* | -0.045* | -0.059** | -0.060** | -0.092*** | -0.099*** | -0.038 | 0.049* | -0.014 | 0.028 |
| Conscientiousness scale | -0.074*** | -0.064** | -0.003 | -0.049* | -0.068** | -0.048* | -0.022 | -0.072** | 0.068** | 0.052* | 0.064** | 0.054* | -0.058** | -0.062** | -0.031 | -0.084*** | -0.061** | 0.011 | -0.031 | -0.035 |
| Anxiety scale | 0.102*** | 0.125*** | 0.096*** | 0.105*** | 0.093*** | 0.121*** | 0.111*** | 0.132*** | -0.105*** | -0.044* | 0.035 | -0.02 | 0.109*** | 0.123*** | 0.097*** | 0.135*** | 0.101*** | -0.025 | -0.007 | 0.037 |
| Individualism vs. Collectivism scale | | | | | | | | | | | | | | | | | | | | |
| Horizontal collectivism | 0.131*** | 0.109*** | 0.144*** | 0.124*** | 0.122*** | 0.116*** | 0.135*** | 0.099*** | 0.008 | 0.078*** | 0.141*** | 0.118*** | 0.086*** | 0.097*** | 0.122*** | 0.084*** | 0.03 | -0.044* | -0.060** | -0.046* |
| Horizontal individualism | 0.006 | 0.002 | 0.022 | 0.043 | 0.017 | 0.02 | 0.039 | 0.005 | 0.113*** | 0.121*** | 0.122*** | 0.153*** | 0.026 | 0.005 | 0.050* | 0.009 | -0.01 | -0.101*** | -0.037 | -0.014 |
| Vertical collectivism | -0.110*** | 0.014 | -0.01 | -0.103*** | -0.114*** | -0.089*** | -0.084*** | -0.142*** | 0.056* | -0.008 | 0.017 | -0.016 | -0.132*** | -0.104*** | -0.091*** | -0.151*** | -0.035 | 0.049* | -0.012 | 0.001 |
| Vertical individualism | -0.146*** | 0.021 | 0.003 | -0.124*** | -0.184*** | -0.151*** | -0.109*** | -0.165*** | 0.049* | -0.105*** | -0.035 | -0.094*** | -0.115*** | -0.112*** | -0.083*** | -0.129*** | -0.045* | 0.094*** | -0.026 | 0.012 |
| Regulatory Focus scale | 0.056* | 0.125*** | 0.155*** | 0.056* | 0.022 | 0.056* | 0.097*** | 0.028 | 0.036 | 0.028 | 0.146*** | 0.097*** | 0.059** | 0.056* | 0.097*** | 0.046* | 0.004 | -0.018 | -0.093*** | -0.021 |
| Tightwads vs. Spendthrift scale | 0.046* | 0.079*** | 0.076*** | 0.043 | 0.042 | 0.069** | 0.052* | 0.065** | -0.081*** | -0.047* | 0.019 | -0.016 | 0.048* | 0.053* | 0.034 | 0.059** | 0.039 | 0.031 | -0.004 | -0.024 |
| Depression scale | 0.108*** | 0.089*** | 0.069** | 0.089*** | 0.111*** | 0.100*** | 0.105*** | 0.104*** | -0.024 | -0.002 | 0.060** | 0.03 | 0.098*** | 0.083*** | 0.097*** | 0.101*** | 0.062** | -0.04 | -0.013 | 0.005 |
| Need for uniqueness scale | 0.079*** | 0.105*** | 0.101*** | 0.058** | 0.038 | 0.089*** | 0.075*** | 0.075*** | -0.081*** | -0.031 | 0.051* | -0.018 | 0.101*** | 0.107*** | 0.105*** | 0.104*** | 0.060** | 0.004 | -0.034 | 0.007 |
| Self-monitoring scale | 0.080*** | 0.110*** | 0.133*** | 0.087*** | 0.094*** | 0.078*** | 0.107*** | 0.054* | 0.055* | 0.019 | 0.128*** | 0.083*** | 0.079*** | 0.099*** | 0.102*** | 0.077*** | -0.005 | -0.031 | -0.068** | -0.014 |
| Self-concept clarity scale | -0.074*** | -0.110*** | -0.074*** | -0.074*** | -0.079*** | -0.096*** | -0.075*** | -0.114*** | 0.100*** | 0.087*** | 0.016 | 0.076*** | -0.108*** | -0.080*** | -0.060** | -0.111*** | -0.111*** | -0.035 | -0.008 | -0.004 |
| Need for closure scale | -0.052* | 0.014 | -0.004 | -0.038 | -0.026 | -0.033 | -0.017 | -0.065** | 0.081*** | 0.017 | 0.039 | -0.002 | -0.071** | -0.043* | -0.046* | -0.065** | -0.028 | 0.016 | -0.046* | 0.074*** |
| Maximization scale | 0.035 | 0.076*** | 0.037 | 0.021 | 0.032 | 0.023 | 0.039 | 0.038 | 0.011 | -0.017 | 0.036 | 0.002 | 0.021 | 0.021 | 0.029 | 0.047* | 0.024 | 0.019 | -0.037 | 0.014 |
| Demographics + Heuristics and biases experiments | | | | | | | | | | | | | | | | | | | | |
| Need for cognition scale | 0.045* | 0.029 | 0.067** | 0.094*** | 0.012 | 0.072** | 0.109*** | 0.073*** | 0.135*** | 0.119*** | 0.143*** | 0.126*** | 0.162*** | 0.093*** | 0.158*** | 0.168*** | -0.075*** | -0.089*** | -0.070** | -0.110*** |
| Agentic vs. Communal Values scale | | | | | | | | | | | | | | | | | | | | |
| Agency | 0.209*** | 0.145*** | 0.199*** | 0.204*** | -0.072** | -0.04 | 0.033 | 0.026 | 0.106*** | 0.075*** | 0.075*** | 0.177*** | 0.035 | -0.036 | -0.005 | 0.022 | -0.126*** | -0.057* | -0.157*** | -0.162*** |
| Communion | 0.047* | 0.014 | 0.049* | 0.062** | 0.211*** | 0.074*** | 0.175*** | 0.132*** | 0.110*** | 0.059** | 0.125*** | 0.107*** | -0.086*** | -0.048* | -0.04 | -0.051* | -0.063** | -0.021 | -0.025 | -0.032 |
| Consumer Minimalism scale | -0.044* | -0.015 | 0.021 | -0.021 | -0.02 | -0.059** | -0.013 | -0.042 | 0.122*** | 0.061** | 0.056* | 0.105*** | -0.031 | -0.061** | -0.028 | -0.053* | -0.082*** | -0.031 | -0.060** | -0.070** |
| Empathy scale | 0.159*** | 0.118*** | 0.061** | 0.149*** | 0.293*** | 0.252*** | 0.237*** | 0.254*** | -0.027 | 0.076*** | 0.099*** | 0.028 | 0.150*** | 0.190*** | 0.181*** | 0.191*** | 0.092*** | -0.046* | 0.128*** | 0.088*** |
| Green values scale | 0.154*** | 0.142*** | 0.125*** | 0.233*** | 0.222*** | 0.335*** | 0.272*** | 0.348*** | 0.033 | 0.053* | 0.181*** | 0.098*** | 0.304*** | 0.326*** | 0.348*** | 0.373*** | 0.038 | -0.077*** | 0.071** | -0.041 |
| Social Desirability scale | -0.078*** | -0.080*** | -0.065** | -0.054* | 0.032 | -0.071** | -0.034 | -0.024 | 0.131*** | 0.058** | 0.046* | 0.062** | -0.142*** | -0.138*** | -0.150*** | -0.136*** | -0.114*** | 0.007 | -0.105*** | -0.089*** |
| Conscientiousness scale | -0.016 | -0.019 | 0.008 | -0.023 | -0.014 | -0.063** | 0.038 | -0.028 | 0.290*** | 0.208*** | 0.167*** | 0.201*** | -0.093*** | -0.136*** | -0.101*** | -0.119*** | -0.246*** | -0.177*** | -0.163*** | -0.192*** |
| Anxiety scale | 0.118*** | 0.112*** | 0.066** | 0.119*** | 0.063** | 0.080*** | 0.036 | 0.100*** | -0.177*** | -0.124*** | -0.060** | -0.080*** | 0.071** | 0.105*** | 0.087*** | 0.103*** | 0.172*** | 0.094*** | 0.154*** | 0.149*** |
| Individualism vs. Collectivism scale | | | | | | | | | | | | | | | | | | | | |
| Horizontal collectivism | 0.103*** | 0.079*** | 0.095*** | 0.145*** | 0.194*** | 0.098*** | 0.202*** | 0.169*** | 0.177*** | 0.144*** | 0.182*** | 0.180*** | 0.001 | -0.006 | 0.023 | 0.038 | -0.112*** | -0.053* | -0.096*** | -0.120*** |
| Horizontal individualism | -0.018 | -0.081*** | -0.038 | -0.044* | 0.015 | -0.022 | -0.009 | -0.037 | 0.097*** | 0.057** | 0.085*** | 0.046* | -0.012 | -0.022 | -0.009 | -0.028 | -0.045* | -0.025 | -0.033 | -0.026 |
| Vertical collectivism | 0.03 | 0.031 | 0.072** | 0.031 | 0.072** | -0.110*** | 0.064** | -0.013 | 0.234*** | 0.099*** | 0.109*** | 0.175*** | -0.239*** | -0.222*** | -0.215*** | -0.228*** | -0.232*** | -0.042 | -0.190*** | -0.191*** |
| Vertical individualism | 0.063** | 0.084*** | 0.104*** | 0.079*** | -0.126*** | -0.141*** | -0.045* | -0.075*** | 0.136*** | 0.055* | 0.025 | 0.130*** | -0.072** | -0.141*** | -0.130*** | -0.103*** | -0.174*** | -0.035 | -0.178*** | -0.154*** |
| Regulatory Focus scale | 0.143*** | 0.096*** | 0.167*** | 0.148*** | 0.069** | 0.045* | 0.111*** | 0.080*** | 0.175*** | 0.129*** | 0.174*** | 0.189*** | 0.031 | -0.008 | 0.039 | 0.048* | -0.130*** | -0.073*** | -0.143*** | -0.138*** |
| Tightwads vs. Spendthrift scale | 0.093*** | 0.061** | 0.057** | 0.092*** | 0.038 | 0.062** | 0.049* | 0.073*** | -0.038 | 0.027 | 0.011 | 0.01 | 0.016 | 0.045* | 0.019 | 0.051* | 0.023 | -0.01 | 0.056* | 0.031 |
| Depression scale | 0.048* | 0.03 | 0.001 | 0.009 | 0.038 | 0.063** | -0.005 | 0.024 | -0.273*** | -0.189*** | -0.128*** | -0.189*** | 0.070** | 0.115*** | 0.096*** | 0.079*** | 0.264*** | 0.138*** | 0.211*** | 0.255*** |
| Need for uniqueness scale | 0.149*** | 0.146*** | 0.124*** | 0.162*** | 0.021 | 0.055* | 0.085*** | 0.101*** | 0.027 | 0.028 | 0.054* | 0.078*** | 0.083*** | 0.074*** | 0.071** | 0.111*** | -0.04 | -0.038 | -0.015 | -0.021 |
| Self-monitoring scale | 0.184 *** | 0.131 *** | 0.178 *** | 0.194 *** | 0.068 ** | 0.039 | 0.106 *** | 0.100 *** | 0.133 *** | 0.077 *** | 0.142 *** | 0.162 *** | 0.028 | 0.005 | 0.045 * | 0.053 * | -0.112 *** | -0.063 ** | -0.107 *** | -0.124 *** |
| Self-concept clarity scale | -0.152 *** | -0.138 *** | -0.116 *** | -0.141 *** | -0.019 | -0.061 ** | -0.031 | -0.080 *** | 0.246 *** | 0.167 *** | 0.103 *** | 0.116 *** | -0.090 *** | -0.127 *** | -0.103 *** | -0.130 *** | -0.180 *** | -0.089 *** | -0.141 *** | -0.143 *** |
| Need for closure scale | -0.006 | -0.016 | -0.017 | -0.008 | 0.046 * | -0.058 ** | 0.018 | -0.025 | 0.022 | 0.008 | 0.029 | 0.037 | -0.129 *** | -0.110 *** | -0.120 *** | -0.118 *** | 0.007 | 0.023 | -0.013 | 0.032 |
| Maximization scale | 0.111 *** | 0.103 *** | 0.095 *** | 0.098 *** | 0.022 | -0.006 | 0.071 ** | 0.036 | -0.011 | -0.005 | 0.027 | 0.042 | 0.02 | 0.01 | 0.024 | 0.028 | -0.015 | -0.003 | -0.015 | -0.023 |
| Demographics + All psychological measures + Heuristics and biases experiments | | | | | | | | | | | | | | | | | | | | |
| Need for cognition scale | 0.258 *** | 0.263 *** | 0.375 *** | 0.308 *** | 0.170 *** | 0.180 *** | 0.290 *** | 0.235 *** | 0.268 *** | 0.252 *** | 0.344 *** | 0.285 *** | 0.830 *** | 0.700 *** | 0.771 *** | 0.785 *** | -0.207 *** | -0.216 *** | -0.222 *** | -0.203 *** |
| Agentic vs. Communal Values scale | | | | | | | | | | | | | | | | | | | | |
| Agency | 0.431 *** | 0.448 *** | 0.441 *** | 0.417 *** | -0.067 ** | -0.122 *** | 0.005 | -0.053 * | 0.132 *** | 0.136 *** | 0.154 *** | 0.156 *** | 0.273 *** | 0.240 *** | 0.268 *** | 0.285 *** | -0.104 *** | -0.065 ** | -0.089 *** | -0.095 *** |
| Communion | 0.156 *** | 0.071 ** | 0.165 *** | 0.176 *** | 0.613 *** | 0.497 *** | 0.560 *** | 0.581 *** | 0.237 *** | 0.198 *** | 0.274 *** | 0.257 *** | 0.119 *** | 0.126 *** | 0.122 *** | 0.106 *** | -0.119 *** | -0.122 *** | -0.130 *** | -0.115 *** |
| Consumer Minimalism scale | 0.048* | 0.037 | 0.044* | 0.060** | 0.156*** | 0.143*** | 0.172*** | 0.178*** | 0.271*** | 0.254*** | 0.292*** | 0.300*** | 0.141*** | 0.086*** | 0.108*** | 0.117*** | -0.156*** | -0.141*** | -0.177*** | -0.169*** |
| Empathy scale | 0.001 | 0.016 | 0.095*** | 0.072** | 0.321*** | 0.309*** | 0.398*** | 0.340*** | 0.072*** | 0.062** | 0.108*** | 0.060** | 0.173*** | 0.155*** | 0.188*** | 0.180*** | 0.163*** | 0.108*** | 0.142*** | 0.145*** |
| Green values scale | 0.150*** | 0.129*** | 0.185*** | 0.215*** | 0.282*** | 0.273*** | 0.334*** | 0.329*** | 0.179*** | 0.158*** | 0.221*** | 0.201*** | 0.393*** | 0.343*** | 0.416*** | 0.428*** | -0.075*** | -0.082*** | -0.071** | -0.081*** |
| Social Desirability scale | 0.241*** | 0.050* | 0.096*** | 0.195*** | 0.507*** | 0.491*** | 0.360*** | 0.483*** | 0.382*** | 0.320*** | 0.342*** | 0.372*** | 0.092*** | 0.039 | 0.049* | 0.085*** | -0.427*** | -0.429*** | -0.405*** | -0.409*** |
| Conscientiousness scale | 0.262*** | 0.144*** | 0.231*** | 0.289*** | 0.243*** | 0.287*** | 0.310*** | 0.323*** | 0.785*** | 0.828*** | 0.667*** | 0.743*** | 0.163*** | 0.104*** | 0.135*** | 0.156*** | -0.397*** | -0.422*** | -0.425*** | -0.433*** |
| Anxiety scale | -0.218*** | -0.03 | -0.085*** | -0.207*** | -0.137*** | -0.145*** | -0.162*** | -0.236*** | -0.319*** | -0.299*** | -0.293*** | -0.351*** | -0.03 | 0.035 | 0.007 | -0.041 | 0.649*** | 0.576*** | 0.632*** | 0.659*** |
| Individualism vs. Collectivism scale | | | | | | | | | | | | | | | | | | | | |
| horizontal_collectivism | 0.350*** | 0.243*** | 0.324*** | 0.396*** | 0.596*** | 0.529*** | 0.637*** | 0.649*** | 0.349*** | 0.314*** | 0.377*** | 0.368*** | 0.269*** | 0.232*** | 0.253*** | 0.279*** | -0.257*** | -0.247*** | -0.266*** | -0.267*** |
| horizontal_individualism | 0.054* | 0.105*** | 0.108*** | 0.032 | -0.066** | -0.041 | 0.062** | -0.038 | 0.248*** | 0.224*** | 0.258*** | 0.227*** | 0.156*** | 0.103*** | 0.154*** | 0.151*** | -0.068** | -0.095*** | -0.129*** | -0.099*** |
| vertical_collectivism | 0.262 *** | 0.209 *** | 0.226 *** | 0.285 *** | 0.328 *** | 0.267 *** | 0.332 *** | 0.348 *** | 0.248 *** | 0.209 *** | 0.248 *** | 0.266 *** | 0.012 | 0.049 * | -0.003 | 0.006 | -0.190 *** | -0.187 *** | -0.202 *** | -0.180 *** |
| vertical_individualism | 0.315 *** | 0.465 *** | 0.329 *** | 0.302 *** | -0.223 *** | -0.250 *** | -0.116 *** | -0.190 *** | 0.070 ** | 0.064 ** | 0.081 *** | 0.086 *** | 0.125 *** | 0.111 *** | 0.123 *** | 0.135 *** | -0.028 | 0.01 | -0.03 | -0.024 |
| Regulatory Focus scale | 0.374 *** | 0.373 *** | 0.478 *** | 0.390 *** | 0.184 *** | 0.157 *** | 0.309 *** | 0.205 *** | 0.289 *** | 0.234 *** | 0.332 *** | 0.285 *** | 0.476 *** | 0.384 *** | 0.451 *** | 0.441 *** | -0.095 *** | -0.084 *** | -0.103 *** | -0.092 *** |
| Tightwads vs. Spendthrift scale | 0.050 * | 0.082 *** | 0.108 *** | 0.042 | -0.053 * | -0.049 * | -0.023 | -0.065 ** | -0.169 *** | -0.143 *** | -0.145 *** | -0.173 *** | -0.024 | 0.014 | -0.015 | -0.002 | 0.149 *** | 0.127 *** | 0.161 *** | 0.156 *** |
| Depression scale | -0.441 *** | -0.184 *** | -0.250 *** | -0.452 *** | -0.264 *** | -0.245 *** | -0.274 *** | -0.412 *** | -0.494 *** | -0.448 *** | -0.439 *** | -0.561 *** | -0.143 *** | -0.054 * | -0.079 *** | -0.172 *** | 0.771 *** | 0.734 *** | 0.765 *** | 0.822 *** |
| Need for uniqueness scale | 0.209 *** | 0.262 *** | 0.299 *** | 0.255 *** | -0.069 ** | -0.071 ** | -0.003 | -0.065 ** | -0.017 | -0.016 | 0.025 | 0.004 | 0.375 *** | 0.366 *** | 0.401 *** | 0.397 *** | 0.073 *** | 0.073 *** | 0.085 *** | 0.082 *** |
| Self-monitoring scale | 0.313 *** | 0.315 *** | 0.355 *** | 0.350 *** | 0.114 *** | 0.089 *** | 0.216 *** | 0.146 *** | 0.201 *** | 0.177 *** | 0.220 *** | 0.219 *** | 0.261 *** | 0.228 *** | 0.273 *** | 0.270 *** | -0.092 *** | -0.085 *** | -0.103 *** | -0.101 *** |
| Self-concept clarity scale | 0.243 *** | 0.011 | 0.106 *** | 0.218 *** | 0.252 *** | 0.247 *** | 0.275 *** | 0.348 *** | 0.470 *** | 0.400 *** | 0.434 *** | 0.473 *** | 0.050 * | -0.033 | 0.002 | 0.056 * | -0.574 *** | -0.544 *** | -0.579 *** | -0.591 *** |
| Need for closure scale | -0.169 *** | -0.031 | -0.130 *** | -0.153 *** | -0.081 *** | -0.089 *** | -0.069 ** | -0.125 *** | -0.017 | -0.032 | -0.038 | -0.046 * | -0.316 *** | -0.244 *** | -0.278 *** | -0.315 *** | 0.256 *** | 0.215 *** | 0.229 *** | 0.249 *** |
| Maximization scale | 0.047 * | 0.187 *** | 0.139 *** | 0.089 *** | -0.063 ** | -0.081 *** | -0.033 | -0.088 *** | -0.015 | -0.017 | 0.03 | -0.004 | 0.120 *** | 0.124 *** | 0.134 *** | 0.126 *** | 0.152 *** | 0.145 *** | 0.130 *** | 0.144 *** |
| Demographics + Heuristics and biases experiments + Features with r < 0.5 | | | | | | | | | | | | | | | | | | | | |
| Need for cognition scale | 0.178 *** | 0.177 *** | 0.131 *** | 0.151 *** | 0.127 *** | 0.173 *** | 0.225 *** | 0.232 *** | 0.160 *** | 0.081 *** | 0.223 *** | 0.212 *** | 0.485 *** | 0.339 *** | 0.388 *** | 0.406 *** | -0.237 *** | -0.275 *** | -0.188 *** | -0.278 *** |
| Agentic vs. Communal Values scale | | | | | | | | | | | | | | | | | | | | |
| Agency | 0.282 *** | 0.239 *** | 0.329 *** | 0.308 *** | -0.001 | -0.021 | 0.043 | 0.047 * | 0.065 ** | 0.016 | 0.105 *** | 0.132 *** | 0.329 *** | 0.297 *** | 0.298 *** | 0.320 *** | -0.112 *** | -0.087 *** | -0.096 *** | -0.117 *** |
| Communion | 0.106 *** | 0.039 | 0.069 ** | 0.105 *** | 0.399 *** | 0.384 *** | 0.374 *** | 0.422 *** | 0.249 *** | 0.200 *** | 0.253 *** | 0.260 *** | 0.133 *** | 0.060 ** | 0.095 *** | 0.110 *** | -0.189 *** | -0.045 * | -0.111 *** | -0.091 *** |
| Consumer Minimalism scale | -0.317 *** | -0.172 *** | -0.108 *** | -0.190 *** | 0.117 *** | 0.086 *** | 0.147 *** | 0.150 *** | 0.236 *** | 0.213 *** | 0.326 *** | 0.297 *** | 0.131 *** | 0.019 | 0.134 *** | 0.077 *** | -0.179 *** | -0.088 *** | -0.216 *** | -0.200 *** |
| Empathy scale | 0.157 *** | 0.175 *** | 0.207 *** | 0.195 *** | 0.207 *** | 0.538 *** | 0.536 *** | 0.420 *** | 0.126 *** | 0.218 *** | 0.213 *** | 0.217 *** | 0.140 *** | 0.112 *** | 0.219 *** | 0.207 *** | 0.190 *** | 0.168 *** | 0.170 *** | 0.199 *** |
| Green values scale | 0.073*** | 0.145*** | 0.129*** | 0.147*** | 0.203*** | 0.350*** | 0.388*** | 0.403*** | 0.128*** | 0.104*** | 0.211*** | 0.219*** | 0.366*** | 0.324*** | 0.424*** | 0.447*** | -0.047* | -0.067** | -0.060** | -0.088*** |
| Social Desirability scale | 0.080*** | -0.055* | -0.055* | -0.015 | 0.696*** | 0.329*** | 0.267*** | 0.472*** | 0.313*** | 0.148*** | 0.217*** | 0.193*** | 0.064** | -0.008 | -0.036 | -0.013 | -0.584*** | -0.267*** | -0.356*** | -0.413*** |
| Conscientiousness scale | 0.054* | -0.026 | 0.026 | 0.011 | 0.240*** | 0.141*** | 0.240*** | 0.208*** | 0.455*** | 0.283*** | 0.397*** | 0.401*** | 0.096*** | 0.008 | 0.050* | 0.03 | -0.369*** | -0.198*** | -0.290*** | -0.340*** |
| Anxiety scale | 0.052* | 0.076*** | 0.112*** | 0.081*** | -0.113*** | -0.011 | -0.054* | -0.095*** | -0.203*** | -0.106*** | -0.135*** | -0.130*** | 0.060** | 0.078*** | 0.101*** | 0.074*** | 0.356*** | 0.237*** | 0.303*** | 0.363*** |
| Individualism vs. Collectivism scale | | | | | | | | | | | | | | | | | | | | |
| Horizontal collectivism | 0.190*** | 0.139*** | 0.192*** | 0.203*** | 0.386*** | 0.396*** | 0.443*** | 0.445*** | 0.291*** | 0.227*** | 0.319*** | 0.296*** | 0.204*** | 0.138*** | 0.191*** | 0.196*** | -0.250*** | -0.116*** | -0.144*** | -0.196*** |
| Horizontal individualism | 0.005 | -0.047* | -0.003 | -0.025 | 0.002 | 0.026 | 0.090*** | 0.022 | 0.275*** | 0.147*** | 0.228*** | 0.251*** | 0.135*** | 0.074*** | 0.113*** | 0.088*** | -0.060** | 0.02 | -0.080*** | -0.057** |
| Vertical collectivism | 0.165*** | 0.059** | 0.124*** | 0.140*** | 0.266*** | 0.197*** | 0.172*** | 0.228*** | 0.229*** | 0.176*** | 0.218*** | 0.206*** | -0.001 | -0.015 | -0.036 | -0.017 | -0.248*** | -0.053* | -0.156*** | -0.166*** |
| Vertical individualism | 0.191*** | 0.129*** | 0.235*** | 0.188*** | -0.144*** | -0.119*** | -0.074*** | -0.123*** | 0.053* | 0.035 | 0.062** | 0.080*** | 0.154*** | 0.133*** | 0.138*** | 0.142*** | -0.002 | 0.017 | -0.034 | -0.041 |
| Regulatory Focus scale | 0.337*** | 0.244*** | 0.341*** | 0.297*** | 0.168*** | 0.213*** | 0.285*** | 0.217*** | 0.301*** | 0.142*** | 0.305*** | 0.340*** | 0.521*** | 0.377*** | 0.426*** | 0.425*** | -0.147*** | -0.065** | -0.108*** | -0.133*** |
| Tightwads vs. Spendthrift scale | 0.242*** | 0.183*** | 0.225*** | 0.176*** | -0.047* | 0.001 | 0.001 | -0.039 | -0.140*** | -0.075*** | -0.129*** | -0.101*** | 0.019 | 0.082*** | 0.063** | 0.066** | 0.112*** | 0.033 | 0.128*** | 0.113*** |
| Depression scale | -0.085*** | -0.016 | 0.007 | -0.054* | -0.214*** | -0.070** | -0.126*** | -0.197*** | -0.249*** | -0.091*** | -0.194*** | -0.187*** | -0.076*** | -0.025 | 0.003 | -0.042 | 0.473*** | 0.309*** | 0.374*** | 0.475*** |
| Need for uniqueness scale | 0.276*** | 0.265*** | 0.318*** | 0.305*** | -0.057** | 0.046* | 0.043* | 0.022 | -0.054* | -0.045* | 0.044* | 0.042 | 0.472*** | 0.552*** | 0.545*** | 0.577*** | 0.051* | -0.018 | 0.005 | -0.002 |
| Self-monitoring scale | 0.349*** | 0.198*** | 0.346*** | 0.388*** | 0.113*** | 0.175*** | 0.304*** | 0.195*** | 0.191*** | 0.145*** | 0.290*** | 0.241*** | 0.320*** | 0.271*** | 0.349*** | 0.331*** | -0.133*** | -0.032 | -0.126*** | -0.116*** |
| Self-concept clarity scale | -0.044* | -0.099*** | -0.145*** | -0.121*** | 0.229*** | 0.115*** | 0.167*** | 0.195*** | 0.330*** | 0.181*** | 0.253*** | 0.229*** | -0.044* | -0.113*** | -0.107*** | -0.096*** | -0.426*** | -0.256*** | -0.323*** | -0.378*** |
| Need for closure scale | -0.168*** | -0.220*** | -0.055* | -0.110*** | -0.070** | -0.017 | -0.007 | -0.143*** | 0.110*** | 0.205*** | 0.072** | 0.072** | -0.290*** | -0.278*** | -0.197*** | -0.272*** | 0.336*** | 0.542*** | 0.276*** | 0.427*** |
| Maximization scale | 0.096*** | 0.080*** | 0.158*** | 0.126*** | -0.045* | -0.017 | 0.016 | -0.052* | 0.050* | 0.055* | 0.067** | 0.080*** | 0.170*** | 0.143*** | 0.195*** | 0.167*** | 0.117*** | 0.128*** | 0.085*** | 0.146*** |
| Demographics + Heuristics and biases experiments + Features with r > 0.5 | | | | | | | | | | | | | | | | | | | | |
| Need for cognition scale | 0.255*** | 0.289*** | 0.366*** | 0.362*** | 0.142*** | 0.204*** | 0.317*** | 0.241*** | 0.274*** | 0.311*** | 0.377*** | 0.310*** | 0.884*** | 0.833*** | 0.849*** | 0.872*** | -0.186*** | -0.214*** | -0.194*** | -0.207*** |
| Agentic vs. Communal Values scale | | | | | | | | | | | | | | | | | | | | |
| Agency | 0.394*** | 0.438*** | 0.406*** | 0.418*** | -0.081*** | -0.127*** | 0.018 | -0.038 | 0.144*** | 0.157*** | 0.161*** | 0.179*** | 0.196*** | 0.190*** | 0.204*** | 0.218*** | -0.085*** | -0.094*** | -0.096*** | -0.099*** |
| Communion | 0.140*** | 0.092*** | 0.147*** | 0.162*** | 0.623*** | 0.534*** | 0.572*** | 0.625*** | 0.216*** | 0.222*** | 0.247*** | 0.224*** | 0.104*** | 0.086*** | 0.095*** | 0.070** | -0.100*** | -0.137*** | -0.120*** | -0.107*** |
| Consumer Minimalism scale | 0.086*** | 0.101*** | 0.115*** | 0.129*** | 0.157*** | 0.115*** | 0.182*** | 0.168*** | 0.260*** | 0.269*** | 0.255*** | 0.279*** | 0.107*** | 0.103*** | 0.118*** | 0.112*** | -0.158*** | -0.133*** | -0.144*** | -0.166*** |
| Empathy scale | -0.028 | -0.019 | 0.03 | 0.036 | 0.326*** | 0.335*** | 0.318*** | 0.345*** | 0.037 | 0.067** | 0.064** | 0.04 | 0.189*** | 0.141*** | 0.181*** | 0.163*** | 0.161*** | 0.103*** | 0.135*** | 0.133*** |
| Green values scale | 0.145*** | 0.169*** | 0.187*** | 0.219*** | 0.270*** | 0.275*** | 0.331*** | 0.331*** | 0.167*** | 0.184*** | 0.208*** | 0.182*** | 0.367*** | 0.328*** | 0.390*** | 0.383*** | -0.061** | -0.083*** | -0.060** | -0.082*** |
| Social Desirability scale | 0.200 *** | 0.116 *** | 0.139 *** | 0.191 *** | 0.333 *** | 0.254 *** | 0.347 *** | 0.341 *** | 0.364 *** | 0.312 *** | 0.342 *** | 0.339 *** | 0.085 *** | 0.068 ** | 0.069 ** | 0.070 ** | -0.373 *** | -0.377 *** | -0.379 *** | -0.362 *** |
| Conscientiousness scale | 0.265 *** | 0.211 *** | 0.259 *** | 0.298 *** | 0.222 *** | 0.242 *** | 0.353 *** | 0.301 *** | 0.800 *** | 0.807 *** | 0.691 *** | 0.749 *** | 0.153 *** | 0.135 *** | 0.155 *** | 0.142 *** | -0.392 *** | -0.413 *** | -0.423 *** | -0.430 *** |
| Anxiety scale | -0.217 *** | -0.112 *** | -0.122 *** | -0.243 *** | -0.138 *** | -0.133 *** | -0.215 *** | -0.203 *** | -0.337 *** | -0.300 *** | -0.332 *** | -0.336 *** | -0.04 | -0.036 | -0.049 * | -0.060 ** | 0.672 *** | 0.582 *** | 0.649 *** | 0.662 *** |
| Individualism vs. Collectivism scale | | | | | | | | | | | | | | | | | | | | |
| Horizontal collectivism | 0.333 *** | 0.290 *** | 0.343 *** | 0.366 *** | 0.615 *** | 0.609 *** | 0.669 *** | 0.675 *** | 0.337 *** | 0.344 *** | 0.363 *** | 0.346 *** | 0.269 *** | 0.244 *** | 0.267 *** | 0.245 *** | -0.236 *** | -0.274 *** | -0.251 *** | -0.258 *** |
| Horizontal individualism | 0.026 | 0.087 *** | 0.110 *** | 0.046 * | -0.039 | -0.050 * | 0.079 *** | -0.032 | 0.244 *** | 0.234 *** | 0.272 *** | 0.240 *** | 0.151 *** | 0.111 *** | 0.170 *** | 0.135 *** | -0.087 *** | -0.079 *** | -0.098 *** | -0.094 *** |
| Vertical collectivism | 0.257 *** | 0.234 *** | 0.237 *** | 0.258 *** | 0.328 *** | 0.318 *** | 0.354 *** | 0.349 *** | 0.242 *** | 0.242 *** | 0.255 *** | 0.242 *** | 0.002 | 0.021 | -0.007 | -0.006 | -0.183 *** | -0.189 *** | -0.197 *** | -0.179 *** |
| Vertical individualism | 0.304 *** | 0.403 *** | 0.318 *** | 0.303 *** | -0.203 *** | -0.224 *** | -0.099 *** | -0.174 *** | 0.087 *** | 0.098 *** | 0.107 *** | 0.107 *** | 0.087 *** | 0.100 *** | 0.095 *** | 0.098 *** | -0.031 | -0.01 | -0.048 * | -0.045 * |
| Regulatory Focus scale | 0.332 *** | 0.363 *** | 0.394 *** | 0.364 *** | 0.182 *** | 0.176 *** | 0.298 *** | 0.224 *** | 0.272 *** | 0.287 *** | 0.313 *** | 0.291 *** | 0.408 *** | 0.378 *** | 0.382 *** | 0.371 *** | -0.062 ** | -0.093 *** | -0.084 *** | -0.092 *** |
| Tightwads vs. Spendthrift scale | 0.02 | 0.031 | 0.014 | 0.002 | -0.047 * | 0.008 | -0.033 | -0.041 | -0.155 *** | -0.144 *** | -0.135 *** | -0.154 *** | -0.04 | -0.04 | -0.048 * | -0.026 | 0.147 *** | 0.140 *** | 0.151 *** | 0.154 *** |
| Depression scale | -0.444 *** | -0.303 *** | -0.312 *** | -0.483 *** | -0.259 *** | -0.228 *** | -0.385 *** | -0.369 *** | -0.519 *** | -0.464 *** | -0.530 *** | -0.548 *** | -0.120 *** | -0.117 *** | -0.126 *** | -0.162 *** | 0.792 *** | 0.803 *** | 0.794 *** | 0.843 *** |
| Need for uniqueness scale | 0.167 *** | 0.233 *** | 0.214 *** | 0.202 *** | -0.063 ** | -0.050 * | -0.013 | -0.028 | -0.01 | 0.007 | 0.015 | 0.015 | 0.248 *** | 0.224 *** | 0.246 *** | 0.251 *** | 0.091 *** | 0.082 *** | 0.089 *** | 0.074 *** |
| Self-monitoring scale | 0.261 *** | 0.280 *** | 0.302 *** | 0.302 *** | 0.121 *** | 0.104 *** | 0.204 *** | 0.155 *** | 0.192 *** | 0.204 *** | 0.219 *** | 0.212 *** | 0.203 *** | 0.178 *** | 0.216 *** | 0.212 *** | -0.083 *** | -0.096 *** | -0.090 *** | -0.094 *** |
| Self-concept clarity scale | 0.255 *** | 0.096 *** | 0.157 *** | 0.252 *** | 0.269 *** | 0.211 *** | 0.366 *** | 0.323 *** | 0.495 *** | 0.422 *** | 0.478 *** | 0.477 *** | 0.066 ** | 0.038 | 0.056 * | 0.076 *** | -0.587 *** | -0.572 *** | -0.626 *** | -0.593 *** |
| Need for closure scale | -0.134 *** | -0.099 *** | -0.119 *** | -0.169 *** | -0.038 | -0.061 ** | -0.091 *** | -0.094 *** | -0.032 | -0.047 * | -0.066 ** | -0.056 * | -0.296 *** | -0.283 *** | -0.292 *** | -0.323 *** | 0.219 *** | 0.224 *** | 0.196 *** | 0.220 *** |
| Maximization scale | 0.046 * | 0.131 *** | 0.153 *** | 0.070 ** | -0.064 ** | -0.059 ** | -0.031 | -0.067 ** | -0.027 | -0.009 | -0.005 | -0.02 | 0.084 *** | 0.088 *** | 0.106 *** | 0.089 *** | 0.152 *** | 0.137 *** | 0.145 *** | 0.142 *** |
| Demographics + Heuristics and biases experiments + Features with r < 0.3 | | | | | | | | | | | | | | | | | | | | |
| Need for cognition scale | 0.048* | 0.015 | 0.101*** | 0.117*** | 0.074*** | 0.177*** | 0.188*** | 0.101*** | 0.227*** | 0.208*** | 0.224*** | 0.233*** | 0.195*** | 0.100*** | 0.214*** | 0.232*** | -0.173*** | -0.121*** | -0.163*** | -0.188*** |
| Agentic vs. Communal Values scale | | | | | | | | | | | | | | | | | | | | |
| Agency | 0.213*** | 0.237*** | 0.283*** | 0.291*** | 0.053* | 0.134*** | 0.156*** | 0.104*** | 0.193*** | 0.177*** | 0.153*** | 0.241*** | 0.179*** | 0.091*** | 0.182*** | 0.201*** | -0.198*** | -0.152*** | -0.160*** | -0.166*** |
| Communion | 0.021 | -0.001 | 0.072** | 0.067** | 0.256*** | 0.173*** | 0.200*** | 0.222*** | 0.210*** | 0.158*** | 0.188*** | 0.180*** | -0.004 | -0.008 | 0.04 | 0.049* | -0.106*** | -0.084*** | -0.060** | -0.094*** |
| Consumer Minimalism scale | -0.387*** | -0.321*** | -0.051* | -0.152*** | 0.027 | 0.145*** | 0.172*** | 0.080*** | 0.437*** | 0.493*** | 0.408*** | 0.435*** | 0.163*** | -0.037 | 0.198*** | 0.152*** | -0.138*** | -0.090*** | -0.204*** | -0.229*** |
| Empathy scale | 0.141*** | 0.086*** | 0.130*** | 0.158*** | 0.302*** | 0.287*** | 0.282*** | 0.265*** | 0.073*** | 0.106*** | 0.161*** | 0.102*** | 0.199*** | 0.186*** | 0.260*** | 0.242*** | 0.013 | -0.056* | 0.027 | 0.03 |
| Green values scale | -0.007 | 0.032 | 0.086*** | 0.141*** | 0.159*** | 0.236*** | 0.232*** | 0.292*** | 0.155*** | 0.190*** | 0.181*** | 0.210*** | 0.313*** | 0.245*** | 0.337*** | 0.398*** | -0.012 | -0.077*** | 0.019 | -0.089*** |
| Social Desirability scale | -0.057** | -0.099*** | -0.051* | -0.047* | 0.095*** | 0.045* | 0.056* | 0.074*** | 0.223*** | 0.172*** | 0.139*** | 0.144*** | -0.096*** | -0.142*** | -0.068** | -0.083*** | -0.184*** | -0.122*** | -0.161*** | -0.186*** |
| Conscientiousness scale | 0.03 | -0.037 | 0.064 ** | 0.064 ** | 0.143 *** | 0.172 *** | 0.187 *** | 0.159 *** | 0.425 *** | 0.337 *** | 0.344 *** | 0.354 *** | -0.001 | -0.073 *** | 0.038 | 0.042 | -0.363 *** | -0.174 *** | -0.253 *** | -0.312 *** |
| Anxiety scale | 0.063 ** | 0.136 *** | 0.091 *** | 0.068 ** | 0.008 | 0.02 | -0.019 | -0.024 | -0.184 *** | -0.107 *** | -0.093 *** | -0.123 *** | 0.081 *** | 0.095 *** | 0.076 *** | 0.057 * | 0.238 *** | 0.131 *** | 0.182 *** | 0.238 *** |
| Individualism vs. Collectivism scale | | | | | | | | | | | | | | | | | | | | |
| Horizontal collectivism | 0.133 *** | 0.053 * | 0.158 *** | 0.200 *** | 0.279 *** | 0.233 *** | 0.310 *** | 0.290 *** | 0.288 *** | 0.223 *** | 0.275 *** | 0.289 *** | 0.087 *** | 0.035 | 0.144 *** | 0.170 *** | -0.250 *** | -0.147 *** | -0.187 *** | -0.227 *** |
| Horizontal individualism | -0.004 | -0.031 | 0.026 | 0.012 | 0.102 *** | 0.134 *** | 0.120 *** | 0.034 | 0.246 *** | 0.224 *** | 0.178 *** | 0.210 *** | 0.109 *** | -0.006 | 0.100 *** | 0.055 * | -0.131 *** | -0.096 *** | -0.096 *** | -0.099 *** |
| Vertical collectivism | 0.102 *** | 0.043 * | 0.143 *** | 0.116 *** | 0.171 *** | 0.101 *** | 0.123 *** | 0.116 *** | 0.240 *** | 0.160 *** | 0.182 *** | 0.214 *** | -0.133 *** | -0.120 *** | -0.078 *** | -0.076 *** | -0.222 *** | -0.115 *** | -0.190 *** | -0.208 *** |
| Vertical individualism | 0.175 *** | 0.190 *** | 0.241 *** | 0.194 *** | 0.026 | 0.098 *** | 0.085 *** | -0.005 | 0.167 *** | 0.134 *** | 0.111 *** | 0.169 *** | 0.043 * | 0.012 | 0.058 ** | 0.035 | -0.190 *** | -0.096 *** | -0.173 *** | -0.129 *** |
| Regulatory Focus scale | 0.228 *** | 0.191 *** | 0.303 *** | 0.293 *** | 0.210 *** | 0.283 *** | 0.294 *** | 0.197 *** | 0.305 *** | 0.289 *** | 0.282 *** | 0.320 *** | 0.210 *** | 0.129 *** | 0.204 *** | 0.222 *** | -0.241 *** | -0.133 *** | -0.189 *** | -0.194 *** |
| Tightwads vs. Spendthrift scale | 0.295*** | 0.271*** | 0.195*** | 0.190*** | 0.071** | -0.009 | -0.001 | 0.024 | -0.227*** | -0.233*** | -0.138*** | -0.115*** | 0.068** | 0.130*** | -0.011 | 0.04 | 0.019 | -0.123*** | 0.105*** | 0.069** |
| Depression scale | -0.018 | 0.039 | 0.007 | -0.059** | -0.056* | -0.073*** | -0.101*** | -0.115*** | -0.310*** | -0.198*** | -0.222*** | -0.228*** | 0.050* | 0.076*** | 0.037 | -0.012 | 0.375*** | 0.215*** | 0.281*** | 0.345*** |
| Need for uniqueness scale | 0.124*** | 0.185*** | 0.217*** | 0.226*** | 0.072** | 0.177*** | 0.153*** | 0.114*** | 0.116*** | 0.138*** | 0.097*** | 0.169*** | 0.178*** | 0.157*** | 0.198*** | 0.224*** | -0.043 | -0.060** | -0.036 | -0.053* |
| Self-monitoring scale | 0.449*** | 0.310*** | 0.468*** | 0.545*** | 0.304*** | 0.418*** | 0.448*** | 0.318*** | 0.338*** | 0.314*** | 0.353*** | 0.432*** | 0.274*** | 0.166*** | 0.299*** | 0.312*** | -0.362*** | -0.247*** | -0.285*** | -0.273*** |
| Self-concept clarity scale | -0.081*** | -0.152*** | -0.108*** | -0.084*** | 0.082*** | 0.037 | 0.072** | 0.068** | 0.302*** | 0.198*** | 0.175*** | 0.221*** | -0.081*** | -0.137*** | -0.073*** | -0.060** | -0.263*** | -0.152*** | -0.208*** | -0.263*** |
| Need for closure scale | -0.004 | 0.009 | 0.016 | -0.017 | 0.107*** | 0.024 | 0.036 | 0.001 | 0.059** | 0.070** | 0.059** | 0.025 | -0.050* | -0.057** | -0.064** | -0.075*** | 0.04 | 0.038 | 0.031 | 0.090*** |
| Maximization scale | 0.050* | 0.160*** | 0.192*** | 0.130*** | 0.062** | 0.088*** | 0.082*** | -0.024 | 0.123*** | 0.181*** | 0.098*** | 0.125*** | 0.132*** | 0.138*** | 0.163*** | 0.145*** | 0.018 | 0.055* | 0.019 | 0.086*** |

*Note.* *** p < .001, ** p < .01, * p < .05.

**Table S27 Recovery of the human Big Five nomothetic span across models and input conditions in Study 3a**

| Model | Input conditions | Pattern similarity | MAE |
|---|---|---|---|
| Gemma-4-31B-it | Demographics + All psychological measures + Heuristics and biases experiments | 0.981 | 0.045 |
| | Demographics | 0.722 | 0.163 |
| | Demographics + Heuristics and biases experiments | 0.692 | 0.159 |
| | Demographics + Heuristics and biases experiments + Features with r > 0.5 | 0.976 | 0.044 |
| | Demographics + Heuristics and biases experiments + Features with r < 0.3 | 0.746 | 0.137 |
| | Demographics + Heuristics and biases experiments + Features with r < 0.5 | 0.904 | 0.078 |
| | Heuristics and biases experiments | 0.256 | 0.202 |
| gpt-oss-120b | Demographics + All psychological measures + Heuristics and biases experiments | 0.961 | 0.051 |
| | Demographics | 0.554 | 0.188 |
| | Demographics + Heuristics and biases experiments | 0.489 | 0.183 |
| | Demographics + Heuristics and biases experiments + Features with r > 0.5 | 0.966 | 0.049 |
| | Demographics + Heuristics and biases experiments + Features with r < 0.3 | 0.592 | 0.162 |
| | Demographics + Heuristics and biases experiments + Features with r < 0.5 | 0.787 | 0.120 |
| | Heuristics and biases experiments | 0.248 | 0.194 |
| DeepSeek V4-Flash | Demographics + All psychological measures + Heuristics and biases experiments | 0.976 | 0.053 |
| | Demographics | 0.710 | 0.167 |
| | Demographics + Heuristics and biases experiments | 0.625 | 0.164 |
| | Demographics + Heuristics and biases experiments + Features with r > 0.5 | 0.976 | 0.052 |
| | Demographics + Heuristics and biases experiments + Features with r < 0.3 | 0.722 | 0.135 |
| | Demographics + Heuristics and biases experiments + Features with r < 0.5 | 0.848 | 0.104 |
| | Heuristics and biases experiments | 0.300 | 0.183 |
| Qwen 3.5 Plus | Demographics + All psychological measures + Heuristics and biases experiments | 0.984 | 0.048 |
| | Demographics | 0.709 | 0.165 |
| | Demographics + Heuristics and biases experiments | 0.594 | 0.169 |
| | Demographics + Heuristics and biases experiments + Features with r > 0.5 | 0.976 | 0.049 |
| | Demographics + Heuristics and biases experiments + Features with r < 0.3 | 0.770 | 0.126 |
| | Demographics + Heuristics and biases experiments + Features with r < 0.5 | 0.876 | 0.090 |
| | Heuristics and biases experiments | 0.233 | 0.200 |

*Note.* Pattern similarity refers to the Pearson correlation between the vectorized human Big Five–external scale correlation matrix and the corresponding model-derived correlation matrix. MAE = Mean absolute error. Mean absolute error represents the average absolute difference between model-derived and human correlations.

**Table S28 Configural invariance results for Big Five responses across input conditions and models in Study 3a**

| Input conditions | Gemma-4-31B-it | gpt-oss-120b | DeepSeek V4-Flash | Qwen 3.5 Plus |
|---|---|---|---|---|
| Demographics | Configural invariance | Configural invariance | Openness 35R | Openness<br>Agreeableness 2R, 12R, 27R, 37R<br>Conscientiousness 8R, 18R, 23R, 43R |
| Heuristics and biases experiments | Configural invariance | Configural invariance | Openness 35R | Configural noninvariance<br>(Agreeableness excluded) |
| Demographics + Heuristics and biases experiments | Configural invariance | Configural invariance | Configural invariance | Openness 35R<br>Agreeableness 12R, 27R, 37R<br>Extraversion 11, 16, 26 |
| Demographics + All psychological measures + Heuristics and biases experiments | Configural invariance | Configural invariance | Openness 35R | Extraversion 11, 16, 26 |
| Demographics + Heuristics and biases experiments + Features with $r < 0.5$ | Configural invariance | Openness 35R<br>Neuroticism 4, 29 | Configural invariance | Configural invariance |
| Demographics + Heuristics and biases experiments + Features with $r > 0.5$ | Configural invariance | Openness: 2 communities | Configural invariance | Agreeableness 2R, 12R, 27R, 37R |
| Demographics + Heuristics and biases experiments + Features with $r < 0.3$ | Configural invariance | Neuroticism 4, 29 | Openness 35R | Openness 35R<br>Extraversion 11, 16, 26 |

*Note.* Listed items are those removed during the iterative invariance procedure, either by the algorithmic procedure or because their community assignment was inconsistent with the expected Big Five domain structure.

Table Item-level configural removals and metric noninvariance across input conditions and models

**Table S29 Network configural invariance of Big Five Inventory responses comparing human and digital twins in Study 3a**

| | | Demographics | | Heuristics and biases experiments | | Demographics + Heuristics and biases experiments | | Demographics + All psychological measures + Heuristics and biases experiments | | Demographics + Heuristics and biases experiments + Features with r > 0.5 | | Demographics + Heuristics and biases experiments + Features with r < 0.5 | | Demographics + Heuristics and biases experiments + Features with r < 0.3 | |
|---|---|---|---|---|---|---|---|---|---|---|---|---|---|---|---|
| Gemma-4-31B-it | | | | | | | | | | | | | | | |
| Item | Dimension | Community | Replication | Community | Replication | Community | Replication | Community | Replication | Community | Replication | Community | Replication | Community | Replication |
| 1 | Extraversion | 1 | 1.000 | 1 | 1.000 | 1 | 1.000 | 1 | 1.000 | 1 | 1.000 | 1 | 1.000 | 1 | 1.000 |
| 2 (R) | Agreeableness | 2 | 0.996 | 2 | 1.000 | 2 | 0.988 | 2 | 1.000 | 2 | 1.000 | 2 | 1.000 | 2 | 0.998 |
| 3 | Conscientiousness | 3 | 1.000 | 3 | 1.000 | 3 | 1.000 | 3 | 1.000 | 3 | 1.000 | 3 | 1.000 | 3 | 1.000 |
| 4 | Neuroticism | 4 | 1.000 | 4 | 1.000 | 4 | 1.000 | 4 | 1.000 | 4 | 1.000 | 4 | 1.000 | 4 | 1.000 |
| 5 | Openness | 5 | 1.000 | 5 | 1.000 | 5 | 1.000 | 5 | 1.000 | 5 | 1.000 | 5 | 1.000 | 5 | 1.000 |
| 6 (R) | Extraversion | 1 | 1.000 | 1 | 1.000 | 1 | 1.000 | 1 | 1.000 | 1 | 1.000 | 1 | 1.000 | 1 | 1.000 |
| 7 | Agreeableness | 2 | 1.000 | 2 | 1.000 | 2 | 1.000 | 2 | 1.000 | 2 | 1.000 | 2 | 1.000 | 2 | 1.000 |
| 8 (R) | Conscientiousness | 3 | 1.000 | 3 | 1.000 | 3 | 1.000 | 3 | 1.000 | 3 | 1.000 | 3 | 1.000 | 3 | 1.000 |
| 9 (R) | Neuroticism | 4 | 1.000 | 4 | 1.000 | 4 | 1.000 | 4 | 1.000 | 4 | 1.000 | 4 | 1.000 | 4 | 1.000 |
| 10 | Openness | 5 | 1.000 | 5 | 1.000 | 5 | 1.000 | 5 | 1.000 | 5 | 1.000 | 5 | 1.000 | 5 | 1.000 |
| 11 | Extraversion | 1 | 0.962 | 1 | 0.860 | 1 | 0.852 | 1 | 1.000 | 1 | 1.000 | 1 | 1.000 | 1 | 0.984 |
| 12 (R) | Agreeableness | 2 | 0.954 | 2 | 0.992 | 2 | 0.988 | 2 | 1.000 | 2 | 1.000 | 2 | 1.000 | 2 | 0.998 |
| 13 | Conscientiousness | 3 | 1.000 | 3 | 1.000 | 3 | 1.000 | 3 | 1.000 | 3 | 1.000 | 3 | 1.000 | 3 | 1.000 |
| 14 | Neuroticism | 4 | 1.000 | 4 | 1.000 | 4 | 1.000 | 4 | 1.000 | 4 | 1.000 | 4 | 1.000 | 4 | 1.000 |
| 15 | Openness | 5 | 1.000 | 5 | 1.000 | 5 | 1.000 | 5 | 1.000 | 5 | 1.000 | 5 | 1.000 | 5 | 1.000 |
| 16 | Extraversion | 1 | 0.962 | 1 | 0.860 | 1 | 0.852 | 1 | 1.000 | 1 | 1.000 | 1 | 1.000 | 1 | 0.984 |
| 17 | Agreeableness | 2 | 1.000 | 2 | 1.000 | 2 | 1.000 | 2 | 1.000 | 2 | 1.000 | 2 | 1.000 | 2 | 1.000 |
| 18 (R) | Conscientiousness | 3 | 1.000 | 3 | 1.000 | 3 | 1.000 | 3 | 1.000 | 3 | 1.000 | 3 | 1.000 | 3 | 1.000 |
| 19 | Neuroticism | 4 | 1.000 | 4 | 1.000 | 4 | 1.000 | 4 | 1.000 | 4 | 1.000 | 4 | 1.000 | 4 | 1.000 |
| 20 | Openness | 5 | 1.000 | 5 | 1.000 | 5 | 1.000 | 5 | 1.000 | 5 | 1.000 | 5 | 1.000 | 5 | 1.000 |
| 21 (R) | Extraversion | 1 | 1.000 | 1 | 1.000 | 1 | 1.000 | 1 | 1.000 | 1 | 1.000 | 1 | 1.000 | 1 | 1.000 |
| 22 | Agreeableness | 2 | 1.000 | 2 | 1.000 | 2 | 1.000 | 2 | 1.000 | 2 | 1.000 | 2 | 1.000 | 2 | 1.000 |
| 23 (R) | Conscientiousness | 3 | 1.000 | 3 | 1.000 | 3 | 1.000 | 3 | 1.000 | 3 | 1.000 | 3 | 1.000 | 3 | 1.000 |
| 24 (R) | Neuroticism | 4 | 1.000 | 4 | 1.000 | 4 | 1.000 | 4 | 1.000 | 4 | 1.000 | 4 | 1.000 | 4 | 1.000 |
| 25 | Openness | 5 | 1.000 | 5 | 1.000 | 5 | 1.000 | 5 | 1.000 | 5 | 1.000 | 5 | 1.000 | 5 | 1.000 |
| 26 | Extraversion | 1 | 0.794 | 1 | 0.860 | 1 | 0.852 | 1 | 1.000 | 1 | 1.000 | 1 | 1.000 | 1 | 0.984 |
| 27 (R) | Agreeableness | 2 | 0.996 | 2 | 1.000 | 2 | 0.988 | 2 | 1.000 | 2 | 1.000 | 2 | 1.000 | 2 | 0.998 |
| 28 | Conscientiousness | 3 | 1.000 | 3 | 1.000 | 3 | 1.000 | 3 | 1.000 | 3 | 1.000 | 3 | 1.000 | 3 | 1.000 |
| 29 | Neuroticism | 4 | 1.000 | 4 | 1.000 | 4 | 1.000 | 4 | 1.000 | 4 | 1.000 | 4 | 1.000 | 4 | 1.000 |
| 30 | Openness | 5 | 1.000 | 5 | 1.000 | 5 | 1.000 | 5 | 1.000 | 5 | 1.000 | 5 | 1.000 | 5 | 1.000 |
| 31 (R) | Extraversion | 1 | 1.000 | 1 | 1.000 | 1 | 1.000 | 1 | 1.000 | 1 | 1.000 | 1 | 1.000 | 1 | 1.000 |
| 32 | Agreeableness | 2 | 1.000 | 2 | 1.000 | 2 | 1.000 | 2 | 1.000 | 2 | 1.000 | 2 | 1.000 | 2 | 1.000 |
| 33 | Conscientiousness | 3 | 1.000 | 3 | 1.000 | 3 | 1.000 | 3 | 1.000 | 3 | 1.000 | 3 | 1.000 | 3 | 1.000 |
| 34 (R) | Neuroticism | 4 | 1.000 | 4 | 1.000 | 4 | 1.000 | 4 | 1.000 | 4 | 1.000 | 4 | 1.000 | 4 | 1.000 |
| 35 (R) | Openness | 5 | 0.976 | 5 | 0.962 | 5 | 0.998 | 5 | 1.000 | 5 | 1.000 | 5 | 1.000 | 5 | 1.000 |
| 36 | Extraversion | 1 | 1.000 | 1 | 1.000 | 1 | 1.000 | 1 | 1.000 | 1 | 1.000 | 1 | 1.000 | 1 | 1.000 |
| 37 (R) | Agreeableness | 2 | 0.996 | 2 | 1.000 | 2 | 0.988 | 2 | 1.000 | 2 | 1.000 | 2 | 1.000 | 2 | 0.998 |
| 38 | Conscientiousness | 3 | 1.000 | 3 | 1.000 | 3 | 1.000 | 3 | 1.000 | 3 | 1.000 | 3 | 1.000 | 3 | 1.000 |
| 39 | Neuroticism | 4 | 1.000 | 4 | 1.000 | 4 | 1.000 | 4 | 1.000 | 4 | 1.000 | 4 | 1.000 | 4 | 1.000 |
| 40 | Openness | 5 | 1.000 | 5 | 1.000 | 5 | 1.000 | 5 | 1.000 | 5 | 1.000 | 5 | 1.000 | 5 | 1.000 |
| 41 (R) | Openness | 5 | 1.000 | 5 | 1.000 | 5 | 1.000 | 5 | 1.000 | 5 | 1.000 | 5 | 1.000 | 5 | 1.000 |
| 42 | Agreeableness | 2 | 1.000 | 2 | 1.000 | 2 | 1.000 | 2 | 1.000 | 2 | 1.000 | 2 | 1.000 | 2 | 1.000 |
| 43 (R) | Conscientiousness | 3 | 1.000 | 3 | 1.000 | 3 | 1.000 | 3 | 1.000 | 3 | 1.000 | 3 | 1.000 | 3 | 1.000 |

| 44 | Openness | 5 | 1.000 | 5 | 1.000 | 5 | 1.000 | 5 | 1.000 | 5 | 1.000 | 5 | 1.000 | 5 | 1.000 |
|---|---|---|---|---|---|---|---|---|---|---|---|---|---|---|---|
| gpt-oss-120b | | | | | | | | | | | | | | | |
| Item | Dimension | Community | Replication | Community | Replication | Community | Replication | Community | Replication | Community | Replication | Community | Replication | Community | Replication |
| 1 | Extraversion | 1 | 1.000 | 1 | 1.000 | 1 | 1.000 | 1 | 1.000 | 1 | 1.000 | 1 | 1.000 | 1 | 1.000 |
| 2 (R) | Agreeableness | 2 | 1.000 | 2 | 1.000 | 2 | 1.000 | 2 | 1.000 | 2 | 0.996 | 2 | 1.000 | 2 | 1.000 |
| 3 | Conscientiousness | 3 | 1.000 | 3 | 1.000 | 3 | 1.000 | 3 | 1.000 | 3 | 1.000 | 3 | 1.000 | 3 | 1.000 |
| 4 | Neuroticism | 4 | 1.000 | 4 | 1.000 | 4 | 1.000 | 4 | 1.000 | 4 | 1.000 | Alg-rem. | | Pre-rem. | |
| 5 | Openness | 5 | 1.000 | 5 | 1.000 | 5 | 1.000 | 5 | 1.000 | 5 | 1.000 | 4 | 1.000 | 4 | 1.000 |
| 6 (R) | Extraversion | 1 | 1.000 | 1 | 1.000 | 1 | 1.000 | 1 | 1.000 | 1 | 1.000 | 1 | 1.000 | 1 | 1.000 |
| 7 | Agreeableness | 2 | 1.000 | 2 | 1.000 | 2 | 1.000 | 2 | 1.000 | 2 | 1.000 | 2 | 1.000 | 2 | 1.000 |
| 8 (R) | Conscientiousness | 3 | 1.000 | 3 | 1.000 | 3 | 1.000 | 3 | 1.000 | 3 | 1.000 | 3 | 1.000 | 3 | 1.000 |
| 9 (R) | Neuroticism | 4 | 1.000 | 4 | 1.000 | 4 | 1.000 | 4 | 1.000 | 4 | 1.000 | 5 | 1.000 | 5 | 1.000 |
| 10 | Openness | 5 | 1.000 | 5 | 1.000 | 5 | 1.000 | 5 | 1.000 | 5 | 1.000 | 4 | 1.000 | 4 | 1.000 |
| 11 | Extraversion | 1 | 1.000 | 1 | 1.000 | 1 | 1.000 | 1 | 1.000 | 1 | 0.878 | 1 | 1.000 | 1 | 1.000 |
| 12 (R) | Agreeableness | 2 | 1.000 | 2 | 1.000 | 2 | 1.000 | 2 | 1.000 | 2 | 0.996 | 2 | 1.000 | 2 | 1.000 |
| 13 | Conscientiousness | 3 | 1.000 | 3 | 1.000 | 3 | 1.000 | 3 | 1.000 | 3 | 1.000 | 3 | 1.000 | 3 | 1.000 |
| 14 | Neuroticism | 4 | 1.000 | 4 | 1.000 | 4 | 1.000 | 4 | 1.000 | 4 | 1.000 | 5 | 1.000 | 5 | 1.000 |
| 15 | Openness | 5 | 1.000 | 5 | 1.000 | 5 | 1.000 | 5 | 1.000 | 5 | 1.000 | 4 | 1.000 | 4 | 1.000 |
| 16 | Extraversion | 1 | 1.000 | 1 | 1.000 | 1 | 1.000 | 1 | 1.000 | 1 | 0.878 | 1 | 1.000 | 1 | 1.000 |
| 17 | Agreeableness | 2 | 1.000 | 2 | 1.000 | 2 | 1.000 | 2 | 1.000 | 2 | 1.000 | 2 | 1.000 | 2 | 1.000 |
| 18 (R) | Conscientiousness | 3 | 1.000 | 3 | 1.000 | 3 | 1.000 | 3 | 1.000 | 3 | 1.000 | 3 | 1.000 | 3 | 1.000 |
| 19 | Neuroticism | 4 | 1.000 | 4 | 1.000 | 4 | 1.000 | 4 | 1.000 | 4 | 1.000 | 5 | 1.000 | 5 | 1.000 |
| 20 | Openness | 5 | 1.000 | 5 | 1.000 | 5 | 1.000 | 5 | 1.000 | 5 | 1.000 | 4 | 1.000 | 4 | 1.000 |
| 21 (R) | Extraversion | 1 | 1.000 | 1 | 1.000 | 1 | 1.000 | 1 | 1.000 | 1 | 1.000 | 1 | 1.000 | 1 | 1.000 |
| 22 | Agreeableness | 2 | 1.000 | 2 | 1.000 | 2 | 1.000 | 2 | 1.000 | 2 | 1.000 | 2 | 1.000 | 2 | 1.000 |
| 23 (R) | Conscientiousness | 3 | 1.000 | 3 | 1.000 | 3 | 1.000 | 3 | 1.000 | 3 | 1.000 | 3 | 1.000 | 3 | 1.000 |
| 24 (R) | Neuroticism | 4 | 1.000 | 4 | 1.000 | 4 | 1.000 | 4 | 1.000 | 4 | 1.000 | 5 | 1.000 | 5 | 1.000 |
| 25 | Openness | 5 | 1.000 | 5 | 1.000 | 5 | 1.000 | 5 | 1.000 | 5 | 1.000 | 4 | 1.000 | 4 | 1.000 |
| 26 | Extraversion | 1 | 1.000 | 1 | 1.000 | 1 | 1.000 | 1 | 1.000 | 1 | 0.878 | 1 | 1.000 | 1 | 1.000 |
| 27 (R) | Agreeableness | 2 | 1.000 | 2 | 1.000 | 2 | 1.000 | 2 | 1.000 | 2 | 0.998 | 2 | 1.000 | 2 | 1.000 |
| 28 | Conscientiousness | 3 | 1.000 | 3 | 1.000 | 3 | 1.000 | 3 | 1.000 | 3 | 1.000 | 3 | 1.000 | 3 | 1.000 |
| 29 | Neuroticism | 4 | 1.000 | 4 | 1.000 | 4 | 1.000 | 4 | 1.000 | 4 | 1.000 | Alg-rem. | | Pre-rem. | |
| 30 | Openness | 5 | 1.000 | 5 | 1.000 | 5 | 1.000 | 5 | 1.000 | 6 | 0.968 | 4 | 1.000 | 4 | 1.000 |
| 31 (R) | Extraversion | 1 | 1.000 | 1 | 1.000 | 1 | 1.000 | 1 | 1.000 | 1 | 1.000 | 1 | 1.000 | 1 | 1.000 |
| 32 | Agreeableness | 2 | 1.000 | 2 | 1.000 | 2 | 1.000 | 2 | 1.000 | 2 | 1.000 | 2 | 1.000 | 2 | 1.000 |
| 33 | Conscientiousness | 3 | 1.000 | 3 | 1.000 | 3 | 1.000 | 3 | 1.000 | 3 | 1.000 | 3 | 1.000 | 3 | 1.000 |
| 34 (R) | Neuroticism | 4 | 1.000 | 4 | 1.000 | 4 | 1.000 | 4 | 1.000 | 4 | 1.000 | 5 | 1.000 | 5 | 1.000 |
| 35 (R) | Openness | 5 | 1.000 | 5 | 1.000 | 5 | 1.000 | 5 | 1.000 | 5 | 1.000 | Alg-rem. | | 4 | 1.000 |
| 36 | Extraversion | 1 | 1.000 | 1 | 1.000 | 1 | 1.000 | 1 | 1.000 | 1 | 1.000 | 1 | 1.000 | 1 | 1.000 |
| 37 (R) | Agreeableness | 2 | 1.000 | 2 | 1.000 | 2 | 1.000 | 2 | 1.000 | 2 | 0.996 | 2 | 1.000 | 2 | 1.000 |
| 38 | Conscientiousness | 3 | 1.000 | 3 | 1.000 | 3 | 1.000 | 3 | 1.000 | 3 | 1.000 | 3 | 1.000 | 3 | 1.000 |
| 39 | Neuroticism | 4 | 1.000 | 4 | 1.000 | 4 | 1.000 | 4 | 1.000 | 4 | 1.000 | 5 | 1.000 | 5 | 1.000 |
| 40 | Openness | 5 | 1.000 | 5 | 1.000 | 5 | 1.000 | 5 | 1.000 | 5 | 1.000 | 4 | 1.000 | 4 | 1.000 |
| 41 (R) | Openness | 5 | 1.000 | 5 | 1.000 | 5 | 1.000 | 5 | 1.000 | 6 | 0.968 | 4 | 1.000 | 4 | 1.000 |
| 42 | Agreeableness | 2 | 1.000 | 2 | 1.000 | 2 | 1.000 | 2 | 1.000 | 2 | 1.000 | 2 | 1.000 | 2 | 1.000 |
| 43 (R) | Conscientiousness | 3 | 1.000 | 3 | 1.000 | 3 | 1.000 | 3 | 1.000 | 3 | 1.000 | 3 | 1.000 | 3 | 1.000 |
| 44 | Openness | 5 | 1.000 | 5 | 1.000 | 5 | 1.000 | 5 | 1.000 | 6 | 0.968 | 4 | 1.000 | 4 | 1.000 |

DeepSeek V4-Flash

| Item | Dimension | Community | Replication | Community | Replication | Community | Replication | Community | Replication | Community | Replication | Community | Replication | Community | Replication |
|---|---|---|---|---|---|---|---|---|---|---|---|---|---|---|---|
| 1 | Extraversion | 1 | 1.000 | 1 | 1.000 | 1 | 1.000 | 1 | 1.000 | 1 | 1.000 | 1 | 1.000 | 1 | 1.000 |

| 2 (R) | Agreeableness | 2 | 0.982 | 2 | 0.998 | 2 | 0.998 | 2 | 1.000 | 2 | 1.000 | 2 | 1.000 | 2 | 1.000 |
|---|---|---|---|---|---|---|---|---|---|---|---|---|---|---|---|
| 3 | Conscientiousness | 3 | 1.000 | 3 | 1.000 | 3 | 1.000 | 3 | 1.000 | 3 | 1.000 | 3 | 1.000 | 3 | 1.000 |
| 4 | Neuroticism | 4 | 1.000 | 4 | 1.000 | 4 | 1.000 | 4 | 1.000 | 4 | 1.000 | 4 | 1.000 | 4 | 1.000 |
| 5 | Openness | 5 | 1.000 | 5 | 1.000 | 5 | 1.000 | 5 | 1.000 | 5 | 1.000 | 5 | 1.000 | 5 | 1.000 |
| 6 (R) | Extraversion | 1 | 1.000 | 1 | 1.000 | 1 | 1.000 | 1 | 1.000 | 1 | 1.000 | 1 | 1.000 | 1 | 1.000 |
| 7 | Agreeableness | 2 | 1.000 | 2 | 1.000 | 2 | 1.000 | 2 | 1.000 | 2 | 1.000 | 2 | 1.000 | 2 | 1.000 |
| 8 (R) | Conscientiousness | 3 | 1.000 | 3 | 1.000 | 3 | 0.998 | 3 | 1.000 | 3 | 1.000 | 3 | 1.000 | 3 | 1.000 |
| 9 (R) | Neuroticism | 4 | 1.000 | 4 | 1.000 | 4 | 1.000 | 4 | 1.000 | 4 | 1.000 | 4 | 1.000 | 4 | 1.000 |
| 10 | Openness | 5 | 1.000 | 5 | 1.000 | 5 | 1.000 | 5 | 1.000 | 5 | 1.000 | 5 | 1.000 | 5 | 1.000 |
| 11 | Extraversion | 1 | 0.984 | 1 | 0.998 | 1 | 1.000 | 1 | 1.000 | 1 | 1.000 | 1 | 1.000 | 1 | 1.000 |
| 12 (R) | Agreeableness | 2 | 0.982 | 2 | 0.998 | 2 | 0.998 | 2 | 1.000 | 2 | 1.000 | 2 | 1.000 | 2 | 1.000 |
| 13 | Conscientiousness | 3 | 1.000 | 3 | 1.000 | 3 | 1.000 | 3 | 1.000 | 3 | 1.000 | 3 | 1.000 | 3 | 1.000 |
| 14 | Neuroticism | 4 | 1.000 | 4 | 1.000 | 4 | 1.000 | 4 | 1.000 | 4 | 1.000 | 4 | 1.000 | 4 | 1.000 |
| 15 | Openness | 5 | 1.000 | 5 | 1.000 | 5 | 1.000 | 5 | 1.000 | 5 | 1.000 | 5 | 1.000 | 5 | 1.000 |
| 16 | Extraversion | 1 | 0.984 | 1 | 0.998 | 1 | 1.000 | 1 | 1.000 | 1 | 1.000 | 1 | 1.000 | 1 | 1.000 |
| 17 | Agreeableness | 2 | 1.000 | 2 | 1.000 | 2 | 1.000 | 2 | 1.000 | 2 | 1.000 | 2 | 1.000 | 2 | 1.000 |
| 18 (R) | Conscientiousness | 3 | 1.000 | 3 | 1.000 | 3 | 0.998 | 3 | 1.000 | 3 | 1.000 | 3 | 1.000 | 3 | 1.000 |
| 19 | Neuroticism | 4 | 1.000 | 4 | 1.000 | 4 | 1.000 | 4 | 1.000 | 4 | 1.000 | 4 | 1.000 | 4 | 1.000 |
| 20 | Openness | 5 | 1.000 | 5 | 1.000 | 5 | 1.000 | 5 | 1.000 | 5 | 1.000 | 5 | 1.000 | 5 | 1.000 |
| 21 (R) | Extraversion | 1 | 1.000 | 1 | 1.000 | 1 | 1.000 | 1 | 1.000 | 1 | 1.000 | 1 | 1.000 | 1 | 1.000 |
| 22 | Agreeableness | 2 | 1.000 | 2 | 1.000 | 2 | 1.000 | 2 | 1.000 | 2 | 1.000 | 2 | 1.000 | 2 | 1.000 |
| 23 (R) | Conscientiousness | 3 | 1.000 | 3 | 1.000 | 3 | 0.998 | 3 | 1.000 | 3 | 1.000 | 3 | 1.000 | 3 | 1.000 |
| 24 (R) | Neuroticism | 4 | 1.000 | 4 | 1.000 | 4 | 1.000 | 4 | 1.000 | 4 | 1.000 | 4 | 1.000 | 4 | 1.000 |
| 25 | Openness | 5 | 1.000 | 5 | 1.000 | 5 | 1.000 | 5 | 1.000 | 5 | 1.000 | 5 | 1.000 | 5 | 1.000 |
| 26 | Extraversion | 1 | 0.984 | 1 | 0.998 | 1 | 1.000 | 1 | 1.000 | 1 | 1.000 | 1 | 1.000 | 1 | 1.000 |
| 27 (R) | Agreeableness | 2 | 0.982 | 2 | 0.998 | 2 | 0.998 | 2 | 1.000 | 2 | 1.000 | 2 | 1.000 | 2 | 1.000 |
| 28 | Conscientiousness | 3 | 1.000 | 3 | 1.000 | 3 | 1.000 | 3 | 1.000 | 3 | 1.000 | 3 | 1.000 | 3 | 1.000 |
| 29 | Neuroticism | 4 | 1.000 | 4 | 1.000 | 4 | 1.000 | 4 | 1.000 | 4 | 1.000 | 4 | 1.000 | 4 | 1.000 |
| 30 | Openness | 5 | 1.000 | 5 | 1.000 | 5 | 1.000 | 5 | 1.000 | 5 | 1.000 | 5 | 1.000 | 5 | 1.000 |
| 31 (R) | Extraversion | 1 | 1.000 | 1 | 1.000 | 1 | 1.000 | 1 | 1.000 | 1 | 1.000 | 1 | 1.000 | 1 | 1.000 |
| 32 | Agreeableness | 2 | 1.000 | 2 | 1.000 | 2 | 1.000 | 2 | 1.000 | 2 | 1.000 | 2 | 1.000 | 2 | 1.000 |
| 33 | Conscientiousness | 3 | 1.000 | 3 | 1.000 | 3 | 1.000 | 3 | 1.000 | 3 | 1.000 | 3 | 1.000 | 3 | 1.000 |
| 34 (R) | Neuroticism | 4 | 1.000 | 4 | 1.000 | 4 | 1.000 | 4 | 1.000 | 4 | 1.000 | 4 | 1.000 | 4 | 1.000 |
| 35 (R) | Openness | Alg-rem. | | Pre-rem. | | 5 | 0.830 | Alg-rem. | | 5 | 1.000 | 5 | 0.994 | Pre-rem. | |
| 36 | Extraversion | 1 | 1.000 | 1 | 1.000 | 1 | 1.000 | 1 | 1.000 | 1 | 1.000 | 1 | 1.000 | 1 | 1.000 |
| 37 (R) | Agreeableness | 2 | 0.982 | 2 | 0.998 | 2 | 0.998 | 2 | 1.000 | 2 | 1.000 | 2 | 1.000 | 2 | 1.000 |
| 38 | Conscientiousness | 3 | 1.000 | 3 | 1.000 | 3 | 1.000 | 3 | 1.000 | 3 | 1.000 | 3 | 1.000 | 3 | 1.000 |
| 39 | Neuroticism | 4 | 1.000 | 4 | 1.000 | 4 | 1.000 | 4 | 1.000 | 4 | 1.000 | 4 | 1.000 | 4 | 1.000 |
| 40 | Openness | 5 | 1.000 | 5 | 1.000 | 5 | 1.000 | 5 | 1.000 | 5 | 1.000 | 5 | 1.000 | 5 | 1.000 |
| 41 (R) | Openness | 5 | 1.000 | 5 | 1.000 | 5 | 1.000 | 5 | 1.000 | 5 | 1.000 | 5 | 1.000 | 5 | 1.000 |
| 42 | Agreeableness | 2 | 1.000 | 2 | 1.000 | 2 | 1.000 | 2 | 1.000 | 2 | 1.000 | 2 | 1.000 | 2 | 1.000 |
| 43 (R) | Conscientiousness | 3 | 1.000 | 3 | 1.000 | 3 | 0.998 | 3 | 1.000 | 3 | 1.000 | 3 | 1.000 | 3 | 1.000 |
| 44 | Openness | 5 | 1.000 | 5 | 1.000 | 5 | 1.000 | 5 | 1.000 | 5 | 1.000 | 5 | 1.000 | 5 | 1.000 |
| Qwen 3.5 Plus | | | | | | | | | | | | | | | |
| Item | Dimension | Community | Replication | Community | Replication | Community | Replication | Community | Replication | Community | Replication | Community | Replication | Community | Replication |
| 1 | Extraversion | 1 | 1.000 | 1 | 1.000 | 1 | 1.000 | 1 | 1.000 | 1 | 1.000 | 1 | 1.000 | 1 | 1.000 |
| 2 (R) | Agreeableness | Alg-rem. | | Alg-rem. | | 2 | 1.000 | 2 | 1.000 | Pre-rem. | | 2 | 1.000 | 2 | 1.000 |
| 3 | Conscientiousness | 2 | 1.000 | 2 | 1.000 | 3 | 1.000 | 3 | 1.000 | 2 | 1.000 | 3 | 1.000 | 3 | 1.000 |
| 4 | Neuroticism | 3 | 1.000 | 3 | 1.000 | 4 | 1.000 | 4 | 1.000 | 3 | 1.000 | 4 | 1.000 | 4 | 1.000 |
| 5 | Openness | 4 | 1.000 | 4 | 1.000 | 5 | 1.000 | 5 | 1.000 | 4 | 1.000 | 5 | 1.000 | 5 | 1.000 |

| 6 (R) | Extraversion | 1 | 1.000 | 1 | 1.000 | 1 | 1.000 | 1 | 1.000 | 1 | 1.000 | 1 | 1.000 | 1 | 1.000 |
|---|---|---|---|---|---|---|---|---|---|---|---|---|---|---|---|
| 7 | Agreeableness | 5 | 1.000 | Alg-rem. | | 2 | 1.000 | 2 | 1.000 | 5 | 1.000 | 2 | 1.000 | 2 | 1.000 |
| 8 (R) | Conscientiousness | Alg-rem. | | 2 | 0.868 | 3 | 1.000 | 3 | 1.000 | 2 | 1.000 | 3 | 1.000 | 3 | 1.000 |
| 9 (R) | Neuroticism | 3 | 1.000 | 3 | 1.000 | 4 | 1.000 | 4 | 1.000 | 3 | 1.000 | 4 | 1.000 | 4 | 1.000 |
| 10 | Openness | 4 | 1.000 | 4 | 1.000 | 5 | 1.000 | 5 | 1.000 | 4 | 1.000 | 5 | 1.000 | 5 | 1.000 |
| 11 | Extraversion | 1 | 1.000 | Alg-rem. | | Pre-rem. | | Pre-rem. | | 1 | 1.000 | 1 | 0.826 | Pre-rem. | |
| 12 (R) | Agreeableness | Alg-rem. | | Alg-rem. | | Pre-rem. | | 2 | 1.000 | Pre-rem. | | 2 | 1.000 | 2 | 1.000 |
| 13 | Conscientiousness | 2 | 1.000 | 2 | 1.000 | 3 | 1.000 | 3 | 1.000 | 2 | 1.000 | 3 | 1.000 | 3 | 1.000 |
| 14 | Neuroticism | 3 | 1.000 | 3 | 1.000 | 4 | 1.000 | 4 | 1.000 | 3 | 1.000 | 4 | 1.000 | 4 | 1.000 |
| 15 | Openness | 4 | 1.000 | 4 | 1.000 | 5 | 1.000 | 5 | 1.000 | 4 | 1.000 | 5 | 1.000 | 5 | 1.000 |
| 16 | Extraversion | 1 | 1.000 | Alg-rem. | | Pre-rem. | | Pre-rem. | | 1 | 1.000 | 1 | 0.826 | Pre-rem. | |
| 17 | Agreeableness | 5 | 1.000 | Alg-rem. | | 2 | 1.000 | 2 | 1.000 | 5 | 1.000 | 2 | 1.000 | 2 | 1.000 |
| 18 (R) | Conscientiousness | Alg-rem. | | 2 | 0.868 | 3 | 1.000 | 3 | 1.000 | 2 | 1.000 | 3 | 1.000 | 3 | 1.000 |
| 19 | Neuroticism | 3 | 1.000 | 3 | 1.000 | 4 | 1.000 | 4 | 1.000 | 3 | 1.000 | 4 | 1.000 | 4 | 1.000 |
| 20 | Openness | 4 | 1.000 | 4 | 1.000 | 5 | 1.000 | 5 | 1.000 | 4 | 1.000 | 5 | 1.000 | 5 | 1.000 |
| 21 (R) | Extraversion | 1 | 1.000 | 1 | 1.000 | 1 | 1.000 | 1 | 1.000 | 1 | 1.000 | 1 | 1.000 | 1 | 1.000 |
| 22 | Agreeableness | 5 | 1.000 | Alg-rem. | | 2 | 1.000 | 2 | 1.000 | 5 | 1.000 | 2 | 1.000 | 2 | 1.000 |
| 23 (R) | Conscientiousness | Alg-rem. | | 2 | 0.868 | 3 | 1.000 | 3 | 1.000 | 2 | 1.000 | 3 | 1.000 | 3 | 1.000 |
| 24 (R) | Neuroticism | 3 | 1.000 | 3 | 1.000 | 4 | 1.000 | 4 | 1.000 | 3 | 1.000 | 4 | 1.000 | 4 | 1.000 |
| 25 | Openness | 4 | 1.000 | 4 | 1.000 | 5 | 1.000 | 5 | 1.000 | 4 | 1.000 | 5 | 1.000 | 5 | 1.000 |
| 26 | Extraversion | 1 | 1.000 | Alg-rem. | | Pre-rem. | | Pre-rem. | | 1 | 1.000 | 1 | 0.826 | Pre-rem. | |
| 27 (R) | Agreeableness | Alg-rem. | | Alg-rem. | | Pre-rem. | | 2 | 1.000 | Pre-rem. | | 2 | 1.000 | 2 | 1.000 |
| 28 | Conscientiousness | 2 | 1.000 | 2 | 1.000 | 3 | 1.000 | 3 | 1.000 | 2 | 1.000 | 3 | 1.000 | 3 | 1.000 |
| 29 | Neuroticism | 3 | 1.000 | 3 | 1.000 | 4 | 1.000 | 4 | 1.000 | 3 | 1.000 | 4 | 1.000 | 4 | 1.000 |
| 30 | Openness | 4 | 1.000 | 4 | 1.000 | 5 | 1.000 | 5 | 1.000 | 4 | 1.000 | 5 | 1.000 | 5 | 1.000 |
| 31 (R) | Extraversion | 1 | 1.000 | 1 | 1.000 | 1 | 1.000 | 1 | 1.000 | 1 | 1.000 | 1 | 1.000 | 1 | 1.000 |
| 32 | Agreeableness | 5 | 1.000 | Alg-rem. | | 2 | 1.000 | 2 | 1.000 | 5 | 1.000 | 2 | 1.000 | 2 | 1.000 |
| 33 | Conscientiousness | 2 | 1.000 | 2 | 1.000 | 3 | 1.000 | 3 | 1.000 | 2 | 1.000 | 3 | 1.000 | 3 | 1.000 |
| 34 (R) | Neuroticism | 3 | 1.000 | 3 | 1.000 | 4 | 1.000 | 4 | 1.000 | 3 | 1.000 | 4 | 1.000 | 4 | 1.000 |
| 35 (R) | Openness | Alg-rem. | | Alg-rem. | removed | Pre-rem. | | 5 | 1.000 | 4 | 1.000 | 5 | 1.000 | Pre-rem. | |
| 36 | Extraversion | 1 | 1.000 | 1 | 1.000 | 1 | 1.000 | 1 | 1.000 | 1 | 1.000 | 1 | 1.000 | 1 | 1.000 |
| 37 (R) | Agreeableness | Alg-rem. | | Alg-rem. | | Pre-rem. | | 2 | 1.000 | Pre-rem. | | 2 | 1.000 | 2 | 1.000 |
| 38 | Conscientiousness | 2 | 1.000 | 2 | 1.000 | 3 | 1.000 | 3 | 1.000 | 2 | 1.000 | 3 | 1.000 | 3 | 1.000 |
| 39 | Neuroticism | 3 | 1.000 | 3 | 1.000 | 4 | 1.000 | 4 | 1.000 | 3 | 1.000 | 4 | 1.000 | 4 | 1.000 |
| 40 | Openness | 4 | 1.000 | 4 | 1.000 | 5 | 1.000 | 5 | 1.000 | 4 | 1.000 | 5 | 1.000 | 5 | 1.000 |
| 41 (R) | Openness | 4 | 1.000 | 4 | 1.000 | 5 | 1.000 | 5 | 1.000 | 4 | 1.000 | 5 | 1.000 | 5 | 1.000 |
| 42 | Agreeableness | 5 | 1.000 | Alg-rem. | | 2 | 1.000 | 2 | 1.000 | 5 | 1.000 | 2 | 1.000 | 2 | 1.000 |
| 43 (R) | Conscientiousness | Alg-rem. | | 2 | 0.868 | 3 | 1.000 | 3 | 1.000 | 2 | 1.000 | 3 | 1.000 | 3 | 1.000 |
| 44 | Openness | 4 | 1.000 | 4 | 1.000 | 5 | 1.000 | 5 | 1.000 | 4 | 1.000 | 5 | 1.000 | 5 | 1.000 |

*Note.* The Big Five personality items were adopted from the Big Five Inventory[29]. Analyses were conducted separately for each human–digital twin pair. (R) denotes reverse-scored items. Community numbers correspond to the clusters identified by the *walktrap* algorithm. "Replication" indicates item-wise replication rates calculated via Bootstrap Exploratory Graph Analysis (BootEGA), where a value of 1.00 represents perfect stability of the item's placement within the detected community structure. Pre-rem. = removed before the invariance test; Alg-rem. = removed during the iterative invariance test.

**Table S30 Network configural invariance of Big Five Inventory responses among digital twins across input conditions in Study 3a**

| | | | Gemma-4-31B-it | | gpt-oss-120b | | DeepSeek V4-Flash | | Qwen 3.5 Plus | |
|---|---|---|---|---|---|---|---|---|---|---|
| **Item** | **Content** | **Dimension** | **Community** | **Replication** | **Community** | **Replication** | **Community** | **Replication** | **Community** | **Replication** |
| 1 | Is talkative | Extraversion | 1 | 1.00 | 1 | 1.00 | 1 | 1.00 | 1 | 1.00 |
| 2 (R) | Tends to find fault with others | Agreeableness | 2 | 1.00 | 2 | 1.00 | 2 | 0.998 | 2 | 1.00 |
| 3 | Does a thorough job | Conscientiousness | 3 | 1.00 | 3 | 1.00 | 3 | 1.00 | 3 | 1.00 |
| 4 | Is depressed, blue | Neuroticism | 4 | 1.00 | 4 | 0.992 | 4 | 1.00 | 4 | 1.00 |
| 5 | Is original, comes up with new ideas | Openness | 5 | 1.00 | 5 | 1.00 | 5 | 1.00 | 5 | 1.00 |
| 6 (R) | Is reserved | Extraversion | 1 | 1.00 | 1 | 1.00 | 1 | 1.00 | 1 | 1.00 |
| 7 | Is helpful and unselfish with others | Agreeableness | 2 | 1.00 | 2 | 1.00 | 2 | 1.00 | 2 | 1.00 |
| 8 (R) | Can be somewhat careless | Conscientiousness | 3 | 1.00 | 3 | 1.00 | 3 | 0.998 | 3 | 1.00 |
| 9 (R) | Is relaxed, handles stress well | Neuroticism | 4 | 1.00 | 4 | 1.00 | 4 | 1.00 | 4 | 1.00 |
| 10 | Is curious about many different things | Openness | 5 | 1.00 | 5 | 1.00 | 5 | 1.00 | 5 | 1.00 |
| 11 | Is full of energy | Extraversion | 1 | 1.00 | 1 | 0.994 | 1 | 1.00 | 1 | 1.00 |
| 12 (R) | Starts quarrels with others | Agreeableness | 2 | 1.00 | 2 | 0.994 | 2 | 0.998 | 2 | 1.00 |
| 13 | Is a reliable worker | Conscientiousness | 3 | 1.00 | 3 | 1.00 | 3 | 1.00 | 3 | 1.00 |
| 14 | Can be tense | Neuroticism | 4 | 1.00 | 4 | 1.00 | 4 | 1.00 | 4 | 1.00 |
| 15 | Is ingenious, a deep thinker | Openness | 5 | 1.00 | 5 | 1.00 | 5 | 1.00 | 5 | 1.00 |
| 16 | Generates a lot of enthusiasm | Extraversion | 1 | 1.00 | 1 | 0.994 | 1 | 1.00 | 1 | 1.00 |
| 17 | Has a forgiving nature | Agreeableness | 2 | 1.00 | 2 | 1.00 | 2 | 1.00 | 2 | 1.00 |
| 18 (R) | Tends to be disorganized | Conscientiousness | 3 | 1.00 | 3 | 1.00 | 3 | 0.998 | 3 | 1.00 |
| 19 | Worries a lot | Neuroticism | 4 | 1.00 | 4 | 1.00 | 4 | 1.00 | 4 | 1.00 |
| 20 | Has an active imagination | Openness | 5 | 1.00 | 5 | 1.00 | 5 | 1.00 | 5 | 1.00 |
| 21 (R) | Tends to be quiet | Extraversion | 1 | 1.00 | 1 | 1.00 | 1 | 1.00 | 1 | 1.00 |
| 22 | Is generally trusting | Agreeableness | 2 | 1.00 | 2 | 1.00 | 2 | 1.00 | 2 | 1.00 |
| 23 (R) | Tends to be lazy | Conscientiousness | 3 | 1.00 | 3 | 1.00 | 3 | 0.998 | 3 | 1.00 |
| 24 (R) | Is emotionally stable, not easily upset | Neuroticism | 4 | 1.00 | 4 | 1.00 | 4 | 1.00 | 4 | 1.00 |
| 25 | Is inventive | Openness | 5 | 1.00 | 5 | 1.00 | 5 | 1.00 | 5 | 1.00 |
| 26 | Has an assertive personality | Extraversion | 1 | 1.00 | 1 | 0.994 | 1 | 1.00 | 1 | 1.00 |
| 27 (R) | Can be cold and aloof | Agreeableness | 2 | 1.00 | 2 | 1.00 | 2 | 0.998 | 2 | 1.00 |
| 28 | Perseveres until the task is finished | Conscientiousness | 3 | 1.00 | 3 | 1.00 | 3 | 1.00 | 3 | 1.00 |
| 29 | Can be moody | Neuroticism | 4 | 1.00 | 4 | 1.00 | 4 | 1.00 | 4 | 1.00 |
| 30 | Values artistic, aesthetic experiences | Openness | 5 | 1.00 | 6 | 1.00 | 5 | 1.00 | 5 | 1.00 |
| 31 (R) | Is sometimes shy, inhibited | Extraversion | 1 | 1.00 | 1 | 1.00 | 1 | 1.00 | 1 | 1.00 |
| 32 | Is considerate and kind to almost everyone | Agreeableness | 2 | 1.00 | 2 | 1.00 | 2 | 1.00 | 2 | 1.00 |
| 33 | Does things efficiently | Conscientiousness | 3 | 1.00 | 3 | 1.00 | 3 | 1.00 | 3 | 1.00 |
| 34 (R) | Remains calm in tense situations | Neuroticism | 4 | 1.00 | 4 | 1.00 | 4 | 1.00 | 4 | 1.00 |
| 35 (R) | Prefers work that is routine | Openness | 5 | 1.00 | 5 | 1.00 | 5 | 1.00 | 5 | 1.00 |
| 36 | Is outgoing, sociable | Extraversion | 1 | 1.00 | 1 | 1.00 | 1 | 1.00 | 1 | 1.00 |
| 37 (R) | Is sometimes rude to others | Agreeableness | 2 | 1.00 | 2 | 0.994 | 2 | 0.998 | 2 | 1.00 |
| 38 | Makes plans and follows through with them | Conscientiousness | 3 | 1.00 | 3 | 1.00 | 3 | 1.00 | 3 | 1.00 |

| 39 | Gets nervous easily | Neuroticism | 4 | 1.00 | 4 | 1.00 | 4 | 1.00 | 4 | 1.00 |
|---|---|---|---|---|---|---|---|---|---|---|
| 40 | Likes to reflect, play with ideas | Openness | 5 | 1.00 | 5 | 1.00 | 5 | 1.00 | 5 | 1.00 |
| 41 (R) | Has few artistic interests | Openness | 5 | 1.00 | 6 | 1.00 | 5 | 1.00 | 5 | 1.00 |
| 42 | Likes to cooperate with others | Agreeableness | 2 | 1.00 | 2 | 1.00 | 2 | 1.00 | 2 | 1.00 |
| 43 (R) | Is easily distracted | Conscientiousness | 3 | 1.00 | 3 | 1.00 | 3 | 0.998 | 3 | 1.00 |
| 44 | Is sophisticated in art, music, or literature | Openness | 5 | 1.00 | 6 | 1.00 | 5 | 1.00 | 5 | 1.00 |

*Note.* The Big Five personality items were adopted from the Big Five Inventory[29]. Analysis was performed on the pooled sample. (R) denotes reverse-scored items. Community numbers correspond to the clusters identified by the *walktrap* algorithm. "Replication" indicates item-wise replication rates calculated via Bootstrap Exploratory Graph Analysis (BootEGA), where a value of 1.00 represents perfect stability of the item's placement within the detected community structure.

**Table S31 Network configural invariance of Big Five Inventory responses among digital-twin models within the same input condition in Study 3a**

| | Demographics | | Heuristics and biases experiments | | Demographics + Heuristics and biases experiments | | Demographics + Heuristics and biases experiments + Features with r < 0.3 | | Demographics + Heuristics and biases experiments + Features with r > 0.5 | | Demographics + Heuristics and biases experiments + Features with r < 0.5 | | Demographics + All psychological measures + Heuristics and biases experiments | |
|---|---|---|---|---|---|---|---|---|---|---|---|---|---|---|
| **Item** | **Community** | **Replication** | **Community** | **Replication** | **Community** | **Replication** | **Community** | **Replication** | **Community** | **Replication** | **Community** | **Replication** | **Community** | **Replication** |
| 1 | 1 | 1.00 | 1 | 1.00 | 1 | 1.00 | 1 | 1.00 | 1 | 1.00 | 1 | 1.00 | 1 | 1.00 |
| 2 (R) | 2 | 1.00 | 2 | 1.00 | 2 | 1.00 | 2 | 0.938 | 2 | 1.00 | 2 | 1.00 | 2 | 1.00 |
| 3 | 3 | 1.00 | 3 | 1.00 | 3 | 1.00 | 3 | 1.00 | 3 | 1.00 | 3 | 1.00 | 3 | 1.00 |
| 4 | 4 | 1.00 | 4 | 1.00 | 4 | 1.00 | 2 | 0.938 | 4 | 1.00 | 4 | 0.968 | 4 | 1.00 |
| 5 | 5 | 1.00 | 5 | 1.00 | 5 | 1.00 | 4 | 1.00 | 5 | 1.00 | 5 | 1.00 | 5 | 1.00 |
| 6 (R) | 1 | 1.00 | 1 | 1.00 | 1 | 1.00 | 1 | 1.00 | 1 | 1.00 | 1 | 1.00 | 1 | 1.00 |
| 7 | 6 | 1.00 | 2 | 1.00 | Removed | | 2 | 0.976 | 2 | 1.00 | 2 | 1.00 | 2 | 1.00 |
| 8 (R) | 3 | 0.778 | 6 | 1.00 | Removed | | 3 | 1.00 | 3 | 1.00 | 4 | 1.00 | 3 | 1.00 |
| 9 (R) | 4 | 1.00 | 4 | 1.00 | 4 | 1.00 | 5 | 1.00 | 4 | 1.00 | 6 | 1.00 | 4 | 1.00 |
| 10 | 5 | 1.00 | 5 | 1.00 | 5 | 1.00 | 4 | 1.00 | 5 | 1.00 | 5 | 1.00 | 5 | 1.00 |
| 11 | 1 | 1.00 | 1 | 1.00 | 1 | 1.00 | 1 | 1.00 | 1 | 1.00 | 1 | 1.00 | 1 | 1.00 |
| 12 (R) | 2 | 1.00 | 2 | 1.00 | 2 | 1.00 | 2 | 0.938 | 2 | 1.00 | 2 | 1.00 | 2 | 1.00 |
| 13 | 3 | 1.00 | 3 | 1.00 | 3 | 1.00 | 3 | 1.00 | 3 | 1.00 | 3 | 1.00 | 3 | 1.00 |
| 14 | 4 | 1.00 | 4 | 1.00 | 4 | 1.00 | 5 | 1.00 | 4 | 1.00 | 6 | 1.00 | 4 | 1.00 |
| 15 | 5 | 1.00 | 5 | 1.00 | 5 | 1.00 | 4 | 1.00 | 5 | 1.00 | 5 | 1.00 | 5 | 1.00 |
| 16 | 1 | 1.00 | 1 | 1.00 | 1 | 1.00 | 1 | 1.00 | 1 | 1.00 | 1 | 1.00 | 1 | 1.00 |
| 17 | 6 | 1.00 | 2 | 1.00 | Removed | | 2 | 0.976 | 2 | 1.00 | 2 | 1.00 | 2 | 1.00 |
| 18 (R) | 3 | 0.778 | 6 | 1.00 | Removed | | 3 | 1.00 | 3 | 1.00 | 4 | 1.00 | 3 | 1.00 |
| 19 | 4 | 1.00 | 4 | 1.00 | 4 | 1.00 | 5 | 1.00 | 4 | 1.00 | 6 | 1.00 | 4 | 1.00 |
| 20 | 5 | 1.00 | 5 | 1.00 | 5 | 1.00 | 4 | 1.00 | 5 | 1.00 | 5 | 1.00 | 5 | 1.00 |
| 21 (R) | 1 | 1.00 | 1 | 1.00 | 1 | 1.00 | 1 | 1.00 | 1 | 1.00 | 1 | 1.00 | 1 | 1.00 |
| 22 | 2 | 1.00 | 2 | 1.00 | 2 | 1.00 | 2 | 0.938 | 2 | 1.00 | 2 | 1.00 | 2 | 1.00 |
| 23 (R) | 3 | 0.778 | 6 | 1.00 | Removed | | 3 | 1.00 | 3 | 1.00 | 4 | 1.00 | 3 | 1.00 |
| 24 (R) | 4 | 1.00 | 4 | 1.00 | 4 | 1.00 | 5 | 1.00 | 4 | 1.00 | 6 | 1.00 | 4 | 1.00 |
| 25 | 5 | 1.00 | 5 | 1.00 | 5 | 1.00 | 4 | 1.00 | 5 | 1.00 | 5 | 1.00 | 5 | 1.00 |
| 26 | 1 | 1.00 | 1 | 1.00 | 1 | 1.00 | 1 | 1.00 | 1 | 1.00 | 1 | 1.00 | 1 | 1.00 |
| 27 (R) | 2 | 1.00 | 2 | 1.00 | 2 | 1.00 | 2 | 0.938 | 2 | 1.00 | 2 | 1.00 | 2 | 1.00 |
| 28 | 3 | 1.00 | 3 | 1.00 | 3 | 1.00 | 3 | 1.00 | 3 | 1.00 | 3 | 1.00 | 3 | 1.00 |
| 29 | 4 | 1.00 | 4 | 1.00 | 4 | 1.00 | 2 | 0.938 | 4 | 1.00 | 4 | 0.968 | 4 | 1.00 |
| 30 | 5 | 1.00 | 5 | 1.00 | 5 | 1.00 | 4 | 1.00 | 6 | 1.00 | 5 | 1.00 | 5 | 0.958 |
| 31 (R) | 1 | 1.00 | 1 | 1.00 | 1 | 1.00 | 1 | 1.00 | 1 | 1.00 | 1 | 1.00 | 1 | 1.00 |
| 32 | 6 | 1.00 | 2 | 1.00 | Removed | | 2 | 0.976 | 2 | 1.00 | 2 | 1.00 | 2 | 1.00 |
| 33 | 3 | 1.00 | 3 | 1.00 | 3 | 1.00 | 3 | 1.00 | 3 | 1.00 | 3 | 1.00 | 3 | 1.00 |
| 34 (R) | 4 | 1.00 | 4 | 1.00 | 4 | 1.00 | 5 | 1.00 | 4 | 1.00 | 4 | 1.00 | 4 | 1.00 |
| 35 (R) | 5 | 1.00 | 5 | 1.00 | 5 | 1.00 | 4 | 0.938 | 5 | 1.00 | Removed | | 5 | 1.00 |
| 36 | 1 | 1.00 | 1 | 1.00 | 1 | 1.00 | 1 | 1.00 | 1 | 1.00 | 1 | 1.00 | 1 | 1.00 |
| 37 (R) | 2 | 1.00 | 2 | 1.00 | 2 | 1.00 | 2 | 0.938 | 2 | 1.00 | 2 | 1.00 | 2 | 1.00 |
| 38 | 3 | 1.00 | 3 | 1.00 | 3 | 1.00 | 3 | 1.00 | 3 | 1.00 | 3 | 1.00 | 3 | 1.00 |
| 39 | 4 | 1.00 | 4 | 1.00 | 4 | 1.00 | 5 | 1.00 | 4 | 1.00 | 6 | 1.00 | 4 | 1.00 |
| 40 | 5 | 1.00 | 5 | 1.00 | 5 | 1.00 | 4 | 1.00 | 5 | 1.00 | 5 | 1.00 | 5 | 1.00 |
| 41 (R) | 5 | 1.00 | 5 | 1.00 | 5 | 1.00 | 4 | 1.00 | 6 | 1.00 | 5 | 1.00 | 5 | 0.958 |
| 42 | 6 | 1.00 | 2 | 1.00 | Removed | | 2 | 0.976 | 2 | 1.00 | 2 | 1.00 | 2 | 1.00 |
| 43 (R) | 3 | 0.778 | 6 | 1.00 | Removed | | 3 | 1.00 | 3 | 1.00 | 4 | 1.00 | 3 | 1.00 |
| 44 | 5 | 1.00 | 5 | 1.00 | 5 | 1.00 | 4 | 1.00 | 6 | 1.00 | 5 | 1.00 | 5 | 0.958 |

*Note.* The Big Five personality items were adopted from the Big Five Inventory[29]. Analysis was performed on the pooled sample. (R) denotes reverse-scored items. Community numbers correspond to the clusters identified by the *walktrap* algorithm. "Replication" indicates item-wise replication rates calculated via Bootstrap Exploratory Graph Analysis (BootEGA), where a value of 1.00 represents perfect stability of the item's placement within the detected community structure. Items with stability below .70 were iteratively removed.

**Table S32 Metric invariance tests for the Big Five personality traits comparing human participants and digital twins in Study 3a**

**Panel A. Gemma-4-31B-it**

| Item | Membership | Difference | *p* | Sig | Direction |
|---|---|---|---|---|---|
| Demographics vs Human | | | | | |
| Q1 | 1 | -0.038 | 0.173 | | |
| Q6 | 1 | -0.051 | 0.129 | | |
| Q11 | 1 | -0.205 | 0.005 | ** | Human < Demographics |
| Q16 | 1 | -0.121 | 0.005 | ** | Human < Demographics |
| Q21 | 1 | 0.094 | 0.005 | ** | Human > Demographics |
| Q26 | 1 | -0.097 | 0.008 | ** | Human < Demographics |
| Q31 | 1 | 0.085 | 0.005 | ** | Human > Demographics |
| Q36 | 1 | 0.073 | 0.039 | * | Human > Demographics |
| Q2 | 2 | 0.025 | 0.614 | | |
| Q7 | 2 | -0.24 | 0.005 | ** | Human < Demographics |
| Q12 | 2 | -0.142 | 0.005 | ** | Human < Demographics |
| Q17 | 2 | 0.158 | 0.005 | ** | Human > Demographics |
| Q22 | 2 | -0.114 | 0.015 | * | Human < Demographics |
| Q27 | 2 | -0.03 | 0.489 | | |
| Q32 | 2 | 0.001 | 0.994 | | |
| Q37 | 2 | -0.027 | 0.533 | | |
| Q42 | 2 | -0.137 | 0.005 | ** | Human < Demographics |
| Q3 | 3 | -0.054 | 0.148 | | |
| Q8 | 3 | -0.121 | 0.005 | ** | Human < Demographics |
| Q13 | 3 | -0.029 | 0.489 | | |
| Q18 | 3 | 0.089 | 0.008 | ** | Human > Demographics |
| Q23 | 3 | 0.03 | 0.421 | | |
| Q28 | 3 | -0.068 | 0.037 | * | Human < Demographics |
| Q33 | 3 | -0.002 | 0.954 | | |
| Q38 | 3 | -0.069 | 0.079 | | |
| Q43 | 3 | 0.005 | 0.921 | | |
| Q4 | 4 | -0.121 | 0.005 | ** | Human < Demographics |
| Q9 | 4 | 0.17 | 0.005 | ** | Human > Demographics |
| Q14 | 4 | -0.021 | 0.523 | | |
| Q19 | 4 | 0.094 | 0.005 | ** | Human > Demographics |
| Q24 | 4 | -0.003 | 0.954 | | |
| Q29 | 4 | -0.159 | 0.005 | ** | Human < Demographics |
| Q34 | 4 | -0.083 | 0.008 | ** | Human < Demographics |
| Q39 | 4 | -0.067 | 0.022 | * | Human < Demographics |
| Q5 | 5 | 0.021 | 0.489 | | |
| Q10 | 5 | -0.087 | 0.005 | ** | Human < Demographics |
| Q15 | 5 | -0.129 | 0.005 | ** | Human < Demographics |
| Q20 | 5 | -0.013 | 0.638 | | |
| Q25 | 5 | -0.068 | 0.025 | * | Human < Demographics |
| Q30 | 5 | 0.168 | 0.005 | ** | Human > Demographics |
| Q35 | 5 | -0.334 | 0.005 | ** | Human < Demographics |
| Q40 | 5 | -0.049 | 0.115 | | |
| Q41 | 5 | -0.176 | 0.005 | ** | Human < Demographics |
| Q44 | 5 | 0.079 | 0.005 | ** | Human > Demographics |
| Heuristics and biases experiments vs Human | | | | | |
| Q1 | 1 | -0.045 | 0.135 | | |
| Q6 | 1 | -0.101 | 0.009 | ** | Human < Heuristics and biases experiments |
| Q11 | 1 | -0.017 | 0.618 | | |
| Q16 | 1 | -0.059 | 0.085 | | |
| Q21 | 1 | 0.125 | 0.005 | ** | Human > Heuristics and biases experiments |
| Q26 | 1 | -0.118 | 0.009 | ** | Human < Heuristics and biases experiments |
| Q31 | 1 | 0.094 | 0.005 | ** | Human > Heuristics and biases experiments |
| Q36 | 1 | 0.078 | 0.036 | * | Human > Heuristics and biases experiments |
| Q2 | 2 | -0.114 | 0.013 | * | Human < Heuristics and biases experiments |
| Q7 | 2 | -0.083 | 0.113 | | |
| Q12 | 2 | -0.065 | 0.098 | | |
| Q17 | 2 | -0.057 | 0.179 | | |
| Q22 | 2 | 0.022 | 0.553 | | |
| Q27 | 2 | -0.13 | 0.005 | ** | Human < Heuristics and biases experiments |
| Q32 | 2 | 0.173 | 0.005 | ** | Human > Heuristics and biases experiments |
| Q37 | 2 | -0.042 | 0.267 | | |
| Q42 | 2 | -0.166 | 0.005 | ** | Human < Heuristics and biases experiments |
| Q3 | 3 | 0.01 | 0.73 | | |
| Q8 | 3 | -0.028 | 0.414 | | |
| Q13 | 3 | -0.023 | 0.486 | | |
| Q18 | 3 | 0.203 | 0.005 | ** | Human > Heuristics and biases experiments |
| Q23 | 3 | 0.064 | 0.053 | | |
| Q28 | 3 | 0.016 | 0.633 | | |
| Q33 | 3 | 0.013 | 0.641 | | |
| Q38 | 3 | -0.086 | 0.017 | * | Human < Heuristics and biases experiments |
| Q43 | 3 | -0.098 | 0.005 | ** | Human < Heuristics and biases experiments |
| Q4 | 4 | 0.144 | 0.005 | ** | Human > Heuristics and biases experiments |
| Q9 | 4 | 0.283 | 0.005 | ** | Human > Heuristics and biases experiments |
| Q14 | 4 | -0.116 | 0.009 | ** | Human < Heuristics and biases experiments |
| Q19 | 4 | 0.112 | 0.005 | ** | Human > Heuristics and biases experiments |
| Q24 | 4 | -0.056 | 0.22 | | |
| Q29 | 4 | -0.096 | 0.005 | ** | Human < Heuristics and biases experiments |
| Q34 | 4 | -0.164 | 0.005 | ** | Human < Heuristics and biases experiments |
| Q39 | 4 | -0.045 | 0.179 | | |
| Q5 | 5 | 0.069 | 0.046 | * | Human > Heuristics and biases experiments |
| Q10 | 5 | 0.016 | 0.702 | | |
| Q15 | 5 | -0.029 | 0.298 | | |
| Q20 | 5 | -0.094 | 0.005 | ** | Human < Heuristics and biases experiments |
| Q25 | 5 | -0.071 | 0.048 | * | Human < Heuristics and biases experiments |
| Q30 | 5 | 0.127 | 0.005 | ** | Human > Heuristics and biases experiments |
| Q35 | 5 | -0.156 | 0.005 | ** | Human < Heuristics and biases experiments |
| Q40 | 5 | -0.059 | 0.13 | | |
| Q41 | 5 | -0.12 | 0.005 | ** | Human < Heuristics and biases experiments |
| Q44 | 5 | -0.039 | 0.179 | | |
| Demographics + Heuristics and biases experiments vs Human | | | | | |
| Q1 | 1 | -0.063 | 0.028 | * | Human < Demographics + Heuristics and biases experiments |
| Q6 | 1 | -0.112 | 0.006 | ** | Human < Demographics + Heuristics and biases experiments |
| Q11 | 1 | -0.053 | 0.067 | | |
| Q16 | 1 | -0.089 | 0.01 | * | Human < Demographics + Heuristics and biases experiments |
| Q21 | 1 | 0.108 | 0.006 | ** | Human > Demographics + Heuristics and biases experiments |
| Q26 | 1 | -0.123 | 0.006 | ** | Human < Demographics + Heuristics and biases experiments |
| Q31 | 1 | 0.038 | 0.165 | | |
| Q36 | 1 | 0.053 | 0.118 | | |
| Q2 | 2 | -0.16 | 0.006 | ** | Human < Demographics + Heuristics and biases experiments |
| Q7 | 2 | -0.185 | 0.006 | ** | Human < Demographics + Heuristics and biases experiments |
| Q12 | 2 | -0.053 | 0.171 | | |
| Q17 | 2 | -0.098 | 0.023 | * | Human < Demographics + Heuristics and biases experiments |
| Q22 | 2 | 0.126 | 0.013 | * | Human > Demographics + Heuristics and biases experiments |
| Q27 | 2 | -0.017 | 0.64 | | |
| Q32 | 2 | 0.073 | 0.108 | | |
| Q37 | 2 | -0.101 | 0.026 | * | Human < Demographics + Heuristics and biases experiments |
| Q42 | 2 | -0.157 | 0.006 | ** | Human < Demographics + Heuristics and biases experiments |
| Q3 | 3 | -0.074 | 0.03 | * | Human < Demographics + Heuristics and biases experiments |
| Q8 | 3 | -0.099 | 0.006 | ** | Human < Demographics + Heuristics and biases experiments |
| Q13 | 3 | -0.039 | 0.266 | | |
| Q18 | 3 | 0.044 | 0.171 | | |
| Q23 | 3 | 0.033 | 0.315 | | |
| Q28 | 3 | -0.022 | 0.465 | | |
| Q33 | 3 | -0.031 | 0.315 | | |
| Q38 | 3 | -0.099 | 0.013 | * | Human < Demographics + Heuristics and biases experiments |
| Q43 | 3 | -0.082 | 0.01 | * | Human < Demographics + Heuristics and biases experiments |
| Q4 | 4 | -0.17 | 0.006 | ** | Human < Demographics + Heuristics and biases experiments |
| Q9 | 4 | 0.157 | 0.006 | ** | Human > Demographics + Heuristics and biases experiments |
| Q14 | 4 | -0.108 | 0.01 | * | Human < Demographics + Heuristics and biases experiments |
| Q19 | 4 | 0.084 | 0.013 | * | Human > Demographics + Heuristics and biases experiments |
| Q24 | 4 | -0.037 | 0.358 | | |
| Q29 | 4 | -0.107 | 0.006 | ** | Human < Demographics + Heuristics and biases experiments |
| Q34 | 4 | -0.12 | 0.006 | ** | Human < Demographics + Heuristics and biases experiments |
| Q39 | 4 | -0.068 | 0.023 | * | Human < Demographics + Heuristics and biases experiments |
| Q5 | 5 | 0.03 | 0.308 | | |
| Q10 | 5 | -0.064 | 0.073 | | |
| Q15 | 5 | -0.074 | 0.013 | * | Human < Demographics + Heuristics and biases experiments |
| Q20 | 5 | -0.016 | 0.516 | | |
| Q25 | 5 | -0.06 | 0.042 | * | Human < Demographics + Heuristics and biases experiments |
| Q30 | 5 | 0.136 | 0.006 | ** | Human > Demographics + Heuristics and biases experiments |
| Q35 | 5 | -0.345 | 0.006 | ** | Human < Demographics + Heuristics and biases experiments |
| Q40 | 5 | -0.063 | 0.063 | | |
| Q41 | 5 | -0.161 | 0.006 | ** | Human < Demographics + Heuristics and biases experiments |
| Q44 | 5 | 0.03 | 0.24 | | |
| Demographics + All psychological measures + Heuristics and biases experiments vs Human | | | | | |
| Q1 | 1 | -0.052 | 0.035 | * | Human < Demographics + All psychological measures + Heuristics and biases experiments |
| Q6 | 1 | -0.103 | 0.003 | ** | Human < Demographics + All psychological measures + Heuristics and biases experiments |
| Q11 | 1 | 0.021 | 0.371 | | |
| Q16 | 1 | -0.132 | 0.003 | ** | Human < Demographics + All psychological measures + Heuristics and biases experiments |
| Q21 | 1 | 0.056 | 0.036 | * | Human > Demographics + All psychological measures + Heuristics and biases experiments |
| Q26 | 1 | -0.057 | 0.024 | * | Human < Demographics + All psychological measures + Heuristics and biases experiments |
| Q31 | 1 | 0.042 | 0.073 | | |
| Q36 | 1 | -0.018 | 0.516 | | |
| Q2 | 2 | -0.224 | 0.003 | ** | Human < Demographics + All psychological measures + Heuristics and biases experiments |
| Q7 | 2 | -0.208 | 0.003 | ** | Human < Demographics + All psychological measures + Heuristics and biases experiments |
| Q12 | 2 | -0.109 | 0.003 | ** | Human < Demographics + All psychological measures + Heuristics and biases experiments |
| Q17 | 2 | 0.157 | 0.003 | ** | Human > Demographics + All psychological measures + Heuristics and biases experiments |
| Q22 | 2 | 0.112 | 0.003 | ** | Human > Demographics + All psychological measures + Heuristics and biases experiments |
| Q27 | 2 | -0.139 | 0.003 | ** | Human < Demographics + All psychological measures + Heuristics and biases experiments |
| Q32 | 2 | -0.054 | 0.111 | | |
| Q37 | 2 | -0.056 | 0.048 | * | Human < Demographics + All psychological measures + Heuristics and biases experiments |
| Q42 | 2 | 0.041 | 0.102 | | |
| Q3 | 3 | -0.037 | 0.17 | | |
| Q8 | 3 | 0.035 | 0.156 | | |
| Q13 | 3 | -0.156 | 0.003 | ** | Human < Demographics + All psychological measures + Heuristics and biases experiments |
| Q18 | 3 | 0.097 | 0.003 | ** | Human > Demographics + All psychological measures + Heuristics and biases experiments |
| Q23 | 3 | 0.041 | 0.102 | | |
| Q28 | 3 | -0.088 | 0.003 | ** | Human < Demographics + All psychological measures + Heuristics and biases experiments |
| Q33 | 3 | 0.085 | 0.003 | ** | Human > Demographics + All psychological measures + Heuristics and biases experiments |
| Q38 | 3 | -0.191 | 0.003 | ** | Human < Demographics + All psychological measures + Heuristics and biases experiments |
| Q43 | 3 | -0.137 | 0.003 | ** | Human < Demographics + All psychological measures + Heuristics and biases experiments |
| Q4 | 4 | 0.012 | 0.604 | | |
| Q9 | 4 | 0.07 | 0.003 | ** | Human > Demographics + All psychological measures + Heuristics and biases experiments |
| Q14 | 4 | -0.184 | 0.003 | ** | Human < Demographics + All psychological measures + Heuristics and biases experiments |
| Q19 | 4 | 0.093 | 0.003 | ** | Human > Demographics + All psychological measures + Heuristics and biases experiments |
| Q24 | 4 | -0.11 | 0.003 | ** | Human < Demographics + All psychological measures + Heuristics and biases experiments |
| Q29 | 4 | -0.081 | 0.003 | ** | Human < Demographics + All psychological measures + Heuristics and biases experiments |
| Q34 | 4 | -0.122 | 0.003 | ** | Human < Demographics + All psychological measures + Heuristics and biases experiments |
| Q39 | 4 | -0.086 | 0.003 | ** | Human < Demographics + All psychological measures + Heuristics and biases experiments |
| Q5 | 5 | 0.052 | 0.036 | * | Human > Demographics + All psychological measures + Heuristics and biases experiments |
| Q10 | 5 | -0.083 | 0.009 | ** | Human < Demographics + All psychological measures + Heuristics and biases experiments |
| Q15 | 5 | -0.114 | 0.003 | ** | Human < Demographics + All psychological measures + Heuristics and biases experiments |
| Q20 | 5 | -0.107 | 0.003 | ** | Human < Demographics + All psychological measures + Heuristics and biases experiments |
| Q25 | 5 | -0.048 | 0.048 | * | Human < Demographics + All psychological measures + Heuristics and biases experiments |
| Q30 | 5 | 0.223 | 0.003 | ** | Human > Demographics + All psychological measures + Heuristics and biases experiments |
| Q35 | 5 | -0.181 | 0.003 | ** | Human < Demographics + All psychological measures + Heuristics and biases experiments |
| Q40 | 5 | -0.093 | 0.003 | ** | Human < Demographics + All psychological measures + Heuristics and biases experiments |
| Q41 | 5 | -0.257 | 0.003 | ** | Human < Demographics + All psychological measures + Heuristics and biases experiments |
| Q44 | 5 | 0.041 | 0.048 | * | Human > Demographics + All psychological measures + Heuristics and biases experiments |
| Demographics + Heuristics and biases experiments + Features with r < 0.5 vs Human | | | | | |
| Q1 | 1 | -0.055 | 0.033 | * | Human < Demographics + Heuristics and biases experiments + Features with r < 0.5 |
| Q6 | 1 | -0.125 | 0.004 | ** | Human < Demographics + Heuristics and biases experiments + Features with r < 0.5 |
| Q11 | 1 | -0.083 | 0.004 | ** | Human < Demographics + Heuristics and biases experiments + Features with r < 0.5 |
| Q16 | 1 | -0.132 | 0.004 | ** | Human < Demographics + Heuristics and biases experiments + Features with r < 0.5 |
| Q21 | 1 | 0.072 | 0.006 | ** | Human > Demographics + Heuristics and biases experiments + Features with r < 0.5 |
| Q26 | 1 | -0.072 | 0.011 | * | Human < Demographics + Heuristics and biases experiments + Features with r < 0.5 |
| Q31 | 1 | 0.085 | 0.004 | ** | Human > Demographics + Heuristics and biases experiments + Features with r < 0.5 |
| Q36 | 1 | 0.008 | 0.821 | | |
| Q2 | 2 | -0.13 | 0.004 | ** | Human < Demographics + Heuristics and biases experiments + Features with r < 0.5 |

| Q7 | 2 | -0.22 | 0.004 | ** | Human < Demographics + Heuristics and biases experiments + Features with r < 0.5 |
|---|---|---|---|---|---|
| Q12 | 2 | -0.122 | 0.004 | ** | Human < Demographics + Heuristics and biases experiments + Features with r < 0.5 |
| Q17 | 2 | 0.105 | 0.004 | ** | Human > Demographics + Heuristics and biases experiments + Features with r < 0.5 |
| Q22 | 2 | -0.082 | 0.006 | ** | Human < Demographics + Heuristics and biases experiments + Features with r < 0.5 |
| Q27 | 2 | -0.101 | 0.004 | ** | Human < Demographics + Heuristics and biases experiments + Features with r < 0.5 |
| Q32 | 2 | 0.091 | 0.006 | ** | Human > Demographics + Heuristics and biases experiments + Features with r < 0.5 |
| Q37 | 2 | 0 | 0.996 | | |
| Q42 | 2 | -0.223 | 0.004 | ** | Human < Demographics + Heuristics and biases experiments + Features with r < 0.5 |
| Q3 | 3 | -0.04 | 0.191 | | |
| Q8 | 3 | -0.051 | 0.081 | | |
| Q13 | 3 | -0.119 | 0.004 | ** | Human < Demographics + Heuristics and biases experiments + Features with r < 0.5 |
| Q18 | 3 | -0.014 | 0.589 | | |
| Q23 | 3 | 0.069 | 0.013 | * | Human > Demographics + Heuristics and biases experiments + Features with r < 0.5 |
| Q28 | 3 | -0.098 | 0.004 | ** | Human < Demographics + Heuristics and biases experiments + Features with r < 0.5 |
| Q33 | 3 | -0.08 | 0.004 | ** | Human < Demographics + Heuristics and biases experiments + Features with r < 0.5 |
| Q38 | 3 | -0.08 | 0.009 | ** | Human < Demographics + Heuristics and biases experiments + Features with r < 0.5 |
| Q43 | 3 | -0.056 | 0.011 | * | Human < Demographics + Heuristics and biases experiments + Features with r < 0.5 |
| Q4 | 4 | 0.109 | 0.004 | ** | Human > Demographics + Heuristics and biases experiments + Features with r < 0.5 |
| Q9 | 4 | 0.119 | 0.004 | ** | Human > Demographics + Heuristics and biases experiments + Features with r < 0.5 |
| Q14 | 4 | -0.227 | 0.004 | ** | Human < Demographics + Heuristics and biases experiments + Features with r < 0.5 |
| Q19 | 4 | 0.107 | 0.004 | ** | Human > Demographics + Heuristics and biases experiments + Features with r < 0.5 |
| Q24 | 4 | -0.086 | 0.009 | ** | Human < Demographics + Heuristics and biases experiments + Features with r < 0.5 |
| Q29 | 4 | -0.05 | 0.048 | * | Human < Demographics + Heuristics and biases experiments + Features with r < 0.5 |
| Q34 | 4 | -0.176 | 0.004 | ** | Human < Demographics + Heuristics and biases experiments + Features with r < 0.5 |
| Q39 | 4 | -0.102 | 0.004 | ** | Human < Demographics + Heuristics and biases experiments + Features with r < 0.5 |
| Q5 | 5 | 0.038 | 0.14 | | |
| Q10 | 5 | 0.03 | 0.275 | | |
| Q15 | 5 | -0.15 | 0.004 | ** | Human < Demographics + Heuristics and biases experiments + Features with r < 0.5 |
| Q20 | 5 | -0.13 | 0.004 | ** | Human < Demographics + Heuristics and biases experiments + Features with r < 0.5 |
| Q25 | 5 | -0.053 | 0.046 | * | Human < Demographics + Heuristics and biases experiments + Features with r < 0.5 |
| Q30 | 5 | 0.194 | 0.004 | ** | Human > Demographics + Heuristics and biases experiments + Features with r < 0.5 |
| Q35 | 5 | -0.045 | 0.016 | * | Human < Demographics + Heuristics and biases experiments + Features with r < 0.5 |
| Q40 | 5 | -0.139 | 0.004 | ** | Human < Demographics + Heuristics and biases experiments + Features with r < 0.5 |
| Q41 | 5 | -0.263 | 0.004 | ** | Human < Demographics + Heuristics and biases experiments + Features with r < 0.5 |
| Q44 | 5 | 0.083 | 0.004 | ** | Human > Demographics + Heuristics and biases experiments + Features with r < 0.5 |
| Demographics + Heuristics and biases experiments + Features with r > 0.5 vs Human | | | | | |
| Q1 | 1 | -0.036 | 0.125 | | |
| Q6 | 1 | -0.113 | 0.003 | ** | Human < Demographics + Heuristics and biases experiments + Features with r > 0.5 |
| Q11 | 1 | 0.029 | 0.214 | | |
| Q16 | 1 | -0.128 | 0.003 | ** | Human < Demographics + Heuristics and biases experiments + Features with r > 0.5 |
| Q21 | 1 | 0.066 | 0.006 | ** | Human > Demographics + Heuristics and biases experiments + Features with r > 0.5 |
| Q26 | 1 | -0.026 | 0.289 | | |
| Q31 | 1 | -0.024 | 0.288 | | |
| Q36 | 1 | -0.001 | 0.97 | | |
| Q2 | 2 | -0.222 | 0.003 | ** | Human < Demographics + Heuristics and biases experiments + Features with r > 0.5 |
| Q7 | 2 | -0.2 | 0.003 | ** | Human < Demographics + Heuristics and biases experiments + Features with r > 0.5 |
| Q12 | 2 | -0.118 | 0.003 | ** | Human < Demographics + Heuristics and biases experiments + Features with r > 0.5 |
| Q17 | 2 | 0.214 | 0.003 | ** | Human > Demographics + Heuristics and biases experiments + Features with r > 0.5 |
| Q22 | 2 | 0.147 | 0.003 | ** | Human > Demographics + Heuristics and biases experiments + Features with r > 0.5 |
| Q27 | 2 | -0.092 | 0.003 | ** | Human < Demographics + Heuristics and biases experiments + Features with r > 0.5 |
| Q32 | 2 | -0.094 | 0.011 | * | Human < Demographics + Heuristics and biases experiments + Features with r > 0.5 |
| Q37 | 2 | -0.069 | 0.024 | * | Human < Demographics + Heuristics and biases experiments + Features with r > 0.5 |
| Q42 | 2 | 0.09 | 0.003 | ** | Human > Demographics + Heuristics and biases experiments + Features with r > 0.5 |
| Q3 | 3 | -0.144 | 0.003 | ** | Human < Demographics + Heuristics and biases experiments + Features with r > 0.5 |
| Q8 | 3 | 0.054 | 0.024 | * | Human > Demographics + Heuristics and biases experiments + Features with r > 0.5 |
| Q13 | 3 | -0.132 | 0.003 | ** | Human < Demographics + Heuristics and biases experiments + Features with r > 0.5 |
| Q18 | 3 | 0.092 | 0.003 | ** | Human > Demographics + Heuristics and biases experiments + Features with r > 0.5 |
| Q23 | 3 | 0.082 | 0.003 | ** | Human > Demographics + Heuristics and biases experiments + Features with r > 0.5 |
| Q28 | 3 | -0.103 | 0.003 | ** | Human < Demographics + Heuristics and biases experiments + Features with r > 0.5 |
| Q33 | 3 | 0.067 | 0.006 | ** | Human > Demographics + Heuristics and biases experiments + Features with r > 0.5 |
| Q38 | 3 | -0.162 | 0.003 | ** | Human < Demographics + Heuristics and biases experiments + Features with r > 0.5 |
| Q43 | 3 | -0.116 | 0.003 | ** | Human < Demographics + Heuristics and biases experiments + Features with r > 0.5 |
| Q4 | 4 | 0.021 | 0.332 | | |
| Q9 | 4 | 0.054 | 0.016 | * | Human > Demographics + Heuristics and biases experiments + Features with r > 0.5 |
| Q14 | 4 | -0.155 | 0.003 | ** | Human < Demographics + Heuristics and biases experiments + Features with r > 0.5 |
| Q19 | 4 | 0.072 | 0.003 | ** | Human > Demographics + Heuristics and biases experiments + Features with r > 0.5 |
| Q24 | 4 | -0.121 | 0.003 | ** | Human < Demographics + Heuristics and biases experiments + Features with r > 0.5 |
| Q29 | 4 | -0.087 | 0.003 | ** | Human < Demographics + Heuristics and biases experiments + Features with r > 0.5 |
| Q34 | 4 | -0.131 | 0.003 | ** | Human < Demographics + Heuristics and biases experiments + Features with r > 0.5 |
| Q39 | 4 | -0.069 | 0.006 | ** | Human < Demographics + Heuristics and biases experiments + Features with r > 0.5 |
| Q5 | 5 | 0.083 | 0.003 | ** | Human > Demographics + Heuristics and biases experiments + Features with r > 0.5 |
| Q10 | 5 | -0.129 | 0.003 | ** | Human < Demographics + Heuristics and biases experiments + Features with r > 0.5 |
| Q15 | 5 | -0.163 | 0.003 | ** | Human < Demographics + Heuristics and biases experiments + Features with r > 0.5 |
| Q20 | 5 | -0.063 | 0.013 | * | Human < Demographics + Heuristics and biases experiments + Features with r > 0.5 |
| Q25 | 5 | -0.037 | 0.109 | | |
| Q30 | 5 | 0.144 | 0.003 | ** | Human > Demographics + Heuristics and biases experiments + Features with r > 0.5 |
| Q35 | 5 | -0.148 | 0.003 | ** | Human < Demographics + Heuristics and biases experiments + Features with r > 0.5 |
| Q40 | 5 | -0.079 | 0.009 | ** | Human < Demographics + Heuristics and biases experiments + Features with r > 0.5 |
| Q41 | 5 | -0.221 | 0.003 | ** | Human < Demographics + Heuristics and biases experiments + Features with r > 0.5 |
| Q44 | 5 | 0.022 | 0.288 | | |
| Demographics + Heuristics and biases experiments + Features with r < 0.3 vs Human | | | | | |
| Q1 | 1 | -0.074 | 0.013 | * | Human < Demographics + Heuristics and biases experiments + Features with r < 0.3 |
| Q6 | 1 | -0.099 | 0.009 | ** | Human < Demographics + Heuristics and biases experiments + Features with r < 0.3 |
| Q11 | 1 | -0.033 | 0.209 | | |
| Q16 | 1 | -0.147 | 0.005 | ** | Human < Demographics + Heuristics and biases experiments + Features with r < 0.3 |
| Q21 | 1 | 0.141 | 0.005 | ** | Human > Demographics + Heuristics and biases experiments + Features with r < 0.3 |
| Q26 | 1 | -0.124 | 0.005 | ** | Human < Demographics + Heuristics and biases experiments + Features with r < 0.3 |
| Q31 | 1 | 0.056 | 0.029 | * | Human > Demographics + Heuristics and biases experiments + Features with r < 0.3 |
| Q36 | 1 | 0.025 | 0.431 | | |
| Q2 | 2 | -0.066 | 0.144 | | |
| Q7 | 2 | -0.193 | 0.005 | ** | Human < Demographics + Heuristics and biases experiments + Features with r < 0.3 |
| Q12 | 2 | -0.097 | 0.019 | * | Human < Demographics + Heuristics and biases experiments + Features with r < 0.3 |
| Q17 | 2 | -0.01 | 0.81 | | |
| Q22 | 2 | -0.055 | 0.144 | | |
| Q27 | 2 | 0.055 | 0.145 | | |
| Q32 | 2 | 0.05 | 0.269 | | |
| Q37 | 2 | -0.037 | 0.341 | | |
| Q42 | 2 | -0.261 | 0.005 | ** | Human < Demographics + Heuristics and biases experiments + Features with r < 0.3 |
| Q3 | 3 | -0.027 | 0.381 | | |
| Q8 | 3 | -0.112 | 0.005 | ** | Human < Demographics + Heuristics and biases experiments + Features with r < 0.3 |
| Q13 | 3 | -0.02 | 0.553 | | |
| Q18 | 3 | 0.077 | 0.017 | * | Human > Demographics + Heuristics and biases experiments + Features with r < 0.3 |
| Q23 | 3 | 0.033 | 0.273 | | |
| Q28 | 3 | -0.059 | 0.066 | | |
| Q33 | 3 | -0.089 | 0.005 | ** | Human < Demographics + Heuristics and biases experiments + Features with r < 0.3 |
| Q38 | 3 | -0.161 | 0.005 | ** | Human < Demographics + Heuristics and biases experiments + Features with r < 0.3 |
| Q43 | 3 | -0.055 | 0.037 | * | Human < Demographics + Heuristics and biases experiments + Features with r < 0.3 |
| Q4 | 4 | -0.072 | 0.049 | * | Human < Demographics + Heuristics and biases experiments + Features with r < 0.3 |
| Q9 | 4 | 0.118 | 0.005 | ** | Human > Demographics + Heuristics and biases experiments + Features with r < 0.3 |
| Q14 | 4 | -0.155 | 0.005 | ** | Human < Demographics + Heuristics and biases experiments + Features with r < 0.3 |
| Q19 | 4 | 0.087 | 0.009 | ** | Human > Demographics + Heuristics and biases experiments + Features with r < 0.3 |
| Q24 | 4 | -0.055 | 0.152 | | |
| Q29 | 4 | -0.043 | 0.146 | | |
| Q34 | 4 | -0.126 | 0.005 | ** | Human < Demographics + Heuristics and biases experiments + Features with r < 0.3 |
| Q39 | 4 | -0.126 | 0.005 | ** | Human < Demographics + Heuristics and biases experiments + Features with r < 0.3 |
| Q5 | 5 | -0.035 | 0.269 | | |
| Q10 | 5 | -0.003 | 0.93 | | |
| Q15 | 5 | -0.107 | 0.005 | ** | Human < Demographics + Heuristics and biases experiments + Features with r < 0.3 |

| Item | Membership | Difference | p | Sig | Direction |
|---|---|---|---|---|---|
| Q20 | 5 | 0.039 | 0.158 | | |
| Q25 | 5 | -0.104 | 0.005 | ** | Human < Demographics + Heuristics and biases experiments + Features with r < 0.3 |
| Q30 | 5 | 0.204 | 0.005 | ** | Human > Demographics + Heuristics and biases experiments + Features with r < 0.3 |
| Q35 | 5 | -0.288 | 0.005 | ** | Human < Demographics + Heuristics and biases experiments + Features with r < 0.3 |
| Q40 | 5 | -0.09 | 0.019 | * | Human < Demographics + Heuristics and biases experiments + Features with r < 0.3 |
| Q41 | 5 | -0.223 | 0.005 | ** | Human < Demographics + Heuristics and biases experiments + Features with r < 0.3 |
| Q44 | 5 | 0.064 | 0.037 | * | Human > Demographics + Heuristics and biases experiments + Features with r < 0.3 |

**Panel B. gpt-oss-120b**

| **Item** | **Membership** | **Difference** | ***p*** | **Sig** | **Direction** |
|---|---|---|---|---|---|
| Demographics vs Human | | | | | |
| Q1 | 1 | 0 | 1 | | |
| Q6 | 1 | -0.086 | 0.005 | ** | Human < Demographics |
| Q11 | 1 | -0.212 | 0.005 | ** | Human < Demographics |
| Q16 | 1 | -0.05 | 0.073 | | |
| Q21 | 1 | 0.06 | 0.019 | * | Human > Demographics |
| Q26 | 1 | -0.091 | 0.005 | ** | Human < Demographics |
| Q31 | 1 | -0.055 | 0.012 | * | Human < Demographics |
| Q36 | 1 | 0.02 | 0.53 | | |
| Q2 | 2 | 0.089 | 0.005 | ** | Human > Demographics |
| Q7 | 2 | -0.172 | 0.005 | ** | Human < Demographics |
| Q12 | 2 | -0.111 | 0.005 | ** | Human < Demographics |
| Q17 | 2 | -0.02 | 0.621 | | |
| Q22 | 2 | 0.149 | 0.005 | ** | Human > Demographics |
| Q27 | 2 | -0.069 | 0.028 | * | Human < Demographics |
| Q32 | 2 | 0.008 | 0.911 | | |
| Q37 | 2 | -0.05 | 0.15 | | |
| Q42 | 2 | -0.172 | 0.005 | ** | Human < Demographics |
| Q3 | 3 | 0.064 | 0.012 | * | Human > Demographics |
| Q8 | 3 | -0.069 | 0.036 | * | Human < Demographics |
| Q13 | 3 | -0.141 | 0.005 | ** | Human < Demographics |
| Q18 | 3 | -0.012 | 0.733 | | |
| Q23 | 3 | 0.139 | 0.005 | ** | Human > Demographics |
| Q28 | 3 | -0.026 | 0.49 | | |
| Q33 | 3 | -0.023 | 0.523 | | |
| Q38 | 3 | -0.145 | 0.005 | ** | Human < Demographics |
| Q43 | 3 | -0.062 | 0.026 | * | Human < Demographics |
| Q4 | 4 | 0.021 | 0.523 | | |
| Q9 | 4 | 0.175 | 0.005 | ** | Human > Demographics |
| Q14 | 4 | -0.064 | 0.012 | * | Human < Demographics |
| Q19 | 4 | 0.026 | 0.338 | | |
| Q24 | 4 | -0.131 | 0.005 | ** | Human < Demographics |
| Q29 | 4 | -0.116 | 0.005 | ** | Human < Demographics |
| Q34 | 4 | -0.052 | 0.034 | * | Human < Demographics |
| Q39 | 4 | -0.081 | 0.005 | ** | Human < Demographics |
| Q5 | 5 | 0.024 | 0.416 | | |
| Q10 | 5 | -0.122 | 0.005 | ** | Human < Demographics |
| Q15 | 5 | -0.159 | 0.005 | ** | Human < Demographics |
| Q20 | 5 | -0.098 | 0.005 | ** | Human < Demographics |
| Q25 | 5 | -0.034 | 0.185 | | |
| Q30 | 5 | 0.041 | 0.06 | | |
| Q35 | 5 | -0.095 | 0.005 | ** | Human < Demographics |
| Q40 | 5 | -0.024 | 0.461 | | |
| Q41 | 5 | -0.001 | 0.995 | | |
| Q44 | 5 | 0.003 | 0.911 | | |
| Heuristics and biases experiments vs Human | | | | | |
| Q1 | 1 | -0.017 | 0.454 | | |
| Q6 | 1 | -0.102 | 0.005 | ** | Human < Heuristics and biases experiments |
| Q11 | 1 | -0.095 | 0.005 | ** | Human < Heuristics and biases experiments |
| Q16 | 1 | 0.016 | 0.574 | | |
| Q21 | 1 | 0.054 | 0.038 | * | Human > Heuristics and biases experiments |
| Q26 | 1 | -0.115 | 0.005 | ** | Human < Heuristics and biases experiments |
| Q31 | 1 | -0.042 | 0.044 | * | Human < Heuristics and biases experiments |
| Q36 | 1 | -0.001 | 0.97 | | |
| Q2 | 2 | 0.041 | 0.2 | | |
| Q7 | 2 | -0.198 | 0.005 | ** | Human < Heuristics and biases experiments |
| Q12 | 2 | 0.097 | 0.005 | ** | Human > Heuristics and biases experiments |
| Q17 | 2 | 0.015 | 0.622 | | |
| Q22 | 2 | 0.054 | 0.022 | * | Human > Heuristics and biases experiments |
| Q27 | 2 | -0.127 | 0.005 | ** | Human < Heuristics and biases experiments |
| Q32 | 2 | 0.036 | 0.359 | | |
| Q37 | 2 | -0.004 | 0.939 | | |
| Q42 | 2 | -0.27 | 0.005 | ** | Human < Heuristics and biases experiments |
| Q3 | 3 | 0.092 | 0.005 | ** | Human > Heuristics and biases experiments |
| Q8 | 3 | -0.049 | 0.15 | | |
| Q13 | 3 | -0.077 | 0.038 | * | Human < Heuristics and biases experiments |
| Q18 | 3 | -0.079 | 0.005 | ** | Human < Heuristics and biases experiments |
| Q23 | 3 | 0.168 | 0.005 | ** | Human > Heuristics and biases experiments |
| Q28 | 3 | -0.056 | 0.091 | | |
| Q33 | 3 | -0.098 | 0.005 | ** | Human < Heuristics and biases experiments |
| Q38 | 3 | -0.162 | 0.005 | ** | Human < Heuristics and biases experiments |
| Q43 | 3 | 0.086 | 0.005 | ** | Human > Heuristics and biases experiments |
| Q4 | 4 | 0.133 | 0.005 | ** | Human > Heuristics and biases experiments |
| Q9 | 4 | 0.135 | 0.005 | ** | Human > Heuristics and biases experiments |
| Q14 | 4 | -0.04 | 0.154 | | |
| Q19 | 4 | 0.051 | 0.053 | | |
| Q24 | 4 | 0.028 | 0.454 | | |
| Q29 | 4 | -0.02 | 0.483 | | |
| Q34 | 4 | -0.039 | 0.112 | | |
| Q39 | 4 | -0.213 | 0.005 | ** | Human < Heuristics and biases experiments |
| Q5 | 5 | 0.052 | 0.098 | | |
| Q10 | 5 | -0.068 | 0.038 | * | Human < Heuristics and biases experiments |
| Q15 | 5 | -0.129 | 0.005 | ** | Human < Heuristics and biases experiments |
| Q20 | 5 | -0.108 | 0.005 | ** | Human < Heuristics and biases experiments |
| Q25 | 5 | -0.049 | 0.112 | | |
| Q30 | 5 | 0.08 | 0.005 | ** | Human > Heuristics and biases experiments |
| Q35 | 5 | 0.022 | 0.295 | | |
| Q40 | 5 | -0.048 | 0.144 | | |
| Q41 | 5 | -0.012 | 0.622 | | |
| Q44 | 5 | -0.023 | 0.359 | | |
| Demographics + Heuristics and biases experiments vs Human | | | | | |
| Q1 | 1 | 0.033 | 0.194 | | |
| Q6 | 1 | -0.123 | 0.004 | ** | Human < Demographics + Heuristics and biases experiments |
| Q11 | 1 | -0.2 | 0.004 | ** | Human < Demographics + Heuristics and biases experiments |
| Q16 | 1 | 0.044 | 0.108 | | |
| Q21 | 1 | 0.049 | 0.097 | | |
| Q26 | 1 | -0.062 | 0.011 | * | Human < Demographics + Heuristics and biases experiments |
| Q31 | 1 | -0.122 | 0.004 | ** | Human < Demographics + Heuristics and biases experiments |
| Q36 | 1 | 0.049 | 0.1 | | |
| Q2 | 2 | 0.097 | 0.004 | ** | Human > Demographics + Heuristics and biases experiments |
| Q7 | 2 | -0.211 | 0.004 | ** | Human < Demographics + Heuristics and biases experiments |
| Q12 | 2 | -0.027 | 0.394 | | |
| Q17 | 2 | 0.088 | 0.004 | ** | Human > Demographics + Heuristics and biases experiments |
| Q22 | 2 | 0.132 | 0.004 | ** | Human > Demographics + Heuristics and biases experiments |
| Q27 | 2 | -0.144 | 0.004 | ** | Human < Demographics + Heuristics and biases experiments |
| Q32 | 2 | 0.018 | 0.666 | | |
| Q37 | 2 | -0.047 | 0.121 | | |
| Q42 | 2 | -0.235 | 0.004 | ** | Human < Demographics + Heuristics and biases experiments |
| Q3 | 3 | 0.067 | 0.008 | ** | Human > Demographics + Heuristics and biases experiments |
| Q8 | 3 | 0.022 | 0.513 | | |
| Q13 | 3 | -0.04 | 0.252 | | |
| Q18 | 3 | -0.117 | 0.004 | ** | Human < Demographics + Heuristics and biases experiments |
| Q23 | 3 | 0.137 | 0.004 | ** | Human > Demographics + Heuristics and biases experiments |
| Q28 | 3 | -0.033 | 0.35 | | |
| Q33 | 3 | -0.102 | 0.004 | ** | Human < Demographics + Heuristics and biases experiments |
| Q38 | 3 | -0.193 | 0.004 | ** | Human < Demographics + Heuristics and biases experiments |
| Q43 | 3 | 0.036 | 0.168 | | |
| Q4 | 4 | 0.076 | 0.014 | * | Human > Demographics + Heuristics and biases experiments |
| Q9 | 4 | 0.082 | 0.004 | ** | Human > Demographics + Heuristics and biases experiments |
| Q14 | 4 | -0.019 | 0.44 | | |
| Q19 | 4 | 0.083 | 0.004 | ** | Human > Demographics + Heuristics and biases experiments |
| Q24 | 4 | 0.009 | 0.81 | | |
| Q29 | 4 | -0.138 | 0.004 | ** | Human < Demographics + Heuristics and biases experiments |
| Q34 | 4 | -0.1 | 0.004 | ** | Human < Demographics + Heuristics and biases experiments |
| Q39 | 4 | -0.15 | 0.004 | ** | Human < Demographics + Heuristics and biases experiments |
| Q5 | 5 | 0.056 | 0.039 | * | Human > Demographics + Heuristics and biases experiments |
| Q10 | 5 | -0.064 | 0.057 | | |
| Q15 | 5 | -0.13 | 0.004 | ** | Human < Demographics + Heuristics and biases experiments |
| Q20 | 5 | -0.1 | 0.004 | ** | Human < Demographics + Heuristics and biases experiments |
| Q25 | 5 | -0.069 | 0.014 | * | Human < Demographics + Heuristics and biases experiments |
| Q30 | 5 | 0.068 | 0.004 | ** | Human > Demographics + Heuristics and biases experiments |
| Q35 | 5 | -0.102 | 0.004 | ** | Human < Demographics + Heuristics and biases experiments |
| Q40 | 5 | -0.002 | 0.934 | | |
| Q41 | 5 | -0.011 | 0.637 | | |
| Q44 | 5 | -0.025 | 0.211 | | |
| Demographics + All psychological measures + Heuristics and biases experiments vs Human | | | | | |
| Q1 | 1 | -0.049 | 0.02 | * | Human < Demographics + All psychological measures + Heuristics and biases experiments |
| Q6 | 1 | -0.094 | 0.004 | ** | Human < Demographics + All psychological measures + Heuristics and biases experiments |
| Q11 | 1 | -0.114 | 0.004 | ** | Human < Demographics + All psychological measures + Heuristics and biases experiments |
| Q16 | 1 | 0.064 | 0.014 | * | Human > Demographics + All psychological measures + Heuristics and biases experiments |
| Q21 | 1 | 0.009 | 0.799 | | |
| Q26 | 1 | -0.023 | 0.337 | | |
| Q31 | 1 | 0.049 | 0.049 | * | Human > Demographics + All psychological measures + Heuristics and biases experiments |
| Q36 | 1 | 0.007 | 0.807 | | |
| Q2 | 2 | -0.068 | 0.014 | * | Human < Demographics + All psychological measures + Heuristics and biases experiments |
| Q7 | 2 | -0.273 | 0.004 | ** | Human < Demographics + All psychological measures + Heuristics and biases experiments |
| Q12 | 2 | -0.069 | 0.007 | ** | Human < Demographics + All psychological measures + Heuristics and biases experiments |
| Q17 | 2 | 0.172 | 0.004 | ** | Human > Demographics + All psychological measures + Heuristics and biases experiments |
| Q22 | 2 | 0.132 | 0.004 | ** | Human > Demographics + All psychological measures + Heuristics and biases experiments |
| Q27 | 2 | -0.163 | 0.004 | ** | Human < Demographics + All psychological measures + Heuristics and biases experiments |
| Q32 | 2 | -0.081 | 0.007 | ** | Human < Demographics + All psychological measures + Heuristics and biases experiments |
| Q37 | 2 | -0.017 | 0.621 | | |
| Q42 | 2 | -0.002 | 0.93 | | |
| Q3 | 3 | -0.02 | 0.535 | | |
| Q8 | 3 | 0.105 | 0.004 | ** | Human > Demographics + All psychological measures + Heuristics and biases experiments |
| Q13 | 3 | -0.125 | 0.004 | ** | Human < Demographics + All psychological measures + Heuristics and biases experiments |
| Q18 | 3 | 0.008 | 0.807 | | |
| Q23 | 3 | 0.102 | 0.004 | ** | Human > Demographics + All psychological measures + Heuristics and biases experiments |
| Q28 | 3 | 0.008 | 0.807 | | |
| Q33 | 3 | -0.029 | 0.336 | | |
| Q38 | 3 | -0.238 | 0.004 | ** | Human < Demographics + All psychological measures + Heuristics and biases experiments |
| Q43 | 3 | -0.138 | 0.004 | ** | Human < Demographics + All psychological measures + Heuristics and biases experiments |
| Q4 | 4 | -0.039 | 0.079 | | |
| Q9 | 4 | 0.142 | 0.004 | ** | Human > Demographics + All psychological measures + Heuristics and biases experiments |
| Q14 | 4 | 0.026 | 0.208 | | |
| Q19 | 4 | 0.097 | 0.004 | ** | Human > Demographics + All psychological measures + Heuristics and biases experiments |
| Q24 | 4 | -0.152 | 0.004 | ** | Human < Demographics + All psychological measures + Heuristics and biases experiments |
| Q29 | 4 | -0.181 | 0.004 | ** | Human < Demographics + All psychological measures + Heuristics and biases experiments |
| Q34 | 4 | -0.147 | 0.004 | ** | Human < Demographics + All psychological measures + Heuristics and biases experiments |
| Q39 | 4 | -0.144 | 0.004 | ** | Human < Demographics + All psychological measures + Heuristics and biases experiments |
| Q5 | 5 | 0.004 | 0.907 | | |
| Q10 | 5 | -0.048 | 0.102 | | |
| Q15 | 5 | -0.086 | 0.004 | ** | Human < Demographics + All psychological measures + Heuristics and biases experiments |
| Q20 | 5 | -0.123 | 0.004 | ** | Human < Demographics + All psychological measures + Heuristics and biases experiments |
| Q25 | 5 | -0.038 | 0.132 | | |
| Q30 | 5 | 0.091 | 0.004 | ** | Human > Demographics + All psychological measures + Heuristics and biases experiments |
| Q35 | 5 | -0.071 | 0.004 | ** | Human < Demographics + All psychological measures + Heuristics and biases experiments |
| Q40 | 5 | -0.069 | 0.031 | * | Human < Demographics + All psychological measures + Heuristics and biases experiments |
| Q41 | 5 | -0.087 | 0.004 | ** | Human < Demographics + All psychological measures + Heuristics and biases experiments |
| Q44 | 5 | -0.014 | 0.526 | | |
| Demographics + Heuristics and biases experiments + Features with r < 0.5 vs Human | | | | | |
| Q1 | 1 | -0.038 | 0.102 | | |
| Q6 | 1 | -0.105 | 0.003 | ** | Human < Demographics + Heuristics and biases experiments + Features with r < 0.5 |
| Q11 | 1 | -0.079 | 0.003 | ** | Human < Demographics + Heuristics and biases experiments + Features with r < 0.5 |
| Q16 | 1 | 0.041 | 0.15 | | |
| Q21 | 1 | 0.038 | 0.162 | | |
| Q26 | 1 | -0.032 | 0.175 | | |
| Q31 | 1 | -0.065 | 0.012 | * | Human < Demographics + Heuristics and biases experiments + Features with r < 0.5 |
| Q36 | 1 | 0.007 | 0.804 | | |
| Q2 | 2 | -0.02 | 0.527 | | |
| Q7 | 2 | -0.214 | 0.003 | ** | Human < Demographics + Heuristics and biases experiments + Features with r < 0.5 |
| Q12 | 2 | 0.108 | 0.003 | ** | Human > Demographics + Heuristics and biases experiments + Features with r < 0.5 |
| Q17 | 2 | 0.207 | 0.003 | ** | Human > Demographics + Heuristics and biases experiments + Features with r < 0.5 |
| Q22 | 2 | 0.101 | 0.003 | ** | Human > Demographics + Heuristics and biases experiments + Features with r < 0.5 |
| Q27 | 2 | -0.073 | 0.003 | ** | Human < Demographics + Heuristics and biases experiments + Features with r < 0.5 |
| Q32 | 2 | -0.087 | 0.023 | * | Human < Demographics + Heuristics and biases experiments + Features with r < 0.5 |
| Q37 | 2 | -0.064 | 0.036 | * | Human < Demographics + Heuristics and biases experiments + Features with r < 0.5 |
| Q42 | 2 | -0.299 | 0.003 | ** | Human < Demographics + Heuristics and biases experiments + Features with r < 0.5 |
| Q3 | 3 | 0.097 | 0.003 | ** | Human > Demographics + Heuristics and biases experiments + Features with r < 0.5 |
| Q8 | 3 | 0.073 | 0.003 | ** | Human > Demographics + Heuristics and biases experiments + Features with r < 0.5 |
| Q13 | 3 | -0.073 | 0.006 | ** | Human < Demographics + Heuristics and biases experiments + Features with r < 0.5 |
| Q18 | 3 | -0.06 | 0.009 | ** | Human < Demographics + Heuristics and biases experiments + Features with r < 0.5 |
| Q23 | 3 | 0.188 | 0.003 | ** | Human > Demographics + Heuristics and biases experiments + Features with r < 0.5 |
| Q28 | 3 | -0.108 | 0.003 | ** | Human < Demographics + Heuristics and biases experiments + Features with r < 0.5 |
| Q33 | 3 | -0.111 | 0.003 | ** | Human < Demographics + Heuristics and biases experiments + Features with r < 0.5 |
| Q38 | 3 | -0.202 | 0.003 | ** | Human < Demographics + Heuristics and biases experiments + Features with r < 0.5 |
| Q43 | 3 | -0.098 | 0.003 | ** | Human < Demographics + Heuristics and biases experiments + Features with r < 0.5 |
| Q5 | 4 | 0.106 | 0.003 | ** | Human > Demographics + Heuristics and biases experiments + Features with r < 0.5 |
| Q10 | 4 | -0.002 | 0.924 | | |
| Q15 | 4 | -0.128 | 0.003 | ** | Human < Demographics + Heuristics and biases experiments + Features with r < 0.5 |
| Q20 | 4 | -0.183 | 0.003 | ** | Human < Demographics + Heuristics and biases experiments + Features with r < 0.5 |
| Q25 | 4 | -0.079 | 0.003 | ** | Human < Demographics + Heuristics and biases experiments + Features with r < 0.5 |
| Q30 | 4 | 0.063 | 0.006 | ** | Human > Demographics + Heuristics and biases experiments + Features with r < 0.5 |
| Q40 | 4 | -0.05 | 0.046 | * | Human < Demographics + Heuristics and biases experiments + Features with r < 0.5 |
| Q41 | 4 | -0.095 | 0.003 | ** | Human < Demographics + Heuristics and biases experiments + Features with r < 0.5 |
| Q44 | 4 | 0.005 | 0.799 | | |
| Q9 | 5 | 0.136 | 0.003 | ** | Human > Demographics + Heuristics and biases experiments + Features with r < 0.5 |
| Q14 | 5 | -0.117 | 0.003 | ** | Human < Demographics + Heuristics and biases experiments + Features with r < 0.5 |
| Q19 | 5 | 0.015 | 0.555 | | |
| Q24 | 5 | 0.056 | 0.04 | * | Human > Demographics + Heuristics and biases experiments + Features with r < 0.5 |
| Q34 | 5 | -0.203 | 0.003 | ** | Human < Demographics + Heuristics and biases experiments + Features with r < 0.5 |
| Q39 | 5 | -0.209 | 0.003 | ** | Human < Demographics + Heuristics and biases experiments + Features with r < 0.5 |
| Demographics + Heuristics and biases experiments + Features with r > 0.5 vs Human | | | | | |
| Q1 | 1 | -0.009 | 0.681 | | |
| Q6 | 1 | -0.122 | 0.004 | ** | Human < Demographics + Heuristics and biases experiments + Features with r > 0.5 |
| Q11 | 1 | -0.093 | 0.004 | ** | Human < Demographics + Heuristics and biases experiments + Features with r > 0.5 |
| Q16 | 1 | 0.073 | 0.009 | ** | Human > Demographics + Heuristics and biases experiments + Features with r > 0.5 |
| Q21 | 1 | 0.023 | 0.42 | | |
| Q26 | 1 | 0.084 | 0.004 | ** | Human > Demographics + Heuristics and biases experiments + Features with r > 0.5 |
| Q31 | 1 | -0.061 | 0.015 | * | Human < Demographics + Heuristics and biases experiments + Features with r > 0.5 |
| Q36 | 1 | 0.015 | 0.547 | | |
| Q2 | 2 | 0.066 | 0.02 | * | Human > Demographics + Heuristics and biases experiments + Features with r > 0.5 |
| Q7 | 2 | -0.26 | 0.004 | ** | Human < Demographics + Heuristics and biases experiments + Features with r > 0.5 |
| Q12 | 2 | -0.039 | 0.186 | | |
| Q17 | 2 | 0.187 | 0.004 | ** | Human > Demographics + Heuristics and biases experiments + Features with r > 0.5 |
| Q22 | 2 | 0.156 | 0.004 | ** | Human > Demographics + Heuristics and biases experiments + Features with r > 0.5 |
| Q27 | 2 | -0.111 | 0.004 | ** | Human < Demographics + Heuristics and biases experiments + Features with r > 0.5 |
| Q32 | 2 | -0.158 | 0.004 | ** | Human < Demographics + Heuristics and biases experiments + Features with r > 0.5 |
| Q37 | 2 | -0.098 | 0.009 | ** | Human < Demographics + Heuristics and biases experiments + Features with r > 0.5 |
| Q42 | 2 | -0.033 | 0.252 | | |
| Q3 | 3 | -0.029 | 0.273 | | |
| Q8 | 3 | 0.024 | 0.329 | | |
| Q13 | 3 | -0.137 | 0.004 | ** | Human < Demographics + Heuristics and biases experiments + Features with r > 0.5 |
| Q18 | 3 | 0.046 | 0.077 | | |
| Q23 | 3 | 0.131 | 0.004 | ** | Human > Demographics + Heuristics and biases experiments + Features with r > 0.5 |
| Q28 | 3 | -0.002 | 0.926 | | |
| Q33 | 3 | 0.02 | 0.446 | | |
| Q38 | 3 | -0.22 | 0.004 | ** | Human < Demographics + Heuristics and biases experiments + Features with r > 0.5 |
| Q43 | 3 | -0.164 | 0.004 | ** | Human < Demographics + Heuristics and biases experiments + Features with r > 0.5 |
| Q4 | 4 | -0.077 | 0.004 | ** | Human < Demographics + Heuristics and biases experiments + Features with r > 0.5 |
| Q9 | 4 | 0.1 | 0.004 | ** | Human > Demographics + Heuristics and biases experiments + Features with r > 0.5 |
| Q14 | 4 | 0.039 | 0.041 | * | Human > Demographics + Heuristics and biases experiments + Features with r > 0.5 |
| Q19 | 4 | 0.076 | 0.009 | ** | Human > Demographics + Heuristics and biases experiments + Features with r > 0.5 |
| Q24 | 4 | -0.108 | 0.004 | ** | Human < Demographics + Heuristics and biases experiments + Features with r > 0.5 |
| Q29 | 4 | -0.154 | 0.004 | ** | Human < Demographics + Heuristics and biases experiments + Features with r > 0.5 |
| Q34 | 4 | -0.152 | 0.004 | ** | Human < Demographics + Heuristics and biases experiments + Features with r > 0.5 |
| Q39 | 4 | -0.143 | 0.004 | ** | Human < Demographics + Heuristics and biases experiments + Features with r > 0.5 |
| Q5 | 5 | 0.035 | 0.191 | | |
| Q10 | 5 | -0.126 | 0.004 | ** | Human < Demographics + Heuristics and biases experiments + Features with r > 0.5 |
| Q15 | 5 | -0.163 | 0.004 | ** | Human < Demographics + Heuristics and biases experiments + Features with r > 0.5 |
| Q20 | 5 | -0.098 | 0.004 | ** | Human < Demographics + Heuristics and biases experiments + Features with r > 0.5 |
| Q25 | 5 | -0.044 | 0.093 | | |
| Q35 | 5 | -0.221 | 0.004 | ** | Human < Demographics + Heuristics and biases experiments + Features with r > 0.5 |
| Q40 | 5 | -0.173 | 0.004 | ** | Human < Demographics + Heuristics and biases experiments + Features with r > 0.5 |
| Q30 | 6 | 0.074 | 0.007 | ** | Human > Demographics + Heuristics and biases experiments + Features with r > 0.5 |
| Q41 | 6 | -0.224 | 0.004 | ** | Human < Demographics + Heuristics and biases experiments + Features with r > 0.5 |
| Q44 | 6 | -0.166 | 0.004 | ** | Human < Demographics + Heuristics and biases experiments + Features with r > 0.5 |

| Demographics + Heuristics and biases experiments + Features with r < 0.3 vs Human | | | | | |
|---|---|---|---|---|---|
| Q1 | 1 | -0.008 | 0.788 | | |
| Q6 | 1 | -0.124 | 0.004 | ** | Human < Demographics + Heuristics and biases experiments + Features with r < 0.3 |
| Q11 | 1 | -0.151 | 0.004 | ** | Human < Demographics + Heuristics and biases experiments + Features with r < 0.3 |
| Q16 | 1 | -0.005 | 0.86 | | |
| Q21 | 1 | 0.043 | 0.137 | | |
| Q26 | 1 | -0.073 | 0.004 | ** | Human < Demographics + Heuristics and biases experiments + Features with r < 0.3 |
| Q31 | 1 | -0.056 | 0.016 | * | Human < Demographics + Heuristics and biases experiments + Features with r < 0.3 |
| Q36 | 1 | 0.046 | 0.137 | | |
| Q2 | 2 | 0.161 | 0.004 | ** | Human > Demographics + Heuristics and biases experiments + Features with r < 0.3 |
| Q7 | 2 | -0.213 | 0.004 | ** | Human < Demographics + Heuristics and biases experiments + Features with r < 0.3 |
| Q12 | 2 | -0.034 | 0.31 | | |
| Q17 | 2 | 0.118 | 0.004 | ** | Human > Demographics + Heuristics and biases experiments + Features with r < 0.3 |
| Q22 | 2 | 0.138 | 0.004 | ** | Human > Demographics + Heuristics and biases experiments + Features with r < 0.3 |
| Q27 | 2 | -0.126 | 0.004 | ** | Human < Demographics + Heuristics and biases experiments + Features with r < 0.3 |
| Q32 | 2 | -0.027 | 0.608 | | |
| Q37 | 2 | 0.008 | 0.853 | | |
| Q42 | 2 | -0.231 | 0.004 | ** | Human < Demographics + Heuristics and biases experiments + Features with r < 0.3 |
| Q3 | 3 | -0.033 | 0.266 | | |
| Q8 | 3 | -0.002 | 0.938 | | |
| Q13 | 3 | 0.014 | 0.718 | | |
| Q18 | 3 | -0.06 | 0.029 | * | Human < Demographics + Heuristics and biases experiments + Features with r < 0.3 |
| Q23 | 3 | 0.176 | 0.004 | ** | Human > Demographics + Heuristics and biases experiments + Features with r < 0.3 |
| Q28 | 3 | -0.046 | 0.162 | | |
| Q33 | 3 | -0.201 | 0.004 | ** | Human < Demographics + Heuristics and biases experiments + Features with r < 0.3 |
| Q38 | 3 | -0.119 | 0.004 | ** | Human < Demographics + Heuristics and biases experiments + Features with r < 0.3 |
| Q43 | 3 | -0.003 | 0.86 | | |
| Q5 | 4 | 0.126 | 0.004 | ** | Human > Demographics + Heuristics and biases experiments + Features with r < 0.3 |
| Q10 | 4 | -0.06 | 0.066 | | |
| Q15 | 4 | -0.18 | 0.004 | ** | Human < Demographics + Heuristics and biases experiments + Features with r < 0.3 |
| Q20 | 4 | -0.135 | 0.004 | ** | Human < Demographics + Heuristics and biases experiments + Features with r < 0.3 |
| Q25 | 4 | -0.037 | 0.293 | | |
| Q30 | 4 | 0.083 | 0.004 | ** | Human > Demographics + Heuristics and biases experiments + Features with r < 0.3 |
| Q35 | 4 | -0.027 | 0.252 | | |
| Q40 | 4 | -0.054 | 0.104 | | |
| Q41 | 4 | 0.005 | 0.86 | | |
| Q44 | 4 | -0.061 | 0.019 | * | Human < Demographics + Heuristics and biases experiments + Features with r < 0.3 |
| Q9 | 5 | 0.132 | 0.004 | ** | Human > Demographics + Heuristics and biases experiments + Features with r < 0.3 |
| Q14 | 5 | -0.059 | 0.016 | * | Human < Demographics + Heuristics and biases experiments + Features with r < 0.3 |
| Q19 | 5 | -0.012 | 0.728 | | |
| Q24 | 5 | -0.02 | 0.647 | | |
| Q34 | 5 | -0.158 | 0.004 | ** | Human < Demographics + Heuristics and biases experiments + Features with r < 0.3 |
| Q39 | 5 | -0.187 | 0.004 | ** | Human < Demographics + Heuristics and biases experiments + Features with r < 0.3 |
| | **Panel C. DeepSeek V4-Flash** | | | | |
| **Item** | **Membership** | **Difference** | ***p*** | **Sig** | **Direction** |
| Demographics vs Human | | | | | |
| Q1 | 1 | -0.026 | 0.319 | | |
| Q6 | 1 | -0.042 | 0.163 | | |
| Q11 | 1 | -0.25 | 0.005 | ** | Human < Demographics |
| Q16 | 1 | -0.031 | 0.319 | | |
| Q21 | 1 | 0.019 | 0.581 | | |
| Q26 | 1 | 0.03 | 0.291 | | |
| Q31 | 1 | 0.072 | 0.005 | ** | Human > Demographics |
| Q36 | 1 | -0.037 | 0.314 | | |
| Q2 | 2 | 0.091 | 0.005 | ** | Human > Demographics |
| Q7 | 2 | -0.194 | 0.005 | ** | Human < Demographics |
| Q12 | 2 | -0.195 | 0.005 | ** | Human < Demographics |
| Q17 | 2 | 0.137 | 0.005 | ** | Human > Demographics |
| Q22 | 2 | 0.099 | 0.011 | * | Human > Demographics |
| Q27 | 2 | -0.014 | 0.736 | | |
| Q32 | 2 | 0.008 | 0.909 | | |
| Q37 | 2 | -0.001 | 0.99 | | |
| Q42 | 2 | -0.301 | 0.005 | ** | Human < Demographics |
| Q3 | 3 | 0.068 | 0.013 | * | Human > Demographics |
| Q8 | 3 | 0.004 | 0.957 | | |
| Q13 | 3 | -0.085 | 0.013 | * | Human < Demographics |
| Q18 | 3 | -0.049 | 0.123 | | |
| Q23 | 3 | 0.063 | 0.045 | * | Human > Demographics |
| Q28 | 3 | -0.088 | 0.007 | ** | Human < Demographics |
| Q33 | 3 | -0.134 | 0.005 | ** | Human < Demographics |
| Q38 | 3 | -0.11 | 0.005 | ** | Human < Demographics |
| Q43 | 3 | 0 | 0.99 | | |
| Q4 | 4 | 0.016 | 0.681 | | |
| Q9 | 4 | 0.173 | 0.005 | ** | Human > Demographics |
| Q14 | 4 | -0.08 | 0.007 | ** | Human < Demographics |
| Q19 | 4 | 0.113 | 0.005 | ** | Human > Demographics |
| Q24 | 4 | -0.11 | 0.007 | ** | Human < Demographics |
| Q29 | 4 | -0.181 | 0.005 | ** | Human < Demographics |
| Q34 | 4 | -0.162 | 0.005 | ** | Human < Demographics |
| Q39 | 4 | -0.107 | 0.005 | ** | Human < Demographics |
| Q5 | 5 | 0.091 | 0.005 | ** | Human > Demographics |
| Q10 | 5 | -0.039 | 0.198 | | |
| Q15 | 5 | -0.12 | 0.005 | ** | Human < Demographics |
| Q20 | 5 | -0.107 | 0.005 | ** | Human < Demographics |
| Q25 | 5 | -0.07 | 0.007 | ** | Human < Demographics |
| Q30 | 5 | 0.026 | 0.371 | | |
| Q40 | 5 | -0.107 | 0.005 | ** | Human < Demographics |
| Q41 | 5 | -0.067 | 0.005 | ** | Human < Demographics |
| Q44 | 5 | -0.002 | 0.986 | | |
| Heuristics and biases experiments vs Human | | | | | |
| Q1 | 1 | 0.023 | 0.396 | | |
| Q6 | 1 | -0.067 | 0.022 | * | Human < Heuristics and biases experiments |
| Q11 | 1 | -0.258 | 0.005 | ** | Human < Heuristics and biases experiments |
| Q16 | 1 | -0.027 | 0.362 | | |
| Q21 | 1 | 0.02 | 0.492 | | |
| Q26 | 1 | 0.081 | 0.009 | ** | Human > Heuristics and biases experiments |
| Q31 | 1 | -0.017 | 0.527 | | |
| Q36 | 1 | 0.007 | 0.808 | | |
| Q2 | 2 | 0.132 | 0.005 | ** | Human > Heuristics and biases experiments |
| Q7 | 2 | -0.191 | 0.005 | ** | Human < Heuristics and biases experiments |
| Q12 | 2 | -0.058 | 0.132 | | |
| Q17 | 2 | 0.06 | 0.033 | * | Human > Heuristics and biases experiments |
| Q22 | 2 | 0.049 | 0.08 | | |
| Q27 | 2 | -0.063 | 0.051 | | |
| Q32 | 2 | 0.009 | 0.808 | | |
| Q37 | 2 | -0.044 | 0.242 | | |
| Q42 | 2 | -0.245 | 0.005 | ** | Human < Heuristics and biases experiments |
| Q3 | 3 | 0.05 | 0.086 | | |
| Q8 | 3 | 0.021 | 0.527 | | |
| Q13 | 3 | -0.072 | 0.02 | * | Human < Heuristics and biases experiments |
| Q18 | 3 | 0.036 | 0.25 | | |
| Q23 | 3 | -0.091 | 0.009 | ** | Human < Heuristics and biases experiments |
| Q28 | 3 | -0.074 | 0.025 | * | Human < Heuristics and biases experiments |
| Q33 | 3 | -0.106 | 0.005 | ** | Human < Heuristics and biases experiments |
| Q38 | 3 | -0.113 | 0.005 | ** | Human < Heuristics and biases experiments |
| Q43 | 3 | -0.091 | 0.005 | ** | Human < Heuristics and biases experiments |
| Q4 | 4 | 0.172 | 0.005 | ** | Human > Heuristics and biases experiments |
| Q9 | 4 | 0.31 | 0.005 | ** | Human > Heuristics and biases experiments |
| Q14 | 4 | -0.073 | 0.012 | * | Human < Heuristics and biases experiments |
| Q19 | 4 | 0.111 | 0.005 | ** | Human > Heuristics and biases experiments |
| Q24 | 4 | -0.174 | 0.005 | ** | Human < Heuristics and biases experiments |
| Q29 | 4 | -0.234 | 0.005 | ** | Human < Heuristics and biases experiments |
| Q34 | 4 | -0.201 | 0.005 | ** | Human < Heuristics and biases experiments |
| Q39 | 4 | -0.11 | 0.005 | ** | Human < Heuristics and biases experiments |
| Q5 | 5 | 0.068 | 0.009 | ** | Human > Heuristics and biases experiments |
| Q10 | 5 | -0.015 | 0.602 | | |
| Q15 | 5 | -0.075 | 0.005 | ** | Human < Heuristics and biases experiments |
| Q20 | 5 | -0.038 | 0.239 | | |
| Q25 | 5 | -0.136 | 0.005 | ** | Human < Heuristics and biases experiments |
| Q30 | 5 | 0.017 | 0.527 | | |
| Q40 | 5 | -0.074 | 0.012 | * | Human < Heuristics and biases experiments |
| Q41 | 5 | -0.048 | 0.028 | * | Human < Heuristics and biases experiments |
| Q44 | 5 | 0.012 | 0.542 | | |
| Demographics + Heuristics and biases experiments vs Human | | | | | |
| Q1 | 1 | 0.012 | 0.578 | | |
| Q6 | 1 | -0.058 | 0.057 | | |
| Q11 | 1 | -0.237 | 0.004 | ** | Human < Demographics + Heuristics and biases experiments |
| Q16 | 1 | -0.016 | 0.578 | | |
| Q21 | 1 | 0.022 | 0.45 | | |
| Q26 | 1 | 0.03 | 0.267 | | |
| Q31 | 1 | 0.038 | 0.168 | | |
| Q36 | 1 | -0.09 | 0.007 | ** | Human < Demographics + Heuristics and biases experiments |
| Q2 | 2 | 0.155 | 0.004 | ** | Human > Demographics + Heuristics and biases experiments |
| Q7 | 2 | -0.169 | 0.004 | ** | Human < Demographics + Heuristics and biases experiments |
| Q12 | 2 | -0.1 | 0.007 | ** | Human < Demographics + Heuristics and biases experiments |
| Q17 | 2 | 0.095 | 0.004 | ** | Human > Demographics + Heuristics and biases experiments |
| Q22 | 2 | 0.109 | 0.004 | ** | Human > Demographics + Heuristics and biases experiments |
| Q27 | 2 | -0.126 | 0.004 | ** | Human < Demographics + Heuristics and biases experiments |
| Q32 | 2 | -0.002 | 0.964 | | |
| Q37 | 2 | -0.06 | 0.111 | | |
| Q42 | 2 | -0.299 | 0.004 | ** | Human < Demographics + Heuristics and biases experiments |
| Q3 | 3 | 0.049 | 0.1 | | |
| Q8 | 3 | 0.029 | 0.382 | | |
| Q13 | 3 | -0.105 | 0.004 | ** | Human < Demographics + Heuristics and biases experiments |
| Q18 | 3 | -0.035 | 0.277 | | |
| Q23 | 3 | 0.023 | 0.475 | | |
| Q28 | 3 | -0.108 | 0.004 | ** | Human < Demographics + Heuristics and biases experiments |
| Q33 | 3 | -0.12 | 0.004 | ** | Human < Demographics + Heuristics and biases experiments |
| Q38 | 3 | -0.104 | 0.004 | ** | Human < Demographics + Heuristics and biases experiments |
| Q43 | 3 | 0.01 | 0.679 | | |
| Q4 | 4 | 0.085 | 0.01 | * | Human > Demographics + Heuristics and biases experiments |
| Q9 | 4 | 0.255 | 0.004 | ** | Human > Demographics + Heuristics and biases experiments |
| Q14 | 4 | -0.053 | 0.076 | | |
| Q19 | 4 | 0.105 | 0.004 | ** | Human > Demographics + Heuristics and biases experiments |
| Q24 | 4 | -0.133 | 0.004 | ** | Human < Demographics + Heuristics and biases experiments |
| Q29 | 4 | -0.227 | 0.004 | ** | Human < Demographics + Heuristics and biases experiments |
| Q34 | 4 | -0.206 | 0.004 | ** | Human < Demographics + Heuristics and biases experiments |
| Q39 | 4 | -0.119 | 0.004 | ** | Human < Demographics + Heuristics and biases experiments |
| Q5 | 5 | 0.097 | 0.004 | ** | Human > Demographics + Heuristics and biases experiments |
| Q10 | 5 | -0.021 | 0.539 | | |
| Q15 | 5 | -0.074 | 0.004 | ** | Human < Demographics + Heuristics and biases experiments |
| Q20 | 5 | -0.096 | 0.004 | ** | Human < Demographics + Heuristics and biases experiments |
| Q25 | 5 | -0.105 | 0.004 | ** | Human < Demographics + Heuristics and biases experiments |
| Q30 | 5 | 0.018 | 0.483 | | |
| Q35 | 5 | -0.18 | 0.004 | ** | Human < Demographics + Heuristics and biases experiments |
| Q40 | 5 | -0.109 | 0.004 | ** | Human < Demographics + Heuristics and biases experiments |
| Q41 | 5 | -0.07 | 0.004 | ** | Human < Demographics + Heuristics and biases experiments |
| Q44 | 5 | 0.018 | 0.416 | | |
| Demographics + All psychological measures + Heuristics and biases experiments vs Human | | | | | |
| Q1 | 1 | 0.06 | 0.012 | * | Human > Demographics + All psychological measures + Heuristics and biases experiments |
| Q6 | 1 | -0.027 | 0.272 | | |
| Q11 | 1 | -0.185 | 0.005 | ** | Human < Demographics + All psychological measures + Heuristics and biases experiments |
| Q16 | 1 | -0.021 | 0.391 | | |
| Q21 | 1 | 0.039 | 0.128 | | |
| Q26 | 1 | 0.022 | 0.366 | | |
| Q31 | 1 | -0.03 | 0.22 | | |
| Q36 | 1 | -0.062 | 0.024 | * | Human < Demographics + All psychological measures + Heuristics and biases experiments |
| Q2 | 2 | 0.122 | 0.005 | ** | Human > Demographics + All psychological measures + Heuristics and biases experiments |
| Q7 | 2 | -0.214 | 0.005 | ** | Human < Demographics + All psychological measures + Heuristics and biases experiments |
| Q12 | 2 | -0.043 | 0.205 | | |
| Q17 | 2 | 0.142 | 0.005 | ** | Human > Demographics + All psychological measures + Heuristics and biases experiments |
| Q22 | 2 | 0.04 | 0.147 | | |
| Q27 | 2 | -0.112 | 0.005 | ** | Human < Demographics + All psychological measures + Heuristics and biases experiments |
| Q32 | 2 | -0.1 | 0.005 | ** | Human < Demographics + All psychological measures + Heuristics and biases experiments |
| Q37 | 2 | -0.091 | 0.009 | ** | Human < Demographics + All psychological measures + Heuristics and biases experiments |
| Q42 | 2 | -0.072 | 0.025 | * | Human < Demographics + All psychological measures + Heuristics and biases experiments |
| Q3 | 3 | 0.027 | 0.328 | | |
| Q8 | 3 | -0.038 | 0.177 | | |
| Q13 | 3 | -0.119 | 0.005 | ** | Human < Demographics + All psychological measures + Heuristics and biases experiments |
| Q18 | 3 | 0.034 | 0.205 | | |
| Q23 | 3 | -0.001 | 0.96 | | |
| Q28 | 3 | -0.066 | 0.018 | * | Human < Demographics + All psychological measures + Heuristics and biases experiments |
| Q33 | 3 | -0.088 | 0.005 | ** | Human < Demographics + All psychological measures + Heuristics and biases experiments |
| Q38 | 3 | -0.105 | 0.005 | ** | Human < Demographics + All psychological measures + Heuristics and biases experiments |
| Q43 | 3 | -0.046 | 0.068 | | |
| Q4 | 4 | -0.003 | 0.934 | | |
| Q9 | 4 | 0.258 | 0.005 | ** | Human > Demographics + All psychological measures + Heuristics and biases experiments |
| Q14 | 4 | -0.108 | 0.005 | ** | Human < Demographics + All psychological measures + Heuristics and biases experiments |
| Q19 | 4 | 0.057 | 0.012 | * | Human > Demographics + All psychological measures + Heuristics and biases experiments |
| Q24 | 4 | -0.07 | 0.015 | * | Human < Demographics + All psychological measures + Heuristics and biases experiments |
| Q29 | 4 | -0.168 | 0.005 | ** | Human < Demographics + All psychological measures + Heuristics and biases experiments |
| Q34 | 4 | -0.221 | 0.005 | ** | Human < Demographics + All psychological measures + Heuristics and biases experiments |
| Q39 | 4 | -0.131 | 0.005 | ** | Human < Demographics + All psychological measures + Heuristics and biases experiments |
| Q5 | 5 | 0.029 | 0.211 | | |
| Q10 | 5 | -0.074 | 0.005 | ** | Human < Demographics + All psychological measures + Heuristics and biases experiments |
| Q15 | 5 | -0.082 | 0.005 | ** | Human < Demographics + All psychological measures + Heuristics and biases experiments |
| Q20 | 5 | -0.077 | 0.005 | ** | Human < Demographics + All psychological measures + Heuristics and biases experiments |
| Q25 | 5 | -0.136 | 0.005 | ** | Human < Demographics + All psychological measures + Heuristics and biases experiments |
| Q30 | 5 | 0.037 | 0.132 | | |
| Q40 | 5 | -0.08 | 0.005 | ** | Human < Demographics + All psychological measures + Heuristics and biases experiments |
| Q41 | 5 | 0.012 | 0.524 | | |
| Q44 | 5 | 0.044 | 0.025 | * | Human > Demographics + All psychological measures + Heuristics and biases experiments |
| Demographics + Heuristics and biases experiments + Features with r < 0.5 vs Human | | | | | |
| Q1 | 1 | 0 | 0.994 | | |
| Q6 | 1 | -0.061 | 0.024 | * | Human < Demographics + Heuristics and biases experiments + Features with r < 0.5 |
| Q11 | 1 | -0.152 | 0.004 | ** | Human < Demographics + Heuristics and biases experiments + Features with r < 0.5 |
| Q16 | 1 | -0.021 | 0.489 | | |
| Q21 | 1 | 0.049 | 0.057 | | |
| Q26 | 1 | 0.004 | 0.903 | | |
| Q31 | 1 | -0.043 | 0.058 | | |
| Q36 | 1 | -0.049 | 0.088 | | |
| Q2 | 2 | 0.137 | 0.004 | ** | Human > Demographics + Heuristics and biases experiments + Features with r < 0.5 |
| Q7 | 2 | -0.149 | 0.004 | ** | Human < Demographics + Heuristics and biases experiments + Features with r < 0.5 |
| Q12 | 2 | -0.087 | 0.007 | ** | Human < Demographics + Heuristics and biases experiments + Features with r < 0.5 |
| Q17 | 2 | 0.032 | 0.297 | | |
| Q22 | 2 | 0.082 | 0.007 | ** | Human > Demographics + Heuristics and biases experiments + Features with $r < 0.5$ |
| Q27 | 2 | -0.14 | 0.004 | ** | Human < Demographics + Heuristics and biases experiments + Features with $r < 0.5$ |
| Q32 | 2 | -0.01 | 0.865 | | |
| Q37 | 2 | -0.061 | 0.075 | | |
| Q42 | 2 | -0.227 | 0.004 | ** | Human < Demographics + Heuristics and biases experiments + Features with $r < 0.5$ |
| Q3 | 3 | 0.071 | 0.007 | ** | Human > Demographics + Heuristics and biases experiments + Features with $r < 0.5$ |
| Q8 | 3 | -0.015 | 0.651 | | |
| Q13 | 3 | -0.12 | 0.004 | ** | Human < Demographics + Heuristics and biases experiments + Features with $r < 0.5$ |
| Q18 | 3 | -0.083 | 0.004 | ** | Human < Demographics + Heuristics and biases experiments + Features with $r < 0.5$ |
| Q23 | 3 | 0.062 | 0.044 | * | Human > Demographics + Heuristics and biases experiments + Features with $r < 0.5$ |
| Q28 | 3 | -0.105 | 0.004 | ** | Human < Demographics + Heuristics and biases experiments + Features with $r < 0.5$ |
| Q33 | 3 | -0.084 | 0.004 | ** | Human < Demographics + Heuristics and biases experiments + Features with $r < 0.5$ |
| Q38 | 3 | -0.119 | 0.004 | ** | Human < Demographics + Heuristics and biases experiments + Features with $r < 0.5$ |
| Q43 | 3 | -0.014 | 0.627 | | |
| Q4 | 4 | 0.188 | 0.004 | ** | Human > Demographics + Heuristics and biases experiments + Features with $r < 0.5$ |
| Q9 | 4 | 0.285 | 0.004 | ** | Human > Demographics + Heuristics and biases experiments + Features with $r < 0.5$ |
| Q14 | 4 | -0.114 | 0.004 | ** | Human < Demographics + Heuristics and biases experiments + Features with $r < 0.5$ |
| Q19 | 4 | 0.155 | 0.004 | ** | Human > Demographics + Heuristics and biases experiments + Features with $r < 0.5$ |
| Q24 | 4 | -0.126 | 0.004 | ** | Human < Demographics + Heuristics and biases experiments + Features with $r < 0.5$ |
| Q29 | 4 | -0.215 | 0.004 | ** | Human < Demographics + Heuristics and biases experiments + Features with $r < 0.5$ |
| Q34 | 4 | -0.204 | 0.004 | ** | Human < Demographics + Heuristics and biases experiments + Features with $r < 0.5$ |
| Q39 | 4 | -0.181 | 0.004 | ** | Human < Demographics + Heuristics and biases experiments + Features with $r < 0.5$ |
| Q5 | 5 | 0.1 | 0.004 | ** | Human > Demographics + Heuristics and biases experiments + Features with $r < 0.5$ |
| Q10 | 5 | -0.022 | 0.489 | | |
| Q15 | 5 | -0.122 | 0.004 | ** | Human < Demographics + Heuristics and biases experiments + Features with r < 0.5 |
| Q20 | 5 | -0.097 | 0.004 | ** | Human < Demographics + Heuristics and biases experiments + Features with r < 0.5 |
| Q25 | 5 | -0.125 | 0.004 | ** | Human < Demographics + Heuristics and biases experiments + Features with r < 0.5 |
| Q30 | 5 | 0.004 | 0.895 | | |
| Q35 | 5 | -0.082 | 0.004 | ** | Human < Demographics + Heuristics and biases experiments + Features with r < 0.5 |
| Q40 | 5 | -0.101 | 0.004 | ** | Human < Demographics + Heuristics and biases experiments + Features with r < 0.5 |
| Q41 | 5 | -0.025 | 0.34 | | |
| Q44 | 5 | 0.05 | 0.019 | * | Human > Demographics + Heuristics and biases experiments + Features with r < 0.5 |
| Demographics + Heuristics and biases experiments + Features with r > 0.5 vs Human | | | | | |
| Q1 | 1 | 0.03 | 0.147 | | |
| Q6 | 1 | -0.017 | 0.576 | | |
| Q11 | 1 | -0.152 | 0.004 | ** | Human < Demographics + Heuristics and biases experiments + Features with r > 0.5 |
| Q16 | 1 | -0.071 | 0.015 | * | Human < Demographics + Heuristics and biases experiments + Features with r > 0.5 |
| Q21 | 1 | 0.045 | 0.065 | | |
| Q26 | 1 | 0.001 | 0.968 | | |
| Q31 | 1 | -0.034 | 0.176 | | |
| Q36 | 1 | -0.052 | 0.065 | | |
| Q2 | 2 | 0.156 | 0.004 | ** | Human > Demographics + Heuristics and biases experiments + Features with r > 0.5 |
| Q7 | 2 | -0.206 | 0.004 | ** | Human < Demographics + Heuristics and biases experiments + Features with r > 0.5 |
| Q12 | 2 | -0.009 | 0.821 | | |
| Q17 | 2 | 0.173 | 0.004 | ** | Human > Demographics + Heuristics and biases experiments + Features with r > 0.5 |
| Q22 | 2 | 0.05 | 0.072 | | |
| Q27 | 2 | -0.113 | 0.004 | ** | Human < Demographics + Heuristics and biases experiments + Features with r > 0.5 |
| Q32 | 2 | -0.132 | 0.004 | ** | Human < Demographics + Heuristics and biases experiments + Features with r > 0.5 |
| Q37 | 2 | -0.107 | 0.004 | ** | Human < Demographics + Heuristics and biases experiments + Features with r > 0.5 |
| Q42 | 2 | -0.079 | 0.011 | * | Human < Demographics + Heuristics and biases experiments + Features with r > 0.5 |
| Q3 | 3 | 0.005 | 0.874 | | |
| Q8 | 3 | -0.016 | 0.588 | | |
| Q13 | 3 | -0.116 | 0.004 | ** | Human < Demographics + Heuristics and biases experiments + Features with r > 0.5 |
| Q18 | 3 | 0.026 | 0.341 | | |
| Q23 | 3 | 0.066 | 0.02 | * | Human > Demographics + Heuristics and biases experiments + Features with r > 0.5 |
| Q28 | 3 | -0.059 | 0.023 | * | Human < Demographics + Heuristics and biases experiments + Features with r > 0.5 |
| Q33 | 3 | -0.04 | 0.136 | | |
| Q38 | 3 | -0.145 | 0.004 | ** | Human < Demographics + Heuristics and biases experiments + Features with r > 0.5 |
| Q43 | 3 | -0.089 | 0.004 | ** | Human < Demographics + Heuristics and biases experiments + Features with r > 0.5 |
| Q4 | 4 | -0.06 | 0.018 | * | Human < Demographics + Heuristics and biases experiments + Features with r > 0.5 |
| Q9 | 4 | 0.21 | 0.004 | ** | Human > Demographics + Heuristics and biases experiments + Features with r > 0.5 |
| Q14 | 4 | -0.115 | 0.004 | ** | Human < Demographics + Heuristics and biases experiments + Features with r > 0.5 |
| Q19 | 4 | 0.071 | 0.008 | ** | Human > Demographics + Heuristics and biases experiments + Features with r > 0.5 |
| Q24 | 4 | -0.053 | 0.058 | | |
| Q29 | 4 | -0.157 | 0.004 | ** | Human < Demographics + Heuristics and biases experiments + Features with r > 0.5 |
| Q34 | 4 | -0.18 | 0.004 | ** | Human < Demographics + Heuristics and biases experiments + Features with r > 0.5 |
| Q39 | 4 | -0.138 | 0.004 | ** | Human < Demographics + Heuristics and biases experiments + Features with r > 0.5 |
| Q5 | 5 | 0.047 | 0.066 | | |
| Q10 | 5 | -0.082 | 0.008 | ** | Human < Demographics + Heuristics and biases experiments + Features with r > 0.5 |
| Q15 | 5 | -0.165 | 0.004 | ** | Human < Demographics + Heuristics and biases experiments + Features with r > 0.5 |
| Q20 | 5 | -0.069 | 0.004 | ** | Human < Demographics + Heuristics and biases experiments + Features with r > 0.5 |
| Q25 | 5 | -0.095 | 0.004 | ** | Human < Demographics + Heuristics and biases experiments + Features with r > 0.5 |
| Q30 | 5 | 0.051 | 0.025 | * | Human > Demographics + Heuristics and biases experiments + Features with r > 0.5 |
| Q35 | 5 | -0.135 | 0.004 | ** | Human < Demographics + Heuristics and biases experiments + Features with r > 0.5 |
| Q40 | 5 | -0.086 | 0.004 | ** | Human < Demographics + Heuristics and biases experiments + Features with $r > 0.5$ |
| Q41 | 5 | 0.021 | 0.324 | | |
| Q44 | 5 | 0.037 | 0.072 | | |
| Demographics + Heuristics and biases experiments + Features with $r < 0.3$ vs Human | | | | | |
| Q1 | 1 | 0.014 | 0.536 | | |
| Q6 | 1 | -0.022 | 0.421 | | |
| Q11 | 1 | -0.175 | 0.003 | ** | Human < Demographics + Heuristics and biases experiments + Features with $r < 0.3$ |
| Q16 | 1 | -0.107 | 0.003 | ** | Human < Demographics + Heuristics and biases experiments + Features with $r < 0.3$ |
| Q21 | 1 | 0.107 | 0.003 | ** | Human > Demographics + Heuristics and biases experiments + Features with $r < 0.3$ |
| Q26 | 1 | 0.016 | 0.499 | | |
| Q31 | 1 | -0.092 | 0.003 | ** | Human < Demographics + Heuristics and biases experiments + Features with $r < 0.3$ |
| Q36 | 1 | -0.041 | 0.172 | | |
| Q2 | 2 | 0.167 | 0.003 | ** | Human > Demographics + Heuristics and biases experiments + Features with $r < 0.3$ |
| Q7 | 2 | -0.077 | 0.049 | * | Human < Demographics + Heuristics and biases experiments + Features with $r < 0.3$ |
| Q12 | 2 | -0.123 | 0.003 | ** | Human < Demographics + Heuristics and biases experiments + Features with $r < 0.3$ |
| Q17 | 2 | 0.113 | 0.003 | ** | Human > Demographics + Heuristics and biases experiments + Features with $r < 0.3$ |
| Q22 | 2 | 0.081 | 0.013 | * | Human > Demographics + Heuristics and biases experiments + Features with $r < 0.3$ |
| Q27 | 2 | -0.049 | 0.169 | | |
| Q32 | 2 | -0.031 | 0.423 | | |
| Q37 | 2 | -0.029 | 0.421 | | |
| Q42 | 2 | -0.321 | 0.003 | ** | Human < Demographics + Heuristics and biases experiments + Features with $r < 0.3$ |
| Q3 | 3 | 0.083 | 0.01 | * | Human > Demographics + Heuristics and biases experiments + Features with $r < 0.3$ |
| Q8 | 3 | -0.008 | 0.776 | | |
| Q13 | 3 | -0.1 | 0.003 | ** | Human < Demographics + Heuristics and biases experiments + Features with $r < 0.3$ |
| Q18 | 3 | -0.035 | 0.268 | | |
| Q23 | 3 | 0.025 | 0.416 | | |
| Q28 | 3 | -0.105 | 0.003 | ** | Human < Demographics + Heuristics and biases experiments + Features with $r < 0.3$ |
| Q33 | 3 | -0.098 | 0.003 | ** | Human < Demographics + Heuristics and biases experiments + Features with r < 0.3 |
| Q38 | 3 | -0.121 | 0.003 | ** | Human < Demographics + Heuristics and biases experiments + Features with r < 0.3 |
| Q43 | 3 | -0.052 | 0.049 | * | Human < Demographics + Heuristics and biases experiments + Features with r < 0.3 |
| Q4 | 4 | 0.152 | 0.003 | ** | Human > Demographics + Heuristics and biases experiments + Features with r < 0.3 |
| Q9 | 4 | 0.32 | 0.003 | ** | Human > Demographics + Heuristics and biases experiments + Features with r < 0.3 |
| Q14 | 4 | -0.066 | 0.022 | * | Human < Demographics + Heuristics and biases experiments + Features with r < 0.3 |
| Q19 | 4 | 0.108 | 0.003 | ** | Human > Demographics + Heuristics and biases experiments + Features with r < 0.3 |
| Q24 | 4 | -0.162 | 0.003 | ** | Human < Demographics + Heuristics and biases experiments + Features with r < 0.3 |
| Q29 | 4 | -0.242 | 0.003 | ** | Human < Demographics + Heuristics and biases experiments + Features with r < 0.3 |
| Q34 | 4 | -0.217 | 0.003 | ** | Human < Demographics + Heuristics and biases experiments + Features with r < 0.3 |
| Q39 | 4 | -0.103 | 0.003 | ** | Human < Demographics + Heuristics and biases experiments + Features with r < 0.3 |
| Q5 | 5 | 0.084 | 0.003 | ** | Human > Demographics + Heuristics and biases experiments + Features with r < 0.3 |
| Q10 | 5 | -0.026 | 0.381 | | |
| Q15 | 5 | -0.092 | 0.003 | ** | Human < Demographics + Heuristics and biases experiments + Features with r < 0.3 |
| Q20 | 5 | -0.1 | 0.003 | ** | Human < Demographics + Heuristics and biases experiments + Features with r < 0.3 |
| Q25 | 5 | -0.107 | 0.003 | ** | Human < Demographics + Heuristics and biases experiments + Features with r < 0.3 |
| Q30 | 5 | 0.025 | 0.352 | | |
| Q40 | 5 | -0.094 | 0.003 | ** | Human < Demographics + Heuristics and biases experiments + Features with r < 0.3 |
| Q41 | 5 | -0.068 | 0.003 | ** | Human < Demographics + Heuristics and biases experiments + Features with r < 0.3 |
| Q44 | 5 | 0.015 | 0.477 | | |
| | **Panel D. Qwen 3.5 Plus** | | | | |
| **Item** | **Membership** | **Difference** | ***p*** | **Sig** | **Direction** |
| Demographics vs Human | | | | | |
| Q1 | 1 | 0.124 | 0.004 | ** | Human < Demographics |
| Q6 | 1 | 0.053 | 0.105 | | |
| Q11 | 1 | 0.157 | 0.004 | ** | Human < Demographics |
| Q16 | 1 | 0.028 | 0.377 | | |
| Q21 | 1 | -0.136 | 0.004 | ** | Human > Demographics |
| Q26 | 1 | 0.089 | 0.004 | ** | Human < Demographics |
| Q31 | 1 | -0.049 | 0.1 | | |
| Q36 | 1 | -0.028 | 0.464 | | |
| Q3 | 2 | 0.051 | 0.192 | | |
| Q13 | 2 | 0.062 | 0.108 | | |
| Q28 | 2 | 0.048 | 0.129 | | |
| Q33 | 2 | -0.037 | 0.167 | | |
| Q38 | 2 | 0.166 | 0.004 | ** | Human < Demographics |
| Q4 | 3 | -0.033 | 0.252 | | |
| Q9 | 3 | -0.093 | 0.004 | ** | Human > Demographics |
| Q14 | 3 | 0.092 | 0.004 | ** | Human < Demographics |
| Q19 | 3 | -0.074 | 0.011 | * | Human > Demographics |
| Q24 | 3 | 0.072 | 0.042 | * | Human < Demographics |
| Q29 | 3 | 0.019 | 0.508 | | |
| Q34 | 3 | 0.16 | 0.004 | ** | Human < Demographics |
| Q39 | 3 | 0.113 | 0.004 | ** | Human < Demographics |
| Q5 | 4 | -0.001 | 0.98 | | |
| Q10 | 4 | 0.084 | 0.004 | ** | Human < Demographics |
| Q15 | 4 | 0.185 | 0.004 | ** | Human < Demographics |
| Q20 | 4 | 0.109 | 0.004 | ** | Human < Demographics |
| Q25 | 4 | 0.043 | 0.108 | | |
| Q30 | 4 | -0.094 | 0.004 | ** | Human > Demographics |
| Q40 | 4 | 0.014 | 0.584 | | |
| Q41 | 4 | 0.16 | 0.004 | ** | Human < Demographics |
| Q44 | 4 | -0.061 | 0.004 | ** | Human > Demographics |
| Q7 | 5 | 0.179 | 0.004 | ** | Human < Demographics |
| Q17 | 5 | 0.013 | 0.719 | | |
| Q22 | 5 | -0.182 | 0.004 | ** | Human > Demographics |
| Q32 | 5 | 0.008 | 0.873 | | |
| Q42 | 5 | 0.324 | 0.004 | ** | Human < Demographics |
| Demographics + Heuristics and biases experiments vs Human | | | | | |
| Q1 | 1 | 0.082 | 0.004 | ** | Human < Demographics + Heuristics and biases experiments |
| Q6 | 1 | 0.043 | 0.219 | | |
| Q21 | 1 | -0.178 | 0.004 | ** | Human > Demographics + Heuristics and biases experiments |
| Q31 | 1 | 0.005 | 0.857 | | |
| Q36 | 1 | 0.117 | 0.004 | ** | Human < Demographics + Heuristics and biases experiments |
| Q2 | 2 | -0.161 | 0.004 | ** | Human > Demographics + Heuristics and biases experiments |
| Q7 | 2 | 0.154 | 0.011 | * | Human < Demographics + Heuristics and biases experiments |
| Q17 | 2 | -0.023 | 0.545 | | |
| Q22 | 2 | -0.021 | 0.545 | | |
| Q32 | 2 | -0.028 | 0.588 | | |
| Q42 | 2 | 0.27 | 0.004 | ** | Human < Demographics + Heuristics and biases experiments |
| Q3 | 3 | 0.054 | 0.174 | | |
| Q8 | 3 | 0.096 | 0.004 | ** | Human < Demographics + Heuristics and biases experiments |
| Q13 | 3 | 0.097 | 0.004 | ** | Human < Demographics + Heuristics and biases experiments |
| Q18 | 3 | -0.021 | 0.545 | | |
| Q23 | 3 | -0.081 | 0.004 | ** | Human > Demographics + Heuristics and biases experiments |
| Q28 | 3 | 0.057 | 0.105 | | |
| Q33 | 3 | -0.059 | 0.032 | * | Human > Demographics + Heuristics and biases experiments |
| Q38 | 3 | 0.172 | 0.004 | ** | Human < Demographics + Heuristics and biases experiments |
| Q43 | 3 | 0.069 | 0.032 | * | Human < Demographics + Heuristics and biases experiments |
| Q4 | 4 | -0.102 | 0.008 | ** | Human > Demographics + Heuristics and biases experiments |
| Q9 | 4 | -0.115 | 0.004 | ** | Human > Demographics + Heuristics and biases experiments |
| Q14 | 4 | 0.156 | 0.004 | ** | Human < Demographics + Heuristics and biases experiments |
| Q19 | 4 | -0.034 | 0.313 | | |
| Q24 | 4 | 0.038 | 0.313 | | |
| Q29 | 4 | 0.114 | 0.004 | ** | Human < Demographics + Heuristics and biases experiments |
| Q34 | 4 | 0.148 | 0.004 | ** | Human < Demographics + Heuristics and biases experiments |
| Q39 | 4 | 0.11 | 0.004 | ** | Human < Demographics + Heuristics and biases experiments |
| Q5 | 5 | 0.004 | 0.872 | | |
| Q10 | 5 | 0.047 | 0.169 | | |
| Q15 | 5 | 0.063 | 0.004 | ** | Human < Demographics + Heuristics and biases experiments |
| Q20 | 5 | 0.133 | 0.004 | ** | Human < Demographics + Heuristics and biases experiments |
| Q25 | 5 | 0.064 | 0.014 | * | Human < Demographics + Heuristics and biases experiments |
| Q30 | 5 | -0.091 | 0.004 | ** | Human > Demographics + Heuristics and biases experiments |
| Q40 | 5 | 0.015 | 0.577 | | |
| Q41 | 5 | 0.082 | 0.004 | ** | Human < Demographics + Heuristics and biases experiments |
| Q44 | 5 | 0.029 | 0.214 | | |
| Demographics + All psychological measures + Heuristics and biases experiments vs Human | | | | | |
| Q1 | 1 | 0.08 | 0.004 | ** | Human < Demographics + All psychological measures + Heuristics and biases experiments |
| Q6 | 1 | 0.113 | 0.004 | ** | Human < Demographics + All psychological measures + Heuristics and biases experiments |
| Q21 | 1 | -0.101 | 0.004 | ** | Human > Demographics + All psychological measures + Heuristics and biases experiments |
| Q31 | 1 | -0.023 | 0.359 | | |
| Q36 | 1 | 0.129 | 0.004 | ** | Human < Demographics + All psychological measures + Heuristics and biases experiments |
| Q2 | 2 | 0.048 | 0.077 | | |
| Q7 | 2 | 0.246 | 0.004 | ** | Human < Demographics + All psychological measures + Heuristics and biases experiments |
| Q12 | 2 | 0.146 | 0.004 | ** | Human < Demographics + All psychological measures + Heuristics and biases experiments |
| Q17 | 2 | -0.084 | 0.013 | * | Human > Demographics + All psychological measures + Heuristics and biases experiments |
| Q22 | 2 | -0.039 | 0.155 | | |
| Q27 | 2 | 0.101 | 0.007 | ** | Human < Demographics + All psychological measures + Heuristics and biases experiments |
| Q32 | 2 | 0.03 | 0.376 | | |
| Q37 | 2 | 0.057 | 0.049 | * | Human < Demographics + All psychological measures + Heuristics and biases experiments |
| Q42 | 2 | 0.067 | 0.023 | * | Human < Demographics + All psychological measures + Heuristics and biases experiments |
| Q3 | 3 | 0.043 | 0.158 | | |
| Q8 | 3 | 0.132 | 0.004 | ** | Human < Demographics + All psychological measures + Heuristics and biases experiments |
| Q13 | 3 | 0.069 | 0.018 | * | Human < Demographics + All psychological measures + Heuristics and biases experiments |
| Q18 | 3 | -0.037 | 0.139 | | |
| Q23 | 3 | -0.117 | 0.004 | ** | Human > Demographics + All psychological measures + Heuristics and biases experiments |
| Q28 | 3 | 0.057 | 0.027 | * | Human < Demographics + All psychological measures + Heuristics and biases experiments |
| Q33 | 3 | 0.089 | 0.007 | ** | Human < Demographics + All psychological measures + Heuristics and biases experiments |
| Q38 | 3 | 0.108 | 0.004 | ** | Human < Demographics + All psychological measures + Heuristics and biases experiments |
| Q43 | 3 | 0.121 | 0.004 | ** | Human < Demographics + All psychological measures + Heuristics and biases experiments |
| Q4 | 4 | 0.042 | 0.105 | | |
| Q9 | 4 | -0.106 | 0.004 | ** | Human > Demographics + All psychological measures + Heuristics and biases experiments |
| Q14 | 4 | 0.179 | 0.004 | ** | Human < Demographics + All psychological measures + Heuristics and biases experiments |
| Q19 | 4 | -0.069 | 0.004 | ** | Human > Demographics + All psychological measures + Heuristics and biases experiments |
| Q24 | 4 | 0.075 | 0.007 | ** | Human < Demographics + All psychological measures + Heuristics and biases experiments |
| Q29 | 4 | 0.082 | 0.004 | ** | Human < Demographics + All psychological measures + Heuristics and biases experiments |
| Q34 | 4 | 0.144 | 0.004 | ** | Human < Demographics + All psychological measures + Heuristics and biases experiments |
| Q39 | 4 | 0.088 | 0.004 | ** | Human < Demographics + All psychological measures + Heuristics and biases experiments |
| Q5 | 5 | -0.048 | 0.084 |  |  |
| Q10 | 5 | 0.102 | 0.004 | ** | Human < Demographics + All psychological measures + Heuristics and biases experiments |
| Q15 | 5 | 0.103 | 0.004 | ** | Human < Demographics + All psychological measures + Heuristics and biases experiments |
| Q20 | 5 | 0.113 | 0.004 | ** | Human < Demographics + All psychological measures + Heuristics and biases experiments |
| Q25 | 5 | 0.065 | 0.023 | * | Human < Demographics + All psychological measures + Heuristics and biases experiments |
| Q30 | 5 | -0.146 | 0.004 | ** | Human > Demographics + All psychological measures + Heuristics and biases experiments |
| Q35 | 5 | 0.265 | 0.004 | ** | Human < Demographics + All psychological measures + Heuristics and biases experiments |
| Q40 | 5 | 0.026 | 0.359 |  |  |
| Q41 | 5 | 0.059 | 0.016 | * | Human < Demographics + All psychological measures + Heuristics and biases experiments |
| Q44 | 5 | 0.042 | 0.042 | * | Human < Demographics + All psychological measures + Heuristics and biases experiments |
| Demographics + Heuristics and biases experiments + Features with r < 0.5 vs Human |  |  |  |  |  |
| Q1 | 1 | -0.092 | 0.004 | ** | Human < Demographics + Heuristics and biases experiments + Features with r < 0.5 |
| Q6 | 1 | -0.121 | 0.004 | ** | Human < Demographics + Heuristics and biases experiments + Features with r < 0.5 |
| Q11 | 1 | -0.239 | 0.004 | ** | Human < Demographics + Heuristics and biases experiments + Features with r < 0.5 |
| Q16 | 1 | 0.052 | 0.08 |  |  |
| Q21 | 1 | 0.081 | 0.009 | ** | Human > Demographics + Heuristics and biases experiments + Features with r < 0.5 |
| Q26 | 1 | -0.089 | 0.004 | ** | Human < Demographics + Heuristics and biases experiments + Features with r < 0.5 |
| Q31 | 1 | 0.039 | 0.143 |  |  |
| Q36 | 1 | 0.019 | 0.522 |  |  |
| Q2 | 2 | -0.018 | 0.566 |  |  |
| Q7 | 2 | -0.2 | 0.004 | ** | Human < Demographics + Heuristics and biases experiments + Features with r < 0.5 |
| Q12 | 2 | -0.119 | 0.004 | ** | Human < Demographics + Heuristics and biases experiments + Features with r < 0.5 |
| Q17 | 2 | 0.019 | 0.555 |  |  |
| Q22 | 2 | 0.02 | 0.504 | | |
| Q27 | 2 | -0.106 | 0.007 | ** | Human < Demographics + Heuristics and biases experiments + Features with r < 0.5 |
| Q32 | 2 | 0.07 | 0.072 | | |
| Q37 | 2 | 0.051 | 0.143 | | |
| Q42 | 2 | -0.285 | 0.004 | ** | Human < Demographics + Heuristics and biases experiments + Features with r < 0.5 |
| Q3 | 3 | -0.031 | 0.359 | | |
| Q8 | 3 | -0.145 | 0.004 | ** | Human < Demographics + Heuristics and biases experiments + Features with r < 0.5 |
| Q13 | 3 | -0.084 | 0.007 | ** | Human < Demographics + Heuristics and biases experiments + Features with r < 0.5 |
| Q18 | 3 | -0.031 | 0.232 | | |
| Q23 | 3 | 0.159 | 0.004 | ** | Human > Demographics + Heuristics and biases experiments + Features with r < 0.5 |
| Q28 | 3 | -0.043 | 0.169 | | |
| Q33 | 3 | -0.083 | 0.009 | ** | Human < Demographics + Heuristics and biases experiments + Features with r < 0.5 |
| Q38 | 3 | -0.125 | 0.004 | ** | Human < Demographics + Heuristics and biases experiments + Features with r < 0.5 |
| Q43 | 3 | -0.091 | 0.007 | ** | Human < Demographics + Heuristics and biases experiments + Features with r < 0.5 |
| Q4 | 4 | 0.1 | 0.004 | ** | Human > Demographics + Heuristics and biases experiments + Features with r < 0.5 |
| Q9 | 4 | 0.141 | 0.004 | ** | Human > Demographics + Heuristics and biases experiments + Features with r < 0.5 |
| Q14 | 4 | -0.212 | 0.004 | ** | Human < Demographics + Heuristics and biases experiments + Features with r < 0.5 |
| Q19 | 4 | 0.06 | 0.045 | * | Human > Demographics + Heuristics and biases experiments + Features with r < 0.5 |
| Q24 | 4 | -0.111 | 0.004 | ** | Human < Demographics + Heuristics and biases experiments + Features with r < 0.5 |
| Q29 | 4 | -0.072 | 0.015 | * | Human < Demographics + Heuristics and biases experiments + Features with r < 0.5 |
| Q34 | 4 | -0.153 | 0.004 | ** | Human < Demographics + Heuristics and biases experiments + Features with r < 0.5 |
| Q39 | 4 | -0.09 | 0.007 | ** | Human < Demographics + Heuristics and biases experiments + Features with r < 0.5 |
| Q5 | 5 | 0.061 | 0.053 | | |
| Q10 | 5 | -0.123 | 0.004 | ** | Human < Demographics + Heuristics and biases experiments + Features with r < 0.5 |
| Q15 | 5 | -0.103 | 0.004 | ** | Human < Demographics + Heuristics and biases experiments + Features with r < 0.5 |
| Q20 | 5 | -0.097 | 0.004 | ** | Human < Demographics + Heuristics and biases experiments + Features with r < 0.5 |
| Q25 | 5 | -0.056 | 0.052 | | |
| Q30 | 5 | 0.128 | 0.004 | ** | Human > Demographics + Heuristics and biases experiments + Features with r < 0.5 |
| Q35 | 5 | -0.124 | 0.004 | ** | Human < Demographics + Heuristics and biases experiments + Features with r < 0.5 |
| Q40 | 5 | -0.064 | 0.044 | * | Human < Demographics + Heuristics and biases experiments + Features with r < 0.5 |
| Q41 | 5 | -0.058 | 0.004 | ** | Human < Demographics + Heuristics and biases experiments + Features with r < 0.5 |
| Q44 | 5 | -0.078 | 0.004 | ** | Human < Demographics + Heuristics and biases experiments + Features with r < 0.5 |
| Demographics + Heuristics and biases experiments + Features with r > 0.5 vs Human | | | | | |
| Q1 | 1 | 0.092 | 0.004 | ** | Human < Demographics + Heuristics and biases experiments + Features with r > 0.5 |
| Q6 | 1 | 0.113 | 0.004 | ** | Human < Demographics + Heuristics and biases experiments + Features with r > 0.5 |
| Q11 | 1 | 0.146 | 0.004 | ** | Human < Demographics + Heuristics and biases experiments + Features with r > 0.5 |
| Q16 | 1 | 0.027 | 0.36 | | |
| Q21 | 1 | -0.094 | 0.004 | ** | Human > Demographics + Heuristics and biases experiments + Features with r > 0.5 |
| Q26 | 1 | 0.007 | 0.752 | | |
| Q31 | 1 | 0.04 | 0.109 | | |
| Q36 | 1 | -0.031 | 0.299 | | |
| Q3 | 2 | 0.052 | 0.068 | | |
| Q8 | 2 | 0.109 | 0.004 | ** | Human < Demographics + Heuristics and biases experiments + Features with r > 0.5 |
| Q13 | 2 | 0.097 | 0.004 | ** | Human < Demographics + Heuristics and biases experiments + Features with r > 0.5 |
| Q18 | 2 | -0.064 | 0.01 | * | Human > Demographics + Heuristics and biases experiments + Features with r > 0.5 |
| Q23 | 2 | -0.109 | 0.004 | ** | Human > Demographics + Heuristics and biases experiments + Features with r > 0.5 |
| Q28 | 2 | 0.054 | 0.032 | * | Human < Demographics + Heuristics and biases experiments + Features with r > 0.5 |
| Q33 | 2 | 0.057 | 0.029 | * | Human < Demographics + Heuristics and biases experiments + Features with r > 0.5 |
| Q38 | 2 | 0.107 | 0.004 | ** | Human < Demographics + Heuristics and biases experiments + Features with r > 0.5 |
| Q43 | 2 | 0.162 | 0.004 | ** | Human < Demographics + Heuristics and biases experiments + Features with r > 0.5 |
| Q4 | 3 | 0.025 | 0.299 | | |
| Q9 | 3 | -0.085 | 0.004 | ** | Human > Demographics + Heuristics and biases experiments + Features with r > 0.5 |
| Q14 | 3 | 0.142 | 0.004 | ** | Human < Demographics + Heuristics and biases experiments + Features with r > 0.5 |
| Q19 | 3 | -0.048 | 0.032 | * | Human > Demographics + Heuristics and biases experiments + Features with r > 0.5 |
| Q24 | 3 | 0.065 | 0.022 | * | Human < Demographics + Heuristics and biases experiments + Features with r > 0.5 |
| Q29 | 3 | 0.057 | 0.007 | ** | Human < Demographics + Heuristics and biases experiments + Features with r > 0.5 |
| Q34 | 3 | 0.151 | 0.004 | ** | Human < Demographics + Heuristics and biases experiments + Features with r > 0.5 |
| Q39 | 3 | 0.086 | 0.004 | ** | Human < Demographics + Heuristics and biases experiments + Features with r > 0.5 |
| Q5 | 4 | -0.02 | 0.506 | | |
| Q10 | 4 | 0.111 | 0.004 | ** | Human < Demographics + Heuristics and biases experiments + Features with r > 0.5 |
| Q15 | 4 | 0.111 | 0.004 | ** | Human < Demographics + Heuristics and biases experiments + Features with r > 0.5 |
| Q20 | 4 | 0.089 | 0.004 | ** | Human < Demographics + Heuristics and biases experiments + Features with r > 0.5 |
| Q25 | 4 | 0.057 | 0.036 | * | Human < Demographics + Heuristics and biases experiments + Features with r > 0.5 |
| Q30 | 4 | -0.143 | 0.004 | ** | Human > Demographics + Heuristics and biases experiments + Features with r > 0.5 |
| Q35 | 4 | 0.3 | 0.004 | ** | Human < Demographics + Heuristics and biases experiments + Features with r > 0.5 |
| Q40 | 4 | 0.01 | 0.724 | | |
| Q41 | 4 | 0.076 | 0.007 | ** | Human < Demographics + Heuristics and biases experiments + Features with r > 0.5 |
| Q44 | 4 | 0.046 | 0.027 | * | Human < Demographics + Heuristics and biases experiments + Features with r > 0.5 |
| Q7 | 5 | 0.3 | 0.004 | ** | Human < Demographics + Heuristics and biases experiments + Features with r > 0.5 |
| Q17 | 5 | -0.17 | 0.004 | ** | Human > Demographics + Heuristics and biases experiments + Features with r > 0.5 |
| Q22 | 5 | -0.115 | 0.004 | ** | Human > Demographics + Heuristics and biases experiments + Features with r > 0.5 |
| Q32 | 5 | 0.21 | 0.004 | ** | Human < Demographics + Heuristics and biases experiments + Features with r > 0.5 |
| Q42 | 5 | 0.01 | 0.724 | | |
| Demographics + Heuristics and biases experiments + Features with r < 0.3 vs Human | | | | | |
| Q1 | 1 | -0.093 | 0.004 | ** | Human < Demographics + Heuristics and biases experiments + Features with r < 0.3 |
| Q6 | 1 | -0.055 | 0.104 | | |
| Q21 | 1 | 0.157 | 0.004 | ** | Human > Demographics + Heuristics and biases experiments + Features with r < 0.3 |
| Q31 | 1 | -0.021 | 0.416 | | |
| Q36 | 1 | -0.13 | 0.004 | ** | Human < Demographics + Heuristics and biases experiments + Features with r < 0.3 |
| Q2 | 2 | 0.053 | 0.147 | | |
| Q7 | 2 | -0.158 | 0.004 | ** | Human < Demographics + Heuristics and biases experiments + Features with r < 0.3 |
| Q12 | 2 | -0.132 | 0.004 | ** | Human < Demographics + Heuristics and biases experiments + Features with r < 0.3 |
| Q17 | 2 | -0.042 | 0.281 | | |
| Q22 | 2 | 0.099 | 0.011 | * | Human > Demographics + Heuristics and biases experiments + Features with r < 0.3 |
| Q27 | 2 | -0.158 | 0.004 | ** | Human < Demographics + Heuristics and biases experiments + Features with r < 0.3 |
| Q32 | 2 | 0.096 | 0.03 | * | Human > Demographics + Heuristics and biases experiments + Features with r < 0.3 |
| Q37 | 2 | 0.042 | 0.301 | | |
| Q42 | 2 | -0.272 | 0.004 | ** | Human < Demographics + Heuristics and biases experiments + Features with r < 0.3 |
| Q3 | 3 | -0.054 | 0.133 | | |
| Q8 | 3 | -0.119 | 0.004 | ** | Human < Demographics + Heuristics and biases experiments + Features with r < 0.3 |
| Q13 | 3 | -0.096 | 0.004 | ** | Human < Demographics + Heuristics and biases experiments + Features with r < 0.3 |
| Q18 | 3 | -0.025 | 0.426 | | |
| Q23 | 3 | 0.141 | 0.004 | ** | Human > Demographics + Heuristics and biases experiments + Features with r < 0.3 |
| Q28 | 3 | -0.055 | 0.105 | | |
| Q33 | 3 | -0.029 | 0.366 | | |
| Q38 | 3 | -0.109 | 0.004 | ** | Human < Demographics + Heuristics and biases experiments + Features with r < 0.3 |
| Q43 | 3 | -0.054 | 0.08 | | |
| Q4 | 4 | 0.113 | 0.004 | ** | Human > Demographics + Heuristics and biases experiments + Features with r < 0.3 |
| Q9 | 4 | 0.096 | 0.004 | ** | Human > Demographics + Heuristics and biases experiments + Features with r < 0.3 |
| Q14 | 4 | -0.198 | 0.004 | ** | Human < Demographics + Heuristics and biases experiments + Features with r < 0.3 |
| Q19 | 4 | 0.044 | 0.15 | | |
| Q24 | 4 | -0.06 | 0.105 | | |
| Q29 | 4 | 0.02 | 0.526 | | |
| Q34 | 4 | -0.152 | 0.004 | ** | Human < Demographics + Heuristics and biases experiments + Features with r < 0.3 |
| Q39 | 4 | -0.121 | 0.004 | ** | Human < Demographics + Heuristics and biases experiments + Features with r < 0.3 |
| Q5 | 5 | 0.01 | 0.734 | | |
| Q10 | 5 | -0.074 | 0.017 | * | Human < Demographics + Heuristics and biases experiments + Features with r < 0.3 |
| Q15 | 5 | -0.071 | 0.008 | ** | Human < Demographics + Heuristics and biases experiments + Features with r < 0.3 |
| Q20 | 5 | -0.13 | 0.004 | ** | Human < Demographics + Heuristics and biases experiments + Features with r < 0.3 |
| Q25 | 5 | -0.073 | 0.011 | * | Human < Demographics + Heuristics and biases experiments + Features with r < 0.3 |
| Q30 | 5 | 0.112 | 0.004 | ** | Human > Demographics + Heuristics and biases experiments + Features with r < 0.3 |
| Q40 | 5 | 0.017 | 0.578 | | |
| Q41 | 5 | -0.049 | 0.064 | | |
| Q44 | 5 | -0.083 | 0.008 | ** | Human < Demographics + Heuristics and biases experiments + Features with r < 0.3 |

*Note.* The Big Five personality items were adopted from the Big Five Inventory[29]. P-values are adjusted for multiple comparisons using the Benjamini-Hochberg (BH) method. *** p < .001, ** p < .01, * p < .05.

## Table S33 Metric invariance results for BFI-44 item loadings comparing digital twins with human participants in Study 3a

| Input conditions | Items retained | Number of invariant items | Number of noninvariant items | Digital twin > Human | Digital twin < Human | Agreeableness | Conscientiousness | Extraversion | Neuroticism | Openness |
|---|---|---|---|---|---|---|---|---|---|---|
| **Gemma-4-31B-it** | | | | | | | | | | |
| Demographics | 44 | 17 (38.64) | 27 (61.36) | 18 (40.91) | 9 (20.45) | 7, 12R, 17, 22, 42 | 8R, 18R, 28 | 11, 16, 21R, 26, 31R, 36 | 4, 9R, 19, 29, 34R, 39 | 10, 15, 25, 30, 35R, 41R, 44 |
| Heuristics and biases experiments | 44 | 20 (45.45) | 24 (54.55) | 14 (31.82) | 10 (22.73) | 2R, 27R, 32, 42 | 18R, 38, 43R | 6R, 21R, 26, 31R, 36 | 4, 9R, 14, 19, 29, 34R | 5, 20, 25, 30, 35R, 41R |
| Demographics + Heuristics and biases experiments | 44 | 17 (38.64) | 27 (61.36) | 22 (50.00) | 5 (11.36) | 2R, 7, 17, 22, 37R, 42 | 3, 8R, 38, 43R | 1, 6R, 16, 21R, 26 | 4, 9R, 14, 19, 29, 34R, 39 | 15, 25, 30, 35R, 41R |
| Demographics + All psychological measures + Heuristics and biases experiments | 44 | 9 (20.45) | 35 (79.55) | 25 (56.82) | 10 (22.73) | 2R, 7, 12R, 17, 22, 27R, 37R | 13, 18R, 28, 33, 38, 43R | 1, 6R, 16, 21R, 26 | 9R, 14, 19, 24R, 29, 34R, 39 | 5, 10, 15, 20, 25, 30, 35R, 40, 41R, 44 |
| Demographics + Heuristics and biases experiments + Features with r < 0.5 | 44 | 7 (15.91) | 37 (84.09) | 27 (61.36) | 10 (22.73) | 2R, 7, 12R, 17, 22, 27R, 32, 42 | 13, 23R, 28, 33, 38, 43R | 1, 6R, 11, 16, 21R, 26, 31R | 4, 9R, 14, 19, 24R, 29, 34R, 39 | 15, 20, 25, 30, 35R, 40, 41R, 44 |
| Demographics + Heuristics and biases experiments + Features with r > 0.5 | 44 | 8 (18.18) | 36 (81.82) | 24 (54.55) | 12 (27.27) | 2R, 7, 12R, 17, 22, 27R, 32, 37R, 42 | 3, 8R, 13, 18R, 23R, 28, 33, 38, 43R | 6R, 16, 21R | 9R, 14, 19, 24R, 29, 34R, 39 | 5, 10, 15, 20, 30, 35R, 40, 41R |
| Demographics + Heuristics and biases experiments + Features with r < 0.3 | 44 | 17 (38.64) | 27 (61.36) | 20 (45.45) | 7 (15.91) | 7, 12R, 42 | 8R, 18R, 33, 38, 43R | 1, 6R, 16, 21R, 26, 31R | 4, 9R, 14, 19, 34R, 39 | 15, 25, 30, 35R, 40, 41R, 44 |
| **gpt-oss-120b** | | | | | | | | | | |
| Demographics | 44 | 17 (38.64) | 27 (61.36) | 21 (47.73) | 6 (13.64) | 2R, 7, 12R, 22, 27R, 42 | 3, 8R, 13, 23R, 38, 43R | 6R, 11, 21R, 26, 31R | 9R, 14, 24R, 29, 34R, 39 | 10, 15, 20, 35R |
| Heuristics and biases experiments | 44 | 20 (45.45) | 24 (54.55) | 15 (34.09) | 9 (20.45) | 7, 12R, 22, 27R, 42 | 3, 13, 18R, 23R, 33, 38, 43R | 6R, 11, 21R, 26, 31R | 4, 9R, 39 | 10, 15, 20, 30 |
| Demographics + Heuristics and biases experiments | 44 | 17 (38.64) | 27 (61.36) | 17 (38.64) | 10 (22.73) | 2R, 7, 17, 22, 27R, 42 | 3, 18R, 23R, 33, 38 | 6R, 11, 26, 31R | 4, 9R, 19, 29, 34R, 39 | 5, 15, 20, 25, 30, 35R |
| Demographics + All psychological measures + Heuristics and biases experiments | 44 | 15 (34.09) | 29 (65.91) | 20 (45.45) | 9 (20.45) | 2R, 7, 12R, 17, 22, 27R, 32 | 8R, 13, 23R, 38, 43R | 1, 6R, 11, 16, 31R | 9R, 19, 24R, 29, 34R, 39 | 15, 20, 30, 35R, 40, 41R |
| Demographics + Heuristics and biases experiments + Features with r < 0.5 | 41 | 9 (20.45) | 32 (72.73) | 22 (50.00) | 10 (22.73) | 7, 12R, 17, 22, 27R, 32, 37R, 42 | 3, 8R, 13, 18R, 23R, 28, 33, 38, 43R | 6R, 11, 31R | 9R, 14, 24R, 34R, 39 | 5, 15, 20, 25, 30, 40, 41R |
| Demographics + Heuristics and biases experiments + Features with r > 0.5 | 44 | 12 (27.27) | 32 (72.73) | 22 (50.00) | 10 (22.73) | 2R, 7, 17, 22, 27R, 32, 37R | 13, 23R, 38, 43R | 6R, 11, 16, 26, 31R | 4, 9R, 14, 19, 24R, 29, 34R, 39 | 10, 15, 20, 30, 35R, 40, 41R, 44 |
| Demographics + Heuristics and biases experiments + Features with r < 0.3 | 42 | 19 (43.18) | 23 (52.27) | 16 (36.36) | 7 (15.91) | 2R, 7, 17, 22, 27R, 42 | 18R, 23R, 33, 38 | 6R, 11, 26, 31R | 9R, 14, 34R, 39 | 5, 15, 20, 30, 44 |
| **DeepSeek V4-Flash** | | | | | | | | | | |
| Demographics | 43 | 16 (36.36) | 27 (61.36) | 18 (40.91) | 9 (20.45) | 2R, 7, 12R, 17, 22, 42 | 3, 13, 23R, 28, 33, 38 | 11, 31R | 9R, 14, 19, 24R, 29, 34R, 39 | 5, 15, 20, 25, 40, 41R |
| Heuristics and biases experiments | 43 | 17 (38.64) | 26 (59.09) | 19 (43.18) | 7 (15.91) | 2R, 7, 17, 42 | 13, 23R, 28, 33, 38, 43R | 6R, 11, 26 | 4, 9R, 14, 19, 24R, 29, 34R, 39 | 5, 15, 25, 40, 41R |
| Demographics + Heuristics and biases experiments | 44 | 17 (38.64) | 27 (61.36) | 20 (45.45) | 7 (15.91) | 2R, 7, 12R, 17, 22, 27R, 42 | 13, 28, 33, 38 | 11, 36 | 4, 9R, 19, 24R, 29, 34R, 39 | 5, 15, 20, 25, 35R, 40, 41R |
| Demographics + All psychological measures + Heuristics and biases experiments | 43 | 16 (36.36) | 27 (61.36) | 21 (47.73) | 6 (13.64) | 2R, 7, 17, 27R, 32, 37R, 42 | 13, 28, 33, 38 | 1, 11, 36 | 9R, 14, 19, 24R, 29, 34R, 39 | 10, 15, 20, 25, 40, 44 |
| Demographics + Heuristics and biases experiments + Features with r < 0.5 | 44 | 14 (31.82) | 30 (68.18) | 21 (47.73) | 9 (20.45) | 2R, 7, 12R, 22, 27R, 42 | 3, 13, 18R, 23R, 28, 33, 38 | 6R, 11 | 4, 9R, 14, 19, 24R, 29, 34R, 39 | 5, 15, 20, 25, 35R, 40, 44 |
| Demographics + Heuristics and biases experiments + Features with r > 0.5 | 44 | 16 (36.36) | 28 (63.64) | 22 (50.00) | 6 (13.64) | 2R, 7, 17, 27R, 32, 37R, 42 | 13, 23R, 28, 38, 43R | 11, 16 | 4, 9R, 14, 19, 29, 34R, 39 | 10, 15, 20, 25, 30, 35R, 40 |
| Demographics + Heuristics and biases experiments + Features with r < 0.3 | 43 | 13 (29.55) | 30 (68.18) | 21 (47.73) | 9 (20.45) | 2R, 7, 12R, 17, 22, 42 | 3, 13, 28, 33, 38, 43R | 11, 16, 21R, 31R | 4, 9R, 14, 19, 24R, 29, 34R, 39 | 5, 15, 20, 25, 40, 41R |

| | | | | | | | | | |
|---|---|---|---|---|---|---|---|---|---|
| **Qwen 3.5 Plus** | | | | | | | | | |
| Demographics | 35 | 15 (34.09) | 20 (45.45) | 14 (31.82) | 6 (13.64) | 7, 22, 42 | 38 | 1, 11, 21R, 26 | 9R, 14, 19, 24R, 34R, 39 | 10, 15, 20, 30, 41R, 44 |
| Demographics + Heuristics and biases experiments | 37 | 14 (31.82) | 23 (52.27) | 16 (36.36) | 7 (15.91) | 2R, 7, 42 | 8R, 13, 23R, 33, 38, 43R | 1, 21R, 36 | 4, 9R, 14, 29, 34R, 39 | 15, 20, 25, 30, 41R |
| Demographics + All psychological measures + Heuristics and biases experiments | 41 | 9 (20.45) | 32 (72.73) | 26 (59.09) | 6 (13.64) | 7, 12R, 17, 27R, 37R, 42 | 8R, 13, 23R, 28, 33, 38, 43R | 1, 6R, 21R, 36 | 9R, 14, 19, 24R, 29, 34R, 39 | 10, 15, 20, 25, 30, 35R, 41R, 44 |
| Demographics + Heuristics and biases experiments + Features with r < 0.5 | 44 | 13 (29.55) | 31 (70.45) | 25 (56.82) | 6 (13.64) | 7, 12R, 27R, 42 | 8R, 13, 23R, 33, 38, 43R | 1, 6R, 11, 21R, 26 | 4, 9R, 14, 19, 24R, 29, 34R, 39 | 10, 15, 20, 30, 35R, 40, 41R, 44 |
| Demographics + Heuristics and biases experiments + Features with r > 0.5 | 40 | 9 (20.45) | 31 (70.45) | 23 (52.27) | 8 (18.18) | 7, 17, 22, 32 | 8R, 13, 18R, 23R, 28, 33, 38, 43R | 1, 6R, 11, 21R | 9R, 14, 19, 24R, 29, 34R, 39 | 10, 15, 20, 25, 30, 35R, 41R, 44 |
| Demographics + Heuristics and biases experiments + Features with r < 0.3 | 40 | 16 (36.36) | 24 (54.55) | 17 (38.64) | 7 (15.91) | 7, 12R, 22, 27R, 32, 42 | 8R, 13, 23R, 38 | 1, 21R, 36 | 4, 9R, 14, 34R, 39 | 10, 15, 20, 25, 30, 44 |

*Note.* Items retained indicates the number of Big Five items retained after the configural-invariance item-removal procedure. Values are counts, with percentages in parentheses calculated relative to the original 44 BFI items. Invariant items are items that did not show evidence of metric noninvariance between the human and digital-twin networks. Noninvariant items are further classified by the direction of the loading difference: Digital twin > Human indicates higher item loadings in the digital-twin responses than in the human responses, whereas Digital twin < Human indicates lower item loadings in the digital-twin responses. Item-level loading differences are reported in Table S33.

**Table S34 Metric invariance results for BFI-44 item loadings among digital twins across input conditions in Study 3a**

| Model | Input condition A | Input condition B | Number of noninvariant items, n | A < B, n | A > B, n |
|---|---|---|---|---|---|
| Gemma-4-31B-it | Demographics + All psychological measures + Heuristics and biases experiments | Demographics + Heuristics and biases experiments | 23 | 10 | 13 |
| | Demographics + All psychological measures + Heuristics and biases experiments | Demographics | 27 | 11 | 16 |
| | Demographics + All psychological measures + Heuristics and biases experiments | Demographics + Heuristics and biases experiments + Features with r > 0.5 | 6 | 4 | 2 |
| | Demographics + All psychological measures + Heuristics and biases experiments | Demographics + Heuristics and biases experiments + Features with r < 0.3 | 24 | 12 | 12 |
| | Demographics + All psychological measures + Heuristics and biases experiments | Demographics + Heuristics and biases experiments + Features with r < 0.5 | 16 | 8 | 8 |
| | Demographics + All psychological measures + Heuristics and biases experiments | Heuristics and biases experiments | 28 | 9 | 19 |
| | Demographics + Heuristics and biases experiments | Demographics + Heuristics and biases experiments + Features with r > 0.5 | 28 | 17 | 11 |
| | Demographics + Heuristics and biases experiments | Demographics + Heuristics and biases experiments + Features with r < 0.3 | 19 | 10 | 9 |
| | Demographics + Heuristics and biases experiments | Demographics + Heuristics and biases experiments + Features with r < 0.5 | 26 | 16 | 10 |
| | Demographics + Heuristics and biases experiments | Heuristics and biases experiments | 17 | 4 | 13 |
| | Demographics | Demographics + Heuristics and biases experiments | 12 | 8 | 4 |
| | Demographics | Demographics + Heuristics and biases experiments + Features with r > 0.5 | 29 | 17 | 12 |
| | Demographics | Demographics + Heuristics and biases experiments + Features with r < 0.3 | 24 | 15 | 9 |
| | Demographics | Demographics + Heuristics and biases experiments + Features with r < 0.5 | 22 | 15 | 7 |
| | Demographics | Heuristics and biases experiments | 30 | 11 | 19 |
| | Demographics + Heuristics and biases experiments + Features with r > 0.5 | Demographics + Heuristics and biases experiments + Features with r < 0.3 | 29 | 14 | 15 |
| | Demographics + Heuristics and biases experiments + Features with r > 0.5 | Demographics + Heuristics and biases experiments + Features with r < 0.5 | 27 | 14 | 13 |
| | Demographics + Heuristics and biases experiments + Features with r > 0.5 | Heuristics and biases experiments | 28 | 8 | 20 |
| | Demographics + Heuristics and biases experiments + Features with r < 0.3 | Demographics + Heuristics and biases experiments + Features with r < 0.5 | 16 | 9 | 7 |
| | Demographics + Heuristics and biases experiments + Features with r < 0.3 | Heuristics and biases experiments | 24 | 4 | 20 |
| | Demographics + Heuristics and biases experiments + Features with r < 0.5 | Heuristics and biases experiments | 24 | 4 | 20 |
| gpt-oss-120b | Demographics + All psychological measures + Heuristics and biases experiments | Demographics + Heuristics and biases experiments | 21 | 9 | 12 |
| | Demographics + All psychological measures + Heuristics and biases experiments | Demographics | 24 | 9 | 15 |
| | Demographics + All psychological measures + Heuristics and biases experiments | Demographics + Heuristics and biases experiments + Features with r > 0.5 | 6 | 4 | 2 |
| | Demographics + All psychological measures + Heuristics and biases experiments | Demographics + Heuristics and biases experiments + Features with r < 0.3 | 20 | 9 | 11 |
| | Demographics + All psychological measures + Heuristics and biases experiments | Demographics + Heuristics and biases experiments + Features with r < 0.5 | 17 | 7 | 10 |
| | Demographics + All psychological measures + Heuristics and biases experiments | Heuristics and biases experiments | 26 | 11 | 15 |
| | Demographics + Heuristics and biases experiments | Demographics + Heuristics and biases experiments + Features with r > 0.5 | 18 | 10 | 8 |
| | Demographics + Heuristics and biases experiments | Demographics + Heuristics and biases experiments + Features with r < 0.3 | 15 | 6 | 9 |
| | Demographics + Heuristics and biases experiments | Demographics + Heuristics and biases experiments + Features with r < 0.5 | 18 | 8 | 10 |
| | Demographics + Heuristics and biases experiments | Heuristics and biases experiments | 14 | 6 | 8 |
| | Demographics | Demographics + Heuristics and biases experiments | 21 | 10 | 11 |
| | Demographics | Demographics + Heuristics and biases experiments + Features with r > 0.5 | 24 | 13 | 11 |
| | Demographics | Demographics + Heuristics and biases experiments + Features with r < 0.3 | 23 | 9 | 14 |
| | Demographics | Demographics + Heuristics and biases experiments + Features with r < 0.5 | 26 | 13 | 13 |
| | Demographics | Heuristics and biases experiments | 15 | 8 | 7 |
| | Demographics + Heuristics and biases experiments + Features with r > 0.5 | Demographics + Heuristics and biases experiments + Features with r < 0.3 | 26 | 11 | 15 |
| | Demographics + Heuristics and biases experiments + Features with r > 0.5 | Demographics + Heuristics and biases experiments + Features with r < 0.5 | 19 | 10 | 9 |
| | Demographics + Heuristics and biases experiments + Features with r > 0.5 | Heuristics and biases experiments | 20 | 9 | 11 |
| | Demographics + Heuristics and biases experiments + Features with r < 0.3 | Demographics + Heuristics and biases experiments + Features with r < 0.5 | 10 | 4 | 6 |
| | Demographics + Heuristics and biases experiments + Features with r < 0.3 | Heuristics and biases experiments | 14 | 6 | 8 |
| | Demographics + Heuristics and biases experiments + Features with r < 0.5 | Heuristics and biases experiments | 13 | 6 | 7 |
| DeepSeek V4-Flash | Demographics + All psychological measures + Heuristics and biases experiments | Demographics + Heuristics and biases experiments | 5 | 2 | 3 |
| | Demographics + All psychological measures + Heuristics and biases experiments | Demographics | 11 | 5 | 6 |
| | Demographics + All psychological measures + Heuristics and biases experiments | Demographics + Heuristics and biases experiments + Features with $r > 0.5$ | 0 | 0 | 0 |
| | Demographics + All psychological measures + Heuristics and biases experiments | Demographics + Heuristics and biases experiments + Features with $r < 0.3$ | 6 | 4 | 2 |
| | Demographics + All psychological measures + Heuristics and biases experiments | Demographics + Heuristics and biases experiments + Features with $r < 0.5$ | 7 | 3 | 4 |
| | Demographics + All psychological measures + Heuristics and biases experiments | Heuristics and biases experiments | 6 | 4 | 2 |
| | Demographics + Heuristics and biases experiments | Demographics + Heuristics and biases experiments + Features with $r > 0.5$ | 13 | 6 | 7 |
| | Demographics + Heuristics and biases experiments | Demographics + Heuristics and biases experiments + Features with $r < 0.3$ | 2 | 2 | 0 |
| | Demographics + Heuristics and biases experiments | Demographics + Heuristics and biases experiments + Features with $r < 0.5$ | 2 | 0 | 2 |
| | Demographics + Heuristics and biases experiments | Heuristics and biases experiments | 4 | 2 | 2 |
| | Demographics | Demographics + Heuristics and biases experiments | 0 | 0 | 0 |
| | Demographics | Demographics + Heuristics and biases experiments + Features with $r > 0.5$ | 14 | 7 | 7 |
| | Demographics | Demographics + Heuristics and biases experiments + Features with $r < 0.3$ | 7 | 2 | 5 |
| | Demographics | Demographics + Heuristics and biases experiments + Features with $r < 0.5$ | 9 | 4 | 5 |
| | Demographics | Heuristics and biases experiments | 9 | 4 | 5 |
| | Demographics + Heuristics and biases experiments + Features with $r > 0.5$ | Demographics + Heuristics and biases experiments + Features with $r < 0.3$ | 12 | 5 | 7 |
| | Demographics + Heuristics and biases experiments + Features with $r > 0.5$ | Demographics + Heuristics and biases experiments + Features with $r < 0.5$ | 11 | 4 | 7 |
| | Demographics + Heuristics and biases experiments + Features with $r > 0.5$ | Heuristics and biases experiments | 12 | 7 | 5 |
| | Demographics + Heuristics and biases experiments + Features with $r < 0.3$ | Demographics + Heuristics and biases experiments + Features with $r < 0.5$ | 0 | 0 | 0 |
| | Demographics + Heuristics and biases experiments + Features with $r < 0.3$ | Heuristics and biases experiments | 4 | 2 | 2 |
| | Demographics + Heuristics and biases experiments + Features with $r < 0.5$ | Heuristics and biases experiments | 4 | 2 | 2 |
| Qwen 3.5 Plus | Demographics + All psychological measures + Heuristics and biases experiments | Demographics + Heuristics and biases experiments | 19 | 8 | 11 |
| | Demographics + All psychological measures + Heuristics and biases experiments | Demographics | 21 | 7 | 14 |
| | Demographics + All psychological measures + Heuristics and biases experiments | Demographics + Heuristics and biases experiments + Features with $r > 0.5$ | 2 | 1 | 1 |
| | Demographics + All psychological measures + Heuristics and biases experiments | Demographics + Heuristics and biases experiments + Features with $r < 0.3$ | 17 | 6 | 11 |

| | | | | |
|---|---|---|---|---|
| Demographics + All psychological measures + Heuristics and biases experiments | Demographics + Heuristics and biases experiments + Features with r < 0.5 | 14 | 8 | 6 |
| Demographics + All psychological measures + Heuristics and biases experiments | Heuristics and biases experiments | 23 | 7 | 16 |
| Demographics + Heuristics and biases experiments | Demographics + Heuristics and biases experiments + Features with r > 0.5 | 22 | 12 | 10 |
| Demographics + Heuristics and biases experiments | Demographics + Heuristics and biases experiments + Features with r < 0.3 | 14 | 8 | 6 |
| Demographics + Heuristics and biases experiments | Demographics + Heuristics and biases experiments + Features with r < 0.5 | 20 | 13 | 7 |
| Demographics + Heuristics and biases experiments | Heuristics and biases experiments | 18 | 3 | 15 |
| Demographics | Demographics + Heuristics and biases experiments | 9 | 5 | 4 |
| Demographics | Demographics + Heuristics and biases experiments + Features with r > 0.5 | 21 | 14 | 7 |
| Demographics | Demographics + Heuristics and biases experiments + Features with r < 0.3 | 12 | 6 | 6 |
| Demographics | Demographics + Heuristics and biases experiments + Features with r < 0.5 | 26 | 15 | 11 |
| Demographics | Heuristics and biases experiments | 24 | 7 | 17 |
| Demographics + Heuristics and biases experiments + Features with r > 0.5 | Demographics + Heuristics and biases experiments + Features with r < 0.3 | 19 | 6 | 13 |
| Demographics + Heuristics and biases experiments + Features with r > 0.5 | Demographics + Heuristics and biases experiments + Features with r < 0.5 | 21 | 11 | 10 |
| Demographics + Heuristics and biases experiments + Features with r > 0.5 | Heuristics and biases experiments | 29 | 9 | 20 |
| Demographics + Heuristics and biases experiments + Features with r < 0.3 | Demographics + Heuristics and biases experiments + Features with r < 0.5 | 11 | 8 | 3 |
| Demographics + Heuristics and biases experiments + Features with r < 0.3 | Heuristics and biases experiments | 28 | 8 | 20 |
| Demographics + Heuristics and biases experiments + Features with r < 0.5 | Heuristics and biases experiments | 23 | 6 | 17 |

*Note.* Entries report the number of BFI-44 items showing significant loading differences between pairs of digital-twin input conditions within each model. “A < B” indicates that the estimated item loading was significantly smaller under input condition A than under input condition B, whereas “A > B” indicates that the estimated item loading was significantly larger under input condition A than under input condition B. Noninvariant items were identified using Benjamini–Hochberg adjusted p values < .05.

**Table S35 Metric invariance results for BFI-44 item loadings among digital-twin models within the same input condition in Study 3a**

| Condition | Model 1 | Model 2 | Number of noninvariant items | Model 1 < Model 2, n | Model 1 > Model 2, n |
|---|---|---|---|---|---|
| Demographics | DeepSeek V4-Flash | Gemma-4-31B-it | 22 | 10 | 12 |
| Demographics | DeepSeek V4-Flash | gpt-oss-120b | 17 | 9 | 8 |
| Demographics | gpt-oss-120b | Gemma-4-31B-it | 18 | 6 | 12 |
| Demographics | Qwen 3.5 Plus | DeepSeek V4-Flash | 18 | 9 | 9 |
| Demographics | Qwen 3.5 Plus | Gemma-4-31B-it | 20 | 6 | 14 |
| Demographics | Qwen 3.5 Plus | gpt-oss-120b | 16 | 9 | 7 |
| Heuristics and biases experiments | DeepSeek V4-Flash | Gemma-4-31B-it | 21 | 4 | 17 |
| Heuristics and biases experiments | DeepSeek V4-Flash | gpt-oss-120b | 15 | 5 | 10 |
| Heuristics and biases experiments | gpt-oss-120b | Gemma-4-31B-it | 23 | 8 | 15 |
| Heuristics and biases experiments | Qwen 3.5 Plus | DeepSeek V4-Flash | 24 | 16 | 8 |
| Heuristics and biases experiments | Qwen 3.5 Plus | Gemma-4-31B-it | 24 | 12 | 12 |
| Heuristics and biases experiments | Qwen 3.5 Plus | gpt-oss-120b | 23 | 15 | 8 |
| Demographics + Heuristics and biases experiments | DeepSeek V4-Flash | Gemma-4-31B-it | 22 | 10 | 12 |
| Demographics + Heuristics and biases experiments | DeepSeek V4-Flash | gpt-oss-120b | 15 | 7 | 8 |
| Demographics + Heuristics and biases experiments | gpt-oss-120b | Gemma-4-31B-it | 17 | 9 | 8 |
| Demographics + Heuristics and biases experiments | Qwen 3.5 Plus | DeepSeek V4-Flash | 16 | 7 | 9 |
| Demographics + Heuristics and biases experiments | Qwen 3.5 Plus | Gemma-4-31B-it | 11 | 8 | 3 |
| Demographics + Heuristics and biases experiments | Qwen 3.5 Plus | gpt-oss-120b | 13 | 6 | 7 |
| Demographics + All psychological measures + Heuristics and biases experiments | DeepSeek V4-Flash | Gemma-4-31B-it | 25 | 13 | 12 |
| Demographics + All psychological measures + Heuristics and biases experiments | DeepSeek V4-Flash | gpt-oss-120b | 28 | 11 | 17 |
| Demographics + All psychological measures + Heuristics and biases experiments | gpt-oss-120b | Gemma-4-31B-it | 18 | 9 | 9 |
| Demographics + All psychological measures + Heuristics and biases experiments | Qwen 3.5 Plus | DeepSeek V4-Flash | 25 | 10 | 15 |

| Demographics + All psychological measures + Heuristics and biases experiments | Qwen 3.5 Plus | Gemma-4-31B-it | 19 | 7 | 12 |
|---|---|---|---|---|---|
| Demographics + All psychological measures + Heuristics and biases experiments | Qwen 3.5 Plus | gpt-oss-120b | 16 | 7 | 9 |
| Demographics + Heuristics and biases experiments + Features with r < 0.5 | DeepSeek V4-Flash | Gemma-4-31B-it | 15 | 10 | 5 |
| Demographics + Heuristics and biases experiments + Features with r < 0.5 | DeepSeek V4-Flash | gpt-oss-120b | 15 | 10 | 5 |
| Demographics + Heuristics and biases experiments + Features with r < 0.5 | gpt-oss-120b | Gemma-4-31B-it | 21 | 10 | 11 |
| Demographics + Heuristics and biases experiments + Features with r < 0.5 | Qwen 3.5 Plus | DeepSeek V4-Flash | 23 | 9 | 14 |
| Demographics + Heuristics and biases experiments + Features with r < 0.5 | Qwen 3.5 Plus | Gemma-4-31B-it | 12 | 5 | 7 |
| Demographics + Heuristics and biases experiments + Features with r < 0.5 | Qwen 3.5 Plus | gpt-oss-120b | 18 | 7 | 11 |
| Demographics + Heuristics and biases experiments + Features with r > 0.5 | DeepSeek V4-Flash | Gemma-4-31B-it | 20 | 10 | 10 |
| Demographics + Heuristics and biases experiments + Features with r > 0.5 | DeepSeek V4-Flash | gpt-oss-120b | 20 | 10 | 10 |
| Demographics + Heuristics and biases experiments + Features with r > 0.5 | gpt-oss-120b | Gemma-4-31B-it | 20 | 9 | 11 |
| Demographics + Heuristics and biases experiments + Features with r > 0.5 | Qwen 3.5 Plus | DeepSeek V4-Flash | 23 | 11 | 12 |
| Demographics + Heuristics and biases experiments + Features with r > 0.5 | Qwen 3.5 Plus | Gemma-4-31B-it | 17 | 8 | 9 |
| Demographics + Heuristics and biases experiments + Features with r > 0.5 | Qwen 3.5 Plus | gpt-oss-120b | 23 | 12 | 11 |
| Demographics + Heuristics and biases experiments + Features with r < 0.3 | DeepSeek V4-Flash | Gemma-4-31B-it | 19 | 11 | 8 |
| Demographics + Heuristics and biases experiments + Features with r < 0.3 | DeepSeek V4-Flash | gpt-oss-120b | 16 | 8 | 8 |
| Demographics + Heuristics and biases experiments + Features with r < 0.3 | gpt-oss-120b | Gemma-4-31B-it | 18 | 8 | 10 |
| Demographics + Heuristics and biases experiments + Features with r < 0.3 | Qwen 3.5 Plus | DeepSeek V4-Flash | 19 | 9 | 10 |
| Demographics + Heuristics and biases experiments + Features with r < 0.3 | Qwen 3.5 Plus | Gemma-4-31B-it | 14 | 8 | 6 |
| Demographics + Heuristics and biases experiments + Features with r < 0.3 | Qwen 3.5 Plus | gpt-oss-120b | 11 | 3 | 8 |

*Note.* Entries report the number of BFI-44 items showing significant loading differences between pairs of digital-twin models within each input condition. Noninvariant items were identified using Benjamini–Hochberg adjusted p values < .05.

**Table S36 Percentile-aligned human–digital-twin correlations across models, input conditions, and calibration proportions**

| Input condition | Calibration proportion | Pearson | Spearman | Δr (Pearson) | Δr (Spearman) | Input condition | Calibration proportion | Pearson | Spearman | Δr (Pearson) | Δr (Spearman) |
|---|---|---|---|---|---|---|---|---|---|---|---|
| DeepSeek V4-Flash | | | | | | Qwen 3.5 Plus | | | | | |
| Demographics | 10% | 0.390 | 0.386 | -0.039 | -0.042 | Demographics | 10% | 0.422 | 0.411 | -0.049 | -0.054 |
| + All | 20% | 0.391 | 0.386 | -0.039 | -0.042 | + All | 20% | 0.422 | 0.411 | -0.048 | -0.054 |
| psychological | 30% | 0.391 | 0.386 | -0.038 | -0.042 | psychological | 30% | 0.422 | 0.411 | -0.048 | -0.054 |
| measures + | 40% | 0.391 | 0.386 | -0.038 | -0.042 | measures + | 40% | 0.422 | 0.412 | -0.048 | -0.054 |
| Heuristics and | 50% | 0.391 | 0.386 | -0.038 | -0.042 | Heuristics and | 50% | 0.422 | 0.411 | -0.048 | -0.054 |
| biases | 60% | 0.391 | 0.387 | -0.038 | -0.042 | biases | 60% | 0.423 | 0.412 | -0.048 | -0.054 |
| experiments | 70% | 0.391 | 0.386 | -0.038 | -0.042 | experiments | 70% | 0.422 | 0.411 | -0.048 | -0.054 |
| | 80% | 0.392 | 0.387 | -0.038 | -0.042 | | 80% | 0.423 | 0.412 | -0.048 | -0.054 |
| | 90% | 0.393 | 0.387 | -0.038 | -0.042 | | 90% | 0.423 | 0.412 | -0.048 | -0.054 |
| Demographics | 10% | 0.091 | 0.089 | -0.016 | -0.016 | Demographics | 10% | 0.096 | 0.094 | -0.026 | -0.026 |
| | 20% | 0.091 | 0.089 | -0.016 | -0.016 | | 20% | 0.095 | 0.094 | -0.026 | -0.026 |
| | 30% | 0.091 | 0.089 | -0.016 | -0.016 | | 30% | 0.096 | 0.094 | -0.026 | -0.025 |
| | 40% | 0.091 | 0.089 | -0.016 | -0.016 | | 40% | 0.096 | 0.094 | -0.026 | -0.025 |
| | 50% | 0.091 | 0.089 | -0.016 | -0.015 | | 50% | 0.096 | 0.094 | -0.026 | -0.025 |
| | 60% | 0.092 | 0.090 | -0.016 | -0.016 | | 60% | 0.096 | 0.094 | -0.026 | -0.026 |
| | 70% | 0.092 | 0.089 | -0.016 | -0.016 | | 70% | 0.096 | 0.094 | -0.026 | -0.026 |
| | 80% | 0.091 | 0.089 | -0.016 | -0.016 | | 80% | 0.096 | 0.094 | -0.026 | -0.026 |
| | 90% | 0.091 | 0.089 | -0.015 | -0.015 | | 90% | 0.096 | 0.094 | -0.026 | -0.025 |
| Demographics | 10% | 0.087 | 0.088 | -0.015 | -0.017 | Demographics | 10% | 0.085 | 0.084 | -0.025 | -0.026 |
| + Heuristics | 20% | 0.087 | 0.088 | -0.015 | -0.016 | + Heuristics | 20% | 0.085 | 0.084 | -0.025 | -0.026 |
| and biases | 30% | 0.087 | 0.088 | -0.015 | -0.016 | and biases | 30% | 0.085 | 0.084 | -0.025 | -0.026 |
| experiments | 40% | 0.087 | 0.088 | -0.015 | -0.016 | experiments | 40% | 0.085 | 0.084 | -0.026 | -0.026 |
| | 50% | 0.087 | 0.088 | -0.015 | -0.017 | | 50% | 0.085 | 0.084 | -0.026 | -0.026 |
| | 60% | 0.087 | 0.088 | -0.015 | -0.016 | | 60% | 0.085 | 0.084 | -0.026 | -0.026 |
| | 70% | 0.087 | 0.088 | -0.015 | -0.016 | | 70% | 0.085 | 0.085 | -0.026 | -0.026 |
| | 80% | 0.087 | 0.088 | -0.015 | -0.016 | | 80% | 0.085 | 0.084 | -0.026 | -0.026 |
| | 90% | 0.085 | 0.087 | -0.015 | -0.017 | | 90% | 0.084 | 0.083 | -0.026 | -0.026 |
| Demographics | 10% | 0.394 | 0.392 | -0.035 | -0.037 | Demographics | 10% | 0.415 | 0.407 | -0.043 | -0.045 |
| + Heuristics | 20% | 0.395 | 0.393 | -0.034 | -0.036 | + Heuristics | 20% | 0.416 | 0.407 | -0.042 | -0.044 |
| and biases | 30% | 0.395 | 0.393 | -0.034 | -0.036 | and biases | 30% | 0.416 | 0.408 | -0.042 | -0.044 |
| experiments + | 40% | 0.395 | 0.393 | -0.034 | -0.036 | experiments + | 40% | 0.416 | 0.408 | -0.042 | -0.044 |
| Features with | 50% | 0.395 | 0.393 | -0.034 | -0.036 | Features with | 50% | 0.416 | 0.408 | -0.041 | -0.044 |
| r > 0.5 | 60% | 0.395 | 0.393 | -0.034 | -0.036 | r > 0.5 | 60% | 0.417 | 0.408 | -0.041 | -0.044 |
| | 70% | 0.395 | 0.393 | -0.034 | -0.036 | | 70% | 0.417 | 0.409 | -0.042 | -0.044 |
| | 80% | 0.396 | 0.393 | -0.034 | -0.036 | | 80% | 0.417 | 0.409 | -0.041 | -0.044 |
| | 90% | 0.396 | 0.393 | -0.034 | -0.036 | | 90% | 0.418 | 0.409 | -0.041 | -0.044 |
| Demographics | 10% | 0.150 | 0.153 | -0.025 | -0.027 | Demographics | 10% | 0.150 | 0.150 | -0.049 | -0.051 |
| + Heuristics | 20% | 0.150 | 0.153 | -0.024 | -0.027 | + Heuristics | 20% | 0.150 | 0.150 | -0.049 | -0.051 |
| and biases experiments + Features with r < 0.3 | 30% | 0.150 | 0.153 | -0.024 | -0.027 | and biases experiments + Features with r < 0.3 | 30% | 0.150 | 0.150 | -0.049 | -0.051 |
| | 40% | 0.150 | 0.153 | -0.024 | -0.027 | | 40% | 0.150 | 0.150 | -0.049 | -0.051 |
| | 50% | 0.150 | 0.153 | -0.024 | -0.027 | | 50% | 0.150 | 0.150 | -0.049 | -0.051 |
| | 60% | 0.150 | 0.153 | -0.024 | -0.027 | | 60% | 0.150 | 0.150 | -0.049 | -0.051 |
| | 70% | 0.150 | 0.153 | -0.025 | -0.027 | | 70% | 0.151 | 0.150 | -0.049 | -0.051 |
| | 80% | 0.150 | 0.153 | -0.025 | -0.027 | | 80% | 0.151 | 0.150 | -0.049 | -0.051 |
| | 90% | 0.151 | 0.153 | -0.025 | -0.027 | | 90% | 0.151 | 0.150 | -0.049 | -0.051 |
| Demographics + Heuristics and biases experiments + Features with r < 0.5 | 10% | 0.233 | 0.233 | -0.033 | -0.037 | Demographics + Heuristics and biases experiments + Features with r < 0.5 | 10% | 0.237 | 0.227 | -0.060 | -0.067 |
| | 20% | 0.233 | 0.233 | -0.033 | -0.037 | | 20% | 0.237 | 0.227 | -0.060 | -0.067 |
| | 30% | 0.234 | 0.233 | -0.033 | -0.037 | | 30% | 0.238 | 0.227 | -0.060 | -0.067 |
| | 40% | 0.234 | 0.233 | -0.033 | -0.036 | | 40% | 0.238 | 0.227 | -0.060 | -0.067 |
| | 50% | 0.234 | 0.233 | -0.033 | -0.037 | | 50% | 0.238 | 0.227 | -0.059 | -0.067 |
| | 60% | 0.234 | 0.233 | -0.033 | -0.036 | | 60% | 0.238 | 0.227 | -0.060 | -0.067 |
| | 70% | 0.234 | 0.233 | -0.033 | -0.036 | | 70% | 0.238 | 0.227 | -0.059 | -0.067 |
| | 80% | 0.234 | 0.233 | -0.033 | -0.036 | | 80% | 0.239 | 0.228 | -0.059 | -0.067 |
| | 90% | 0.234 | 0.233 | -0.033 | -0.036 | | 90% | 0.238 | 0.227 | -0.059 | -0.067 |
| Heuristics and biases experiments | 10% | 0.055 | 0.061 | -0.005 | -0.010 | Heuristics and biases experiments | 10% | 0.039 | 0.042 | -0.010 | -0.009 |
| | 20% | 0.055 | 0.062 | -0.005 | -0.009 | | 20% | 0.039 | 0.042 | -0.010 | -0.009 |
| | 30% | 0.055 | 0.062 | -0.005 | -0.009 | | 30% | 0.039 | 0.042 | -0.010 | -0.009 |
| | 40% | 0.055 | 0.062 | -0.005 | -0.009 | | 40% | 0.039 | 0.043 | -0.010 | -0.009 |
| | 50% | 0.055 | 0.061 | -0.005 | -0.009 | | 50% | 0.039 | 0.042 | -0.010 | -0.009 |
| | 60% | 0.055 | 0.061 | -0.005 | -0.009 | | 60% | 0.039 | 0.042 | -0.010 | -0.009 |
| | 70% | 0.057 | 0.063 | -0.005 | -0.009 | | 70% | 0.040 | 0.043 | -0.010 | -0.009 |
| | 80% | 0.055 | 0.062 | -0.005 | -0.009 | | 80% | 0.040 | 0.043 | -0.010 | -0.009 |
| | 90% | 0.056 | 0.062 | -0.005 | -0.009 | | 90% | 0.040 | 0.043 | -0.010 | -0.009 |

| **Input condition** | **Calibration proportion** | **Pearson** | **Spearman** | **Δr (Pearson)** | **Δr (Spearman)** | **Input condition** | **Calibration proportion** | **Pearson** | **Spearman** | **Δr (Pearson)** | **Δr (Spearman)** |
|---|---|---|---|---|---|---|---|---|---|---|---|
| Gemma-4-31B-it | | | | | | gpt-oss-120b | | | | | |
| Demographics + All psychological measures + Heuristics and biases experiments | 10% | 0.435 | 0.422 | -0.050 | -0.058 | Demographics + All psychological measures + Heuristics and biases experiments | 10% | 0.343 | 0.328 | -0.035 | -0.046 |
| | 20% | 0.436 | 0.423 | -0.049 | -0.058 | | 20% | 0.343 | 0.328 | -0.035 | -0.046 |
| | 30% | 0.436 | 0.423 | -0.049 | -0.058 | | 30% | 0.343 | 0.328 | -0.034 | -0.046 |
| | 40% | 0.436 | 0.423 | -0.049 | -0.058 | | 40% | 0.344 | 0.328 | -0.034 | -0.046 |
| | 50% | 0.436 | 0.423 | -0.049 | -0.058 | | 50% | 0.343 | 0.328 | -0.034 | -0.046 |
| | 60% | 0.437 | 0.423 | -0.049 | -0.058 | | 60% | 0.344 | 0.328 | -0.034 | -0.046 |
| | 70% | 0.436 | 0.423 | -0.049 | -0.058 | | 70% | 0.343 | 0.328 | -0.034 | -0.046 |
| | 80% | 0.437 | 0.423 | -0.049 | -0.058 | | 80% | 0.344 | 0.328 | -0.034 | -0.046 |
| | 90% | 0.437 | 0.424 | -0.049 | -0.058 | | 90% | 0.344 | 0.328 | -0.034 | -0.046 |
| Demographics | 10% | 0.096 | 0.096 | -0.033 | -0.032 | Demographics | 10% | 0.071 | 0.066 | -0.014 | -0.011 |
| | 20% | 0.095 | 0.096 | -0.033 | -0.032 | | 20% | 0.071 | 0.066 | -0.014 | -0.011 |
| | 30% | 0.096 | 0.096 | -0.033 | -0.032 | | 30% | 0.071 | 0.066 | -0.014 | -0.011 |
| | 40% | 0.096 | 0.096 | -0.033 | -0.032 | | 40% | 0.071 | 0.066 | -0.014 | -0.011 |
| | 50% | 0.096 | 0.096 | -0.033 | -0.032 | | 50% | 0.071 | 0.066 | -0.014 | -0.011 |
| | 60% | 0.096 | 0.096 | -0.033 | -0.032 | | 60% | 0.071 | 0.066 | -0.014 | -0.011 |
| | 70% | 0.096 | 0.096 | -0.033 | -0.032 | | 70% | 0.071 | 0.066 | -0.014 | -0.011 |
| | 80% | 0.095 | 0.096 | -0.034 | -0.032 | | 80% | 0.071 | 0.067 | -0.014 | -0.011 |
| | 90% | 0.096 | 0.097 | -0.035 | -0.033 | | 90% | 0.071 | 0.067 | -0.014 | -0.011 |
| Demographics | 10% | 0.085 | 0.085 | -0.037 | -0.038 | Demographics | 10% | 0.063 | 0.063 | -0.012 | -0.011 |
| + Heuristics | 20% | 0.085 | 0.086 | -0.037 | -0.038 | + Heuristics | 20% | 0.064 | 0.063 | -0.012 | -0.011 |
| and biases | 30% | 0.085 | 0.086 | -0.037 | -0.038 | and biases | 30% | 0.063 | 0.063 | -0.012 | -0.011 |
| experiments | 40% | 0.085 | 0.086 | -0.037 | -0.038 | experiments | 40% | 0.064 | 0.063 | -0.012 | -0.011 |
| | 50% | 0.084 | 0.085 | -0.037 | -0.038 | | 50% | 0.063 | 0.063 | -0.012 | -0.012 |
| | 60% | 0.085 | 0.086 | -0.037 | -0.038 | | 60% | 0.064 | 0.063 | -0.012 | -0.011 |
| | 70% | 0.085 | 0.086 | -0.037 | -0.038 | | 70% | 0.064 | 0.063 | -0.012 | -0.011 |
| | 80% | 0.085 | 0.086 | -0.038 | -0.038 | | 80% | 0.064 | 0.063 | -0.012 | -0.011 |
| | 90% | 0.084 | 0.085 | -0.038 | -0.038 | | 90% | 0.063 | 0.062 | -0.012 | -0.012 |
| Demographics | 10% | 0.432 | 0.424 | -0.047 | -0.050 | Demographics | 10% | 0.358 | 0.342 | -0.038 | -0.050 |
| + Heuristics | 20% | 0.433 | 0.425 | -0.046 | -0.049 | + Heuristics | 20% | 0.359 | 0.343 | -0.038 | -0.049 |
| and biases | 30% | 0.433 | 0.425 | -0.046 | -0.049 | and biases | 30% | 0.359 | 0.343 | -0.038 | -0.049 |
| experiments + | 40% | 0.433 | 0.425 | -0.046 | -0.049 | experiments + | 40% | 0.359 | 0.343 | -0.038 | -0.049 |
| Features with | 50% | 0.433 | 0.425 | -0.046 | -0.049 | Features with | 50% | 0.359 | 0.343 | -0.038 | -0.049 |
| r > 0.5 | 60% | 0.433 | 0.425 | -0.046 | -0.049 | r > 0.5 | 60% | 0.359 | 0.343 | -0.037 | -0.049 |
| | 70% | 0.434 | 0.425 | -0.046 | -0.049 | | 70% | 0.360 | 0.343 | -0.037 | -0.049 |
| | 80% | 0.434 | 0.425 | -0.046 | -0.050 | | 80% | 0.360 | 0.343 | -0.038 | -0.049 |
| | 90% | 0.434 | 0.425 | -0.046 | -0.049 | | 90% | 0.360 | 0.343 | -0.037 | -0.049 |
| Demographics | 10% | 0.156 | 0.157 | -0.049 | -0.051 | Demographics | 10% | 0.117 | 0.119 | -0.021 | -0.022 |
| + Heuristics | 20% | 0.156 | 0.158 | -0.049 | -0.051 | + Heuristics | 20% | 0.117 | 0.119 | -0.021 | -0.022 |
| and biases | 30% | 0.156 | 0.157 | -0.049 | -0.051 | and biases | 30% | 0.117 | 0.119 | -0.021 | -0.022 |
| experiments + | 40% | 0.156 | 0.157 | -0.049 | -0.051 | experiments + | 40% | 0.117 | 0.119 | -0.021 | -0.022 |
| Features with r | 50% | 0.156 | 0.157 | -0.049 | -0.051 | Features with r | 50% | 0.118 | 0.119 | -0.021 | -0.022 |
| < 0.3 | 60% | 0.156 | 0.157 | -0.049 | -0.051 | < 0.3 | 60% | 0.117 | 0.119 | -0.021 | -0.022 |
| | 70% | 0.157 | 0.158 | -0.049 | -0.051 | | 70% | 0.118 | 0.120 | -0.021 | -0.022 |
| | 80% | 0.157 | 0.158 | -0.049 | -0.051 | | 80% | 0.118 | 0.119 | -0.021 | -0.023 |
| | 90% | 0.157 | 0.158 | -0.049 | -0.051 | | 90% | 0.118 | 0.120 | -0.021 | -0.023 |
| Demographics | 10% | 0.291 | 0.281 | -0.053 | -0.059 | Demographics | 10% | 0.214 | 0.214 | -0.029 | -0.034 |
| + Heuristics | 20% | 0.291 | 0.281 | -0.053 | -0.059 | + Heuristics | 20% | 0.214 | 0.214 | -0.029 | -0.033 |
| and biases | 30% | 0.291 | 0.282 | -0.052 | -0.059 | and biases | 30% | 0.214 | 0.214 | -0.029 | -0.033 |
| experiments + | 40% | 0.291 | 0.282 | -0.052 | -0.059 | experiments + | 40% | 0.215 | 0.214 | -0.029 | -0.033 |
| Features with r | 50% | 0.291 | 0.282 | -0.052 | -0.059 | Features with r | 50% | 0.215 | 0.214 | -0.029 | -0.033 |
| < 0.5 | 60% | 0.291 | 0.282 | -0.052 | -0.059 | < 0.5 | 60% | 0.215 | 0.214 | -0.029 | -0.033 |
| | 70% | 0.292 | 0.282 | -0.052 | -0.059 | | 70% | 0.215 | 0.214 | -0.029 | -0.033 |
| | 80% | 0.292 | 0.282 | -0.052 | -0.059 | | 80% | 0.214 | 0.214 | -0.029 | -0.033 |
| | 90% | 0.292 | 0.282 | -0.052 | -0.059 | | 90% | 0.215 | 0.214 | -0.029 | -0.033 |
| | 10% | 0.037 | 0.038 | -0.017 | -0.021 | | 10% | 0.041 | 0.046 | -0.006 | -0.010 |
| | 20% | 0.037 | 0.038 | -0.017 | -0.021 | | 20% | 0.041 | 0.045 | -0.006 | -0.010 |
| Heuristics and biases experiments | 30% | 0.037 | 0.037 | -0.017 | -0.021 | Heuristics and biases experiments | 30% | 0.041 | 0.046 | -0.006 | -0.010 |
| | 40% | 0.037 | 0.038 | -0.017 | -0.021 | | 40% | 0.041 | 0.046 | -0.006 | -0.010 |
| | 50% | 0.037 | 0.038 | -0.017 | -0.021 | | 50% | 0.041 | 0.045 | -0.006 | -0.010 |
| | 60% | 0.037 | 0.037 | -0.017 | -0.021 | | 60% | 0.041 | 0.046 | -0.006 | -0.010 |
| | 70% | 0.038 | 0.038 | -0.017 | -0.021 | | 70% | 0.041 | 0.046 | -0.006 | -0.010 |
| | 80% | 0.037 | 0.038 | -0.017 | -0.021 | | 80% | 0.042 | 0.046 | -0.006 | -0.010 |
| | 90% | 0.038 | 0.038 | -0.018 | -0.022 | | 90% | 0.041 | 0.046 | -0.006 | -0.010 |

*Note.* Calibration proportion refers to the proportion of participants used to estimate the item-level randomized ordinal equipercentile mapping between digital-twin and human response distributions; correlations were computed on the held-out participants. Item-level Pearson and Spearman correlations were first computed separately for each item and then summarized using Fisher-z averages across items and repeated random splits. Δr denotes the difference between equated and raw, unequated item-level correlations computed on the same held-out participants, such that negative values indicate lower correlations after equating.

**Table S37 Alignment between human and digital-twin Big Five responses after statistical calibration**

| LLM | Calibration proportion | Calibration method | Item-level Pearson r | Item-level Spearman ρ | Profile Pearson r | Profile Spearman ρ | Extraversion Disattenuated r | Agreeableness Disattenuated r | Conscientiousness Disattenuated r | Neuroticism Disattenuated r | Openness Disattenuated r |
|---|---|---|---|---|---|---|---|---|---|---|---|
| DeepSeek V4-Flash | 10% | Digital twin | 0.429 [0.425, 0.433] | 0.428 [0.425, 0.432] | 0.521 [0.518, 0.525] | 0.504 [0.501, 0.507] | 0.504 [0.491, 0.517] | 0.753 [0.743, 0.763] | 0.782 [0.774, 0.790] | 0.787 [0.780, 0.795] | 0.716 [0.707, 0.726] |
| | 10% | Item FE | 0.429 [0.425, 0.433] | 0.428 [0.425, 0.432] | 0.589 [0.584, 0.595] | 0.563 [0.555, 0.570] | 0.504 [0.491, 0.517] | 0.753 [0.743, 0.763] | 0.782 [0.774, 0.790] | 0.787 [0.780, 0.795] | 0.716 [0.707, 0.726] |
| | 10% | Item FE + LLM | 0.429 [0.425, 0.433] | 0.428 [0.425, 0.432] | 0.622 [0.619, 0.625] | 0.599 [0.594, 0.603] | 0.504 [0.491, 0.517] | 0.753 [0.743, 0.763] | 0.782 [0.774, 0.790] | 0.787 [0.780, 0.795] | 0.716 [0.707, 0.726] |
| | 10% | Item FE + demographics | 0.423 [0.418, 0.428] | 0.399 [0.394, 0.405] | 0.584 [0.577, 0.590] | 0.556 [0.545, 0.564] | 0.503 [0.490, 0.518] | 0.758 [0.748, 0.768] | 0.785 [0.777, 0.793] | 0.788 [0.781, 0.796] | 0.715 [0.704, 0.727] |
| | 10% | Item FE + LLM + demographics | 0.408 [0.397, 0.417] | 0.395 [0.386, 0.402] | 0.615 [0.609, 0.619] | 0.592 [0.585, 0.597] | 0.504 [0.489, 0.519] | 0.773 [0.759, 0.788] | 0.794 [0.784, 0.807] | 0.792 [0.784, 0.801] | 0.706 [0.690, 0.722] |
| | 25% | Digital twin | 0.429 [0.422, 0.436] | 0.428 [0.421, 0.435] | 0.521 [0.515, 0.527] | 0.504 [0.499, 0.510] | 0.504 [0.481, 0.525] | 0.753 [0.735, 0.771] | 0.782 [0.768, 0.796] | 0.787 [0.775, 0.801] | 0.716 [0.701, 0.732] |
| | 25% | Item FE | 0.429 [0.422, 0.436] | 0.428 [0.421, 0.435] | 0.590 [0.585, 0.596] | 0.564 [0.558, 0.570] | 0.504 [0.481, 0.525] | 0.753 [0.735, 0.771] | 0.782 [0.768, 0.796] | 0.787 [0.775, 0.801] | 0.716 [0.701, 0.732] |
| | 25% | Item FE + LLM | 0.429 [0.422, 0.436] | 0.428 [0.421, 0.435] | 0.624 [0.620, 0.628] | 0.601 [0.596, 0.606] | 0.504 [0.481, 0.525] | 0.753 [0.735, 0.771] | 0.782 [0.768, 0.796] | 0.787 [0.775, 0.801] | 0.716 [0.701, 0.732] |
| | 25% | Item FE + demographics | 0.427 [0.420, 0.435] | 0.400 [0.392, 0.407] | 0.589 [0.583, 0.594] | 0.562 [0.556, 0.568] | 0.503 [0.479, 0.525] | 0.754 [0.737, 0.772] | 0.782 [0.769, 0.796] | 0.788 [0.775, 0.801] | 0.717 [0.701, 0.734] |
| | 25% | Item FE + LLM + demographics | 0.422 [0.414, 0.429] | 0.401 [0.393, 0.409] | 0.622 [0.618, 0.626] | 0.599 [0.594, 0.604] | 0.502 [0.478, 0.525] | 0.760 [0.742, 0.779] | 0.785 [0.772, 0.800] | 0.790 [0.777, 0.803] | 0.713 [0.695, 0.730] |
| | 50% | Digital twin | 0.429 [0.416, 0.441] | 0.429 [0.416, 0.440] | 0.521 [0.511, 0.532] | 0.504 [0.495, 0.514] | 0.503 [0.465, 0.542] | 0.753 [0.722, 0.783] | 0.782 [0.756, 0.806] | 0.788 [0.765, 0.809] | 0.716 [0.688, 0.744] |
| | 50% | Item FE | 0.429 [0.416, 0.441] | 0.429 [0.416, 0.440] | 0.591 [0.582, 0.599] | 0.565 [0.558, 0.573] | 0.503 [0.465, 0.542] | 0.753 [0.722, 0.783] | 0.782 [0.756, 0.806] | 0.788 [0.765, 0.809] | 0.716 [0.688, 0.744] |
| | 50% | Item FE + LLM | 0.429 [0.416, 0.441] | 0.429 [0.416, 0.440] | 0.625 [0.617, 0.632] | 0.601 [0.594, 0.608] | 0.503 [0.465, 0.542] | 0.753 [0.722, 0.783] | 0.782 [0.756, 0.806] | 0.788 [0.765, 0.809] | 0.716 [0.688, 0.744] |
| | 50% | Item FE + demographics | 0.429 [0.415, 0.440] | 0.401 [0.388, 0.412] | 0.591 [0.582, 0.599] | 0.565 [0.557, 0.572] | 0.502 [0.464, 0.540] | 0.754 [0.722, 0.783] | 0.782 [0.757, 0.806] | 0.788 [0.765, 0.809] | 0.717 [0.689, 0.745] |
| | 50% | Item FE + LLM + demographics | 0.426 [0.413, 0.438] | 0.403 [0.391, 0.414] | 0.625 [0.617, 0.632] | 0.601 [0.594, 0.608] | 0.501 [0.462, 0.540] | 0.757 [0.726, 0.787] | 0.783 [0.757, 0.806] | 0.789 [0.767, 0.811] | 0.714 [0.686, 0.742] |
| Gemma -4-31B-it | 10% | Digital twin | 0.485 [0.481, 0.489] | 0.481 [0.477, 0.484] | 0.605 [0.602, 0.608] | 0.588 [0.585, 0.590] | 0.601 [0.591, 0.612] | 0.820 [0.812, 0.829] | 0.861 [0.856, 0.867] | 0.772 [0.765, 0.779] | 0.762 [0.754, 0.771] |
| | 10% | Item FE | 0.485 [0.481, 0.489] | 0.481 [0.477, 0.484] | 0.636 [0.631, 0.640] | 0.603 [0.597, 0.609] | 0.601 [0.591, 0.612] | 0.820 [0.812, 0.829] | 0.861 [0.856, 0.867] | 0.772 [0.765, 0.779] | 0.762 [0.754, 0.771] |
| | 10% | Item FE + LLM | 0.485 [0.481, 0.489] | 0.481 [0.477, 0.484] | 0.657 [0.654, 0.659] | 0.628 [0.624, 0.631] | 0.601 [0.591, 0.612] | 0.820 [0.812, 0.829] | 0.861 [0.856, 0.867] | 0.772 [0.765, 0.779] | 0.762 [0.754, 0.771] |
| | 10% | Item FE + demographics | 0.478 [0.473, 0.483] | 0.439 [0.433, 0.444] | 0.631 [0.625, 0.636] | 0.597 [0.590, 0.603] | 0.601 [0.591, 0.612] | 0.825 [0.817, 0.834] | 0.864 [0.858, 0.870] | 0.773 [0.766, 0.781] | 0.759 [0.749, 0.770] |
| | 10% | Item FE + LLM + demographics | 0.467 [0.457, 0.474] | 0.437 [0.430, 0.443] | 0.649 [0.644, 0.653] | 0.621 [0.615, 0.625] | 0.604 [0.592, 0.616] | 0.838 [0.826, 0.854] | 0.870 [0.863, 0.880] | 0.776 [0.768, 0.785] | 0.753 [0.740, 0.768] |
| | 25% | Digital twin | 0.485 [0.478, 0.492] | 0.481 [0.474, 0.487] | 0.605 [0.600, 0.610] | 0.588 [0.583, 0.592] | 0.602 [0.583, 0.621] | 0.821 [0.806, 0.835] | 0.861 [0.850, 0.871] | 0.772 [0.759, 0.784] | 0.762 [0.747, 0.776] |
| | 25% | Item FE | 0.485 [0.478, 0.492] | 0.481 [0.474, 0.487] | 0.637 [0.632, 0.642] | 0.605 [0.600, 0.610] | 0.602 [0.583, 0.621] | 0.821 [0.806, 0.835] | 0.861 [0.850, 0.871] | 0.772 [0.759, 0.784] | 0.762 [0.747, 0.776] |
| | 25% | Item FE + LLM | 0.485 [0.478, 0.492] | 0.481 [0.474, 0.487] | 0.658 [0.654, 0.662] | 0.630 [0.625, 0.634] | 0.602 [0.583, 0.621] | 0.821 [0.806, 0.835] | 0.861 [0.850, 0.871] | 0.772 [0.759, 0.784] | 0.762 [0.747, 0.776] |
| | 25% | Item FE + demographics | 0.483 [0.476, 0.490] | 0.440 [0.432, 0.447] | 0.635 [0.630, 0.640] | 0.603 [0.598, 0.607] | 0.601 [0.583, 0.620] | 0.822 [0.807, 0.836] | 0.862 [0.851, 0.872] | 0.772 [0.760, 0.784] | 0.761 [0.746, 0.776] |
| | 25% | Item FE + LLM + demographics | 0.479 [0.471, 0.486] | 0.440 [0.432, 0.448] | 0.656 [0.652, 0.660] | 0.627 [0.623, 0.632] | 0.601 [0.582, 0.620] | 0.826 [0.810, 0.840] | 0.864 [0.853, 0.874] | 0.774 [0.761, 0.786] | 0.759 [0.742, 0.774] |
| | 50% | Digital twin | 0.485 [0.473, 0.497] | 0.481 [0.470, 0.492] | 0.605 [0.597, 0.613] | 0.588 [0.579, 0.596] | 0.601 [0.568, 0.634] | 0.821 [0.797, 0.846] | 0.860 [0.843, 0.876] | 0.773 [0.751, 0.793] | 0.762 [0.737, 0.787] |
| | 50% | Item FE | 0.485 [0.473, 0.497] | 0.481 [0.470, 0.492] | 0.637 [0.630, 0.645] | 0.606 [0.598, 0.613] | 0.601 [0.568, 0.634] | 0.821 [0.797, 0.846] | 0.860 [0.843, 0.876] | 0.773 [0.751, 0.793] | 0.762 [0.737, 0.787] |
| | 50% | Item FE + LLM | 0.485 [0.473, 0.497] | 0.481 [0.470, 0.492] | 0.659 [0.651, 0.666] | 0.630 [0.623, 0.637] | 0.601 [0.568, 0.634] | 0.821 [0.797, 0.846] | 0.860 [0.843, 0.876] | 0.773 [0.751, 0.793] | 0.762 [0.737, 0.787] |
| | 50% | Item FE + demographics | 0.484 [0.472, 0.496] | 0.441 [0.430, 0.452] | 0.637 [0.629, 0.644] | 0.604 [0.597, 0.611] | 0.600 [0.567, 0.633] | 0.821 [0.797, 0.847] | 0.861 [0.843, 0.877] | 0.773 [0.751, 0.794] | 0.762 [0.737, 0.787] |
| | 50% | Item FE + LLM + demographics | 0.482 [0.470, 0.494] | 0.442 [0.431, 0.453] | 0.658 [0.651, 0.665] | 0.629 [0.622, 0.636] | 0.600 [0.566, 0.631] | 0.823 [0.799, 0.849] | 0.862 [0.845, 0.878] | 0.774 [0.752, 0.795] | 0.761 [0.735, 0.786] |
| gpt-oss-120b | 10% | Digital twin | 0.378 [0.373, 0.382] | 0.374 [0.370, 0.378] | 0.531 [0.528, 0.534] | 0.512 [0.509, 0.515] | 0.344 [0.330, 0.358] | 0.731 [0.720, 0.742] | 0.811 [0.805, 0.819] | 0.716 [0.707, 0.725] | 0.607 [0.595, 0.620] |
| | 10% | Item FE | 0.378 [0.373, 0.382] | 0.374 [0.370, 0.378] | 0.537 [0.529, 0.546] | 0.501 [0.489, 0.512] | 0.344 [0.330, 0.358] | 0.731 [0.720, 0.742] | 0.811 [0.805, 0.819] | 0.716 [0.707, 0.725] | 0.607 [0.595, 0.620] |
| | 10% | Item FE + LLM | 0.378 [0.373, 0.382] | 0.374 [0.370, 0.378] | 0.605 [0.602, 0.608] | 0.585 [0.580, 0.588] | 0.344 [0.330, 0.358] | 0.731 [0.720, 0.742] | 0.811 [0.805, 0.819] | 0.716 [0.707, 0.725] | 0.607 [0.595, 0.620] |
| | 10% | Item FE + demographics | 0.373 [0.368, 0.377] | 0.338 [0.331, 0.344] | 0.532 [0.523, 0.541] | 0.497 [0.485, 0.509] | 0.344 [0.329, 0.359] | 0.735 [0.724, 0.746] | 0.814 [0.807, 0.822] | 0.717 [0.708, 0.726] | 0.606 [0.591, 0.620] |
| | 10% | Item FE + LLM + demographics | 0.352 [0.339, 0.362] | 0.327 [0.315, 0.336] | 0.597 [0.591, 0.602] | 0.576 [0.570, 0.582] | 0.345 [0.327, 0.366] | 0.756 [0.739, 0.774] | 0.829 [0.818, 0.843] | 0.723 [0.712, 0.735] | 0.590 [0.567, 0.613] |
| | 25% | Digital twin | 0.378 [0.370, 0.385] | 0.374 [0.368, 0.381] | 0.531 [0.526, 0.537] | 0.512 [0.507, 0.518] | 0.344 [0.321, 0.371] | 0.731 [0.711, 0.750] | 0.812 [0.799, 0.825] | 0.716 [0.700, 0.731] | 0.607 [0.586, 0.627] |
| | 25% | Item FE | 0.378 [0.370, 0.385] | 0.374 [0.368, 0.381] | 0.538 [0.531, 0.545] | 0.502 [0.494, 0.511] | 0.344 [0.321, 0.371] | 0.731 [0.711, 0.750] | 0.812 [0.799, 0.825] | 0.716 [0.700, 0.731] | 0.607 [0.586, 0.627] |
| | 25% | Item FE + LLM | 0.378 [0.370, 0.385] | 0.374 [0.368, 0.381] | 0.607 [0.603, 0.611] | 0.587 [0.582, 0.592] | 0.344 [0.321, 0.371] | 0.731 [0.711, 0.750] | 0.812 [0.799, 0.825] | 0.716 [0.700, 0.731] | 0.607 [0.586, 0.627] |
| | 25% | Item FE + demographics | 0.376 [0.369, 0.383] | 0.339 [0.331, 0.346] | 0.537 [0.529, 0.544] | 0.501 [0.493, 0.510] | 0.344 [0.320, 0.369] | 0.732 [0.713, 0.751] | 0.812 [0.800, 0.825] | 0.716 [0.701, 0.732] | 0.607 [0.586, 0.628] |
| | 25% | Item FE + LLM + demographics | 0.368 [0.361, 0.376] | 0.339 [0.331, 0.346] | 0.605 [0.601, 0.609] | 0.584 [0.580, 0.589] | 0.344 [0.320, 0.370] | 0.741 [0.721, 0.761] | 0.818 [0.805, 0.831] | 0.719 [0.704, 0.735] | 0.602 [0.578, 0.625] |
| | 50% | Digital twin | 0.378 [0.365, 0.390] | 0.375 [0.362, 0.386] | 0.531 [0.521, 0.541] | 0.512 [0.503, 0.521] | 0.344 [0.302, 0.385] | 0.733 [0.701, 0.764] | 0.811 [0.790, 0.831] | 0.716 [0.688, 0.743] | 0.607 [0.571, 0.642] |
| | 50% | Item FE | 0.378 [0.365, 0.390] | 0.375 [0.362, 0.386] | 0.539 [0.529, 0.549] | 0.503 [0.494, 0.512] | 0.344 [0.302, 0.385] | 0.733 [0.701, 0.764] | 0.811 [0.790, 0.831] | 0.716 [0.688, 0.743] | 0.607 [0.571, 0.642] |
| | 50% | Item FE + LLM | 0.378 [0.365, 0.390] | 0.375 [0.362, 0.386] | 0.608 [0.600, 0.615] | 0.587 [0.580, 0.595] | 0.344 [0.302, 0.385] | 0.733 [0.701, 0.764] | 0.811 [0.790, 0.831] | 0.716 [0.688, 0.743] | 0.607 [0.571, 0.642] |
| | 50% | Item FE + demographics | 0.377 [0.365, 0.389] | 0.340 [0.328, 0.351] | 0.538 [0.528, 0.548] | 0.502 [0.493, 0.511] | 0.343 [0.301, 0.385] | 0.733 [0.701, 0.765] | 0.812 [0.790, 0.832] | 0.717 [0.689, 0.744] | 0.609 [0.572, 0.643] |
| | 50% | Item FE + LLM + demographics | 0.373 [0.361, 0.386] | 0.342 [0.330, 0.354] | 0.608 [0.600, 0.615] | 0.587 [0.579, 0.595] | 0.342 [0.299, 0.385] | 0.738 [0.706, 0.770] | 0.814 [0.793, 0.834] | 0.719 [0.691, 0.746] | 0.606 [0.569, 0.640] |
| Qwen 3.5 Plus | 10% | Digital twin | 0.470 [0.467, 0.474] | 0.466 [0.462, 0.470] | 0.574 [0.571, 0.577] | 0.559 [0.556, 0.562] | 0.543 [0.532, 0.554] | 0.801 [0.792, 0.810] | 0.822 [0.816, 0.829] | 0.777 [0.771, 0.784] | 0.707 [0.698, 0.717] |

| | | | | | | | | | | |
|---|---|---|---|---|---|---|---|---|---|---|
| 10% | Item FE | 0.470 [0.467, 0.474] | 0.466 [0.462, 0.470] | 0.634 [0.629, 0.638] | 0.604 [0.599, 0.609] | 0.543 [0.532, 0.554] | 0.801 [0.792, 0.810] | 0.822 [0.816, 0.829] | 0.777 [0.771, 0.784] | 0.707 [0.698, 0.717] |
| 10% | Item FE + LLM | 0.470 [0.467, 0.474] | 0.466 [0.462, 0.470] | 0.649 [0.646, 0.652] | 0.625 [0.621, 0.628] | 0.543 [0.532, 0.554] | 0.801 [0.792, 0.810] | 0.822 [0.816, 0.829] | 0.777 [0.771, 0.784] | 0.707 [0.698, 0.717] |
| 10% | Item FE + demographics | 0.462 [0.456, 0.467] | 0.427 [0.421, 0.433] | 0.628 [0.621, 0.633] | 0.598 [0.590, 0.603] | 0.545 [0.533, 0.557] | 0.807 [0.798, 0.816] | 0.826 [0.819, 0.833] | 0.779 [0.772, 0.786] | 0.708 [0.694, 0.720] |
| 10% | Item FE + LLM + demographics | 0.452 [0.441, 0.460] | 0.426 [0.419, 0.432] | 0.641 [0.636, 0.646] | 0.616 [0.610, 0.621] | 0.545 [0.530, 0.560] | 0.815 [0.804, 0.827] | 0.831 [0.824, 0.840] | 0.780 [0.772, 0.788] | 0.701 [0.683, 0.717] |
| 25% | Digital twin | 0.470 [0.463, 0.477] | 0.466 [0.459, 0.472] | 0.574 [0.569, 0.579] | 0.559 [0.554, 0.564] | 0.543 [0.522, 0.562] | 0.801 [0.786, 0.816] | 0.822 [0.811, 0.834] | 0.777 [0.764, 0.788] | 0.707 [0.692, 0.723] |
| 25% | Item FE | 0.470 [0.463, 0.477] | 0.466 [0.459, 0.472] | 0.635 [0.630, 0.639] | 0.605 [0.601, 0.610] | 0.543 [0.522, 0.562] | 0.801 [0.786, 0.816] | 0.822 [0.811, 0.834] | 0.777 [0.764, 0.788] | 0.707 [0.692, 0.723] |
| 25% | Item FE + LLM | 0.470 [0.463, 0.477] | 0.466 [0.459, 0.472] | 0.651 [0.647, 0.655] | 0.626 [0.622, 0.631] | 0.543 [0.522, 0.562] | 0.801 [0.786, 0.816] | 0.822 [0.811, 0.834] | 0.777 [0.764, 0.788] | 0.707 [0.692, 0.723] |
| 25% | Item FE + demographics | 0.468 [0.460, 0.475] | 0.429 [0.420, 0.437] | 0.633 [0.628, 0.638] | 0.603 [0.599, 0.608] | 0.543 [0.522, 0.562] | 0.804 [0.788, 0.819] | 0.823 [0.812, 0.835] | 0.779 [0.765, 0.790] | 0.710 [0.693, 0.727] |
| 25% | Item FE + LLM + demographics | 0.464 [0.457, 0.471] | 0.429 [0.421, 0.438] | 0.649 [0.644, 0.653] | 0.624 [0.619, 0.628] | 0.543 [0.523, 0.563] | 0.807 [0.791, 0.823] | 0.825 [0.813, 0.837] | 0.779 [0.766, 0.791] | 0.707 [0.689, 0.725] |
| 50% | Digital twin | 0.470 [0.457, 0.482] | 0.466 [0.453, 0.477] | 0.574 [0.565, 0.583] | 0.559 [0.550, 0.568] | 0.542 [0.506, 0.575] | 0.802 [0.776, 0.828] | 0.822 [0.803, 0.841] | 0.778 [0.756, 0.798] | 0.707 [0.680, 0.734] |
| 50% | Item FE | 0.470 [0.457, 0.482] | 0.466 [0.453, 0.477] | 0.635 [0.627, 0.643] | 0.606 [0.599, 0.613] | 0.542 [0.505, 0.575] | 0.802 [0.776, 0.828] | 0.822 [0.803, 0.841] | 0.778 [0.756, 0.798] | 0.707 [0.680, 0.734] |
| 50% | Item FE + LLM | 0.470 [0.457, 0.482] | 0.466 [0.453, 0.477] | 0.651 [0.644, 0.659] | 0.627 [0.620, 0.634] | 0.542 [0.506, 0.575] | 0.802 [0.776, 0.828] | 0.822 [0.803, 0.841] | 0.778 [0.756, 0.798] | 0.707 [0.680, 0.734] |
| 50% | Item FE + demographics | 0.469 [0.456, 0.481] | 0.430 [0.418, 0.442] | 0.635 [0.627, 0.642] | 0.605 [0.598, 0.612] | 0.541 [0.505, 0.575] | 0.803 [0.778, 0.829] | 0.822 [0.803, 0.842] | 0.778 [0.757, 0.798] | 0.711 [0.682, 0.738] |
| 50% | Item FE + LLM + demographics | 0.467 [0.455, 0.479] | 0.431 [0.418, 0.442] | 0.651 [0.644, 0.658] | 0.626 [0.619, 0.633] | 0.541 [0.505, 0.575] | 0.806 [0.780, 0.831] | 0.823 [0.804, 0.843] | 0.779 [0.757, 0.799] | 0.709 [0.680, 0.737] |

*Note.* LLM denotes the digital twin model. Calibration proportion denotes the proportion of participants used to estimate the calibration model; all reported metrics were computed among held out participants. Calibration method denotes the response source or bias correction specification. Digital twin refers to the uncalibrated digital twin response. Item FE estimates item specific mean bias using BFI-44 item fixed effects. Item FE + LLM additionally includes the raw digital twin response. Item FE + demographics additionally includes participant demographic covariates. Item FE + LLM + demographics includes item fixed effects, the raw digital twin response and demographic covariates. Item level Pearson r and Spearman ρ compare human and digital twin BFI-44 responses across participant and item observations in the held out sample. Profile Pearson r and Spearman ρ were computed within each held out participant by correlating the participant's 44 item human BFI-44 response profile with the corresponding 44 item digital twin response profile. Disattenuated r denotes the Pearson correlation between human and digital twin trait scores, computed across participants and corrected for measurement reliability. Item level correlations, within participant profile correlations and trait score correlations were summarized across 1,000 repeated participant level splits for each calibration proportion using Fisher z transformations. Values in brackets are empirical 95% intervals across the 1,000 repeated splits. Higher values indicate stronger correspondence with held out human responses. FE, fixed effects.

**Table S38 Trait-level distributional accuracy and prediction error of digital-twin Big Five scores after statistical calibration**

| LLM | Calibration proportion | Calibration method | Human mean | Predicted mean | Wasserstein distance | MAE | RMSE |
|---|---|---|---|---|---|---|---|
| Extraversion | | | | | | | |
| DeepSeek V4-Flash | 10% | Digital twin | 2.874 [2.860, 2.887] | 3.233 [3.222, 3.245] | 0.359 [0.345, 0.372] | 0.791 [0.782, 0.799] | 0.992 [0.981, 1.002] |
| | 10% | Item FE | 2.874 [2.860, 2.887] | 2.873 [2.745, 2.994] | 0.126 [0.097, 0.184] | 0.734 [0.725, 0.744] | 0.927 [0.916, 0.940] |
| | 10% | Item FE + LLM | 2.874 [2.860, 2.887] | 2.871 [2.757, 2.980] | 0.434 [0.400, 0.469] | 0.677 [0.668, 0.687] | 0.834 [0.825, 0.846] |
| | 10% | Item FE + demographics | 2.874 [2.860, 2.887] | 2.873 [2.746, 2.996] | 0.121 [0.096, 0.182] | 0.735 [0.726, 0.746] | 0.929 [0.917, 0.943] |
| | 10% | Item FE + LLM + demographics | 2.874 [2.860, 2.887] | 2.871 [2.755, 2.980] | 0.433 [0.399, 0.469] | 0.679 [0.670, 0.689] | 0.836 [0.826, 0.848] |
| | 25% | Digital twin | 2.874 [2.852, 2.895] | 3.233 [3.212, 3.253] | 0.359 [0.338, 0.382] | 0.791 [0.775, 0.806] | 0.992 [0.974, 1.010] |
| | 25% | Item FE | 2.874 [2.852, 2.895] | 2.873 [2.793, 2.952] | 0.116 [0.091, 0.159] | 0.733 [0.717, 0.747] | 0.926 [0.907, 0.942] |
| | 25% | Item FE + LLM | 2.874 [2.852, 2.895] | 2.872 [2.812, 2.936] | 0.432 [0.406, 0.457] | 0.675 [0.663, 0.687] | 0.832 [0.819, 0.845] |
| | 25% | Item FE + demographics | 2.874 [2.852, 2.895] | 2.873 [2.793, 2.952] | 0.114 [0.091, 0.156] | 0.734 [0.718, 0.748] | 0.927 [0.908, 0.944] |
| | 25% | Item FE + LLM + demographics | 2.874 [2.852, 2.895] | 2.872 [2.812, 2.936] | 0.431 [0.405, 0.457] | 0.677 [0.664, 0.689] | 0.834 [0.820, 0.847] |
| | 50% | Digital twin | 2.875 [2.833, 2.916] | 3.233 [3.195, 3.269] | 0.358 [0.318, 0.400] | 0.790 [0.764, 0.817] | 0.992 [0.961, 1.021] |
| | 50% | Item FE | 2.875 [2.833, 2.916] | 2.872 [2.807, 2.939] | 0.113 [0.084, 0.157] | 0.732 [0.707, 0.756] | 0.925 [0.894, 0.953] |
| | 50% | Item FE + LLM | 2.875 [2.833, 2.916] | 2.872 [2.831, 2.914] | 0.431 [0.403, 0.458] | 0.675 [0.655, 0.694] | 0.832 [0.809, 0.854] |
| | 50% | Item FE + demographics | 2.875 [2.833, 2.916] | 2.872 [2.806, 2.939] | 0.112 [0.084, 0.154] | 0.733 [0.708, 0.757] | 0.927 [0.895, 0.955] |
| | 50% | Item FE + LLM + demographics | 2.875 [2.833, 2.916] | 2.872 [2.831, 2.914] | 0.431 [0.403, 0.457] | 0.676 [0.656, 0.695] | 0.833 [0.810, 0.856] |
| Gemma-4-31B-it | 10% | Digital twin | 2.874 [2.860, 2.887] | 2.792 [2.782, 2.802] | 0.232 [0.223, 0.240] | 0.635 [0.628, 0.642] | 0.798 [0.790, 0.805] |
| | 10% | Item FE | 2.874 [2.860, 2.887] | 2.871 [2.763, 2.973] | 0.239 [0.229, 0.262] | 0.637 [0.630, 0.647] | 0.796 [0.787, 0.807] |
| | 10% | Item FE + LLM | 2.874 [2.860, 2.887] | 2.871 [2.767, 2.975] | 0.442 [0.417, 0.469] | 0.643 [0.634, 0.653] | 0.791 [0.782, 0.802] |
| | 10% | Item FE + demographics | 2.874 [2.860, 2.887] | 2.871 [2.762, 2.975] | 0.233 [0.221, 0.257] | 0.639 [0.631, 0.648] | 0.798 [0.789, 0.808] |
| | 10% | Item FE + LLM + demographics | 2.874 [2.860, 2.887] | 2.871 [2.766, 2.975] | 0.441 [0.413, 0.468] | 0.644 [0.635, 0.655] | 0.793 [0.783, 0.804] |
| | 25% | Digital twin | 2.874 [2.852, 2.895] | 2.793 [2.775, 2.809] | 0.232 [0.219, 0.247] | 0.635 [0.622, 0.646] | 0.798 [0.784, 0.811] |
| | 25% | Item FE | 2.874 [2.852, 2.895] | 2.873 [2.807, 2.933] | 0.237 [0.224, 0.253] | 0.636 [0.624, 0.648] | 0.794 [0.780, 0.807] |
| | 25% | Item FE + LLM | 2.874 [2.852, 2.895] | 2.872 [2.814, 2.928] | 0.441 [0.419, 0.463] | 0.641 [0.629, 0.653] | 0.789 [0.777, 0.802] |
| | 25% | Item FE + demographics | 2.874 [2.852, 2.895] | 2.873 [2.807, 2.934] | 0.232 [0.219, 0.249] | 0.637 [0.625, 0.649] | 0.796 [0.781, 0.809] |
| | 25% | Item FE + LLM + demographics | 2.874 [2.852, 2.895] | 2.872 [2.815, 2.928] | 0.439 [0.418, 0.462] | 0.643 [0.630, 0.655] | 0.791 [0.778, 0.804] |
| | 50% | Digital twin | 2.875 [2.833, 2.916] | 2.793 [2.764, 2.821] | 0.234 [0.210, 0.260] | 0.635 [0.614, 0.655] | 0.798 [0.773, 0.822] |
| | 50% | Item FE | 2.875 [2.833, 2.916] | 2.873 [2.827, 2.920] | 0.237 [0.217, 0.261] | 0.636 [0.616, 0.656] | 0.794 [0.770, 0.818] |
| | 50% | Item FE + LLM | 2.875 [2.833, 2.916] | 2.872 [2.835, 2.908] | 0.440 [0.415, 0.465] | 0.641 [0.622, 0.661] | 0.789 [0.767, 0.810] |
| | 50% | Item FE + demographics | 2.875 [2.833, 2.916] | 2.873 [2.826, 2.921] | 0.233 [0.212, 0.257] | 0.637 [0.617, 0.657] | 0.795 [0.771, 0.819] |
| | 50% | Item FE + LLM + demographics | 2.875 [2.833, 2.916] | 2.872 [2.836, 2.908] | 0.439 [0.414, 0.465] | 0.642 [0.623, 0.662] | 0.790 [0.768, 0.812] |
| gpt-oss-120b | 10% | Digital twin | 2.874 [2.860, 2.887] | 2.645 [2.632, 2.658] | 0.275 [0.261, 0.290] | 0.864 [0.855, 0.874] | 1.082 [1.071, 1.093] |
| | 10% | Item FE | 2.874 [2.860, 2.887] | 2.872 [2.718, 3.029] | 0.220 [0.207, 0.253] | 0.852 [0.842, 0.864] | 1.061 [1.049, 1.076] |
| | 10% | Item FE + LLM | 2.874 [2.860, 2.887] | 2.871 [2.751, 2.985] | 0.526 [0.496, 0.558] | 0.731 [0.722, 0.741] | 0.892 [0.882, 0.904] |
| | 10% | Item FE + demographics | 2.874 [2.860, 2.887] | 2.872 [2.718, 3.030] | 0.218 [0.205, 0.248] | 0.852 [0.842, 0.865] | 1.062 [1.049, 1.078] |
| | 10% | Item FE + LLM + demographics | 2.874 [2.860, 2.887] | 2.871 [2.752, 2.987] | 0.523 [0.492, 0.555] | 0.732 [0.723, 0.743] | 0.894 [0.884, 0.906] |
| | 25% | Digital twin | 2.874 [2.852, 2.895] | 2.645 [2.624, 2.668] | 0.276 [0.255, 0.299] | 0.864 [0.849, 0.880] | 1.082 [1.063, 1.101] |
| | 25% | Item FE | 2.874 [2.852, 2.895] | 2.872 [2.776, 2.960] | 0.216 [0.201, 0.236] | 0.850 [0.835, 0.866] | 1.059 [1.040, 1.078] |
| | 25% | Item FE + LLM | 2.874 [2.852, 2.895] | 2.872 [2.804, 2.935] | 0.524 [0.500, 0.548] | 0.729 [0.716, 0.742] | 0.890 [0.876, 0.904] |
| | 25% | Item FE + demographics | 2.874 [2.852, 2.895] | 2.872 [2.775, 2.961] | 0.215 [0.201, 0.235] | 0.850 [0.835, 0.867] | 1.059 [1.039, 1.079] |
| | 25% | Item FE + LLM + demographics | 2.874 [2.852, 2.895] | 2.872 [2.803, 2.935] | 0.522 [0.498, 0.546] | 0.730 [0.717, 0.744] | 0.891 [0.877, 0.905] |
| | 50% | Digital twin | 2.875 [2.833, 2.916] | 2.646 [2.606, 2.682] | 0.278 [0.241, 0.320] | 0.865 [0.837, 0.891] | 1.082 [1.049, 1.116] |
| | 50% | Item FE | 2.875 [2.833, 2.916] | 2.873 [2.791, 2.945] | 0.217 [0.193, 0.245] | 0.850 [0.823, 0.877] | 1.059 [1.027, 1.091] |
| | 50% | Item FE + LLM | 2.875 [2.833, 2.916] | 2.872 [2.829, 2.915] | 0.523 [0.493, 0.554] | 0.729 [0.707, 0.750] | 0.889 [0.865, 0.913] |
| | 50% | Item FE + demographics | 2.875 [2.833, 2.916] | 2.873 [2.790, 2.945] | 0.216 [0.193, 0.244] | 0.850 [0.823, 0.877] | 1.059 [1.027, 1.091] |
| | 50% | Item FE + LLM + demographics | 2.875 [2.833, 2.916] | 2.872 [2.829, 2.915] | 0.522 [0.491, 0.552] | 0.730 [0.707, 0.752] | 0.890 [0.866, 0.914] |
| Qwen 3.5 Plus | 10% | Digital twin | 2.874 [2.860, 2.887] | 2.750 [2.740, 2.759] | 0.283 [0.274, 0.292] | 0.668 [0.661, 0.675] | 0.839 [0.830, 0.847] |
| | 10% | Item FE | 2.874 [2.860, 2.887] | 2.872 [2.756, 2.984] | 0.256 [0.241, 0.287] | 0.669 [0.660, 0.679] | 0.832 [0.823, 0.844] |
| | 10% | Item FE + LLM | 2.874 [2.860, 2.887] | 2.871 [2.764, 2.976] | 0.436 [0.408, 0.465] | 0.663 [0.655, 0.673] | 0.815 [0.807, 0.826] |
| | 10% | Item FE + demographics | 2.874 [2.860, 2.887] | 2.872 [2.755, 2.982] | 0.253 [0.237, 0.287] | 0.670 [0.662, 0.681] | 0.834 [0.824, 0.846] |
| | 10% | Item FE + LLM + demographics | 2.874 [2.860, 2.887] | 2.871 [2.764, 2.973] | 0.434 [0.406, 0.462] | 0.665 [0.657, 0.676] | 0.818 [0.809, 0.829] |
| | 25% | Digital twin | 2.874 [2.852, 2.895] | 2.750 [2.734, 2.766] | 0.283 [0.268, 0.298] | 0.668 [0.656, 0.680] | 0.839 [0.825, 0.852] |
| | 25% | Item FE | 2.874 [2.852, 2.895] | 2.872 [2.804, 2.939] | 0.251 [0.235, 0.274] | 0.667 [0.654, 0.679] | 0.831 [0.816, 0.844] |
| | 25% | Item FE + LLM | 2.874 [2.852, 2.895] | 2.872 [2.812, 2.929] | 0.434 [0.412, 0.455] | 0.662 [0.650, 0.674] | 0.814 [0.800, 0.826] |
| | 25% | Item FE + demographics | 2.874 [2.852, 2.895] | 2.872 [2.805, 2.936] | 0.250 [0.234, 0.274] | 0.669 [0.656, 0.681] | 0.832 [0.817, 0.845] |
| | 25% | Item FE + LLM + demographics | 2.874 [2.852, 2.895] | 2.872 [2.812, 2.929] | 0.433 [0.410, 0.455] | 0.664 [0.651, 0.676] | 0.816 [0.803, 0.828] |
| | 50% | Digital twin | 2.875 [2.833, 2.916] | 2.750 [2.723, 2.779] | 0.284 [0.255, 0.311] | 0.668 [0.646, 0.691] | 0.839 [0.814, 0.865] |
| | 50% | Item FE | 2.875 [2.833, 2.916] | 2.873 [2.828, 2.923] | 0.251 [0.227, 0.278] | 0.667 [0.645, 0.688] | 0.830 [0.807, 0.855] |
| | 50% | Item FE + LLM | 2.875 [2.833, 2.916] | 2.872 [2.835, 2.912] | 0.434 [0.408, 0.461] | 0.662 [0.641, 0.682] | 0.813 [0.791, 0.835] |
| | 50% | Item FE + demographics | 2.875 [2.833, 2.916] | 2.873 [2.827, 2.923] | 0.250 [0.226, 0.278] | 0.668 [0.646, 0.690] | 0.832 [0.808, 0.856] |
| | 50% | Item FE + LLM + demographics | 2.875 [2.833, 2.916] | 2.872 [2.835, 2.912] | 0.433 [0.407, 0.461] | 0.664 [0.643, 0.684] | 0.815 [0.793, 0.837] |
| Agreeableness | | | | | | | |
| DeepSeek V4-Flash | 10% | Digital twin | 3.902 [3.892, 3.912] | 4.232 [4.225, 4.241] | 0.331 [0.323, 0.338] | 0.483 [0.477, 0.489] | 0.636 [0.629, 0.644] |
| | 10% | Item FE | 3.902 [3.892, 3.912] | 3.903 [3.831, 3.975] | 0.112 [0.099, 0.139] | 0.424 [0.418, 0.432] | 0.546 [0.538, 0.553] |
| | 10% | Item FE + LLM | 3.902 [3.892, 3.912] | 3.902 [3.832, 3.973] | 0.327 [0.305, 0.350] | 0.439 [0.431, 0.449] | 0.546 [0.536, 0.557] |
| | 10% | Item FE + demographics | 3.902 [3.892, 3.912] | 3.903 [3.831, 3.976] | 0.110 [0.098, 0.138] | 0.424 [0.418, 0.432] | 0.546 [0.538, 0.554] |
| | 10% | Item FE + LLM + demographics | 3.902 [3.892, 3.912] | 3.902 [3.831, 3.972] | 0.328 [0.306, 0.351] | 0.441 [0.432, 0.451] | 0.547 [0.537, 0.558] |
| | 25% | Digital twin | 3.902 [3.885, 3.919] | 4.232 [4.217, 4.247] | 0.331 [0.317, 0.344] | 0.483 [0.472, 0.493] | 0.637 [0.622, 0.649] |
| | 25% | Item FE | 3.902 [3.885, 3.919] | 3.901 [3.853, 3.948] | 0.108 [0.095, 0.129] | 0.423 [0.415, 0.432] | 0.545 [0.534, 0.556] |
| | 25% | Item FE + LLM | 3.902 [3.885, 3.919] | 3.902 [3.861, 3.940] | 0.326 [0.309, 0.343] | 0.439 [0.430, 0.448] | 0.545 [0.534, 0.556] |
| | 25% | Item FE + demographics | 3.902 [3.885, 3.919] | 3.901 [3.853, 3.948] | 0.107 [0.095, 0.128] | 0.423 [0.415, 0.432] | 0.545 [0.534, 0.556] |
| | 25% | Item FE + LLM + demographics | 3.902 [3.885, 3.919] | 3.901 [3.861, 3.941] | 0.327 [0.310, 0.343] | 0.440 [0.431, 0.449] | 0.546 [0.535, 0.557] |
| | 50% | Digital twin | 3.902 [3.873, 3.932] | 4.232 [4.208, 4.258] | 0.331 [0.308, 0.355] | 0.483 [0.465, 0.501] | 0.636 [0.614, 0.659] |
| | 50% | Item FE | 3.902 [3.873, 3.932] | 3.902 [3.864, 3.940] | 0.107 [0.091, 0.131] | 0.422 [0.408, 0.437] | 0.545 [0.524, 0.565] |
| | 50% | Item FE + LLM | 3.902 [3.873, 3.932] | 3.902 [3.879, 3.926] | 0.326 [0.306, 0.345] | 0.439 [0.424, 0.453] | 0.545 [0.528, 0.563] |
| | 50% | Item FE + demographics | 3.902 [3.873, 3.932] | 3.902 [3.865, 3.941] | 0.106 [0.090, 0.131] | 0.423 [0.409, 0.437] | 0.545 [0.525, 0.565] |
| | 50% | Item FE + LLM + demographics | 3.902 [3.873, 3.932] | 3.902 [3.879, 3.926] | 0.327 [0.306, 0.345] | 0.439 [0.425, 0.454] | 0.545 [0.529, 0.563] |
| Gemma-4-31B-it | 10% | Digital twin | 3.902 [3.892, 3.912] | 4.257 [4.249, 4.266] | 0.355 [0.347, 0.362] | 0.476 [0.470, 0.481] | 0.608 [0.601, 0.614] |
| | 10% | Item FE | 3.902 [3.892, 3.912] | 3.902 [3.833, 3.969] | 0.120 [0.112, 0.138] | 0.390 [0.384, 0.397] | 0.496 [0.489, 0.502] |
| | 10% | Item FE + LLM | 3.902 [3.892, 3.912] | 3.901 [3.833, 3.966] | 0.280 [0.259, 0.300] | 0.401 [0.393, 0.410] | 0.496 [0.487, 0.505] |
| | 10% | Item FE + demographics | 3.902 [3.892, 3.912] | 3.902 [3.833, 3.967] | 0.118 [0.110, 0.134] | 0.390 [0.385, 0.397] | 0.496 [0.490, 0.503] |
| | 10% | Item FE + LLM + demographics | 3.902 [3.892, 3.912] | 3.901 [3.833, 3.967] | 0.280 [0.258, 0.300] | 0.402 [0.394, 0.411] | 0.497 [0.488, 0.507] |
| | 25% | Digital twin | 3.902 [3.885, 3.919] | 4.256 [4.242, 4.272] | 0.354 [0.342, 0.366] | 0.476 [0.466, 0.485] | 0.608 [0.597, 0.619] |
| | 25% | Item FE | 3.902 [3.885, 3.919] | 3.901 [3.858, 3.944] | 0.118 [0.110, 0.128] | 0.389 [0.381, 0.398] | 0.495 [0.485, 0.505] |
| | 25% | Item FE + LLM | 3.902 [3.885, 3.919] | 3.901 [3.864, 3.936] | 0.279 [0.264, 0.295] | 0.401 [0.393, 0.409] | 0.495 [0.486, 0.504] |
| | 25% | Item FE + demographics | 3.902 [3.885, 3.919] | 3.901 [3.858, 3.944] | 0.116 [0.108, 0.127] | 0.390 [0.382, 0.398] | 0.495 [0.485, 0.505] |
| | 25% | Item FE + LLM + demographics | 3.902 [3.885, 3.919] | 3.901 [3.864, 3.936] | 0.279 [0.264, 0.295] | 0.401 [0.394, 0.409] | 0.496 [0.486, 0.505] |
| | 50% | Digital twin | 3.902 [3.873, 3.932] | 4.256 [4.230, 4.283] | 0.354 [0.335, 0.375] | 0.476 [0.461, 0.492] | 0.608 [0.589, 0.627] |
| | 50% | Item FE | 3.902 [3.873, 3.932] | 3.901 [3.864, 3.942] | 0.119 [0.104, 0.133] | 0.389 [0.375, 0.402] | 0.495 [0.476, 0.512] |
| | 50% | Item FE + LLM | 3.902 [3.873, 3.932] | 3.902 [3.880, 3.925] | 0.279 [0.261, 0.298] | 0.401 [0.388, 0.413] | 0.494 [0.479, 0.510] |
| | 50% | Item FE + demographics | 3.902 [3.873, 3.932] | 3.901 [3.863, 3.941] | 0.117 [0.103, 0.132] | 0.389 [0.376, 0.403] | 0.495 [0.477, 0.512] |
| | 50% | Item FE + LLM + demographics | 3.902 [3.873, 3.932] | 3.902 [3.880, 3.925] | 0.279 [0.260, 0.298] | 0.401 [0.389, 0.414] | 0.495 [0.480, 0.511] |
| gpt-oss-120b | 10% | Digital twin | 3.902 [3.892, 3.912] | 4.232 [4.222, 4.243] | 0.338 [0.330, 0.346] | 0.545 [0.539, 0.552] | 0.698 [0.689, 0.705] |
| | 10% | Item FE | 3.902 [3.892, 3.912] | 3.900 [3.817, 3.985] | 0.155 [0.143, 0.183] | 0.471 [0.464, 0.478] | 0.617 [0.608, 0.625] |
| | 10% | Item FE + LLM | 3.902 [3.892, 3.912] | 3.901 [3.829, 3.974] | 0.343 [0.318, 0.367] | 0.446 [0.437, 0.456] | 0.556 [0.545, 0.568] |
| | 10% | Item FE + demographics | 3.902 [3.892, 3.912] | 3.900 [3.816, 3.986] | 0.153 [0.139, 0.181] | 0.471 [0.464, 0.478] | 0.617 [0.608, 0.626] |
| | 10% | Item FE + LLM + demographics | 3.902 [3.892, 3.912] | 3.901 [3.829, 3.973] | 0.344 [0.319, 0.368] | 0.447 [0.437, 0.457] | 0.557 [0.546, 0.569] |
| | 25% | Digital twin | 3.902 [3.885, 3.919] | 4.231 [4.213, 4.250] | 0.338 [0.324, 0.352] | 0.545 [0.534, 0.556] | 0.698 [0.684, 0.711] |
| | 25% | Item FE | 3.902 [3.885, 3.919] | 3.900 [3.844, 3.956] | 0.153 [0.140, 0.174] | 0.469 [0.459, 0.480] | 0.616 [0.602, 0.629] |
| | 25% | Item FE + LLM | 3.902 [3.885, 3.919] | 3.901 [3.862, 3.942] | 0.342 [0.324, 0.360] | 0.445 [0.436, 0.455] | 0.555 [0.543, 0.566] |
| | 25% | Item FE + demographics | 3.902 [3.885, 3.919] | 3.900 [3.844, 3.956] | 0.152 [0.138, 0.173] | 0.470 [0.460, 0.480] | 0.616 [0.602, 0.629] |
| | 25% | Item FE + LLM + demographics | 3.902 [3.885, 3.919] | 3.901 [3.862, 3.942] | 0.343 [0.324, 0.360] | 0.446 [0.436, 0.455] | 0.555 [0.544, 0.567] |
| | 50% | Digital twin | 3.902 [3.873, 3.932] | 4.231 [4.200, 4.263] | 0.339 [0.315, 0.362] | 0.545 [0.527, 0.563] | 0.697 [0.674, 0.720] |
| | 50% | Item FE | 3.902 [3.873, 3.932] | 3.901 [3.852, 3.953] | 0.152 [0.132, 0.175] | 0.468 [0.451, 0.486] | 0.615 [0.591, 0.638] |
| | 50% | Item FE + LLM | 3.902 [3.873, 3.932] | 3.901 [3.878, 3.925] | 0.342 [0.321, 0.363] | 0.445 [0.430, 0.459] | 0.554 [0.536, 0.573] |
| | 50% | Item FE + demographics | 3.902 [3.873, 3.932] | 3.901 [3.852, 3.953] | 0.151 [0.131, 0.174] | 0.469 [0.451, 0.486] | 0.615 [0.592, 0.638] |
| | 50% | Item FE + LLM + demographics | 3.902 [3.873, 3.932] | 3.901 [3.878, 3.925] | 0.342 [0.322, 0.363] | 0.445 [0.431, 0.460] | 0.555 [0.536, 0.573] |
| Qwen 3.5 Plus | 10% | Digital twin | 3.902 [3.892, 3.912] | 4.141 [4.133, 4.150] | 0.246 [0.239, 0.253] | 0.424 [0.419, 0.430] | 0.556 [0.549, 0.562] |
| | 10% | Item FE | 3.902 [3.892, 3.912] | 3.901 [3.831, 3.969] | 0.108 [0.100, 0.135] | 0.392 [0.386, 0.399] | 0.503 [0.496, 0.511] |
| | 10% | Item FE + LLM | 3.902 [3.892, 3.912] | 3.901 [3.833, 3.962] | 0.261 [0.237, 0.285] | 0.403 [0.395, 0.412] | 0.503 [0.495, 0.512] |
| | 10% | Item FE + demographics | 3.902 [3.892, 3.912] | 3.901 [3.830, 3.969] | 0.105 [0.095, 0.133] | 0.393 [0.387, 0.399] | 0.504 [0.497, 0.511] |
| | 10% | Item FE + LLM + demographics | 3.902 [3.892, 3.912] | 3.901 [3.832, 3.962] | 0.260 [0.237, 0.284] | 0.404 [0.396, 0.413] | 0.504 [0.495, 0.513] |
| | 25% | Digital twin | 3.902 [3.885, 3.919] | 4.141 [4.126, 4.156] | 0.246 [0.233, 0.258] | 0.424 [0.415, 0.434] | 0.556 [0.544, 0.567] |
| | 25% | Item FE | 3.902 [3.885, 3.919] | 3.901 [3.857, 3.944] | 0.106 [0.098, 0.123] | 0.392 [0.384, 0.400] | 0.503 [0.491, 0.513] |
| | 25% | Item FE + LLM | 3.902 [3.885, 3.919] | 3.901 [3.861, 3.937] | 0.260 [0.243, 0.277] | 0.402 [0.394, 0.410] | 0.502 [0.493, 0.512] |
| | 25% | Item FE + demographics | 3.902 [3.885, 3.919] | 3.901 [3.857, 3.944] | 0.104 [0.095, 0.122] | 0.392 [0.384, 0.400] | 0.503 [0.492, 0.513] |
| | 25% | Item FE + LLM + demographics | 3.902 [3.885, 3.919] | 3.901 [3.861, 3.937] | 0.260 [0.242, 0.277] | 0.403 [0.395, 0.411] | 0.503 [0.493, 0.512] |
| | 50% | Digital twin | 3.902 [3.873, 3.932] | 4.141 [4.117, 4.168] | 0.246 [0.226, 0.266] | 0.424 [0.409, 0.440] | 0.555 [0.537, 0.575] |
| | 50% | Item FE | 3.902 [3.873, 3.932] | 3.901 [3.867, 3.940] | 0.106 [0.094, 0.123] | 0.391 [0.377, 0.404] | 0.502 [0.483, 0.520] |
| | 50% | Item FE + LLM | 3.902 [3.873, 3.932] | 3.902 [3.879, 3.925] | 0.260 [0.241, 0.279] | 0.402 [0.389, 0.415] | 0.501 [0.486, 0.517] |
| | 50% | Item FE + demographics | 3.902 [3.873, 3.932] | 3.901 [3.867, 3.940] | 0.105 [0.091, 0.123] | 0.391 [0.378, 0.404] | 0.502 [0.483, 0.520] |
| | 50% | Item FE + LLM + demographics | 3.902 [3.873, 3.932] | 3.901 [3.878, 3.924] | 0.260 [0.241, 0.279] | 0.402 [0.390, 0.415] | 0.502 [0.487, 0.518] |
| Conscientiousness | | | | | | | |
| DeepSeek V4-Flash | 10% | Digital twin | 3.935 [3.924, 3.946] | 4.325 [4.316, 4.336] | 0.394 [0.386, 0.401] | 0.513 [0.507, 0.519] | 0.676 [0.668, 0.683] |
| | 10% | Item FE | 3.935 [3.924, 3.946] | 3.935 [3.857, 4.007] | 0.172 [0.159, 0.190] | 0.432 [0.426, 0.440] | 0.554 [0.546, 0.561] |
| | 10% | Item FE + LLM | 3.935 [3.924, 3.946] | 3.934 [3.863, 4.004] | 0.369 [0.345, 0.393] | 0.465 [0.453, 0.477] | 0.565 [0.552, 0.578] |
| | 10% | Item FE + demographics | 3.935 [3.924, 3.946] | 3.935 [3.857, 4.007] | 0.170 [0.158, 0.189] | 0.432 [0.426, 0.440] | 0.554 [0.546, 0.562] |
| | 10% | Item FE + LLM + demographics | 3.935 [3.924, 3.946] | 3.934 [3.863, 4.005] | 0.370 [0.345, 0.396] | 0.467 [0.455, 0.480] | 0.566 [0.553, 0.581] |
| | 25% | Digital twin | 3.935 [3.917, 3.954] | 4.325 [4.308, 4.343] | 0.394 [0.381, 0.407] | 0.513 [0.502, 0.523] | 0.676 [0.662, 0.689] |
| | 25% | Item FE | 3.935 [3.917, 3.954] | 3.934 [3.883, 3.983] | 0.171 [0.157, 0.187] | 0.432 [0.423, 0.441] | 0.553 [0.540, 0.565] |
| | 25% | Item FE + LLM | 3.935 [3.917, 3.954] | 3.934 [3.893, 3.972] | 0.368 [0.350, 0.387] | 0.465 [0.455, 0.475] | 0.564 [0.552, 0.576] |
| | 25% | Item FE + demographics | 3.935 [3.917, 3.954] | 3.934 [3.883, 3.984] | 0.170 [0.156, 0.186] | 0.432 [0.423, 0.441] | 0.553 [0.541, 0.565] |
| | 25% | Item FE + LLM + demographics | 3.935 [3.917, 3.954] | 3.934 [3.893, 3.972] | 0.369 [0.351, 0.387] | 0.466 [0.456, 0.477] | 0.565 [0.553, 0.577] |
| | 50% | Digital twin | 3.935 [3.904, 3.968] | 4.325 [4.298, 4.356] | 0.394 [0.372, 0.416] | 0.513 [0.495, 0.531] | 0.676 [0.652, 0.699] |
| | 50% | Item FE | 3.935 [3.904, 3.968] | 3.934 [3.894, 3.977] | 0.171 [0.153, 0.192] | 0.432 [0.417, 0.446] | 0.553 [0.531, 0.573] |
| | 50% | Item FE + LLM | 3.935 [3.904, 3.968] | 3.934 [3.911, 3.957] | 0.368 [0.347, 0.390] | 0.464 [0.450, 0.479] | 0.563 [0.546, 0.581] |
| | 50% | Item FE + demographics | 3.935 [3.904, 3.968] | 3.934 [3.894, 3.977] | 0.171 [0.152, 0.191] | 0.432 [0.417, 0.446] | 0.553 [0.531, 0.573] |
| | 50% | Item FE + LLM + demographics | 3.935 [3.904, 3.968] | 3.934 [3.910, 3.957] | 0.369 [0.349, 0.391] | 0.465 [0.451, 0.480] | 0.564 [0.547, 0.583] |
| Gemma-4-31B-it | 10% | Digital twin | 3.935 [3.924, 3.946] | 4.261 [4.250, 4.271] | 0.326 [0.319, 0.333] | 0.445 [0.439, 0.451] | 0.584 [0.577, 0.590] |
| | 10% | Item FE | 3.935 [3.924, 3.946] | 3.934 [3.869, 4.000] | 0.161 [0.152, 0.178] | 0.384 [0.379, 0.392] | 0.486 [0.480, 0.492] |
| | 10% | Item FE + LLM | 3.935 [3.924, 3.946] | 3.934 [3.872, 3.998] | 0.306 [0.284, 0.331] | 0.409 [0.399, 0.420] | 0.498 [0.487, 0.510] |
| | 10% | Item FE + demographics | 3.935 [3.924, 3.946] | 3.934 [3.870, 3.999] | 0.160 [0.151, 0.176] | 0.384 [0.379, 0.392] | 0.486 [0.480, 0.493] |
| | 10% | Item FE + LLM + demographics | 3.935 [3.924, 3.946] | 3.934 [3.872, 3.997] | 0.308 [0.285, 0.333] | 0.411 [0.399, 0.422] | 0.499 [0.488, 0.512] |
| | 25% | Digital twin | 3.935 [3.917, 3.954] | 4.261 [4.244, 4.279] | 0.326 [0.314, 0.338] | 0.445 [0.436, 0.455] | 0.584 [0.572, 0.595] |
| | 25% | Item FE | 3.935 [3.917, 3.954] | 3.934 [3.889, 3.980] | 0.160 [0.149, 0.175] | 0.383 [0.376, 0.391] | 0.486 [0.475, 0.495] |
| | 25% | Item FE + LLM | 3.935 [3.917, 3.954] | 3.934 [3.897, 3.968] | 0.305 [0.287, 0.324] | 0.409 [0.400, 0.419] | 0.497 [0.487, 0.508] |
| | 25% | Item FE + demographics | 3.935 [3.917, 3.954] | 3.934 [3.889, 3.980] | 0.159 [0.148, 0.174] | 0.384 [0.376, 0.392] | 0.486 [0.475, 0.496] |
| | 25% | Item FE + LLM + demographics | 3.935 [3.917, 3.954] | 3.934 [3.897, 3.968] | 0.307 [0.288, 0.326] | 0.410 [0.401, 0.420] | 0.498 [0.488, 0.509] |
| | 50% | Digital twin | 3.935 [3.904, 3.968] | 4.261 [4.233, 4.291] | 0.326 [0.306, 0.347] | 0.446 [0.429, 0.462] | 0.584 [0.565, 0.604] |
| | 50% | Item FE | 3.935 [3.904, 3.968] | 3.935 [3.897, 3.977] | 0.160 [0.142, 0.179] | 0.383 [0.371, 0.396] | 0.486 [0.469, 0.502] |
| | 50% | Item FE + LLM | 3.935 [3.904, 3.968] | 3.935 [3.911, 3.958] | 0.305 [0.286, 0.326] | 0.409 [0.396, 0.423] | 0.497 [0.482, 0.513] |
| | 50% | Item FE + demographics | 3.935 [3.904, 3.968] | 3.935 [3.897, 3.977] | 0.159 [0.142, 0.179] | 0.383 [0.371, 0.396] | 0.486 [0.469, 0.502] |
| | 50% | Item FE + LLM + demographics | 3.935 [3.904, 3.968] | 3.935 [3.911, 3.958] | 0.306 [0.287, 0.328] | 0.409 [0.396, 0.423] | 0.498 [0.482, 0.514] |
| gpt-oss-120b | 10% | Digital twin | 3.935 [3.924, 3.946] | 4.322 [4.311, 4.334] | 0.417 [0.409, 0.424] | 0.534 [0.528, 0.541] | 0.690 [0.683, 0.697] |
| | 10% | Item FE | 3.935 [3.924, 3.946] | 3.931 [3.849, 4.014] | 0.213 [0.198, 0.236] | 0.442 [0.436, 0.449] | 0.573 [0.565, 0.581] |
| | 10% | Item FE + LLM | 3.935 [3.924, 3.946] | 3.933 [3.860, 4.007] | 0.402 [0.378, 0.428] | 0.470 [0.456, 0.484] | 0.569 [0.554, 0.586] |
| | 10% | Item FE + demographics | 3.935 [3.924, 3.946] | 3.931 [3.849, 4.013] | 0.210 [0.197, 0.232] | 0.442 [0.436, 0.449] | 0.573 [0.565, 0.582] |
| | 10% | Item FE + LLM + demographics | 3.935 [3.924, 3.946] | 3.933 [3.860, 4.006] | 0.403 [0.378, 0.429] | 0.472 [0.458, 0.487] | 0.571 [0.555, 0.589] |
| | 25% | Digital twin | 3.935 [3.917, 3.954] | 4.323 [4.303, 4.342] | 0.417 [0.404, 0.430] | 0.534 [0.524, 0.545] | 0.690 [0.677, 0.704] |
| | 25% | Item FE | 3.935 [3.917, 3.954] | 3.933 [3.878, 3.990] | 0.213 [0.199, 0.231] | 0.441 [0.432, 0.451] | 0.572 [0.560, 0.585] |
| | 25% | Item FE + LLM | 3.935 [3.917, 3.954] | 3.934 [3.892, 3.974] | 0.401 [0.383, 0.420] | 0.470 [0.458, 0.482] | 0.568 [0.554, 0.581] |
| | 25% | Item FE + demographics | 3.935 [3.917, 3.954] | 3.933 [3.877, 3.989] | 0.211 [0.196, 0.229] | 0.441 [0.432, 0.451] | 0.572 [0.560, 0.585] |
| | 25% | Item FE + LLM + demographics | 3.935 [3.917, 3.954] | 3.934 [3.892, 3.974] | 0.402 [0.384, 0.421] | 0.471 [0.459, 0.483] | 0.570 [0.556, 0.583] |
| | 50% | Digital twin | 3.935 [3.904, 3.968] | 4.322 [4.290, 4.356] | 0.417 [0.395, 0.438] | 0.534 [0.517, 0.553] | 0.690 [0.668, 0.713] |
| | 50% | Item FE | 3.935 [3.904, 3.968] | 3.933 [3.885, 3.983] | 0.213 [0.194, 0.233] | 0.441 [0.425, 0.457] | 0.572 [0.550, 0.593] |
| | 50% | Item FE + LLM | 3.935 [3.904, 3.968] | 3.934 [3.912, 3.956] | 0.400 [0.381, 0.421] | 0.469 [0.455, 0.485] | 0.568 [0.551, 0.586] |
| | 50% | Item FE + demographics | 3.935 [3.904, 3.968] | 3.933 [3.885, 3.984] | 0.212 [0.193, 0.232] | 0.441 [0.425, 0.457] | 0.572 [0.550, 0.593] |
| | 50% | Item FE + LLM + demographics | 3.935 [3.904, 3.968] | 3.934 [3.912, 3.956] | 0.402 [0.382, 0.422] | 0.471 [0.456, 0.486] | 0.569 [0.552, 0.587] |
| Qwen 3.5 Plus | 10% | Digital twin | 3.935 [3.924, 3.946] | 4.237 [4.226, 4.248] | 0.337 [0.329, 0.343] | 0.464 [0.458, 0.469] | 0.612 [0.604, 0.618] |
| | 10% | Item FE | 3.935 [3.924, 3.946] | 3.934 [3.862, 4.007] | 0.205 [0.196, 0.218] | 0.415 [0.409, 0.423] | 0.533 [0.526, 0.541] |
| | 10% | Item FE + LLM | 3.935 [3.924, 3.946] | 3.934 [3.867, 3.997] | 0.289 [0.265, 0.318] | 0.410 [0.402, 0.418] | 0.500 [0.492, 0.509] |
| | 10% | Item FE + demographics | 3.935 [3.924, 3.946] | 3.934 [3.861, 4.006] | 0.204 [0.195, 0.217] | 0.415 [0.409, 0.423] | 0.533 [0.526, 0.541] |
| | 10% | Item FE + LLM + demographics | 3.935 [3.924, 3.946] | 3.934 [3.868, 3.997] | 0.289 [0.264, 0.318] | 0.411 [0.403, 0.420] | 0.501 [0.493, 0.510] |
| | 25% | Digital twin | 3.935 [3.917, 3.954] | 4.237 [4.218, 4.255] | 0.337 [0.326, 0.349] | 0.464 [0.455, 0.475] | 0.612 [0.600, 0.624] |
| | 25% | Item FE | 3.935 [3.917, 3.954] | 3.934 [3.886, 3.982] | 0.205 [0.192, 0.219] | 0.415 [0.406, 0.423] | 0.533 [0.521, 0.545] |
| | 25% | Item FE + LLM | 3.935 [3.917, 3.954] | 3.934 [3.898, 3.973] | 0.289 [0.268, 0.308] | 0.409 [0.401, 0.417] | 0.499 [0.489, 0.508] |
| | 25% | Item FE + demographics | 3.935 [3.917, 3.954] | 3.934 [3.887, 3.982] | 0.204 [0.191, 0.218] | 0.415 [0.406, 0.423] | 0.533 [0.521, 0.544] |
| | 25% | Item FE + LLM + demographics | 3.935 [3.917, 3.954] | 3.934 [3.898, 3.973] | 0.289 [0.269, 0.309] | 0.410 [0.402, 0.418] | 0.500 [0.490, 0.509] |
| | 50% | Digital twin | 3.935 [3.904, 3.968] | 4.236 [4.206, 4.270] | 0.337 [0.317, 0.357] | 0.464 [0.447, 0.480] | 0.612 [0.591, 0.631] |
| | 50% | Item FE | 3.935 [3.904, 3.968] | 3.935 [3.892, 3.982] | 0.205 [0.186, 0.222] | 0.415 [0.400, 0.429] | 0.533 [0.513, 0.552] |
| | 50% | Item FE + LLM | 3.935 [3.904, 3.968] | 3.935 [3.907, 3.963] | 0.289 [0.269, 0.311] | 0.409 [0.397, 0.422] | 0.499 [0.483, 0.514] |
| | 50% | Item FE + demographics | 3.935 [3.904, 3.968] | 3.934 [3.892, 3.981] | 0.204 [0.186, 0.222] | 0.415 [0.400, 0.429] | 0.533 [0.513, 0.552] |
| | 50% | Item FE + LLM + demographics | 3.935 [3.904, 3.968] | 3.935 [3.907, 3.963] | 0.290 [0.270, 0.312] | 0.410 [0.397, 0.423] | 0.499 [0.484, 0.515] |
| Neuroticism | | | | | | | |
| DeepSeek V4-Flash | 10% | Digital twin | 2.715 [2.700, 2.729] | 2.800 [2.786, 2.816] | 0.100 [0.092, 0.107] | 0.567 [0.560, 0.575] | 0.737 [0.726, 0.745] |
| | 10% | Item FE | 2.715 [2.700, 2.729] | 2.716 [2.617, 2.817] | 0.099 [0.088, 0.136] | 0.567 [0.560, 0.575] | 0.733 [0.723, 0.744] |
| | 10% | Item FE + LLM | 2.715 [2.700, 2.729] | 2.715 [2.628, 2.811] | 0.418 [0.379, 0.461] | 0.593 [0.578, 0.611] | 0.721 [0.705, 0.740] |
| | 10% | Item FE + demographics | 2.715 [2.700, 2.729] | 2.715 [2.616, 2.816] | 0.095 [0.082, 0.131] | 0.568 [0.560, 0.576] | 0.735 [0.725, 0.745] |
| | 10% | Item FE + LLM + demographics | 2.715 [2.700, 2.729] | 2.715 [2.627, 2.811] | 0.418 [0.378, 0.461] | 0.594 [0.578, 0.612] | 0.722 [0.706, 0.741] |
| | 25% | Digital twin | 2.714 [2.688, 2.738] | 2.801 [2.777, 2.826] | 0.101 [0.087, 0.114] | 0.567 [0.556, 0.579] | 0.737 [0.721, 0.751] |
| | 25% | Item FE | 2.714 [2.688, 2.738] | 2.716 [2.646, 2.785] | 0.097 [0.087, 0.113] | 0.566 [0.555, 0.577] | 0.732 [0.717, 0.747] |
| | 25% | Item FE + LLM | 2.714 [2.688, 2.738] | 2.716 [2.666, 2.766] | 0.417 [0.386, 0.444] | 0.592 [0.578, 0.606] | 0.720 [0.705, 0.734] |
| | 25% | Item FE + demographics | 2.714 [2.688, 2.738] | 2.716 [2.646, 2.784] | 0.092 [0.081, 0.110] | 0.566 [0.555, 0.577] | 0.733 [0.717, 0.748] |
| | 25% | Item FE + LLM + demographics | 2.714 [2.688, 2.738] | 2.716 [2.666, 2.765] | 0.416 [0.385, 0.443] | 0.592 [0.577, 0.607] | 0.720 [0.705, 0.734] |
| | 50% | Digital twin | 2.714 [2.671, 2.754] | 2.800 [2.758, 2.843] | 0.103 [0.080, 0.126] | 0.567 [0.547, 0.589] | 0.736 [0.710, 0.763] |
| | 50% | Item FE | 2.714 [2.671, 2.754] | 2.715 [2.654, 2.773] | 0.098 [0.084, 0.117] | 0.566 [0.545, 0.587] | 0.732 [0.705, 0.758] |
| | 50% | Item FE + LLM | 2.714 [2.671, 2.754] | 2.715 [2.683, 2.747] | 0.417 [0.386, 0.446] | 0.592 [0.573, 0.612] | 0.719 [0.699, 0.740] |
| | 50% | Item FE + demographics | 2.714 [2.671, 2.754] | 2.715 [2.653, 2.774] | 0.095 [0.079, 0.115] | 0.566 [0.545, 0.587] | 0.732 [0.706, 0.759] |
| | 50% | Item FE + LLM + demographics | 2.714 [2.671, 2.754] | 2.715 [2.683, 2.747] | 0.416 [0.385, 0.444] | 0.591 [0.572, 0.611] | 0.719 [0.698, 0.740] |
| Gemma-4-31B-it | 10% | Digital twin | 2.715 [2.700, 2.729] | 3.209 [3.194, 3.225] | 0.494 [0.483, 0.505] | 0.705 [0.696, 0.713] | 0.900 [0.890, 0.910] |
| | 10% | Item FE | 2.715 [2.700, 2.729] | 2.717 [2.613, 2.819] | 0.191 [0.182, 0.206] | 0.586 [0.578, 0.595] | 0.754 [0.745, 0.765] |
| | 10% | Item FE + LLM | 2.715 [2.700, 2.729] | 2.716 [2.630, 2.804] | 0.328 [0.287, 0.372] | 0.563 [0.553, 0.574] | 0.698 [0.689, 0.710] |
| | 10% | Item FE + demographics | 2.715 [2.700, 2.729] | 2.717 [2.611, 2.820] | 0.188 [0.180, 0.203] | 0.587 [0.579, 0.596] | 0.756 [0.746, 0.766] |
| | 10% | Item FE + LLM + demographics | 2.715 [2.700, 2.729] | 2.716 [2.629, 2.805] | 0.327 [0.286, 0.371] | 0.564 [0.554, 0.575] | 0.699 [0.690, 0.711] |
| | 25% | Digital twin | 2.714 [2.688, 2.738] | 3.208 [3.184, 3.234] | 0.494 [0.476, 0.513] | 0.705 [0.691, 0.718] | 0.900 [0.883, 0.915] |
| | 25% | Item FE | 2.714 [2.688, 2.738] | 2.715 [2.648, 2.782] | 0.189 [0.177, 0.202] | 0.585 [0.574, 0.597] | 0.753 [0.738, 0.768] |
| | 25% | Item FE + LLM | 2.714 [2.688, 2.738] | 2.716 [2.665, 2.767] | 0.327 [0.296, 0.359] | 0.562 [0.551, 0.573] | 0.697 [0.685, 0.709] |
| | 25% | Item FE + demographics | 2.714 [2.688, 2.738] | 2.715 [2.648, 2.782] | 0.187 [0.175, 0.199] | 0.586 [0.574, 0.597] | 0.754 [0.739, 0.768] |
| | 25% | Item FE + LLM + demographics | 2.714 [2.688, 2.738] | 2.716 [2.664, 2.767] | 0.326 [0.294, 0.358] | 0.562 [0.551, 0.574] | 0.697 [0.685, 0.710] |
| | 50% | Digital twin | 2.714 [2.671, 2.754] | 3.208 [3.165, 3.248] | 0.494 [0.463, 0.524] | 0.704 [0.678, 0.727] | 0.899 [0.869, 0.929] |
| | 50% | Item FE | 2.714 [2.671, 2.754] | 2.714 [2.653, 2.772] | 0.188 [0.171, 0.208] | 0.584 [0.564, 0.606] | 0.752 [0.725, 0.777] |
| | 50% | Item FE + LLM | 2.714 [2.671, 2.754] | 2.715 [2.678, 2.748] | 0.326 [0.296, 0.358] | 0.561 [0.543, 0.579] | 0.696 [0.675, 0.716] |
| | 50% | Item FE + demographics | 2.714 [2.671, 2.754] | 2.714 [2.653, 2.772] | 0.187 [0.170, 0.206] | 0.585 [0.565, 0.606] | 0.753 [0.726, 0.778] |
| | 50% | Item FE + LLM + demographics | 2.714 [2.671, 2.754] | 2.715 [2.678, 2.748] | 0.325 [0.295, 0.358] | 0.561 [0.543, 0.580] | 0.696 [0.675, 0.716] |
| gpt-oss-120b | 10% | Digital twin | 2.715 [2.700, 2.729] | 2.541 [2.523, 2.557] | 0.265 [0.256, 0.274] | 0.713 [0.704, 0.721] | 0.919 [0.908, 0.928] |
| | 10% | Item FE | 2.715 [2.700, 2.729] | 2.719 [2.591, 2.840] | 0.238 [0.227, 0.256] | 0.704 [0.695, 0.714] | 0.904 [0.893, 0.915] |
| | 10% | Item FE + LLM | 2.715 [2.700, 2.729] | 2.717 [2.620, 2.815] | 0.453 [0.408, 0.498] | 0.640 [0.625, 0.656] | 0.781 [0.766, 0.798] |
| | 10% | Item FE + demographics | 2.715 [2.700, 2.729] | 2.719 [2.592, 2.839] | 0.238 [0.226, 0.256] | 0.706 [0.696, 0.715] | 0.906 [0.894, 0.917] |
| | 10% | Item FE + LLM + demographics | 2.715 [2.700, 2.729] | 2.717 [2.617, 2.811] | 0.453 [0.409, 0.497] | 0.641 [0.625, 0.656] | 0.781 [0.766, 0.800] |
| | 25% | Digital twin | 2.714 [2.688, 2.738] | 2.540 [2.512, 2.569] | 0.265 [0.250, 0.281] | 0.713 [0.698, 0.728] | 0.919 [0.899, 0.936] |
| | 25% | Item FE | 2.714 [2.688, 2.738] | 2.717 [2.631, 2.801] | 0.237 [0.222, 0.252] | 0.703 [0.689, 0.717] | 0.903 [0.885, 0.920] |
| | 25% | Item FE + LLM | 2.714 [2.688, 2.738] | 2.716 [2.663, 2.773] | 0.452 [0.421, 0.483] | 0.639 [0.626, 0.654] | 0.780 [0.765, 0.794] |
| | 25% | Item FE + demographics | 2.714 [2.688, 2.738] | 2.717 [2.630, 2.800] | 0.237 [0.222, 0.253] | 0.704 [0.690, 0.718] | 0.904 [0.886, 0.921] |
| | 25% | Item FE + LLM + demographics | 2.714 [2.688, 2.738] | 2.716 [2.662, 2.773] | 0.451 [0.421, 0.483] | 0.639 [0.625, 0.653] | 0.779 [0.764, 0.793] |
| | 50% | Digital twin | 2.714 [2.671, 2.754] | 2.541 [2.493, 2.588] | 0.265 [0.236, 0.292] | 0.713 [0.687, 0.738] | 0.918 [0.885, 0.951] |
| | 50% | Item FE | 2.714 [2.671, 2.754] | 2.716 [2.637, 2.791] | 0.237 [0.211, 0.262] | 0.703 [0.679, 0.727] | 0.902 [0.871, 0.934] |
| | 50% | Item FE + LLM | 2.714 [2.671, 2.754] | 2.715 [2.682, 2.750] | 0.451 [0.418, 0.485] | 0.639 [0.618, 0.659] | 0.779 [0.757, 0.801] |
| | 50% | Item FE + demographics | 2.714 [2.671, 2.754] | 2.716 [2.637, 2.791] | 0.238 [0.210, 0.263] | 0.703 [0.679, 0.728] | 0.903 [0.871, 0.935] |
| | 50% | Item FE + LLM + demographics | 2.714 [2.671, 2.754] | 2.715 [2.682, 2.750] | 0.451 [0.419, 0.485] | 0.638 [0.617, 0.659] | 0.778 [0.756, 0.801] |
| Qwen 3.5 Plus | 10% | Digital twin | 2.715 [2.700, 2.729] | 3.078 [3.064, 3.092] | 0.364 [0.353, 0.374] | 0.635 [0.627, 0.642] | 0.813 [0.804, 0.822] |
| | 10% | Item FE | 2.715 [2.700, 2.729] | 2.716 [2.610, 2.810] | 0.158 [0.148, 0.191] | 0.568 [0.560, 0.577] | 0.729 [0.719, 0.739] |
| | 10% | Item FE + LLM | 2.715 [2.700, 2.729] | 2.716 [2.626, 2.801] | 0.303 [0.264, 0.345] | 0.561 [0.551, 0.572] | 0.691 [0.681, 0.703] |
| | 10% | Item FE + demographics | 2.715 [2.700, 2.729] | 2.716 [2.612, 2.812] | 0.145 [0.134, 0.179] | 0.570 [0.562, 0.578] | 0.731 [0.721, 0.741] |
| | 10% | Item FE + LLM + demographics | 2.715 [2.700, 2.729] | 2.716 [2.626, 2.803] | 0.299 [0.258, 0.341] | 0.561 [0.550, 0.572] | 0.691 [0.682, 0.702] |
| | 25% | Digital twin | 2.714 [2.688, 2.738] | 3.078 [3.054, 3.102] | 0.364 [0.347, 0.382] | 0.635 [0.623, 0.647] | 0.813 [0.799, 0.828] |
| | 25% | Item FE | 2.714 [2.688, 2.738] | 2.716 [2.653, 2.780] | 0.154 [0.144, 0.173] | 0.567 [0.556, 0.579] | 0.728 [0.715, 0.743] |
| | 25% | Item FE + LLM | 2.714 [2.688, 2.738] | 2.716 [2.667, 2.767] | 0.302 [0.273, 0.330] | 0.560 [0.549, 0.571] | 0.690 [0.677, 0.702] |
| | 25% | Item FE + demographics | 2.714 [2.688, 2.738] | 2.716 [2.652, 2.780] | 0.147 [0.136, 0.166] | 0.568 [0.557, 0.580] | 0.730 [0.716, 0.744] |
| | 25% | Item FE + LLM + demographics | 2.714 [2.688, 2.738] | 2.716 [2.668, 2.767] | 0.298 [0.268, 0.326] | 0.559 [0.548, 0.570] | 0.690 [0.676, 0.702] |
| | 50% | Digital twin | 2.714 [2.671, 2.754] | 3.078 [3.038, 3.120] | 0.364 [0.332, 0.395] | 0.635 [0.613, 0.655] | 0.813 [0.786, 0.839] |
| | 50% | Item FE | 2.714 [2.671, 2.754] | 2.715 [2.656, 2.774] | 0.154 [0.140, 0.173] | 0.567 [0.547, 0.585] | 0.728 [0.703, 0.751] |
| | 50% | Item FE + LLM | 2.714 [2.671, 2.754] | 2.715 [2.678, 2.750] | 0.302 [0.274, 0.330] | 0.559 [0.541, 0.577] | 0.689 [0.669, 0.709] |
| | 50% | Item FE + demographics | 2.714 [2.671, 2.754] | 2.715 [2.656, 2.774] | 0.148 [0.134, 0.167] | 0.568 [0.549, 0.586] | 0.729 [0.704, 0.753] |
| | 50% | Item FE + LLM + demographics | 2.714 [2.671, 2.754] | 2.715 [2.677, 2.750] | 0.298 [0.270, 0.327] | 0.559 [0.540, 0.576] | 0.689 [0.669, 0.708] |
| Openness | | | | | | | |
| DeepSeek V4-Flash | 10% | Digital twin | 3.750 [3.740, 3.760] | 3.404 [3.389, 3.418] | 0.364 [0.354, 0.374] | 0.621 [0.613, 0.629] | 0.814 [0.804, 0.823] |
| | 10% | Item FE | 3.750 [3.740, 3.760] | 3.750 [3.631, 3.860] | 0.254 [0.245, 0.270] | 0.585 [0.577, 0.596] | 0.739 [0.730, 0.749] |
| | 10% | Item FE + LLM | 3.750 [3.740, 3.760] | 3.750 [3.678, 3.816] | 0.164 [0.130, 0.200] | 0.418 [0.414, 0.424] | 0.530 [0.523, 0.536] |
| | 10% | Item FE + demographics | 3.750 [3.740, 3.760] | 3.750 [3.633, 3.860] | 0.255 [0.242, 0.271] | 0.587 [0.575, 0.601] | 0.743 [0.730, 0.756] |
| | 10% | Item FE + LLM + demographics | 3.750 [3.740, 3.760] | 3.750 [3.677, 3.817] | 0.156 [0.123, 0.193] | 0.424 [0.415, 0.433] | 0.537 [0.527, 0.548] |
| | 25% | Digital twin | 3.751 [3.733, 3.769] | 3.404 [3.379, 3.427] | 0.365 [0.349, 0.382] | 0.622 [0.609, 0.634] | 0.815 [0.798, 0.830] |
| | 25% | Item FE | 3.751 [3.733, 3.769] | 3.749 [3.676, 3.822] | 0.253 [0.240, 0.267] | 0.584 [0.573, 0.594] | 0.738 [0.723, 0.751] |
| | 25% | Item FE + LLM | 3.751 [3.733, 3.769] | 3.749 [3.706, 3.790] | 0.161 [0.138, 0.185] | 0.417 [0.409, 0.425] | 0.528 [0.518, 0.538] |
| | 25% | Item FE + demographics | 3.751 [3.733, 3.769] | 3.748 [3.676, 3.822] | 0.252 [0.239, 0.267] | 0.583 [0.571, 0.594] | 0.738 [0.723, 0.752] |
| | 25% | Item FE + LLM + demographics | 3.751 [3.733, 3.769] | 3.748 [3.705, 3.790] | 0.157 [0.133, 0.181] | 0.419 [0.410, 0.428] | 0.531 [0.520, 0.542] |
| | 50% | Digital twin | 3.751 [3.722, 3.780] | 3.404 [3.363, 3.443] | 0.365 [0.337, 0.393] | 0.622 [0.599, 0.644] | 0.815 [0.787, 0.842] |
| | 50% | Item FE | 3.751 [3.722, 3.780] | 3.750 [3.684, 3.816] | 0.253 [0.231, 0.274] | 0.584 [0.565, 0.603] | 0.738 [0.714, 0.762] |
| | 50% | Item FE + LLM | 3.751 [3.722, 3.780] | 3.750 [3.719, 3.781] | 0.161 [0.138, 0.183] | 0.417 [0.403, 0.431] | 0.528 [0.510, 0.546] |
| | 50% | Item FE + demographics | 3.751 [3.722, 3.780] | 3.750 [3.683, 3.816] | 0.251 [0.229, 0.273] | 0.582 [0.563, 0.601] | 0.736 [0.712, 0.761] |
| | 50% | Item FE + LLM + demographics | 3.751 [3.722, 3.780] | 3.750 [3.720, 3.782] | 0.158 [0.135, 0.181] | 0.418 [0.404, 0.432] | 0.529 [0.512, 0.547] |
| Gemma-4-31B-it | 10% | Digital twin | 3.750 [3.740, 3.760] | 3.754 [3.742, 3.767] | 0.165 [0.158, 0.172] | 0.480 [0.474, 0.486] | 0.628 [0.620, 0.635] |
| | 10% | Item FE | 3.750 [3.740, 3.760] | 3.751 [3.664, 3.842] | 0.168 [0.141, 0.208] | 0.482 [0.475, 0.493] | 0.630 [0.621, 0.638] |
| | 10% | Item FE + LLM | 3.750 [3.740, 3.760] | 3.750 [3.687, 3.813] | 0.149 [0.123, 0.179] | 0.400 [0.395, 0.406] | 0.505 [0.498, 0.512] |
| | 10% | Item FE + demographics | 3.750 [3.740, 3.760] | 3.751 [3.661, 3.843] | 0.167 [0.137, 0.209] | 0.487 [0.477, 0.499] | 0.635 [0.623, 0.648] |
| | 10% | Item FE + LLM + demographics | 3.750 [3.740, 3.760] | 3.750 [3.687, 3.815] | 0.141 [0.115, 0.172] | 0.406 [0.397, 0.415] | 0.512 [0.502, 0.523] |
| | 25% | Digital twin | 3.751 [3.733, 3.769] | 3.754 [3.734, 3.777] | 0.165 [0.155, 0.176] | 0.480 [0.470, 0.490] | 0.628 [0.615, 0.641] |
| | 25% | Item FE | 3.751 [3.733, 3.769] | 3.749 [3.688, 3.813] | 0.166 [0.148, 0.192] | 0.481 [0.471, 0.491] | 0.629 [0.616, 0.642] |
| | 25% | Item FE + LLM | 3.751 [3.733, 3.769] | 3.749 [3.708, 3.790] | 0.147 [0.130, 0.168] | 0.399 [0.393, 0.407] | 0.504 [0.493, 0.513] |
| | 25% | Item FE + demographics | 3.751 [3.733, 3.769] | 3.749 [3.690, 3.813] | 0.164 [0.141, 0.193] | 0.483 [0.472, 0.494] | 0.631 [0.617, 0.645] |
| | 25% | Item FE + LLM + demographics | 3.751 [3.733, 3.769] | 3.748 [3.707, 3.790] | 0.142 [0.123, 0.164] | 0.402 [0.394, 0.410] | 0.506 [0.495, 0.517] |
| | 50% | Digital twin | 3.751 [3.722, 3.780] | 3.755 [3.717, 3.790] | 0.166 [0.147, 0.185] | 0.480 [0.463, 0.497] | 0.628 [0.606, 0.651] |
| | 50% | Item FE | 3.751 [3.722, 3.780] | 3.750 [3.692, 3.806] | 0.166 [0.146, 0.189] | 0.481 [0.464, 0.497] | 0.629 [0.606, 0.651] |
| | 50% | Item FE + LLM | 3.751 [3.722, 3.780] | 3.750 [3.718, 3.782] | 0.147 [0.126, 0.168] | 0.399 [0.386, 0.413] | 0.503 [0.486, 0.520] |
| | 50% | Item FE + demographics | 3.751 [3.722, 3.780] | 3.750 [3.693, 3.806] | 0.164 [0.140, 0.189] | 0.481 [0.465, 0.499] | 0.629 [0.606, 0.652] |
| | 50% | Item FE + LLM + demographics | 3.751 [3.722, 3.780] | 3.750 [3.718, 3.783] | 0.143 [0.122, 0.165] | 0.400 [0.386, 0.414] | 0.504 [0.487, 0.521] |
| gpt-oss-120b | 10% | Digital twin | 3.750 [3.740, 3.760] | 3.719 [3.706, 3.731] | 0.206 [0.197, 0.215] | 0.615 [0.606, 0.622] | 0.805 [0.794, 0.814] |
| | 10% | Item FE | 3.750 [3.740, 3.760] | 3.750 [3.639, 3.859] | 0.224 [0.188, 0.277] | 0.620 [0.610, 0.634] | 0.806 [0.795, 0.817] |
| | 10% | Item FE + LLM | 3.750 [3.740, 3.760] | 3.750 [3.676, 3.827] | 0.279 [0.248, 0.310] | 0.469 [0.464, 0.476] | 0.584 [0.577, 0.592] |
| | 10% | Item FE + demographics | 3.750 [3.740, 3.760] | 3.750 [3.642, 3.860] | 0.222 [0.184, 0.275] | 0.623 [0.610, 0.637] | 0.809 [0.795, 0.824] |
| | 10% | Item FE + LLM + demographics | 3.750 [3.740, 3.760] | 3.750 [3.677, 3.826] | 0.266 [0.233, 0.296] | 0.474 [0.466, 0.484] | 0.591 [0.580, 0.602] |
| | 25% | Digital twin | 3.751 [3.733, 3.769] | 3.718 [3.695, 3.741] | 0.206 [0.192, 0.222] | 0.615 [0.602, 0.629] | 0.805 [0.789, 0.822] |
| | 25% | Item FE | 3.751 [3.733, 3.769] | 3.747 [3.673, 3.822] | 0.221 [0.194, 0.254] | 0.619 [0.605, 0.632] | 0.806 [0.789, 0.823] |
| | 25% | Item FE + LLM | 3.751 [3.733, 3.769] | 3.748 [3.704, 3.792] | 0.278 [0.256, 0.298] | 0.469 [0.460, 0.477] | 0.583 [0.572, 0.593] |
| | 25% | Item FE + demographics | 3.751 [3.733, 3.769] | 3.747 [3.673, 3.822] | 0.218 [0.188, 0.252] | 0.618 [0.605, 0.632] | 0.804 [0.786, 0.822] |
| | 25% | Item FE + LLM + demographics | 3.751 [3.733, 3.769] | 3.748 [3.704, 3.792] | 0.270 [0.249, 0.292] | 0.470 [0.461, 0.479] | 0.585 [0.573, 0.596] |
| | 50% | Digital twin | 3.751 [3.722, 3.780] | 3.719 [3.679, 3.757] | 0.206 [0.181, 0.231] | 0.615 [0.591, 0.636] | 0.805 [0.772, 0.832] |
| | 50% | Item FE | 3.751 [3.722, 3.780] | 3.751 [3.685, 3.814] | 0.222 [0.192, 0.254] | 0.619 [0.596, 0.639] | 0.805 [0.774, 0.831] |
| | 50% | Item FE + LLM | 3.751 [3.722, 3.780] | 3.750 [3.722, 3.777] | 0.278 [0.255, 0.300] | 0.468 [0.453, 0.483] | 0.583 [0.564, 0.601] |
| | 50% | Item FE + demographics | 3.751 [3.722, 3.780] | 3.750 [3.684, 3.813] | 0.219 [0.188, 0.250] | 0.617 [0.594, 0.637] | 0.802 [0.771, 0.828] |
| | 50% | Item FE + LLM + demographics | 3.751 [3.722, 3.780] | 3.750 [3.722, 3.777] | 0.272 [0.249, 0.295] | 0.469 [0.453, 0.483] | 0.583 [0.564, 0.602] |
| Qwen 3.5 Plus | 10% | Digital twin | 3.750 [3.740, 3.760] | 3.461 [3.450, 3.472] | 0.292 [0.283, 0.300] | 0.531 [0.524, 0.536] | 0.682 [0.674, 0.689] |
| | 10% | Item FE | 3.750 [3.740, 3.760] | 3.751 [3.667, 3.837] | 0.117 [0.104, 0.138] | 0.488 [0.482, 0.495] | 0.619 [0.612, 0.627] |
| | 10% | Item FE + LLM | 3.750 [3.740, 3.760] | 3.750 [3.684, 3.816] | 0.164 [0.135, 0.196] | 0.423 [0.418, 0.430] | 0.534 [0.527, 0.541] |
| | 10% | Item FE + demographics | 3.750 [3.740, 3.760] | 3.751 [3.667, 3.837] | 0.110 [0.098, 0.132] | 0.489 [0.480, 0.500] | 0.622 [0.610, 0.635] |
| | 10% | Item FE + LLM + demographics | 3.750 [3.740, 3.760] | 3.750 [3.683, 3.817] | 0.154 [0.125, 0.186] | 0.428 [0.419, 0.438] | 0.540 [0.530, 0.552] |
| | 25% | Digital twin | 3.751 [3.733, 3.769] | 3.461 [3.442, 3.480] | 0.292 [0.277, 0.307] | 0.531 [0.520, 0.542] | 0.682 [0.668, 0.695] |
| | 25% | Item FE | 3.751 [3.733, 3.769] | 3.749 [3.692, 3.805] | 0.113 [0.102, 0.127] | 0.487 [0.477, 0.497] | 0.619 [0.606, 0.630] |
| | 25% | Item FE + LLM | 3.751 [3.733, 3.769] | 3.749 [3.707, 3.790] | 0.162 [0.140, 0.183] | 0.422 [0.414, 0.430] | 0.533 [0.523, 0.542] |
| | 25% | Item FE + demographics | 3.751 [3.733, 3.769] | 3.748 [3.693, 3.804] | 0.109 [0.099, 0.120] | 0.485 [0.474, 0.496] | 0.616 [0.602, 0.629] |
| | 25% | Item FE + LLM + demographics | 3.751 [3.733, 3.769] | 3.748 [3.706, 3.790] | 0.157 [0.134, 0.180] | 0.423 [0.414, 0.432] | 0.534 [0.523, 0.545] |
| | 50% | Digital twin | 3.751 [3.722, 3.780] | 3.461 [3.429, 3.492] | 0.292 [0.268, 0.318] | 0.531 [0.512, 0.548] | 0.683 [0.659, 0.704] |
| | 50% | Item FE | 3.751 [3.722, 3.780] | 3.750 [3.699, 3.803] | 0.113 [0.097, 0.130] | 0.487 [0.471, 0.503] | 0.618 [0.598, 0.638] |

| | | | | | | |
|---|---|---|---|---|---|---|
| 50% | Item FE + LLM | 3.751 [3.722, 3.780] | 3.750 [3.719, 3.781] | 0.162 [0.141, 0.183] | 0.422 [0.408, 0.436] | 0.533 [0.515, 0.550] |
| 50% | Item FE + demographics | 3.751 [3.722, 3.780] | 3.750 [3.700, 3.802] | 0.110 [0.095, 0.125] | 0.484 [0.468, 0.501] | 0.615 [0.594, 0.635] |
| 50% | Item FE + LLM + demographics | 3.751 [3.722, 3.780] | 3.750 [3.719, 3.781] | 0.158 [0.136, 0.180] | 0.421 [0.407, 0.436] | 0.532 [0.514, 0.550] |

*Note.* Values are means across 1,000 repeated participant level training and test splits for each calibration proportion; values in brackets are empirical 95% intervals across the 1,000 repeated splits. Human mean denotes the average held out human Big Five trait score. Predicted mean denotes the average raw or calibrated digital twin trait score, depending on the calibration method. Trait scores were computed after BFI-44 reverse coding and aggregation. Wasserstein distance quantifies the discrepancy between the marginal distributions of human and predicted trait scores. MAE and RMSE quantify prediction errors between held out human trait scores and the corresponding raw or bias corrected digital twin trait scores. Lower Wasserstein distance, MAE and RMSE indicate better distributional and predictive accuracy. MAE, mean absolute error; RMSE, root mean square error.

**Table S39 Nomothetic-span recovery of human big five associations with external psychological constructs after statistical calibration**

| LLM | Calibration proportion | Calibration method | Nomothetic matrix Pearson r | Nomothetic matrix MAE | Nomothetic matrix RMSE | LLM-as-DV Pearson r | LLM-as-DV corrected Pearson r | LLM-as-DV MAE | LLM-as-DV corrected MAE | LLM-as-covariate Pearson r | LLM-as-covariate corrected Pearson r | LLM-as-covariate MAE | LLM-as-covariate corrected MAE |
|---|---|---|---|---|---|---|---|---|---|---|---|---|---|
| DeepSeek V4-Flash | 10% | Digital twin | 0.976 [0.974, 0.978] | 0.053 [0.051, 0.056] | 0.066 [0.064, 0.069] | 0.958 [0.955, 0.961] | 0.968 [0.950, 0.980] | 0.050 [0.048, 0.052] | 0.043 [0.035, 0.052] | 0.927 [0.922, 0.933] | 0.904 [0.820, 0.942] | 0.088 [0.082, 0.092] | 0.094 [0.067, 0.139] |
| | 10% | Item FE | 0.976 [0.974, 0.978] | 0.053 [0.051, 0.056] | 0.066 [0.064, 0.069] | 0.958 [0.955, 0.961] | 0.968 [0.949, 0.979] | 0.050 [0.048, 0.052] | 0.043 [0.035, 0.053] | 0.927 [0.922, 0.933] | 0.903 [0.817, 0.942] | 0.088 [0.082, 0.092] | 0.095 [0.068, 0.142] |
| | 10% | Item FE + LLM | 0.976 [0.974, 0.978] | 0.053 [0.051, 0.056] | 0.066 [0.064, 0.069] | 0.958 [0.955, 0.961] | 0.975 [0.961, 0.984] | 0.084 [0.075, 0.093] | 0.038 [0.031, 0.046] | 0.927 [0.922, 0.933] | 0.936 [0.906, 0.957] | 0.271 [0.234, 0.316] | 0.072 [0.058, 0.087] |
| | 10% | Item FE + demographics | 0.976 [0.974, 0.978] | 0.053 [0.051, 0.056] | 0.066 [0.064, 0.069] | 0.958 [0.954, 0.961] | 0.968 [0.949, 0.979] | 0.050 [0.048, 0.053] | 0.043 [0.035, 0.053] | 0.927 [0.921, 0.933] | 0.902 [0.812, 0.941] | 0.088 [0.083, 0.093] | 0.096 [0.069, 0.145] |
| | 10% | Item FE + LLM + demographics | 0.975 [0.972, 0.977] | 0.054 [0.051, 0.056] | 0.066 [0.064, 0.070] | 0.956 [0.952, 0.960] | 0.974 [0.960, 0.984] | 0.085 [0.076, 0.093] | 0.039 [0.031, 0.047] | 0.925 [0.918, 0.931] | 0.936 [0.906, 0.957] | 0.271 [0.234, 0.312] | 0.072 [0.058, 0.088] |
| | 25% | Digital twin | 0.976 [0.972, 0.979] | 0.054 [0.050, 0.057] | 0.067 [0.063, 0.071] | 0.957 [0.953, 0.962] | 0.985 [0.977, 0.990] | 0.050 [0.047, 0.053] | 0.030 [0.024, 0.036] | 0.926 [0.916, 0.934] | 0.954 [0.931, 0.970] | 0.088 [0.082, 0.095] | 0.060 [0.048, 0.076] |
| | 25% | Item FE | 0.976 [0.972, 0.979] | 0.054 [0.050, 0.057] | 0.067 [0.063, 0.071] | 0.957 [0.953, 0.962] | 0.984 [0.977, 0.990] | 0.050 [0.047, 0.053] | 0.030 [0.024, 0.036] | 0.926 [0.916, 0.934] | 0.954 [0.931, 0.969] | 0.088 [0.082, 0.095] | 0.061 [0.048, 0.077] |
|  | 25% | Item FE + LLM | 0.976 [0.972, 0.979] | 0.054 [0.050, 0.057] | 0.067 [0.063, 0.071] | 0.957 [0.953, 0.962] | 0.988 [0.982, 0.992] | 0.084 [0.078, 0.090] | 0.026 [0.021, 0.032] | 0.926 [0.916, 0.934] | 0.969 [0.954, 0.978] | 0.271 [0.247, 0.297] | 0.049 [0.040, 0.059] |
|  | 25% | Item FE + demographics | 0.975 [0.971, 0.979] | 0.054 [0.050, 0.058] | 0.067 [0.063, 0.071] | 0.957 [0.952, 0.962] | 0.984 [0.977, 0.990] | 0.051 [0.047, 0.054] | 0.030 [0.024, 0.036] | 0.926 [0.916, 0.934] | 0.954 [0.930, 0.969] | 0.089 [0.082, 0.095] | 0.061 [0.048, 0.077] |
|  | 25% | Item FE + LLM + demographics | 0.974 [0.970, 0.978] | 0.054 [0.050, 0.058] | 0.068 [0.063, 0.072] | 0.956 [0.951, 0.960] | 0.988 [0.982, 0.992] | 0.084 [0.078, 0.091] | 0.026 [0.021, 0.033] | 0.924 [0.915, 0.933] | 0.969 [0.954, 0.978] | 0.271 [0.247, 0.298] | 0.049 [0.040, 0.059] |
|  | 50% | Digital twin | 0.974 [0.968, 0.979] | 0.055 [0.049, 0.061] | 0.069 [0.062, 0.075] | 0.956 [0.948, 0.963] | 0.988 [0.983, 0.992] | 0.052 [0.047, 0.056] | 0.025 [0.021, 0.031] | 0.922 [0.904, 0.935] | 0.966 [0.952, 0.977] | 0.091 [0.082, 0.102] | 0.051 [0.042, 0.062] |
|  | 50% | Item FE | 0.974 [0.968, 0.979] | 0.055 [0.049, 0.061] | 0.069 [0.062, 0.075] | 0.956 [0.948, 0.963] | 0.988 [0.983, 0.992] | 0.052 [0.047, 0.056] | 0.026 [0.021, 0.031] | 0.922 [0.904, 0.935] | 0.966 [0.952, 0.977] | 0.091 [0.082, 0.102] | 0.051 [0.042, 0.063] |
|  | 50% | Item FE + LLM | 0.974 [0.968, 0.979] | 0.055 [0.049, 0.061] | 0.069 [0.062, 0.075] | 0.956 [0.948, 0.963] | 0.991 [0.987, 0.994] | 0.084 [0.077, 0.091] | 0.023 [0.019, 0.028] | 0.922 [0.904, 0.935] | 0.976 [0.966, 0.983] | 0.274 [0.254, 0.298] | 0.042 [0.035, 0.050] |
|  | 50% | Item FE + demographics | 0.974 [0.967, 0.979] | 0.055 [0.049, 0.061] | 0.069 [0.062, 0.076] | 0.956 [0.947, 0.963] | 0.988 [0.983, 0.992] | 0.052 [0.047, 0.057] | 0.026 [0.021, 0.031] | 0.922 [0.903, 0.936] | 0.966 [0.952, 0.977] | 0.091 [0.082, 0.102] | 0.051 [0.042, 0.063] |
|  | 50% | Item FE + LLM + demographics | 0.973 [0.966, 0.978] | 0.056 [0.049, 0.062] | 0.069 [0.063, 0.076] | 0.954 [0.945, 0.961] | 0.991 [0.987, 0.994] | 0.085 [0.078, 0.091] | 0.023 [0.019, 0.028] | 0.920 [0.902, 0.935] | 0.976 [0.966, 0.983] | 0.275 [0.254, 0.299] | 0.042 [0.035, 0.050] |
| Gemma-4-31B-it | 10% | Digital twin | 0.981 [0.980, 0.982] | 0.045 [0.043, 0.046] | 0.059 [0.057, 0.060] | 0.967 [0.964, 0.969] | 0.974 [0.961, 0.983] | 0.039 [0.038, 0.041] | 0.038 [0.031, 0.046] | 0.944 [0.939, 0.949] | 0.925 [0.875, 0.953] | 0.078 [0.074, 0.082] | 0.080 [0.061, 0.106] |

| | | | | | | | | | | | | |
|---|---|---|---|---|---|---|---|---|---|---|---|---|
| 10% | Item FE | 0.981 [0.980, 0.982] | 0.045 [0.043, 0.046] | 0.059 [0.057, 0.060] | 0.967 [0.964, 0.969] | 0.974 [0.961, 0.983] | 0.039 [0.038, 0.041] | 0.038 [0.031, 0.046] | 0.944 [0.939, 0.949] | 0.923 [0.874, 0.952] | 0.078 [0.074, 0.082] | 0.081 [0.062, 0.108] |
| 10% | Item FE + LLM | 0.981 [0.980, 0.982] | 0.045 [0.043, 0.046] | 0.059 [0.057, 0.060] | 0.967 [0.964, 0.969] | 0.977 [0.966, 0.985] | 0.069 [0.062, 0.076] | 0.036 [0.029, 0.043] | 0.944 [0.939, 0.949] | 0.942 [0.915, 0.963] | 0.205 [0.179, 0.234] | 0.068 [0.055, 0.083] |
| 10% | Item FE + demographics | 0.981 [0.979, 0.982] | 0.045 [0.043, 0.047] | 0.059 [0.057, 0.060] | 0.966 [0.964, 0.969] | 0.974 [0.960, 0.983] | 0.040 [0.038, 0.042] | 0.038 [0.031, 0.047] | 0.944 [0.939, 0.949] | 0.923 [0.874, 0.952] | 0.077 [0.074, 0.082] | 0.082 [0.062, 0.109] |
| 10% | Item FE + LLM + demographics | 0.981 [0.979, 0.983] | 0.045 [0.042, 0.047] | 0.058 [0.055, 0.060] | 0.966 [0.963, 0.968] | 0.977 [0.966, 0.985] | 0.070 [0.063, 0.077] | 0.036 [0.029, 0.044] | 0.944 [0.938, 0.949] | 0.942 [0.915, 0.962] | 0.203 [0.177, 0.232] | 0.069 [0.055, 0.083] |
| 25% | Digital twin | 0.980 [0.978, 0.983] | 0.045 [0.043, 0.048] | 0.059 [0.056, 0.062] | 0.966 [0.963, 0.969] | 0.987 [0.982, 0.991] | 0.040 [0.037, 0.042] | 0.026 [0.022, 0.032] | 0.943 [0.936, 0.950] | 0.963 [0.945, 0.976] | 0.079 [0.074, 0.084] | 0.053 [0.043, 0.067] |
| 25% | Item FE | 0.980 [0.978, 0.983] | 0.045 [0.043, 0.048] | 0.059 [0.056, 0.062] | 0.966 [0.963, 0.969] | 0.987 [0.982, 0.991] | 0.040 [0.037, 0.042] | 0.026 [0.022, 0.032] | 0.943 [0.936, 0.950] | 0.963 [0.944, 0.976] | 0.079 [0.074, 0.084] | 0.054 [0.043, 0.067] |
| 25% | Item FE + LLM | 0.980 [0.978, 0.983] | 0.045 [0.043, 0.048] | 0.059 [0.056, 0.062] | 0.966 [0.963, 0.969] | 0.989 [0.984, 0.992] | 0.069 [0.064, 0.074] | 0.025 [0.020, 0.031] | 0.943 [0.936, 0.950] | 0.972 [0.959, 0.981] | 0.205 [0.188, 0.224] | 0.047 [0.038, 0.056] |
| 25% | Item FE + demographics | 0.980 [0.978, 0.983] | 0.045 [0.043, 0.048] | 0.059 [0.057, 0.062] | 0.966 [0.963, 0.969] | 0.987 [0.982, 0.991] | 0.040 [0.038, 0.042] | 0.027 [0.022, 0.032] | 0.943 [0.935, 0.950] | 0.963 [0.944, 0.976] | 0.079 [0.073, 0.084] | 0.054 [0.043, 0.067] |
| 25% | Item FE + LLM + demographics | 0.980 [0.978, 0.983] | 0.045 [0.043, 0.048] | 0.059 [0.056, 0.062] | 0.965 [0.962, 0.969] | 0.989 [0.983, 0.992] | 0.070 [0.065, 0.075] | 0.025 [0.020, 0.031] | 0.943 [0.935, 0.950] | 0.972 [0.959, 0.981] | 0.205 [0.188, 0.223] | 0.047 [0.038, 0.056] |

| | | | | | | | | | | | | | |
|---|---|---|---|---|---|---|---|---|---|---|---|---|---|
| | 50% | Digital twin | 0.979 [0.975, 0.983] | 0.046 [0.042, 0.051] | 0.061 [0.056, 0.066] | 0.965 [0.959, 0.970] | 0.990 [0.986, 0.994] | 0.041 [0.038, 0.044] | 0.023 [0.019, 0.027] | 0.940 [0.926, 0.951] | 0.973 [0.959, 0.981] | 0.081 [0.073, 0.090] | 0.046 [0.037, 0.056] |
| | 50% | Item FE | 0.979 [0.975, 0.983] | 0.046 [0.042, 0.051] | 0.061 [0.056, 0.066] | 0.965 [0.959, 0.970] | 0.990 [0.986, 0.994] | 0.041 [0.038, 0.044] | 0.023 [0.019, 0.027] | 0.940 [0.926, 0.951] | 0.973 [0.959, 0.981] | 0.081 [0.073, 0.090] | 0.046 [0.037, 0.057] |
| | 50% | Item FE + LLM | 0.979 [0.975, 0.983] | 0.046 [0.042, 0.051] | 0.061 [0.056, 0.066] | 0.965 [0.959, 0.970] | 0.992 [0.988, 0.994] | 0.070 [0.063, 0.075] | 0.021 [0.018, 0.026] | 0.940 [0.926, 0.951] | 0.978 [0.969, 0.985] | 0.208 [0.193, 0.224] | 0.040 [0.033, 0.048] |
| | 50% | Item FE + demographics | 0.979 [0.975, 0.983] | 0.047 [0.042, 0.051] | 0.061 [0.056, 0.066] | 0.965 [0.959, 0.970] | 0.990 [0.986, 0.994] | 0.041 [0.038, 0.045] | 0.023 [0.019, 0.027] | 0.939 [0.925, 0.950] | 0.972 [0.959, 0.981] | 0.081 [0.073, 0.090] | 0.046 [0.038, 0.057] |
| | 50% | Item FE + LLM + demographics | 0.979 [0.975, 0.983] | 0.047 [0.042, 0.051] | 0.061 [0.056, 0.066] | 0.964 [0.958, 0.969] | 0.992 [0.988, 0.994] | 0.070 [0.064, 0.076] | 0.021 [0.018, 0.026] | 0.939 [0.925, 0.950] | 0.978 [0.969, 0.985] | 0.208 [0.192, 0.224] | 0.040 [0.033, 0.048] |
| gpt-oss-120b | 10% | Digital twin | 0.960 [0.958, 0.963] | 0.051 [0.049, 0.053] | 0.069 [0.067, 0.071] | 0.955 [0.951, 0.958] | 0.960 [0.939, 0.974] | 0.051 [0.048, 0.053] | 0.048 [0.039, 0.058] | 0.933 [0.927, 0.939] | 0.877 [0.762, 0.930] | 0.063 [0.060, 0.067] | 0.110 [0.077, 0.177] |
| | 10% | Item FE | 0.960 [0.958, 0.963] | 0.051 [0.049, 0.053] | 0.069 [0.067, 0.071] | 0.955 [0.951, 0.958] | 0.960 [0.939, 0.974] | 0.051 [0.048, 0.053] | 0.048 [0.039, 0.058] | 0.933 [0.927, 0.939] | 0.874 [0.753, 0.928] | 0.063 [0.060, 0.067] | 0.113 [0.078, 0.183] |
| | 10% | Item FE + LLM | 0.960 [0.958, 0.963] | 0.051 [0.049, 0.053] | 0.069 [0.067, 0.071] | 0.955 [0.951, 0.958] | 0.972 [0.958, 0.982] | 0.105 [0.097, 0.113] | 0.040 [0.032, 0.049] | 0.933 [0.927, 0.939] | 0.935 [0.905, 0.956] | 0.270 [0.226, 0.321] | 0.073 [0.059, 0.088] |
| | 10% | Item FE + demographics | 0.960 [0.957, 0.963] | 0.051 [0.049, 0.053] | 0.069 [0.067, 0.072] | 0.955 [0.951, 0.958] | 0.960 [0.939, 0.974] | 0.051 [0.048, 0.053] | 0.048 [0.039, 0.058] | 0.933 [0.927, 0.939] | 0.873 [0.751, 0.928] | 0.064 [0.060, 0.067] | 0.115 [0.078, 0.187] |

| | | | | | | | | | | | | |
|---|---|---|---|---|---|---|---|---|---|---|---|---|
| 10% | Item FE + LLM + demographics | 0.959 [0.954, 0.963] | 0.052 [0.049, 0.055] | 0.070 [0.066, 0.074] | 0.952 [0.947, 0.957] | 0.972 [0.957, 0.981] | 0.105 [0.097, 0.113] | 0.041 [0.032, 0.049] | 0.929 [0.922, 0.936] | 0.935 [0.905, 0.956] | 0.263 [0.220, 0.310] | 0.073 [0.059, 0.089] |
| 25% | Digital twin | 0.960 [0.955, 0.965] | 0.051 [0.048, 0.055] | 0.070 [0.065, 0.074] | 0.954 [0.949, 0.959] | 0.981 [0.971, 0.987] | 0.051 [0.048, 0.054] | 0.033 [0.027, 0.040] | 0.932 [0.924, 0.940] | 0.942 [0.911, 0.963] | 0.065 [0.060, 0.069] | 0.069 [0.054, 0.089] |
| 25% | Item FE | 0.960 [0.955, 0.965] | 0.051 [0.048, 0.055] | 0.070 [0.065, 0.074] | 0.954 [0.949, 0.959] | 0.980 [0.971, 0.987] | 0.051 [0.048, 0.054] | 0.033 [0.027, 0.040] | 0.932 [0.924, 0.940] | 0.942 [0.911, 0.963] | 0.065 [0.060, 0.069] | 0.070 [0.054, 0.091] |
| 25% | Item FE + LLM | 0.960 [0.955, 0.965] | 0.051 [0.048, 0.055] | 0.070 [0.065, 0.074] | 0.954 [0.949, 0.959] | 0.986 [0.980, 0.991] | 0.105 [0.099, 0.111] | 0.028 [0.023, 0.034] | 0.932 [0.924, 0.940] | 0.968 [0.953, 0.978] | 0.270 [0.241, 0.302] | 0.050 [0.040, 0.060] |
| 25% | Item FE + demographics | 0.960 [0.954, 0.965] | 0.052 [0.048, 0.055] | 0.070 [0.066, 0.074] | 0.954 [0.948, 0.959] | 0.980 [0.971, 0.987] | 0.051 [0.048, 0.054] | 0.033 [0.027, 0.040] | 0.932 [0.924, 0.940] | 0.942 [0.911, 0.963] | 0.065 [0.060, 0.069] | 0.070 [0.054, 0.090] |
| 25% | Item FE + LLM + demographics | 0.958 [0.953, 0.964] | 0.052 [0.049, 0.056] | 0.071 [0.066, 0.075] | 0.952 [0.946, 0.957] | 0.986 [0.980, 0.990] | 0.105 [0.099, 0.112] | 0.028 [0.023, 0.034] | 0.930 [0.922, 0.938] | 0.968 [0.953, 0.977] | 0.267 [0.239, 0.298] | 0.050 [0.040, 0.060] |
| 50% | Digital twin | 0.958 [0.949, 0.966] | 0.053 [0.048, 0.058] | 0.071 [0.064, 0.078] | 0.953 [0.944, 0.961] | 0.985 [0.979, 0.990] | 0.053 [0.048, 0.057] | 0.028 [0.023, 0.034] | 0.928 [0.912, 0.940] | 0.957 [0.936, 0.971] | 0.068 [0.061, 0.075] | 0.058 [0.046, 0.073] |
| 50% | Item FE | 0.958 [0.949, 0.966] | 0.053 [0.048, 0.058] | 0.071 [0.064, 0.078] | 0.953 [0.944, 0.961] | 0.985 [0.978, 0.990] | 0.053 [0.048, 0.057] | 0.028 [0.023, 0.034] | 0.928 [0.912, 0.940] | 0.957 [0.936, 0.971] | 0.068 [0.061, 0.075] | 0.058 [0.047, 0.073] |
| 50% | Item FE + LLM | 0.958 [0.949, 0.966] | 0.053 [0.048, 0.058] | 0.071 [0.064, 0.078] | 0.953 [0.944, 0.961] | 0.990 [0.985, 0.993] | 0.105 [0.097, 0.112] | 0.024 [0.020, 0.029] | 0.928 [0.912, 0.940] | 0.976 [0.966, 0.983] | 0.274 [0.252, 0.298] | 0.043 [0.035, 0.052] |

| | | | | | | | | | | | | | |
|---|---|---|---|---|---|---|---|---|---|---|---|---|---|
| | 50% | Item FE + demographics | 0.958 [0.949, 0.966] | 0.053 [0.048, 0.059] | 0.071 [0.064, 0.079] | 0.952 [0.943, 0.961] | 0.985 [0.979, 0.990] | 0.053 [0.048, 0.057] | 0.028 [0.023, 0.034] | 0.927 [0.912, 0.941] | 0.957 [0.936, 0.971] | 0.068 [0.061, 0.075] | 0.058 [0.047, 0.073] |
| | 50% | Item FE + LLM + demographics | 0.957 [0.947, 0.965] | 0.054 [0.049, 0.059] | 0.072 [0.065, 0.080] | 0.950 [0.941, 0.959] | 0.990 [0.985, 0.993] | 0.106 [0.097, 0.113] | 0.024 [0.020, 0.029] | 0.926 [0.910, 0.939] | 0.976 [0.966, 0.983] | 0.273 [0.252, 0.296] | 0.043 [0.035, 0.052] |
| Qwen 3.5 Plus | 10% | Digital twin | 0.984 [0.982, 0.985] | 0.048 [0.046, 0.050] | 0.063 [0.061, 0.064] | 0.978 [0.977, 0.980] | 0.973 [0.960, 0.983] | 0.034 [0.032, 0.035] | 0.039 [0.032, 0.048] | 0.918 [0.912, 0.924] | 0.928 [0.884, 0.954] | 0.117 [0.111, 0.122] | 0.078 [0.061, 0.101] |
| | 10% | Item FE | 0.984 [0.982, 0.985] | 0.048 [0.046, 0.050] | 0.063 [0.061, 0.064] | 0.978 [0.977, 0.980] | 0.973 [0.959, 0.982] | 0.034 [0.032, 0.035] | 0.039 [0.032, 0.048] | 0.918 [0.912, 0.924] | 0.927 [0.881, 0.954] | 0.117 [0.111, 0.122] | 0.079 [0.061, 0.102] |
| | 10% | Item FE + LLM | 0.984 [0.982, 0.985] | 0.048 [0.046, 0.050] | 0.063 [0.061, 0.064] | 0.978 [0.977, 0.980] | 0.976 [0.964, 0.984] | 0.061 [0.053, 0.070] | 0.037 [0.030, 0.044] | 0.918 [0.912, 0.924] | 0.941 [0.912, 0.960] | 0.252 [0.222, 0.285] | 0.070 [0.056, 0.085] |
| | 10% | Item FE + demographics | 0.984 [0.982, 0.985] | 0.047 [0.045, 0.049] | 0.062 [0.059, 0.064] | 0.978 [0.977, 0.980] | 0.973 [0.958, 0.982] | 0.034 [0.032, 0.035] | 0.039 [0.032, 0.048] | 0.917 [0.911, 0.924] | 0.926 [0.878, 0.953] | 0.116 [0.110, 0.121] | 0.079 [0.061, 0.102] |
| | 10% | Item FE + LLM + demographics | 0.984 [0.982, 0.985] | 0.047 [0.044, 0.049] | 0.061 [0.058, 0.064] | 0.978 [0.976, 0.980] | 0.976 [0.964, 0.984] | 0.062 [0.054, 0.070] | 0.037 [0.030, 0.044] | 0.916 [0.910, 0.923] | 0.940 [0.911, 0.959] | 0.249 [0.220, 0.280] | 0.070 [0.056, 0.085] |
| | 25% | Digital twin | 0.983 [0.981, 0.985] | 0.049 [0.046, 0.051] | 0.063 [0.060, 0.066] | 0.978 [0.975, 0.980] | 0.987 [0.981, 0.991] | 0.034 [0.032, 0.036] | 0.027 [0.022, 0.032] | 0.917 [0.906, 0.926] | 0.965 [0.949, 0.976] | 0.117 [0.111, 0.124] | 0.052 [0.042, 0.064] |
| | 25% | Item FE | 0.983 [0.981, 0.985] | 0.049 [0.046, 0.051] | 0.063 [0.060, 0.066] | 0.978 [0.975, 0.980] | 0.987 [0.981, 0.991] | 0.034 [0.032, 0.036] | 0.027 [0.022, 0.032] | 0.917 [0.906, 0.926] | 0.965 [0.948, 0.976] | 0.117 [0.111, 0.124] | 0.052 [0.042, 0.064] |

| | | | | | | | | | | | | |
|---|---|---|---|---|---|---|---|---|---|---|---|---|
| 25% | Item FE + LLM | 0.983 [0.981, 0.985] | 0.049 [0.046, 0.051] | 0.063 [0.060, 0.066] | 0.978 [0.975, 0.980] | 0.988 [0.983, 0.992] | 0.061 [0.055, 0.067] | 0.025 [0.021, 0.031] | 0.917 [0.906, 0.926] | 0.971 [0.958, 0.979] | 0.252 [0.233, 0.273] | 0.047 [0.039, 0.057] |
| 25% | Item FE + demographics | 0.983 [0.981, 0.985] | 0.048 [0.045, 0.051] | 0.063 [0.060, 0.066] | 0.978 [0.975, 0.980] | 0.987 [0.981, 0.991] | 0.034 [0.032, 0.036] | 0.027 [0.022, 0.032] | 0.916 [0.906, 0.925] | 0.965 [0.948, 0.976] | 0.117 [0.111, 0.123] | 0.052 [0.043, 0.064] |
| 25% | Item FE + LLM + demographics | 0.983 [0.981, 0.985] | 0.048 [0.045, 0.051] | 0.062 [0.059, 0.065] | 0.978 [0.975, 0.980] | 0.988 [0.983, 0.992] | 0.062 [0.055, 0.068] | 0.025 [0.021, 0.031] | 0.916 [0.905, 0.925] | 0.970 [0.958, 0.979] | 0.251 [0.232, 0.271] | 0.047 [0.039, 0.057] |
| 50% | Digital twin | 0.982 [0.978, 0.985] | 0.050 [0.045, 0.055] | 0.065 [0.060, 0.070] | 0.976 [0.972, 0.980] | 0.990 [0.986, 0.993] | 0.036 [0.033, 0.039] | 0.023 [0.019, 0.028] | 0.913 [0.895, 0.928] | 0.974 [0.962, 0.981] | 0.119 [0.109, 0.129] | 0.045 [0.037, 0.053] |
| 50% | Item FE | 0.982 [0.978, 0.985] | 0.050 [0.045, 0.055] | 0.065 [0.060, 0.070] | 0.976 [0.972, 0.980] | 0.990 [0.986, 0.993] | 0.036 [0.033, 0.039] | 0.023 [0.019, 0.028] | 0.913 [0.895, 0.928] | 0.973 [0.962, 0.981] | 0.119 [0.109, 0.129] | 0.045 [0.037, 0.053] |
| 50% | Item FE + LLM | 0.982 [0.978, 0.985] | 0.050 [0.045, 0.055] | 0.065 [0.060, 0.070] | 0.976 [0.972, 0.980] | 0.991 [0.988, 0.994] | 0.062 [0.055, 0.068] | 0.022 [0.018, 0.026] | 0.913 [0.895, 0.928] | 0.978 [0.969, 0.984] | 0.255 [0.238, 0.273] | 0.041 [0.034, 0.049] |
| 50% | Item FE + demographics | 0.982 [0.978, 0.985] | 0.050 [0.045, 0.055] | 0.064 [0.059, 0.070] | 0.976 [0.972, 0.980] | 0.990 [0.986, 0.993] | 0.036 [0.033, 0.039] | 0.023 [0.019, 0.028] | 0.913 [0.894, 0.928] | 0.973 [0.962, 0.982] | 0.119 [0.109, 0.129] | 0.045 [0.037, 0.053] |
| 50% | Item FE + LLM + demographics | 0.982 [0.978, 0.985] | 0.050 [0.045, 0.055] | 0.064 [0.059, 0.069] | 0.976 [0.971, 0.980] | 0.991 [0.988, 0.994] | 0.062 [0.055, 0.069] | 0.022 [0.018, 0.026] | 0.913 [0.894, 0.927] | 0.978 [0.969, 0.984] | 0.254 [0.237, 0.272] | 0.041 [0.034, 0.049] |

*Note.* Nomothetic matrix metrics compare the human Big Five–external-scale association matrix with the corresponding matrix estimated from digital-twin-derived Big Five trait scores. LLM-as-DV metrics treat digital-twin-derived Big Five traits as dependent variables predicted by standardized external scales; LLM-as-DV corrected metrics subtract calibration-sample estimates of the association between prediction error and external scales. LLM-as-covariate metrics treat the external scales as dependent variables

jointly predicted by the five digital-twin-derived Big Five traits; LLM-as-covariate corrected metrics adjust the predictor and predictor–outcome covariance matrices using calibration-sample estimates of measurement-error bias. Higher Pearson r and lower MAE or RMSE indicate closer recovery of the human nomothetic-span structure. Values are means across repeated participant-level splits, with empirical 95% intervals in brackets.

**Table S40 Accuracy of digital twin responses to the NEO Five-Factor Personality Inventory across input conditions and models**

| Input condition | Model | Extraversion | Agreeableness | Conscientiousness | Neuroticism | Openness | Overall |
|---|---|---|---|---|---|---|---|
| Demographics | Gemma-4-31B-it | 0.773 | 0.806 | 0.805 | 0.747 | 0.763 | 0.779 |
| Demographics | gpt-oss-120b | 0.754 | 0.797 | 0.804 | 0.735 | 0.748 | 0.768 |
| Demographics | DeepSeek V4-Flash | 0.752 | 0.795 | 0.793 | 0.743 | 0.730 | 0.762 |
| Demographics | Qwen 3.5 Plus | 0.777 | 0.805 | 0.804 | 0.758 | 0.764 | 0.782 |
| COVID Memories | Gemma-4-31B-it | 0.747 | 0.794 | 0.800 | 0.720 | 0.762 | 0.764 |
| COVID Memories | gpt-oss-120b | 0.731 | 0.782 | 0.786 | 0.693 | 0.737 | 0.746 |
| COVID Memories | DeepSeek V4-Flash | 0.742 | 0.791 | 0.782 | 0.706 | 0.729 | 0.750 |
| COVID Memories | Qwen 3.5 Plus | 0.742 | 0.787 | 0.79 | 0.699 | 0.735 | 0.750 |
| Demographics + COVID Memories | Gemma-4-31B-it | 0.776 | 0.805 | 0.808 | 0.745 | 0.773 | 0.781 |
| Demographics + COVID Memories | gpt-oss-120b | 0.755 | 0.796 | 0.799 | 0.729 | 0.746 | 0.765 |
| Demographics + COVID Memories | DeepSeek V4-Flash | 0.752 | 0.795 | 0.793 | 0.735 | 0.736 | 0.762 |
| Demographics + COVID Memories | Qwen 3.5 Plus | 0.773 | 0.801 | 0.805 | 0.739 | 0.763 | 0.776 |
| Demographics + All psychological measures + COVID Memories | Gemma-4-31B-it | 0.795 | 0.819 | 0.814 | 0.811 | 0.792 | 0.806 |
| Demographics + All psychological measures + COVID Memories | gpt-oss-120b | 0.766 | 0.805 | 0.776 | 0.792 | 0.751 | 0.778 |
| Demographics + All psychological measures + COVID Memories | DeepSeek V4-Flash | 0.779 | 0.809 | 0.807 | 0.802 | 0.761 | 0.792 |
| Demographics + All psychological measures + COVID Memories | Qwen 3.5 Plus | 0.788 | 0.818 | 0.807 | 0.806 | 0.778 | 0.799 |
| Demographics + Features with $r < 0.5$ + COVID contexts | Gemma-4-31B-it | 0.777 | 0.808 | 0.804 | 0.744 | 0.777 | 0.782 |
| Demographics + Features with $r < 0.5$ + COVID contexts | gpt-oss-120b | 0.751 | 0.798 | 0.802 | 0.738 | 0.757 | 0.769 |
| Demographics + Features with $r < 0.5$ + COVID contexts | DeepSeek V4-Flash | 0.749 | 0.795 | 0.786 | 0.733 | 0.746 | 0.762 |
| Demographics + Features with $r < 0.5$ + COVID contexts | Qwen 3.5 Plus | 0.774 | 0.803 | 0.804 | 0.745 | 0.771 | 0.779 |
| Demographics + Features with $r > 0.5$ + COVID contexts | Gemma-4-31B-it | 0.794 | 0.818 | 0.814 | 0.810 | 0.788 | 0.805 |
| Demographics + Features with $r > 0.5$ + COVID contexts | gpt-oss-120b | 0.764 | 0.807 | 0.781 | 0.794 | 0.753 | 0.780 |
| Demographics + Features with $r > 0.5$ + COVID contexts | DeepSeek V4-Flash | 0.778 | 0.807 | 0.804 | 0.804 | 0.755 | 0.790 |
| Demographics + Features with $r > 0.5$ + COVID contexts | Qwen 3.5 Plus | 0.787 | 0.819 | 0.811 | 0.806 | 0.777 | 0.800 |
| Demographics + COVID Memories + COVID contexts | Gemma-4-31B-it | 0.776 | 0.807 | 0.807 | 0.747 | 0.776 | 0.782 |
| Demographics + COVID Memories + COVID contexts | gpt-oss-120b | 0.755 | 0.801 | 0.804 | 0.738 | 0.753 | 0.770 |
| Demographics + COVID Memories + COVID contexts | DeepSeek V4-Flash | 0.755 | 0.799 | 0.791 | 0.734 | 0.743 | 0.764 |
| Demographics + COVID Memories + COVID contexts | Qwen 3.5 Plus | 0.771 | 0.805 | 0.807 | 0.744 | 0.766 | 0.779 |

**Table S41 Correlations of digital twin responses to the NEO Five-Factor Personality Inventory across input conditions and models**

| **Feature** | **Model** | **Extraversion** | | **Agreeableness** | | **Conscientiousness** | | **Neuroticism** | | **Openness** | | **Overall** | |
|---|---|---|---|---|---|---|---|---|---|---|---|---|---|
| Item-level correlation | | Pearson | Spearman | Pearson | Spearman | Pearson | Spearman | Pearson | Spearman | Pearson | Spearman | Pearson | Spearman |
| Demographics | Gemma-4-31B-it | 0.272 | 0.273 | 0.109 | 0.104 | 0.204 | 0.213 | 0.326 | 0.313 | 0.195 | 0.191 | 0.223 | 0.22 |
| | gpt-oss-120b | 0.244 | 0.239 | 0.093 | 0.095 | 0.19 | 0.193 | 0.299 | 0.293 | 0.152 | 0.149 | 0.197 | 0.195 |
| | DeepSeek V4-Flash | 0.202 | 0.199 | 0.079 | 0.076 | 0.183 | 0.19 | 0.283 | 0.277 | 0.144 | 0.141 | 0.179 | 0.177 |
| | Qwen 3.5 Plus | 0.271 | 0.27 | 0.103 | 0.099 | 0.21 | 0.209 | 0.317 | 0.3 | 0.177 | 0.171 | 0.217 | 0.211 |
| COVID Memories | Gemma-4-31B-it | 0.207 | 0.206 | 0.093 | 0.088 | 0.147 | 0.14 | 0.213 | 0.206 | 0.142 | 0.146 | 0.161 | 0.158 |
| | gpt-oss-120b | 0.151 | 0.151 | 0.075 | 0.074 | 0.114 | 0.113 | 0.157 | 0.158 | 0.053 | 0.055 | 0.11 | 0.111 |
| | DeepSeek V4-Flash | 0.151 | 0.153 | 0.082 | 0.081 | 0.091 | 0.095 | 0.201 | 0.199 | 0.12 | 0.122 | 0.129 | 0.13 |
| | Qwen 3.5 Plus | 0.182 | 0.178 | 0.086 | 0.084 | 0.145 | 0.139 | 0.184 | 0.18 | 0.135 | 0.136 | 0.146 | 0.144 |
| Demographics + COVID Memories | Gemma-4-31B-it | 0.308 | 0.308 | 0.137 | 0.131 | 0.235 | 0.236 | 0.352 | 0.341 | 0.236 | 0.234 | 0.255 | 0.251 |
| | gpt-oss-120b | 0.246 | 0.247 | 0.085 | 0.085 | 0.196 | 0.196 | 0.304 | 0.301 | 0.159 | 0.149 | 0.199 | 0.197 |
| | DeepSeek V4-Flash | 0.209 | 0.209 | 0.093 | 0.098 | 0.192 | 0.196 | 0.313 | 0.31 | 0.179 | 0.176 | 0.198 | 0.199 |
| | Qwen 3.5 Plus | 0.295 | 0.293 | 0.124 | 0.118 | 0.252 | 0.257 | 0.322 | 0.314 | 0.214 | 0.21 | 0.242 | 0.24 |
| Demographics + All psychological measures + COVID Memories | Gemma-4-31B-it | 0.46 | 0.459 | 0.334 | 0.334 | 0.398 | 0.398 | 0.66 | 0.653 | 0.308 | 0.31 | 0.443 | 0.441 |
| | gpt-oss-120b | 0.38 | 0.372 | 0.253 | 0.247 | 0.266 | 0.267 | 0.586 | 0.569 | 0.184 | 0.177 | 0.343 | 0.335 |
| | DeepSeek V4-Flash | 0.383 | 0.377 | 0.301 | 0.3 | 0.353 | 0.377 | 0.595 | 0.588 | 0.288 | 0.298 | 0.391 | 0.394 |
| | Qwen 3.5 Plus | 0.437 | 0.431 | 0.287 | 0.284 | 0.336 | 0.34 | 0.6 | 0.592 | 0.286 | 0.289 | 0.397 | 0.394 |
| Demographics + Features with r < 0.5 + COVID contexts | Gemma-4-31B-it | 0.316 | 0.314 | 0.166 | 0.168 | 0.219 | 0.224 | 0.349 | 0.34 | 0.252 | 0.253 | 0.262 | 0.261 |
| | gpt-oss-120b | 0.227 | 0.225 | 0.106 | 0.1 | 0.171 | 0.169 | 0.308 | 0.299 | 0.176 | 0.165 | 0.198 | 0.193 |
| | DeepSeek V4-Flash | 0.217 | 0.219 | 0.134 | 0.137 | 0.188 | 0.193 | 0.316 | 0.312 | 0.206 | 0.221 | 0.213 | 0.217 |
| | Qwen 3.5 Plus | 0.279 | 0.275 | 0.121 | 0.12 | 0.236 | 0.242 | 0.328 | 0.32 | 0.248 | 0.25 | 0.244 | 0.242 |
| Demographics + Features with r > 0.5 + COVID contexts | Gemma-4-31B-it | 0.459 | 0.457 | 0.341 | 0.34 | 0.4 | 0.407 | 0.656 | 0.651 | 0.294 | 0.301 | 0.44 | 0.441 |
| | gpt-oss-120b | 0.375 | 0.369 | 0.269 | 0.262 | 0.289 | 0.294 | 0.601 | 0.588 | 0.202 | 0.193 | 0.357 | 0.35 |
| | DeepSeek V4-Flash | 0.386 | 0.381 | 0.267 | 0.269 | 0.361 | 0.376 | 0.597 | 0.593 | 0.262 | 0.272 | 0.383 | 0.386 |
| | Qwen 3.5 Plus | 0.441 | 0.435 | 0.293 | 0.289 | 0.347 | 0.356 | 0.598 | 0.591 | 0.273 | 0.274 | 0.398 | 0.396 |
| Demographics + COVID Memories + COVID contexts | Gemma-4-31B-it | 0.307 | 0.308 | 0.154 | 0.148 | 0.237 | 0.241 | 0.36 | 0.35 | 0.234 | 0.234 | 0.26 | 0.258 |
| | gpt-oss-120b | 0.245 | 0.243 | 0.104 | 0.111 | 0.196 | 0.206 | 0.32 | 0.316 | 0.16 | 0.152 | 0.206 | 0.207 |
| | DeepSeek V4-Flash | 0.208 | 0.207 | 0.11 | 0.119 | 0.216 | 0.218 | 0.313 | 0.304 | 0.192 | 0.197 | 0.209 | 0.21 |
| | Qwen 3.5 Plus | 0.269 | 0.261 | 0.131 | 0.127 | 0.243 | 0.246 | 0.327 | 0.318 | 0.22 | 0.219 | 0.239 | 0.235 |
| Profile correlation | | | | | | | | | | | | | |
| Demographics | Gemma-4-31B-it | 0.392 | 0.383 | 0.749 | 0.734 | 0.715 | 0.649 | 0.442 | 0.412 | 0.524 | 0.507 | 0.548 | 0.545 |
| | gpt-oss-120b | 0.347 | 0.341 | 0.7 | 0.683 | 0.713 | 0.659 | 0.446 | 0.418 | 0.483 | 0.472 | 0.478 | 0.476 |
| | DeepSeek V4-Flash | 0.28 | 0.282 | 0.702 | 0.692 | 0.678 | 0.617 | 0.388 | 0.364 | 0.37 | 0.372 | 0.467 | 0.467 |
| | Qwen 3.5 Plus | 0.39 | 0.383 | 0.737 | 0.733 | 0.745 | 0.686 | 0.46 | 0.439 | 0.519 | 0.511 | 0.555 | 0.554 |
| COVID Memories | Gemma-4-31B-it | 0.252 | 0.252 | 0.717 | 0.699 | 0.702 | 0.649 | 0.225 | 0.205 | 0.529 | 0.525 | 0.486 | 0.485 |
| | gpt-oss-120b | 0.205 | 0.207 | 0.661 | 0.648 | 0.613 | 0.561 | 0.176 | 0.156 | 0.382 | 0.381 | 0.392 | 0.39 |
| | DeepSeek V4-Flash | 0.224 | 0.226 | 0.703 | 0.695 | 0.663 | 0.604 | 0.194 | 0.187 | 0.351 | 0.347 | 0.438 | 0.435 |
| | Qwen 3.5 Plus | 0.286 | 0.279 | 0.692 | 0.682 | 0.667 | 0.619 | 0.152 | 0.136 | 0.357 | 0.351 | 0.436 | 0.432 |
| Demographics + COVID Memories | Gemma-4-31B-it | 0.421 | 0.415 | 0.737 | 0.715 | 0.705 | 0.652 | 0.392 | 0.362 | 0.589 | 0.574 | 0.557 | 0.554 |
| | gpt-oss-120b | 0.347 | 0.343 | 0.695 | 0.677 | 0.673 | 0.616 | 0.378 | 0.351 | 0.467 | 0.46 | 0.462 | 0.46 |
| | DeepSeek V4-Flash | 0.285 | 0.286 | 0.701 | 0.69 | 0.651 | 0.596 | 0.366 | 0.346 | 0.401 | 0.396 | 0.469 | 0.467 |
| | Qwen 3.5 Plus | 0.385 | 0.379 | 0.716 | 0.711 | 0.721 | 0.667 | 0.322 | 0.299 | 0.512 | 0.501 | 0.529 | 0.528 |
| Demographics + All psychological measures + COVID Memories | Gemma-4-31B-it | 0.541 | 0.528 | 0.767 | 0.746 | 0.709 | 0.645 | 0.692 | 0.637 | 0.646 | 0.642 | 0.648 | 0.644 |
| | gpt-oss-120b | 0.428 | 0.418 | 0.722 | 0.696 | 0.52 | 0.468 | 0.677 | 0.633 | 0.478 | 0.473 | 0.524 | 0.52 |
| | DeepSeek V4-Flash | 0.431 | 0.429 | 0.76 | 0.737 | 0.742 | 0.678 | 0.702 | 0.665 | 0.525 | 0.518 | 0.596 | 0.594 |
| | Qwen 3.5 Plus | 0.489 | 0.482 | 0.749 | 0.733 | 0.699 | 0.646 | 0.647 | 0.612 | 0.56 | 0.55 | 0.61 | 0.608 |
| Demographics + Features with $r < 0.5$ + COVID contexts | Gemma-4-31B-it | 0.435 | 0.436 | 0.747 | 0.728 | 0.701 | 0.639 | 0.388 | 0.364 | 0.618 | 0.606 | 0.565 | 0.562 |
| | gpt-oss-120b | 0.309 | 0.312 | 0.708 | 0.688 | 0.708 | 0.65 | 0.397 | 0.365 | 0.52 | 0.513 | 0.484 | 0.483 |
| | DeepSeek V4-Flash | 0.312 | 0.316 | 0.719 | 0.698 | 0.689 | 0.634 | 0.376 | 0.357 | 0.494 | 0.49 | 0.503 | 0.501 |
| | Qwen 3.5 Plus | 0.375 | 0.376 | 0.727 | 0.718 | 0.733 | 0.676 | 0.335 | 0.314 | 0.549 | 0.536 | 0.541 | 0.539 |
| Demographics + Features with $r > 0.5$ + COVID contexts | Gemma-4-31B-it | 0.555 | 0.542 | 0.767 | 0.746 | 0.712 | 0.649 | 0.688 | 0.635 | 0.631 | 0.62 | 0.648 | 0.643 |
| | gpt-oss-120b | 0.408 | 0.398 | 0.722 | 0.694 | 0.536 | 0.482 | 0.691 | 0.637 | 0.487 | 0.482 | 0.528 | 0.524 |
| | DeepSeek V4-Flash | 0.442 | 0.444 | 0.751 | 0.724 | 0.745 | 0.688 | 0.693 | 0.661 | 0.497 | 0.49 | 0.591 | 0.588 |
| | Qwen 3.5 Plus | 0.488 | 0.48 | 0.749 | 0.732 | 0.696 | 0.647 | 0.648 | 0.61 | 0.551 | 0.545 | 0.606 | 0.605 |
| Demographics + COVID Memories + COVID contexts | Gemma-4-31B-it | 0.429 | 0.425 | 0.744 | 0.726 | 0.704 | 0.648 | 0.391 | 0.374 | 0.604 | 0.592 | 0.564 | 0.561 |
| | gpt-oss-120b | 0.324 | 0.324 | 0.709 | 0.693 | 0.679 | 0.624 | 0.398 | 0.365 | 0.495 | 0.49 | 0.472 | 0.471 |
| | DeepSeek V4-Flash | 0.303 | 0.304 | 0.72 | 0.7 | 0.703 | 0.637 | 0.364 | 0.34 | 0.463 | 0.458 | 0.504 | 0.501 |
| | Qwen 3.5 Plus | 0.362 | 0.363 | 0.729 | 0.723 | 0.729 | 0.672 | 0.314 | 0.303 | 0.519 | 0.513 | 0.533 | 0.531 |
| Trait-score correlation | | Observed | Disattenuated | Observed | Disattenuated | Observed | Disattenuated | Observed | Disattenuated | Observed | Disattenuated | | |
| Demographics | Gemma-4-31B-it | 0.49 | 0.536 | 0.214 | 0.26 | 0.315 | 0.337 | 0.458 | 0.479 | 0.317 | 0.364 | | |
| | gpt-oss-120b | 0.449 | 0.493 | 0.217 | 0.272 | 0.321 | 0.35 | 0.452 | 0.475 | 0.268 | 0.321 | | |
| | DeepSeek V4-Flash | 0.336 | 0.367 | 0.159 | 0.188 | 0.295 | 0.316 | 0.432 | 0.456 | 0.228 | 0.266 | | |
| | Qwen 3.5 Plus | 0.467 | 0.509 | 0.221 | 0.266 | 0.332 | 0.355 | 0.471 | 0.495 | 0.289 | 0.341 | | |
| COVID Memories | Gemma-4-31B-it | 0.313 | 0.341 | 0.191 | 0.223 | 0.225 | 0.242 | 0.31 | 0.329 | 0.262 | 0.304 | | |
| | gpt-oss-120b | 0.246 | 0.27 | 0.146 | 0.169 | 0.199 | 0.215 | 0.244 | 0.258 | 0.107 | 0.129 | | |
| | DeepSeek V4-Flash | 0.238 | 0.26 | 0.164 | 0.191 | 0.149 | 0.159 | 0.312 | 0.33 | 0.252 | 0.295 | | |
| | Qwen 3.5 Plus | 0.273 | 0.298 | 0.171 | 0.195 | 0.201 | 0.214 | 0.269 | 0.284 | 0.242 | 0.283 | | |
| Demographics + COVID Memories | Gemma-4-31B-it | 0.501 | 0.546 | 0.28 | 0.336 | 0.364 | 0.389 | 0.499 | 0.525 | 0.362 | 0.416 | | |
| | gpt-oss-120b | 0.438 | 0.483 | 0.187 | 0.231 | 0.335 | 0.365 | 0.486 | 0.516 | 0.253 | 0.305 | | |
| | DeepSeek V4-Flash | 0.344 | 0.376 | 0.207 | 0.246 | 0.312 | 0.334 | 0.474 | 0.5 | 0.31 | 0.361 | | |
| | Qwen 3.5 Plus | 0.48 | 0.523 | 0.261 | 0.305 | 0.381 | 0.406 | 0.477 | 0.503 | 0.329 | 0.384 | | |
| Demographics + All psychological measures + COVID Memories | Gemma-4-31B-it | 0.736 | 0.808 | 0.586 | 0.682 | 0.587 | 0.623 | 0.854 | 0.894 | 0.48 | 0.555 | | |
| | gpt-oss-120b | 0.68 | 0.754 | 0.503 | 0.598 | 0.47 | 0.508 | 0.827 | 0.876 | 0.365 | 0.453 | | |
| | DeepSeek V4-Flash | 0.67 | 0.737 | 0.534 | 0.626 | 0.555 | 0.594 | 0.83 | 0.873 | 0.496 | 0.58 | | |
| | Qwen 3.5 Plus | 0.708 | 0.771 | 0.487 | 0.566 | 0.501 | 0.532 | 0.808 | 0.846 | 0.466 | 0.545 | | |
| Demographics + Features with r < 0.5 + COVID contexts | Gemma-4-31B-it | 0.511 | 0.557 | 0.312 | 0.379 | 0.355 | 0.38 | 0.503 | 0.53 | 0.386 | 0.442 | | |
| | gpt-oss-120b | 0.415 | 0.459 | 0.171 | 0.209 | 0.303 | 0.329 | 0.477 | 0.508 | 0.316 | 0.375 | | |
| | DeepSeek V4-Flash | 0.361 | 0.398 | 0.235 | 0.282 | 0.305 | 0.329 | 0.48 | 0.507 | 0.401 | 0.464 | | |

| | | | | | | | | | | | |
|---|---|---|---|---|---|---|---|---|---|---|---|
| | Qwen 3.5 Plus | 0.457 | 0.498 | 0.196 | 0.23 | 0.366 | 0.391 | 0.481 | 0.508 | 0.396 | 0.458 |
| Demographics + Features with r > 0.5 + COVID contexts | Gemma-4-31B-it | 0.738 | 0.81 | 0.586 | 0.679 | 0.589 | 0.625 | 0.855 | 0.895 | 0.456 | 0.527 |
| | gpt-oss-120b | 0.666 | 0.737 | 0.505 | 0.596 | 0.49 | 0.53 | 0.844 | 0.894 | 0.373 | 0.467 |
| | DeepSeek V4-Flash | 0.67 | 0.735 | 0.493 | 0.575 | 0.568 | 0.607 | 0.835 | 0.877 | 0.465 | 0.543 |
| | Qwen 3.5 Plus | 0.713 | 0.777 | 0.511 | 0.593 | 0.508 | 0.54 | 0.806 | 0.844 | 0.443 | 0.52 |
| Demographics + COVID Memories + COVID contexts | Gemma-4-31B-it | 0.502 | 0.547 | 0.328 | 0.398 | 0.369 | 0.395 | 0.512 | 0.54 | 0.362 | 0.415 |
| | gpt-oss-120b | 0.455 | 0.505 | 0.191 | 0.238 | 0.342 | 0.375 | 0.502 | 0.536 | 0.269 | 0.328 |
| | DeepSeek V4-Flash | 0.372 | 0.409 | 0.237 | 0.281 | 0.336 | 0.362 | 0.472 | 0.498 | 0.341 | 0.396 |
| | Qwen 3.5 Plus | 0.447 | 0.489 | 0.235 | 0.276 | 0.361 | 0.386 | 0.478 | 0.505 | 0.346 | 0.403 |

*Note.* Item-level correlations were calculated by correlating human and digital-twin responses across participants for each item and then summarized within each trait. Profile correlations were calculated by correlating human and digital-twin response profiles across items within participants and then summarized by trait. Trait-score correlations were calculated using trait scores. Observed trait-score correlations are Pearson correlations between human and digital-twin trait scores. Disattenuated correlations correct the observed Pearson correlations for measurement unreliability using McDonald's ω. Overall correlations were computed using Fisher's z transformation.

**Table S42 Nomothetic span of NEO Five-Factor personality trait across input conditions and models**

| | Extraversion | | | | Agreeableness | | | | Conscientiousness | | | | Openness | | | | Neuroticism | | | |
|---|---|---|---|---|---|---|---|---|---|---|---|---|---|---|---|---|---|---|---|---|
| | Gemma-4-31B-it | gpt-oss-120b | DeepSeek V4-Flash | Qwen 3.5 Plus | Gemma-4-31B-it | gpt-oss-120b | DeepSeek V4-Flash | Qwen 3.5 Plus | Gemma-4-31B-it | gpt-oss-120b | DeepSeek V4-Flash | Qwen 3.5 Plus | Gemma-4-31B-it | gpt-oss-120b | DeepSeek V4-Flash | Qwen 3.5 Plus | Gemma-4-31B-it | gpt-oss-120b | DeepSeek V4-Flash | Qwen 3.5 Plus |
| Demographics | | | | | | | | | | | | | | | | | | | | |
| BDI-II | -0.354 *** | -0.287 *** | -0.188 *** | -0.327 *** | -0.233 *** | -0.158 *** | -0.178 *** | -0.217 *** | -0.389 *** | -0.289 *** | -0.278 *** | -0.362 *** | 0.146 *** | 0.024 | -0.003 | -0.013 | 0.441 *** | 0.411 *** | 0.390 *** | 0.440 *** |
| CD-RISC-10 | 0.350 *** | 0.308 *** | 0.234 *** | 0.318 *** | 0.204 *** | 0.188 *** | 0.217 *** | 0.217 *** | 0.345 *** | 0.304 *** | 0.275 *** | 0.341 *** | -0.137 *** | 0.05 | 0.014 | 0.029 | -0.327 *** | -0.349 *** | -0.304 *** | -0.333 *** |
| IPDS | -0.365 *** | -0.300 *** | -0.165 *** | -0.315 *** | -0.227 *** | -0.208 *** | -0.181 *** | -0.237 *** | -0.315 *** | -0.307 *** | -0.300 *** | -0.335 *** | 0.100 ** | -0.005 | -0.038 | -0.07 | 0.333 *** | 0.345 *** | 0.308 *** | 0.345 *** |
| PSS | -0.295 *** | -0.264 *** | -0.173 *** | -0.265 *** | -0.161 *** | -0.138 *** | -0.143 *** | -0.162 *** | -0.336 *** | -0.261 *** | -0.272 *** | -0.306 *** | 0.184 *** | 0.043 | 0.03 | 0.046 | 0.378 *** | 0.366 *** | 0.351 *** | 0.385 *** |
| PANAS | | | | | | | | | | | | | | | | | | | | |
| Positive affect | 0.318 *** | 0.272 *** | 0.171 *** | 0.273 *** | 0.199 *** | 0.171 *** | 0.186 *** | 0.166 *** | 0.301 *** | 0.248 *** | 0.276 *** | 0.300 *** | -0.176 *** | -0.01 | -0.044 | -0.068 | -0.260 *** | -0.292 *** | -0.303 *** | -0.272 *** |
| Negative affect | -0.108 ** | -0.096 * | -0.032 | -0.073 | -0.083 * | -0.047 | -0.057 | -0.05 | -0.141 *** | -0.06 | -0.090 * | -0.084 * | 0.081 * | 0.047 | 0.036 | 0.066 | 0.210 *** | 0.167 *** | 0.181 *** | 0.186 *** |
| SEE | | | | | | | | | | | | | | | | | | | | |
| Empathic feeling | 0.141*** | 0.166*** | 0.170*** | 0.180*** | 0.272*** | 0.264*** | 0.232*** | 0.227*** | -0.138*** | -0.108** | -0.004 | -0.063 | 0.368*** | 0.305*** | 0.293*** | 0.347*** | 0.204*** | 0.151*** | 0.199*** | 0.194*** |
| Perspective taking | 0.072 | 0.089* | 0.097* | 0.103** | 0.137*** | 0.112** | 0.076* | 0.107** | -0.023 | 0.009 | -0.006 | -0.018 | 0.147*** | 0.129*** | 0.161*** | 0.165*** | 0.022 | 0.051 | 0.086* | 0.032 |
| Acceptance of differences | 0.058 | 0.083* | 0.126*** | 0.083* | 0.145*** | 0.121** | 0.106** | 0.108** | -0.153*** | -0.154*** | -0.059 | -0.127*** | 0.312*** | 0.236*** | 0.253*** | 0.283*** | 0.163*** | 0.154*** | 0.191*** | 0.152*** |
| Empathic awareness | 0.032 | 0.093* | 0.115** | 0.098** | 0.205*** | 0.224*** | 0.162*** | 0.151*** | -0.185*** | -0.181*** | -0.140*** | -0.176*** | 0.466*** | 0.366*** | 0.361*** | 0.382*** | 0.193*** | 0.210*** | 0.253*** | 0.213*** |
| STAI | | | | | | | | | | | | | | | | | | | | |
| State anxiety | -0.255*** | -0.239*** | -0.126*** | -0.218*** | -0.183*** | -0.130*** | -0.143*** | -0.126*** | -0.277*** | -0.216*** | -0.224*** | -0.236*** | 0.126*** | 0.022 | 0.017 | 0.064 | 0.319*** | 0.301*** | 0.313*** | 0.317*** |
| Trait anxiety | -0.405*** | -0.326*** | -0.242*** | -0.360*** | -0.244*** | -0.179*** | -0.228*** | -0.221*** | -0.432*** | -0.370*** | -0.366*** | -0.415*** | 0.187*** | 0.017 | -0.006 | 0.014 | 0.445*** | 0.446*** | 0.434*** | 0.451*** |
| VSA | 0.101** | 0.043 | 0.035 | 0.067 | -0.072 | -0.114** | -0.053 | 0.007 | 0.256*** | 0.212*** | 0.244*** | 0.263*** | -0.583*** | -0.426*** | -0.376*** | -0.435*** | -0.246*** | -0.229*** | -0.278*** | -0.242*** |
| COVID Memories | | | | | | | | | | | | | | | | | | | | |
| BDI-II | -0.298*** | -0.218*** | -0.187*** | -0.218*** | -0.188*** | -0.150*** | -0.152*** | -0.154*** | -0.256*** | -0.279*** | -0.227*** | -0.249*** | 0.039 | -0.048 | 0.046 | 0.01 | 0.368*** | 0.304*** | 0.335*** | 0.308*** |
| CD-RISC-10 | 0.241*** | 0.182*** | 0.183*** | 0.226*** | 0.113** | 0.090* | 0.104** | 0.119** | 0.203*** | 0.229*** | 0.181*** | 0.232*** | -0.007 | 0.054 | 0.007 | 0.03 | -0.273*** | -0.224*** | -0.240*** | -0.248*** |
| IPDS | -0.221*** | -0.184*** | -0.127*** | -0.191*** | -0.167*** | -0.115** | -0.137*** | -0.140*** | -0.192*** | -0.193*** | -0.178*** | -0.208*** | 0.017 | -0.052 | 0.007 | -0.018 | 0.247*** | 0.212*** | 0.228*** | 0.207*** |
| PSS | -0.281*** | -0.220*** | -0.176*** | -0.219*** | -0.142*** | -0.121** | -0.131*** | -0.138*** | -0.251*** | -0.274*** | -0.221*** | -0.262*** | 0.079* | -0.04 | 0.066 | 0.03 | 0.384*** | 0.316*** | 0.371*** | 0.341*** |
| PANAS | | | | | | | | | | | | | | | | | | | | |
| Positive affect | 0.187*** | 0.153*** | 0.123** | 0.135*** | 0.118** | 0.071 | 0.129*** | 0.099** | 0.204*** | 0.188*** | 0.217*** | 0.204*** | -0.109** | -0.027 | -0.075* | -0.077* | -0.210*** | -0.199*** | -0.201*** | -0.182*** |
| Negative affect | -0.137*** | -0.103** | -0.091* | -0.133*** | -0.095* | -0.090* | -0.102** | -0.097* | -0.105** | -0.169*** | -0.151*** | -0.128*** | 0.021 | 0.02 | 0.007 | 0.001 | 0.191*** | 0.175*** | 0.217*** | 0.200*** |
| SEE | | | | | | | | | | | | | | | | | | | | |
| Empathic feeling | 0.009 | 0.021 | 0.045 | 0.028 | 0.089* | 0.077* | 0.117** | 0.103** | -0.07 | 0.007 | -0.013 | 0.002 | 0.219*** | 0.108** | 0.218*** | 0.208*** | 0.137*** | 0.079* | 0.135*** | 0.091* |
| Perspective taking | -0.054 | -0.002 | -0.026 | -0.009 | -0.041 | -0.018 | -0.018 | -0.013 | -0.035 | -0.024 | -0.006 | -0.007 | 0.106** | 0.028 | 0.104** | 0.084* | 0.072 | 0.043 | 0.073 | 0.045 |
| Acceptance of differences | 0.028 | 0.074 | 0.02 | 0.027 | 0.092* | 0.117** | 0.075* | 0.072 | -0.067 | 0.033 | -0.036 | -0.012 | 0.268*** | 0.154*** | 0.242*** | 0.212*** | 0.058 | -0.003 | 0.039 | 0.041 |
| Empathic awareness | -0.044 | 0 | 0.045 | 0.027 | 0.095* | 0.123** | 0.129*** | 0.141*** | -0.129*** | 0.002 | -0.069 | -0.033 | 0.282*** | 0.155*** | 0.276*** | 0.292*** | 0.157*** | 0.081* | 0.147*** | 0.100** |
| STAI | | | | | | | | | | | | | | | | | | | | |
| State anxiety | -0.241*** | -0.200*** | -0.183*** | -0.205*** | -0.147*** | -0.133*** | -0.166*** | -0.147*** | -0.207*** | -0.234*** | -0.229*** | -0.213*** | 0.049 | -0.01 | 0.041 | 0.01 | 0.312*** | 0.261*** | 0.314*** | 0.284*** |
| Trait anxiety | -0.284*** | -0.216*** | -0.169*** | -0.230*** | -0.149*** | -0.117** | -0.119** | -0.123** | -0.251*** | -0.254*** | -0.236*** | -0.258*** | 0.05 | -0.035 | 0.051 | 0.02 | 0.341*** | 0.277*** | 0.316*** | 0.296*** |
| VSA | 0.148*** | 0.042 | 0.035 | 0.072 | 0.06 | -0.023 | 0.024 | 0 | 0.133*** | 0.048 | 0.122** | 0.078* | -0.307*** | -0.149*** | -0.259*** | -0.257*** | -0.143*** | -0.087* | -0.105** | -0.07 |
| Demographics + COVID Memories | | | | | | | | | | | | | | | | | | | | |
| BDI-II | -0.377*** | -0.289*** | -0.284*** | -0.363*** | -0.219*** | -0.162*** | -0.249*** | -0.210*** | -0.368*** | -0.388*** | -0.381*** | -0.394*** | 0.167*** | -0.002 | -0.012 | 0.017 | 0.509*** | 0.457*** | 0.463*** | 0.493*** |
| CD-RISC-10 | 0.359*** | 0.294*** | 0.265*** | 0.326*** | 0.198*** | 0.141*** | 0.175*** | 0.176*** | 0.340*** | 0.346*** | 0.312*** | 0.376*** | -0.137*** | 0.036 | 0.029 | 0.03 | -0.386*** | -0.389*** | -0.363*** | -0.368*** |
| IPDS | -0.352*** | -0.298*** | -0.199*** | -0.300*** | -0.239*** | -0.201*** | -0.197*** | -0.217*** | -0.296*** | -0.376*** | -0.313*** | -0.373*** | 0.106** | -0.039 | -0.02 | -0.025 | 0.364*** | 0.372*** | 0.338*** | 0.355*** |
| PSS | -0.341*** | -0.251*** | -0.264*** | -0.324*** | -0.180*** | -0.141*** | -0.210*** | -0.183*** | -0.342*** | -0.331*** | -0.351*** | -0.358*** | 0.207*** | 0.023 | 0.021 | 0.051 | 0.498*** | 0.442*** | 0.463*** | 0.485*** |
| PANAS | | | | | | | | | | | | | | | | | | | | |
| Positive affect | 0.321*** | 0.247*** | 0.199*** | 0.251*** | 0.198*** | 0.103** | 0.173*** | 0.163*** | 0.303*** | 0.307*** | 0.305*** | 0.351*** | -0.196*** | -0.016 | -0.074* | -0.067 | -0.312*** | -0.323*** | -0.285*** | -0.305*** |
| Negative affect | -0.124*** | -0.112** | -0.111** | -0.129*** | -0.074* | -0.058 | -0.095* | -0.086* | -0.121** | -0.138*** | -0.141*** | -0.125*** | 0.085* | 0.013 | -0.009 | 0.045 | 0.242*** | 0.202*** | 0.223*** | 0.235*** |
| SEE | | | | | | | | | | | | | | | | | | | | |
| Empathic feeling | 0.090* | 0.139*** | 0.110** | 0.079* | 0.292*** | 0.222*** | 0.198*** | 0.211*** | -0.119** | -0.092* | -0.107** | -0.051 | 0.393*** | 0.306*** | 0.330*** | 0.346*** | 0.223*** | 0.157*** | 0.263*** | 0.233*** |
| Perspective taking | 0.001 | 0.035 | 0.027 | 0.021 | 0.080* | 0.071 | 0.019 | 0.05 | -0.023 | -0.043 | -0.073 | 0.004 | 0.149*** | 0.099** | 0.113** | 0.148*** | 0.076* | 0.071 | 0.113** | 0.074* |
| Acceptance of differences | 0.018 | 0.076* | 0.05 | 0.04 | 0.164*** | 0.152*** | 0.107** | 0.124*** | -0.156*** | -0.122** | -0.089* | -0.103** | 0.352*** | 0.278*** | 0.263*** | 0.306*** | 0.150*** | 0.108** | 0.180*** | 0.136*** |
| Empathic awareness | -0.01 | 0.072 | 0.043 | 0 | 0.250*** | 0.225*** | 0.134*** | 0.187*** | -0.199*** | -0.166*** | -0.236*** | -0.164*** | 0.491*** | 0.386*** | 0.390*** | 0.411*** | 0.251*** | 0.211*** | 0.312*** | 0.276*** |
| STAI | | | | | | | | | | | | | | | | | | | | |
| State anxiety | -0.289*** | -0.227*** | -0.215*** | -0.275*** | -0.170*** | -0.121** | -0.183*** | -0.163*** | -0.276*** | -0.285*** | -0.297*** | -0.303*** | 0.148*** | 0.001 | 0.018 | 0.035 | 0.402*** | 0.371*** | 0.373*** | 0.385*** |
| Trait anxiety | -0.412*** | -0.336*** | -0.310*** | -0.374*** | -0.219*** | -0.163*** | -0.252*** | -0.204*** | -0.406*** | -0.419*** | -0.427*** | -0.442*** | 0.196*** | 0.007 | -0.011 | 0.02 | 0.506*** | 0.483*** | 0.480*** | 0.491*** |
| VSA | 0.161*** | 0.044 | 0.071 | 0.122** | -0.043 | -0.108** | -0.022 | 0.018 | 0.263*** | 0.235*** | 0.274*** | 0.231*** | -0.584*** | -0.441*** | -0.439*** | -0.437*** | -0.290*** | -0.254*** | -0.310*** | -0.276*** |
| Demographics + All psychological measures + COVID Memories | | | | | | | | | | | | | | | | | | | | |
| BDI-II | -0.710*** | -0.540*** | -0.540*** | -0.695*** | -0.553*** | -0.400*** | -0.407*** | -0.555*** | -0.757*** | -0.723*** | -0.691*** | -0.793*** | 0.145*** | -0.103** | -0.062 | -0.016 | 0.840*** | 0.831*** | 0.842*** | 0.862*** |
| CD-RISC-10 | 0.686*** | 0.543*** | 0.584*** | 0.633*** | 0.539*** | 0.394*** | 0.464*** | 0.495*** | 0.749*** | 0.635*** | 0.723*** | 0.677*** | -0.03 | 0.169*** | 0.236*** | 0.153*** | -0.722*** | -0.684*** | -0.715*** | -0.683*** |
| IPDS | -0.560*** | -0.534*** | -0.426*** | -0.521*** | -0.727*** | -0.712*** | -0.619*** | -0.590*** | -0.604*** | -0.624*** | -0.534*** | -0.581*** | 0.059 | -0.119** | -0.113** | -0.066 | 0.632*** | 0.652*** | 0.631*** | 0.606*** |
| PSS | -0.694*** | -0.543*** | -0.515*** | -0.669*** | -0.591*** | -0.435*** | -0.451*** | -0.517*** | -0.754*** | -0.673*** | -0.692*** | -0.710*** | 0.149*** | -0.079* | -0.081* | -0.011 | 0.861*** | 0.806*** | 0.851*** | 0.839*** |
| PANAS | | | | | | | | | | | | | | | | | | | | |
| Positive affect | 0.615*** | 0.521*** | 0.656*** | 0.532*** | 0.396*** | 0.282*** | 0.361*** | 0.362*** | 0.547*** | 0.447*** | 0.563*** | 0.509*** | -0.072 | 0.134*** | 0.238*** | 0.024 | -0.511*** | -0.501*** | -0.495*** | -0.490*** |
| Negative affect | -0.403*** | -0.282*** | -0.274*** | -0.387*** | -0.393*** | -0.285*** | -0.351*** | -0.386*** | -0.415*** | -0.401*** | -0.384*** | -0.403*** | 0.01 | -0.093* | -0.093* | -0.058 | 0.550*** | 0.514*** | 0.549*** | 0.533*** |
| SEE | | | | | | | | | | | | | | | | | | | | |
| Empathic feeling | 0.087* | 0.184*** | 0.186*** | 0.07 | 0.331*** | 0.284*** | 0.496*** | 0.379*** | -0.022 | 0.02 | 0.103** | 0.015 | 0.626*** | 0.466*** | 0.644*** | 0.574*** | 0.077* | 0.022 | 0.052 | 0.099** |
| Perspective taking | 0.074* | 0.127*** | 0.113** | 0.047 | 0.138*** | 0.108** | 0.265*** | 0.136*** | 0.05 | 0.047 | 0.117** | 0.03 | 0.313*** | 0.251*** | 0.410*** | 0.300*** | -0.038 | -0.074* | -0.052 | 0.001 |
| Acceptance of differences | 0.028 | 0.124*** | 0.055 | 0.04 | 0.321*** | 0.293*** | 0.476*** | 0.320*** | -0.066 | 0.022 | 0.038 | -0.034 | 0.524*** | 0.413*** | 0.492*** | 0.465*** | 0.031 | -0.01 | -0.006 | 0.039 |
| Empathic awareness | -0.063 | 0.03 | 0.068 | -0.059 | 0.194*** | 0.156*** | 0.346*** | 0.312*** | -0.145*** | -0.082* | -0.033 | -0.114** | 0.668*** | 0.453*** | 0.630*** | 0.615*** | 0.180*** | 0.125*** | 0.145*** | 0.190*** |
| STAI | | | | | | | | | | | | | | | | | | | | |
| State anxiety | -0.637*** | -0.487*** | -0.505*** | -0.603*** | -0.519*** | -0.366*** | -0.441*** | -0.481*** | -0.639*** | -0.560*** | -0.617*** | -0.601*** | 0.062 | -0.122** | -0.152*** | -0.057 | 0.755*** | 0.694*** | 0.743*** | 0.718*** |
| Trait anxiety | -0.778*** | -0.624*** | -0.586*** | -0.729*** | -0.626*** | -0.460*** | -0.493*** | -0.566*** | -0.828*** | -0.744*** | -0.760*** | -0.788*** | 0.154*** | -0.105** | -0.105** | -0.033 | 0.913*** | 0.865*** | 0.887*** | 0.871*** |
| VSA | 0.145*** | 0.069 | 0.057 | 0.130*** | -0.102** | -0.087* | -0.269*** | -0.194*** | 0.194*** | 0.142*** | 0.135*** | 0.187*** | -0.703*** | -0.456*** | -0.583*** | -0.665*** | -0.193*** | -0.148*** | -0.164*** | -0.191*** |
| Demographics + Features with $r < 0.5$ + COVID contexts | | | | | | | | | | | | | | | | | | | | |
| BDI-II | -0.384*** | -0.264*** | -0.264*** | -0.361*** | -0.207*** | -0.180*** | -0.153*** | -0.189*** | -0.377*** | -0.299*** | -0.266*** | -0.371*** | 0.160*** | -0.01 | 0.077* | 0.067 | 0.520*** | 0.468*** | 0.448*** | 0.505*** |
| CD-RISC-10 | 0.382*** | 0.291*** | 0.250*** | 0.344*** | 0.193*** | 0.156*** | 0.146*** | 0.108** | 0.335*** | 0.278*** | 0.291*** | 0.333*** | -0.112** | 0.052 | 0.008 | 0.017 | -0.377*** | -0.356*** | -0.345*** | -0.359*** |
| IPDS | -0.368*** | -0.264*** | -0.229*** | -0.307*** | -0.254*** | -0.173*** | -0.203*** | -0.177*** | -0.328*** | -0.252*** | -0.286*** | -0.333*** | 0.109** | -0.035 | -0.001 | 0.022 | 0.393*** | 0.370*** | 0.332*** | 0.367*** |
| PSS | -0.360*** | -0.255*** | -0.238*** | -0.335*** | -0.188*** | -0.138*** | -0.128*** | -0.148*** | -0.349*** | -0.254*** | -0.261*** | -0.333*** | 0.193*** | 0.017 | 0.111** | 0.100** | 0.526*** | 0.471*** | 0.465*** | 0.527*** |
| PANAS | | | | | | | | | | | | | | | | | | | | |
| Positive affect | 0.356*** | 0.227*** | 0.213*** | 0.260*** | 0.189*** | 0.106** | 0.147*** | 0.083* | 0.346*** | 0.199*** | 0.312*** | 0.349*** | -0.179*** | -0.033 | -0.056 | -0.080* | -0.267*** | -0.225*** | -0.234*** | -0.270*** |
| Negative affect | -0.149*** | -0.087* | -0.072 | -0.141*** | -0.115** | -0.067 | -0.075* | -0.119** | -0.120** | -0.089* | -0.059 | -0.098** | 0.06 | -0.022 | 0.034 | 0.018 | 0.246*** | 0.219*** | 0.252*** | 0.260*** |
| SEE | | | | | | | | | | | | | | | | | | | | |
| Empathic feeling | 0.056 | 0.141*** | 0.129*** | 0.054 | 0.320*** | 0.281*** | 0.321*** | 0.284*** | -0.126*** | -0.016 | 0.011 | -0.038 | 0.432*** | 0.377*** | 0.460*** | 0.440*** | 0.315*** | 0.254*** | 0.294*** | 0.332*** |
| Perspective taking | -0.038 | 0.033 | 0.032 | -0.023 | 0.058 | 0.080* | 0.102** | 0.057 | -0.062 | -0.027 | 0.006 | -0.012 | 0.171*** | 0.146*** | 0.221*** | 0.187*** | 0.143*** | 0.116** | 0.135*** | 0.154*** |
| Acceptance of differences | 0.014 | 0.089* | 0.083* | 0.054 | 0.226*** | 0.194*** | 0.262*** | 0.226*** | -0.175*** | -0.06 | -0.074 | -0.112** | 0.410*** | 0.313*** | 0.369*** | 0.394*** | 0.208*** | 0.171*** | 0.203*** | 0.186*** |
| Empathic awareness | -0.042 | 0.113** | 0.115** | -0.007 | 0.323*** | 0.323*** | 0.321*** | 0.359*** | -0.235*** | -0.068 | -0.160*** | -0.163*** | 0.564*** | 0.477*** | 0.547*** | 0.530*** | 0.354*** | 0.275*** | 0.353*** | 0.363*** |
| STAI | | | | | | | | | | | | | | | | | | | | |
| State anxiety | -0.324*** | -0.216*** | -0.198*** | -0.294*** | -0.198*** | -0.133*** | -0.162*** | -0.154*** | -0.300*** | -0.178*** | -0.251*** | -0.288*** | 0.120** | 0.007 | 0.041 | 0.041 | 0.415*** | 0.366*** | 0.371*** | 0.414*** |
| Trait anxiety | -0.404*** | -0.293*** | -0.249*** | -0.358*** | -0.211*** | -0.166*** | -0.167*** | -0.133*** | -0.413*** | -0.302*** | -0.345*** | -0.396*** | 0.197*** | 0.027 | 0.080* | 0.097* | 0.520*** | 0.497*** | 0.472*** | 0.500*** |
| VSA | 0.145*** | -0.025 | -0.059 | 0.052 | -0.196*** | -0.226*** | -0.289*** | -0.284*** | 0.306*** | 0.167*** | 0.243*** | 0.256*** | -0.735*** | -0.553*** | -0.679*** | -0.677*** | -0.312*** | -0.261*** | -0.295*** | -0.297*** |
| Demographics + Features with r > 0.5 + COVID contexts | | | | | | | | | | | | | | | | | | | | |
| BDI-II | -0.712*** | -0.557*** | -0.554*** | -0.679*** | -0.537*** | -0.390*** | -0.421*** | -0.540*** | -0.760*** | -0.698*** | -0.669*** | -0.790*** | 0.141*** | -0.076* | -0.098** | -0.054 | 0.837*** | 0.827*** | 0.838*** | 0.863*** |
| CD-RISC-10 | 0.690*** | 0.568*** | 0.585*** | 0.626*** | 0.543*** | 0.389*** | 0.510*** | 0.485*** | 0.747*** | 0.638*** | 0.746*** | 0.679*** | -0.037 | 0.181*** | 0.276*** | 0.148*** | -0.726*** | -0.716*** | -0.705*** | -0.678*** |
| IPDS | -0.557*** | -0.548*** | -0.439*** | -0.524*** | -0.735*** | -0.716*** | -0.639*** | -0.602*** | -0.607*** | -0.593*** | -0.529*** | -0.583*** | 0.067 | -0.109** | -0.178*** | -0.073 | 0.642*** | 0.671*** | 0.629*** | 0.607*** |
| PSS | -0.706*** | -0.548*** | -0.530*** | -0.664*** | -0.574*** | -0.436*** | -0.464*** | -0.517*** | -0.742*** | -0.648*** | -0.677*** | -0.718*** | 0.152*** | -0.068 | -0.081* | -0.03 | 0.859*** | 0.811*** | 0.853*** | 0.833*** |
| PANAS | | | | | | | | | | | | | | | | | | | | |
| Positive affect | 0.604*** | 0.508*** | 0.678*** | 0.552*** | 0.385*** | 0.289*** | 0.389*** | 0.365*** | 0.537*** | 0.472*** | 0.558*** | 0.516*** | -0.063 | 0.136*** | 0.274*** | 0.05 | -0.512*** | -0.503*** | -0.478*** | -0.503*** |
| Negative affect | -0.411*** | -0.308*** | -0.285*** | -0.379*** | -0.384*** | -0.265*** | -0.345*** | -0.362*** | -0.431*** | -0.348*** | -0.365*** | -0.409*** | 0.033 | -0.066 | -0.07 | -0.057 | 0.544*** | 0.519*** | 0.554*** | 0.530*** |
| SEE | | | | | | | | | | | | | | | | | | | | |
| Empathic feeling | 0.100** | 0.191*** | 0.224*** | 0.078* | 0.341*** | 0.282*** | 0.472*** | 0.371*** | -0.027 | 0.013 | 0.123** | 0.025 | 0.613*** | 0.490*** | 0.653*** | 0.560*** | 0.067 | 0.027 | 0.076* | 0.100** |
| Perspective taking | 0.094* | 0.109** | 0.135*** | 0.048 | 0.145*** | 0.072 | 0.269*** | 0.138*** | 0.05 | 0.023 | 0.160*** | 0.043 | 0.312*** | 0.251*** | 0.411*** | 0.309*** | -0.045 | -0.058 | -0.044 | -0.002 |
| Acceptance of differences | 0.044 | 0.126*** | 0.053 | 0.041 | 0.313*** | 0.256*** | 0.374*** | 0.333*** | -0.06 | -0.037 | 0.015 | -0.005 | 0.504*** | 0.387*** | 0.468*** | 0.487*** | 0.025 | 0.002 | 0.008 | 0.032 |
| Empathic awareness | -0.049 | 0.048 | 0.075* | -0.053 | 0.197*** | 0.161*** | 0.318*** | 0.291*** | -0.129*** | -0.098** | -0.012 | -0.094* | 0.660*** | 0.498*** | 0.599*** | 0.555*** | 0.178*** | 0.152*** | 0.160*** | 0.197*** |
| STAI | | | | | | | | | | | | | | | | | | | | |
| State anxiety | -0.646*** | -0.510*** | -0.518*** | -0.599*** | -0.512*** | -0.352*** | -0.442*** | -0.477*** | -0.639*** | -0.539*** | -0.620*** | -0.613*** | 0.066 | -0.109** | -0.142*** | -0.072 | 0.750*** | 0.693*** | 0.746*** | 0.720*** |
| Trait anxiety | -0.783*** | -0.638*** | -0.604*** | -0.717*** | -0.623*** | -0.462*** | -0.503*** | -0.551*** | -0.821*** | -0.744*** | -0.759*** | -0.780*** | 0.156*** | -0.085* | -0.140*** | -0.058 | 0.913*** | 0.882*** | 0.882*** | 0.861*** |
| VSA | 0.140*** | 0.036 | 0.091* | 0.130*** | -0.086* | -0.112** | -0.187*** | -0.175*** | 0.191*** | 0.166*** | 0.116** | 0.150*** | -0.644*** | -0.486*** | -0.468*** | -0.556*** | -0.196*** | -0.166*** | -0.160*** | -0.193*** |
| Demographics + COVID Memories + COVID contexts | | | | | | | | | | | | | | | | | | | | |
| BDI-II | -0.401*** | -0.281*** | -0.275*** | -0.356*** | -0.221*** | -0.189*** | -0.233*** | -0.172*** | -0.364*** | -0.356*** | -0.353*** | -0.337*** | 0.145*** | 0.008 | 0.029 | 0.054 | 0.528*** | 0.479*** | 0.484*** | 0.499*** |
| CD-RISC-10 | 0.379*** | 0.306*** | 0.239*** | 0.335*** | 0.181*** | 0.150*** | 0.193*** | 0.123** | 0.366*** | 0.340*** | 0.336*** | 0.322*** | -0.120** | 0.052 | 0.026 | 0.037 | -0.378*** | -0.360*** | -0.349*** | -0.355*** |
| IPDS | -0.374*** | -0.308*** | -0.177*** | -0.315*** | -0.257*** | -0.205*** | -0.213*** | -0.187*** | -0.321*** | -0.325*** | -0.264*** | -0.335*** | 0.103** | -0.018 | -0.02 | 0.001 | 0.393*** | 0.408*** | 0.337*** | 0.365*** |
| PSS | -0.369*** | -0.274*** | -0.224*** | -0.353*** | -0.182*** | -0.149*** | -0.185*** | -0.142*** | -0.344*** | -0.312*** | -0.292*** | -0.308*** | 0.187*** | 0.031 | 0.096* | 0.086* | 0.535*** | 0.501*** | 0.497*** | 0.520*** |
| PANAS | | | | | | | | | | | | | | | | | | | | |
| Positive affect | 0.345*** | 0.266*** | 0.179*** | 0.250*** | 0.178*** | 0.141*** | 0.163*** | 0.090* | 0.323*** | 0.306*** | 0.303*** | 0.340*** | -0.176*** | -0.03 | -0.089* | -0.073 | -0.284*** | -0.255*** | -0.264*** | -0.258*** |
| Negative affect | -0.141*** | -0.091* | -0.099** | -0.150*** | -0.119** | -0.081* | -0.128*** | -0.125*** | -0.125*** | -0.133*** | -0.159*** | -0.095* | 0.045 | -0.012 | -0.001 | 0.021 | 0.254*** | 0.239*** | 0.235*** | 0.267*** |
| SEE | | | | | | | | | | | | | | | | | | | | |
| Empathic feeling | 0.055 | 0.168*** | 0.084* | 0.033 | 0.330*** | 0.301*** | 0.259*** | 0.261*** | -0.096* | 0.001 | 0.021 | -0.008 | 0.434*** | 0.357*** | 0.424*** | 0.399*** | 0.310*** | 0.245*** | 0.357*** | 0.343*** |
| Perspective taking | -0.033 | 0.022 | -0.034 | -0.039 | 0.071 | 0.082* | 0.072 | 0.043 | -0.025 | 0.002 | 0.028 | 0.026 | 0.183*** | 0.127*** | 0.158*** | 0.179*** | 0.137*** | 0.102** | 0.179*** | 0.160*** |
| Acceptance of differences | 0.008 | 0.099** | 0.025 | 0.029 | 0.208*** | 0.182*** | 0.182*** | 0.196*** | -0.153*** | -0.073 | -0.071 | -0.081* | 0.408*** | 0.289*** | 0.335*** | 0.356*** | 0.209*** | 0.158*** | 0.235*** | 0.197*** |
| Empathic awareness | -0.05 | 0.080* | 0.056 | -0.042 | 0.306*** | 0.307*** | 0.247*** | 0.300*** | -0.204*** | -0.044 | -0.093* | -0.120** | 0.552*** | 0.412*** | 0.491*** | 0.479*** | 0.347*** | 0.290*** | 0.381*** | 0.375*** |
| STAI | | | | | | | | | | | | | | | | | | | | |
| State anxiety | -0.324*** | -0.227*** | -0.218*** | -0.303*** | -0.192*** | -0.158*** | -0.207*** | -0.170*** | -0.296*** | -0.275*** | -0.275*** | -0.278*** | 0.109** | 0.003 | 0.022 | 0.021 | 0.418*** | 0.398*** | 0.408*** | 0.420*** |
| Trait anxiety | -0.426*** | -0.319*** | -0.257*** | -0.369*** | -0.217*** | -0.179*** | -0.213*** | -0.141*** | -0.417*** | -0.394*** | -0.372*** | -0.378*** | 0.186*** | 0.023 | 0.058 | 0.06 | 0.530*** | 0.515*** | 0.485*** | 0.502*** |
| VSA | 0.162*** | 0.046 | 0.025 | 0.112** | -0.131*** | -0.178*** | -0.137*** | -0.154*** | 0.262*** | 0.170*** | 0.194*** | 0.216*** | -0.636*** | -0.429*** | -0.520*** | -0.525*** | -0.325*** | -0.257*** | -0.314*** | -0.309*** |

Note. Beck Depression Inventory-II = BDI-II; Connor-Davidson Resilience Scale = CD-RISC-10; Iowa Personality Disorder Screen = IPDS; Perceived Stress Scale = PSS; Positive and Negative Affect Scales = PANAS; Scale of Ethnocultural Empathy = SEE; State-Trait Anxiety Inventory = STAI; Very Short Authoritarianism Scale = VSA. *** $p < .001$, ** $p < .01$, * $p < .05$.

**Table S43 Recovery of the human Big Five nomothetic span across models and input conditions in Study 3b**

| Model | Input conditions | Pattern similarity | MAE |
|---|---|---|---|
| Gemma-4-31B-it | Demographics | 0.903 | 0.152 |
| | COVID Memories | 0.934 | 0.184 |
| | Demographics + COVID Memories | 0.909 | 0.149 |
| | Demographics + All psychological measures + COVID Memories | 0.975 | 0.127 |
| | Demographics + Features with $r < 0.5$ + COVID contexts | 0.887 | 0.153 |
| | Demographics + Features with $r > 0.5$ + COVID contexts | 0.975 | 0.125 |
| | Demographics + COVID Memories + COVID contexts | 0.901 | 0.146 |
| gpt-oss-120b | Demographics | 0.928 | 0.157 |
| | COVID Memories | 0.95 | 0.199 |
| | Demographics + COVID Memories | 0.933 | 0.149 |
| | Demographics + All psychological measures + COVID Memories | 0.982 | 0.078 |
| | Demographics + Features with $r < 0.5$ + COVID contexts | 0.908 | 0.162 |
| | Demographics + Features with $r > 0.5$ + COVID contexts | 0.984 | 0.08 |
| | Demographics + COVID Memories + COVID contexts | 0.934 | 0.14 |
| DeepSeek V4-Flash | Demographics | 0.904 | 0.171 |
| | COVID Memories | 0.934 | 0.193 |
| | Demographics + COVID Memories | 0.906 | 0.157 |
| | Demographics + All psychological measures + COVID Memories | 0.977 | 0.101 |
| | Demographics + Features with $r < 0.5$ + COVID contexts | 0.877 | 0.168 |
| | Demographics + Features with $r > 0.5$ + COVID contexts | 0.976 | 0.103 |
| | Demographics + COVID Memories + COVID contexts | 0.891 | 0.156 |
| Qwen 3.5 Plus | Demographics | 0.928 | 0.15 |
| | COVID Memories | 0.955 | 0.186 |
| | Demographics + COVID Memories | 0.936 | 0.137 |
| | Demographics + All psychological measures + COVID Memories | 0.978 | 0.107 |
| | Demographics + Features with $r < 0.5$ + COVID contexts | 0.893 | 0.153 |
| | Demographics + Features with $r > 0.5$ + COVID contexts | 0.979 | 0.103 |
| | Demographics + COVID Memories + COVID contexts | 0.909 | 0.147 |

*Note.* Pattern similarity refers to the Pearson correlation between the vectorized human Big Five–external scale correlation matrix and the corresponding model-derived correlation matrix. MAE = Mean absolute error. Mean absolute error represents the average absolute difference between model-derived and human correlations.

Table S44 Accuracy of digital twin responses to the NEO Five-Factor Personality Inventory across input conditions and temporal manipulations

| Input condition | Date cue | Covid context cue | Extraversion | Agreeableness | Conscientiousness | Neuroticism | Openness | Overall |
|---|---|---|---|---|---|---|---|---|
| Demographics | Absent | Absent | 0.753 | 0.798 | 0.799 | 0.744 | 0.725 | 0.764 |
| | Absent | Present | 0.754 | 0.799 | 0.798 | 0.742 | 0.730 | 0.765 |
| | Present | Present | 0.756 | 0.796 | 0.798 | 0.741 | 0.735 | 0.765 |
| | Present | Absent | 0.753 | 0.798 | 0.792 | 0.739 | 0.734 | 0.763 |
| COVID Memories | Absent | Absent | 0.746 | 0.795 | 0.790 | 0.705 | 0.728 | 0.753 |
| | Absent | Present | 0.746 | 0.797 | 0.784 | 0.702 | 0.732 | 0.752 |
| | Present | Present | 0.747 | 0.794 | 0.784 | 0.707 | 0.737 | 0.754 |
| | Present | Absent | 0.749 | 0.795 | 0.791 | 0.710 | 0.737 | 0.756 |
| Demographics + COVID Memories | Absent | Absent | 0.762 | 0.797 | 0.802 | 0.733 | 0.733 | 0.765 |
| | Absent | Present | 0.762 | 0.801 | 0.800 | 0.739 | 0.739 | 0.768 |
| | Present | Present | 0.763 | 0.797 | 0.798 | 0.735 | 0.737 | 0.766 |
| | Present | Absent | 0.765 | 0.797 | 0.799 | 0.730 | 0.740 | 0.766 |
| Demographics + All psychological measures + COVID Memories | Absent | Absent | 0.782 | 0.808 | 0.807 | 0.810 | 0.772 | 0.796 |
| | Absent | Present | 0.781 | 0.808 | 0.803 | 0.806 | 0.766 | 0.793 |
| | Present | Present | 0.781 | 0.808 | 0.801 | 0.804 | 0.768 | 0.792 |
| | Present | Absent | 0.78 | 0.809 | 0.807 | 0.805 | 0.770 | 0.794 |
| Demographics + Features with r < 0.5 + COVID contexts | Absent | Absent | 0.751 | 0.795 | 0.790 | 0.737 | 0.758 | 0.766 |
| | Absent | Present | 0.753 | 0.795 | 0.787 | 0.732 | 0.757 | 0.765 |
| | Present | Present | 0.755 | 0.794 | 0.785 | 0.732 | 0.757 | 0.764 |
| | Present | Absent | 0.746 | 0.794 | 0.789 | 0.733 | 0.761 | 0.764 |
| Demographics + Features with r > 0.5 + COVID contexts | Absent | Absent | 0.779 | 0.811 | 0.806 | 0.810 | 0.763 | 0.794 |
| | Absent | Present | 0.781 | 0.811 | 0.804 | 0.807 | 0.763 | 0.793 |
| | Present | Present | 0.784 | 0.809 | 0.808 | 0.804 | 0.759 | 0.793 |
| | Present | Absent | 0.777 | 0.809 | 0.803 | 0.801 | 0.761 | 0.790 |
| Demographics + COVID Memories + COVID contexts | Absent | Absent | 0.758 | 0.801 | 0.789 | 0.739 | 0.751 | 0.768 |
| | Absent | Present | 0.758 | 0.801 | 0.789 | 0.736 | 0.750 | 0.767 |
| | Present | Present | 0.756 | 0.799 | 0.791 | 0.742 | 0.748 | 0.767 |
| | Present | Absent | 0.752 | 0.800 | 0.792 | 0.739 | 0.751 | 0.767 |

**Table S45 Correlations of digital twin responses to the NEO Five-Factor Personality Inventory across input conditions and models**

| Feature | Date cue | Covid context cue | Extraversion | | Agreeableness | | Conscientiousness | | Neuroticism | | Openness | | Overall | |
|---|---|---|---|---|---|---|---|---|---|---|---|---|---|---|
| Item-level correlation | | | Pearson | Spearman | Pearson | Spearman | Pearson | Spearman | Pearson | Spearman | Pearson | Spearman | Pearson | Spearman |
| Demographics | Absent | Absent | 0.193 | 0.192 | 0.098 | 0.094 | 0.197 | 0.2 | 0.296 | 0.286 | 0.153 | 0.149 | 0.188 | 0.185 |
| | Absent | Present | 0.18 | 0.175 | 0.086 | 0.086 | 0.182 | 0.186 | 0.271 | 0.256 | 0.15 | 0.151 | 0.174 | 0.171 |
| | Present | Present | 0.197 | 0.195 | 0.081 | 0.084 | 0.189 | 0.193 | 0.272 | 0.264 | 0.156 | 0.153 | 0.18 | 0.178 |
| | Present | Absent | 0.21 | 0.214 | 0.091 | 0.097 | 0.146 | 0.154 | 0.276 | 0.271 | 0.17 | 0.171 | 0.18 | 0.182 |
| COVID Memories | Absent | Absent | 0.147 | 0.146 | 0.098 | 0.097 | 0.117 | 0.121 | 0.199 | 0.194 | 0.124 | 0.13 | 0.137 | 0.138 |
| | Absent | Present | 0.123 | 0.123 | 0.085 | 0.087 | 0.089 | 0.082 | 0.153 | 0.145 | 0.117 | 0.117 | 0.113 | 0.111 |
| | Present | Present | 0.139 | 0.138 | 0.089 | 0.083 | 0.078 | 0.072 | 0.165 | 0.16 | 0.114 | 0.118 | 0.117 | 0.114 |
| | Present | Absent | 0.159 | 0.158 | 0.099 | 0.095 | 0.129 | 0.124 | 0.194 | 0.191 | 0.12 | 0.123 | 0.14 | 0.138 |
| Demographics + COVID Memories | Absent | Absent | 0.243 | 0.243 | 0.111 | 0.11 | 0.215 | 0.219 | 0.296 | 0.29 | 0.191 | 0.19 | 0.212 | 0.211 |
| | Absent | Present | 0.216 | 0.214 | 0.127 | 0.128 | 0.2 | 0.205 | 0.312 | 0.307 | 0.191 | 0.189 | 0.21 | 0.209 |
| | Present | Present | 0.225 | 0.224 | 0.104 | 0.104 | 0.206 | 0.212 | 0.297 | 0.292 | 0.166 | 0.169 | 0.2 | 0.201 |
| | Present | Absent | 0.247 | 0.245 | 0.114 | 0.113 | 0.199 | 0.202 | 0.287 | 0.281 | 0.188 | 0.186 | 0.208 | 0.206 |
| Demographics + All psychological measures + COVID Memories | Absent | Absent | 0.405 | 0.408 | 0.276 | 0.285 | 0.343 | 0.375 | 0.604 | 0.601 | 0.305 | 0.312 | 0.394 | 0.403 |
| | Absent | Present | 0.402 | 0.395 | 0.279 | 0.293 | 0.316 | 0.339 | 0.587 | 0.584 | 0.272 | 0.278 | 0.379 | 0.385 |
| | Present | Present | 0.398 | 0.394 | 0.283 | 0.284 | 0.32 | 0.337 | 0.591 | 0.583 | 0.281 | 0.294 | 0.382 | 0.385 |
| | Present | Absent | 0.39 | 0.387 | 0.278 | 0.282 | 0.337 | 0.357 | 0.59 | 0.588 | 0.291 | 0.3 | 0.384 | 0.389 |
| Demographics + Features with r < 0.5 + COVID contexts | Absent | Absent | 0.232 | 0.232 | 0.103 | 0.103 | 0.189 | 0.199 | 0.317 | 0.319 | 0.245 | 0.252 | 0.218 | 0.222 |
| | Absent | Present | 0.207 | 0.204 | 0.112 | 0.11 | 0.167 | 0.172 | 0.284 | 0.279 | 0.212 | 0.222 | 0.197 | 0.198 |
| | Present | Present | 0.236 | 0.227 | 0.12 | 0.128 | 0.172 | 0.174 | 0.303 | 0.302 | 0.23 | 0.235 | 0.213 | 0.214 |
| | Present | Absent | 0.193 | 0.194 | 0.099 | 0.108 | 0.183 | 0.183 | 0.288 | 0.283 | 0.222 | 0.227 | 0.198 | 0.2 |
| | Absent | Absent | 0.388 | 0.39 | 0.299 | 0.303 | 0.342 | 0.359 | 0.601 | 0.599 | 0.277 | 0.282 | 0.389 | 0.394 |
| | Absent | Present | 0.379 | 0.375 | 0.275 | 0.275 | 0.348 | 0.359 | 0.592 | 0.585 | 0.266 | 0.279 | 0.38 | 0.381 |
| Demographics + Features with r > 0.5 + COVID contexts | Present | Present | 0.403 | 0.401 | 0.272 | 0.277 | 0.354 | 0.365 | 0.589 | 0.584 | 0.241 | 0.257 | 0.38 | 0.384 |
| | Present | Absent | 0.379 | 0.379 | 0.282 | 0.278 | 0.328 | 0.351 | 0.58 | 0.575 | 0.266 | 0.275 | 0.374 | 0.378 |
| Demographics + COVID Memories + COVID contexts | Absent | Absent | 0.232 | 0.233 | 0.128 | 0.13 | 0.185 | 0.195 | 0.31 | 0.3 | 0.229 | 0.233 | 0.218 | 0.219 |
| | Absent | Present | 0.227 | 0.224 | 0.125 | 0.132 | 0.184 | 0.194 | 0.299 | 0.292 | 0.196 | 0.199 | 0.207 | 0.209 |
| | Present | Present | 0.216 | 0.214 | 0.112 | 0.114 | 0.189 | 0.191 | 0.316 | 0.306 | 0.199 | 0.202 | 0.207 | 0.206 |
| | Present | Absent | 0.198 | 0.199 | 0.122 | 0.124 | 0.196 | 0.197 | 0.315 | 0.309 | 0.195 | 0.2 | 0.206 | 0.207 |
| Profile correlation | | | | | | | | | | | | | | |
| Demographics | Absent | Absent | 0.263 | 0.266 | 0.706 | 0.693 | 0.698 | 0.641 | 0.387 | 0.358 | 0.313 | 0.312 | 0.461 | 0.46 |
| | Absent | Present | 0.292 | 0.294 | 0.717 | 0.707 | 0.705 | 0.644 | 0.345 | 0.33 | 0.34 | 0.337 | 0.475 | 0.474 |
| | Present | Present | 0.285 | 0.285 | 0.704 | 0.689 | 0.695 | 0.64 | 0.389 | 0.371 | 0.374 | 0.367 | 0.479 | 0.477 |
| | Present | Absent | 0.285 | 0.287 | 0.713 | 0.705 | 0.674 | 0.619 | 0.38 | 0.358 | 0.398 | 0.394 | 0.479 | 0.478 |
| COVID Memories | Absent | Absent | 0.236 | 0.24 | 0.705 | 0.695 | 0.668 | 0.604 | 0.177 | 0.173 | 0.33 | 0.332 | 0.436 | 0.432 |
| | Absent | Present | 0.258 | 0.261 | 0.718 | 0.705 | 0.676 | 0.605 | 0.155 | 0.152 | 0.356 | 0.351 | 0.447 | 0.443 |
| | Present | Present | 0.267 | 0.272 | 0.719 | 0.708 | 0.684 | 0.616 | 0.165 | 0.166 | 0.379 | 0.381 | 0.451 | 0.447 |
| | Present | Absent | 0.259 | 0.263 | 0.709 | 0.698 | 0.674 | 0.607 | 0.173 | 0.173 | 0.38 | 0.38 | 0.45 | 0.446 |
| Demographics + COVID Memories | Absent | Absent | 0.333 | 0.334 | 0.705 | 0.694 | 0.702 | 0.651 | 0.327 | 0.31 | 0.366 | 0.363 | 0.476 | 0.475 |
| | Absent | Present | 0.333 | 0.333 | 0.715 | 0.703 | 0.691 | 0.633 | 0.334 | 0.321 | 0.404 | 0.401 | 0.488 | 0.487 |
| | Present | Present | 0.331 | 0.331 | 0.709 | 0.7 | 0.69 | 0.63 | 0.344 | 0.332 | 0.39 | 0.384 | 0.48 | 0.479 |
| | Present | Absent | 0.337 | 0.339 | 0.705 | 0.7 | 0.671 | 0.615 | 0.311 | 0.3 | 0.407 | 0.403 | 0.479 | 0.477 |
| Demographics + All psychological measures + COVID Memories | Absent | Absent | 0.45 | 0.445 | 0.762 | 0.737 | 0.756 | 0.698 | 0.709 | 0.672 | 0.547 | 0.535 | 0.608 | 0.604 |
| | Absent | Present | 0.431 | 0.433 | 0.755 | 0.725 | 0.759 | 0.696 | 0.678 | 0.635 | 0.543 | 0.534 | 0.603 | 0.599 |
| | Present | Present | 0.436 | 0.437 | 0.758 | 0.737 | 0.75 | 0.69 | 0.687 | 0.649 | 0.554 | 0.546 | 0.603 | 0.599 |
| | Present | Absent | 0.417 | 0.418 | 0.754 | 0.731 | 0.745 | 0.681 | 0.701 | 0.664 | 0.553 | 0.542 | 0.603 | 0.6 |
| Demographics + Features with r < 0.5 + COVID contexts | Absent | Absent | 0.314 | 0.314 | 0.729 | 0.708 | 0.734 | 0.67 | 0.384 | 0.359 | 0.534 | 0.528 | 0.519 | 0.516 |
| | Absent | Present | 0.317 | 0.32 | 0.723 | 0.699 | 0.739 | 0.685 | 0.323 | 0.299 | 0.508 | 0.505 | 0.513 | 0.511 |
| | Present | Present | 0.318 | 0.321 | 0.721 | 0.697 | 0.729 | 0.667 | 0.361 | 0.333 | 0.523 | 0.519 | 0.515 | 0.512 |
| | Present | Absent | 0.275 | 0.283 | 0.726 | 0.702 | 0.714 | 0.654 | 0.347 | 0.337 | 0.534 | 0.528 | 0.511 | 0.508 |
| Demographics + Features with r > 0.5 + COVID contexts | Absent | Absent | 0.433 | 0.437 | 0.765 | 0.738 | 0.768 | 0.714 | 0.692 | 0.659 | 0.511 | 0.501 | 0.598 | 0.596 |
| | Absent | Present | 0.435 | 0.436 | 0.765 | 0.742 | 0.758 | 0.699 | 0.704 | 0.667 | 0.496 | 0.489 | 0.595 | 0.593 |
| | Present | Present | 0.455 | 0.458 | 0.759 | 0.732 | 0.761 | 0.703 | 0.699 | 0.666 | 0.504 | 0.496 | 0.598 | 0.595 |
| | Present | Absent | 0.437 | 0.439 | 0.755 | 0.734 | 0.753 | 0.692 | 0.688 | 0.649 | 0.519 | 0.507 | 0.594 | 0.591 |
| Demographics + COVID Memories + COVID contexts | Absent | Absent | 0.324 | 0.326 | 0.734 | 0.716 | 0.723 | 0.654 | 0.377 | 0.357 | 0.477 | 0.471 | 0.514 | 0.511 |
| | Absent | Present | 0.321 | 0.323 | 0.737 | 0.72 | 0.718 | 0.653 | 0.349 | 0.327 | 0.476 | 0.476 | 0.513 | 0.51 |
| | Present | Present | 0.31 | 0.314 | 0.73 | 0.709 | 0.727 | 0.663 | 0.379 | 0.362 | 0.476 | 0.472 | 0.51 | 0.508 |
| | Present | Absent | 0.293 | 0.297 | 0.74 | 0.721 | 0.731 | 0.668 | 0.368 | 0.345 | 0.473 | 0.47 | 0.51 | 0.507 |
| Trait-score correlation | | | Observed | Disattenuated | Observed | Disattenuated | Observed | Disattenuated | Observed | Disattenuated | Observed | Disattenuated | | |
| Demographics | Absent | Absent | 0.338 | 0.37 | 0.202 | 0.24 | 0.316 | 0.338 | 0.448 | 0.472 | 0.266 | 0.31 | | |
| | Absent | Present | 0.323 | 0.353 | 0.161 | 0.19 | 0.297 | 0.318 | 0.42 | 0.444 | 0.26 | 0.302 | | |
| | Present | Present | 0.338 | 0.37 | 0.205 | 0.243 | 0.297 | 0.318 | 0.418 | 0.441 | 0.291 | 0.337 | | |
| | Present | Absent | 0.365 | 0.399 | 0.186 | 0.219 | 0.24 | 0.257 | 0.428 | 0.451 | 0.267 | 0.31 | | |
| COVID Memories | Absent | Absent | 0.234 | 0.256 | 0.225 | 0.264 | 0.202 | 0.218 | 0.298 | 0.316 | 0.268 | 0.313 | | |
| | Absent | Present | 0.197 | 0.216 | 0.166 | 0.197 | 0.156 | 0.168 | 0.242 | 0.256 | 0.254 | 0.298 | | |
| | Present | Present | 0.219 | 0.241 | 0.173 | 0.204 | 0.122 | 0.132 | 0.254 | 0.269 | 0.249 | 0.291 | | |
| | Present | Absent | 0.253 | 0.278 | 0.197 | 0.231 | 0.214 | 0.231 | 0.287 | 0.304 | 0.224 | 0.263 | | |
| Demographics + COVID Memories | Absent | Absent | 0.412 | 0.452 | 0.235 | 0.278 | 0.344 | 0.368 | 0.451 | 0.477 | 0.306 | 0.357 | | |
| | Absent | Present | 0.36 | 0.395 | 0.257 | 0.305 | 0.321 | 0.345 | 0.478 | 0.505 | 0.315 | 0.367 | | |
| | Present | Present | 0.386 | 0.424 | 0.203 | 0.239 | 0.316 | 0.339 | 0.453 | 0.479 | 0.268 | 0.312 | | |
| | Present | Absent | 0.427 | 0.47 | 0.236 | 0.281 | 0.321 | 0.344 | 0.447 | 0.473 | 0.336 | 0.391 | | |
| Demographics + All psychological measures + COVID Memories | Absent | Absent | 0.7 | 0.766 | 0.487 | 0.567 | 0.532 | 0.568 | 0.831 | 0.872 | 0.512 | 0.599 | | |
| | Absent | Present | 0.694 | 0.762 | 0.527 | 0.616 | 0.494 | 0.529 | 0.817 | 0.858 | 0.47 | 0.55 | | |
| | Present | Present | 0.679 | 0.746 | 0.508 | 0.594 | 0.518 | 0.553 | 0.818 | 0.859 | 0.49 | 0.572 | | |
| | Present | Absent | 0.679 | 0.748 | 0.507 | 0.593 | 0.539 | 0.577 | 0.823 | 0.865 | 0.5 | 0.582 | | |

| | | | | | | | | | | | | |
|---|---|---|---|---|---|---|---|---|---|---|---|---|
| Demographics + | Absent | Absent | 0.386 | 0.426 | 0.201 | 0.241 | 0.309 | 0.333 | 0.479 | 0.505 | 0.416 | 0.48 |
| Features with r < 0.5 + | Absent | Present | 0.351 | 0.388 | 0.212 | 0.254 | 0.288 | 0.311 | 0.425 | 0.449 | 0.386 | 0.447 |
| COVID contexts | Present | Present | 0.399 | 0.438 | 0.235 | 0.279 | 0.282 | 0.305 | 0.462 | 0.487 | 0.397 | 0.458 |
| | Present | Absent | 0.349 | 0.389 | 0.179 | 0.213 | 0.308 | 0.333 | 0.438 | 0.464 | 0.386 | 0.448 |
| Demographics + | Absent | Absent | 0.669 | 0.734 | 0.55 | 0.64 | 0.533 | 0.57 | 0.826 | 0.868 | 0.48 | 0.561 |
| Features with r > 0.5 + | Absent | Present | 0.661 | 0.725 | 0.493 | 0.573 | 0.538 | 0.576 | 0.821 | 0.862 | 0.459 | 0.536 |
| COVID contexts | Present | Present | 0.674 | 0.739 | 0.481 | 0.561 | 0.547 | 0.585 | 0.816 | 0.856 | 0.417 | 0.486 |
| | Present | Absent | 0.658 | 0.724 | 0.515 | 0.602 | 0.537 | 0.576 | 0.821 | 0.864 | 0.453 | 0.529 |
| Demographics + COVID | Absent | Absent | 0.4 | 0.442 | 0.242 | 0.287 | 0.294 | 0.316 | 0.474 | 0.5 | 0.388 | 0.451 |
| Memories + COVID | Absent | Present | 0.394 | 0.435 | 0.214 | 0.255 | 0.315 | 0.34 | 0.452 | 0.478 | 0.326 | 0.38 |
| contexts | Present | Present | 0.374 | 0.414 | 0.204 | 0.245 | 0.309 | 0.335 | 0.48 | 0.507 | 0.344 | 0.402 |
| | Present | Absent | 0.36 | 0.4 | 0.221 | 0.264 | 0.318 | 0.344 | 0.476 | 0.504 | 0.342 | 0.399 |

*Note.* Item-level correlations were calculated by correlating human and digital-twin responses across participants for each item and then summarized within each trait. Profile correlations were calculated by correlating human and digital-twin response profiles across items within participants and then summarized by trait. Trait-score correlations were calculated using trait scores. Observed trait-score correlations are Pearson correlations between human and digital-twin trait scores. Disattenuated correlations correct the observed Pearson correlations for measurement unreliability using McDonald’s ω. Overall correlations were computed using Fisher’s z transformation.

**Table S46 Date manipulation check for Study 3b: model date responses to date and COVID-19 context cues**

| Input condition | Date cue | Covid context cue | Mean | SD | Exact target-date responses, n (%) |
|---|---|---|---|---|---|
| Demographics + All psychological measures + COVID Memories | Absent | Absent | 83.69 | 208.98 | 0 (0.00) |
| Demographics + All psychological measures + COVID Memories | Absent | Present | 20.66 | 78.26 | 13 (1.85) |
| Demographics + All psychological measures + COVID Memories | Present | Present | 0.00 | 0.00 | 702 (100.00) |
| Demographics + All psychological measures + COVID Memories | Present | Absent | 0.00 | 0.00 | 702 (100.00) |
| Demographics | Absent | Absent | 173.62 | 254.31 | 0 (0.00) |
| Demographics | Absent | Present | 7.11 | 7.37 | 3 (0.43) |
| Demographics | Present | Present | 0.00 | 0.00 | 702 (100.00) |
| Demographics | Present | Absent | 0.00 | 0.00 | 702 (100.00) |
| Demographics + COVID Memories | Absent | Absent | 127.62 | 297.84 | 2 (0.28) |
| Demographics + COVID Memories | Absent | Present | 13.38 | 17.73 | 15 (2.14) |
| Demographics + COVID Memories | Present | Present | 0.00 | 0.04 | 701 (99.86) |
| Demographics + COVID Memories | Present | Absent | 0.00 | 0.00 | 702 (100.00) |
| Demographics + COVID Memories + COVID contexts | Absent | Absent | 89.15 | 215.38 | 0 (0.00) |
| Demographics + COVID Memories + COVID contexts | Absent | Present | 15.88 | 57.98 | 10 (1.42) |
| Demographics + COVID Memories + COVID contexts | Present | Present | 0.00 | 0.00 | 702 (100.00) |
| Demographics + COVID Memories + COVID contexts | Present | Absent | 0.00 | 0.00 | 702 (100.00) |
| Demographics + Features with $r > 0.5$ + COVID contexts | Absent | Absent | 84.93 | 204.04 | 1 (0.14) |
| Demographics + Features with $r > 0.5$ + COVID contexts | Absent | Present | 19.55 | 83.06 | 18 (2.56) |
| Demographics + Features with $r > 0.5$ + COVID contexts | Present | Present | 0.00 | 0.00 | 702 (100.00) |
| Demographics + Features with $r > 0.5$ + COVID contexts | Present | Absent | 0.00 | 0.00 | 702 (100.00) |
| Demographics + Features with $r < 0.5$ + COVID contexts | Absent | Absent | 106.64 | 576.08 | 1 (0.14) |
| Demographics + Features with $r < 0.5$ + COVID contexts | Absent | Present | 16.95 | 60.89 | 15 (2.14) |
| Demographics + Features with $r < 0.5$ + COVID contexts | Present | Present | 0.00 | 0.00 | 702 (100.00) |
| Demographics + Features with $r < 0.5$ + COVID contexts | Present | Absent | 0.00 | 0.00 | 702 (100.00) |
| COVID Memories | Absent | Absent | 91.42 | 203.25 | 1 (0.14) |
| COVID Memories | Absent | Present | 24.78 | 61.66 | 2 (0.28) |
| COVID Memories | Present | Present | 0.00 | 0.00 | 702 (100.00) |
| COVID Memories | Present | Absent | 0.00 | 0.00 | 702 (100.00) |

*Note.* All results were generated using DeepSeek V4-Flash. "Date cue" indicates whether the wave-specific target date was explicitly provided in the prompt. "COVID context cue" indicates whether contemporaneous COVID-19 contextual information was included in the prompt. Mean and SD represent the absolute difference, in days, between the model-returned date and the wave-specific target date. "Exact target-date responses, n (%)" reports the number and percentage of responses that exactly matched the target date, which was 2020-11-07. The total sample size was 702 responses.

**Table S47 Event manipulation check for Study 3b: model event responses to date and COVID-19 context cues**

| Input condition | Date cue | Covid context cue | Target-event responses, n (%) | Target-plus-other event responses, n (%) | Non-target event responses, n (%) | Unspecified event responses, n (%) | Distinct event categories, n | Event response distribution |
|---|---|---|---|---|---|---|---|---|
| Demographics + All psychological measures + COVID Memories | Absent | Absent | 655 (93.30) | 8 (1.14) | 38 (5.41) | 1 (0.14) | 5 | COVID-19 pandemic (n=663, 94.44%) \| 2020 US presidential election/Election Day (n=41, 5.84%) \| Natural disaster/wildfire/hurricane (n=2, 0.28%) \| Other protests/social unrest (n=2, 0.28%) \| Personal discovery about biological father (n=1, 0.14%) |
| Demographics + All psychological measures + COVID Memories | Absent | Present | 269 (38.32) | 43 (6.13) | 390 (55.56) | 0 (0.00) | 10 | COVID-19 pandemic (n=308, 43.87%) \| COVID-19 case/death toll or surge context (n=14, 1.99%) \| COVID-19 public-health restrictions/guidance (n=2, 0.28%) \| COVID-19 economic context/recession (n=1, 0.14%) \| COVID-19 vaccine authorization/distribution (n=2, 0.28%) \| Trump COVID-19 diagnosis (n=2, 0.28%) \| George Floyd/Black Lives Matter protests (n=7, 1.00%) \| 2020 US presidential election/Election Day (n=416, 59.26%) \| Trump presidency/administration (n=1, 0.14%) \| Natural disaster/wildfire/hurricane (n=2, 0.28%) |
| Demographics + All psychological measures + | Present | Present | 193 (27.49) | 29 (4.13) | 480 (68.38) | 0 (0.00) | 5 | COVID-19 pandemic (n=222, 31.62%) \| COVID-19 case/death toll or surge context (n=4, 0.57%) \| George Floyd/Black Lives Matter protests (n=1, 0.14%) \| 2020 US presidential election/Election Day (n=508, 72.36%) \| Justice Ruth Bader Ginsburg death (n=1, 0.14%) |
| COVID Memories | | | | | | | | |
| Demographics + All psychological measures + COVID Memories | Present | Absent | 32 (4.56) | 11 (1.57) | 659 (93.87) | 0 (0.00) | 2 | COVID-19 pandemic (n=43, 6.13%) \| 2020 US presidential election/Election Day (n=670, 95.44%) |
| Demographics | Absent | Absent | 655 (93.30) | 2 (0.28) | 44 (6.27) | 1 (0.14) | 9 | COVID-19 pandemic (n=657, 93.59%) \| COVID-19 public-health restrictions/guidance (n=1, 0.14%) \| COVID-19 economic context/recession (n=1, 0.14%) \| 2020 US presidential election/Election Day (n=36, 5.13%) \| Trump presidency/administration (n=5, 0.71%) \| 2019 federal government response (n=1, 0.14%) \| 2019 general political climate (n=1, 0.14%) \| 2019 U.S. political climate (n=1, 0.14%) \| 2019 US political climate (n=1, 0.14%) |
| Demographics | Absent | Present | 83 (11.82) | 18 (2.56) | 601 (85.61) | 0 (0.00) | 9 | COVID-19 pandemic (n=98, 13.96%) \| COVID-19 case/death toll or surge context (n=1, 0.14%) \| COVID-19 public-health restrictions/guidance (n=1, 0.14%) \| COVID-19 economic context/recession (n=1, 0.14%) \| COVID-19 vaccine authorization/distribution (n=1, 0.14%) \| Trump COVID-19 diagnosis (n=6, 0.85%) \| George Floyd/Black Lives Matter protests (n=13, 1.85%) \| 2020 US presidential election/Election Day (n=597, 85.04%) \| Trump presidency/administration (n=2, 0.28%) |
| Demographics | Present | Present | 417 (59.40) | 20 (2.85) | 265 (37.75) | 0 (0.00) | 5 | COVID-19 pandemic (n=437, 62.25%) \| COVID-19 case/death toll or surge context (n=2, 0.28%) \| George Floyd/Black Lives Matter |
| | | | | | | | | protests (n=4, 0.57%) \| 2020 US presidential election/Election Day (n=279, 39.74%) \| Current events (n=1, 0.14%) |
| Demographics | Present | Absent | 0 (0.00) | 3 (0.43) | 699 (99.57) | 0 (0.00) | 2 | COVID-19 pandemic (n=3, 0.43%) \| 2020 US presidential election/Election Day (n=702, 100.00%) |
| Demographics + COVID Memories | Absent | Absent | 653 (93.02) | 6 (0.85) | 42 (5.98) | 1 (0.14) | 6 | COVID-19 pandemic (n=659, 93.87%) \| George Floyd/Black Lives Matter protests (n=2, 0.28%) \| 2020 US presidential election/Election Day (n=39, 5.56%) \| Natural disaster/wildfire/hurricane (n=4, 0.57%) \| Thanksgiving/holiday context (n=2, 0.28%) \| Other protests/social unrest (n=1, 0.14%) |
| Demographics + COVID Memories | Absent | Present | 285 (40.60) | 16 (2.28) | 401 (57.12) | 0 (0.00) | 10 | COVID-19 pandemic (n=300, 42.74%) \| COVID-19 case/death toll or surge context (n=3, 0.43%) \| COVID-19 economic context/recession (n=2, 0.28%) \| COVID-19 vaccine authorization/distribution (n=1, 0.14%) \| Trump COVID-19 diagnosis (n=1, 0.14%) \| George Floyd/Black Lives Matter protests (n=21, 2.99%) \| 2020 US presidential election/Election Day (n=385, 54.84%) \| Trump presidency/administration (n=1, 0.14%) \| Natural disaster/wildfire/hurricane (n=2, 0.28%) \| Thanksgiving/holiday context (n=4, 0.57%) |
| Demographics + COVID Memories | Present | Present | 337 (48.01) | 22 (3.13) | 343 (48.86) | 0 (0.00) | 5 | COVID-19 pandemic (n=359, 51.14%) \| COVID-19 case/death toll or surge context (n=2, 0.28%) \| Trump COVID-19 diagnosis (n=1, 0.14%) \| George Floyd/Black Lives Matter protests (n=6, 0.85%) \| 2020 US presidential election/Election Day (n=358, 51.00%) |
| Demographics + COVID Memories | Present | Absent | 121 (17.24) | 14 (1.99) | 567 (80.77) | 0 (0.00) | 5 | COVID-19 pandemic (n=135, 19.23%) \| COVID-19 public-health restrictions/guidance (n=1, 0.14%) \| 2020 US presidential election/Election Day (n=578, 82.34%) \| Natural disaster/wildfire/hurricane (n=1, 0.14%) \| Great Reset (n=1, 0.14%) |

<table>
<tr><td>Demographics + COVID Memories + COVID contexts</td><td>Absent</td><td>Absent</td><td>659 (93.87)</td><td>6 (0.85)</td><td>37 (5.27)</td><td>0 (0.00)</td><td>4</td><td>COVID-19 pandemic (n=665, 94.73%) | 2020 US presidential election/Election Day (n=40, 5.70%) | Natural disaster/wildfire/hurricane (n=2, 0.28%) | Other protests/social unrest (n=1, 0.14%)</td></tr>
<tr><td>Demographics + COVID Memories + COVID contexts</td><td>Absent</td><td>Present</td><td>273 (38.89)</td><td>28 (3.99)</td><td>401 (57.12)</td><td>0 (0.00)</td><td>6</td><td>COVID-19 pandemic (n=301, 42.88%) | COVID-19 case/death toll or surge context (n=6, 0.85%) | George Floyd/Black Lives Matter protests (n=5, 0.71%) | 2020 US presidential election/Election Day (n=422, 60.11%) | Natural disaster/wildfire/hurricane (n=1, 0.14%) | Thanksgiving/holiday context (n=1, 0.14%)</td></tr>
<tr><td>Demographics + COVID Memories + COVID contexts</td><td>Present</td><td>Present</td><td>208 (29.63)</td><td>19 (2.71)</td><td>475 (67.66)</td><td>0 (0.00)</td><td>5</td><td>COVID-19 pandemic (n=227, 32.34%) | COVID-19 case/death toll or surge context (n=5, 0.71%) | George Floyd/Black Lives Matter protests (n=3, 0.43%) | 2020 US presidential election/Election Day (n=490, 69.80%) | Natural disaster/wildfire/hurricane (n=1, 0.14%)</td></tr>
<tr><td>Demographics + COVID Memories + COVID contexts</td><td>Present</td><td>Absent</td><td>33 (4.70)</td><td>17 (2.42)</td><td>652 (92.88)</td><td>0 (0.00)</td><td>3</td><td>COVID-19 pandemic (n=50, 7.12%) | 2020 US presidential election/Election Day (n=668, 95.16%) | Natural disaster/wildfire/hurricane (n=1, 0.14%)</td></tr>
<tr><td>Demographics + Features with r > 0.5 + COVID contexts</td><td>Absent</td><td>Absent</td><td>661 (94.16)</td><td>11 (1.57)</td><td>29 (4.13)</td><td>1 (0.14)</td><td>3</td><td>COVID-19 pandemic (n=672, 95.73%) | 2020 US presidential election/Election Day (n=37, 5.27%) | Natural disaster/wildfire/hurricane (n=3, 0.43%)</td></tr>
<tr><td>Demographics + Features</td><td>Absent</td><td>Present</td><td>248 (35.33)</td><td>40 (5.70)</td><td>414 (58.97)</td><td>0 (0.00)</td><td>8</td><td>COVID-19 pandemic (n=286, 40.74%) | COVID-19 case/death toll or surge context (n=8, 1.14%) | COVID-19 public-health</td></tr>
</table>

| | | | | | | | | |
|---|---|---|---|---|---|---|---|---|
| with r > 0.5 + COVID contexts | | | | | | | | restrictions/guidance (n=2, 0.28%) \| COVID-19 vaccine authorization/distribution (n=3, 0.43%) \| Trump COVID-19 diagnosis (n=1, 0.14%) \| George Floyd/Black Lives Matter protests (n=5, 0.71%) \| 2020 US presidential election/Election Day (n=444, 63.25%) \| Natural disaster/wildfire/hurricane (n=1, 0.14%) |
| Demographics + Features with r > 0.5 + COVID contexts | Present | Present | 185 (26.35) | 15 (2.14) | 502 (71.51) | 0 (0.00) | 6 | COVID-19 pandemic (n=200, 28.49%) \| COVID-19 case/death toll or surge context (n=3, 0.43%) \| Trump COVID-19 diagnosis (n=1, 0.14%) \| George Floyd/Black Lives Matter protests (n=2, 0.28%) \| 2020 US presidential election/Election Day (n=513, 73.08%) \| Natural disaster/wildfire/hurricane (n=1, 0.14%) |
| Demographics + Features with r > 0.5 + COVID contexts | Present | Absent | 22 (3.13) | 7 (1.00) | 673 (95.87) | 0 (0.00) | 5 | COVID-19 pandemic (n=29, 4.13%) \| COVID-19 case/death toll or surge context (n=1, 0.14%) \| COVID-19 public-health restrictions/guidance (n=1, 0.14%) \| 2020 US presidential election/Election Day (n=679, 96.72%) \| Natural disaster/wildfire/hurricane (n=1, 0.14%) |
| Demographics + Features with r < 0.5 + COVID contexts | Absent | Absent | 655 (93.30) | 12 (1.71) | 35 (4.99) | 0 (0.00) | 10 | COVID-19 pandemic (n=666, 94.87%) \| COVID-19 case/death toll or surge context (n=1, 0.14%) \| COVID-19 public-health restrictions/guidance (n=1, 0.14%) \| COVID-19 vaccine authorization/distribution (n=1, 0.14%) \| George Floyd/Black Lives Matter protests (n=1, 0.14%) \| 2020 US presidential election/Election Day (n=39, 5.56%) \| Natural disaster/wildfire/hurricane (n=2, 0.28%) \| Thanksgiving/holiday context (n=1, 0.14%) \| Other protests/social unrest (n=1, 0.14%) \| Homelessness (n=1, 0.14%) |
| Demographics + Features with r < 0.5 + | Absent | Present | 285 (40.60) | 22 (3.13) | 395 (56.27) | 0 (0.00) | 7 | COVID-19 pandemic (n=307, 43.73%) \| COVID-19 case/death toll or surge context (n=10, 1.42%) \| COVID-19 vaccine authorization/distribution (n=2, 0.28%) \| George Floyd/Black Lives Matter protests (n=6, 0.85%) \| 2020 US presidential election/Election |

<table>
<tr><td>COVID contexts</td><td></td><td></td><td></td><td></td><td></td><td></td><td></td><td>Day (n=407, 57.98%) | Natural disaster/wildfire/hurricane (n=1, 0.14%) | Thanksgiving/holiday context (n=1, 0.14%)</td></tr>
<tr><td>Demographics + Features with r < 0.5 + COVID contexts</td><td>Present</td><td>Present</td><td>208 (29.63)</td><td>16 (2.28)</td><td>478 (68.09)</td><td>0 (0.00)</td><td>3</td><td>COVID-19 pandemic (n=224, 31.91%) | COVID-19 case/death toll or surge context (n=1, 0.14%) | 2020 US presidential election/Election Day (n=494, 70.37%)</td></tr>
<tr><td>Demographics + Features with r < 0.5 + COVID contexts</td><td>Present</td><td>Absent</td><td>34 (4.84)</td><td>7 (1.00)</td><td>661 (94.16)</td><td>0 (0.00)</td><td>3</td><td>COVID-19 pandemic (n=41, 5.84%) | 2020 US presidential election/Election Day (n=667, 95.01%) | Natural disaster/wildfire/hurricane (n=1, 0.14%)</td></tr>
<tr><td>COVID Memories</td><td>Absent</td><td>Absent</td><td>654 (93.16)</td><td>10 (1.42)</td><td>38 (5.41)</td><td>0 (0.00)</td><td>9</td><td>COVID-19 pandemic (n=664, 94.59%) | George Floyd/Black Lives Matter protests (n=3, 0.43%) | 2020 US presidential election/Election Day (n=35, 4.99%) | Natural disaster/wildfire/hurricane (n=2, 0.28%) | Thanksgiving/holiday context (n=4, 0.57%) | Other protests/social unrest (n=1, 0.14%) | Discord convention planning (n=1, 0.14%) | High school graduation (n=1, 0.14%) | Reuniting with boyfriend after his sobriety (n=1, 0.14%)</td></tr>
<tr><td>COVID Memories</td><td>Absent</td><td>Present</td><td>506 (72.08)</td><td>11 (1.57)</td><td>185 (26.35)</td><td>0 (0.00)</td><td>8</td><td>COVID-19 pandemic (n=517, 73.65%) | COVID-19 economic context/recession (n=2, 0.28%) | Trump COVID-19 diagnosis (n=1, 0.14%) | George Floyd/Black Lives Matter protests (n=17, 2.42%) | 2020 US presidential election/Election Day (n=173, 24.64%) | Thanksgiving/holiday context (n=2, 0.28%) | A dark year (n=1, 0.14%) | Good Friday Agreement (n=1, 0.14%)</td></tr>
<tr><td>COVID Memories</td><td>Present</td><td>Present</td><td>548 (78.06)</td><td>11 (1.57)</td><td>143 (20.37)</td><td>0 (0.00)</td><td>8</td><td>COVID-19 pandemic (n=558, 79.49%) | COVID-19 case/death toll or surge context (n=1, 0.14%) | COVID-19 economic context/recession</td></tr>
</table>

| | | | | | | | | (n=1, 0.14%) \| COVID-19 vaccine authorization/distribution (n=1, 0.14%) \| Trump COVID-19 diagnosis (n=1, 0.14%) \| George Floyd/Black Lives Matter protests (n=3, 0.43%) \| 2020 US presidential election/Election Day (n=150, 21.37%) \| Trump presidency/administration (n=2, 0.28%) |
|---|---|---|---|---|---|---|---|---|
| COVID Memories | Present | Absent | 522 (74.36) | 6 (0.85) | 174 (24.79) | 0 (0.00) | 7 | COVID-19 pandemic (n=528, 75.21%) \| COVID-19 economic context/recession (n=1, 0.14%) \| 2020 US presidential election/Election Day (n=173, 24.64%) \| Natural disaster/wildfire/hurricane (n=1, 0.14%) \| Thanksgiving/holiday context (n=3, 0.43%) \| Other protests/social unrest (n=1, 0.14%) \| NASCAR Cup Series championship (n=1, 0.14%) |

*Note.* All results were generated using DeepSeek V4-Flash. "Date cue" indicates whether the wave-specific target date was explicitly provided in the prompt. "COVID context cue" indicates whether contemporaneous COVID-19 contextual information was included in the prompt. The target event was the COVID-19 pandemic. "Target-event responses" refers to responses coded as mentioning the COVID-19 pandemic only. "Target-plus-other event responses" refers to responses that mentioned the COVID-19 pandemic together with at least one additional codable event or contextual category. "Non-target event responses" refers to responses that mentioned codable events other than the COVID-19 pandemic. "Unspecified event responses" refers to responses that did not identify a specific codable event or could not be coded. The four response-type columns are mutually exclusive and sum to N within each wave-condition combination. "Distinct event categories" and "Event response distribution" are based on normalized multi-label coding; therefore, a single response could be assigned to multiple event or contextual categories, and percentages in the final column are not mutually exclusive and may sum to more than 100%. For example, "COVID-19 pandemic (n = 1498, 100.00%) | COVID-19 case/death toll or surge context (n = 2, 0.13%)" indicates that all 1498 responses mentioned the COVID-19 pandemic, and two of those responses also mentioned case counts, death counts, or surge-related contextual information.

**Table S48 Configural invariance results for Big Five responses across input conditions and models in Study 3b**

| Model | Input conditions | Agreeableness | Conscientiousness | Extraversion | Neuroticism | Openness | N |
|---|---|---|---|---|---|---|---|
| Gemma-4-31B-it | Demographics | 4, 29R, 44R | | 7, 12R, 37, 42R | | 3R, 8R | 9 |
| | COVID Memories | 24R, 29R | | 7, 12R, 37, 42R | 36 | 8R | 8 |
| | Demographics + COVID Memories | 24R, 29R, 44R | | 7, 12R, 37, 42R | | 3R, 8R | 9 |
| | Demographics + All psychological measures + COVID Memories | 34, 44R | | 32, 47, 52 | 36 | 3R, 8R | 8 |
| | Demographics + Features with r > 0.5 + COVID contexts | 34, 44R | | 2, 17, 22, 27R, 57R | 36 | 3R, 8R | 10 |
| | Demographics + Features with r < 0.5 + COVID contexts | 24R, 29R, 44R | 15R | 7, 12R, 37, 42R | 36 | 3R, 8R | 11 |
| | Demographics + COVID Memories + COVID contexts | 24R, 29R, 44R | | 7, 12R, 37, 42R | | 3R, 8R | 9 |
| gpt-oss-120b | Demographics | 24R, 29R, 44R | 20 | | | 3R, 8R, 18R, 28, 33R, 38R | 10 |
| | COVID Memories | 24R, 29R, 34 | | | 36 | 3R, 8R, 18R, 28, 38R | 9 |
| | Demographics + COVID Memories | 24R, 29R, 34, 44R | 20 | | 36 | 3R, 8R, 18R, 28, 38R | 11 |
| | Demographics + All psychological measures + COVID Memories | 4, 9R, 14R, 19, 24R, 29R, 34, 39R, 44R, 49, 54R, 59R | | | 36 | 3R, 8R, 18R, 28, 38R | 18 |
| | Demographics + Features with r > 0.5 + COVID contexts | 34 | | | 36 | 3R, 8R, 18R, 28, 38R | 7 |
| | Demographics + Features with r < 0.5 + COVID contexts | 24R, 29R, 34, 44R | | | 36 | 3R, 8R, 18R, 28, 33R, 38R | 11 |
| | Demographics + COVID Memories + COVID contexts | 24R, 29R, 34, 44R | 20 | 7, 12R, 37, 42R | 1R, 11, 21, 31R, 36 | 3R, 8R, 18R, 38R | 18 |
| DeepSeek V4-Flash | Demographics | | | | | 3R, 8R, 18R, 38R | 4 |
| | COVID Memories | | | | 36 | 8R | 2 |
| | Demographics + COVID Memories | | | | 36 | 3R, 8R | 3 |
| | Demographics + All psychological measures + COVID Memories | 34 | | | | 3R, 8R, 18R, 38R | 5 |
| | Demographics + Features with r > 0.5 + COVID contexts | 34 | | | | 3R, 8R, 18R, 38R | 5 |
| | Demographics + Features with r < 0.5 + COVID contexts | 44R | | | 36 | 3R, 8R | 4 |
| | Demographics + COVID Memories + COVID contexts | | | | 36 | 3R, 8R | 3 |
| Qwen 3.5 Plus | Demographics | 4, 49 | | 7, 12R, 37, 42R | | 3R, 8R, 18R, 38R | 10 |
| | COVID Memories | 24R, 29R, 34 | | 7, 12R, 37, 42R | 36 | 3R, 8R, 18R, 28, 38R | 13 |

| | | | | | |
|---|---|---|---|---|---|
| Demographics + COVID Memories | 4, 49 | 7, 12R, 37, 42R | 36 | 3R, 8R, 18R, 28, 38R | 12 |
| Demographics + All psychological measures + COVID Memories | 4, 34, 49 | 7, 12R, 37, 42R | 36 | 3R, 8R, 18R, 28, 38R, 53 | 14 |
| Demographics + Features with r > 0.5 + COVID contexts | 4, 34, 49 | 7, 12R, 37, 42R | 36 | 3R, 8R, 18R, 28, 38R, 53 | 14 |
| Demographics + Features with r < 0.5 + COVID contexts | 4, 34, 49 | 7, 12R, 37, 42R | 36 | 3R, 8R, 18R, 28, 38R | 13 |
| Demographics + COVID Memories + COVID contexts | 4, 34, 49 | | 36 | 3R, 8R, 18R, 28, 38R | 9 |

*Note.* Listed items are those removed during the iterative invariance procedure, either by the algorithmic procedure or because their community assignment was inconsistent with the expected Big Five domain structure.

**Table S49 Network configural invariance of NEO Five-Factor Personality Inventory responses comparing human and digital twins in Study 3b**

| | | Demographics | | COVID Memories | | Demographics + COVID Memories | | Demographics + Features with r < 0.5 + COVID contexts | | Demographics + Features with r > 0.5 + COVID contexts | | Demographics + COVID Memories + COVID contexts | | Demographics + All psychological measures + COVID Memories | |
|---|---|---|---|---|---|---|---|---|---|---|---|---|---|---|---|
| Gemma-4-31B-it | | | | | | | | | | | | | | | |
| Item | Dimension | Community | Replication | Community | Replication | Community | Replication | Community | Replication | Community | Replication | Community | Replication | Community | Replication |
| 1 (R) | Neuroticism | 1 | 1.000 | 1 | 1.000 | 1 | 1.000 | 1 | 1.000 | 1 | 1.000 | 1 | 1.000 | 1 | 1.000 |
| 2 | Extraversion | 2 | 1.000 | 2 | 1.000 | 2 | 1.000 | 2 | 1.000 | Alg-rem. | | 2 | 1.000 | 2 | 1.000 |
| 3 (R) | Openness | Alg-rem. | | 3 | 0.980 | Alg-rem. | | Alg-rem. | | Alg-rem. | | Alg-rem. | | Alg-rem. | |
| 4 | Agreeableness | Pre-rem. | | 4 | 1.000 | 3 | 1.000 | 3 | 1.000 | 2 | 1.000 | 3 | 1.000 | 3 | 1.000 |
| 5 | Conscientiousness | 3 | 1.000 | 5 | 1.000 | 4 | 1.000 | 4 | 1.000 | 3 | 1.000 | 4 | 1.000 | 4 | 1.000 |
| 6 | Neuroticism | 1 | 1.000 | 1 | 1.000 | 1 | 1.000 | 1 | 1.000 | 1 | 1.000 | 1 | 1.000 | 1 | 1.000 |
| 7 | Extraversion | Alg-rem. | | Alg-rem. | | Alg-rem. | | Alg-rem. | | 4 | 0.974 | Alg-rem. | | 5 | 0.824 |
| 8 (R) | Openness | Pre-rem. | | Alg-rem. | | Alg-rem. | | Alg-rem. | | Alg-rem. | | Alg-rem. | | Alg-rem. | |
| 9 (R) | Agreeableness | 4 | 0.998 | 4 | 1.000 | 3 | 0.998 | 3 | 1.000 | 2 | 1.000 | 3 | 0.998 | 3 | 1.000 |
| 10 | Conscientiousness | 3 | 1.000 | 5 | 1.000 | 4 | 1.000 | 4 | 1.000 | 3 | 1.000 | 4 | 1.000 | 4 | 1.000 |
| 11 | Neuroticism | 1 | 1.000 | 1 | 1.000 | 1 | 1.000 | 1 | 1.000 | 1 | 1.000 | 1 | 1.000 | 1 | 1.000 |
| 12 (R) | Extraversion | Alg-rem. | | Alg-rem. | | Alg-rem. | | Alg-rem. | | 4 | 0.974 | Alg-rem. | | 5 | 0.824 |
| 13 | Openness | 5 | 1.000 | 3 | 1.000 | 5 | 1.000 | 5 | 1.000 | 5 | 1.000 | 5 | 1.000 | 6 | 1.000 |
| 14 (R) | Agreeableness | 4 | 1.000 | 4 | 1.000 | 3 | 1.000 | 3 | 1.000 | 2 | 1.000 | 3 | 1.000 | 3 | 1.000 |
| 15 (R) | Conscientiousness | 3 | 1.000 | 5 | 1.000 | 4 | 1.000 | Alg-rem. | | 3 | 1.000 | 4 | 1.000 | 4 | 1.000 |
| 16 (R) | Neuroticism | 1 | 1.000 | 1 | 1.000 | 1 | 1.000 | 1 | 1.000 | 1 | 1.000 | 1 | 1.000 | 1 | 1.000 |
| 17 | Extraversion | 2 | 1.000 | 2 | 1.000 | 2 | 1.000 | 2 | 1.000 | Alg-rem. | | 2 | 1.000 | 2 | 1.000 |
| 18 (R) | Openness | 5 | 1.000 | 3 | 0.980 | 5 | 1.000 | 5 | 1.000 | 5 | 1.000 | 5 | 1.000 | 6 | 1.000 |
| 19 | Agreeableness | 4 | 1.000 | 4 | 1.000 | 3 | 1.000 | 3 | 1.000 | 2 | 1.000 | 3 | 1.000 | 3 | 1.000 |
| 20 | Conscientiousness | 3 | 0.996 | 5 | 1.000 | 4 | 1.000 | 4 | 1.000 | 3 | 1.000 | 4 | 0.996 | 4 | 1.000 |
| 21 | Neuroticism | 1 | 1.000 | 1 | 1.000 | 1 | 1.000 | 1 | 1.000 | 1 | 1.000 | 1 | 1.000 | 1 | 1.000 |
| 22 | Extraversion | 2 | 1.000 | 2 | 1.000 | 2 | 1.000 | 2 | 1.000 | Alg-rem. | | 2 | 1.000 | 2 | 1.000 |
| 23 (R) | Openness | 5 | 1.000 | 3 | 1.000 | 5 | 1.000 | 5 | 1.000 | 5 | 1.000 | 5 | 1.000 | 6 | 1.000 |
| 24 (R) | Agreeableness | 4 | 1.000 | Pre-rem. | | Alg-rem. | | Alg-rem. | | 2 | 0.968 | Alg-rem. | | 3 | 0.946 |
| 25 | Conscientiousness | 3 | 1.000 | 5 | 1.000 | 4 | 1.000 | 4 | 1.000 | 3 | 1.000 | 4 | 1.000 | 4 | 1.000 |
| 26 | Neuroticism | 1 | 1.000 | 1 | 1.000 | 1 | 1.000 | 1 | 1.000 | 1 | 1.000 | 1 | 1.000 | 1 | 1.000 |
| 27 (R) | Extraversion | 2 | 1.000 | 2 | 1.000 | 2 | 1.000 | 2 | 1.000 | Alg-rem. | | 2 | 1.000 | 2 | 1.000 |
| 28 | Openness | 5 | 1.000 | 3 | 0.980 | 5 | 1.000 | 5 | 1.000 | 5 | 1.000 | 5 | 1.000 | 6 | 1.000 |
| 29 (R) | Agreeableness | Pre-rem. | | Pre-rem. | | Alg-rem. | | Alg-rem. | | 2 | 0.968 | Alg-rem. | | 3 | 0.946 |
| 30 (R) | Conscientiousness | 3 | 1.000 | 5 | 1.000 | 4 | 1.000 | 4 | 1.000 | 3 | 1.000 | 4 | 1.000 | 4 | 1.000 |
| 31 (R) | Neuroticism | 1 | 1.000 | 1 | 1.000 | 1 | 1.000 | 1 | 1.000 | 1 | 1.000 | 1 | 1.000 | 1 | 1.000 |
| 32 | Extraversion | 2 | 1.000 | 2 | 1.000 | 2 | 1.000 | 2 | 1.000 | 4 | 0.830 | 2 | 1.000 | Alg-rem. | |
| 33 (R) | Openness | 5 | 1.000 | 3 | 1.000 | 5 | 1.000 | 5 | 1.000 | 5 | 1.000 | 5 | 1.000 | 6 | 1.000 |
| 34 | Agreeableness | 4 | 0.996 | 4 | 0.992 | 3 | 0.998 | 3 | 0.990 | Alg-rem. | | 3 | 1.000 | Alg-rem. | |
| 35 | Conscientiousness | 3 | 1.000 | 5 | 1.000 | 4 | 1.000 | 4 | 1.000 | 3 | 1.000 | 4 | 1.000 | 4 | 1.000 |
| 36 | Neuroticism | 1 | 0.996 | Pre-rem. | | 1 | 0.944 | Alg-rem. | | Pre-rem. | | 1 | 0.854 | Pre-rem. | |
| 37 | Extraversion | Alg-rem. | | Alg-rem. | | Alg-rem. | | Alg-rem. | | 4 | 0.974 | Alg-rem. | | 5 | 0.824 |
| 38 (R) | Openness | 5 | 1.000 | 3 | 0.980 | 5 | 1.000 | 5 | 1.000 | 5 | 1.000 | 5 | 1.000 | 6 | 1.000 |
| 39 (R) | Agreeableness | 4 | 1.000 | 4 | 1.000 | 3 | 1.000 | 3 | 1.000 | 2 | 1.000 | 3 | 1.000 | 3 | 1.000 |
| 40 | Conscientiousness | 3 | 0.996 | 5 | 1.000 | 4 | 1.000 | 4 | 1.000 | 3 | 1.000 | 4 | 0.996 | 4 | 1.000 |
| 41 | Neuroticism | 1 | 1.000 | 1 | 1.000 | 1 | 1.000 | 1 | 1.000 | 1 | 1.000 | 1 | 1.000 | 1 | 1.000 |
| 42 (R) | Extraversion | Alg-rem. | | Alg-rem. | | Alg-rem. | | Alg-rem. | | 4 | 0.974 | Alg-rem. | | 5 | 0.824 |
| 43 | Openness | 5 | 1.000 | 3 | 1.000 | 5 | 1.000 | 5 | 1.000 | 5 | 1.000 | 5 | 1.000 | 6 | 1.000 |
| 44 (R) | Agreeableness | Alg-rem. | | 4 | 1.000 | Alg-rem. | | Pre-rem. | | Alg-rem. | | Alg-rem. | | Alg-rem. | |
| 45 (R) | Conscientiousness | 3 | 1.000 | 5 | 1.000 | 4 | 1.000 | 4 | 1.000 | 3 | 1.000 | 4 | 1.000 | 4 | 1.000 |
| 46 (R) | Neuroticism | 1 | 1.000 | 1 | 1.000 | 1 | 1.000 | 1 | 1.000 | 1 | 1.000 | 1 | 1.000 | 1 | 1.000 |
| 47 | Extraversion | 2 | 1.000 | 2 | 1.000 | 2 | 1.000 | 2 | 1.000 | 4 | 0.830 | 2 | 1.000 | Alg-rem. | |
| 48 (R) | Openness | 5 | 1.000 | 3 | 1.000 | 5 | 1.000 | 5 | 1.000 | 5 | 1.000 | 5 | 1.000 | 6 | 1.000 |
| 49 | Agreeableness | 4 | 0.982 | 4 | 1.000 | 3 | 1.000 | 3 | 1.000 | 2 | 1.000 | 3 | 1.000 | 3 | 1.000 |
| 50 | Conscientiousness | 3 | 1.000 | 5 | 1.000 | 4 | 1.000 | 4 | 1.000 | 3 | 1.000 | 4 | 1.000 | 4 | 1.000 |
| 51 | Neuroticism | 1 | 0.998 | 1 | 1.000 | 1 | 1.000 | 1 | 1.000 | 1 | 1.000 | 1 | 1.000 | 1 | 1.000 |
| 52 | Extraversion | 2 | 1.000 | 2 | 1.000 | 2 | 1.000 | 2 | 1.000 | 4 | 0.830 | 2 | 1.000 | Alg-rem. | |
| 53 | Openness | 5 | 1.000 | 3 | 1.000 | 5 | 1.000 | 5 | 1.000 | 5 | 1.000 | 5 | 1.000 | 6 | 1.000 |
| 54 (R) | Agreeableness | 4 | 1.000 | 4 | 1.000 | 3 | 1.000 | 3 | 1.000 | 2 | 1.000 | 3 | 1.000 | 3 | 1.000 |
| 55 (R) | Conscientiousness | 3 | 1.000 | 5 | 1.000 | 4 | 1.000 | 4 | 1.000 | 3 | 1.000 | 4 | 1.000 | 4 | 1.000 |
| 56 | Neuroticism | 1 | 0.998 | 1 | 1.000 | 1 | 1.000 | 1 | 1.000 | 1 | 1.000 | 1 | 1.000 | 1 | 1.000 |
| 57 (R) | Extraversion | 2 | 1.000 | 2 | 1.000 | 2 | 1.000 | 2 | 1.000 | Alg-rem. | | 2 | 1.000 | 2 | 1.000 |
| 58 | Openness | 5 | 1.000 | 3 | 1.000 | 5 | 1.000 | 5 | 1.000 | 5 | 1.000 | 5 | 1.000 | 6 | 1.000 |
| 59 (R) | Agreeableness | 4 | 1.000 | 4 | 1.000 | 3 | 1.000 | 3 | 1.000 | 2 | 1.000 | 3 | 1.000 | 3 | 1.000 |
| 60 | Conscientiousness | 3 | 1.000 | 5 | 1.000 | 4 | 1.000 | 4 | 1.000 | 3 | 1.000 | 4 | 1.000 | 4 | 1.000 |
| gpt-oss-120b | | | | | | | | | | | | | | | |
| Item | Dimension | Community | Replication | Community | Replication | Community | Replication | Community | Replication | Community | Replication | Community | Replication | Community | Replication |
| 1 (R) | Neuroticism | 1 | 1 | 1 | 1 | 1 | 1 | 1 | 1 | 1 | 1 | Alg-rem. | | 1 | 1 |
| 2 | Extraversion | 2 | 1 | 2 | 1 | 2 | 1 | 2 | 1 | 2 | 1 | 1 | 1 | 2 | 0.99 |
| 3 (R) | Openness | Pre-rem. | | Pre-rem. | | Alg-rem. | | Alg-rem. | | Alg-rem. | | Alg-rem. | | Alg-rem. | |
| 4 | Agreeableness | 3 | 1 | 3 | 1 | 3 | 1 | 3 | 0.992 | 3 | 0.998 | 2 | 1 | Alg-rem. | |
| 5 | Conscientiousness | 4 | 1 | 4 | 1 | 4 | 1 | 4 | 1 | 4 | 1 | 3 | 1 | 3 | 1 |
| 6 | Neuroticism | 1 | 1 | 1 | 1 | 1 | 1 | 1 | 1 | 1 | 1 | 4 | 1 | 1 | 1 |
| 7 | Extraversion | 2 | 0.978 | 2 | 0.998 | 2 | 1 | 2 | 0.996 | 2 | 1 | Alg-rem. | | 2 | 0.824 |
| 8 (R) | Openness | Pre-rem. | | Pre-rem. | | Pre-rem. | | Pre-rem. | | Pre-rem. | | Pre-rem. | | Pre-rem. | |
| 9 (R) | Agreeableness | 3 | 1 | 3 | 1 | 3 | 1 | 3 | 1 | 3 | 1 | 2 | 1 | Alg-rem. | |
| 10 | Conscientiousness | 4 | 1 | 4 | 1 | 4 | 1 | 4 | 1 | 4 | 1 | 3 | 1 | 3 | 1 |
| 11 | Neuroticism | 1 | 1 | 1 | 1 | 1 | 1 | 1 | 1 | 1 | 1 | Alg-rem. | | 1 | 1 |
| 12 (R) | Extraversion | 2 | 0.978 | 2 | 0.998 | 2 | 1 | 2 | 0.996 | 2 | 1 | Alg-rem. | | 2 | 0.824 |
| 13 | Openness | 5 | 1 | 5 | 1 | 5 | 1 | 5 | 1 | 5 | 1 | 5 | 1 | 4 | 1 |
| 14 (R) | Agreeableness | 3 | 1 | 3 | 1 | 3 | 1 | 3 | 1 | 3 | 1 | 2 | 1 | Alg-rem. | |
| 15 (R) | Conscientiousness | 4 | 1 | 4 | 1 | 4 | 1 | 4 | 1 | 4 | 1 | 3 | 1 | 3 | 1 |
| 16 (R) | Neuroticism | 1 | 1 | 1 | 1 | 1 | 1 | 1 | 1 | 1 | 1 | 4 | 1 | 1 | 1 |
| 17 | Extraversion | 2 | 1 | 2 | 1 | 2 | 1 | 2 | 1 | 2 | 1 | 1 | 1 | 2 | 0.99 |
| 18 (R) | Openness | Pre-rem. | | Pre-rem. | | Alg-rem. | | Alg-rem. | | Alg-rem. | | Alg-rem. | | Alg-rem. | |
| 19 | Agreeableness | 3 | 1 | 3 | 1 | 3 | 1 | 3 | 1 | 3 | 1 | 2 | 1 | Alg-rem. | |
| 20 | Conscientiousness | Alg-rem. | | 4 | 0.97 | Alg-rem. | | 4 | 0.766 | 4 | 0.984 | Alg-rem. | | 3 | 1 |
| 21 | Neuroticism | 1 | 1 | 1 | 1 | 1 | 1 | 1 | 1 | 1 | 1 | Alg-rem. | | 1 | 1 |
| 22 | Extraversion | 2 | 1 | 2 | 1 | 2 | 1 | 2 | 1 | 2 | 1 | 1 | 1 | 2 | 0.99 |
| 23 (R) | Openness | 5 | 1 | 5 | 1 | 5 | 1 | 5 | 1 | 5 | 1 | 5 | 1 | 4 | 1 |
| 24 (R) | Agreeableness | Pre-rem. | | Alg-rem. | | Alg-rem. | | Alg-rem. | | 3 | 1 | Pre-rem. | | Alg-rem. | |
| 25 | Conscientiousness | 4 | 1 | 4 | 1 | 4 | 1 | 4 | 1 | 4 | 1 | 3 | 1 | 3 | 1 |
| 26 | Neuroticism | 1 | 1 | 1 | 1 | 1 | 1 | 1 | 1 | 1 | 1 | 4 | 1 | 1 | 1 |
| 27 (R) | Extraversion | 2 | 1 | 2 | 1 | 2 | 1 | 2 | 1 | 2 | 1 | 1 | 1 | 2 | 0.99 |
| 28 | Openness | Pre-rem. | | Pre-rem. | | Alg-rem. | | Alg-rem. | | Alg-rem. | | 5 | 0.988 | Alg-rem. | |
| 29 (R) | Agreeableness | Pre-rem. | | Alg-rem. | | Alg-rem. | | Alg-rem. | | 3 | 1 | Pre-rem. | | Alg-rem. | |
| 30 (R) | Conscientiousness | 4 | 1 | 4 | 1 | 4 | 1 | 4 | 1 | 4 | 1 | 3 | 1 | 3 | 1 |
| 31 (R) | Neuroticism | 1 | 1 | 1 | 1 | 1 | 1 | 1 | 1 | 1 | 1 | Alg-rem. | | 1 | 1 |
| 32 | Extraversion | 2 | 0.998 | 2 | 1 | 2 | 1 | 2 | 1 | 2 | 1 | 1 | 1 | 2 | 1 |
| 33 (R) | Openness | Pre-rem. |  | 5 | 1 | 5 | 1 | Alg-rem. |  | 5 | 0.998 | 5 | 1 | 4 | 1 |
| 34 | Agreeableness | 3 | 0.9 | Alg-rem. |  | Alg-rem. |  | Alg-rem. |  | Alg-rem. |  | Alg-rem. |  | Alg-rem. |  |
| 35 | Conscientiousness | 4 | 1 | 4 | 1 | 4 | 1 | 4 | 1 | 4 | 1 | 3 | 1 | 3 | 1 |
| 36 | Neuroticism | 1 | 0.884 | Alg-rem. |  | Alg-rem. |  | Alg-rem. |  | Pre-rem. |  | Alg-rem. |  | Pre-rem. |  |
| 37 | Extraversion | 2 | 0.978 | 2 | 0.998 | 2 | 1 | 2 | 0.996 | 2 | 1 | Alg-rem. |  | 2 | 0.824 |
| 38 (R) | Openness | Pre-rem. |  | Pre-rem. |  | Alg-rem. |  | Alg-rem. |  | Alg-rem. |  | Alg-rem. |  | Alg-rem. |  |
| 39 (R) | Agreeableness | 3 | 1 | 3 | 1 | 3 | 1 | 3 | 1 | 3 | 1 | 2 | 1 | Alg-rem. |  |
| 40 | Conscientiousness | 4 | 1 | 4 | 1 | 4 | 1 | 4 | 1 | 4 | 1 | 3 | 1 | 3 | 1 |
| 41 | Neuroticism | 1 | 1 | 1 | 1 | 1 | 1 | 1 | 1 | 1 | 1 | 4 | 1 | 1 | 1 |
| 42 (R) | Extraversion | 2 | 0.978 | 2 | 0.998 | 2 | 1 | 2 | 0.996 | 2 | 1 | Alg-rem. |  | 2 | 0.824 |
| 43 | Openness | 5 | 1 | 5 | 1 | 5 | 1 | 5 | 1 | 5 | 1 | 5 | 1 | 4 | 1 |
| 44 (R) | Agreeableness | Pre-rem. |  | 3 | 1 | Alg-rem. |  | Alg-rem. |  | 3 | 1 | Pre-rem. |  | Alg-rem. |  |
| 45 (R) | Conscientiousness | 4 | 1 | 4 | 1 | 4 | 1 | 4 | 1 | 4 | 1 | 3 | 1 | 3 | 1 |
| 46 (R) | Neuroticism | 1 | 1 | 1 | 1 | 1 | 1 | 1 | 1 | 1 | 1 | 4 | 1 | 1 | 1 |
| 47 | Extraversion | 2 | 0.998 | 2 | 1 | 2 | 1 | 2 | 1 | 2 | 1 | 1 | 1 | 2 | 1 |
| 48 (R) | Openness | 5 | 1 | 5 | 1 | 5 | 1 | 5 | 1 | 5 | 1 | 5 | 1 | 4 | 1 |
| 49 | Agreeableness | 3 | 1 | 3 | 0.99 | 3 | 1 | 3 | 0.992 | 3 | 0.998 | 2 | 1 | Alg-rem. |  |
| 50 | Conscientiousness | 4 | 1 | 4 | 1 | 4 | 1 | 4 | 1 | 4 | 1 | 3 | 1 | 3 | 1 |
| 51 | Neuroticism | 1 | 0.914 | 1 | 1 | 1 | 1 | 1 | 1 | 1 | 1 | 4 | 1 | 1 | 1 |
| 52 | Extraversion | 2 | 0.998 | 2 | 1 | 2 | 1 | 2 | 1 | 2 | 1 | 1 | 1 | 2 | 1 |
| 53 | Openness | 5 | 1 | 5 | 1 | 5 | 1 | 5 | 1 | 5 | 1 | 5 | 1 | 4 | 1 |
| 54 (R) | Agreeableness | 3 | 1 | 3 | 1 | 3 | 1 | 3 | 1 | 3 | 1 | 2 | 1 | Alg-rem. |  |
| 55 (R) | Conscientiousness | 4 | 1 | 4 | 1 | 4 | 1 | 4 | 1 | 4 | 1 | 3 | 1 | 3 | 1 |
| 56 | Neuroticism | 1 | 1 | 1 | 1 | 1 | 1 | 1 | 1 | 1 | 1 | 4 | 1 | 1 | 1 |
| 57 (R) | Extraversion | 2 | 1 | 2 | 1 | 2 | 1 | 2 | 1 | 2 | 1 | 1 | 1 | 2 | 0.99 |
| 58 | Openness | 5 | 1 | 5 | 1 | 5 | 1 | 5 | 1 | 5 | 1 | 5 | 1 | 4 | 1 |
| 59 (R) | Agreeableness | 3 | 1 | 3 | 1 | 3 | 1 | 3 | 1 | 3 | 1 | 2 | 1 | Alg-rem. | |
| 60 | Conscientiousness | 4 | 1 | 4 | 1 | 4 | 1 | 4 | 0.996 | 4 | 1 | 3 | 1 | 3 | 1 |
| DeepSeek V4-Flash | | | | | | | | | | | | | | | |
| Item | Dimension | Community | Replication | Community | Replication | Community | Replication | Community | Replication | Community | Replication | Community | Replication | Community | Replication |
| 1 (R) | Neuroticism | 1 | 1 | 1 | 1 | 1 | 1 | 1 | 1 | 1 | 1 | 1 | 1 | 1 | 1 |
| 2 | Extraversion | 2 | 1 | 2 | 1 | 2 | 1 | 2 | 1 | 2 | 1 | 2 | 1 | 2 | 1 |
| 3 (R) | Openness | Alg-rem. | | 3 | 0.814 | Alg-rem. | | Pre-rem. | | Alg-rem. | | Pre-rem. | | Alg-rem. | |
| 4 | Agreeableness | 3 | 0.996 | 4 | 0.976 | 3 | 1 | 3 | 0.928 | 3 | 0.996 | 3 | 0.788 | 3 | 1 |
| 5 | Conscientiousness | 4 | 1 | 5 | 1 | 4 | 1 | 4 | 1 | 4 | 1 | 4 | 1 | 4 | 1 |
| 6 | Neuroticism | 1 | 1 | 1 | 1 | 1 | 1 | 1 | 1 | 1 | 1 | 1 | 1 | 1 | 1 |
| 7 | Extraversion | 2 | 1 | 2 | 0.83 | 2 | 0.972 | 2 | 0.912 | 2 | 1 | 2 | 1 | 2 | 1 |
| 8 (R) | Openness | Pre-rem. | | Pre-rem. | | Pre-rem. | | Pre-rem. | | Alg-rem. | | Pre-rem. | | Alg-rem. | |
| 9 (R) | Agreeableness | 3 | 0.998 | 4 | 0.994 | 3 | 1 | 3 | 1 | 3 | 1 | 3 | 1 | 3 | 1 |
| 10 | Conscientiousness | 4 | 1 | 5 | 1 | 4 | 1 | 4 | 1 | 4 | 1 | 4 | 1 | 4 | 1 |
| 11 | Neuroticism | 1 | 1 | 1 | 1 | 1 | 1 | 1 | 1 | 1 | 1 | 1 | 1 | 1 | 1 |
| 12 (R) | Extraversion | 2 | 1 | 2 | 0.83 | 2 | 0.972 | 2 | 0.912 | 2 | 1 | 2 | 1 | 2 | 1 |
| 13 | Openness | 5 | 1 | 3 | 1 | 5 | 1 | 5 | 1 | 5 | 1 | 5 | 1 | 5 | 1 |
| 14 (R) | Agreeableness | 3 | 1 | 4 | 1 | 3 | 1 | 3 | 1 | 3 | 1 | 3 | 1 | 3 | 1 |
| 15 (R) | Conscientiousness | 4 | 1 | 5 | 1 | 4 | 1 | 4 | 1 | 4 | 1 | 4 | 1 | 4 | 1 |
| 16 (R) | Neuroticism | 1 | 1 | 1 | 1 | 1 | 1 | 1 | 1 | 1 | 1 | 1 | 1 | 1 | 1 |
| 17 | Extraversion | 2 | 1 | 2 | 1 | 2 | 1 | 2 | 1 | 2 | 1 | 2 | 1 | 2 | 1 |
| 18 (R) | Openness | Alg-rem. | | 3 | 0.814 | 5 | 0.976 | 5 | 1 | Alg-rem. | | 5 | 0.994 | Alg-rem. | |
| 19 | Agreeableness | 3 | 0.998 | 4 | 0.994 | 3 | 1 | 3 | 1 | 3 | 0.998 | 3 | 0.998 | 3 | 1 |
| 20 | Conscientiousness | 4 | 1 | 5 | 1 | 4 | 1 | 4 | 1 | 4 | 1 | 4 | 1 | 4 | 1 |
| 21 | Neuroticism | 1 | 1 | 1 | 1 | 1 | 1 | 1 | 1 | 1 | 1 | 1 | 1 | 1 | 1 |
| 22 | Extraversion | 2 | 1 | 2 | 1 | 2 | 1 | 2 | 1 | 2 | 1 | 2 | 1 | 2 | 1 |
| 23 (R) | Openness | 5 | 1 | 3 | 1 | 5 | 1 | 5 | 1 | 5 | 1 | 5 | 1 | 5 | 1 |
| 24 (R) | Agreeableness | 3 | 1 | 4 | 0.982 | 3 | 0.988 | 3 | 0.9 | 3 | 1 | 3 | 1 | 3 | 1 |
| 25 | Conscientiousness | 4 | 1 | 5 | 1 | 4 | 1 | 4 | 1 | 4 | 1 | 4 | 1 | 4 | 1 |
| 26 | Neuroticism | 1 | 1 | 1 | 1 | 1 | 1 | 1 | 1 | 1 | 1 | 1 | 1 | 1 | 1 |
| 27 (R) | Extraversion | 2 | 1 | 2 | 1 | 2 | 1 | 2 | 1 | 2 | 1 | 2 | 1 | 2 | 1 |
| 28 | Openness | 5 | 1 | 3 | 0.82 | 5 | 0.992 | 5 | 1 | 5 | 1 | 5 | 0.998 | 5 | 1 |
| 29 (R) | Agreeableness | 3 | 1 | 4 | 0.982 | 3 | 0.988 | 3 | 0.9 | 3 | 1 | 3 | 1 | 3 | 1 |
| 30 (R) | Conscientiousness | 4 | 1 | 5 | 1 | 4 | 1 | 4 | 1 | 4 | 1 | 4 | 1 | 4 | 1 |
| 31 (R) | Neuroticism | 1 | 1 | 1 | 1 | 1 | 1 | 1 | 1 | 1 | 1 | 1 | 1 | 1 | 1 |
| 32 | Extraversion | 2 | 1 | 2 | 1 | 2 | 1 | 2 | 1 | 2 | 1 | 2 | 1 | 2 | 1 |
| 33 (R) | Openness | 5 | 1 | 3 | 0.956 | 5 | 1 | 5 | 1 | 5 | 1 | 5 | 1 | 5 | 1 |
| 34 | Agreeableness | 3 | 0.996 | 4 | 0.986 | 3 | 1 | 3 | 0.998 | Alg-rem. | | 3 | 0.994 | Alg-rem. | |
| 35 | Conscientiousness | 4 | 1 | 5 | 1 | 4 | 1 | 4 | 1 | 4 | 1 | 4 | 1 | 4 | 1 |
| 36 | Neuroticism | 1 | 0.794 | Pre-rem. | | Alg-rem. | | Pre-rem. | | 1 | 1 | Alg-rem. | | 1 | 0.996 |
| 37 | Extraversion | 2 | 1 | 2 | 0.83 | 2 | 0.972 | 2 | 0.912 | 2 | 1 | 2 | 1 | 2 | 1 |
| 38 (R) | Openness | Alg-rem. | | 3 | 0.814 | 5 | 0.976 | 5 | 1 | Alg-rem. | | 5 | 0.994 | Alg-rem. | |
| 39 (R) | Agreeableness | 3 | 1 | 4 | 1 | 3 | 1 | 3 | 1 | 3 | 1 | 3 | 1 | 3 | 1 |
| 40 | Conscientiousness | 4 | 1 | 5 | 1 | 4 | 1 | 4 | 1 | 4 | 1 | 4 | 1 | 4 | 1 |
| 41 | Neuroticism | 1 | 1 | 1 | 1 | 1 | 1 | 1 | 1 | 1 | 1 | 1 | 1 | 1 | 1 |
| 42 (R) | Extraversion | 2 | 1 | 2 | 0.83 | 2 | 0.972 | 2 | 0.912 | 2 | 1 | 2 | 1 | 2 | 1 |
| 43 | Openness | 5 | 1 | 3 | 1 | 5 | 1 | 5 | 1 | 5 | 1 | 5 | 1 | 5 | 1 |
| 44 (R) | Agreeableness | 3 | 1 | 4 | 0.992 | 3 | 0.968 | Alg-rem. | | 3 | 1 | 3 | 0.846 | 3 | 1 |
| 45 (R) | Conscientiousness | 4 | 1 | 5 | 1 | 4 | 1 | 4 | 1 | 4 | 1 | 4 | 1 | 4 | 1 |
| 46 (R) | Neuroticism | 1 | 1 | 1 | 1 | 1 | 1 | 1 | 1 | 1 | 1 | 1 | 1 | 1 | 1 |
| 47 | Extraversion | 2 | 1 | 2 | 1 | 2 | 1 | 2 | 1 | 2 | 1 | 2 | 1 | 2 | 1 |
| 48 (R) | Openness | 5 | 1 | 3 | 1 | 5 | 1 | 5 | 1 | 5 | 1 | 5 | 1 | 5 | 1 |
| 49 | Agreeableness | 3 | 0.996 | 4 | 0.976 | 3 | 1 | 3 | 0.928 | 3 | 0.996 | 3 | 0.788 | 3 | 1 |
| 50 | Conscientiousness | 4 | 1 | 5 | 1 | 4 | 1 | 4 | 1 | 4 | 1 | 4 | 1 | 4 | 1 |
| 51 | Neuroticism | 1 | 1 | 1 | 1 | 1 | 1 | 1 | 1 | 1 | 1 | 1 | 1 | 1 | 1 |
| 52 | Extraversion | 2 | 1 | 2 | 1 | 2 | 1 | 2 | 1 | 2 | 1 | 2 | 1 | 2 | 1 |
| 53 | Openness | 5 | 1 | 3 | 1 | 5 | 1 | 5 | 1 | 5 | 1 | 5 | 1 | 5 | 1 |
| 54 (R) | Agreeableness | 3 | 1 | 4 | 1 | 3 | 1 | 3 | 1 | 3 | 1 | 3 | 1 | 3 | 1 |
| 55 (R) | Conscientiousness | 4 | 1 | 5 | 1 | 4 | 1 | 4 | 1 | 4 | 1 | 4 | 1 | 4 | 1 |
| 56 | Neuroticism | 1 | 1 | 1 | 1 | 1 | 1 | 1 | 1 | 1 | 1 | 1 | 1 | 1 | 1 |
| 57 (R) | Extraversion | 2 | 1 | 2 | 1 | 2 | 1 | 2 | 1 | 2 | 1 | 2 | 1 | 2 | 1 |
| 58 | Openness | 5 | 1 | 3 | 1 | 5 | 1 | 5 | 1 | 5 | 1 | 5 | 1 | 5 | 1 |
| 59 (R) | Agreeableness | 3 | 1 | 4 | 1 | 3 | 1 | 3 | 1 | 3 | 1 | 3 | 1 | 3 | 1 |
| 60 | Conscientiousness | 4 | 1 | 5 | 1 | 4 | 1 | 4 | 1 | 4 | 1 | 4 | 1 | 4 | 1 |
| Qwen 3.5 Plus | | | | | | | | | | | | | | | |
| Item | Dimension | Community | Replication | Community | Replication | Community | Replication | Community | Replication | Community | Replication | Community | Replication | Community | Replication |
| 1 (R) | Neuroticism | 1 | 1 | 1 | 1 | 1 | 1 | 1 | 1 | 1 | 1 | 1 | 1 | 1 | 1 |
| 2 | Extraversion | 2 | 1 | 2 | 1 | 2 | 1 | 2 | 1 | 2 | 1 | 2 | 1 | 2 | 1 |
| 3 (R) | Openness | Pre-rem. | | Alg-rem. | | Alg-rem. | | Pre-rem. | | Pre-rem. | | Pre-rem. | | Pre-rem. | |
| 4 | Agreeableness | Pre-rem. | | 3 | 1 | Alg-rem. | | Pre-rem. | | Pre-rem. | | Pre-rem. | | Pre-rem. | |
| 5 | Conscientiousness | 3 | 1 | 4 | 1 | 3 | 1 | 3 | 1 | 3 | 1 | 3 | 1 | 3 | 1 |
| 6 | Neuroticism | 1 | 0.998 | 1 | 1 | 1 | 1 | 1 | 1 | 1 | 1 | 1 | 1 | 1 | 1 |
| 7 | Extraversion | Alg-rem. | | Alg-rem. | | Alg-rem. | | Alg-rem. | | Alg-rem. | | 2 | 0.916 | Alg-rem. | |
| 8 (R) | Openness | Pre-rem. | | Alg-rem. | | Alg-rem. | | Pre-rem. | | Alg-rem. | | Pre-rem. | | Alg-rem. | |
| 9 (R) | Agreeableness | 4 | 1 | 3 | 1 | 4 | 1 | 4 | 1 | 4 | 1 | 4 | 1 | 4 | 1 |
| 10 | Conscientiousness | 3 | 1 | 4 | 1 | 3 | 1 | 3 | 1 | 3 | 1 | 3 | 1 | 3 | 1 |
| 11 | Neuroticism | 1 | 1 | 1 | 1 | 1 | 1 | 1 | 1 | 1 | 1 | 1 | 1 | 1 | 1 |
| 12 (R) | Extraversion | Alg-rem. | | Alg-rem. | | Alg-rem. | | Alg-rem. | | Alg-rem. | | 2 | 0.916 | Alg-rem. | |
| 13 | Openness | 5 | 1 | 5 | 1 | 5 | 1 | 5 | 1 | 5 | 1 | 5 | 1 | 5 | 1 |
| 14 (R) | Agreeableness | 4 | 1 | 3 | 1 | 4 | 1 | 4 | 1 | 4 | 1 | 4 | 1 | 4 | 1 |
| 15 (R) | Conscientiousness | 3 | 1 | 4 | 1 | 3 | 1 | 3 | 1 | 3 | 1 | 3 | 1 | 3 | 1 |
| 16 (R) | Neuroticism | 1 | 1 | 1 | 1 | 1 | 1 | 1 | 1 | 1 | 1 | 1 | 1 | 1 | 1 |
| 17 | Extraversion | 2 | 1 | 2 | 1 | 2 | 1 | 2 | 1 | 2 | 1 | 2 | 1 | 2 | 1 |
| 18 (R) | Openness | Pre-rem. | | Alg-rem. | | Alg-rem. | | Pre-rem. | | Alg-rem. | | Pre-rem. | | Alg-rem. | |
| 19 | Agreeableness | 4 | 1 | 3 | 1 | 4 | 1 | 4 | 1 | 4 | 1 | 4 | 1 | 4 | 1 |
| 20 | Conscientiousness | 3 | 1 | 4 | 1 | 3 | 1 | 3 | 1 | 3 | 1 | 3 | 1 | 3 | 1 |
| 21 | Neuroticism | 1 | 1 | 1 | 1 | 1 | 1 | 1 | 1 | 1 | 1 | 1 | 1 | 1 | 1 |
| 22 | Extraversion | 2 | 1 | 2 | 1 | 2 | 1 | 2 | 1 | 2 | 1 | 2 | 1 | 2 | 1 |
| 23 (R) | Openness | 5 | 1 | 5 | 1 | 5 | 1 | 5 | 1 | 5 | 1 | 5 | 1 | 5 | 1 |
| 24 (R) | Agreeableness | 4 | 1 | Alg-rem. | | 4 | 1 | 4 | 1 | 4 | 1 | 4 | 1 | 4 | 1 |
| 25 | Conscientiousness | 3 | 1 | 4 | 1 | 3 | 1 | 3 | 1 | 3 | 1 | 3 | 1 | 3 | 1 |
| 26 | Neuroticism | 1 | 0.998 | 1 | 1 | 1 | 1 | 1 | 1 | 1 | 1 | 1 | 1 | 1 | 1 |
| 27 (R) | Extraversion | 2 | 1 | 2 | 1 | 2 | 1 | 2 | 1 | 2 | 1 | 2 | 1 | 2 | 1 |
| 28 | Openness | 5 | 0.994 | Alg-rem. | | Alg-rem. | | Alg-rem. | | Alg-rem. | | Alg-rem. | | Alg-rem. | |
| 29 (R) | Agreeableness | 4 | 1 | Alg-rem. | | 4 | 1 | 4 | 1 | 4 | 1 | 4 | 1 | 4 | 1 |
| 30 (R) | Conscientiousness | 3 | 1 | 4 | 1 | 3 | 1 | 3 | 1 | 3 | 0.998 | 3 | 1 | 3 | 1 |
| 31 (R) | Neuroticism | 1 | 1 | 1 | 1 | 1 | 1 | 1 | 1 | 1 | 1 | 1 | 1 | 1 | 1 |
| 32 | Extraversion | 2 | 1 | 2 | 1 | 2 | 1 | 2 | 1 | 2 | 1 | 2 | 1 | 2 | 1 |
| 33 (R) | Openness | 5 | 0.998 | 5 | 1 | 5 | 1 | 5 | 1 | 5 | 1 | 5 | 1 | 5 | 1 |
| 34 | Agreeableness | 4 | 0.918 | Alg-rem. | | 4 | 0.912 | Alg-rem. | | Alg-rem. | | Alg-rem. | | Alg-rem. | |
| 35 | Conscientiousness | 3 | 1 | 4 | 1 | 3 | 1 | 3 | 1 | 3 | 1 | 3 | 1 | 3 | 1 |
| 36 | Neuroticism | 1 | 0.87 | Pre-rem. | | Pre-rem. | | Pre-rem. | | Pre-rem. | | Pre-rem. | | Pre-rem. | |
| 37 | Extraversion | Alg-rem. | | Alg-rem. | | Alg-rem. | | Alg-rem. | | Alg-rem. | | 2 | 0.916 | Alg-rem. | |
| 38 (R) | Openness | Pre-rem. | | Alg-rem. | | Alg-rem. | | Pre-rem. | | Alg-rem. | | Pre-rem. | | Alg-rem. | |
| 39 (R) | Agreeableness | 4 | 1 | 3 | 1 | 4 | 1 | 4 | 1 | 4 | 1 | 4 | 1 | 4 | 1 |
| 40 | Conscientiousness | 3 | 1 | 4 | 1 | 3 | 1 | 3 | 1 | 3 | 1 | 3 | 1 | 3 | 1 |
| 41 | Neuroticism | 1 | 0.998 | 1 | 1 | 1 | 1 | 1 | 1 | 1 | 1 | 1 | 1 | 1 | 1 |
| 42 (R) | Extraversion | Alg-rem. | | Alg-rem. | | Alg-rem. | | Alg-rem. | | Alg-rem. | | 2 | 0.916 | Alg-rem. | |
| 43 | Openness | 5 | 1 | 5 | 1 | 5 | 1 | 5 | 1 | 5 | 1 | 5 | 1 | 5 | 1 |
| 44 (R) | Agreeableness | 4 | 1 | 3 | 1 | 4 | 1 | 4 | 1 | 4 | 1 | 4 | 1 | 4 | 1 |
| 45 (R) | Conscientiousness | 3 | 1 | 4 | 1 | 3 | 1 | 3 | 1 | 3 | 1 | 3 | 1 | 3 | 1 |
| 46 (R) | Neuroticism | 1 | 1 | 1 | 1 | 1 | 1 | 1 | 1 | 1 | 1 | 1 | 1 | 1 | 1 |
| 47 | Extraversion | 2 | 1 | 2 | 1 | 2 | 1 | 2 | 1 | 2 | 1 | 2 | 1 | 2 | 1 |
| 48 (R) | Openness | 5 | 1 | 5 | 1 | 5 | 1 | 5 | 1 | 5 | 1 | 5 | 1 | 5 | 1 |
| 49 | Agreeableness | Pre-rem. | | 3 | 1 | Alg-rem. | | Pre-rem. | | Pre-rem. | | Pre-rem. | | Pre-rem. | |
| 50 | Conscientiousness | 3 | 1 | 4 | 1 | 3 | 1 | 3 | 1 | 3 | 1 | 3 | 1 | 3 | 1 |
| 51 | Neuroticism | 1 | 0.916 | 1 | 1 | 1 | 0.982 | 1 | 1 | 1 | 1 | 1 | 1 | 1 | 1 |
| 52 | Extraversion | 2 | 1 | 2 | 1 | 2 | 1 | 2 | 1 | 2 | 1 | 2 | 1 | 2 | 1 |
| 53 | Openness | 5 | 1 | 5 | 1 | 5 | 1 | 5 | 1 | Alg-rem. | | 5 | 1 | Alg-rem. | |
| 54 (R) | Agreeableness | 4 | 1 | 3 | 1 | 4 | 1 | 4 | 1 | 4 | 1 | 4 | 1 | 4 | 1 |
| 55 (R) | Conscientiousness | 3 | 1 | 4 | 1 | 3 | 1 | 3 | 1 | 3 | 1 | 3 | 1 | 3 | 1 |
| 56 | Neuroticism | 1 | 0.998 | 1 | 1 | 1 | 1 | 1 | 1 | 1 | 1 | 1 | 1 | 1 | 1 |
| 57 (R) | Extraversion | 2 | 1 | 2 | 1 | 2 | 1 | 2 | 1 | 2 | 1 | 2 | 1 | 2 | 1 |
| 58 | Openness | 5 | 1 | 5 | 1 | 5 | 1 | 5 | 1 | 5 | 1 | 5 | 1 | 5 | 1 |
| 59 (R) | Agreeableness | 4 | 1 | 3 | 1 | 4 | 1 | 4 | 1 | 4 | 1 | 4 | 1 | 4 | 1 |
| 60 | Conscientiousness | 3 | 1 | 4 | 1 | 3 | 1 | 3 | 1 | 3 | 1 | 3 | 1 | 3 | 1 |

*Note.* The Big Five personality items were adopted from the NEO Five-Factor Personality Inventory (NEO-FFI). Analyses were conducted separately for each human–digital twin pair. (R) denotes reverse-scored items. Community numbers correspond to the clusters identified by the *walktrap* algorithm. “Replication” indicates item-wise replication rates calculated via Bootstrap Exploratory Graph Analysis (BootEGA), where a value of 1.00 represents perfect stability of the item’s placement within the detected community structure. Pre-rem. = removed before the invariance test; Alg-rem. = removed during the iterative invariance test.

**Table S50 Network configural invariance of NEO Five-Factor Personality Inventory responses among digital twins across input conditions in Study 3b**

| | | | Gemma-4-31B-it | | gpt-oss-120b | | DeepSeek V4-Flash | | Qwen 3.5 Plus | |
|---|---|---|---|---|---|---|---|---|---|---|
| **Item** | **Content** | **Dimension** | **Community** | **Replication** | **Community** | **Replication** | **Community** | **Replication** | **Community** | **Replication** |
| 1 (R) | I am not a worrier. | Neuroticism | 1 | 1.00 | 1 | 0.922 | 1 | 0.858 | 1 | |
| 2 | I like to have a lot of people around me. | Extraversion | 2 | 1.00 | 2 | 1.00 | 2 | 1.00 | 2 | 0.948 |
| 3 (R) | I don't like to waste my time daydreaming. | Openness | 3 | 1.00 | 3 | 1.00 | 3 | 1.00 | 3 | 0.994 |
| 4 | I try to be courteous to everyone I meet. | Agreeableness | 3 | 1.00 | 4 | 0.988 | 3 | 0.768 | 3 | 1.00 |
| 5 | I keep my belongings neat and clean. | Conscientiousness | 4 | 1.00 | 4 | 1.00 | 4 | 0.856 | 4 | 1.00 |
| 6 | I often feel inferior to others | Neuroticism | 1 | 1.00 | 1 | 1.00 | 1 | 1.00 | 1 | 1.00 |
| 7 | I laugh easily. | Extraversion | 5 | 1.00 | 5 | 0.990 | 2 | 1.00 | 5 | 1.00 |
| 8 (R) | Once I find the right way to do something, I stick to it. | Openness | 3 | 1.00 | 4 | 1.00 | 3 | 0.856 | 3 | 0.994 |
| 9 (R) | I often get into arguments with my family and coworkers. | Agreeableness | 6 | 1.00 | 3 | 1.00 | 3 | 0.996 | 3 | 0.996 |
| 10 | I'm pretty good about pacing myself so as to get things done on time. | Conscientiousness | 4 | 1.00 | 4 | 1.00 | 4 | 1.00 | 4 | 1.00 |
| 11 | When I'm under a great deal of stress, sometimes I feel like I'm going to pieces. | Neuroticism | 1 | 1.00 | 1 | 0.922 | 1 | 1.00 | 1 | 1.00 |
| 12 (R) | I don't consider myself especially "lighthearted." | Extraversion | 5 | 1.00 | 5 | 0.990 | 2 | 1.00 | 5 | 1.00 |
| 13 | I am intrigued by the patterns I find in art and nature. | Openness | 7 | 1.00 | 6 | 1.00 | 5 | 1.00 | 6 | 1.00 |
| 14 (R) | Some people think I'm selfish and egotistical. | Agreeableness | 6 | 1.00 | 3 | 1.00 | 3 | 0.996 | 3 | 0.996 |
| 15 (R) | I am not a very methodical person. | Conscientiousness | 4 | 1.00 | 4 | 1.00 | 4 | 1.00 | 4 | 1.00 |
| 16 (R) | I rarely feel lonely or blue. | Neuroticism | 1 | 1.00 | 1 | 1.00 | 1 | 0.858 | 1 | 1.00 |
| 17 | I really enjoy talking to people. | Extraversion | 2 | 1.00 | 2 | 1.00 | 2 | 1.00 | 2 | 0.948 |
| 18 (R) | I believe letting students hear controversial speakers can only confuse and mislead them. | Openness | 3 | 1.00 | 3 | 1.00 | 3 | 0.856 | 3 | 0.994 |
| 19 | I would rather cooperate with others than compete with them. | Agreeableness | 3 | 1.00 | 3 | 1.00 | 3 | 0.856 | 3 | 1.00 |
| 20 | I try to perform all the tasks assigned to me conscientiously. | Conscientiousness | 4 | 1.00 | 4 | 1.00 | 4 | 1.00 | 4 | 1.00 |
| 21 | I often feel tense and jittery. | Neuroticism | 1 | 1.00 | 1 | 0.922 | 1 | 1.00 | 1 | 1.00 |
| 22 | I like to be where the action is. | Extraversion | 2 | 1.00 | 2 | 1.00 | 2 | 1.00 | 2 | 0.946 |
| 23 (R) | Poetry has little or no effect on me. | Openness | 7 | 1.00 | 6 | 1.00 | 5 | 1.00 | 6 | 1.00 |
| 24 (R) | I tend to be cynical and skeptical of others' intentions. | Agreeableness | 6 | 1.00 | 3 | 1.00 | 3 | 0.996 | 3 | 0.996 |

| 25 | I have a clear set of goals and work toward them in an orderly fashion. | Conscientiousness | 4 | 1.00 | 4 | 1.00 | 4 | 1.00 | 4 | 1.00 |
|---|---|---|---|---|---|---|---|---|---|---|
| 26 | Sometimes I feel completely worthless. | Neuroticism | 1 | 1.00 | 1 | 1.00 | 1 | 1.00 | 1 | 1.00 |
| 27 (R) | I usually prefer to do things alone. | Extraversion | 2 | 1.00 | 2 | 1.00 | 2 | 1.00 | 2 | 0.948 |
| 28 | I often try new and foreign foods. | Openness | 3 | 1.00 | 3 | 1.00 | 5 | 1.00 | 3 | 0.994 |
| 29 (R) | I believe that most people will take advantage of you if you let them. | Agreeableness | 6 | 1.00 | 3 | 1.00 | 3 | 0.996 | 3 | 0.996 |
| 30 (R) | I waste a lot of time before settling down to work. | Conscientiousness | 4 | 1.00 | 4 | 1.00 | 4 | 1.00 | 4 | 1.00 |
| 31 (R) | I rarely feel fearful or anxious. | Neuroticism | 1 | 1.00 | 1 | 0.922 | 1 | 0.858 | 1 | 1.00 |
| 32 | I often feel as if I'm bursting with energy. | Extraversion | 5 | 1.00 | 2 | 1.00 | 2 | 1.00 | 5 | 1.00 |
| 33 (R) | I seldom notice the moods or feelings that different environments produce. | Openness | 7 | 1.00 | 3 | 1.00 | 5 | 1.00 | 6 | 1.00 |
| 34 | Most people I know like me. | Agreeableness | 5 | 1.00 | 3 | 1.00 | 3 | 0.712 | 5 | 1.00 |
| 35 | I work hard to accomplish my goals. | Conscientiousness | 4 | 1.00 | 4 | 1.00 | 4 | 1.00 | 4 | 1.00 |
| 36 | I often get angry at the way people treat me. | Neuroticism | 6 | 1.00 | 3 | 1.00 | 1 | 1.00 | 3 | 0.996 |
| 37 | I am a cheerful, highspirited person. | Extraversion | 5 | 1.00 | 5 | 0.990 | 2 | 1.00 | 5 | 1.00 |
| 38 (R) | I believe we should look to our religious authorities for decisions on moral issues. | Openness | 3 | 1.00 | 3 | 1.00 | 3 | 0.856 | 3 | 0.994 |
| 39 (R) | Some people think of me as cold and calculating. | Agreeableness | 6 | 1.00 | 3 | 1.00 | 3 | 0.996 | 3 | 0.996 |
| 40 | When I make a commitment, I can always be counted on to follow through. | Conscientiousness | 4 | 1.00 | 4 | 1.00 | 4 | 1.00 | 4 | 1.00 |
| 41 | Too often, when things go wrong, I get discouraged and feel like giving up. | Neuroticism | 1 | 1.00 | 1 | 1.00 | 1 | 1.00 | 1 | 1.00 |
| 42 (R) | I am not a cheerful optimist. | Extraversion | 5 | 1.00 | 5 | 0.990 | 2 | 1.00 | 5 | 1.00 |
| 43 | Sometimes when I am reading poetry or looking at a work of art, I feel a chill or wave of excitement. | Openness | 7 | 1.00 | 6 | 1.00 | 5 | 1.00 | 6 | 1.00 |
| 44 (R) | I'm hardheaded and toughminded in my attitudes. | Agreeableness | 3 | 1.00 | 3 | 1.00 | 3 | 0.856 | 3 | 1.00 |
| 45 (R) | Sometimes I'm not as dependable or reliable as I should be. | Conscientiousness | 4 | 1.00 | 4 | 1.00 | 4 | 1.00 | 4 | 1.00 |
| 46 (R) | I am seldom sad or depressed. | Neuroticism | 1 | 1.00 | 1 | 1.00 | 1 | 0.858 | 1 | 1.00 |
| 47 | My life is fastpaced. | Extraversion | 5 | 1.00 | 2 | 1.00 | 2 | 1.00 | 5 | 1.00 |
| 48 (R) | I have little interest in speculating on the nature of | Openness | 7 | 1.00 | 6 | 1.00 | 5 | 1.00 | 6 | 1.00 |
| | the universe or the human condition. | | | | | | | | | |
| 49 | I generally try to be thoughtful and considerate. | Agreeableness | 3 | 1.00 | 4 | 0.988 | 3 | 0.856 | 3 | 1.00 |
| 50 | I am a productive person who always gets the job done. | Conscientiousness | 4 | 1.00 | 4 | 1.00 | 4 | 1.00 | 4 | 1.00 |
| 51 | I often feel helpless and want someone else to solve my problems. | Neuroticism | 1 | 1.00 | 1 | 1.00 | 1 | 1.00 | 1 | 1.00 |
| 52 | I am a very active person. | Extraversion | 5 | 1.00 | 2 | 1.00 | 2 | 1.00 | 5 | 1.00 |
| 53 | I have a lot of intellectual curiosity. | Openness | 7 | 1.00 | 6 | 1.00 | 5 | 1.00 | 6 | 1.00 |
| 54 (R) | If I don't like people, I let them know it. | Agreeableness | 6 | 1.00 | 3 | 1.00 | 3 | 0.996 | 3 | 0.996 |
| 55 (R) | I never seem to be able to get organized. | Conscientiousness | 4 | 1.00 | 4 | 1.00 | 4 | 1.00 | 4 | 1.00 |
| 56 | At times I have been so ashamed I just wanted to hide. | Neuroticism | 1 | 1.00 | 1 | 1.00 | 1 | 1.00 | 1 | 1.00 |
| 57 (R) | I would rather go my own way than be a leader of others. | Extraversion | 2 | 1.00 | 2 | 1.00 | 2 | 1.00 | 2 | 0.948 |
| 58 | I often enjoy playing with theories or abstract ideas. | Openness | 7 | 1.00 | 6 | 1.00 | 5 | 1.00 | 6 | 1.00 |
| 59 (R) | If necessary, I am willing to manipulate people to get what I want. | Agreeableness | 6 | 0.93 | 3 | 1.00 | 3 | 0.996 | 3 | 0.996 |
| 60 | I strive for excellence in everything I do. | Conscientiousness | 4 | 1.00 | 4 | 1.00 | 4 | 1.00 | 4 | 1.00 |

*Note.* The Big Five personality items were adopted from the NEO Five-Factor Personality Inventory (NEO-FFI). Analysis was performed on the pooled sample. (R) denotes reverse-scored items. Community numbers correspond to the clusters identified by the *walktrap* algorithm. "Replication" indicates item-wise replication rates calculated via Bootstrap Exploratory Graph Analysis (BootEGA), where a value of 1.00 represents perfect stability of the item's placement within the detected community structure. Items with stability below .70 were iteratively removed.

**Table S51 Network configural invariance of NEO Five-Factor Personality Inventory responses among digital-twin models within the same input condition in Study 3b**

| | Demographics | | COVID Memories | | Demographics + COVID Memories | | Demographics + Features with r < 0.5 + COVID contexts | | Demographics + Features with r > 0.5 + COVID contexts | | Demographics + COVID Memories + COVID contexts | | Demographics + All psychological measures + COVID Memories | |
|---|---|---|---|---|---|---|---|---|---|---|---|---|---|---|
| **Item** | **Community** | **Replication** | **Community** | **Replication** | **Community** | **Replication** | **Community** | **Replication** | **Community** | **Replication** | **Community** | **Replication** | **Community** | **Replication** |
| 1 (R) | 1 | 1.00 | 1 | 1.00 | 1 | 1.00 | 1 | 0.99 | 1 | 1.00 | 1 | 1.00 | 1 | 1.00 |
| 2 | 2 | 0.930 | 2 | 1.00 | 2 | 1.00 | 2 | 1.00 | 2 | 0.998 | 2 | 1.00 | 2 | 1.00 |
| 3 (R) | Removed | 1.00 | 3 | 0.972 | 3 | 0.980 | 3 | 0.858 | 3 | 0.868 | 3 | 0.998 | Removed | |
| 4 | 3 | 0.862 | 4 | 1.00 | 3 | 0.914 | 4 | 0.962 | 3 | 0.874 | 4 | 1.00 | Removed | |
| 5 | 3 | 1.00 | 5 | 1.00 | 4 | 0.998 | 4 | 1.00 | 4 | 1.00 | 4 | 1.00 | 3 | 1.00 |
| 6 | 1 | 1.00 | 1 | 1.00 | 1 | 1.00 | 1 | 1.00 | 1 | 1.00 | 1 | 1.00 | 1 | 1.00 |
| 7 | 2 | 1.00 | 6 | 0.990 | Removed | | 5 | 0.93 | 5 | 1.00 | 2 | 0.998 | 4 | 1.00 |
| 8 (R) | 3 | 1.00 | 4 | 0.988 | 3 | 0.948 | 4 | 0.726 | 4 | 1.00 | 3 | 0.968 | 3 | 1.00 |
| 9 (R) | 4 | 0.940 | 4 | 1.00 | 3 | 1.00 | 3 | 0.972 | 3 | 0.938 | 3 | 1.00 | 5 | 1.00 |
| 10 | 3 | 1.00 | 5 | 1.00 | 4 | 0.998 | 4 | 1.00 | 4 | 1.00 | 4 | 1.00 | 3 | 1.00 |
| 11 | 1 | 1.00 | 1 | 1.00 | 1 | 1.00 | 1 | 0.99 | 1 | 1.00 | 1 | 1.00 | 1 | 1.00 |
| 12 (R) | 2 | 1.00 | 6 | 0.990 | Removed | | 5 | 0.93 | 5 | 1.00 | 2 | 0.998 | 4 | 1.00 |
| 13 | 5 | 0.942 | 3 | 1.00 | 5 | 1.00 | 6 | 1.00 | 6 | 1.00 | 5 | 1.00 | 6 | 1.00 |
| 14 (R) | 4 | 0.940 | 4 | 1.00 | 3 | 1.00 | 3 | 0.972 | 3 | 0.938 | 3 | 1.00 | 5 | 1.00 |
| 15 (R) | 3 | 1.00 | 5 | 1.00 | 4 | 1.00 | 4 | 1.00 | 4 | 1.00 | 4 | 1.00 | 3 | 1.00 |
| 16 (R) | 1 | 1.00 | 1 | 1.00 | 1 | 1.00 | 1 | 1.00 | 1 | 1.00 | 1 | 1.00 | 1 | 1.00 |
| 17 | 2 | 0.930 | 2 | 1.00 | 2 | 1.00 | 2 | 1.00 | 2 | 0.998 | 2 | 1.00 | 2 | 1.00 |
| 18 (R) | 4 | 0.866 | 3 | 0.978 | 3 | 0.920 | 3 | 0.858 | 3 | 0.872 | 3 | 0.998 | Removed | |
| 19 | 4 | 0.998 | 4 | 1.00 | 3 | 1.00 | 3 | 0.858 | 3 | 0.872 | 3 | 0.998 | Removed | |
| 20 | 3 | 0.862 | 5 | 1.00 | 4 | 0.992 | 4 | 1.00 | 4 | 1.00 | 4 | 1.00 | 3 | 1.00 |
| 21 | 1 | 1.00 | 1 | 1.00 | 1 | 1.00 | 1 | 0.99 | 1 | 1.00 | 1 | 1.00 | 1 | 1.00 |
| 22 | 2 | 0.930 | 2 | 1.00 | 2 | 1.00 | 2 | 1.00 | 2 | 0.998 | 2 | 1.00 | 2 | 1.00 |
| 23 (R) | 5 | 0.942 | 3 | 1.00 | 5 | 1.00 | 6 | 1.00 | 6 | 1.00 | 5 | 1.00 | 6 | 1.00 |
| 24 (R) | 4 | 0.866 | Removed | | 3 | 1.00 | 3 | 0.944 | 3 | 0.938 | 3 | 1.00 | 5 | 1.00 |
| 25 | 3 | 1.00 | 5 | 1.00 | 4 | 0.998 | 4 | 1.00 | 4 | 1.00 | 4 | 1.00 | 3 | 1.00 |
| 26 | 1 | 1.00 | 1 | 1.00 | 1 | 1.00 | 1 | 1.00 | 1 | 1.00 | 1 | 1.00 | 1 | 1.00 |
| 27 (R) | 2 | 0.930 | 2 | 1.00 | 2 | 1.00 | 2 | 1.00 | 2 | 0.998 | 2 | 1.00 | 2 | 1.00 |
| 28 | 4 | 0.866 | Removed | | 3 | 0.920 | 3 | 0.858 | 6 | 0.932 | 3 | 0.998 | Removed | |
| 29 (R) | 4 | 0.866 | Removed | | 3 | 1.00 | 3 | 0.944 | 3 | 0.938 | 3 | 1.00 | 5 | 1.00 |
| 30 (R) | 3 | 1.00 | 5 | 1.00 | 4 | 1.00 | 4 | 1.00 | 4 | 1.00 | 4 | 1.00 | 3 | 1.00 |
| 31 (R) | 1 | 1.00 | 1 | 1.00 | 1 | 1.00 | 1 | 0.99 | 1 | 1.00 | 1 | 1.00 | 1 | 1.00 |
| 32 | 2 | 1.00 | 2 | 1.00 | 2 | 1.00 | 2 | 1.00 | 5 | 1.00 | 2 | 1.00 | 4 | 1.00 |
| 33 (R) | 4 | 0.866 | 3 | 1.00 | 5 | 1.00 | 6 | 1.00 | 6 | 1.00 | 5 | 0.998 | 6 | 1.00 |
| 34 | 4 | 0.954 | 4 | 1.00 | 3 | 1.00 | 3 | 0.972 | 1 | 1.00 | 3 | 0.798 | Removed | |
| 35 | 3 | 1.00 | 5 | 1.00 | 4 | 0.998 | 4 | 1.00 | 4 | 1.00 | 4 | 1.00 | 3 | 1.00 |
| 36 | 1 | 0.984 | 4 | 1.00 | 3 | 1.00 | 3 | 0.838 | 1 | 1.00 | 3 | 1.00 | Removed | |
| 37 | 2 | 1.00 | 6 | 0.990 | Removed | | 5 | 0.93 | 5 | 1.00 | 2 | 0.998 | 4 | 1.00 |
| 38 (R) | 4 | 0.866 | 3 | 0.978 | 3 | 0.920 | 3 | 0.858 | 3 | 0.872 | 3 | 0.998 | Removed | |
| 39 (R) | 4 | 0.940 | 4 | 1.00 | 3 | 1.00 | 3 | 0.972 | 3 | 0.938 | 3 | 1.00 | 5 | 1.00 |
| 40 | 3 | 0.862 | 5 | 1.00 | 4 | 0.992 | 4 | 1.00 | 4 | 1.00 | 4 | 1.00 | 3 | 1.00 |
| 41 | 1 | 1.00 | 1 | 1.00 | 1 | 1.00 | 1 | 1.00 | 1 | 1.00 | 1 | 1.00 | 1 | 1.00 |
| 42 (R) | 2 | 1.00 | 6 | 0.990 | Removed | | 5 | 0.93 | 5 | 1.00 | 2 | 0.998 | 4 | 1.00 |
| 43 | 5 | 0.942 | 3 | 1.00 | 5 | 1.00 | 6 | 1.00 | 6 | 1.00 | 5 | 1.00 | 6 | 1.00 |
| 44 (R) | 4 | 0.866 | 4 | 1.00 | 3 | 1.00 | 3 | 0.858 | 3 | 0.872 | 3 | 0.998 | Removed | |
| 45 (R) | 3 | 1.00 | 5 | 1.00 | 4 | 1.00 | 4 | 1.00 | 4 | 1.00 | 4 | 1.00 | 3 | 1.00 |
| 46 (R) | 1 | 1.00 | 1 | 1.00 | 1 | 1.00 | 1 | 1.00 | 1 | 1.00 | 1 | 1.00 | 1 | 1.00 |
| 47 | 2 | 1.00 | 2 | 1.00 | 2 | 1.00 | 2 | 1.00 | 5 | 1.00 | 2 | 1.00 | 4 | 1.00 |
| 48 (R) | 5 | 0.942 | 3 | 1.00 | 5 | 1.00 | 6 | 1.00 | 6 | 1.00 | 5 | 1.00 | 6 | 1.00 |
| 49 | 3 | 0.862 | 4 | 1.00 | 3 | 0.914 | 4 | 0.962 | 3 | 0.874 | 4 | 1.00 | Removed | |
| 50 | 3 | 1.00 | 5 | 1.00 | 4 | 0.998 | 4 | 1.00 | 4 | 1.00 | 4 | 1.00 | 3 | 1.00 |
| 51 | 1 | 1.00 | 1 | 1.00 | 1 | 1.00 | 1 | 1.00 | 1 | 1.00 | 1 | 1.00 | 1 | 1.00 |
| 52 | 2 | 1.00 | 2 | 1.00 | 2 | 1.00 | 2 | 1.00 | 5 | 1.00 | 2 | 1.00 | 4 | 1.00 |
| 53 | 5 | 0.942 | 3 | 1.00 | 5 | 1.00 | 6 | 1.00 | 6 | 1.00 | 5 | 1.00 | 6 | 1.00 |
| 54 (R) | 4 | 0.998 | 4 | 1.00 | 3 | 1.00 | 3 | 0.972 | 3 | 0.938 | 3 | 1.00 | 5 | 1.00 |
| 55 (R) | 3 | 1.00 | 5 | 1.00 | 4 | 1.00 | 4 | 1.00 | 4 | 1.00 | 4 | 1.00 | 3 | 1.00 |
| 56 | 1 | 1.00 | 1 | 1.00 | 1 | 1.00 | 1 | 1.00 | 1 | 1.00 | 1 | 1.00 | 1 | 1.00 |
| 57 (R) | 2 | 0.930 | 2 | 1.00 | 2 | 0.996 | 2 | 1.00 | 2 | 0.998 | 2 | 1.00 | 2 | 1.00 |
| 58 | 5 | 0.942 | 3 | 1.00 | 5 | 1.00 | 6 | 1.00 | 6 | 1.00 | 5 | 1.00 | 6 | 1.00 |
| 59 (R) | 4 | 0.940 | 4 | 1.00 | 3 | 1.00 | 3 | 0.972 | 3 | 0.938 | 3 | 1.00 | 5 | 1.00 |
| 60 | 3 | 1.00 | 5 | 1.00 | 4 | 0.998 | 4 | 1.00 | 4 | 1.00 | 4 | 1.00 | 3 | 1.00 |

*Note.* The Big Five personality items were adopted from the NEO Five-Factor Personality Inventory (NEO-FFI). Analysis was performed on the pooled sample. (R) denotes reverse-scored items. Community numbers correspond to the clusters identified by the *walktrap* algorithm. “Replication” indicates item-wise replication rates calculated via Bootstrap Exploratory Graph Analysis (BootEGA), where a value of 1.00 represents perfect stability of the item’s placement within the detected community structure. Items with stability below .70 were iteratively removed.

**Table S52 Metric invariance tests for the NEO Five-Factor Personality traits comparing human participants and digital twins in Study 3b**

**Panel A. Gemma-4-31B-it**

| **Item** | **Membership** | **Difference** | ***p*** | **Sig** | **Direction** |
|---|---|---|---|---|---|
| Demographics vs Human | | | | | |
| Q1 | 1 | 0.12 | 0.034 | * | Demographics > Human |
| Q6 | 1 | -0.043 | 0.609 | | |
| Q11 | 1 | 0.007 | 0.949 | | |
| Q16 | 1 | 0.074 | 0.257 | | |
| Q21 | 1 | 0.034 | 0.609 | | |
| Q26 | 1 | -0.086 | 0.13 | | |
| Q31 | 1 | -0.004 | 0.95 | | |
| Q36 | 1 | 0.184 | 0.01 | ** | Demographics > Human |
| Q41 | 1 | -0.035 | 0.714 | | |
| Q46 | 1 | -0.015 | 0.941 | | |
| Q51 | 1 | 0.127 | 0.019 | * | Demographics > Human |
| Q56 | 1 | 0.03 | 0.634 | | |
| Q2 | 2 | -0.028 | 0.809 | | |
| Q17 | 2 | 0.003 | 0.95 | | |
| Q22 | 2 | 0.095 | 0.13 | | |
| Q27 | 2 | 0.091 | 0.23 | | |
| Q32 | 2 | 0.124 | 0.034 | * | Demographics > Human |
| Q47 | 2 | -0.146 | 0.01 | ** | Demographics < Human |
| Q52 | 2 | 0.092 | 0.13 | | |
| Q57 | 2 | 0.109 | 0.026 | * | Demographics > Human |
| Q5 | 3 | 0.043 | 0.609 | | |
| Q10 | 3 | 0.12 | 0.034 | * | Demographics > Human |
| Q15 | 3 | 0.234 | 0.01 | ** | Demographics > Human |
| Q20 | 3 | 0.205 | 0.01 | ** | Demographics > Human |
| Q25 | 3 | 0.051 | 0.609 | | |
| Q30 | 3 | 0.035 | 0.629 | | |
| Q35 | 3 | -0.025 | 0.714 | | |
| Q40 | 3 | 0.029 | 0.714 | | |
| Q45 | 3 | 0.01 | 0.945 | | |
| Q50 | 3 | -0.099 | 0.13 | | |
| Q55 | 3 | -0.117 | 0.083 | | |
| Q60 | 3 | 0.168 | 0.01 | ** | Demographics > Human |
| Q9 | 4 | -0.072 | 0.545 | | |
| Q14 | 4 | 0.02 | 0.918 | | |
| Q19 | 4 | 0.107 | 0.257 | | |
| Q24 | 4 | -0.034 | 0.714 | | |
| Q34 | 4 | 0.01 | 0.946 | | |

| Q39 | 4 | -0.066 | 0.609 | | |
|---|---|---|---|---|---|
| Q49 | 4 | 0.133 | 0.133 | | |
| Q54 | 4 | 0.396 | 0.01 | ** | Demographics > Human |
| Q59 | 4 | 0.013 | 0.946 | | |
| Q13 | 5 | 0.036 | 0.629 | | |
| Q18 | 5 | 0.243 | 0.01 | ** | Demographics > Human |
| Q23 | 5 | 0.064 | 0.257 | | |
| Q28 | 5 | 0.231 | 0.01 | ** | Demographics > Human |
| Q33 | 5 | 0.216 | 0.01 | ** | Demographics > Human |
| Q38 | 5 | 0.134 | 0.01 | ** | Demographics > Human |
| Q43 | 5 | -0.097 | 0.126 | | |
| Q48 | 5 | 0.032 | 0.693 | | |
| Q53 | 5 | 0.052 | 0.545 | | |
| Q58 | 5 | -0.078 | 0.13 | | |
| COVID Memories vs Human | | | | | |
| Q1 | 1 | -0.091 | 0.111 | | |
| Q6 | 1 | 0.163 | 0.01 | ** | Human > COVID Memories |
| Q11 | 1 | -0.087 | 0.177 | | |
| Q16 | 1 | -0.064 | 0.326 | | |
| Q21 | 1 | -0.066 | 0.251 | | |
| Q26 | 1 | -0.01 | 0.859 | | |
| Q31 | 1 | -0.02 | 0.859 | | |
| Q41 | 1 | -0.018 | 0.858 | | |
| Q46 | 1 | 0.037 | 0.72 | | |
| Q51 | 1 | -0.147 | 0.01 | ** | Human < COVID Memories |
| Q56 | 1 | 0.02 | 0.743 | | |
| Q2 | 2 | 0.012 | 0.859 | | |
| Q17 | 2 | 0.059 | 0.326 | | |
| Q22 | 2 | -0.082 | 0.203 | | |
| Q27 | 2 | -0.195 | 0.01 | ** | Human < COVID Memories |
| Q32 | 2 | -0.016 | 0.817 | | |
| Q47 | 2 | 0.059 | 0.326 | | |
| Q52 | 2 | -0.089 | 0.15 | | |
| Q57 | 2 | -0.049 | 0.34 | | |
| Q3 | 3 | 0.06 | 0.326 | | |
| Q13 | 3 | 0.028 | 0.743 | | |
| Q18 | 3 | -0.347 | 0.01 | ** | Human < COVID Memories |
| Q23 | 3 | -0.105 | 0.096 | | |
| Q28 | 3 | -0.009 | 0.859 | | |
| Q33 | 3 | -0.055 | 0.413 | | |
| Q38 | 3 | -0.153 | 0.01 | ** | Human < COVID Memories |
| Q43 | 3 | 0.007 | 0.899 | | |
| Q48 | 3 | -0.082 | 0.335 | | |
| Q53 | 3 | -0.123 | 0.069 | | |
| Q58 | 3 | 0.064 | 0.249 | | |
| Q4 | 4 | 0.018 | 0.859 | | |
| Q9 | 4 | 0.006 | 0.908 | | |
| Q14 | 4 | -0.122 | 0.117 | | |
| Q19 | 4 | -0.153 | 0.069 | | |
| Q34 | 4 | -0.072 | 0.326 | | |
| Q39 | 4 | 0.058 | 0.54 | | |
| Q44 | 4 | -0.163 | 0.01 | ** | Human < COVID Memories |
| Q49 | 4 | -0.12 | 0.232 | | |
| Q54 | 4 | -0.264 | 0.01 | ** | Human < COVID Memories |
| Q59 | 4 | -0.015 | 0.859 | | |
| Q5 | 5 | -0.024 | 0.763 | | |
| Q10 | 5 | 0.028 | 0.743 | | |
| Q15 | 5 | -0.262 | 0.01 | ** | Human < COVID Memories |
| Q20 | 5 | -0.225 | 0.01 | ** | Human < COVID Memories |
| Q25 | 5 | -0.044 | 0.593 | | |
| Q30 | 5 | -0.07 | 0.326 | | |
| Q35 | 5 | 0.055 | 0.413 | | |
| Q40 | 5 | -0.053 | 0.335 | | |
| Q45 | 5 | -0.094 | 0.117 | | |
| Q50 | 5 | 0.093 | 0.17 | | |
| Q55 | 5 | 0.11 | 0.135 | | |
| Q60 | 5 | -0.172 | 0.01 | ** | Human < COVID Memories |
| Demographics + COVID Memories vs Human | | | | | |
| Q1 | 1 | 0.085 | 0.156 | | |
| Q6 | 1 | -0.098 | 0.102 | | |
| Q11 | 1 | 0.016 | 0.832 | | |
| Q16 | 1 | 0.085 | 0.176 | | |
| Q21 | 1 | 0.1 | 0.102 | | |
| Q26 | 1 | -0.072 | 0.212 | | |
| Q31 | 1 | 0.011 | 0.912 | | |
| Q36 | 1 | -0.06 | 0.187 | | |
| Q41 | 1 | -0.007 | 0.947 | | |
| Q46 | 1 | 0.021 | 0.832 | | |
| Q51 | 1 | 0.138 | 0.023 | * | Demographics + COVID Memories > Human |
| Q56 | 1 | 0.076 | 0.13 | | |
| Q2 | 2 | -0.054 | 0.498 | | |
| Q17 | 2 | -0.002 | 0.968 | | |
| Q22 | 2 | 0.086 | 0.165 | | |
| Q27 | 2 | 0.169 | 0.013 | * | Demographics + COVID Memories > Human |
| Q32 | 2 | 0.037 | 0.505 | | |
| Q47 | 2 | -0.103 | 0.102 | | |
| Q52 | 2 | 0.113 | 0.102 | | |
| Q57 | 2 | 0.099 | 0.046 | * | Demographics + COVID Memories > Human |
| Q4 | 3 | 0.036 | 0.823 | | |
| Q9 | 3 | 0.018 | 0.853 | | |
| Q14 | 3 | 0.071 | 0.473 | | |
| Q19 | 3 | 0.063 | 0.449 | | |
| Q34 | 3 | 0.019 | 0.832 | | |
| Q39 | 3 | 0.003 | 0.968 | | |
| Q49 | 3 | 0.099 | 0.411 | | |
| Q54 | 3 | 0.052 | 0.505 | | |
| Q59 | 3 | 0.049 | 0.619 | | |
| Q5 | 4 | 0.035 | 0.597 | | |
| Q10 | 4 | 0.104 | 0.102 | | |
| Q15 | 4 | 0.257 | 0.013 | * | Demographics + COVID Memories > Human |
| Q20 | 4 | 0.228 | 0.013 | * | Demographics + COVID Memories > Human |
| Q25 | 4 | 0.058 | 0.449 | | |
| Q30 | 4 | 0.047 | 0.492 | | |
| Q35 | 4 | -0.069 | 0.247 | | |
| Q40 | 4 | 0.018 | 0.825 | | |
| Q45 | 4 | 0.091 | 0.158 | | |
| Q50 | 4 | -0.106 | 0.108 | | |
| Q55 | 4 | -0.156 | 0.041 | * | Demographics + COVID Memories < Human |
| Q60 | 4 | 0.181 | 0.013 | * | Demographics + COVID Memories > Human |
| Q13 | 5 | -0.035 | 0.578 | | |
| Q18 | 5 | 0.271 | 0.013 | * | Demographics + COVID Memories > Human |
| Q23 | 5 | 0.058 | 0.306 | | |
| Q28 | 5 | 0.118 | 0.013 | * | Demographics + COVID Memories > Human |
| Q33 | 5 | 0.175 | 0.013 | * | Demographics + COVID Memories > Human |
| Q38 | 5 | 0.141 | 0.013 | * | Demographics + COVID Memories > Human |
| Q43 | 5 | -0.042 | 0.505 | | |
| Q48 | 5 | 0.076 | 0.288 | | |
| Q53 | 5 | 0.086 | 0.187 | | |
| Q58 | 5 | -0.062 | 0.287 | | |
| Demographics + All psychological measures + COVID Memories vs Human | | | | | |
| Q1 | 1 | 0.095 | 0.052 | | |
| Q6 | 1 | -0.11 | 0.04 | * | Demographics + All psychological measures + COVID Memories < Human |
| Q11 | 1 | -0.05 | 0.348 | | |
| Q16 | 1 | 0.06 | 0.212 | | |
| Q21 | 1 | -0.002 | 0.958 | | |
| Q26 | 1 | -0.095 | 0.071 | | |
| Q31 | 1 | 0.012 | 0.957 | | |
| Q41 | 1 | -0.091 | 0.071 | | |
| Q46 | 1 | -0.084 | 0.071 | | |
| Q51 | 1 | 0.119 | 0.017 | * | Demographics + All psychological measures + COVID Memories > Human |
| Q56 | 1 | -0.008 | 0.957 | | |
| Q2 | 2 | 0.003 | 0.958 | | |
| Q17 | 2 | 0.088 | 0.221 | | |
| Q22 | 2 | 0.116 | 0.066 | | |
| Q27 | 2 | 0.157 | 0.01 | ** | Demographics + All psychological measures + COVID Memories > Human |
| Q57 | 2 | 0.104 | 0.066 | | |
| Q4 | 3 | -0.022 | 0.939 | | |
| Q9 | 3 | -0.004 | 0.958 | | |
| Q14 | 3 | 0.008 | 0.958 | | |
| Q19 | 3 | 0.041 | 0.572 | | |
| Q24 | 3 | 0.092 | 0.071 | | |
| Q29 | 3 | -0.022 | 0.765 | | |
| Q39 | 3 | 0.041 | 0.572 | | |
| Q49 | 3 | 0.103 | 0.212 | | |
| Q54 | 3 | 0.169 | 0.01 | ** | Demographics + All psychological measures + COVID Memories > Human |
| Q59 | 3 | 0.185 | 0.017 | * | Demographics + All psychological measures + COVID Memories > Human |
| Q5 | 4 | 0.033 | 0.565 | | |
| Q10 | 4 | -0.01 | 0.957 | | |
| Q15 | 4 | 0.263 | 0.01 | ** | Demographics + All psychological measures + COVID Memories > Human |
| Q20 | 4 | 0.252 | 0.01 | ** | Demographics + All psychological measures + COVID Memories > Human |
| Q25 | 4 | -0.004 | 0.958 | | |
| Q30 | 4 | 0.044 | 0.566 | | |
| Q35 | 4 | -0.092 | 0.066 | | |
| Q40 | 4 | -0.012 | 0.957 | | |
| Q45 | 4 | 0.075 | 0.168 | | |
| Q50 | 4 | -0.098 | 0.071 | | |
| Q55 | 4 | -0.13 | 0.071 | | |
| Q60 | 4 | 0.167 | 0.01 | ** | Demographics + All psychological measures + COVID Memories > Human |
| Q7 | 5 | 0.187 | 0.01 | ** | Demographics + All psychological measures + COVID Memories > Human |
| Q12 | 5 | 0.141 | 0.066 | | |
| Q37 | 5 | -0.076 | 0.218 | | |
| Q42 | 5 | -0.006 | 0.958 | | |
| Q13 | 6 | -0.003 | 0.958 | | |
| Q18 | 6 | 0.352 | 0.01 | ** | Demographics + All psychological measures + COVID Memories > Human |
| Q23 | 6 | 0.105 | 0.066 | | |
| Q28 | 6 | 0.148 | 0.01 | ** | Demographics + All psychological measures + COVID Memories > Human |
| Q33 | 6 | 0.142 | 0.01 | ** | Demographics + All psychological measures + COVID Memories > Human |
| Q38 | 6 | 0.123 | 0.01 | ** | Demographics + All psychological measures + COVID Memories > Human |
| Q43 | 6 | -0.083 | 0.193 | | |
| Q48 | 6 | 0.022 | 0.863 | | |
| Q53 | 6 | 0.06 | 0.359 | | |
| Q58 | 6 | -0.04 | 0.563 | | |
| Demographics + Features with r < 0.5 + COVID contexts vs Human | | | | | |
| Q1 | 1 | -0.045 | 0.395 | | |
| Q6 | 1 | 0.098 | 0.077 | | |
| Q11 | 1 | -0.074 | 0.226 | | |
| Q16 | 1 | -0.038 | 0.536 | | |
| Q21 | 1 | -0.111 | 0.03 | * | Human < Demographics + Features with r < 0.5 + COVID contexts |
| Q26 | 1 | 0.043 | 0.441 | | |
| Q31 | 1 | -0.003 | 0.974 | | |
| Q41 | 1 | 0.003 | 0.974 | | |
| Q46 | 1 | 0.007 | 0.971 | | |
| Q51 | 1 | -0.173 | 0.011 | * | Human < Demographics + Features with r < 0.5 + COVID contexts |
| Q56 | 1 | -0.054 | 0.237 | | |
| Q2 | 2 | 0.021 | 0.766 | | |
| Q17 | 2 | 0.036 | 0.585 | | |
| Q22 | 2 | -0.144 | 0.011 | * | Human < Demographics + Features with r < 0.5 + COVID contexts |
| Q27 | 2 | -0.16 | 0.03 | * | Human < Demographics + Features with r < 0.5 + COVID contexts |
| Q32 | 2 | 0.005 | 0.971 | | |
| Q47 | 2 | 0.111 | 0.042 | * | Human > Demographics + Features with r < 0.5 + COVID contexts |
| Q52 | 2 | -0.145 | 0.02 | * | Human < Demographics + Features with r < 0.5 + COVID contexts |
| Q57 | 2 | -0.03 | 0.66 | | |
| Q4 | 3 | 0.01 | 0.971 | | |
| Q9 | 3 | -0.028 | 0.739 | | |
| Q14 | 3 | -0.068 | 0.487 | | |
| Q19 | 3 | -0.113 | 0.147 | | |
| Q34 | 3 | -0.051 | 0.554 | | |
| Q39 | 3 | 0.156 | 0.065 | | |
| Q49 | 3 | -0.104 | 0.304 | | |
| Q54 | 3 | -0.215 | 0.011 | * | Human < Demographics + Features with r < 0.5 + COVID contexts |
| Q59 | 3 | -0.081 | 0.355 | | |
| Q5 | 4 | 0.019 | 0.766 | | |
| Q10 | 4 | -0.124 | 0.03 | * | Human < Demographics + Features with r < 0.5 + COVID contexts |
| Q20 | 4 | -0.215 | 0.011 | * | Human < Demographics + Features with r < 0.5 + COVID contexts |
| Q25 | 4 | -0.078 | 0.294 | | |
| Q30 | 4 | -0.026 | 0.681 | | |
| Q35 | 4 | 0.025 | 0.737 | | |
| Q40 | 4 | -0.062 | 0.294 | | |
| Q45 | 4 | -0.06 | 0.294 | | |
| Q50 | 4 | 0.114 | 0.058 | | |
| Q55 | 4 | 0.093 | 0.098 | | |
| Q60 | 4 | -0.197 | 0.011 | * | Human < Demographics + Features with r < 0.5 + COVID contexts |
| Q13 | 5 | 0.025 | 0.721 | | |
| Q18 | 5 | -0.319 | 0.011 | * | Human < Demographics + Features with r < 0.5 + COVID contexts |
| Q23 | 5 | -0.102 | 0.058 | | |
| Q28 | 5 | -0.142 | 0.011 | * | Human < Demographics + Features with r < 0.5 + COVID contexts |
| Q33 | 5 | -0.195 | 0.011 | * | Human < Demographics + Features with r < 0.5 + COVID contexts |
| Q38 | 5 | -0.167 | 0.011 | * | Human < Demographics + Features with r < 0.5 + COVID contexts |
| Q43 | 5 | 0.106 | 0.058 | | |
| Q48 | 5 | -0.002 | 0.976 | | |
| Q53 | 5 | -0.092 | 0.135 | | |
| Q58 | 5 | 0.047 | 0.439 | | |
| Demographics + Features with r > 0.5 + COVID contexts vs Human | | | | | |
| Q1 | 1 | 0.092 | 0.058 | | |
| Q6 | 1 | -0.079 | 0.164 | | |
| Q11 | 1 | -0.018 | 0.726 | | |
| Q16 | 1 | 0.06 | 0.182 | | |
| Q21 | 1 | -0.039 | 0.44 | | |
| Q26 | 1 | -0.105 | 0.047 | * | Demographics + Features with r > 0.5 + COVID contexts < Human |
| Q31 | 1 | 0.053 | 0.486 | | |
| Q41 | 1 | -0.083 | 0.117 | | |
| Q46 | 1 | -0.076 | 0.129 | | |
| Q51 | 1 | 0.103 | 0.058 | | |
| Q56 | 1 | 0.017 | 0.719 | | |
| Q4 | 2 | -0.015 | 0.845 | | |
| Q9 | 2 | -0.037 | 0.603 | | |
| Q14 | 2 | 0.041 | 0.603 | | |
| Q19 | 2 | 0.028 | 0.673 | | |
| Q24 | 2 | 0.118 | 0.031 | * | Demographics + Features with r > 0.5 + COVID contexts > Human |
| Q29 | 2 | -0.018 | 0.722 | | |
| Q39 | 2 | 0.074 | 0.277 | | |
| Q49 | 2 | 0.107 | 0.177 | | |
| Q54 | 2 | 0.149 | 0.018 | * | Demographics + Features with r > 0.5 + COVID contexts > Human |
| Q59 | 2 | 0.158 | 0.031 | * | Demographics + Features with r > 0.5 + COVID contexts > Human |
| Q5 | 3 | 0.019 | 0.722 | | |
| Q10 | 3 | 0.03 | 0.603 | | |
| Q15 | 3 | 0.264 | 0.01 | ** | Demographics + Features with r > 0.5 + COVID contexts > Human |
| Q20 | 3 | 0.216 | 0.01 | ** | Demographics + Features with r > 0.5 + COVID contexts > Human |
| Q25 | 3 | -0.014 | 0.845 | | |
| Q30 | 3 | 0.057 | 0.382 | | |
| Q35 | 3 | -0.089 | 0.086 | | |
| Q40 | 3 | -0.035 | 0.603 | | |
| Q45 | 3 | 0.089 | 0.1 | | |
| Q50 | 3 | -0.077 | 0.183 | | |
| Q55 | 3 | -0.14 | 0.058 | | |
| Q60 | 3 | 0.198 | 0.01 | ** | Demographics + Features with r > 0.5 + COVID contexts > Human |
| Q7 | 4 | 0.197 | 0.01 | ** | Demographics + Features with r > 0.5 + COVID contexts > Human |
| Q12 | 4 | 0.157 | 0.01 | ** | Demographics + Features with r > 0.5 + COVID contexts > Human |
| Q32 | 4 | 0.107 | 0.065 | | |
| Q37 | 4 | -0.216 | 0.01 | ** | Demographics + Features with r > 0.5 + COVID contexts < Human |
| Q42 | 4 | 0.003 | 0.96 | | |
| Q47 | 4 | -0.114 | 0.01 | ** | Demographics + Features with r > 0.5 + COVID contexts < Human |
| Q52 | 4 | 0.061 | 0.312 | | |
| Q13 | 5 | -0.02 | 0.726 | | |
| Q18 | 5 | 0.331 | 0.01 | ** | Demographics + Features with r > 0.5 + COVID contexts > Human |
| Q23 | 5 | 0.108 | 0.056 | | |
| Q28 | 5 | 0.164 | 0.01 | ** | Demographics + Features with r > 0.5 + COVID contexts > Human |
| Q33 | 5 | 0.19 | 0.01 | ** | Demographics + Features with r > 0.5 + COVID contexts > Human |
| Q38 | 5 | 0.099 | 0.036 | * | Demographics + Features with r > 0.5 + COVID contexts > Human |
| Q43 | 5 | -0.059 | 0.379 | | |
| Q48 | 5 | 0.027 | 0.719 | | |
| Q53 | 5 | 0.072 | 0.26 | | |
| Q58 | 5 | -0.07 | 0.182 | | |
| Demographics + COVID Memories + COVID contexts vs Human | | | | | |
| Q1 | 1 | 0.055 | 0.331 | | |
| Q6 | 1 | -0.101 | 0.07 | | |
| Q11 | 1 | 0.052 | 0.435 | | |
| Q16 | 1 | 0.087 | 0.167 | | |
| Q21 | 1 | 0.1 | 0.07 | . | |
| Q26 | 1 | -0.059 | 0.326 | | |
| Q31 | 1 | 0.015 | 0.859 | | |
| Q36 | 1 | -0.056 | 0.233 | | |
| Q41 | 1 | -0.016 | 0.837 | | |
| Q46 | 1 | 0.011 | 0.859 | | |
| Q51 | 1 | 0.139 | 0.026 | * | Demographics + COVID Memories + COVID contexts > Human |
| Q56 | 1 | 0.041 | 0.435 | | |
| Q2 | 2 | -0.028 | 0.709 | | |
| Q17 | 2 | -0.044 | 0.495 | | |
| Q22 | 2 | 0.091 | 0.167 | | |
| Q27 | 2 | 0.203 | 0.015 | * | Demographics + COVID Memories + COVID contexts > Human |
| Q32 | 2 | 0.044 | 0.423 | | |
| Q47 | 2 | -0.112 | 0.07 | | |
| Q52 | 2 | 0.068 | 0.331 | | |
| Q57 | 2 | 0.092 | 0.07 | | |
| Q4 | 3 | 0.03 | 0.825 | | |
| Q9 | 3 | 0.073 | 0.354 | | |
| Q14 | 3 | -0.055 | 0.529 | | |
| Q19 | 3 | 0.073 | 0.354 | | |
| Q34 | 3 | -0.015 | 0.837 | | |
| Q39 | 3 | -0.066 | 0.435 | | |
| Q49 | 3 | 0.134 | 0.233 | | |
| Q54 | 3 | 0.122 | 0.13 | | |
| Q59 | 3 | 0.136 | 0.14 | | |
| Q5 | 4 | 0.038 | 0.517 | | |
| Q10 | 4 | 0.086 | 0.167 | | |
| Q15 | 4 | 0.26 | 0.015 | * | Demographics + COVID Memories + COVID contexts > Human |
| Q20 | 4 | 0.225 | 0.015 | * | Demographics + COVID Memories + COVID contexts > Human |
| Q25 | 4 | 0.071 | 0.334 | | |
| Q30 | 4 | 0.03 | 0.61 | | |
| Q35 | 4 | -0.055 | 0.396 | | |
| Q40 | 4 | 0.041 | 0.469 | | |
| Q45 | 4 | 0.046 | 0.447 | | |
| Q50 | 4 | -0.116 | 0.07 | | |
| Q55 | 4 | -0.153 | 0.045 | * | Demographics + COVID Memories + COVID contexts < Human |
| Q60 | 4 | 0.196 | 0.015 | * | Demographics + COVID Memories + COVID contexts > Human |
| Q13 | 5 | -0.06 | 0.354 | | |
| Q18 | 5 | 0.32 | 0.015 | * | Demographics + COVID Memories + COVID contexts > Human |
| Q23 | 5 | 0.097 | 0.07 | | |
| Q28 | 5 | 0.08 | 0.07 | | |
| Q33 | 5 | 0.231 | 0.015 | * | Demographics + COVID Memories + COVID contexts > Human |
| Q38 | 5 | 0.126 | 0.015 | * | Demographics + COVID Memories + COVID contexts > Human |
| Q43 | 5 | -0.054 | 0.396 | | |
| Q48 | 5 | 0.005 | 0.924 | | |
| Q53 | 5 | 0.123 | 0.078 | | |
| Q58 | 5 | -0.067 | 0.233 | | |
| **Panel B. gpt-oss-120b** | | | | | |
| **Item** | **Membership** | **Difference** | ***p*** | **Sig** | **Direction** |
| Demographics vs Human | | | | | |
| Q1 | 1 | -0.01 | 0.924 | | |
| Q6 | 1 | 0.009 | 0.924 | | |
| Q11 | 1 | -0.049 | 0.488 | | |
| Q16 | 1 | -0.108 | 0.083 | | |
| Q21 | 1 | 0.144 | 0.008 | ** | Demographics > Human |
| Q26 | 1 | -0.019 | 0.853 | | |
| Q31 | 1 | 0.066 | 0.488 | | |
| Q36 | 1 | 0.069 | 0.133 | | |
| Q41 | 1 | -0.002 | 0.978 | | |
| Q46 | 1 | -0.067 | 0.387 | | |
| Q51 | 1 | 0.19 | 0.008 | ** | Demographics > Human |
| Q56 | 1 | 0.089 | 0.082 | | |
| Q2 | 2 | -0.127 | 0.014 | * | Demographics < Human |
| Q7 | 2 | -0.043 | 0.488 | | |
| Q12 | 2 | 0.182 | 0.008 | ** | Demographics > Human |
| Q17 | 2 | -0.033 | 0.635 | | |
| Q22 | 2 | 0.067 | 0.261 | | |
| Q27 | 2 | 0.277 | 0.008 | ** | Demographics > Human |
| Q32 | 2 | 0.024 | 0.692 | | |
| Q37 | 2 | -0.055 | 0.376 | | |
| Q42 | 2 | 0.145 | 0.027 | * | Demographics > Human |
| Q47 | 2 | -0.218 | 0.008 | ** | Demographics < Human |
| Q52 | 2 | 0.155 | 0.014 | * | Demographics > Human |
| Q57 | 2 | 0.041 | 0.537 | | |
| Q4 | 3 | -0.227 | 0.008 | ** | Demographics < Human |
| Q9 | 3 | 0.002 | 0.978 | | |
| Q14 | 3 | 0.107 | 0.376 | | |
| Q19 | 3 | 0.11 | 0.25 | | |
| Q34 | 3 | -0.038 | 0.687 | | |
| Q39 | 3 | -0.106 | 0.348 | | |
| Q49 | 3 | -0.078 | 0.488 | | |
| Q54 | 3 | 0.209 | 0.008 | ** | Demographics > Human |
| Q59 | 3 | 0.015 | 0.924 | | |
| Q5 | 4 | -0.084 | 0.084 | | |
| Q10 | 4 | 0.101 | 0.13 | | |
| Q15 | 4 | 0.16 | 0.008 | ** | Demographics > Human |

| Q25 | 4 | 0.065 | 0.376 | | |
|---|---|---|---|---|---|
| Q30 | 4 | 0.015 | 0.924 | | |
| Q35 | 4 | -0.149 | 0.008 | ** | Demographics < Human |
| Q40 | 4 | -0.105 | 0.038 | * | Demographics < Human |
| Q45 | 4 | -0.066 | 0.354 | | |
| Q50 | 4 | 0.006 | 0.973 | | |
| Q55 | 4 | -0.007 | 0.973 | | |
| Q60 | 4 | 0.262 | 0.008 | ** | Demographics > Human |
| Q13 | 5 | -0.144 | 0.008 | ** | Demographics < Human |
| Q23 | 5 | 0.069 | 0.329 | | |
| Q43 | 5 | -0.005 | 0.973 | | |
| Q48 | 5 | 0.039 | 0.597 | | |
| Q53 | 5 | 0.182 | 0.008 | ** | Demographics > Human |
| Q58 | 5 | 0.01 | 0.924 | | |
| COVID Memories vs Human | | | | | |
| Q1 | 1 | -0.017 | 0.794 | | |
| Q6 | 1 | 0.106 | 0.089 | | |
| Q11 | 1 | -0.031 | 0.628 | | |
| Q16 | 1 | 0.019 | 0.768 | | |
| Q21 | 1 | -0.081 | 0.204 | | |
| Q26 | 1 | 0.025 | 0.734 | | |
| Q31 | 1 | 0.023 | 0.792 | | |
| Q41 | 1 | -0.023 | 0.738 | | |
| Q46 | 1 | -0.008 | 0.928 | | |
| Q51 | 1 | -0.171 | 0.015 | * | Human < COVID Memories |
| Q56 | 1 | 0.044 | 0.422 | | |
| Q2 | 2 | 0.033 | 0.641 | | |
| Q7 | 2 | -0.091 | 0.091 | | |
| Q12 | 2 | -0.187 | 0.015 | * | Human < COVID Memories |
| Q17 | 2 | 0.011 | 0.833 | | |
| Q22 | 2 | -0.068 | 0.245 | | |
| Q27 | 2 | -0.117 | 0.043 | * | Human < COVID Memories |
| Q32 | 2 | 0.024 | 0.738 | | |
| Q37 | 2 | 0.087 | 0.089 | | |
| Q42 | 2 | -0.047 | 0.573 | | |
| Q47 | 2 | 0.143 | 0.026 | * | Human > COVID Memories |
| Q52 | 2 | -0.106 | 0.09 | | |
| Q57 | 2 | -0.049 | 0.507 | | |
| Q4 | 3 | 0.078 | 0.422 | | |
| Q9 | 3 | 0.014 | 0.833 | | |
| Q14 | 3 | -0.057 | 0.573 | | |
| Q19 | 3 | -0.132 | 0.091 | | |
| Q39 | 3 | 0.16 | 0.041 | * | Human > COVID Memories |
| Q44 | 3 | -0.148 | 0.015 | * | Human < COVID Memories |
| Q49 | 3 | 0.057 | 0.573 | | |
| Q54 | 3 | -0.287 | 0.015 | * | Human < COVID Memories |
| Q59 | 3 | -0.288 | 0.015 | * | Human < COVID Memories |
| Q5 | 4 | 0.053 | 0.336 | | |
| Q10 | 4 | -0.088 | 0.222 | | |
| Q15 | 4 | -0.105 | 0.063 | | |
| Q20 | 4 | -0.127 | 0.034 | * | Human < COVID Memories |
| Q25 | 4 | -0.036 | 0.628 | | |
| Q30 | 4 | -0.06 | 0.336 | | |
| Q35 | 4 | 0.047 | 0.573 | | |
| Q40 | 4 | -0.004 | 0.93 | | |
| Q45 | 4 | 0.079 | 0.168 | | |
| Q50 | 4 | 0.037 | 0.618 | | |
| Q55 | 4 | -0.045 | 0.6 | | |
| Q60 | 4 | -0.148 | 0.043 | * | Human < COVID Memories |
| Q13 | 5 | 0.183 | 0.015 | * | Human > COVID Memories |
| Q23 | 5 | -0.025 | 0.734 | | |
| Q33 | 5 | 0.071 | 0.222 | | |
| Q43 | 5 | -0.022 | 0.768 | | |
| Q48 | 5 | -0.126 | 0.08 | | |
| Q53 | 5 | -0.196 | 0.015 | * | Human < COVID Memories |
| Q58 | 5 | 0.056 | 0.245 | | |
| Demographics + COVID Memories vs Human | | | | | |
| Q1 | 1 | 0.018 | 0.887 | | |
| Q6 | 1 | -0.08 | 0.239 | | |
| Q11 | 1 | -0.033 | 0.677 | | |
| Q16 | 1 | -0.063 | 0.27 | | |
| Q21 | 1 | 0.105 | 0.065 | | |
| Q26 | 1 | -0.016 | 0.887 | | |
| Q31 | 1 | 0.004 | 0.958 | | |
| Q41 | 1 | 0.06 | 0.412 | | |
| Q46 | 1 | -0.027 | 0.858 | | |
| Q51 | 1 | 0.226 | 0.011 | * | Demographics + COVID Memories > Human |
| Q56 | 1 | -0.011 | 0.9 | | |
| Q2 | 2 | -0.017 | 0.887 | | |
| Q7 | 2 | -0.051 | 0.364 | | |
| Q12 | 2 | 0.168 | 0.011 | * | Demographics + COVID Memories > Human |
| Q17 | 2 | -0.009 | 0.906 | | |
| Q22 | 2 | 0.012 | 0.893 | | |
| Q27 | 2 | 0.137 | 0.023 | * | Demographics + COVID Memories > Human |

| Q32 | 2 | 0.015 | 0.887 | | |
|---|---|---|---|---|---|
| Q37 | 2 | -0.078 | 0.192 | | |
| Q42 | 2 | 0.182 | 0.011 | * | Demographics + COVID Memories > Human |
| Q47 | 2 | -0.126 | 0.023 | * | Demographics + COVID Memories < Human |
| Q52 | 2 | 0.151 | 0.018 | * | Demographics + COVID Memories > Human |
| Q57 | 2 | 0.054 | 0.457 | | |
| Q4 | 3 | -0.176 | 0.043 | * | Demographics + COVID Memories < Human |
| Q9 | 3 | 0.063 | 0.573 | | |
| Q14 | 3 | 0.101 | 0.364 | | |
| Q19 | 3 | -0.005 | 0.949 | | |
| Q39 | 3 | -0.183 | 0.033 | * | Demographics + COVID Memories < Human |
| Q49 | 3 | -0.104 | 0.364 | | |
| Q54 | 3 | 0.192 | 0.018 | * | Demographics + COVID Memories > Human |
| Q59 | 3 | 0.325 | 0.011 | * | Demographics + COVID Memories > Human |
| Q5 | 4 | -0.085 | 0.077 | | |
| Q10 | 4 | -0.009 | 0.948 | | |
| Q15 | 4 | 0.184 | 0.011 | * | Demographics + COVID Memories > Human |
| Q25 | 4 | 0.097 | 0.172 | | |
| Q30 | 4 | 0.082 | 0.187 | | |
| Q35 | 4 | -0.151 | 0.011 | * | Demographics + COVID Memories < Human |
| Q40 | 4 | -0.014 | 0.887 | | |
| Q45 | 4 | -0.046 | 0.566 | | |
| Q50 | 4 | -0.043 | 0.659 | | |
| Q55 | 4 | 0.005 | 0.949 | | |
| Q60 | 4 | 0.192 | 0.011 | * | Demographics + COVID Memories > Human |
| Q13 | 5 | -0.197 | 0.011 | * | Demographics + COVID Memories < Human |
| Q23 | 5 | 0.123 | 0.063 | | |
| Q33 | 5 | -0.127 | 0.011 | * | Demographics + COVID Memories < Human |
| Q43 | 5 | 0.014 | 0.887 | | |
| Q48 | 5 | 0.023 | 0.858 | | |
| Q53 | 5 | 0.116 | 0.033 | * | Demographics + COVID Memories > Human |
| Q58 | 5 | 0.052 | 0.364 | | |
| Demographics + Features with r < 0.5 + COVID contexts vs Human | | | | | |
| Q1 | 1 | -0.109 | 0.049 | * | Human < Demographics + Features with r < 0.5 + COVID contexts |
| Q6 | 1 | 0.049 | 0.496 | | |
| Q11 | 1 | 0.049 | 0.428 | | |
| Q16 | 1 | 0.074 | 0.23 | | |
| Q21 | 1 | -0.117 | 0.02 | * | Human < Demographics + Features with r < 0.5 + COVID contexts |
| Q26 | 1 | 0.044 | 0.502 | | |
| Q31 | 1 | 0.075 | 0.307 | | |
| Q41 | 1 | -0.011 | 0.856 | | |
| Q46 | 1 | -0.037 | 0.615 | | |
| Q51 | 1 | -0.282 | 0.02 | * | Human < Demographics + Features with r < 0.5 + COVID contexts |
| Q56 | 1 | -0.026 | 0.691 | | |
| Q2 | 2 | 0.055 | 0.441 | | |
| Q7 | 2 | -0.075 | 0.219 | | |
| Q12 | 2 | -0.104 | 0.098 | | |
| Q17 | 2 | -0.004 | 0.919 | | |
| Q22 | 2 | -0.057 | 0.427 | | |
| Q27 | 2 | -0.228 | 0.02 | * | Human < Demographics + Features with r < 0.5 + COVID contexts |
| Q32 | 2 | 0.025 | 0.708 | | |
| Q37 | 2 | -0.049 | 0.428 | | |
| Q42 | 2 | 0.056 | 0.484 | | |
| Q47 | 2 | 0.076 | 0.223 | | |
| Q52 | 2 | -0.09 | 0.219 | | |
| Q57 | 2 | 0.029 | 0.691 | | |
| Q4 | 3 | 0.223 | 0.02 | * | Human > Demographics + Features with r < 0.5 + COVID contexts |
| Q9 | 3 | -0.034 | 0.753 | | |
| Q14 | 3 | -0.186 | 0.065 | | |
| Q19 | 3 | 0 | 1 | | |
| Q39 | 3 | 0.105 | 0.307 | | |
| Q49 | 3 | 0.061 | 0.646 | | |
| Q54 | 3 | -0.025 | 0.753 | | |
| Q59 | 3 | -0.09 | 0.404 | | |
| Q5 | 4 | 0.171 | 0.02 | * | Human > Demographics + Features with r < 0.5 + COVID contexts |
| Q10 | 4 | -0.05 | 0.529 | | |
| Q15 | 4 | -0.118 | 0.033 | * | Human < Demographics + Features with r < 0.5 + COVID contexts |
| Q20 | 4 | -0.025 | 0.719 | | |
| Q25 | 4 | -0.099 | 0.217 | | |
| Q30 | 4 | -0.096 | 0.163 | | |
| Q35 | 4 | 0.031 | 0.708 | | |
| Q40 | 4 | -0.046 | 0.529 | | |
| Q45 | 4 | 0.086 | 0.196 | | |
| Q50 | 4 | 0.059 | 0.484 | | |
| Q55 | 4 | -0.015 | 0.808 | | |
| Q60 | 4 | -0.178 | 0.042 | * | Human < Demographics + Features with r < 0.5 + COVID contexts |
| Q13 | 5 | 0.096 | 0.098 | | |
| Q23 | 5 | -0.082 | 0.219 | | |
| Q43 | 5 | -0.009 | 0.899 | | |
| Q48 | 5 | -0.079 | 0.23 | | |
| Q53 | 5 | -0.056 | 0.349 | | |
| Q58 | 5 | -0.04 | 0.484 | | |
| Demographics + Features with r > 0.5 + COVID contexts vs Human | | | | | |
| Q1 | 1 | 0.132 | 0.012 | * | Demographics + Features with r > 0.5 + COVID contexts > Human |
| Q6 | 1 | -0.052 | 0.445 | | |
| Q11 | 1 | 0.036 | 0.559 | | |
| Q16 | 1 | 0.039 | 0.533 | | |
| Q21 | 1 | -0.067 | 0.167 | | |
| Q26 | 1 | -0.135 | 0.012 | * | Demographics + Features with r > 0.5 + COVID contexts < Human |
| Q31 | 1 | -0.025 | 0.749 | | |
| Q41 | 1 | -0.025 | 0.717 | | |
| Q46 | 1 | -0.053 | 0.335 | | |
| Q51 | 1 | 0.283 | 0.012 | * | Demographics + Features with r > 0.5 + COVID contexts > Human |
| Q56 | 1 | -0.035 | 0.561 | | |
| Q2 | 2 | -0.028 | 0.708 | | |
| Q7 | 2 | 0.006 | 0.902 | | |
| Q12 | 2 | 0.146 | 0.019 | * | Demographics + Features with r > 0.5 + COVID contexts > Human |
| Q17 | 2 | -0.055 | 0.445 | | |
| Q22 | 2 | 0.052 | 0.445 | | |
| Q27 | 2 | 0.161 | 0.012 | * | Demographics + Features with r > 0.5 + COVID contexts > Human |
| Q32 | 2 | -0.071 | 0.18 | | |
| Q37 | 2 | -0.021 | 0.749 | | |
| Q42 | 2 | 0.047 | 0.609 | | |
| Q47 | 2 | -0.114 | 0.041 | * | Demographics + Features with r > 0.5 + COVID contexts < Human |
| Q52 | 2 | 0.161 | 0.012 | * | Demographics + Features with r > 0.5 + COVID contexts > Human |
| Q57 | 2 | 0.013 | 0.827 | | |
| Q4 | 3 | -0.131 | 0.124 | | |
| Q9 | 3 | -0.082 | 0.338 | | |
| Q14 | 3 | 0.018 | 0.827 | | |
| Q19 | 3 | 0.108 | 0.237 | | |
| Q24 | 3 | 0.074 | 0.335 | | |
| Q29 | 3 | 0.01 | 0.877 | | |
| Q39 | 3 | 0.018 | 0.827 | | |
| Q44 | 3 | 0.215 | 0.012 | * | Demographics + Features with r > 0.5 + COVID contexts > Human |
| Q49 | 3 | -0.043 | 0.708 | | |
| Q54 | 3 | 0.293 | 0.012 | * | Demographics + Features with r > 0.5 + COVID contexts > Human |
| Q59 | 3 | 0.04 | 0.708 | | |
| Q5 | 4 | -0.018 | 0.8 | | |
| Q10 | 4 | 0.026 | 0.717 | | |
| Q15 | 4 | 0.092 | 0.093 | | |
| Q20 | 4 | 0.169 | 0.012 | * | Demographics + Features with r > 0.5 + COVID contexts > Human |
| Q25 | 4 | 0.015 | 0.819 | | |
| Q30 | 4 | 0.051 | 0.445 | | |
| Q35 | 4 | -0.109 | 0.068 | | |
| Q40 | 4 | 0.022 | 0.796 | | |
| Q45 | 4 | 0.047 | 0.523 | | |

| Q50 | 4 | -0.041 | 0.665 | | |
|---|---|---|---|---|---|
| Q55 | 4 | -0.072 | 0.323 | | |
| Q60 | 4 | 0.12 | 0.085 | | |
| Q13 | 5 | -0.224 | 0.012 | * | Demographics + Features with r > 0.5 + COVID contexts < Human |
| Q23 | 5 | 0.062 | 0.323 | | |
| Q33 | 5 | -0.136 | 0.019 | * | Demographics + Features with r > 0.5 + COVID contexts < Human |
| Q43 | 5 | -0.015 | 0.814 | | |
| Q48 | 5 | -0.045 | 0.624 | | |
| Q53 | 5 | 0.129 | 0.035 | * | Demographics + Features with r > 0.5 + COVID contexts > Human |
| Q58 | 5 | 0.096 | 0.124 | | |
| Demographics + COVID Memories + COVID contexts vs Human | | | | | |
| Q2 | 1 | -0.049 | 0.423 | | |
| Q17 | 1 | 0.087 | 0.187 | | |
| Q22 | 1 | 0.099 | 0.115 | | |
| Q27 | 1 | 0.166 | 0.034 | * | Demographics + COVID Memories + COVID contexts > Human |
| Q32 | 1 | -0.069 | 0.197 | | |
| Q47 | 1 | -0.103 | 0.054 | | |
| Q52 | 1 | 0.066 | 0.291 | | |
| Q57 | 1 | -0.028 | 0.646 | | |
| Q4 | 2 | -0.24 | 0.017 | * | Demographics + COVID Memories + COVID contexts < Human |
| Q9 | 2 | -0.101 | 0.265 | | |
| Q14 | 2 | 0.195 | 0.046 | * | Demographics + COVID Memories + COVID contexts > Human |
| Q19 | 2 | 0.011 | 0.874 | | |
| Q39 | 2 | -0.118 | 0.206 | | |
| Q49 | 2 | -0.208 | 0.049 | * | Demographics + COVID Memories + COVID contexts < Human |
| Q54 | 2 | 0.246 | 0.017 | * | Demographics + COVID Memories + COVID contexts > Human |
| Q59 | 2 | 0.184 | 0.068 | | |
| Q5 | 3 | -0.099 | 0.028 | * | Demographics + COVID Memories + COVID contexts < Human |
| Q10 | 3 | 0.067 | 0.291 | | |
| Q15 | 3 | 0.121 | 0.017 | * | Demographics + COVID Memories + COVID contexts > Human |
| Q25 | 3 | 0.104 | 0.107 | | |
| Q30 | 3 | 0.05 | 0.391 | | |
| Q35 | 3 | -0.161 | 0.028 | * | Demographics + COVID Memories + COVID contexts < Human |
| Q40 | 3 | -0.07 | 0.187 | | |
| Q45 | 3 | -0.112 | 0.056 | | |
| Q50 | 3 | -0.024 | 0.729 | | |
| Q55 | 3 | 0.048 | 0.44 | | |
| Q60 | 3 | 0.099 | 0.104 | | |
| Q6 | 4 | -0.061 | 0.291 | | |
| Q16 | 4 | -0.026 | 0.653 | | |
| Q26 | 4 | -0.069 | 0.187 | | |
| Q41 | 4 | 0.053 | 0.325 | | |
| Q46 | 4 | 0.009 | 0.867 | | |
| Q51 | 4 | 0.217 | 0.017 | * | Demographics + COVID Memories + COVID contexts > Human |
| Q56 | 4 | 0.101 | 0.054 | | |
| Q13 | 5 | -0.281 | 0.017 | * | Demographics + COVID Memories + COVID contexts < Human |
| Q23 | 5 | 0.144 | 0.028 | * | Demographics + COVID Memories + COVID contexts > Human |
| Q28 | 5 | 0.007 | 0.867 | | |
| Q33 | 5 | -0.058 | 0.291 | | |
| Q43 | 5 | -0.015 | 0.867 | | |
| Q48 | 5 | 0.187 | 0.028 | * | Demographics + COVID Memories + COVID contexts > Human |
| Q53 | 5 | -0.029 | 0.67 | | |
| Q58 | 5 | 0.055 | 0.291 | | |

**Panel C. DeepSeek V4-Flash**

| **Item** | **Membership** | **Difference** | ***p*** | **Sig** | **Direction** |
|---|---|---|---|---|---|
| Demographics vs Human | | | | | |
| Q1 | 1 | -0.259 | 0.01 | ** | Demographics < Human |
| Q6 | 1 | -0.091 | 0.139 | | |
| Q11 | 1 | 0.015 | 0.812 | | |
| Q16 | 1 | -0.015 | 0.811 | | |
| Q21 | 1 | 0.162 | 0.017 | * | Demographics > Human |
| Q26 | 1 | 0.026 | 0.655 | | |
| Q31 | 1 | 0.036 | 0.656 | | |
| Q36 | 1 | 0.122 | 0.01 | ** | Demographics > Human |
| Q41 | 1 | 0.043 | 0.589 | | |
| Q46 | 1 | 0.017 | 0.825 | | |
| Q51 | 1 | 0.097 | 0.076 | | |
| Q56 | 1 | 0.114 | 0.01 | ** | Demographics > Human |
| Q2 | 2 | -0.136 | 0.01 | ** | Demographics < Human |
| Q7 | 2 | 0.151 | 0.01 | ** | Demographics > Human |
| Q12 | 2 | 0.098 | 0.04 | * | Demographics > Human |
| Q17 | 2 | 0.032 | 0.589 | | |
| Q22 | 2 | 0.061 | 0.25 | | |
| Q27 | 2 | 0.159 | 0.01 | ** | Demographics > Human |
| Q32 | 2 | 0.095 | 0.069 | | |
| Q37 | 2 | -0.101 | 0.035 | * | Demographics < Human |
| Q42 | 2 | 0.13 | 0.03 | * | Demographics > Human |
| Q47 | 2 | 0.028 | 0.655 | | |
| Q52 | 2 | 0.056 | 0.294 | | |
| Q57 | 2 | -0.088 | 0.053 | | |
| Q4 | 3 | -0.104 | 0.172 | | |
| Q9 | 3 | -0.093 | 0.285 | | |
| Q14 | 3 | 0.043 | 0.655 | | |
| Q19 | 3 | -0.09 | 0.294 | | |
| Q24 | 3 | 0.167 | 0.01 | ** | Demographics > Human |
| Q29 | 3 | 0.091 | 0.216 | | |
| Q34 | 3 | -0.017 | 0.812 | | |
| Q39 | 3 | -0.042 | 0.656 | | |
| Q44 | 3 | 0.146 | 0.05 | * | Demographics > Human |
| Q49 | 3 | 0.099 | 0.308 | | |
| Q54 | 3 | -0.06 | 0.439 | | |
| Q59 | 3 | 0.231 | 0.017 | * | Demographics > Human |
| Q5 | 4 | -0.048 | 0.387 | | |
| Q10 | 4 | 0.007 | 0.862 | | |
| Q15 | 4 | 0.271 | 0.01 | ** | Demographics > Human |
| Q20 | 4 | 0.203 | 0.01 | ** | Demographics > Human |
| Q25 | 4 | -0.013 | 0.825 | | |
| Q30 | 4 | 0.105 | 0.056 | | |
| Q35 | 4 | -0.011 | 0.853 | | |
| Q40 | 4 | 0.099 | 0.112 | | |
| Q45 | 4 | -0.065 | 0.262 | | |
| Q50 | 4 | -0.036 | 0.589 | | |
| Q55 | 4 | -0.135 | 0.03 | * | Demographics < Human |
| Q60 | 4 | 0.196 | 0.01 | ** | Demographics > Human |
| Q13 | 5 | 0.023 | 0.766 | | |
| Q23 | 5 | 0.027 | 0.589 | | |
| Q28 | 5 | 0.18 | 0.01 | ** | Demographics > Human |
| Q33 | 5 | 0.059 | 0.262 | | |
| Q43 | 5 | -0.069 | 0.196 | | |
| Q48 | 5 | 0.077 | 0.195 | | |
| Q53 | 5 | 0.088 | 0.131 | | |
| Q58 | 5 | -0.031 | 0.589 | | |
| COVID Memories vs Human | | | | | |
| Q1 | 1 | 0.168 | 0.013 | * | Human > COVID Memories |
| Q6 | 1 | 0.1 | 0.068 | | |
| Q11 | 1 | 0.04 | 0.438 | | |
| Q16 | 1 | -0.088 | 0.128 | | |
| Q21 | 1 | -0.114 | 0.019 | * | Human < COVID Memories |
| Q26 | 1 | -0.039 | 0.438 | | |
| Q31 | 1 | -0.063 | 0.438 | | |
| Q41 | 1 | -0.016 | 0.87 | | |
| Q46 | 1 | -0.039 | 0.673 | | |
| Q51 | 1 | -0.18 | 0.013 | * | Human < COVID Memories |
| Q56 | 1 | -0.04 | 0.361 | | |
| Q2 | 2 | 0.089 | 0.107 | | |

| Q7 | 2 | -0.14 | 0.031 | * | Human < COVID Memories |
|---|---|---|---|---|---|
| Q12 | 2 | -0.095 | 0.039 | * | Human < COVID Memories |
| Q17 | 2 | 0.029 | 0.638 | | |
| Q22 | 2 | -0.066 | 0.239 | | |
| Q27 | 2 | -0.217 | 0.013 | * | Human < COVID Memories |
| Q32 | 2 | -0.067 | 0.225 | | |
| Q37 | 2 | 0.111 | 0.031 | * | Human > COVID Memories |
| Q42 | 2 | -0.137 | 0.032 | * | Human < COVID Memories |
| Q47 | 2 | -0.064 | 0.255 | | |
| Q52 | 2 | -0.034 | 0.597 | | |
| Q57 | 2 | 0.08 | 0.068 | | |
| Q3 | 3 | 0.134 | 0.013 | * | Human > COVID Memories |
| Q13 | 3 | 0.012 | 0.905 | | |
| Q18 | 3 | -0.151 | 0.027 | * | Human < COVID Memories |
| Q23 | 3 | -0.103 | 0.039 | * | Human < COVID Memories |
| Q28 | 3 | -0.084 | 0.068 | | |
| Q33 | 3 | -0.185 | 0.019 | * | Human < COVID Memories |
| Q38 | 3 | 0.013 | 0.905 | | |
| Q43 | 3 | 0.107 | 0.019 | * | Human > COVID Memories |
| Q48 | 3 | -0.005 | 0.942 | | |
| Q53 | 3 | -0.118 | 0.039 | * | Human < COVID Memories |
| Q58 | 3 | -0.055 | 0.284 | | |
| Q4 | 4 | 0.098 | 0.166 | | |
| Q9 | 4 | -0.047 | 0.604 | | |
| Q14 | 4 | 0.01 | 0.942 | | |
| Q19 | 4 | -0.048 | 0.604 | | |
| Q24 | 4 | -0.109 | 0.07 | | |
| Q29 | 4 | -0.032 | 0.676 | | |
| Q34 | 4 | -0.139 | 0.032 | * | Human < COVID Memories |
| Q39 | 4 | 0.006 | 0.942 | | |
| Q44 | 4 | -0.169 | 0.013 | * | Human < COVID Memories |
| Q49 | 4 | -0.145 | 0.103 | | |
| Q54 | 4 | 0.032 | 0.679 | | |
| Q59 | 4 | -0.161 | 0.032 | * | Human < COVID Memories |
| Q5 | 5 | -0.012 | 0.917 | | |
| Q10 | 5 | -0.005 | 0.942 | | |
| Q15 | 5 | -0.267 | 0.013 | * | Human < COVID Memories |
| Q20 | 5 | -0.249 | 0.013 | * | Human < COVID Memories |
| Q25 | 5 | 0.006 | 0.942 | | |
| Q30 | 5 | -0.068 | 0.232 | | |
| Q35 | 5 | -0.005 | 0.942 | | |
| Q40 | 5 | -0.064 | 0.232 | | |
| Q45 | 5 | 0.069 | 0.136 | | |
| Q50 | 5 | 0.14 | 0.013 | * | Human > COVID Memories |
| Q55 | 5 | 0.009 | 0.942 | | |
| Q60 | 5 | -0.231 | 0.013 | * | Human < COVID Memories |
| Demographics + COVID Memories vs Human | | | | | |
| Q1 | 1 | -0.257 | 0.011 | * | Demographics + COVID Memories < Human |
| Q6 | 1 | -0.074 | 0.175 | | |
| Q11 | 1 | -0.013 | 0.798 | | |
| Q16 | 1 | 0.038 | 0.456 | | |
| Q21 | 1 | 0.125 | 0.018 | * | Demographics + COVID Memories > Human |
| Q26 | 1 | 0.005 | 0.951 | | |
| Q31 | 1 | 0.043 | 0.57 | | |
| Q41 | 1 | 0.025 | 0.609 | | |
| Q46 | 1 | 0.066 | 0.403 | | |
| Q51 | 1 | 0.137 | 0.011 | * | Demographics + COVID Memories > Human |
| Q56 | 1 | 0.117 | 0.018 | * | Demographics + COVID Memories > Human |
| Q2 | 2 | -0.149 | 0.011 | * | Demographics + COVID Memories < Human |
| Q7 | 2 | 0.171 | 0.011 | * | Demographics + COVID Memories > Human |
| Q12 | 2 | 0.084 | 0.105 | | |
| Q17 | 2 | 0.004 | 0.951 | | |
| Q22 | 2 | 0.068 | 0.196 | | |
| Q27 | 2 | 0.098 | 0.06 | | |
| Q32 | 2 | 0.065 | 0.196 | | |
| Q37 | 2 | -0.049 | 0.378 | | |
| Q42 | 2 | 0.212 | 0.011 | * | Demographics + COVID Memories > Human |
| Q47 | 2 | 0.073 | 0.195 | | |
| Q52 | 2 | 0.044 | 0.403 | | |
| Q57 | 2 | -0.105 | 0.023 | * | Demographics + COVID Memories < Human |
| Q4 | 3 | -0.115 | 0.105 | | |
| Q9 | 3 | -0.115 | 0.19 | | |
| Q14 | 3 | -0.103 | 0.19 | | |
| Q19 | 3 | -0.121 | 0.122 | | |
| Q24 | 3 | 0.175 | 0.018 | * | Demographics + COVID Memories > Human |
| Q29 | 3 | 0.086 | 0.196 | | |
| Q34 | 3 | -0.008 | 0.951 | | |
| Q39 | 3 | -0.056 | 0.53 | | |
| Q44 | 3 | 0.042 | 0.609 | | |
| Q49 | 3 | 0.156 | 0.081 | | |
| Q54 | 3 | -0.089 | 0.196 | | |
| Q59 | 3 | 0.229 | 0.011 | * | Demographics + COVID Memories > Human |
| Q5 | 4 | -0.068 | 0.196 | | |
| Q10 | 4 | 0.038 | 0.453 | | |
| Q15 | 4 | 0.228 | 0.011 | * | Demographics + COVID Memories > Human |
| Q20 | 4 | 0.218 | 0.011 | * | Demographics + COVID Memories > Human |
| Q25 | 4 | -0.038 | 0.527 | | |
| Q30 | 4 | 0.104 | 0.06 | | |
| Q35 | 4 | -0.003 | 0.966 | | |
| Q40 | 4 | 0.089 | 0.105 | | |
| Q45 | 4 | -0.045 | 0.403 | | |
| Q50 | 4 | -0.057 | 0.32 | | |
| Q55 | 4 | -0.097 | 0.098 | | |
| Q60 | 4 | 0.168 | 0.011 | * | Demographics + COVID Memories > Human |
| Q13 | 5 | 0.042 | 0.49 | | |
| Q18 | 5 | 0.126 | 0.023 | * | Demographics + COVID Memories > Human |
| Q23 | 5 | 0.025 | 0.587 | | |
| Q28 | 5 | 0.083 | 0.076 | | |
| Q33 | 5 | 0.149 | 0.011 | * | Demographics + COVID Memories > Human |
| Q38 | 5 | 0.095 | 0.081 | | |
| Q43 | 5 | -0.077 | 0.114 | | |
| Q48 | 5 | -0.004 | 0.951 | | |
| Q53 | 5 | 0.116 | 0.081 | | |
| Q58 | 5 | -0.021 | 0.715 | | |
| Demographics + All psychological measures + COVID Memories vs Human | | | | | |
| Q1 | 1 | -0.104 | 0.012 | * | Demographics + All psychological measures + COVID Memories < Human |
| Q6 | 1 | -0.04 | 0.446 | | |
| Q11 | 1 | 0.005 | 0.93 | | |
| Q16 | 1 | 0.087 | 0.073 | | |
| Q21 | 1 | 0.034 | 0.454 | | |
| Q26 | 1 | -0.029 | 0.555 | | |
| Q31 | 1 | 0.036 | 0.562 | | |
| Q36 | 1 | 0.142 | 0.012 | * | Demographics + All psychological measures + COVID Memories > Human |
| Q41 | 1 | -0.012 | 0.856 | | |
| Q46 | 1 | 0.039 | 0.555 | | |
| Q51 | 1 | 0.143 | 0.012 | * | Demographics + All psychological measures + COVID Memories > Human |
| Q56 | 1 | 0.045 | 0.33 | | |
| Q2 | 2 | -0.267 | 0.012 | * | Demographics + All psychological measures + COVID Memories < Human |
| Q7 | 2 | 0.14 | 0.022 | * | Demographics + All psychological measures + COVID Memories > Human |
| Q12 | 2 | 0.081 | 0.103 | | |
| Q17 | 2 | -0.026 | 0.61 | | |
| Q22 | 2 | 0.046 | 0.409 | | |
| Q27 | 2 | 0.081 | 0.185 | | |
| Q32 | 2 | 0.063 | 0.304 | | |
| Q37 | 2 | -0.056 | 0.33 | | |
| Q42 | 2 | 0.14 | 0.042 | * | Demographics + All psychological measures + COVID Memories > Human |
| Q47 | 2 | 0.101 | 0.116 | | |
| Q52 | 2 | 0.069 | 0.21 | | |
| Q57 | 2 | -0.061 | 0.305 | | |
| Q4 | 3 | -0.082 | 0.295 | | |
| Q9 | 3 | 0.028 | 0.651 | | |
| Q14 | 3 | 0.127 | 0.127 | | |
| Q19 | 3 | -0.033 | 0.61 | | |
| Q24 | 3 | 0.118 | 0.03 | * | Demographics + All psychological measures + COVID Memories > Human |
| Q29 | 3 | -0.011 | 0.856 | | |
| Q39 | 3 | -0.095 | 0.211 | | |
| Q44 | 3 | 0.058 | 0.33 | | |
| Q49 | 3 | 0.116 | 0.183 | | |
| Q54 | 3 | -0.037 | 0.565 | | |
| Q59 | 3 | 0.267 | 0.012 | * | Demographics + All psychological measures + COVID Memories > Human |
| Q5 | 4 | -0.052 | 0.313 | | |
| Q10 | 4 | -0.007 | 0.882 | | |
| Q15 | 4 | 0.211 | 0.012 | * | Demographics + All psychological measures + COVID Memories > Human |
| Q20 | 4 | 0.187 | 0.012 | * | Demographics + All psychological measures + COVID Memories > Human |
| Q25 | 4 | -0.07 | 0.228 | | |
| Q30 | 4 | 0.113 | 0.047 | * | Demographics + All psychological measures + COVID Memories > Human |
| Q35 | 4 | 0.032 | 0.558 | | |
| Q40 | 4 | 0.067 | 0.244 | | |
| Q45 | 4 | -0.008 | 0.882 | | |
| Q50 | 4 | -0.089 | 0.104 | | |
| Q55 | 4 | -0.126 | 0.037 | * | Demographics + All psychological measures + COVID Memories < Human |
| Q60 | 4 | 0.175 | 0.012 | * | Demographics + All psychological measures + COVID Memories > Human |
| Q13 | 5 | -0.044 | 0.481 | | |
| Q23 | 5 | 0.008 | 0.882 | | |

| | | | | | Demographics + All psychological measures + COVID |
|---|---|---|---|---|---|
| Q28 | 5 | 0.151 | 0.012 | * | Memories > Human |
| Q33 | 5 | 0.073 | 0.16 | | |
| Q43 | 5 | -0.056 | 0.304 | | |
| Q48 | 5 | 0.026 | 0.61 | | |
| Q53 | 5 | 0.073 | 0.211 | | |
| Q58 | 5 | 0.041 | 0.422 | | |
| Demographics + Features with r < 0.5 + COVID contexts vs Human | | | | | |
| Q1 | 1 | 0.154 | 0.01 | ** | Human > Demographics + Features with r < 0.5 + COVID contexts |
| Q6 | 1 | 0.107 | 0.028 | * | Human > Demographics + Features with r < 0.5 + COVID contexts |
| Q11 | 1 | 0.006 | 0.888 | | |
| Q16 | 1 | -0.049 | 0.367 | | |
| Q21 | 1 | -0.139 | 0.01 | ** | Human < Demographics + Features with r < 0.5 + COVID contexts |
| Q26 | 1 | -0.054 | 0.238 | | |
| Q31 | 1 | -0.024 | 0.784 | | |
| Q41 | 1 | 0 | 0.996 | | |
| Q46 | 1 | -0.042 | 0.542 | | |
| Q51 | 1 | -0.205 | 0.01 | ** | Human < Demographics + Features with r < 0.5 + COVID contexts |
| Q56 | 1 | -0.075 | 0.045 | * | Human < Demographics + Features with r < 0.5 + COVID contexts |
| Q2 | 2 | 0.188 | 0.01 | ** | Human > Demographics + Features with r < 0.5 + COVID contexts |
| Q7 | 2 | -0.149 | 0.01 | ** | Human < Demographics + Features with r < 0.5 + COVID contexts |
| Q12 | 2 | -0.121 | 0.026 | * | Human < Demographics + Features with r < 0.5 + COVID contexts |
| Q17 | 2 | -0.026 | 0.643 | | |
| Q22 | 2 | -0.058 | 0.253 | | |
| Q27 | 2 | -0.113 | 0.045 | * | Human < Demographics + Features with r < 0.5 + COVID contexts |
| Q32 | 2 | -0.065 | 0.238 | | |
| Q37 | 2 | 0.084 | 0.072 | | |
| Q42 | 2 | -0.182 | 0.01 | ** | Human < Demographics + Features with r < 0.5 + COVID contexts |
| Q47 | 2 | -0.101 | 0.044 | * | Human < Demographics + Features with r < 0.5 + COVID contexts |
| Q52 | 2 | -0.056 | 0.3 | | |
| Q57 | 2 | 0.072 | 0.152 | | |
| Q4 | 3 | 0.157 | 0.028 | * | Human > Demographics + Features with r < 0.5 + COVID contexts |
| Q9 | 3 | 0.044 | 0.589 | | |
| Q14 | 3 | 0.016 | 0.883 | | |
| Q19 | 3 | 0.15 | 0.04 | * | Human > Demographics + Features with r < 0.5 + COVID contexts |
| Q24 | 3 | -0.096 | 0.068 | | |
| Q29 | 3 | -0.04 | 0.497 | | |
| Q34 | 3 | -0.012 | 0.888 | | |
| Q39 | 3 | 0.023 | 0.821 | | |
| Q49 | 3 | -0.145 | 0.07 | | |
| Q54 | 3 | 0.084 | 0.174 | | |
| Q59 | 3 | -0.176 | 0.019 | * | Human < Demographics + Features with r < 0.5 + COVID contexts |
| Q5 | 4 | 0.055 | 0.336 | | |
| Q10 | 4 | 0.039 | 0.467 | | |
| Q15 | 4 | -0.211 | 0.01 | ** | Human < Demographics + Features with r < 0.5 + COVID contexts |
| Q20 | 4 | -0.215 | 0.01 | ** | Human < Demographics + Features with r < 0.5 + COVID contexts |
| Q25 | 4 | 0.003 | 0.987 | | |
| Q30 | 4 | -0.101 | 0.048 | * | Human < Demographics + Features with r < 0.5 + COVID contexts |
| Q35 | 4 | 0.001 | 0.996 | | |
| Q40 | 4 | -0.08 | 0.124 | | |
| Q45 | 4 | 0.047 | 0.396 | | |
| Q50 | 4 | 0.058 | 0.261 | | |
| Q55 | 4 | 0.054 | 0.367 | | |
| Q60 | 4 | -0.18 | 0.01 | ** | Human < Demographics + Features with r < 0.5 + COVID contexts |
| Q13 | 5 | 0.062 | 0.247 | | |
| Q18 | 5 | -0.091 | 0.099 | | |
| Q23 | 5 | -0.066 | 0.187 | | |
| Q28 | 5 | -0.107 | 0.01 | ** | Human < Demographics + Features with r < 0.5 + COVID contexts |
| Q33 | 5 | -0.1 | 0.056 | | |
| Q38 | 5 | -0.173 | 0.01 | ** | Human < Demographics + Features with r < 0.5 + COVID contexts |
| Q43 | 5 | 0.058 | 0.247 | | |
| Q48 | 5 | -0.039 | 0.545 | | |
| Q53 | 5 | -0.128 | 0.028 | * | Human < Demographics + Features with r < 0.5 + COVID contexts |
| Q58 | 5 | -0.014 | 0.822 | | |
| Demographics + Features with r > 0.5 + COVID contexts vs Human | | | | | |
| Q1 | 1 | -0.106 | 0.01 | ** | Demographics + Features with r > 0.5 + COVID contexts < Human |
| Q6 | 1 | -0.023 | 0.689 | | |
| Q11 | 1 | -0.007 | 0.917 | | |
| Q16 | 1 | 0.091 | 0.041 | * | Demographics + Features with r > 0.5 + COVID contexts > Human |
| Q21 | 1 | 0.057 | 0.141 | | |
| Q26 | 1 | -0.019 | 0.699 | | |
| Q31 | 1 | 0.034 | 0.6 | | |
| Q36 | 1 | 0.113 | 0.01 | ** | Demographics + Features with r > 0.5 + COVID contexts > Human |
| Q41 | 1 | -0.042 | 0.6 | | |
| Q46 | 1 | 0.039 | 0.6 | | |
| Q51 | 1 | 0.162 | 0.01 | ** | Demographics + Features with r > 0.5 + COVID contexts > Human |
| Q56 | 1 | 0.079 | 0.067 | | |
| Q2 | 2 | -0.249 | 0.01 | ** | Demographics + Features with r > 0.5 + COVID contexts < Human |
| Q7 | 2 | 0.231 | 0.01 | ** | Demographics + Features with r > 0.5 + COVID contexts > Human |
| Q12 | 2 | 0.067 | 0.175 | | |
| Q17 | 2 | -0.047 | 0.408 | | |
| Q22 | 2 | 0.093 | 0.084 | | |
| Q27 | 2 | 0.078 | 0.152 | | |
| Q32 | 2 | 0.06 | 0.286 | | |
| Q37 | 2 | -0.086 | 0.081 | | |
| Q42 | 2 | 0.176 | 0.01 | ** | Demographics + Features with r > 0.5 + COVID contexts > Human |
| Q47 | 2 | 0.047 | 0.461 | | |
| Q52 | 2 | 0.085 | 0.105 | | |
| Q57 | 2 | -0.054 | 0.33 | | |
| Q4 | 3 | -0.108 | 0.124 | | |
| Q9 | 3 | -0.005 | 0.927 | | |
| Q14 | 3 | 0.075 | 0.412 | | |
| Q19 | 3 | -0.031 | 0.697 | | |
| Q24 | 3 | 0.113 | 0.041 | * | Demographics + Features with r > 0.5 + COVID contexts > Human |
| Q29 | 3 | -0.037 | 0.586 | | |
| Q39 | 3 | -0.044 | 0.6 | | |
| Q44 | 3 | 0.07 | 0.258 | | |
| Q49 | 3 | 0.14 | 0.11 | | |
| Q54 | 3 | 0.036 | 0.6 | | |
| Q59 | 3 | 0.25 | 0.01 | ** | Demographics + Features with r > 0.5 + COVID contexts > Human |
| Q5 | 4 | -0.025 | 0.65 | | |
| Q10 | 4 | 0.049 | 0.33 | | |
| Q15 | 4 | 0.184 | 0.01 | ** | Demographics + Features with r > 0.5 + COVID contexts > Human |
| Q20 | 4 | 0.192 | 0.01 | ** | Demographics + Features with r > 0.5 + COVID contexts > Human |
| Q25 | 4 | -0.016 | 0.831 | | |
| Q30 | 4 | 0.095 | 0.067 | | |
| Q35 | 4 | -0.005 | 0.927 | | |
| Q40 | 4 | 0.032 | 0.6 | | |
| Q45 | 4 | -0.054 | 0.33 | | |
| Q50 | 4 | -0.102 | 0.041 | * | Demographics + Features with r > 0.5 + COVID contexts < Human |
| Q55 | 4 | -0.139 | 0.018 | * | Demographics + Features with r > 0.5 + COVID contexts < Human |
| Q60 | 4 | 0.255 | 0.01 | ** | Demographics + Features with r > 0.5 + COVID contexts > Human |
| Q13 | 5 | -0.074 | 0.258 | | |
| Q23 | 5 | -0.005 | 0.927 | | |
| Q28 | 5 | 0.131 | 0.01 | ** | Demographics + Features with r > 0.5 + COVID contexts > Human |
| Q33 | 5 | 0.111 | 0.025 | * | Demographics + Features with r > 0.5 + COVID contexts > Human |
| Q43 | 5 | -0.003 | 0.948 | | |
| Q48 | 5 | -0.012 | 0.86 | | |
| Q53 | 5 | 0.102 | 0.082 | | |
| Q58 | 5 | 0.03 | 0.6 | | |
| Demographics + COVID Memories + COVID contexts vs Human | | | | | |
| Q1 | 1 | -0.164 | 0.011 | * | Demographics + COVID Memories + COVID contexts < Human |
| Q6 | 1 | -0.102 | 0.04 | * | Demographics + COVID Memories + COVID contexts < Human |
| Q11 | 1 | 0.004 | 0.943 | | |
| Q16 | 1 | 0.034 | 0.575 | | |
| Q21 | 1 | 0.104 | 0.028 | * | Demographics + COVID Memories + COVID contexts > Human |
| Q26 | 1 | 0.039 | 0.426 | | |
| Q31 | 1 | 0.063 | 0.394 | | |
| Q41 | 1 | 0.052 | 0.342 | | |
| Q46 | 1 | 0.043 | 0.575 | | |
| Q51 | 1 | 0.195 | 0.011 | * | Demographics + COVID Memories + COVID contexts > Human |
| Q56 | 1 | 0.019 | 0.776 | | |
| Q2 | 2 | -0.233 | 0.011 | * | Demographics + COVID Memories + COVID contexts < Human |
| Q7 | 2 | 0.157 | 0.011 | * | Demographics + COVID Memories + COVID contexts > Human |
| Q12 | 2 | 0.123 | 0.021 | * | Demographics + COVID Memories + COVID contexts > Human |
| Q17 | 2 | 0.01 | 0.855 | | |
| Q22 | 2 | 0.112 | 0.04 | * | Demographics + COVID Memories + COVID contexts > Human |
| Q27 | 2 | 0.114 | 0.04 | * | Demographics + COVID Memories + COVID contexts > Human |
| Q32 | 2 | 0.029 | 0.65 | | |
| Q37 | 2 | -0.034 | 0.542 | | |
| Q42 | 2 | 0.204 | 0.011 | * | Demographics + COVID Memories + COVID contexts > Human |
| Q47 | 2 | 0.06 | 0.311 | | |
| Q52 | 2 | 0.074 | 0.209 | | |
| Q57 | 2 | -0.066 | 0.178 | | |
| Q4 | 3 | -0.122 | 0.136 | | |
| Q9 | 3 | -0.021 | 0.855 | | |
| Q14 | 3 | -0.024 | 0.817 | | |
| Q19 | 3 | -0.124 | 0.161 | | |
| Q24 | 3 | 0.152 | 0.033 | * | Demographics + COVID Memories + COVID contexts > Human |
| Q29 | 3 | 0.076 | 0.327 | | |
| Q34 | 3 | -0.046 | 0.575 | | |
| Q39 | 3 | -0.01 | 0.906 | | |
| Q44 | 3 | 0.078 | 0.327 | | |
| Q49 | 3 | 0.063 | 0.575 | | |
| Q54 | 3 | -0.043 | 0.615 | | |
| Q59 | 3 | 0.187 | 0.033 | * | Demographics + COVID Memories + COVID contexts > Human |
| Q5 | 4 | -0.088 | 0.131 | | |
| Q10 | 4 | -0.008 | 0.901 | | |
| Q15 | 4 | 0.179 | 0.011 | * | Demographics + COVID Memories + COVID contexts > Human |
| Q20 | 4 | 0.216 | 0.011 | * | Demographics + COVID Memories + COVID contexts > Human |
| Q25 | 4 | -0.017 | 0.837 | | |
| Q30 | 4 | 0.089 | 0.131 | | |
| Q35 | 4 | -0.017 | 0.805 | | |
| Q40 | 4 | 0.065 | 0.305 | | |
| Q45 | 4 | -0.039 | 0.575 | | |
| Q50 | 4 | -0.03 | 0.659 | | |
| Q55 | 4 | -0.046 | 0.542 | | |
| Q60 | 4 | 0.179 | 0.011 | * | Demographics + COVID Memories + COVID contexts > Human |
| Q13 | 5 | -0.02 | 0.805 | | |
| Q18 | 5 | 0.067 | 0.242 | | |
| Q23 | 5 | 0.036 | 0.542 | | |
| Q28 | 5 | 0.121 | 0.011 | * | Demographics + COVID Memories + COVID contexts > Human |
| Q33 | 5 | 0.14 | 0.011 | * | Demographics + COVID Memories + COVID contexts > Human |
| Q38 | 5 | 0.06 | 0.281 | | |
| Q43 | 5 | -0.058 | 0.287 | | |
| Q48 | 5 | 0.011 | 0.901 | | |
| Q53 | 5 | 0.118 | 0.051 | | |
| Q58 | 5 | -0.002 | 0.964 | | |

**Panel D. Qwen 3.5 Plus**

| **Item** | **Membership** | **Difference** | ***p*** | **Sig** | **Direction** |
|---|---|---|---|---|---|
| Demographics vs Human | | | | | |
| Q1 | 1 | -0.06 | 0.209 | | |
| Q6 | 1 | -0.072 | 0.353 | | |
| Q11 | 1 | 0.055 | 0.497 | | |
| Q16 | 1 | 0.062 | 0.336 | | |
| Q21 | 1 | 0.108 | 0.071 | | |
| Q26 | 1 | -0.14 | 0.02 | * | Demographics < Human |
| Q31 | 1 | 0.075 | 0.374 | | |
| Q36 | 1 | 0.085 | 0.081 | | |
| Q41 | 1 | -0.071 | 0.406 | | |
| Q46 | 1 | -0.057 | 0.519 | | |
| Q51 | 1 | 0 | 0.994 | | |
| Q56 | 1 | 0.129 | 0.033 | * | Demographics > Human |
| Q2 | 2 | -0.139 | 0.033 | * | Demographics < Human |
| Q17 | 2 | -0.001 | 0.994 | | |
| Q22 | 2 | 0.07 | 0.324 | | |
| Q27 | 2 | 0.215 | 0.02 | * | Demographics > Human |
| Q32 | 2 | 0.096 | 0.129 | | |
| Q47 | 2 | -0.072 | 0.292 | | |
| Q52 | 2 | 0.067 | 0.336 | | |
| Q57 | 2 | -0.013 | 0.843 | | |
| Q5 | 3 | 0.123 | 0.033 | * | Demographics > Human |
| Q10 | 3 | 0.13 | 0.033 | * | Demographics > Human |
| Q15 | 3 | 0.285 | 0.02 | * | Demographics > Human |
| Q20 | 3 | 0.136 | 0.033 | * | Demographics > Human |
| Q25 | 3 | 0.074 | 0.353 | | |
| Q30 | 3 | -0.101 | 0.083 | | |
| Q35 | 3 | -0.052 | 0.406 | | |
| Q40 | 3 | -0.03 | 0.627 | | |
| Q45 | 3 | -0.055 | 0.394 | | |
| Q50 | 3 | -0.115 | 0.083 | | |
| Q55 | 3 | -0.146 | 0.038 | * | Demographics < Human |
| Q60 | 3 | 0.132 | 0.033 | * | Demographics > Human |
| Q9 | 4 | -0.034 | 0.745 | | |
| Q14 | 4 | -0.276 | 0.02 | * | Demographics < Human |
| Q19 | 4 | -0.045 | 0.627 | | |
| Q24 | 4 | 0.14 | 0.08 | | |
| Q29 | 4 | 0.044 | 0.668 | | |
| Q34 | 4 | -0.046 | 0.611 | | |
| Q39 | 4 | -0.115 | 0.336 | | |
| Q44 | 4 | 0.27 | 0.02 | * | Demographics > Human |
| Q54 | 4 | 0.12 | 0.209 | | |
| Q59 | 4 | 0.212 | 0.033 | * | Demographics > Human |
| Q13 | 5 | -0.024 | 0.782 | | |
| Q23 | 5 | 0.043 | 0.457 | | |
| Q28 | 5 | 0.073 | 0.11 | | |
| Q33 | 5 | -0.008 | 0.894 | | |
| Q43 | 5 | -0.03 | 0.688 | | |
| Q48 | 5 | 0.029 | 0.676 | | |
| Q53 | 5 | 0.108 | 0.1 | | |
| Q58 | 5 | -0.008 | 0.894 | | |
| COVID Memories vs Human | | | | | |
| Q1 | 1 | -0.054 | 0.247 | | |
| Q6 | 1 | 0.037 | 0.526 | | |
| Q11 | 1 | -0.07 | 0.247 | | |
| Q16 | 1 | -0.06 | 0.346 | | |
| Q21 | 1 | -0.103 | 0.06 | | |
| Q26 | 1 | -0.002 | 0.958 | | |
| Q31 | 1 | 0.035 | 0.693 | | |
| Q41 | 1 | 0.053 | 0.346 | | |
| Q46 | 1 | -0.008 | 0.936 | | |
| Q51 | 1 | -0.093 | 0.06 | | |
| Q56 | 1 | -0.053 | 0.247 | | |
| Q2 | 2 | 0.045 | 0.517 | | |
| Q17 | 2 | -0.006 | 0.936 | | |
| Q22 | 2 | -0.109 | 0.094 | | |
| Q27 | 2 | -0.183 | 0.035 | * | Human < COVID Memories |
| Q32 | 2 | -0.057 | 0.346 | | |
| Q47 | 2 | 0.093 | 0.094 | | |
| Q52 | 2 | -0.041 | 0.473 | | |
| Q57 | 2 | -0.122 | 0.035 | * | Human < COVID Memories |
| Q4 | 3 | -0.017 | 0.936 | | |
| Q9 | 3 | 0.031 | 0.733 | | |
| Q14 | 3 | 0.048 | 0.473 | | |
| Q19 | 3 | -0.054 | 0.473 | | |
| Q39 | 3 | 0.103 | 0.151 | | |
| Q44 | 3 | -0.154 | 0.016 | * | Human < COVID Memories |
| Q49 | 3 | -0.096 | 0.346 | | |
| Q54 | 3 | -0.377 | 0.016 | * | Human < COVID Memories |
| Q59 | 3 | -0.216 | 0.016 | * | Human < COVID Memories |
| Q5 | 4 | -0.115 | 0.094 | | |
| Q10 | 4 | -0.094 | 0.122 | | |
| Q15 | 4 | -0.295 | 0.016 | * | Human < COVID Memories |
| Q20 | 4 | -0.178 | 0.016 | * | Human < COVID Memories |
| Q25 | 4 | -0.013 | 0.919 | | |
| Q30 | 4 | -0.005 | 0.936 | | |
| Q35 | 4 | 0.043 | 0.473 | | |
| Q40 | 4 | 0.004 | 0.936 | | |
| Q45 | 4 | -0.083 | 0.151 | | |
| Q50 | 4 | 0.103 | 0.094 | | |
| Q55 | 4 | 0.09 | 0.151 | | |
| Q60 | 4 | -0.116 | 0.06 | | |
| Q13 | 5 | -0.028 | 0.721 | | |
| Q23 | 5 | -0.149 | 0.016 | * | Human < COVID Memories |
| Q33 | 5 | -0.078 | 0.122 | | |
| Q43 | 5 | 0.098 | 0.087 | | |
| Q48 | 5 | -0.109 | 0.094 | | |
| Q53 | 5 | -0.1 | 0.07 | | |
| Q58 | 5 | 0.056 | 0.247 | | |
| Demographics + COVID Memories vs Human | | | | | |
| Q1 | 1 | -0.021 | 0.684 | | |
| Q6 | 1 | -0.065 | 0.299 | | |
| Q11 | 1 | 0.061 | 0.331 | | |
| Q16 | 1 | 0.085 | 0.167 | | |
| Q21 | 1 | 0.124 | 0.021 | * | Demographics + COVID Memories > Human |
| Q26 | 1 | -0.032 | 0.597 | | |
| Q31 | 1 | 0.047 | 0.602 | | |
| Q41 | 1 | -0.049 | 0.422 | | |
| Q46 | 1 | 0.002 | 0.972 | | |
| Q51 | 1 | 0.048 | 0.301 | | |
| Q56 | 1 | 0.065 | 0.221 | | |
| Q2 | 2 | -0.074 | 0.299 | | |
| Q17 | 2 | 0.016 | 0.886 | | |
| Q22 | 2 | 0.057 | 0.415 | | |
| Q27 | 2 | 0.178 | 0.011 | * | Demographics + COVID Memories > Human |
| Q32 | 2 | 0.15 | 0.017 | * | Demographics + COVID Memories > Human |
| Q47 | 2 | -0.107 | 0.056 | | |
| Q52 | 2 | 0.051 | 0.415 | | |
| Q57 | 2 | 0.064 | 0.299 | | |
| Q5 | 3 | 0.096 | 0.056 | | |
| Q10 | 3 | 0.124 | 0.011 | * | Demographics + COVID Memories > Human |
| Q15 | 3 | 0.311 | 0.011 | * | Demographics + COVID Memories > Human |
| Q20 | 3 | 0.186 | 0.011 | * | Demographics + COVID Memories > Human |
| Q25 | 3 | 0.08 | 0.299 | | |
| Q30 | 3 | -0.011 | 0.886 | | |
| Q35 | 3 | -0.066 | 0.275 | | |
| Q40 | 3 | 0.009 | 0.89 | | |
| Q45 | 3 | 0.021 | 0.758 | | |
| Q50 | 3 | -0.154 | 0.011 | * | Demographics + COVID Memories < Human |
| Q55 | 3 | -0.096 | 0.107 | | |
| Q60 | 3 | 0.109 | 0.021 | * | Demographics + COVID Memories > Human |
| Q9 | 4 | 0.019 | 0.886 | | |
| Q14 | 4 | 0.011 | 0.927 | | |
| Q19 | 4 | 0.038 | 0.684 | | |
| Q24 | 4 | 0.075 | 0.301 | | |
| Q29 | 4 | 0.068 | 0.39 | | |
| Q34 | 4 | -0.005 | 0.972 | | |
| Q39 | 4 | -0.096 | 0.301 | | |
| Q44 | 4 | 0.223 | 0.017 | * | Demographics + COVID Memories > Human |
| Q54 | 4 | 0.235 | 0.011 | * | Demographics + COVID Memories > Human |
| Q59 | 4 | 0.203 | 0.021 | * | Demographics + COVID Memories > Human |
| Q13 | 5 | 0.01 | 0.89 | | |
| Q23 | 5 | 0.146 | 0.011 | * | Demographics + COVID Memories > Human |
| Q33 | 5 | 0.1 | 0.056 | | |
| Q43 | 5 | -0.071 | 0.274 | | |
| Q48 | 5 | 0.071 | 0.301 | | |
| Q53 | 5 | 0.149 | 0.011 | * | Demographics + COVID Memories > Human |
| Q58 | 5 | -0.101 | 0.011 | * | Demographics + COVID Memories < Human |
| Demographics + All psychological measures + COVID Memories vs Human | | | | | |
| Q1 | 1 | 0.063 | 0.184 | | |
| Q6 | 1 | -0.028 | 0.675 | | |
| Q11 | 1 | 0.055 | 0.414 | | |
| Q16 | 1 | 0.095 | 0.099 | | |
| Q21 | 1 | 0.023 | 0.675 | | |
| Q26 | 1 | -0.115 | 0.018 | * | Demographics + All psychological measures + COVID Memories < Human |
| Q31 | 1 | -0.022 | 0.787 | | |
| Q41 | 1 | -0.006 | 0.882 | | |
| Q46 | 1 | 0.009 | 0.879 | | |
| Q51 | 1 | 0.053 | 0.322 | | |
| Q56 | 1 | 0.016 | 0.745 | | |
| Q2 | 2 | -0.05 | 0.54 | | |
| Q17 | 2 | 0.04 | 0.649 | | |
| Q22 | 2 | 0.146 | 0.023 | * | Demographics + All psychological measures + COVID Memories > Human |
| Q27 | 2 | 0.172 | 0.018 | * | Demographics + All psychological measures + COVID Memories > Human |
| Q32 | 2 | -0.051 | 0.414 | | |
| Q47 | 2 | -0.054 | 0.422 | | |
| Q52 | 2 | 0.079 | 0.208 | | |
| Q57 | 2 | 0.07 | 0.276 | | |
| Q5 | 3 | 0.093 | 0.099 | | |
| Q10 | 3 | 0.107 | 0.061 | | |
| Q15 | 3 | 0.3 | 0.018 | * | Demographics + All psychological measures + COVID Memories > Human |
| Q20 | 3 | 0.159 | 0.018 | * | Demographics + All psychological measures + COVID Memories > Human |
| Q25 | 3 | 0.042 | 0.601 | | |
| Q30 | 3 | -0.042 | 0.507 | | |
| Q35 | 3 | -0.081 | 0.172 | | |
| Q40 | 3 | 0.028 | 0.675 | | |
| Q45 | 3 | -0.045 | 0.5 | | |
| Q50 | 3 | -0.174 | 0.018 | * | Demographics + All psychological measures + COVID Memories < Human |
| Q55 | 3 | -0.086 | 0.184 | | |
| Q60 | 3 | 0.099 | 0.099 | | |
| Q9 | 4 | -0.02 | 0.79 | | |
| Q14 | 4 | -0.118 | 0.184 | | |
| Q19 | 4 | 0.09 | 0.276 | | |
| Q24 | 4 | 0.151 | 0.074 | | |
| Q29 | 4 | -0.046 | 0.601 | | |
| Q39 | 4 | 0.039 | 0.698 | | |
| Q44 | 4 | 0.237 | 0.023 | * | Demographics + All psychological measures + COVID Memories > Human |
| Q54 | 4 | 0.124 | 0.184 | | |
| Q59 | 4 | 0.119 | 0.184 | | |
| Q13 | 5 | 0.023 | 0.758 | | |
| Q23 | 5 | 0.076 | 0.184 | | |
| Q33 | 5 | 0.133 | 0.023 | * | Demographics + All psychological measures + COVID Memories > Human |
| Q43 | 5 | -0.026 | 0.701 | | |
| Q48 | 5 | 0.111 | 0.147 | | |
| Q58 | 5 | -0.084 | 0.118 | | |
| Demographics + Features with r < 0.5 + COVID contexts vs Human | | | | | |
| Q1 | 1 | -0.009 | 0.871 | | |
| Q6 | 1 | 0.045 | 0.508 | | |
| Q11 | 1 | -0.077 | 0.217 | | |
| Q16 | 1 | -0.084 | 0.188 | | |
| Q21 | 1 | -0.115 | 0.031 | * | Human < Demographics + Features with r < 0.5 + COVID contexts |
| Q26 | 1 | 0.018 | 0.743 | | |
| Q31 | 1 | 0.02 | 0.842 | | |
| Q41 | 1 | 0.033 | 0.663 | | |
| Q46 | 1 | 0.015 | 0.871 | | |
| Q51 | 1 | -0.066 | 0.188 | | |
| Q56 | 1 | -0.082 | 0.101 | | |
| Q2 | 2 | 0.09 | 0.188 | | |
| Q17 | 2 | -0.055 | 0.447 | | |
| Q22 | 2 | -0.054 | 0.444 | | |
| Q27 | 2 | -0.19 | 0.016 | * | Human < Demographics + Features with r < 0.5 + COVID contexts |
| Q32 | 2 | -0.121 | 0.031 | * | Human < Demographics + Features with r < 0.5 + COVID contexts |
| Q47 | 2 | 0.081 | 0.188 | | |
| Q52 | 2 | -0.085 | 0.188 | | |
| Q57 | 2 | -0.001 | 0.982 | | |
| Q5 | 3 | -0.075 | 0.188 | | |
| Q10 | 3 | -0.143 | 0.024 | * | Human < Demographics + Features with r < 0.5 + COVID contexts |
| Q15 | 3 | -0.311 | 0.016 | * | Human < Demographics + Features with r < 0.5 + COVID contexts |
| Q20 | 3 | -0.218 | 0.016 | * | Human < Demographics + Features with r < 0.5 + COVID contexts |
| Q25 | 3 | -0.08 | 0.229 | | |
| Q30 | 3 | 0.056 | 0.279 | | |
| Q35 | 3 | 0.092 | 0.107 | | |
| Q40 | 3 | -0.023 | 0.743 | | |
| Q45 | 3 | -0.011 | 0.871 | | |
| Q50 | 3 | 0.125 | 0.031 | * | Human > Demographics + Features with r < 0.5 + COVID contexts |
| Q55 | 3 | 0.116 | 0.072 | | |
| Q60 | 3 | -0.128 | 0.024 | * | Human < Demographics + Features with r < 0.5 + COVID contexts |
| Q9 | 4 | -0.088 | 0.279 | | |
| Q14 | 4 | 0.039 | 0.737 | | |
| Q19 | 4 | -0.009 | 0.909 | | |
| Q24 | 4 | -0.05 | 0.498 | | |
| Q29 | 4 | -0.084 | 0.275 | | |
| Q39 | 4 | 0.097 | 0.275 | | |
| Q44 | 4 | -0.213 | 0.016 | * | Human < Demographics + Features with r < 0.5 + COVID contexts |
| Q54 | 4 | -0.27 | 0.016 | * | Human < Demographics + Features with r < 0.5 + COVID contexts |
| Q59 | 4 | -0.121 | 0.141 | | |
| Q13 | 5 | -0.021 | 0.812 | | |
| Q23 | 5 | -0.069 | 0.219 | | |
| Q33 | 5 | -0.046 | 0.443 | | |
| Q43 | 5 | 0.029 | 0.714 | | |
| Q48 | 5 | -0.07 | 0.285 | | |
| Q53 | 5 | -0.173 | 0.016 | * | Human < Demographics + Features with r < 0.5 + COVID contexts |
| Q58 | 5 | 0.105 | 0.031 | * | Human > Demographics + Features with r < 0.5 + COVID contexts |
| Demographics + Features with r > 0.5 + COVID contexts vs Human | | | | | |
| Q1 | 1 | 0.067 | 0.174 | | |
| Q6 | 1 | -0.027 | 0.63 | | |
| Q11 | 1 | 0.035 | 0.574 | | |
| Q16 | 1 | 0.088 | 0.126 | | |
| Q21 | 1 | 0.027 | 0.609 | | |
| Q26 | 1 | -0.096 | 0.075 | | |
| Q31 | 1 | -0.027 | 0.736 | | |
| Q41 | 1 | -0.016 | 0.738 | | |
| Q46 | 1 | -0.017 | 0.773 | | |
| Q51 | 1 | 0.05 | 0.357 | | |
| Q56 | 1 | 0.044 | 0.365 | | |
| Q2 | 2 | -0.057 | 0.436 | | |
| Q17 | 2 | 0.069 | 0.365 | | |
| Q22 | 2 | 0.15 | 0.015 | * | Demographics + Features with r > 0.5 + COVID contexts > Human |
| Q27 | 2 | 0.185 | 0.015 | * | Demographics + Features with r > 0.5 + COVID contexts > Human |
| Q32 | 2 | -0.049 | 0.429 | | |
| Q47 | 2 | -0.073 | 0.289 | | |
| Q52 | 2 | 0.071 | 0.289 | | |
| Q57 | 2 | 0.037 | 0.522 | | |
| Q5 | 3 | 0.09 | 0.084 | | |
| Q10 | 3 | 0.115 | 0.031 | * | Demographics + Features with r > 0.5 + COVID contexts > Human |
| Q15 | 3 | 0.287 | 0.015 | * | Demographics + Features with r > 0.5 + COVID contexts > Human |
| Q20 | 3 | 0.177 | 0.015 | * | Demographics + Features with r > 0.5 + COVID contexts > Human |
| Q25 | 3 | 0.045 | 0.511 | | |
| Q30 | 3 | -0.057 | 0.357 | | |
| Q35 | 3 | -0.09 | 0.104 | | |
| Q40 | 3 | 0.042 | 0.443 | | |
| Q45 | 3 | -0.062 | 0.345 | | |
| Q50 | 3 | -0.155 | 0.015 | * | Demographics + Features with r > 0.5 + COVID contexts < Human |
| Q55 | 3 | -0.104 | 0.104 | | |
| Q60 | 3 | 0.096 | 0.104 | | |
| Q9 | 4 | 0.017 | 0.796 | | |
| Q14 | 4 | -0.032 | 0.736 | | |
| Q19 | 4 | 0.024 | 0.736 | | |
| Q24 | 4 | 0.154 | 0.046 | * | Demographics + Features with r > 0.5 + COVID contexts > Human |
| Q29 | 4 | -0.054 | 0.463 | | |
| Q39 | 4 | -0.069 | 0.463 | | |
| Q44 | 4 | 0.273 | 0.015 | * | Demographics + Features with r > 0.5 + COVID contexts > Human |
| Q54 | 4 | 0.211 | 0.023 | * | Demographics + Features with r > 0.5 + COVID contexts > Human |
| Q59 | 4 | 0.058 | 0.463 | | |
| Q13 | 5 | 0.072 | 0.365 | | |
| Q23 | 5 | 0.084 | 0.152 | | |
| Q33 | 5 | 0.137 | 0.023 | * | Demographics + Features with r > 0.5 + COVID contexts > Human |
| Q43 | 5 | -0.087 | 0.266 | | |
| Q48 | 5 | 0.078 | 0.3 | | |
| Q58 | 5 | -0.059 | 0.289 | | |
| Demographics + COVID Memories + COVID contexts vs Human | | | | | |
| Q1 | 1 | 0.032 | 0.489 | | |
| Q6 | 1 | -0.021 | 0.784 | | |
| Q11 | 1 | 0.056 | 0.393 | | |
| Q16 | 1 | 0.076 | 0.22 | | |
| Q21 | 1 | 0.095 | 0.096 | | |
| Q26 | 1 | -0.012 | 0.894 | | |
| Q31 | 1 | -0.06 | 0.489 | | |
| Q41 | 1 | -0.04 | 0.525 | | |
| Q46 | 1 | -0.007 | 0.926 | | |
| Q51 | 1 | 0.068 | 0.186 | | |
| Q56 | 1 | 0.065 | 0.22 | | |
| Q2 | 2 | -0.007 | 0.926 | | |
| Q7 | 2 | 0.254 | 0.009 | ** | Demographics + COVID Memories + COVID contexts > Human |
| Q12 | 2 | 0.2 | 0.009 | ** | Demographics + COVID Memories + COVID contexts > Human |
| Q17 | 2 | -0.104 | 0.07 | | |
| Q22 | 2 | 0.101 | 0.118 | | |
| Q27 | 2 | 0.149 | 0.016 | * | Demographics + COVID Memories + COVID contexts > Human |
| Q32 | 2 | 0.116 | 0.058 | | |
| Q37 | 2 | -0.189 | 0.009 | ** | Demographics + COVID Memories + COVID contexts < Human |
| Q42 | 2 | 0.033 | 0.657 | | |
| Q47 | 2 | -0.087 | 0.165 | | |
| Q52 | 2 | 0.05 | 0.411 | | |
| Q57 | 2 | 0.06 | 0.333 | | |
| Q5 | 3 | 0.058 | 0.321 | | |
| Q10 | 3 | 0.157 | 0.009 | ** | Demographics + COVID Memories + COVID contexts > Human |
| Q15 | 3 | 0.313 | 0.009 | ** | Demographics + COVID Memories + COVID contexts > Human |
| Q20 | 3 | 0.193 | 0.009 | ** | Demographics + COVID Memories + COVID contexts > Human |
| Q25 | 3 | 0.062 | 0.393 | | |
| Q30 | 3 | -0.053 | 0.333 | | |
| Q35 | 3 | -0.079 | 0.163 | | |
| Q40 | 3 | 0.012 | 0.894 | | |
| Q45 | 3 | 0.002 | 0.97 | | |
| Q50 | 3 | -0.132 | 0.009 | ** | Demographics + COVID Memories + COVID contexts < Human |
| Q55 | 3 | -0.084 | 0.186 | | |
| Q60 | 3 | 0.138 | 0.009 | ** | Demographics + COVID Memories + COVID contexts > Human |
| Q9 | 4 | 0.042 | 0.648 | | |
| Q14 | 4 | -0.085 | 0.336 | | |
| Q19 | 4 | -0.009 | 0.926 | | |
| Q24 | 4 | 0.051 | 0.518 | | |
| Q29 | 4 | 0.04 | 0.647 | | |
| Q39 | 4 | -0.01 | 0.926 | | |
| Q44 | 4 | 0.268 | 0.009 | ** | Demographics + COVID Memories + COVID contexts > Human |
| Q54 | 4 | 0.255 | 0.009 | ** | Demographics + COVID Memories + COVID contexts > Human |
| Q59 | 4 | 0.141 | 0.118 | | |
| Q13 | 5 | 0.051 | 0.475 | | |
| Q23 | 5 | 0.039 | 0.485 | | |
| Q33 | 5 | 0.071 | 0.196 | | |
| Q43 | 5 | -0.063 | 0.333 | | |
| Q48 | 5 | 0.125 | 0.07 | | |
| Q53 | 5 | 0.141 | 0.016 | * | Demographics + COVID Memories + COVID contexts > Human |
| Q58 | 5 | -0.104 | 0.009 | ** | Demographics + COVID Memories + COVID contexts < Human |

*Note.* The p reports Benjamini–Hochberg adjusted p values. *** $p < .001$, ** $p < .01$, and * $p < .05$. Direction labels are expressed from the human-reference perspective; the original difference values were not reversed.

**Table S53 Metric invariance results for NEO Five-Factor Personality Inventory item loadings comparing digital twins with human participants in Study 3b**

| Input conditions | Items retained | Number of invariant items | Number of noninvariant items | Digital twin > Human | Digital twin < Human | Agreeableness | Conscientiousness | Extraversion | Neuroticism | Openness |
|---|---|---|---|---|---|---|---|---|---|---|
| Gemma-4-31B-it | | | | | | | | | | |
| Demographics | 51 | 36 (60.00) | 15 (25.00) | 14 (23.33) | 1 (1.67) | 54R | 10, 15R, 20, 60 | 32, 47, 57R | 1R, 36, 51 | 18R, 28, 33R, 38R |
| COVID Memories | 52 | 42 (70.00) | 10 (16.67) | 9 (15.00) | 1 (1.67) | 44R, 54R | 15R, 20, 60 | 27R | 6, 51 | 18R, 38R |
| Demographics + COVID Memories | 51 | 40 (66.67) | 11 (18.33) | 10 (16.67) | 1 (1.67) | | 15R, 20, 55R, 60 | 27R, 57R | 51 | 18R, 28, 33R, 38R |
| Demographics + All psychological measures + COVID Memories | 52 | 39 (65.00) | 13 (21.67) | 12 (20.00) | 1 (1.67) | 54R, 59R | 15R, 20, 60 | 7, 27R | 6, 51 | 18R, 28, 33R, 38R |
| Demographics + Features with $r < 0.5$ + COVID contexts | 49 | 35 (58.33) | 14 (23.33) | 13 (21.67) | 1 (1.67) | 54R | 10, 20, 60 | 22, 27R, 47, 52 | 21, 51 | 18R, 28, 33R, 38R |
| Demographics + Features with $r > 0.5$ + COVID contexts | 50 | 35 (58.33) | 15 (25.00) | 12 (20.00) | 3 (5.00) | 24R, 54R, 59R | 15R, 20, 60 | 7, 12R, 37, 47 | 26 | 18R, 28, 33R, 38R |
| Demographics + COVID Memories + COVID contexts | 51 | 42 (70.00) | 9 (15.00) | 8 (13.33) | 1 (1.67) | | 15R, 20, 55R, 60 | 27R | 51 | 18R, 33R, 38R |
| gpt-oss-120b | | | | | | | | | | |
| Demographics | 50 | 34 (56.67) | 16 (26.67) | 10 (16.67) | 6 (10.00) | 4, 54R | 15R, 35, 40, 60 | 2, 12R, 27R, 42R, 47, 52 | 21, 51 | 13, 53 |
| COVID Memories | 51 | 39 (65.00) | 12 (20.00) | 9 (15.00) | 3 (5.00) | 39R, 44R, 54R, 59R | 20, 60 | 12R, 27R, 47 | 51 | 13, 53 |
| Demographics + COVID Memories | 49 | 33 (55.00) | 16 (26.67) | 10 (16.67) | 6 (10.00) | 4, 39R, 54R, 59R | 15R, 35, 60 | 12R, 27R, 42R, 47, 52 | 51 | 13, 33R, 53 |
| Demographics + Features with $r < 0.5$ + COVID contexts | 49 | 41 (68.33) | 8 (13.33) | 6 (10.00) | 2 (3.33) | 4 | 5, 15R, 60 | 27R | 1R, 21, 51 | |
| Demographics + Features with $r > 0.5$ + COVID contexts | 53 | 40 (66.67) | 13 (21.67) | 9 (15.00) | 4 (6.67) | 44R, 54R | 20 | 12R, 27R, 47, 52 | 1R, 26, 51 | 13, 33R, 53 |
| Demographics + COVID Memories + COVID contexts | 42 | 30 (50.00) | 12 (20.00) | 7 (11.67) | 5 (8.33) | 4, 14R, 49, 54R | 5, 15R, 35 | 27R | 51 | 13, 23R, 48R |
| DeepSeek V4-Flash | | | | | | | | | | |
| Demographics | 56 | 38 (63.33) | 18 (30.00) | 14 (23.33) | 4 (6.67) | 24R, 44R, 59R | 15R, 20, 55R, 60 | 2, 7, 12R, 27R, 37, 42R | 1R, 21, 36, 56 | 28 |
| COVID Memories | 58 | 37 (61.67) | 21 (35.00) | 16 (26.67) | 5 (8.33) | 34, 44R, 59R | 15R, 20, 50, 60 | 7, 12R, 27R, 37, 42R | 1R, 21, 51 | 3R, 18R, 23R, 33R, 43, 53 |
| Demographics + COVID Memories | 57 | 42 (70.00) | 15 (25.00) | 12 (20.00) | 3 (5.00) | 24R, 59R | 15R, 20, 60 | 2, 7, 42R, 57R | 1R, 21, 51, 56 | 18R, 33R |
| Demographics + All psychological measures + COVID Memories | 55 | 41 (68.33) | 14 (23.33) | 11 (18.33) | 3 (5.00) | 24R, 59R | 15R, 20, 30R, 55R, 60 | 2, 7, 42R | 1R, 36, 51 | 28 |
| Demographics + Features with r < 0.5 + COVID contexts | 56 | 35 (58.33) | 21 (35.00) | 16 (26.67) | 5 (8.33) | 4, 19, 59R | 15R, 20, 30R, 60 | 2, 7, 12R, 27R, 42R, 47 | 1R, 6, 21, 51, 56 | 28, 38R, 53 |
| Demographics + Features with r > 0.5 + COVID contexts | 55 | 39 (65.00) | 16 (26.67) | 12 (20.00) | 4 (6.67) | 24R, 59R | 15R, 20, 50, 55R, 60 | 2, 7, 42R | 1R, 16R, 36, 51 | 28, 33R |
| Demographics + COVID Memories + COVID contexts | 57 | 40 (66.67) | 17 (28.33) | 14 (23.33) | 3 (5.00) | 24R, 59R | 15R, 20, 60 | 2, 7, 12R, 22, 27R, 42R | 1R, 6, 21, 51 | 28, 33R |
| Qwen 3.5 Plus | | | | | | | | | | |
| Demographics | 50 | 37 (61.67) | 13 (21.67) | 9 (15.00) | 4 (6.67) | 14R, 44R, 59R | 5, 10, 15R, 20, 55R, 60 | 2, 27R | 26, 56 | |
| COVID Memories | 47 | 39 (65.00) | 8 (13.33) | 8 (13.33) | 0 (0.00) | 44R, 54R, 59R | 15R, 20 | 27R, 57R | | 23R |
| Demographics + COVID Memories | 48 | 34 (56.67) | 14 (23.33) | 12 (20.00) | 2 (3.33) | 44R, 54R, 59R | 10, 15R, 20, 50, 60 | 27R, 32 | 21 | 23R, 53, 58 |
| Demographics + All psychological measures + COVID Memories | 46 | 38 (63.33) | 8 (13.33) | 6 (10.00) | 2 (3.33) | 44R | 15R, 20, 50 | 22, 27R | 26 | 33R |
| Demographics + Features with r < 0.5 + COVID contexts | 47 | 35 (58.33) | 12 (20.00) | 10 (16.67) | 2 (3.33) | 44R, 54R | 10, 15R, 20, 50, 60 | 27R, 32 | 21 | 53, 58 |
| Demographics + Features with r > 0.5 + COVID contexts | 46 | 36 (60.00) | 10 (16.67) | 9 (15.00) | 1 (1.67) | 24R, 44R, 54R | 10, 15R, 20, 50 | 22, 27R | | 33R |
| Demographics + COVID Memories + COVID contexts | 51 | 38 (63.33) | 13 (21.67) | 10 (16.67) | 3 (5.00) | 44R, 54R | 10, 15R, 20, 50, 60 | 7, 12R, 27R, 37 | | 53, 58 |

*Note.* Items retained indicates the number of items retained after the configural-invariance item-removal procedure. Values are counts, with percentages in parentheses calculated relative to the original 60 items. Invariant items are items that did not show evidence of metric noninvariance between the human and digital-twin networks. Noninvariant items are further classified by the direction of the loading difference: Digital twin > Human indicates higher item loadings in the digital-twin responses than in the human responses, whereas Digital twin < Human indicates lower item loadings in the digital-twin responses. Item-level loading differences are reported in Table S53.

**Table S54 Metric invariance results for NEO Five-Factor Personality Inventory item loadings among digital twins across input conditions in Study 3b**

| Model | Input condition A | Input condition B | Number of noninvariant items, n | A < B, n | A > B, n |
|---|---|---|---|---|---|
| Gemma-4-31B-it | Demographics | COVID Memories | 14 | 8 | 6 |
| | Demographics | Demographics + COVID Memories | 4 | 2 | 2 |
| | Demographics | Demographics + All psychological measures + COVID Memories | 10 | 4 | 6 |
| | Demographics | Demographics + Features with r < 0.5 + COVID contexts | 8 | 6 | 2 |
| | Demographics | Demographics + Features with r > 0.5 + COVID contexts | 5 | 3 | 2 |
| | Demographics | Demographics + COVID Memories + COVID contexts | 8 | 5 | 3 |
| | COVID Memories | Demographics + COVID Memories | 5 | 3 | 2 |
| | COVID Memories | Demographics + All psychological measures + COVID Memories | 13 | 6 | 7 |
| | COVID Memories | Demographics + Features with r < 0.5 + COVID contexts | 4 | 3 | 1 |
| | COVID Memories | Demographics + Features with r > 0.5 + COVID contexts | 11 | 4 | 7 |
| | COVID Memories | Demographics + COVID Memories + COVID contexts | 5 | 3 | 2 |
| | Demographics + COVID Memories | Demographics + All psychological measures + COVID Memories | 7 | 4 | 3 |
| | Demographics + COVID Memories | Demographics + Features with r < 0.5 + COVID contexts | 0 | 0 | 0 |
| | Demographics + COVID Memories | Demographics + Features with r > 0.5 + COVID contexts | 6 | 4 | 2 |
| | Demographics + COVID Memories | Demographics + COVID Memories + COVID contexts | 0 | 0 | 0 |
| | Demographics + All psychological measures + COVID Memories | Demographics + Features with r < 0.5 + COVID contexts | 5 | 2 | 3 |
| | Demographics + All psychological measures + COVID Memories | Demographics + Features with r > 0.5 + COVID contexts | 0 | 0 | 0 |
| | Demographics + All psychological measures + COVID Memories | Demographics + COVID Memories + COVID contexts | 13 | 7 | 6 |
| | Demographics + Features with r < 0.5 + COVID contexts | Demographics + Features with r > 0.5 + COVID contexts | 8 | 2 | 6 |
| | Demographics + Features with r < 0.5 + COVID contexts | Demographics + COVID Memories + COVID contexts | 0 | 0 | 0 |
| | Demographics + Features with r > 0.5 + COVID contexts | Demographics + COVID Memories + COVID contexts | 9 | 6 | 3 |
| gpt-oss-120b | Demographics | COVID Memories | 0 | 0 | 0 |
| | Demographics | Demographics + COVID Memories | 0 | 0 | 0 |
| | Demographics | Demographics + All psychological measures + COVID Memories | 10 | 2 | 8 |
| | Demographics | Demographics + Features with r < 0.5 + COVID contexts | 5 | 3 | 2 |
| | Demographics | Demographics + Features with r > 0.5 + COVID contexts | 4 | 2 | 2 |
| | Demographics | Demographics + COVID Memories + COVID contexts | 0 | 0 | 0 |
| | COVID Memories | Demographics + COVID Memories | 0 | 0 | 0 |
| | COVID Memories | Demographics + All psychological measures + COVID Memories | 3 | 1 | 2 |
| | COVID Memories | Demographics + Features with r < 0.5 + COVID contexts | 5 | 4 | 1 |
| | COVID Memories | Demographics + Features with r > 0.5 + COVID contexts | 0 | 0 | 0 |
| | COVID Memories | Demographics + COVID Memories + COVID contexts | 5 | 4 | 1 |
| | Demographics + COVID Memories | Demographics + All psychological measures + COVID Memories | 5 | 1 | 4 |
| | Demographics + COVID Memories | Demographics + Features with r < 0.5 + COVID contexts | 0 | 0 | 0 |
| | Demographics + COVID Memories | Demographics + Features with r > 0.5 + COVID contexts | 3 | 1 | 2 |
| | Demographics + COVID Memories | Demographics + COVID Memories + COVID contexts | 0 | 0 | 0 |
| | Demographics + All psychological measures + COVID Memories | Demographics + Features with r < 0.5 + COVID contexts | 8 | 6 | 2 |
| | Demographics + All psychological measures + COVID Memories | Demographics + Features with r > 0.5 + COVID contexts | 0 | 0 | 0 |
| | Demographics + All psychological measures + COVID Memories | Demographics + COVID Memories + COVID contexts | 5 | 5 | 0 |
| | Demographics + Features with r < 0.5 + COVID contexts | Demographics + Features with r > 0.5 + COVID contexts | 3 | 1 | 2 |
| | Demographics + Features with r < 0.5 + COVID contexts | Demographics + COVID Memories + COVID contexts | 0 | 0 | 0 |
| | Demographics + Features with r > 0.5 + COVID contexts | Demographics + COVID Memories + COVID contexts | 0 | 0 | 0 |
| DeepSeek V4-Flash | Demographics | COVID Memories | 8 | 4 | 4 |
| | Demographics | Demographics + COVID Memories | 0 | 0 | 0 |
| | Demographics | Demographics + All psychological measures + COVID Memories | 4 | 2 | 2 |
| | Demographics | Demographics + Features with r < 0.5 + COVID contexts | 0 | 0 | 0 |
| | Demographics | Demographics + Features with r > 0.5 + COVID contexts | 3 | 1 | 2 |
| | Demographics | Demographics + COVID Memories + COVID contexts | 0 | 0 | 0 |
| | COVID Memories | Demographics + COVID Memories | 3 | 1 | 2 |
| | COVID Memories | Demographics + All psychological measures + COVID Memories | 8 | 3 | 5 |
| | COVID Memories | Demographics + Features with r < 0.5 + COVID contexts | 0 | 0 | 0 |
| | COVID Memories | Demographics + Features with r > 0.5 + COVID contexts | 11 | 5 | 6 |
| | COVID Memories | Demographics + COVID Memories + COVID contexts | 6 | 2 | 4 |
| | Demographics + COVID Memories | Demographics + All psychological measures + COVID Memories | 3 | 2 | 1 |
| | Demographics + COVID Memories | Demographics + Features with r < 0.5 + COVID contexts | 0 | 0 | 0 |
| | Demographics + COVID Memories | Demographics + Features with r > 0.5 + COVID contexts | 0 | 0 | 0 |
| | Demographics + COVID Memories | Demographics + COVID Memories + COVID contexts | 0 | 0 | 0 |
| | Demographics + All psychological measures + COVID Memories | Demographics + Features with r < 0.5 + COVID contexts | 0 | 0 | 0 |
| | Demographics + All psychological measures + COVID Memories | Demographics + Features with r > 0.5 + COVID contexts | 0 | 0 | 0 |
| | Demographics + All psychological measures + COVID Memories | Demographics + COVID Memories + COVID contexts | 0 | 0 | 0 |
| | Demographics + Features with r < 0.5 + COVID contexts | Demographics + Features with r > 0.5 + COVID contexts | 0 | 0 | 0 |
| | Demographics + Features with r < 0.5 + COVID contexts | Demographics + COVID Memories + COVID contexts | 0 | 0 | 0 |
| | Demographics + Features with r > 0.5 + COVID contexts | Demographics + COVID Memories + COVID contexts | 0 | 0 | 0 |
| Qwen 3.5 Plus | Demographics | COVID Memories | 28 | 21 | 7 |
| | Demographics | Demographics + COVID Memories | 14 | 12 | 2 |
| | Demographics | Demographics + All psychological measures + COVID Memories | 19 | 10 | 9 |
| | Demographics | Demographics + Features with r < 0.5 + COVID contexts | 15 | 13 | 2 |
| | Demographics | Demographics + Features with r > 0.5 + COVID contexts | 17 | 11 | 6 |
| | Demographics | Demographics + COVID Memories + COVID contexts | 20 | 16 | 4 |
| | COVID Memories | Demographics + COVID Memories | 0 | 0 | 0 |
| | COVID Memories | Demographics + All psychological measures + COVID Memories | 10 | 2 | 8 |

| | | | | |
|---|---|---|---|---|
| COVID Memories | Demographics + Features with r < 0.5 + COVID contexts | 0 | 0 | 0 |
| COVID Memories | Demographics + Features with r > 0.5 + COVID contexts | 9 | 3 | 6 |
| COVID Memories | Demographics + COVID Memories + COVID contexts | 3 | 2 | 1 |
| Demographics + COVID Memories | Demographics + All psychological measures + COVID Memories | 9 | 3 | 6 |
| Demographics + COVID Memories | Demographics + Features with r < 0.5 + COVID contexts | 0 | 0 | 0 |
| Demographics + COVID Memories | Demographics + Features with r > 0.5 + COVID contexts | 9 | 4 | 5 |
| Demographics + COVID Memories | Demographics + COVID Memories + COVID contexts | 0 | 0 | 0 |
| Demographics + All psychological measures + COVID Memories | Demographics + Features with r < 0.5 + COVID contexts | 10 | 7 | 3 |
| Demographics + All psychological measures + COVID Memories | Demographics + Features with r > 0.5 + COVID contexts | 0 | 0 | 0 |
| Demographics + All psychological measures + COVID Memories | Demographics + COVID Memories + COVID contexts | 5 | 4 | 1 |
| Demographics + Features with r < 0.5 + COVID contexts | Demographics + Features with r > 0.5 + COVID contexts | 9 | 2 | 7 |
| Demographics + Features with r < 0.5 + COVID contexts | Demographics + COVID Memories + COVID contexts | 0 | 0 | 0 |
| Demographics + Features with r > 0.5 + COVID contexts | Demographics + COVID Memories + COVID contexts | 7 | 6 | 1 |

*Note.* Entries report the number of items showing significant loading differences between pairs of digital-twin input conditions within each model. "A < B" indicates that the estimated item loading was significantly smaller under input condition A than under input condition B, whereas "A > B" indicates that the estimated item loading was significantly larger under input condition A than under input condition B. Noninvariant items were identified using Benjamini–Hochberg adjusted p values < .05.

**Table S55 Metric invariance results for NEO Five-Factor Personality Inventory item loadings among digital-twin models within the same input condition in Study 3b**

| Condition | Model 1 | Model 2 | Number of noninvariant items | Model 1 < Model 2, n | Model 1 > Model 2, n |
|---|---|---|---|---|---|
| Demographics | Gemma-4-31B-it | gpt-oss-120b | 17 | 8 | 9 |
| Demographics | Gemma-4-31B-it | DeepSeek V4-Flash | 12 | 8 | 4 |
| Demographics | Gemma-4-31B-it | Qwen 3.5 Plus | 5 | 2 | 3 |
| Demographics | gpt-oss-120b | DeepSeek V4-Flash | 18 | 10 | 8 |
| Demographics | gpt-oss-120b | Qwen 3.5 Plus | 15 | 7 | 8 |
| Demographics | DeepSeek V4-Flash | Qwen 3.5 Plus | 16 | 4 | 12 |
| COVID Memories | Gemma-4-31B-it | gpt-oss-120b | 15 | 4 | 11 |
| COVID Memories | Gemma-4-31B-it | DeepSeek V4-Flash | 17 | 8 | 9 |
| COVID Memories | Gemma-4-31B-it | Qwen 3.5 Plus | 3 | 2 | 1 |
| COVID Memories | gpt-oss-120b | DeepSeek V4-Flash | 23 | 15 | 8 |
| COVID Memories | gpt-oss-120b | Qwen 3.5 Plus | 11 | 9 | 2 |
| COVID Memories | DeepSeek V4-Flash | Qwen 3.5 Plus | 23 | 12 | 11 |
| Demographics + COVID Memories | Gemma-4-31B-it | gpt-oss-120b | 17 | 3 | 14 |
| Demographics + COVID Memories | Gemma-4-31B-it | DeepSeek V4-Flash | 8 | 2 | 6 |
| Demographics + COVID Memories | Gemma-4-31B-it | Qwen 3.5 Plus | 0 | 0 | 0 |
| Demographics + COVID Memories | gpt-oss-120b | DeepSeek V4-Flash | 23 | 14 | 9 |
| Demographics + COVID Memories | gpt-oss-120b | Qwen 3.5 Plus | 16 | 9 | 7 |
| Demographics + COVID Memories | DeepSeek V4-Flash | Qwen 3.5 Plus | 11 | 7 | 4 |
| Demographics + All psychological measures + COVID Memories | Gemma-4-31B-it | gpt-oss-120b | 14 | 5 | 9 |
| Demographics + All psychological measures + COVID Memories | Gemma-4-31B-it | DeepSeek V4-Flash | 17 | 9 | 8 |
| Demographics + All psychological measures + COVID Memories | Gemma-4-31B-it | Qwen 3.5 Plus | 0 | 0 | 0 |
| Demographics + All psychological measures + COVID Memories | gpt-oss-120b | DeepSeek V4-Flash | 15 | 10 | 5 |
| Demographics + All psychological measures + COVID Memories | gpt-oss-120b | Qwen 3.5 Plus | 18 | 9 | 9 |
| Demographics + All psychological measures + COVID Memories | DeepSeek V4-Flash | Qwen 3.5 Plus | 21 | 9 | 12 |
| Demographics + Features with $r < 0.5$ + COVID contexts | Gemma-4-31B-it | gpt-oss-120b | 14 | 6 | 8 |
| Demographics + Features with $r < 0.5$ + COVID contexts | Gemma-4-31B-it | DeepSeek V4-Flash | 10 | 4 | 6 |
| Demographics + Features with $r < 0.5$ + COVID contexts | Gemma-4-31B-it | Qwen 3.5 Plus | 0 | 0 | 0 |
| Demographics + Features with $r < 0.5$ + COVID contexts | gpt-oss-120b | DeepSeek V4-Flash | 20 | 13 | 7 |
| Demographics + Features with $r < 0.5$ + COVID contexts | gpt-oss-120b | Qwen 3.5 Plus | 22 | 13 | 9 |
| Demographics + Features with $r < 0.5$ + COVID contexts | DeepSeek V4-Flash | Qwen 3.5 Plus | 13 | 5 | 8 |
| Demographics + Features with $r > 0.5$ + COVID contexts | Gemma-4-31B-it | gpt-oss-120b | 13 | 3 | 10 |
| Demographics + Features with $r > 0.5$ + COVID contexts | Gemma-4-31B-it | DeepSeek V4-Flash | 15 | 8 | 7 |
| Demographics + Features with $r > 0.5$ + COVID contexts | Gemma-4-31B-it | Qwen 3.5 Plus | 0 | 0 | 0 |
| Demographics + Features with $r > 0.5$ + COVID contexts | gpt-oss-120b | DeepSeek V4-Flash | 22 | 14 | 8 |
| Demographics + Features with $r > 0.5$ + COVID contexts | gpt-oss-120b | Qwen 3.5 Plus | 10 | 6 | 4 |
| Demographics + Features with $r > 0.5$ + COVID contexts | DeepSeek V4-Flash | Qwen 3.5 Plus | 18 | 8 | 10 |
| Demographics + COVID Memories + COVID contexts | Gemma-4-31B-it | gpt-oss-120b | 17 | 5 | 12 |
| Demographics + COVID Memories + COVID contexts | Gemma-4-31B-it | DeepSeek V4-Flash | 16 | 11 | 5 |
| Demographics + COVID Memories + COVID contexts | Gemma-4-31B-it | Qwen 3.5 Plus | 0 | 0 | 0 |
| Demographics + COVID Memories + COVID contexts | gpt-oss-120b | DeepSeek V4-Flash | 21 | 11 | 10 |
| Demographics + COVID Memories + COVID contexts | gpt-oss-120b | Qwen 3.5 Plus | 19 | 12 | 7 |
| Demographics + COVID Memories + COVID contexts | DeepSeek V4-Flash | Qwen 3.5 Plus | 18 | 8 | 10 |

*Note.* Entries report the number of items showing significant loading differences between pairs of digital-twin models within each input condition. Noninvariant items were identified using Benjamini–Hochberg adjusted p values $< .05$.

**Table S56 Linguistic metrics for human and digital twin texts**

| | Model | Sentences (n) | Average sentence length | Normalized mean dependency distance | Mean dependency depth | HD-D lexical diversity | Named entity density |
|---|---|---|---|---|---|---|---|
| Study 4a | | | | | | | |
| | Human | 3.426 (3.432) | 17.638 (9.158) | 0.146 (0.054) | 2.620 (0.780) | 0.861 (0.078) | 0.066 (0.052) |
| Gemma-4-31B-it | Digital twin (feature-rich) | 2.435 (0.501) | 17.140 (2.589) | 0.130 (0.021) | 2.975 (0.513) | 0.875 (0.041) | 0.012 (0.016) |
| | Digital twin (demographic) | 2.151 (0.366) | 18.255 (2.560) | 0.123 (0.019) | 2.638 (0.453) | 0.896 (0.044) | 0.015 (0.017) |
| gpt-oss-120b | Digital twin (feature-rich) | 1.203 (0.471) | 40.309 (10.048) | 0.068 (0.019) | 3.615 (0.741) | 0.882 (0.041) | 0.047 (0.021) |
| | Digital twin (demographic) | 1.163 (0.417) | 38.520 (9.117) | 0.071 (0.019) | 3.378 (0.674) | 0.883 (0.043) | 0.051 (0.024) |
| DeepSeek V4-Flash | Digital twin (feature-rich) | 1.899 (0.653) | 17.414 (5.470) | 0.136 (0.035) | 2.565 (0.573) | 0.939 (0.044) | 0.051 (0.029) |
| | Digital twin (demographic) | 2.047 (0.620) | 16.087 (4.997) | 0.146 (0.036) | 2.424 (0.471) | 0.937 (0.043) | 0.055 (0.031) |
| Qwen 3.5 Plus | Digital twin (feature-rich) | 3.236 (0.588) | 14.264 (2.200) | 0.150 (0.021) | 2.414 (0.352) | 0.927 (0.033) | 0.035 (0.021) |
| | Digital twin (demographic) | 3.174 (0.556) | 14.843 (2.252) | 0.147 (0.021) | 2.381 (0.335) | 0.933 (0.031) | 0.048 (0.023) |
| Study 4b | | | | | | | |
| | Human | 9.595 (5.365) | 20.778 (11.434) | 0.180 (0.070) | 2.275 (0.375) | 0.867 (0.041) | 0.064 (0.037) |
| Gemma-4-31B-it | Digital twin (feature-rich) | 7.357 (1.322) | 20.346 (3.180) | 0.161 (0.020) | 2.525 (0.189) | 0.894 (0.015) | 0.021 (0.013) |
| | Digital twin (demographic) | 5.238 (0.850) | 22.905 (3.164) | 0.154 (0.018) | 2.739 (0.225) | 0.878 (0.016) | 0.001 (0.002) |
| gpt-oss-120b | Digital twin (feature-rich) | 9.643 (4.310) | 23.392 (6.622) | 0.153 (0.029) | 2.580 (0.279) | 0.886 (0.016) | 0.034 (0.011) |
| | Digital twin (demographic) | 7.405 (2.509) | 20.325 (4.411) | 0.174 (0.027) | 2.518 (0.206) | 0.884 (0.016) | 0.002 (0.004) |
| DeepSeek V4-Flash | Digital twin (feature-rich) | 10.214 (2.007) | 18.760 (2.116) | 0.154 (0.014) | 2.359 (0.199) | 0.915 (0.017) | 0.050 (0.018) |
| | Digital twin (demographic) | 9.595 (1.926) | 18.933 (2.942) | 0.171 (0.020) | 2.442 (0.209) | 0.895 (0.017) | 0.005 (0.006) |
| Qwen 3.5 Plus | Digital twin (feature-rich) | 10.000 (2.209) | 17.003 (3.557) | 0.161 (0.022) | 2.274 (0.207) | 0.927 (0.016) | 0.038 (0.010) |
| | Digital twin (demographic) | 7.214 (1.661) | 18.014 (2.829) | 0.161 (0.018) | 2.499 (0.210) | 0.908 (0.016) | 0.002 (0.003) |

Note: Values are reported as Mean (SD). Normalized mean dependency distance was defined as MDD/ASL, where MDD is the mean absolute linear head–dependent distance (span, adjacent = 1) computed within sentence only, and ASL is the mean sentence length in the same token units; named-entity density was the number of recognized entity spans divided by the same non-punctuation token count.

**Table S57 Tests of differences in linguistic metrics across human and digital twin texts**

| | Metric | df | χ² | Kendall's W | *p* | |
|---|---|---|---|---|---|---|
| Study 4a | | | | | | |
| Gemma-4-31B-it | Sentences (n) | 2 | 80.324 | 0.049 | <.001 | *** |
| | Average sentence length | 2 | 78.471 | 0.048 | <.001 | *** |
| | Normalized dependency distance | 2 | 72.123 | 0.044 | <.001 | *** |
| | Mean dependency depth | 2 | 169.146 | 0.104 | <.001 | *** |
| | HD-D lexical diversity | 2 | 137.655 | 0.085 | <.001 | *** |
| | Named entity density | 2 | 777.19 | 0.479 | <.001 | *** |
| gpt-oss-120b | Sentences (n) | 2 | 767.117 | 0.475 | <.001 | *** |
| | Average sentence length | 2 | 977.394 | 0.606 | <.001 | *** |
| | Normalized dependency distance | 2 | 1017.021 | 0.63 | <.001 | *** |
| | Mean dependency depth | 2 | 513.281 | 0.318 | <.001 | *** |
| | HD-D lexical diversity | 2 | 79.518 | 0.049 | <.001 | *** |
| | Named entity density | 2 | 53.439 | 0.033 | <.001 | *** |
| DeepSeek V4-Flash | Sentences (n) | 2 | 111.993 | 0.069 | <.001 | *** |
| | Average sentence length | 2 | 13.93 | 0.009 | <.001 | *** |
| | Normalized dependency distance | 2 | 26.69 | 0.016 | <.001 | *** |
| | Mean dependency depth | 2 | 25.687 | 0.016 | <.001 | *** |
| | HD-D lexical diversity | 2 | 422.731 | 0.26 | <.001 | *** |
| | Named entity density | 2 | 24.048 | 0.015 | <.001 | *** |
| Qwen 3.5 Plus | Sentences (n) | 2 | 72.241 | 0.044 | <.001 | *** |
| | Average sentence length | 2 | 75.784 | 0.047 | <.001 | *** |
| | Normalized dependency distance | 2 | 40.494 | 0.025 | <.001 | *** |
| | Mean dependency depth | 2 | 42.169 | 0.026 | <.001 | *** |
| | HD-D lexical diversity | 2 | 390.392 | 0.24 | <.001 | *** |
| | Named entity density | 2 | 180.108 | 0.111 | <.001 | *** |
| Study 4b | | | | | | |
| Gemma-4-31B-it | Sentences (n) | 2 | 39.569 | 0.471 | <.001 | *** |
| | Average sentence length | 2 | 7.19 | 0.086 | 0.031 | * |
| | Normalized dependency distance | 2 | 3.571 | 0.043 | 0.168 | |
| | Mean dependency depth | 2 | 32.333 | 0.385 | <.001 | *** |
| | HD-D lexical diversity | 2 | 14.333 | 0.171 | <.001 | *** |
| | Named entity density | 2 | 72.898 | 0.868 | <.001 | *** |
| gpt-oss-120b | Sentences (n) | 2 | 5.057 | 0.06 | 0.108 | |
| | Average sentence length | 2 | 4.333 | 0.052 | 0.115 | |
| | Normalized dependency distance | 2 | 8.048 | 0.096 | 0.03 | * |
| | Mean dependency depth | 2 | 12.905 | 0.154 | 0.003 | ** |
| | HD-D lexical diversity | 2 | 4.619 | 0.055 | 0.11 | |
| | Named entity density | 2 | 67.476 | 0.803 | <.001 | *** |
| DeepSeek V4-Flash | Sentences (n) | 2 | 3.434 | 0.041 | 0.225 | |
| | Average sentence length | 2 | 0.19 | 0.002 | 0.909 | |
| | Normalized dependency distance | 2 | 12.333 | 0.147 | 0.003 | ** |
| | Mean dependency depth | 2 | 3.19 | 0.038 | 0.225 | |
| | HD-D lexical diversity | 2 | 41.762 | 0.497 | <.001 | *** |
| | Named entity density | 2 | 60.048 | 0.715 | <.001 | *** |
| Qwen 3.5 Plus | Sentences (n) | 2 | 17.478 | 0.208 | <.001 | *** |
| | Average sentence length | 2 | 0.429 | 0.005 | 0.807 | |
| | Normalized dependency distance | 2 | 0.619 | 0.007 | 0.807 | |
| | Mean dependency depth | 2 | 13.476 | 0.16 | 0.001 | ** |
| | HD-D lexical diversity | 2 | 47.286 | 0.563 | <.001 | *** |
| | Named entity density | 2 | 67.413 | 0.803 | <.001 | *** |

*Note:* Differences across the three conditions (Human, Digital Twin–feature-rich, and Digital Twin–demographic) were tested using the Friedman test. Kendall's W is reported as the effect size for the Friedman test. p reports Benjamini-Hochberg (BH)-adjusted p-values. *** $p < .001$, ** $p < .01$, * $p < .05$.

**Table S58 Paired comparisons of linguistic metrics between human and digital twin texts**

| Model | Metric | Group 1 | Group 2 | Median (Group 1) | Median (Group 2) | r | p | |
|---|---|---|---|---|---|---|---|---|
| Study 4a | | | | | | | | |
| Gemma-4-31B-it | Sentences (n) | Digital twin (feature-rich) | Digital twin (demographic) | 2 | 2 | -0.613 | <.001 | *** |
| | Sentences (n) | Human | Digital twin (demographic) | 3 | 2 | -0.416 | <.001 | *** |
| | Sentences (n) | Human | Digital twin (feature-rich) | 3 | 2 | -0.256 | <.001 | *** |
| | Average sentence length | Digital twin (feature-rich) | Digital twin (demographic) | 16.667 | 18 | 0.308 | <.001 | *** |
| | Average sentence length | Human | Digital twin (demographic) | 16 | 18 | 0.203 | <.001 | *** |
| | Average sentence length | Human | Digital twin (feature-rich) | 16 | 16.667 | -0.065 | 0.067 | |
| | Normalized dependency distance | Digital twin (feature-rich) | Digital twin (demographic) | 0.13 | 0.121 | -0.235 | <.001 | *** |
| | Normalized dependency distance | Human | Digital twin (demographic) | 0.136 | 0.121 | -0.348 | <.001 | *** |
| | Normalized dependency distance | Human | Digital twin (feature-rich) | 0.136 | 0.13 | -0.224 | <.001 | *** |
| | Mean dependency depth | Digital twin (feature-rich) | Digital twin (demographic) | 2.902 | 2.562 | -0.453 | <.001 | *** |
| | Mean dependency depth | Human | Digital twin (demographic) | 2.528 | 2.562 | 0.045 | 0.199 | |
| | Mean dependency depth | Human | Digital twin (feature-rich) | 2.528 | 2.902 | 0.409 | <.001 | *** |
| | HD-D lexical diversity | Digital twin (feature-rich) | Digital twin (demographic) | 0.872 | 0.897 | 0.331 | <.001 | *** |
| | HD-D lexical diversity | Human | Digital twin (demographic) | 0.846 | 0.897 | 0.363 | <.001 | *** |
| | HD-D lexical diversity | Human | Digital twin (feature-rich) | 0.846 | 0.872 | 0.159 | <.001 | *** |
| | Named entity density | Digital twin (feature-rich) | Digital twin (demographic) | 0 | 0.009 | 0.207 | <.001 | *** |
| | Named entity density | Human | Digital twin (demographic) | 0.057 | 0.009 | -0.789 | <.001 | *** |
| | Named entity density | Human | Digital twin (feature-rich) | 0.057 | 0 | -0.812 | <.001 | *** |
| gpt-oss-120b | Sentences (n) | Digital twin (feature-rich) | Digital twin (demographic) | 1 | 1 | -0.119 | 0.08 | |
| | Sentences (n) | Human | Digital twin (demographic) | 3 | 1 | -0.824 | <.001 | *** |
| | Sentences (n) | Human | Digital twin (feature-rich) | 3 | 1 | -0.82 | <.001 | *** |
| | Average sentence length | Digital twin (feature-rich) | Digital twin (demographic) | 42 | 41 | -0.134 | <.001 | *** |
| | Average sentence length | Human | Digital twin (demographic) | 16 | 41 | 0.827 | <.001 | *** |
| | Average sentence length | Human | Digital twin (feature-rich) | 16 | 42.5 | 0.836 | <.001 | *** |
| | Normalized dependency distance | Digital twin (feature-rich) | Digital twin (demographic) | 0.062 | 0.065 | 0.14 | <.001 | *** |
| | Normalized dependency distance | Human | Digital twin (demographic) | 0.136 | 0.065 | -0.84 | <.001 | *** |
| | Normalized dependency distance | Human | Digital twin (feature-rich) | 0.136 | 0.062 | -0.856 | <.001 | *** |
| | Mean dependency depth | Digital twin (feature-rich) | Digital twin (demographic) | 3.513 | 3.286 | -0.225 | <.001 | *** |
| | Mean dependency depth | Human | Digital twin (demographic) | 2.529 | 3.289 | 0.628 | <.001 | *** |
| | Mean dependency depth | Human | Digital twin (feature-rich) | 2.528 | 3.512 | 0.724 | <.001 | *** |
| | HD-D lexical diversity | Digital twin (feature-rich) | Digital twin (demographic) | 0.881 | 0.882 | 0.02 | 0.593 | |
| | HD-D lexical diversity | Human | Digital twin (demographic) | 0.846 | 0.882 | 0.254 | <.001 | *** |
| | HD-D lexical diversity | Human | Digital twin (feature-rich) | 0.846 | 0.881 | 0.236 | <.001 | *** |
| | Named entity density | Digital twin (feature-rich) | Digital twin (demographic) | 0.043 | 0.048 | 0.147 | <.001 | *** |
| | Named entity density | Human | Digital twin (demographic) | 0.057 | 0.048 | -0.2 | <.001 | *** |
| | Named entity density | Human | Digital twin (feature-rich) | 0.057 | 0.043 | -0.297 | <.001 | *** |
| DeepSeek V4-Flash | Sentences (n) | Digital twin (feature-rich) | Digital twin (demographic) | 2 | 2 | 0.217 | <.001 | *** |
| | Sentences (n) | Human | Digital twin (demographic) | 3 | 2 | -0.472 | <.001 | *** |
| | Sentences (n) | Human | Digital twin (feature-rich) | 3 | 2 | -0.546 | <.001 | *** |
| | Average sentence length | Digital twin (feature-rich) | Digital twin (demographic) | 16 | 15 | -0.178 | <.001 | *** |
| | Average sentence length | Human | Digital twin (demographic) | 16 | 15 | -0.107 | 0.003 | ** |
| | Average sentence length | Human | Digital twin (feature-rich) | 16 | 16 | -0.046 | 0.221 | |
| | Normalized dependency distance | Digital twin (feature-rich) | Digital twin (demographic) | 0.134 | 0.145 | 0.201 | <.001 | *** |
| | Normalized dependency distance | Human | Digital twin (demographic) | 0.136 | 0.145 | -0.062 | 0.094 | |
| | Normalized dependency distance | Human | Digital twin (feature-rich) | 0.136 | 0.134 | -0.117 | 0.001 | ** |
| | Mean dependency depth | Digital twin (feature-rich) | Digital twin (demographic) | 2.464 | 2.364 | -0.166 | <.001 | *** |
| | Mean dependency depth | Human | Digital twin (demographic) | 2.528 | 2.364 | -0.186 | <.001 | *** |
| | Mean dependency depth | Human | Digital twin (feature-rich) | 2.528 | 2.464 | -0.047 | 0.21 | |

| | HD-D lexical diversity | Digital twin (feature-rich) | Digital twin (demographic) | 0.941 | 0.938 | -0.037 | 0.319 | |
|---|---|---|---|---|---|---|---|---|
| | HD-D lexical diversity | Human | Digital twin (demographic) | 0.846 | 0.938 | 0.674 | <.001 | *** |
| | HD-D lexical diversity | Human | Digital twin (feature-rich) | 0.846 | 0.941 | 0.691 | <.001 | *** |
| | Named entity density | Digital twin (feature-rich) | Digital twin (demographic) | 0.043 | 0.053 | 0.1 | 0.006 | ** |
| | Named entity density | Human | Digital twin (demographic) | 0.057 | 0.053 | -0.14 | <.001 | *** |
| | Named entity density | Human | Digital twin (feature-rich) | 0.057 | 0.043 | -0.204 | <.001 | *** |
| Qwen 3.5 Plus | Sentences (n) | Digital twin (feature-rich) | Digital twin (demographic) | 3 | 3 | -0.107 | 0.034 | * |
| | Sentences (n) | Human | Digital twin (demographic) | 3 | 3 | -0.112 | 0.004 | ** |
| | Sentences (n) | Human | Digital twin (feature-rich) | 3 | 3 | -0.144 | <.001 | *** |
| | Average sentence length | Digital twin (feature-rich) | Digital twin (demographic) | 14.333 | 15 | 0.185 | <.001 | *** |
| | Average sentence length | Human | Digital twin (demographic) | 16 | 15 | -0.26 | <.001 | *** |
| | Average sentence length | Human | Digital twin (feature-rich) | 16 | 14.333 | -0.338 | <.001 | *** |
| | Normalized dependency distance | Digital twin (feature-rich) | Digital twin (demographic) | 0.147 | 0.145 | -0.079 | 0.03 | * |
| | Normalized dependency distance | Human | Digital twin (demographic) | 0.136 | 0.145 | 0.121 | <.001 | *** |
| | Normalized dependency distance | Human | Digital twin (feature-rich) | 0.136 | 0.147 | 0.166 | <.001 | *** |
| | Mean dependency depth | Digital twin (feature-rich) | Digital twin (demographic) | 2.386 | 2.333 | -0.067 | 0.06 | |
| | Mean dependency depth | Human | Digital twin (demographic) | 2.528 | 2.333 | -0.252 | <.001 | *** |
| | Mean dependency depth | Human | Digital twin (feature-rich) | 2.528 | 2.386 | -0.217 | <.001 | *** |
| | HD-D lexical diversity | Digital twin (feature-rich) | Digital twin (demographic) | 0.928 | 0.935 | 0.112 | 0.002 | ** |
| | HD-D lexical diversity | Human | Digital twin (demographic) | 0.846 | 0.935 | 0.663 | <.001 | *** |
| | HD-D lexical diversity | Human | Digital twin (feature-rich) | 0.846 | 0.928 | 0.62 | <.001 | *** |
| | Named entity density | Digital twin (feature-rich) | Digital twin (demographic) | 0.029 | 0.044 | 0.37 | <.001 | *** |
| | Named entity density | Human | Digital twin (demographic) | 0.057 | 0.044 | -0.271 | <.001 | *** |
| | Named entity density | Human | Digital twin (feature-rich) | 0.057 | 0.029 | -0.499 | <.001 | *** |
| Study 4b | | | | | | | | |
| Gemma-4-31B-it | Sentences (n) | Digital twin (feature-rich) | Digital twin (demographic) | 7 | 5 | -0.869 | <.001 | *** |
| | Sentences (n) | Human | Digital twin (demographic) | 8 | 5 | -0.818 | <.001 | *** |
| | Sentences (n) | Human | Digital twin (feature-rich) | 8 | 7 | -0.348 | 0.038 | * |
| | Average sentence length | Digital twin (feature-rich) | Digital twin (demographic) | 20.643 | 22.75 | 0.572 | <.001 | *** |
| | Average sentence length | Human | Digital twin (demographic) | 17.477 | 22.75 | 0.237 | 0.149 | |
| | Average sentence length | Human | Digital twin (feature-rich) | 17.477 | 20.643 | -0.02 | 0.896 | |
| | Normalized dependency distance | Digital twin (feature-rich) | Digital twin (demographic) | 0.158 | 0.151 | -0.261 | 0.113 | |
| | Normalized dependency distance | Human | Digital twin (demographic) | 0.166 | 0.151 | -0.331 | 0.042 | * |
| | Normalized dependency distance | Human | Digital twin (feature-rich) | 0.166 | 0.158 | -0.138 | 0.398 | |
| | Mean dependency depth | Digital twin (feature-rich) | Digital twin (demographic) | 2.558 | 2.72 | 0.638 | <.001 | *** |
| | Mean dependency depth | Human | Digital twin (demographic) | 2.339 | 2.72 | 0.794 | <.001 | *** |
| | Mean dependency depth | Human | Digital twin (feature-rich) | 2.339 | 2.558 | 0.487 | 0.002 | ** |
| | HD-D lexical diversity | Digital twin (feature-rich) | Digital twin (demographic) | 0.894 | 0.88 | -0.605 | <.001 | *** |
| | HD-D lexical diversity | Human | Digital twin (demographic) | 0.881 | 0.88 | 0.107 | 0.505 | |
| | HD-D lexical diversity | Human | Digital twin (feature-rich) | 0.881 | 0.894 | 0.524 | 0.001 | ** |
| | Named entity density | Digital twin (feature-rich) | Digital twin (demographic) | 0.019 | 0 | -0.865 | <.001 | *** |
| | Named entity density | Human | Digital twin (demographic) | 0.06 | 0 | -0.871 | <.001 | *** |
| | Named entity density | Human | Digital twin (feature-rich) | 0.06 | 0.019 | -0.773 | <.001 | *** |
| gpt-oss-120b | Sentences (n) | Digital twin (feature-rich) | Digital twin (demographic) | 9 | 7 | -0.437 | 0.016 | * |
| | Sentences (n) | Human | Digital twin (demographic) | 8 | 7 | -0.35 | 0.054 | |
| | Sentences (n) | Human | Digital twin (feature-rich) | 8 | 9 | 0.042 | 0.821 | |
| | Average sentence length | Digital twin (feature-rich) | Digital twin (demographic) | 22.545 | 20.583 | -0.329 | 0.055 | |
| | Average sentence length | Human | Digital twin (demographic) | 17.477 | 20.583 | -0.082 | 0.661 | |
| | Average sentence length | Human | Digital twin (feature-rich) | 17.477 | 22.545 | 0.232 | 0.172 | |
| | Normalized dependency distance | Digital twin (feature-rich) | Digital twin (demographic) | 0.151 | 0.17 | 0.501 | 0.003 | ** |
| | Normalized dependency distance | Human | Digital twin (demographic) | 0.166 | 0.17 | -0.007 | 0.965 | |
| | Normalized dependency distance | Human | Digital twin (feature-rich) | 0.166 | 0.151 | -0.271 | 0.113 | |
| | Mean dependency depth | Digital twin (feature-rich) | Digital twin (demographic) | 2.558 | 2.543 | -0.099 | 0.6 | |
| | Mean dependency depth | Human | Digital twin (demographic) | 2.339 | 2.543 | 0.514 | 0.003 | ** |
| | Mean dependency depth | Human | Digital twin (feature-rich) | 2.339 | 2.558 | 0.566 | <.001 | *** |
| | HD-D lexical diversity | Digital twin (feature-rich) | Digital twin (demographic) | 0.886 | 0.884 | -0.121 | 0.521 | |
| | HD-D lexical diversity | Human | Digital twin (demographic) | 0.881 | 0.884 | 0.312 | 0.069 | |
| | HD-D lexical diversity | Human | Digital twin (feature-rich) | 0.881 | 0.886 | 0.375 | 0.032 | * |
| | Named entity density | Digital twin (feature-rich) | Digital twin (demographic) | 0.033 | 0 | -0.871 | <.001 | *** |
| | Named entity density | Human | Digital twin (demographic) | 0.06 | 0 | -0.869 | <.001 | *** |
| | Named entity density | Human | Digital twin (feature-rich) | 0.06 | 0.033 | -0.651 | <.001 | *** |
| DeepSeek V4-Flash | Sentences (n) | Digital twin (feature-rich) | Digital twin (demographic) | 10 | 9 | -0.188 | 0.343 | |
| | Sentences (n) | Human | Digital twin (demographic) | 8 | 9 | 0 | 0.431 | |
| | Sentences (n) | Human | Digital twin (feature-rich) | 8 | 10 | 0.213 | 0.252 | |
| | Average sentence length | Digital twin (feature-rich) | Digital twin (demographic) | 18.697 | 18.812 | 0.011 | 0.945 | |
| | Average sentence length | Human | Digital twin (demographic) | 17.477 | 18.812 | -0.049 | 0.803 | |
| | Average sentence length | Human | Digital twin (feature-rich) | 17.477 | 18.697 | -0.065 | 0.75 | |
| | Normalized dependency distance | Digital twin (feature-rich) | Digital twin (demographic) | 0.153 | 0.172 | 0.613 | <.001 | *** |
| | Normalized dependency distance | Human | Digital twin (demographic) | 0.166 | 0.172 | -0.018 | 0.937 | |
| | Normalized dependency distance | Human | Digital twin (feature-rich) | 0.166 | 0.153 | -0.302 | 0.094 | |
| | Mean dependency depth | Digital twin (feature-rich) | Digital twin (demographic) | 2.314 | 2.395 | 0.306 | 0.094 | |
| | Mean dependency depth | Human | Digital twin (demographic) | 2.339 | 2.395 | 0.331 | 0.069 | |
| | Mean dependency depth | Human | Digital twin (feature-rich) | 2.339 | 2.314 | 0.196 | 0.292 | |
| | HD-D lexical diversity | Digital twin (feature-rich) | Digital twin (demographic) | 0.914 | 0.896 | -0.68 | <.001 | *** |
| | HD-D lexical diversity | Human | Digital twin (demographic) | 0.881 | 0.896 | 0.532 | 0.002 | ** |
| | HD-D lexical diversity | Human | Digital twin (feature-rich) | 0.881 | 0.914 | 0.811 | <.001 | *** |
| | Named entity density | Digital twin (feature-rich) | Digital twin (demographic) | 0.049 | 0.005 | -0.871 | <.001 | *** |
| | Named entity density | Human | Digital twin (demographic) | 0.06 | 0.005 | -0.863 | <.001 | *** |
| | Named entity density | Human | Digital twin (feature-rich) | 0.06 | 0.049 | -0.254 | 0.173 | |
| Qwen 3.5 Plus | Sentences (n) | Digital twin (feature-rich) | Digital twin (demographic) | 10 | 8 | -0.783 | <.001 | *** |
| | Sentences (n) | Human | Digital twin (demographic) | 8 | 8 | -0.34 | 0.047 | * |
| | Sentences (n) | Human | Digital twin (feature-rich) | 8 | 10 | 0.152 | 0.41 | |
| | Average sentence length | Digital twin (feature-rich) | Digital twin (demographic) | 16.25 | 17.771 | 0.195 | 0.27 | |
| | Average sentence length | Human | Digital twin (demographic) | 17.477 | 17.771 | -0.117 | 0.499 | |
| | Average sentence length | Human | Digital twin (feature-rich) | 17.477 | 16.25 | -0.225 | 0.198 | |
| | Normalized dependency distance | Digital twin (feature-rich) | Digital twin (demographic) | 0.16 | 0.161 | 0.059 | 0.737 | |
| | Normalized dependency distance | Human | Digital twin (demographic) | 0.166 | 0.161 | -0.124 | 0.485 | |
| | Normalized dependency distance | Human | Digital twin (feature-rich) | 0.166 | 0.16 | -0.177 | 0.316 | |
| | Mean dependency depth | Digital twin (feature-rich) | Digital twin (demographic) | 2.277 | 2.467 | 0.613 | <.001 | *** |
| | Mean dependency depth | Human | Digital twin (demographic) | 2.339 | 2.467 | 0.383 | 0.021 | * |
| | Mean dependency depth | Human | Digital twin (feature-rich) | 2.339 | 2.277 | -0.057 | 0.737 | |
| | HD-D lexical diversity | Digital twin (feature-rich) | Digital twin (demographic) | 0.927 | 0.908 | -0.647 | <.001 | *** |
| | HD-D lexical diversity | Human | Digital twin (demographic) | 0.881 | 0.908 | 0.711 | <.001 | *** |
| | HD-D lexical diversity | Human | Digital twin (feature-rich) | 0.881 | 0.927 | 0.848 | <.001 | *** |
| | Named entity density | Digital twin (feature-rich) | Digital twin (demographic) | 0.039 | 0 | -0.871 | <.001 | *** |
| | Named entity density | Human | Digital twin (demographic) | 0.06 | 0 | -0.871 | <.001 | *** |
| | Named entity density | Human | Digital twin (feature-rich) | 0.06 | 0.039 | -0.584 | <.001 | *** |

*Note:* Pairwise differences between conditions were tested using two-sided paired Wilcoxon signed-rank tests. p reports Benjamini-Hochberg (BH)-adjusted p-values across all pairwise tests reported in this table. *** $p < .001$, ** $p < .01$, * $p < .05$.

**Table S59 Pairwise Jensen–Shannon divergence of NER label and POS bigram distributions between human and digital twin texts**

| | Subject | Comparison | N | Divergence | *p* |
|---|---|---|---|---|---|
| Study 4a | | | | | |
| Gemma-4-31B-it | NER | Human vs Digital twin (feature-rich) | 18 | 0.12 | <.001 |
| | NER | Human vs Digital twin (demographic) | 18 | 0.084 | <.001 |
| | NER | Digital twin (feature-rich) vs Digital twin (demographic) | 10 | 0.076 | <.001 |
| | POS Bigram | Human vs Digital twin (feature-rich) | 202 | 0.106 | <.001 |
| | POS Bigram | Human vs Digital twin (demographic) | 203 | 0.09 | <.001 |
| | POS Bigram | Digital twin (feature-rich) vs Digital twin (demographic) | 136 | 0.044 | <.001 |
| gpt-oss-120b | NER | Human vs Digital twin (feature-rich) | 18 | 0.1 | <.001 |
| | NER | Human vs Digital twin (demographic) | 18 | 0.089 | <.001 |
| | NER | Digital twin (feature-rich) vs Digital twin (demographic) | 16 | 0.015 | <.001 |
| | POS Bigram | Human vs Digital twin (feature-rich) | 207 | 0.114 | <.001 |
| | POS Bigram | Human vs Digital twin (demographic) | 204 | 0.119 | <.001 |
| | POS Bigram | Digital twin (feature-rich) vs Digital twin (demographic) | 156 | 0.01 | <.001 |
| DeepSeek V4-Flash | NER | Human vs Digital twin (feature-rich) | 18 | 0.066 | <.001 |
| | NER | Human vs Digital twin (demographic) | 18 | 0.06 | <.001 |
| | NER | Digital twin (feature-rich) vs Digital twin (demographic) | 17 | 0.034 | <.001 |
| | POS Bigram | Human vs Digital twin (feature-rich) | 202 | 0.062 | <.001 |
| | POS Bigram | Human vs Digital twin (demographic) | 203 | 0.054 | <.001 |
| | POS Bigram | Digital twin (feature-rich) vs Digital twin (demographic) | 155 | 0.014 | <.001 |
| Qwen 3.5 Plus | NER | Human vs Digital twin (feature-rich) | 18 | 0.137 | <.001 |
| | NER | Human vs Digital twin (demographic) | 18 | 0.103 | <.001 |
| | NER | Digital twin (feature-rich) vs Digital twin (demographic) | 13 | 0.024 | <.001 |
| | POS Bigram | Human vs Digital twin (feature-rich) | 202 | 0.114 | <.001 |
| | POS Bigram | Human vs Digital twin (demographic) | 203 | 0.11 | <.001 |
| | POS Bigram | Digital twin (feature-rich) vs Digital twin (demographic) | 145 | 0.017 | <.001 |
| Study 4b | | | | | |
| Gemma-4-31B-it | NER | Human vs Digital twin (feature-rich) | 11 | 0.158 | <.001 |
| | NER | Human vs Digital twin (demographic) | 11 | 0.373 | 0.51 |
| | NER | Digital twin (feature-rich) vs Digital twin (demographic) | 4 | 0.595 | 0.149 |
| | POS Bigram | Human vs Digital twin (feature-rich) | 365 | 0.108 | <.001 |
| | POS Bigram | Human vs Digital twin (demographic) | 359 | 0.152 | <.001 |
| | POS Bigram | Digital twin (feature-rich) vs Digital twin (demographic) | 254 | 0.081 | <.001 |
| gpt-oss-120b | NER | Human vs Digital twin (feature-rich) | 12 | 0.183 | <.001 |
| | NER | Human vs Digital twin (demographic) | 12 | 0.377 | 0.014 |
| | NER | Digital twin (feature-rich) vs Digital twin (demographic) | 10 | 0.685 | 0.006 |
| | POS Bigram | Human vs Digital twin (feature-rich) | 366 | 0.104 | <.001 |
| | POS Bigram | Human vs Digital twin (demographic) | 354 | 0.147 | <.001 |
| | POS Bigram | Digital twin (feature-rich) vs Digital twin (demographic) | 247 | 0.055 | <.001 |
| DeepSeek V4-Flash | NER | Human vs Digital twin (feature-rich) | 12 | 0.114 | <.001 |
| | NER | Human vs Digital twin (demographic) | 12 | 0.293 | <.001 |
| | NER | Digital twin (feature-rich) vs Digital twin (demographic) | 9 | 0.414 | <.001 |
| | POS Bigram | Human vs Digital twin (feature-rich) | 377 | 0.084 | <.001 |
| | POS Bigram | Human vs Digital twin (demographic) | 377 | 0.113 | <.001 |
| | POS Bigram | Digital twin (feature-rich) vs Digital twin (demographic) | 325 | 0.08 | <.001 |
| Qwen 3.5 Plus | NER | Human vs Digital twin (feature-rich) | 12 | 0.112 | <.001 |
| | NER | Human vs Digital twin (demographic) | 12 | 0.2 | 0.268 |
| | NER | Digital twin (feature-rich) vs Digital twin (demographic) | 9 | 0.484 | 0.035 |
| | POS Bigram | Human vs Digital twin (feature-rich) | 375 | 0.093 | <.001 |
| | POS Bigram | Human vs Digital twin (demographic) | 365 | 0.141 | <.001 |
| | POS Bigram | Digital twin (feature-rich) vs Digital twin (demographic) | 281 | 0.097 | <.001 |

Note: Divergence is the Jensen–Shannon divergence computed between the two conditions' mean category-proportion vectors (i.e., counts were converted to within-text proportions and then averaged across matched pairs). N denotes the number of categories included in the distribution for that block in the given comparison (i.e., the number of NER labels or POS-bigram types). p-values were obtained via a paired permutation test that randomly exchanges condition labels within each matched pairs and compares the observed divergence to the permutation distribution. *** $p < .001$, ** $p < .01$, * $p < .05$.

**Table S60 Results of Kolmogorov–Smirnov tests comparing empirical cumulative distribution functions (ECDFs) across dimensions among human participants and digital twins**

| | **Dimension** | **Group1** | **Group2** | **D** | ***p*** | |
|---|---|---|---|---|---|---|
| Study 4a | | | | | | |
| Gemma-4-31B-it | Nostalgia | Human | Digital twin (demographic) | 0.267 | <.001 | *** |
| | Nostalgia | Human | Digital twin (feature-rich) | 0.153 | <.001 | *** |
| | Nostalgia | Digital twin (demographic) | Digital twin (feature-rich) | 0.346 | <.001 | *** |
| | Agency | Human | Digital twin (demographic) | 0.267 | <.001 | *** |
| | Agency | Human | Digital twin (feature-rich) | 0.147 | <.001 | *** |
| | Agency | Digital twin (demographic) | Digital twin (feature-rich) | 0.345 | <.001 | *** |
| | Communion | Human | Digital twin (demographic) | 0.271 | <.001 | *** |
| | Communion | Human | Digital twin (feature-rich) | 0.145 | <.001 | *** |
| | Communion | Digital twin (demographic) | Digital twin (feature-rich) | 0.34 | <.001 | *** |
| | Social ties | Human | Digital twin (demographic) | 0.259 | <.001 | *** |
| | Social ties | Human | Digital twin (feature-rich) | 0.15 | <.001 | *** |
| | Social ties | Digital twin (demographic) | Digital twin (feature-rich) | 0.333 | <.001 | *** |
| | Threat | Human | Digital twin (demographic) | 0.268 | <.001 | *** |
| | Threat | Human | Digital twin (feature-rich) | 0.151 | <.001 | *** |
| | Threat | Digital twin (demographic) | Digital twin (feature-rich) | 0.35 | <.001 | *** |
| gpt-oss-120b | Nostalgia | Human | Digital twin (demographic) | 0.689 | <.001 | *** |
| | Nostalgia | Human | Digital twin (feature-rich) | 0.67 | <.001 | *** |
| | Nostalgia | Digital twin (demographic) | Digital twin (feature-rich) | 0.109 | <.001 | *** |
| | Agency | Human | Digital twin (demographic) | 0.691 | <.001 | *** |
| | Agency | Human | Digital twin (feature-rich) | 0.67 | <.001 | *** |
| | Agency | Digital twin (demographic) | Digital twin (feature-rich) | 0.109 | <.001 | *** |
| | Communion | Human | Digital twin (demographic) | 0.688 | <.001 | *** |
| | Communion | Human | Digital twin (feature-rich) | 0.671 | <.001 | *** |
| | Communion | Digital twin (demographic) | Digital twin (feature-rich) | 0.108 | <.001 | *** |
| | Social ties | Human | Digital twin (demographic) | 0.683 | <.001 | *** |
| | Social ties | Human | Digital twin (feature-rich) | 0.665 | <.001 | *** |
| | Social ties | Digital twin (demographic) | Digital twin (feature-rich) | 0.112 | <.001 | *** |
| | Threat | Human | Digital twin (demographic) | 0.693 | <.001 | *** |
| | Threat | Human | Digital twin (feature-rich) | 0.673 | <.001 | *** |
| | Threat | Digital twin (demographic) | Digital twin (feature-rich) | 0.11 | <.001 | *** |
| DeepSeek V4-Flash | Nostalgia | Human | Digital twin (demographic) | 0.401 | <.001 | *** |
| | Nostalgia | Human | Digital twin (feature-rich) | 0.305 | <.001 | *** |
| | Nostalgia | Digital twin (demographic) | Digital twin (feature-rich) | 0.132 | <.001 | *** |
| | Agency | Human | Digital twin (demographic) | 0.405 | <.001 | *** |
| | Agency | Human | Digital twin (feature-rich) | 0.309 | <.001 | *** |
| | Agency | Digital twin (demographic) | Digital twin (feature-rich) | 0.128 | <.001 | *** |
| | Communion | Human | Digital twin (demographic) | 0.405 | <.001 | *** |
| | Communion | Human | Digital twin (feature-rich) | 0.308 | <.001 | *** |
| | Communion | Digital twin (demographic) | Digital twin (feature-rich) | 0.129 | <.001 | *** |
| | Social ties | Human | Digital twin (demographic) | 0.401 | <.001 | *** |
| | Social ties | Human | Digital twin (feature-rich) | 0.299 | <.001 | *** |
| | Social ties | Digital twin (demographic) | Digital twin (feature-rich) | 0.134 | <.001 | *** |
| | Threat | Human | Digital twin (demographic) | 0.406 | <.001 | *** |
| | Threat | Human | Digital twin (feature-rich) | 0.308 | <.001 | *** |
| | Threat | Digital twin (demographic) | Digital twin (feature-rich) | 0.132 | <.001 | *** |
| Qwen 3.5 Plus | Nostalgia | Human | Digital twin (demographic) | 0.387 | <.001 | *** |
| | Nostalgia | Human | Digital twin (feature-rich) | 0.22 | <.001 | *** |
| | Nostalgia | Digital twin (demographic) | Digital twin (feature-rich) | 0.27 | <.001 | *** |
| | Agency | Human | Digital twin (demographic) | 0.392 | <.001 | *** |
| | Agency | Human | Digital twin (feature-rich) | 0.227 | <.001 | *** |
| | Agency | Digital twin (demographic) | Digital twin (feature-rich) | 0.268 | <.001 | *** |
| | Communion | Human | Digital twin (demographic) | 0.394 | <.001 | *** |
| | Communion | Human | Digital twin (feature-rich) | 0.229 | <.001 | *** |
| | Communion | Digital twin (demographic) | Digital twin (feature-rich) | 0.27 | <.001 | *** |
| | Social ties | Human | Digital twin (demographic) | 0.382 | <.001 | *** |
| | Social ties | Human | Digital twin (feature-rich) | 0.222 | <.001 | *** |
| | Social ties | Digital twin (demographic) | Digital twin (feature-rich) | 0.273 | <.001 | *** |
| | Threat | Human | Digital twin (demographic) | 0.389 | <.001 | *** |
| | Threat | Human | Digital twin (feature-rich) | 0.22 | <.001 | *** |
| | Threat | Digital twin (demographic) | Digital twin (feature-rich) | 0.267 | <.001 | *** |
| Study 4b | | | | | | |
| Gemma-4-31B-it | Socioeconomic status | Human | Digital twin (demographic) | 0.452 | 0.002 | ** |
| | Socioeconomic status | Human | Digital twin (feature-rich) | 0.143 | 0.78 | |
| | Socioeconomic status | Digital twin (demographic) | Digital twin (feature-rich) | 0.31 | 0.054 | |
| | Sociability | Human | Digital twin (demographic) | 0.548 | <.001 | *** |
| | Sociability | Human | Digital twin (feature-rich) | 0.262 | 0.11 | |
| | Sociability | Digital twin (demographic) | Digital twin (feature-rich) | 0.381 | 0.007 | ** |
| | Competence | Human | Digital twin (demographic) | 0.548 | <.001 | *** |
| | Competence | Human | Digital twin (feature-rich) | 0.19 | 0.407 | |
| | Competence | Digital twin (demographic) | Digital twin (feature-rich) | 0.429 | <.001 | *** |
| | Morality | Human | Digital twin (demographic) | 0.381 | 0.029 | * |
| | Morality | Human | Digital twin (feature-rich) | 0.214 | 0.295 | |
| | Morality | Digital twin (demographic) | Digital twin (feature-rich) | 0.262 | 0.169 | |
| | Appearance | Human | Digital twin (demographic) | 0.452 | 0.002 | ** |
| | Appearance | Human | Digital twin (feature-rich) | 0.19 | 0.4 | |
| | Appearance | Digital twin (demographic) | Digital twin (feature-rich) | 0.31 | 0.071 | |
| gpt-oss-120b | Socioeconomic status | Human | Digital twin (demographic) | 0.381 | 0.005 | ** |
| | Socioeconomic status | Human | Digital twin (feature-rich) | 0.643 | <.001 | *** |
| | Socioeconomic status | Digital twin (demographic) | Digital twin (feature-rich) | 0.381 | 0.01 | * |
| | Sociability | Human | Digital twin (demographic) | 0.357 | 0.008 | ** |
| | Sociability | Human | Digital twin (feature-rich) | 0.643 | <.001 | *** |
| | Sociability | Digital twin (demographic) | Digital twin (feature-rich) | 0.381 | 0.008 | ** |
| | Competence | Human | Digital twin (demographic) | 0.333 | 0.013 | * |
| | Competence | Human | Digital twin (feature-rich) | 0.69 | <.001 | *** |
| | Competence | Digital twin (demographic) | Digital twin (feature-rich) | 0.381 | 0.012 | * |
| | Morality | Human | Digital twin (demographic) | 0.31 | 0.029 | * |
| | Morality | Human | Digital twin (feature-rich) | 0.619 | <.001 | *** |
| | Morality | Digital twin (demographic) | Digital twin (feature-rich) | 0.381 | 0.009 | ** |
| | Appearance | Human | Digital twin (demographic) | 0.286 | 0.053 | |
| | Appearance | Human | Digital twin (feature-rich) | 0.667 | <.001 | *** |
| | Appearance | Digital twin (demographic) | Digital twin (feature-rich) | 0.452 | 0.001 | ** |
| DeepSeek V4-Flash | Socioeconomic status | Human | Digital twin (demographic) | 0.476 | 0.002 | ** |
| | Socioeconomic status | Human | Digital twin (feature-rich) | 0.262 | 0.109 | |
| | Socioeconomic status | Digital twin (demographic) | Digital twin (feature-rich) | 0.262 | 0.109 | |
| | Sociability | Human | Digital twin (demographic) | 0.5 | 0.002 | ** |
| | Sociability | Human | Digital twin (feature-rich) | 0.357 | 0.013 | * |
| | Sociability | Digital twin (demographic) | Digital twin (feature-rich) | 0.262 | 0.091 | |
| | Competence | Human | Digital twin (demographic) | 0.5 | <.001 | *** |
| | Competence | Human | Digital twin (feature-rich) | 0.262 | 0.102 | |
| | Competence | Digital twin (demographic) | Digital twin (feature-rich) | 0.333 | 0.025 | * |
| | Morality | Human | Digital twin (demographic) | 0.429 | 0.011 | * |
| | Morality | Human | Digital twin (feature-rich) | 0.286 | 0.096 | |
| | Morality | Digital twin (demographic) | Digital twin (feature-rich) | 0.238 | 0.146 | |
| | Appearance | Human | Digital twin (demographic) | 0.452 | 0.005 | ** |
| | Appearance | Human | Digital twin (feature-rich) | 0.31 | 0.041 | * |
| | Appearance | Digital twin (demographic) | Digital twin (feature-rich) | 0.214 | 0.264 | |
| Qwen 3.5 Plus | Socioeconomic status | Human | Digital twin (demographic) | 0.476 | 0.001 | ** |
| | Socioeconomic status | Human | Digital twin (feature-rich) | 0.262 | 0.097 | |
| | Socioeconomic status | Digital twin (demographic) | Digital twin (feature-rich) | 0.286 | 0.097 | |
| | Sociability | Human | Digital twin (demographic) | 0.405 | 0.005 | ** |
| | Sociability | Human | Digital twin (feature-rich) | 0.381 | 0.005 | ** |
| | Sociability | Digital twin (demographic) | Digital twin (feature-rich) | 0.119 | 0.935 | |
| | Competence | Human | Digital twin (demographic) | 0.381 | 0.013 | * |
| | Competence | Human | Digital twin (feature-rich) | 0.357 | 0.013 | * |
| | Competence | Digital twin (demographic) | Digital twin (feature-rich) | 0.19 | 0.449 | |
| | Morality | Human | Digital twin (demographic) | 0.381 | 0.011 | * |
| | Morality | Human | Digital twin (feature-rich) | 0.405 | 0.003 | ** |
| | Morality | Digital twin (demographic) | Digital twin (feature-rich) | 0.143 | 0.794 | |
| | Appearance | Human | Digital twin (demographic) | 0.31 | 0.154 | |
| | Appearance | Human | Digital twin (feature-rich) | 0.238 | 0.217 | |
| | Appearance | Digital twin (demographic) | Digital twin (feature-rich) | 0.119 | 0.935 | |

Note. D denotes the Kolmogorov–Smirnov test statistic. p-values were obtained via within-participant permutation tests. *** p < .001, ** p < .01, * p < .05.

**Table S61 Construct-validity recommendations for LLM-based digital-twin studies.**

| | Guideline and rationale | Key references |
|---|---|---|
| **Step 1** | **Intended inference** | |
| **Guideline** | **State the role first.** Before generating any data, declare what the simulated responses are meant to do, and raise the evidence bar as the use moves from exploratory construct work, to predicting broad individual profiles, to estimating associations among constructs, to serving as calibrated proxy data, to substituting for human measurement. | Anthis et al. (2025): LLM social simulations can be used cautiously for exploratory studies, with broader use requiring iterative evaluation.<br>Abramski et al. (2026): Semantic networks approximated System 1 and System 2 structures rather than full dual-process reasoning.<br>Kolluri et al. (2025)： Fine-tuned models were positioned as simulations for screening social-science experimental hypotheses.<br>Ludwig et al. (2025): The framework separates prediction from estimation, assigning distinct assumptions to each empirical use. |
| **Rationale** | The validity evidence required scales with the claim: exploration tolerates approximation, whereas measurement substitution demands equivalence. | Meincke et al. (2026): The study tests whether LLMs show humanlike susceptibility to persuasion principles.<br>Pataranutaporn et al. (2025): Well-being prediction was framed as a diagnostic stress test, not policy substitution.<br>Tjuatja et al. (2024): LLM response-distribution shifts were compared with established human survey-bias patterns.<br>Toubia et al. (2025): Twin-2K-500 is introduced as a public testbed for developing and benchmarking digital twins.<br>Wright et al. (2026): LLMs were used as zero-shot scorers of personality from open-ended narratives.<br>Xu et al. (2025)： The study treats LLMs as probes of concept formation from language prediction.<br>Xie et al. (2026): SSDataBench defines validity for LLM-generated social science data as statistical realism.<br>Zewail et al. (2026): LLMs were tested as estimators of average national moral values. |
| **Step 2** | **Level of comparability** | |
| **Guideline** | **Name the level(s) evaluated.** Report which forms of Human–digital-twin correspondence are tested | Abramski et al. (2026): Comparable human, Mistral, and Llama3 networks were analyzed using identical reducibility and bias procedures. |
| | — population/aggregate, item-level (between-person), profile (within-person), construct-score, nomothetic, and measurement — and present each separately. | Gao et al. (2025): LLM response distributions diverged significantly from human participants in the 11-20 money request game.<br>Kim and Lee (2024): The study reports personal-level AUC and public-opinion correlations across missing-data tasks. |
| **Rationale** | A digital twin can pass at one level and fail at another; strong within-person profiles do not by themselves establish item-level (between-person) validity or measurement equivalence. | Li et al. (2025): Simulated distributions were compared with real election outcomes and OpinionQA ground-truth responses.<br>Loru et al. (2025): Outputs often matched expert classifications, especially for unreliable sources, but diverged from human judgments.<br>Michelmann et al. (2025): GPT-3 event boundaries were closer to human consensus than individual annotators, on average.<br>Pataranutaporn et al. (2025): LLM predictions were benchmarked against self-reports and OLS/Lasso estimates across life-satisfaction scales.<br>Peng et al. (2025): The study separately measured matched accuracy, correlations, population means, and response dispersion<br>Strachan et al. (2024): Human and LLM performance was compared across multiple theory-of-mind abilities using shared scoring protocols.<br>Toubia et al. (2025): The dataset supports individual- and aggregate-level comparison between digital twins and human responses.<br>A. Wang et al. (2025): The study compares LLM outputs with in-group representations and out-group imitations across identity groups.<br>Wright et al. (2026): LLM trait scores were compared with self-reports, benchmarks, and nomological networks.<br>Xie et al. (2026): Assessment centres on population-level statistical patterns rather than individual predictability.<br>N. Xu et al. (2025): LLM representations were compared with human judgments and neural activity patterns. |
| | | W. Xu et al. (2025): The study compares human-written and LLM-generated stories at distributional, prompt-level plot diversity.<br>Zewail et al. (2026): GPT estimates were compared with country-level human survey means across six moral foundations. |
| **Step 3** | **Task-demand matching** | |
| **Guideline** | **Match the twin to what the task requires.**<br>**Lower-risk demands** — broad semantic organization, stable profile prediction under rich inputs, and calibrated nomothetic estimation — are defensible with standard validation.<br>**Higher-risk demands** — fine-grained item variation, bounded rationality and cognitive biases, temporal or historical change, human heterogeneity, and individual writing style — require explicit, demand-specific evidence or should be avoided. | Bail (2024): Generative AI may support surveys, experiments, simulations, and text analysis, but task-specific limitations remain.<br>Chen et al. (2025): GPT mirrors human biases in almost half of experiments.<br>Gui and Toubia (2025): Blind LLM simulations violated unconfoundedness by letting unspecified variables vary with treatment.<br>Kozlowski et al. (2024): The study used a model trained through 2019 to test whether later COVID-19 polarization was prefigured.<br>Loru et al. (2025): LLMs operationalized credibility through lexical associations and statistical priors rather than contextual reasoning.<br>Michelmann et al. (2025): Zero-shot GPT-3 segmented continuous narrative text into events significantly above chance. |
| **Rationale** | Digital twins are not universally interchangeable substitutes for human respondents; an inference that hinges on a capacity they lack is invalid regardless of surface accuracy, so the evidence required should be set before validation begins. | Strachan et al. (2024): GPT-4 matched humans on indirect requests, false beliefs and misdirection but struggled with faux pas.<br>Xie et al. (2026): Sparse conditioning compressed real-world heterogeneity into simplified typological structures.<br>N. Xu et al. (2025): The task probes semantic concept inference from linguistic descriptions across contexts.<br>W. Xu et al. (2025): Creative story generation is evaluated as a high-demand task requiring diverse narrative alternatives.<br>Zewail et al. (2026): Models poorly captured global moral diversity and stereotyped cultural populations predictably. |

| **Step 4** | **Person-specific inputs** | |
|---|---|---|
| **Guideline** | **Document and justify every input.** List all information used to condition each twin (demographic, behavioral, questionnaire, textual, and contextual), justify why each should predict the target, confirm that no target item, paraphrase, or criterion score leaks into the prompt, and benchmark feature-rich prompts against reduced-input ones. | Argyle et al. (2023): GPT-3 was conditioned on sociodemographic backstories from real survey participants to generate silicon samples.<br>Kim and Lee (2024): Fine-tuned LLMs used survey questions, individual responses, and years to personalize opinion prediction.<br>Li et al. (2025): Increasing LLM-generated persona content exacerbated skew in simulated opinions and preferences.<br>Park et al. (2024): Two-hour qualitative interviews, averaging 6,491 words, formed each participant-specific agent knowledge base. |
| **Rationale** | Construct-relevance, not sheer volume, governs validity, and a single cue can silently activate correlated but unobserved attributes. | Pataranutaporn et al. (2025): Natural-language profiles encoded socioeconomic, attitudinal, and psychological covariates without observed self-reports.<br>Peng et al. (2025): Each twin used 500 prior answers covering demographics, personality, cognition, preferences, biases, and pricing.<br>A. Wang et al. (2025): The paper evaluates demographic identity prompts and alternatives such as identity-coded names.<br>Westwood et al. (2025): Synthetic respondents used demographic personas and memory of prior answers.<br>Wright et al. (2026): Person-specific inputs were spontaneous thought streams and daily video-diary narratives. |
| **Step 5** | **Generation procedure** | |
| **Guideline** | **Report the full instrument.** Specify the model name, version, and provider; system and user prompt structure; persona-construction rules; item format and response options; output schema; decoding settings (temperature and top-p); parsing, exclusion, and reasoning procedures; and whether the N simulated respondents were | Gui and Toubia (2025): Ambiguous prompts mapped distinct causal questions to identical inputs, limiting fine-tuning and RAG.<br>Kolluri et al. (2025): The study compared direct prompting, reasoning prompts, SFT, reasoning-trace SFT, and DPO.<br>Lyman et al. (2025): Prompt adaptations were required because aligned and base models responded differently to identical prompts. |

| | generated through separate API calls or as multiple answers within a single chat exchange. | Meincke et al. (2026): Treatment and control prompts varied seven persuasion principles across models and request categories. |
|---|---|---|
| **Rationale** | Prompts and decoding settings are part of the measurement instrument, and drawing the N respondents as separate calls captures between-person variance rather than within-call stochasticity; undocumented settings cannot be reproduced or audited. | Park et al. (2024): The architecture combined interview transcripts, memory retrieval, and expert reflection to generate responses.<br>Peng et al. (2025): Prompts embedded persona profiles, randomized survey questions, JSON formatting, post-processing, and retry on validation failure.<br>Xie et al. (2026): Llama-3.1-8B was fine-tuned with LoRA using structured prompts and JSON survey outputs.<br>W. Xu et al. (2025): GPT-4 and LLaMA-3 generated continuations under specified prompts, sampling settings, and segmentation rules. |
| **Step 6** | **Validation battery** | |
| **Guideline** | **Evaluate beyond accuracy.** Report descriptive and prediction-error agreement (means, standard deviations, and mean absolute error); item-level (between-person) and profile (within-person) correlations with human data; and distributional fidelity (variance, under-dispersion, floor and ceiling effects, and a distributional distance such as the Wasserstein distance). Also report raw and disattenuated construct-score correlations and nomothetic span; for multicomponent psychological measures, assess measurement invariance (Step 7). For free text, evaluate surface linguistic form and construct-level content as separate targets. | Bisbee et al. (2024): ChatGPT's average scores matched ANES, but responses varied less and regression coefficients often differed significantly.<br>Kolluri et al. (2025): Validation used individual accuracy, Wasserstein distributional alignment, empirical bounds, and demographic parity.<br>Li et al. (2025): Validation combined election forecasts, 500 OpinionQA questions, alignment scores, and semantic profile analysis.<br>Loru et al. (2025): Validation combined expert benchmarks, a controlled human experiment, confusion matrices, keyword analyses, and criterion rankings.<br>Lyman et al. (2025): Benchmarking assessed task completion, AI commentary, persona consistency, and pattern correspondence before human-data comparison.<br>Michelmann et al. (2025): Validation used Hamming distances and continuous boundary probabilities against human annotations.<br>Pataranutaporn et al. (2025): Validation assessed MAE, scale extremes, predictor coefficients, country effects, semantic gradients, and injection tests. |
| **Rationale** | Accuracy alone can mask systematic error, response compression toward the mean, and measurement nonequivalence. | Wright et al. (2026): Scores showed convergence and predictive validity for daily behaviours and mental health outcomes.<br>Xie et al. (2026): The benchmark tests distributions, associations, predictions, event sequences, and sequence-covariate associations.<br>W. Xu et al. (2025): Sui Generis scores were checked against human surprise judgments and alternative similarity metrics. |
| **Step 7** | **Measurement invariance** | |
| **Guideline** | **Test invariance before score substitution.** Before treating digital twin and human scores as the same measure, evaluate — for network psychometric models, network configural then network metric invariance — whether a common structure holds, items occupy comparable roles, and dispersions match, and whether achieving partial invariance forced dropping so many items that the retained set no longer represents the construct. Where invariance is unsupported, label outputs model-generated proxy estimates. | Russell-Lasalandra et al. (2024): Network psychometrics tested theoretical structure, local independence, and item stability in AI-generated scales.<br>Sühr et al. (2025): Using exploratory factor analysis (EFA), confirmatory factor analysis (CFA), and reliability measures, the authors show the BFI-2 lacks measurement invariance between humans and LLMs, invalidating personality scores. |
| **Rationale** | Without invariance, any Human–digital-twin comparison may reflect measurement differences rather than the construct itself. | |
| **Step 8** | **Statistical calibration** | |
| **Guideline** | **Calibrate cautiously.** When digital twins are used as proxy data, treat calibration as an empirical validation step: anchor calibration | Angelopoulos et al. (2023): Gold-standard data quantify prediction error, rectifying imputed estimates into statistically valid confidence intervals. |

| | models in a gold-standard human sample, evaluate performance in held-out data, and specify which targets calibration improves and which it does not. | Broska et al. (2025): PPI combines human observations with LLM predictions to produce valid estimates and narrower confidence intervals.<br>Hullman et al. (2026): The paper distinguishes heuristic substitution from statistical calibration, which corrects LLM predictions using human validation data. |
|---|---|---|
| **Rationale** | Calibration corrects aggregate or relational estimates but not the full human data-generating process. | Kolluri et al. (2025): Human response data were used to fine-tune models and evaluate held-out prediction improvements.<br>Ludwig et al. (2025): Validation samples debias LLM outputs, restoring consistency, coverage, and often tighter standard errors. |
| **Step 9** | **Generalizability** | |
| **Guideline** | **Replicate before generalizing.** Test whether the primary validation results hold across the model families, datasets, populations, languages, tasks, and input conditions to which the claim is intended to generalize, and limit claims to the settings in which such robustness has been demonstrated. | De Deyne et al. (2024): Analyses included English and Spanish semantic datasets with human relation annotations.<br>Gao et al. (2025): Prompts in English, Chinese, Spanish, and German produced different response distributions.<br>Pataranutaporn et al. (2025): LLM well-being predictions varied across 64 countries, with larger errors in underrepresented settings.<br>Tao et al. (2024): Cultural alignment varied across 107 countries/territories, even under country-specific prompting. |
| **Rationale** | Human–digital-twin correspondence can vary across models, constructs, populations, languages, and tasks; evidence from a single model or setting cannot support general claims about LLM-based digital twins. | C. Wang et al. (2025): LLM knowledge learning and transfer differed across high-, medium-, and low-resource languages.<br>Xie et al. (2026): Benchmark datasets covered the United States, China, and the United Kingdom populations.<br>Zewail et al. (2026): GPT moral-value estimates diverged across 48 countries, showing limited cross-cultural generalizability. |

*Note.* Digital twins should be judged not by surface humanlikeness but by whether their outputs support the same psychometric and substantive inferences as human data. The steps proceed from planning the intended use (Steps 1–3), to constructing the twins (Steps 4–5), to evaluating them (Steps 6–9), and operationalize the construct-validity framework developed across the four studies. Evidentiary requirements increase as the intended use moves from exploration toward measurement substitution. References listed under each step are intended as representative anchors for that recommendation rather than an exhaustive or mutually exclusive classification of their relevance.

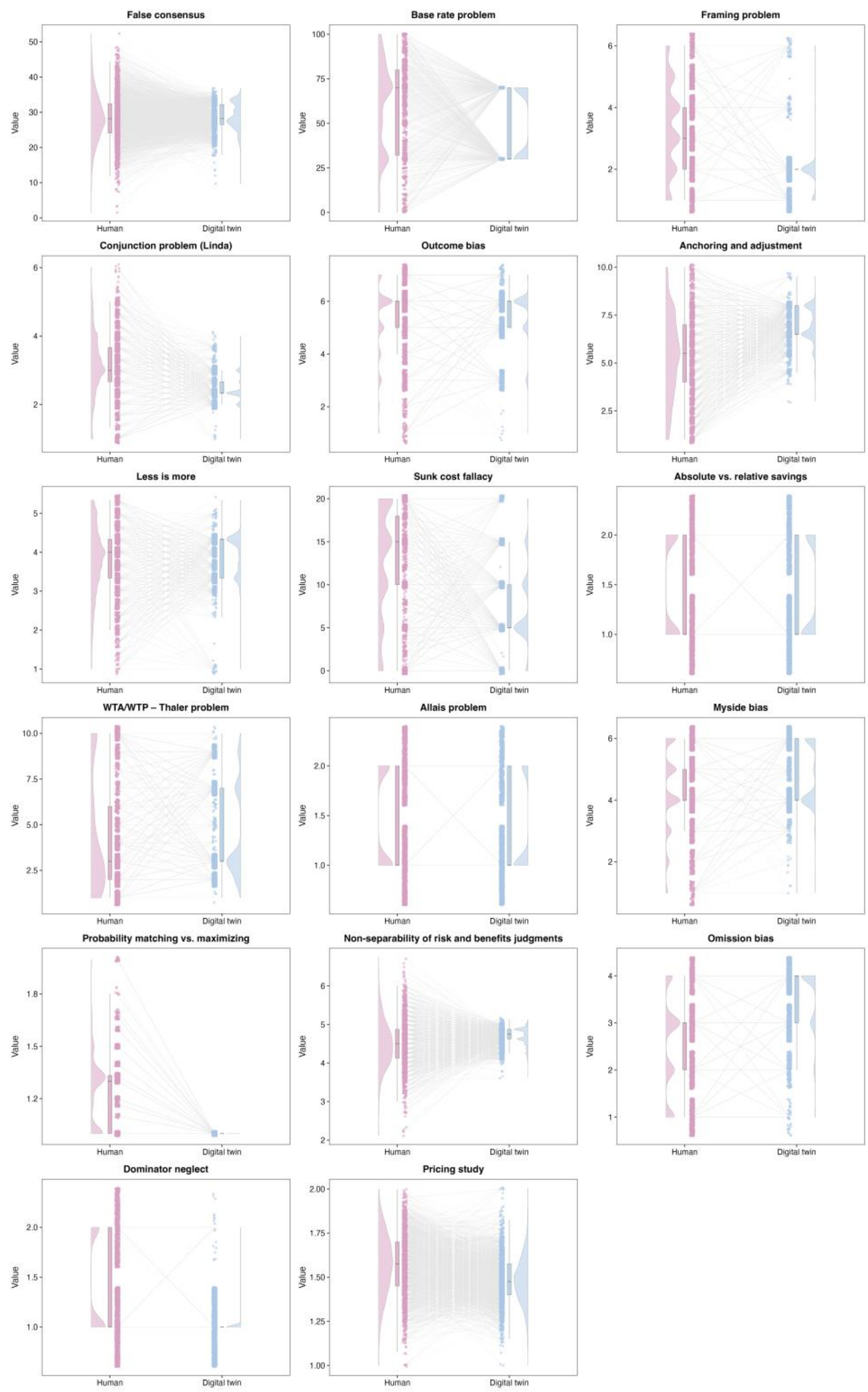

**Supplementary Fig 1. Distribution of responses for humans and digital twins on the heuristics and biases experiments**

The figure visualizes the distribution of responses for Humans (left, pink) and Digital twins (right, blue) for each task. Each panel consists of a half-violin plot showing the density distribution, a boxplot indicating the median and interquartile range, and jittered raw data points. Grey lines connect the responses to the same specific problem item, illustrating the item-level correspondence between Human and LLM responses.

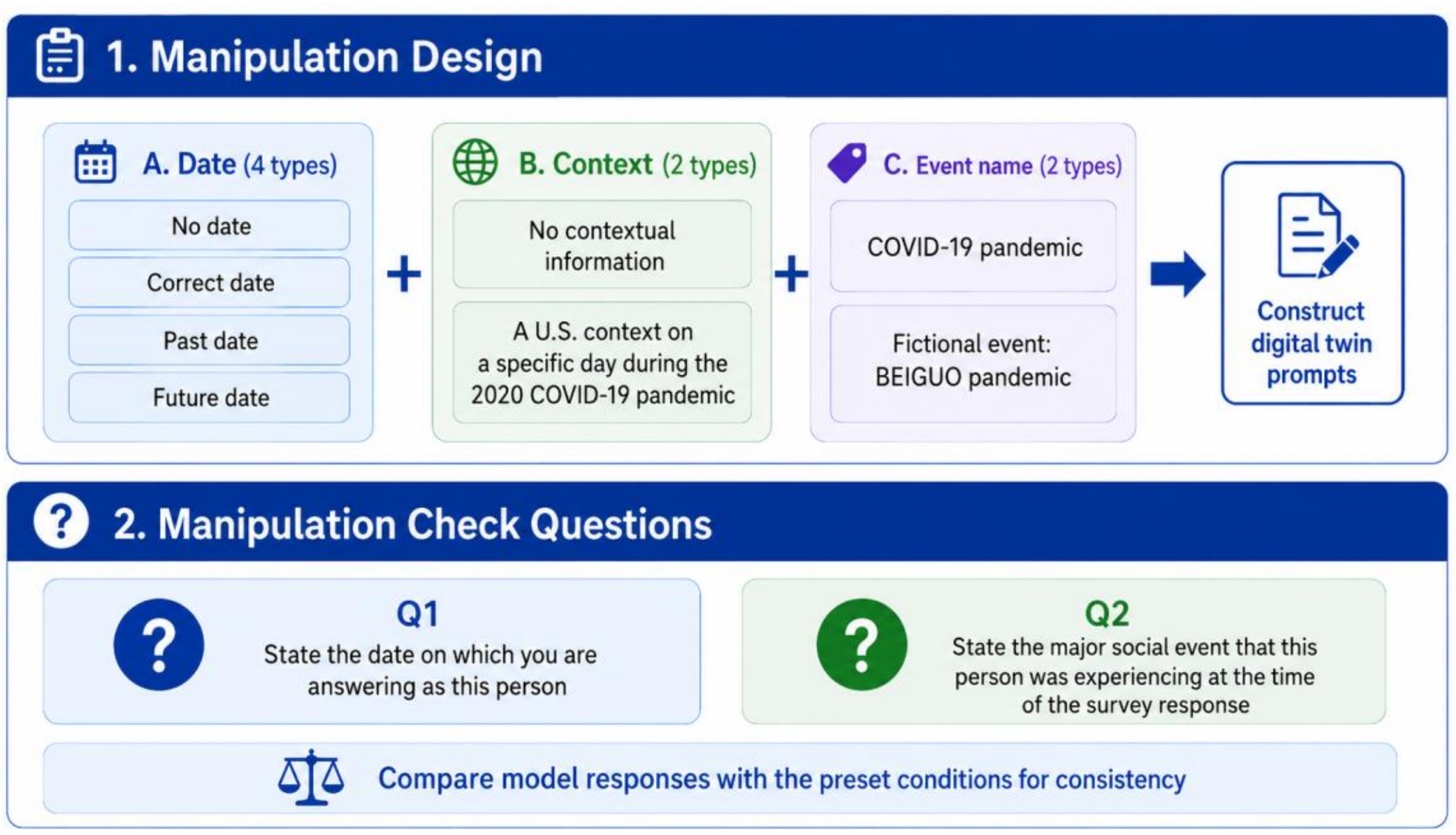

**Supplementary Fig 2 (a). Temporal prompt manipulation design and manipulation check workflow**

| | Accuracy | | | | Item correlation | | | | Profile correlation | | | | Slope | | | |
|---|---|---|---|---|---|---|---|---|---|---|---|---|---|---|---|---|
| | No date No context | No date COVID ctx | Target date COVID ctx | Target date No context | No date No context | No date COVID ctx | Target date COVID ctx | Target date No context | No date No context | No date COVID ctx | Target date COVID ctx | Target date No context | No date No context | No date COVID ctx | Target date COVID ctx | Target date No context |
| **W1** | 0.74 | 0.75 | 0.75 | 0.75 | 0.19 | 0.21 | 0.22 | 0.21 | 0.43 | 0.45 | 0.44 | 0.45 | -0.63 | -0.62 | -0.64 | -0.63 |
| **W2** | 0.75 | 0.75 | 0.75 | 0.76 | 0.27 | 0.26 | 0.27 | 0.26 | 0.48 | 0.48 | 0.49 | 0.49 | -0.55 | -0.55 | -0.55 | -0.55 |
| **W3** | 0.75 | 0.74 | 0.75 | 0.75 | 0.22 | 0.21 | 0.22 | 0.22 | 0.46 | 0.44 | 0.45 | 0.45 | -0.63 | -0.65 | -0.65 | -0.64 |
| **W4** | 0.76 | 0.75 | 0.75 | 0.75 | 0.33 | 0.31 | 0.32 | 0.32 | 0.52 | 0.50 | 0.50 | 0.51 | -0.55 | -0.58 | -0.58 | -0.57 |
| **W5** | 0.75 | 0.75 | 0.75 | 0.75 | 0.27 | 0.27 | 0.27 | 0.28 | 0.48 | 0.49 | 0.49 | 0.49 | -0.55 | -0.55 | -0.54 | -0.54 |
| **W6** | 0.74 | 0.74 | 0.74 | 0.74 | 0.23 | 0.23 | 0.22 | 0.24 | 0.47 | 0.46 | 0.46 | 0.47 | -0.63 | -0.64 | -0.64 | -0.63 |
| **W7** | 0.73 | 0.72 | 0.73 | 0.72 | 0.26 | 0.24 | 0.26 | 0.25 | 0.36 | 0.34 | 0.36 | 0.34 | -0.67 | -0.69 | -0.68 | -0.69 |
| **W8** | 0.73 | 0.73 | 0.73 | 0.73 | 0.22 | 0.23 | 0.21 | 0.23 | 0.46 | 0.46 | 0.46 | 0.46 | -0.56 | -0.56 | -0.56 | -0.56 |
| **W9** | 0.74 | 0.74 | 0.74 | 0.75 | 0.23 | 0.21 | 0.21 | 0.24 | 0.48 | 0.46 | 0.46 | 0.49 | -0.63 | -0.65 | -0.64 | -0.62 |
| **W10** | 0.77 | 0.77 | 0.77 | 0.77 | 0.34 | 0.34 | 0.34 | 0.34 | 0.58 | 0.57 | 0.59 | 0.58 | -0.44 | -0.45 | -0.44 | -0.43 |
| **W11** | 0.75 | 0.75 | 0.75 | 0.75 | 0.27 | 0.28 | 0.28 | 0.29 | 0.48 | 0.49 | 0.50 | 0.50 | -0.54 | -0.53 | -0.52 | -0.52 |
| **W12** | 0.74 | 0.74 | 0.74 | 0.74 | 0.31 | 0.27 | 0.28 | 0.30 | 0.50 | 0.45 | 0.47 | 0.49 | -0.58 | -0.63 | -0.61 | -0.59 |
| **W13** | 0.73 | 0.73 | 0.73 | 0.73 | 0.25 | 0.25 | 0.27 | 0.26 | 0.46 | 0.45 | 0.47 | 0.47 | -0.60 | -0.60 | -0.58 | -0.59 |
| **W14** | 0.74 | 0.74 | 0.74 | 0.74 | 0.25 | 0.23 | 0.25 | 0.25 | 0.47 | 0.44 | 0.45 | 0.45 | -0.61 | -0.64 | -0.64 | -0.63 |
| **W15** | 0.73 | 0.72 | 0.73 | 0.73 | 0.26 | 0.25 | 0.26 | 0.27 | 0.47 | 0.44 | 0.46 | 0.47 | -0.60 | -0.61 | -0.59 | -0.59 |
| **W16** | 0.74 | 0.74 | 0.74 | 0.75 | 0.28 | 0.27 | 0.30 | 0.29 | 0.48 | 0.46 | 0.48 | 0.47 | -0.58 | -0.60 | -0.58 | -0.60 |

**Supplementary Fig 2 (b). Digital-twin performance under temporal prompt manipulations across 16 survey waves**

Heatmaps show digital-twin performance across 16 survey waves under four temporal-prompt conditions: no date with no contextual information, no date with COVID-19 contextual information, target date with COVID-19 contextual information, and target date with no contextual information. Performance was evaluated using accuracy, item-level correlation, profile correlation, and prediction-error slope, with human responses used as the gold standard. Across waves, performance remained highly similar across temporal-prompt conditions, indicating that adding the target date or COVID-19 context did not systematically improve digital-twin alignment with human responses.

(a)

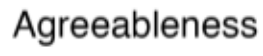

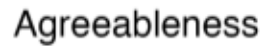

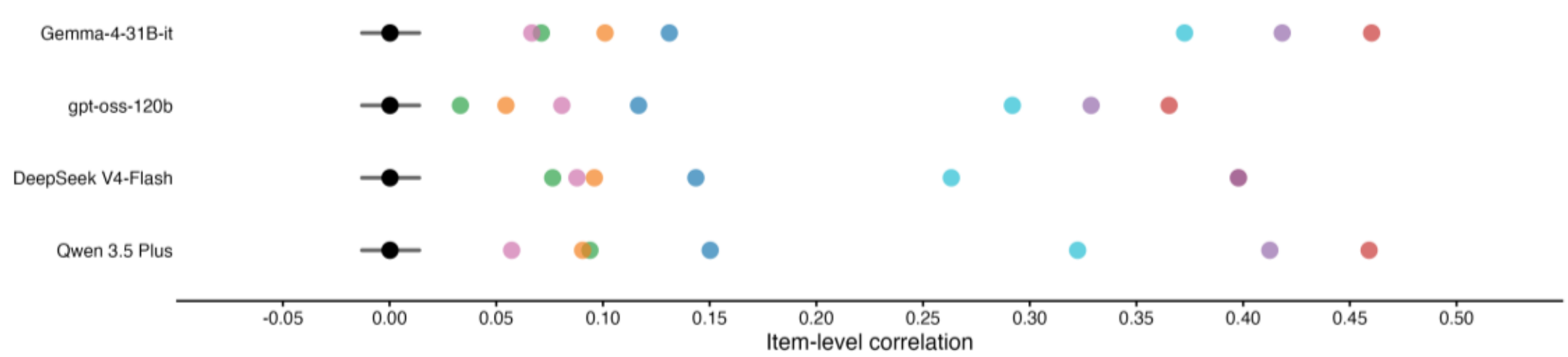

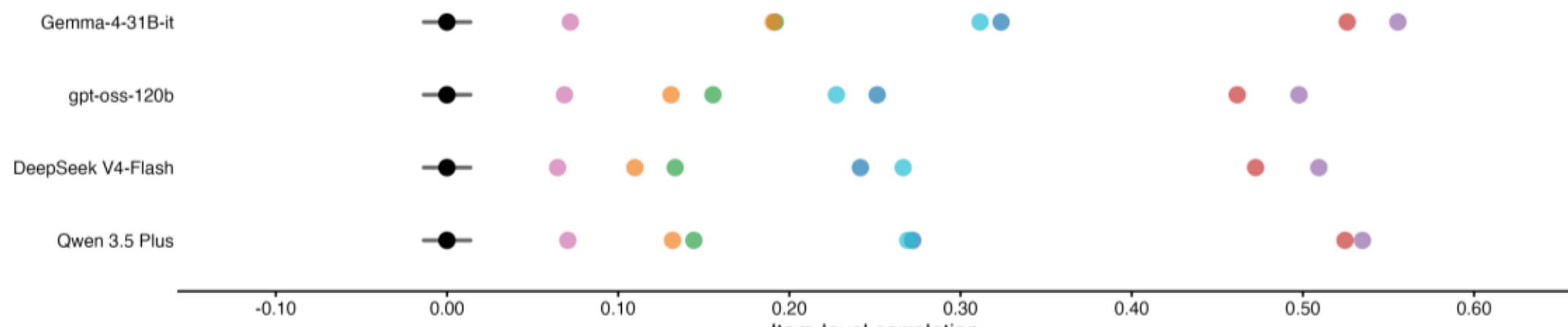

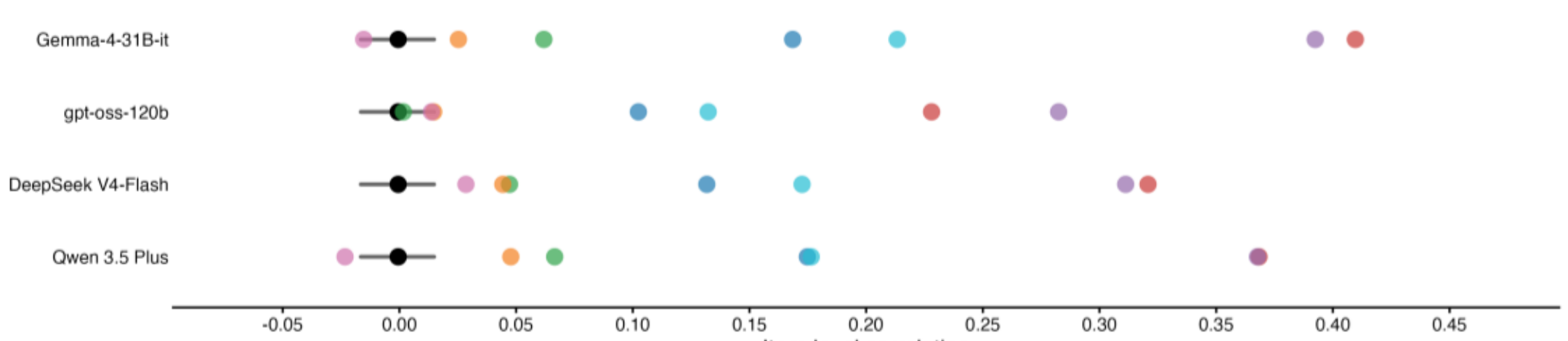

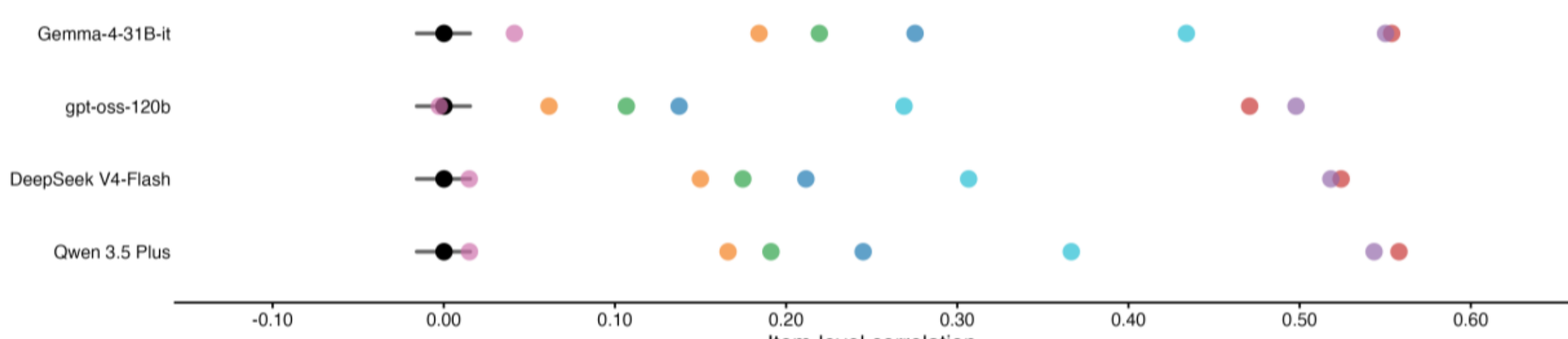

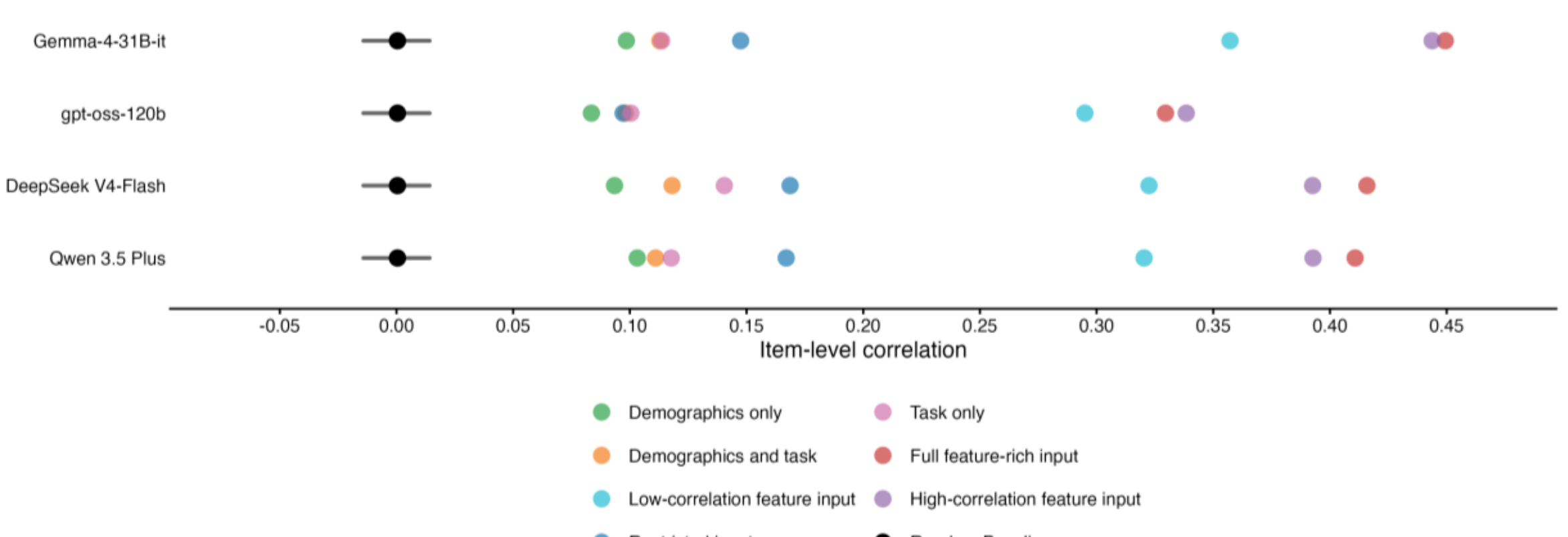

(c)

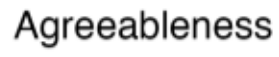

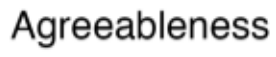
(d)

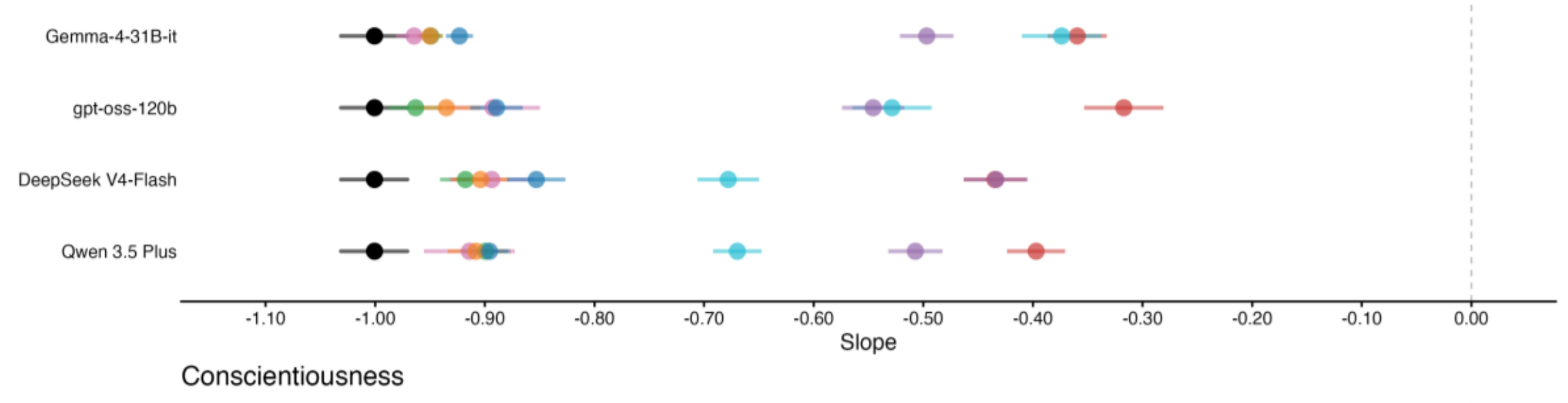

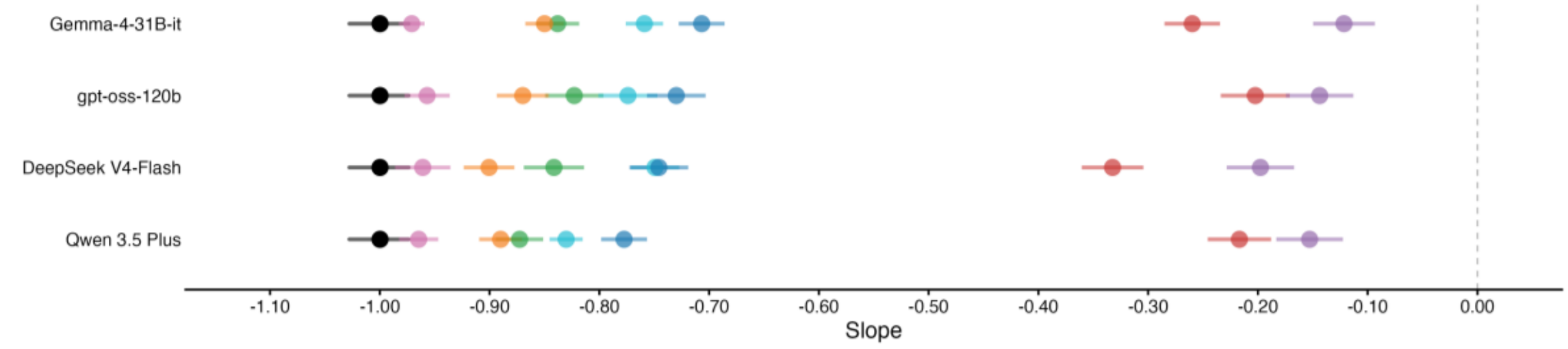

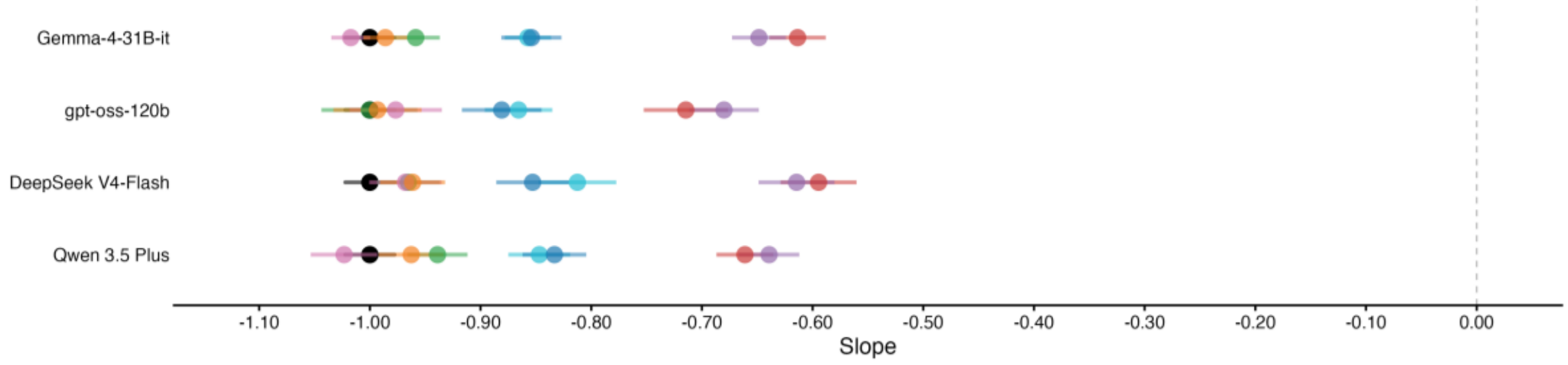

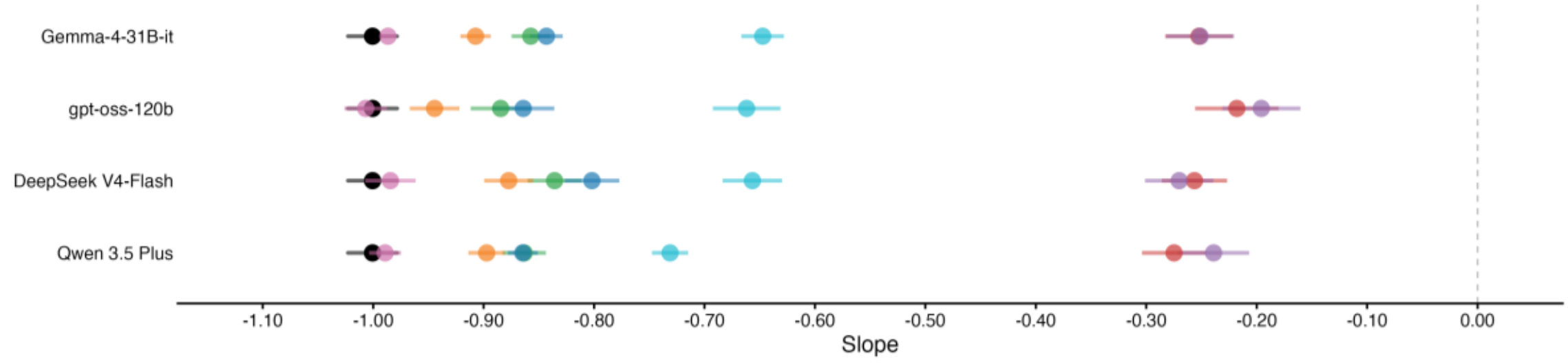

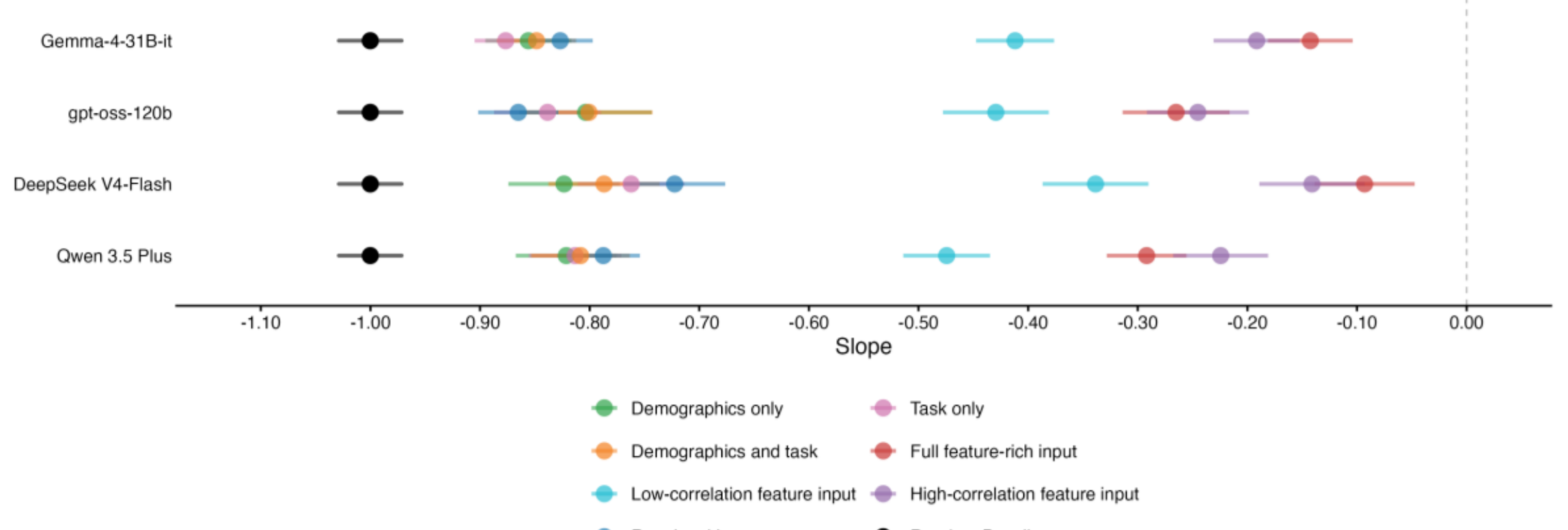

(e)

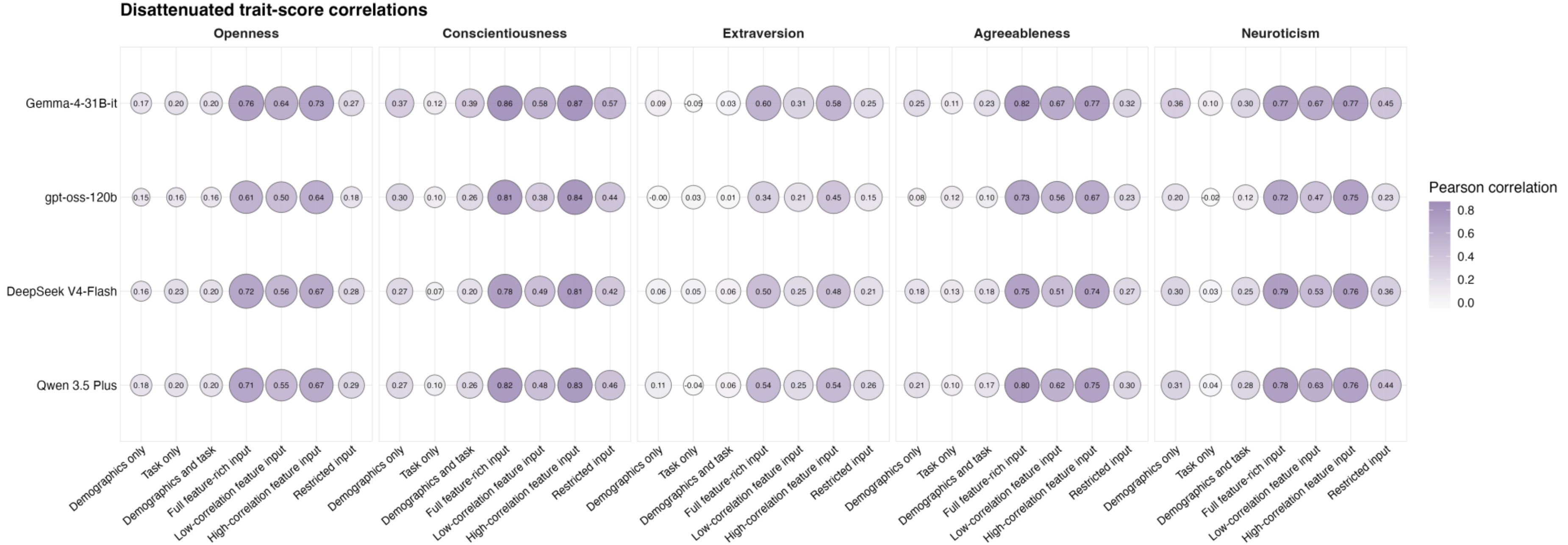

(f)

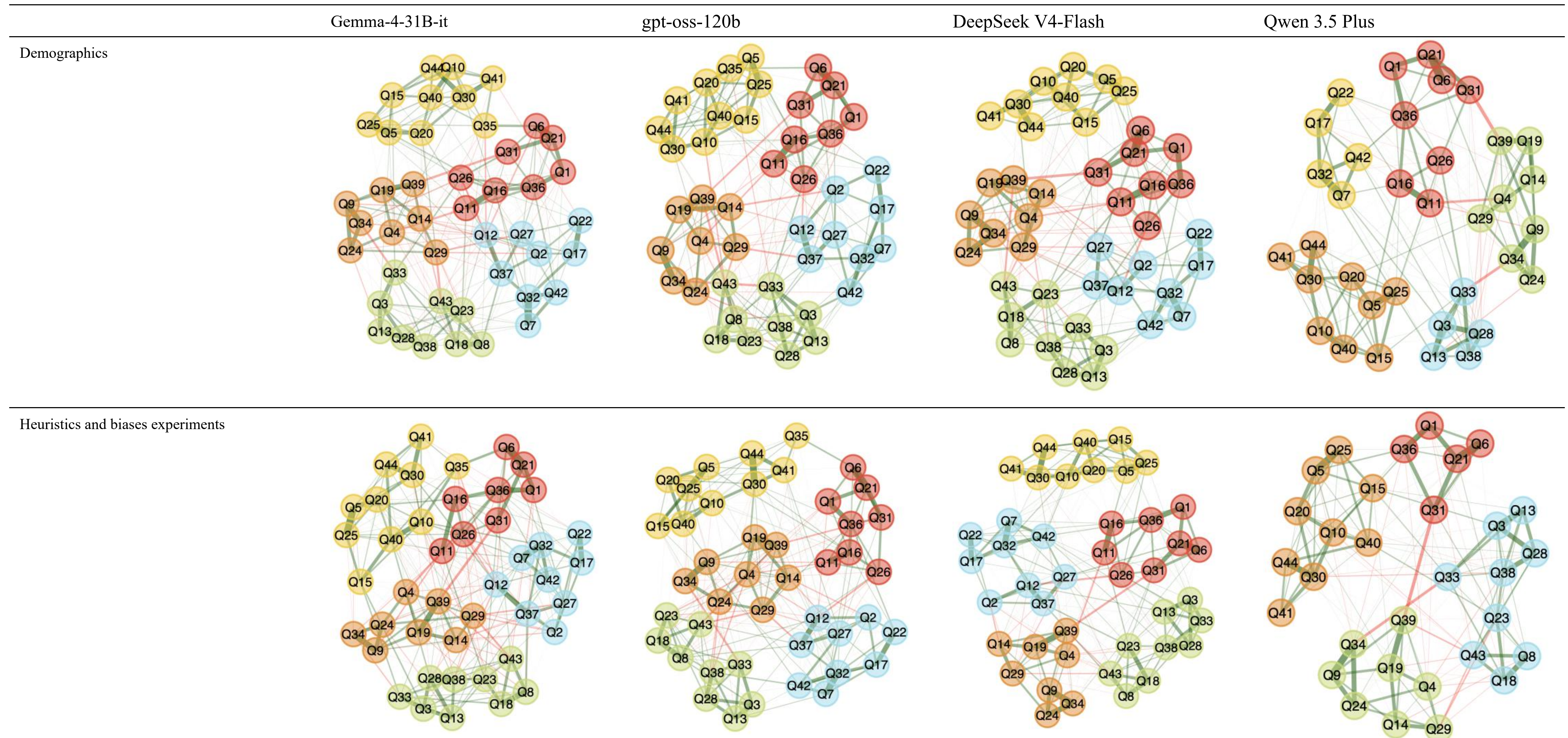

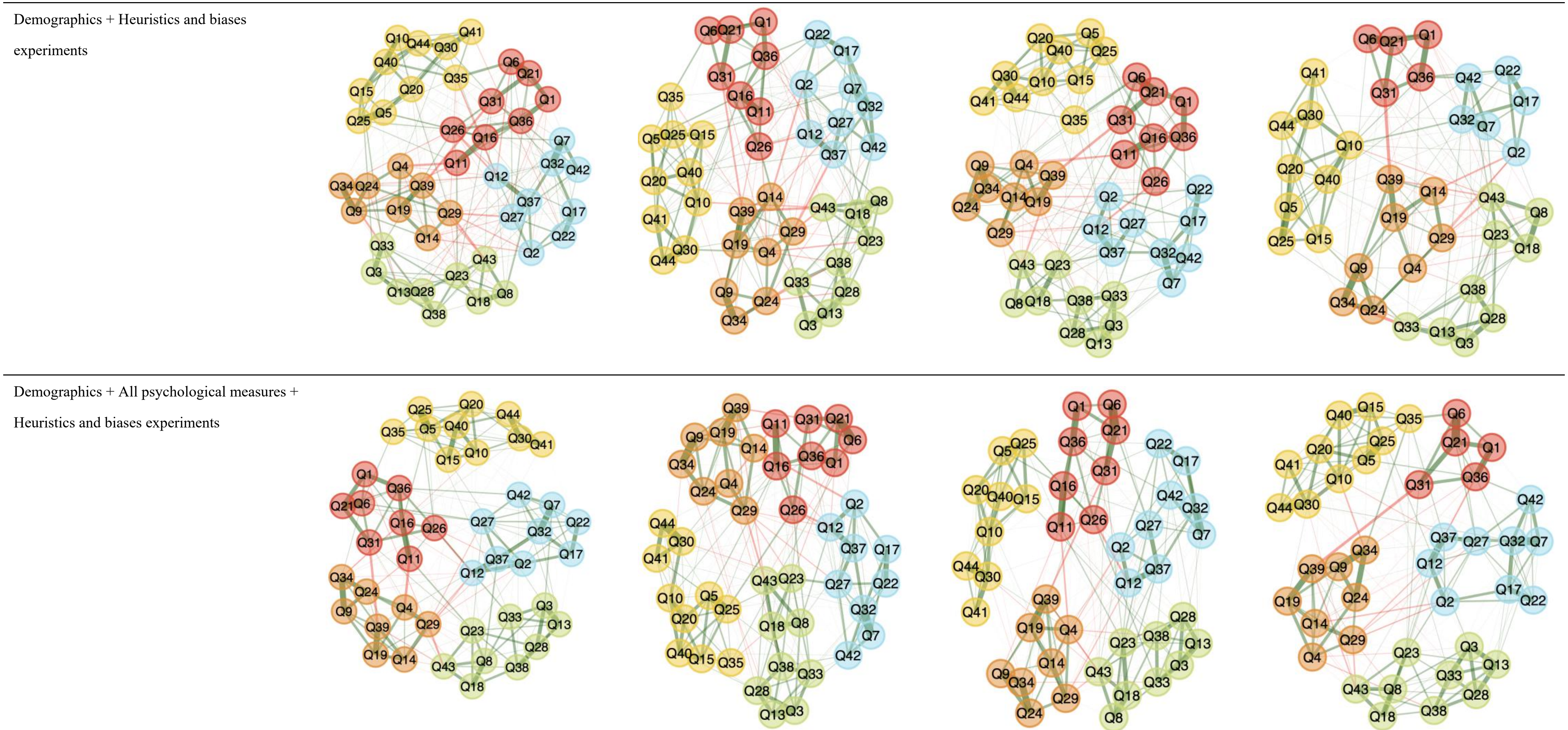

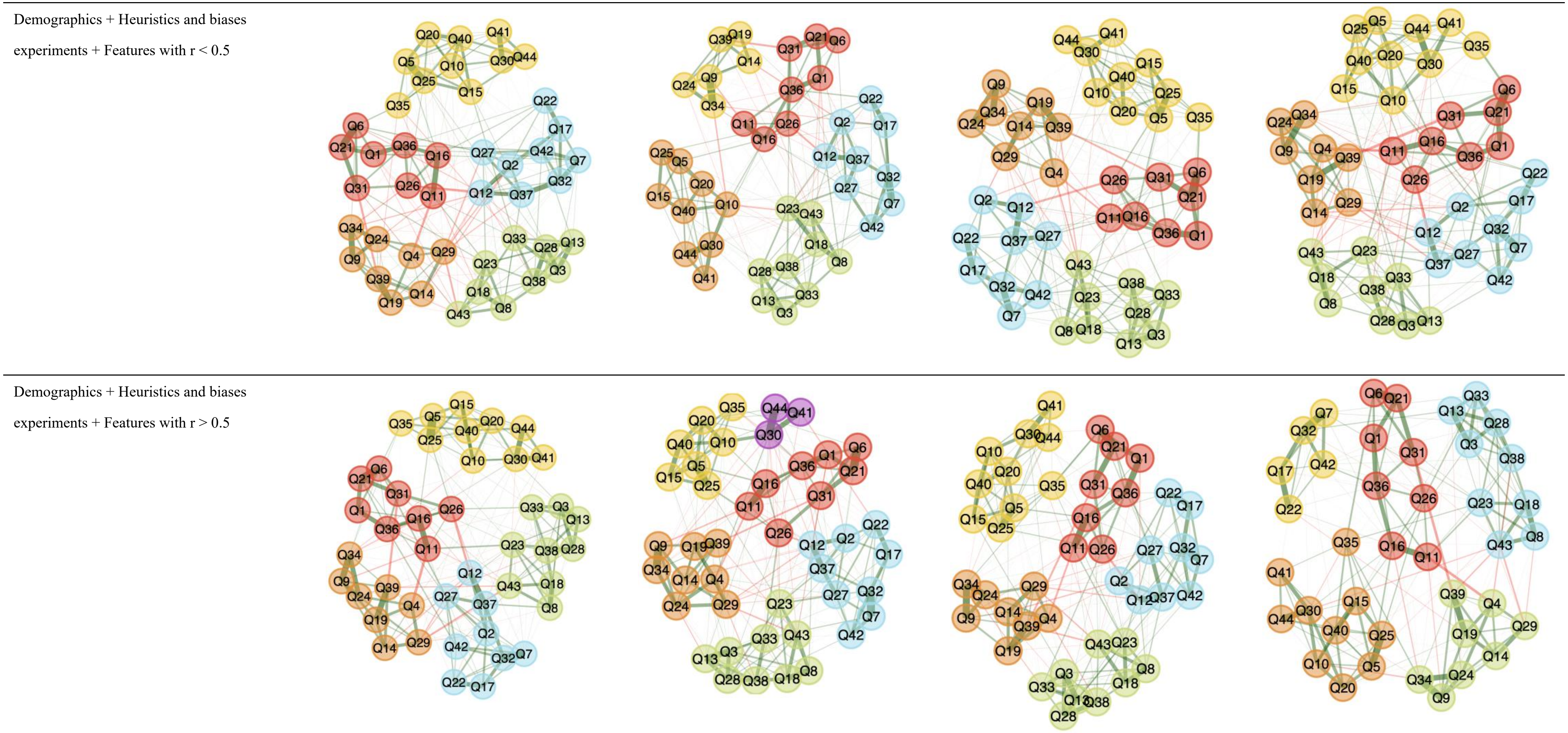

| | |
|---|---|
| Demographics + Heuristics and biases experiments + Features with $r < 0.3$ | 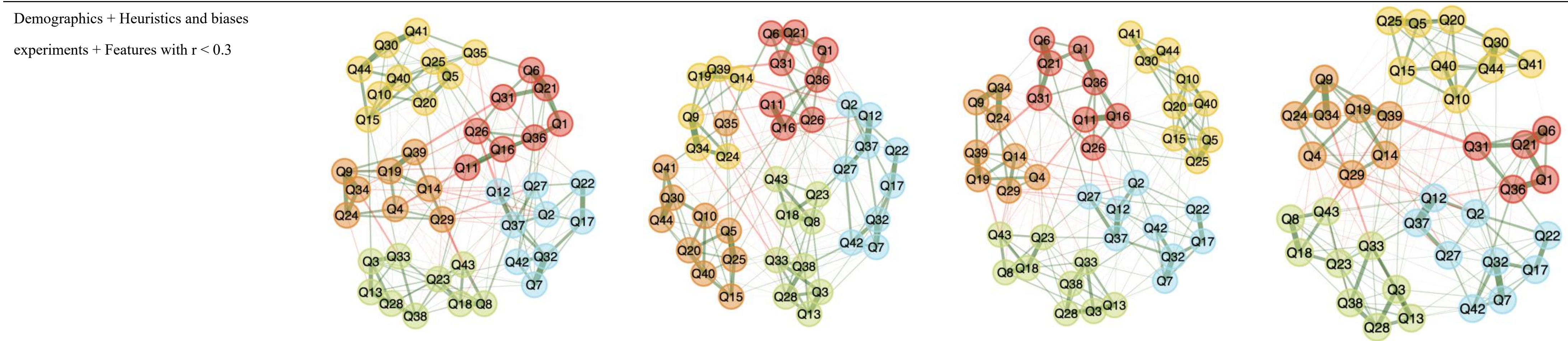 |

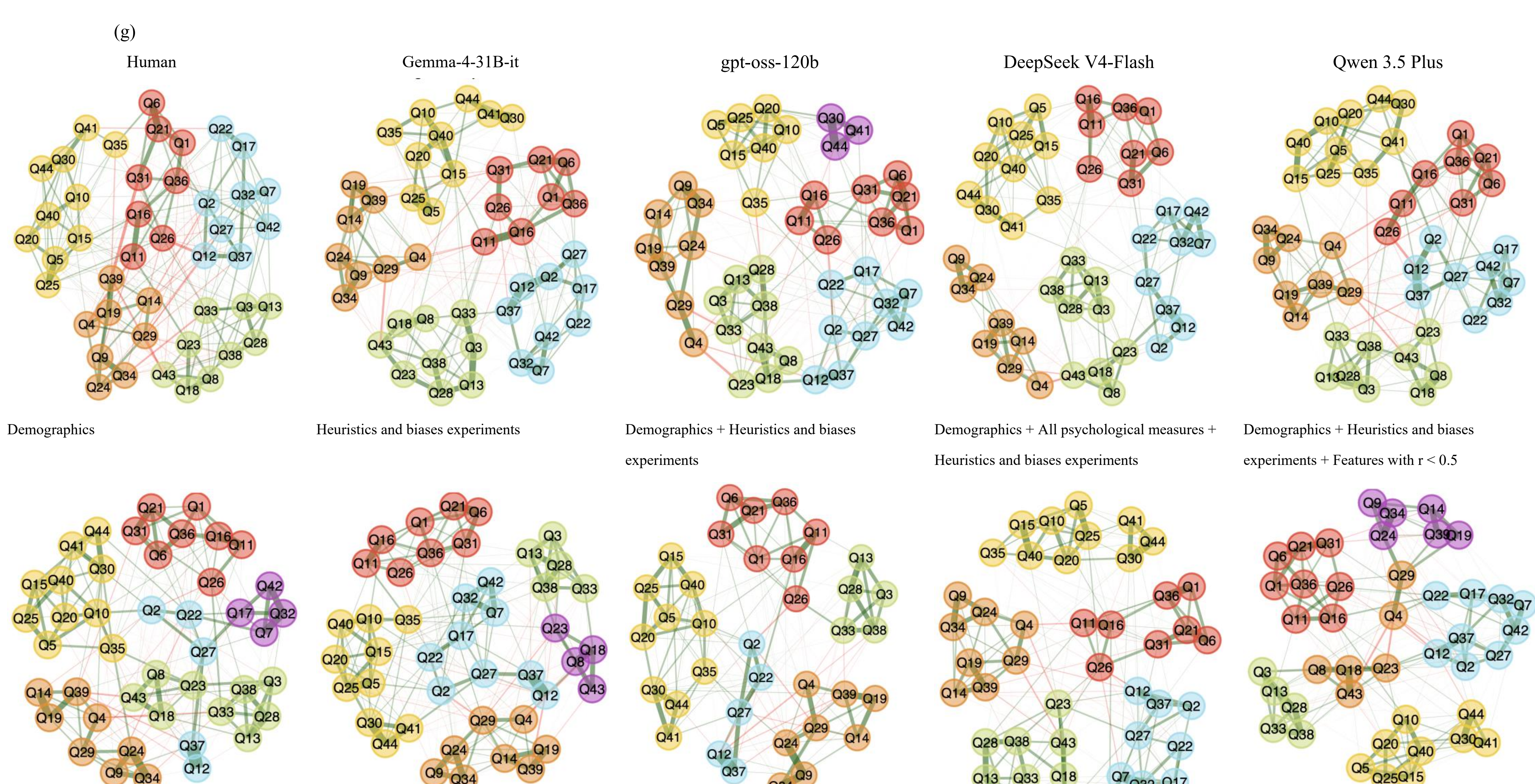

Demographics + Heuristics and biases experiments + Features with $r > 0.5$

Demographics + Heuristics and biases experiments + Features with $r < 0.3$

**Supplementary Fig 3. Digital-twin performance against human responses in Study 3a**

(a) Accuracy, (b) item-level correlation, (c) profile correlation, (d) prediction-error slope and (e) disattenuated trait-score correlation for four LLM-based digital twins across seven feature-input conditions, shown separately for the five Big Five domains. Human BFI-44 responses were treated as the gold standard. Random baseline intervals indicate 95% confidence intervals from 1,000 random simulations. (f–g) Network configural invariance solutions. Nodes represent BFI-44 items and colors indicate EGA-derived communities. (f) Pairwise comparisons between human responses and each digital-twin model under each input condition. (g) The human EGA solution, within-model comparisons across the seven input conditions, and between-model comparisons among the four models within each input condition.

(a)

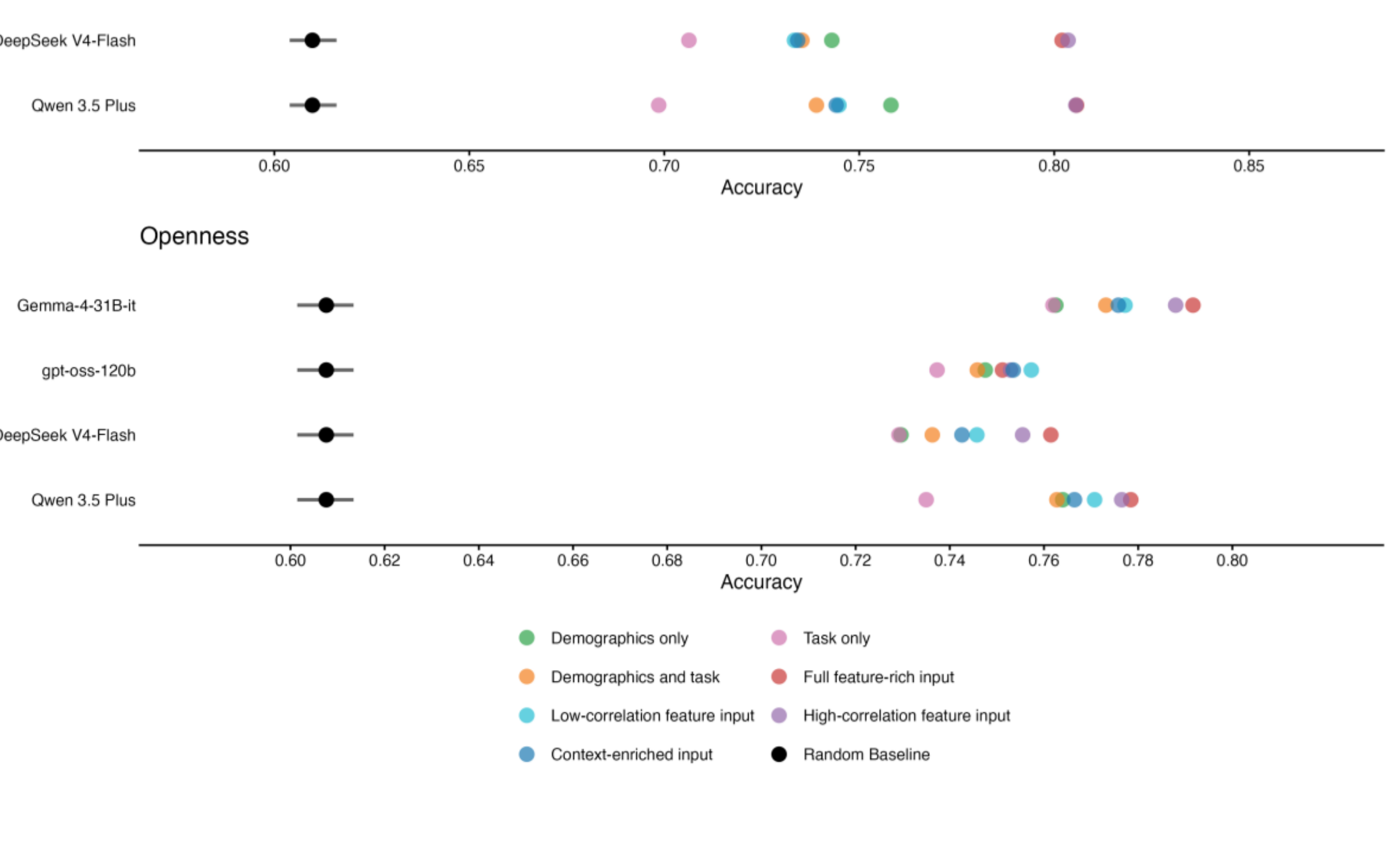

(b)

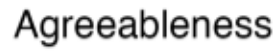

(c)

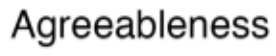

(d)

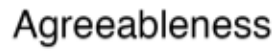

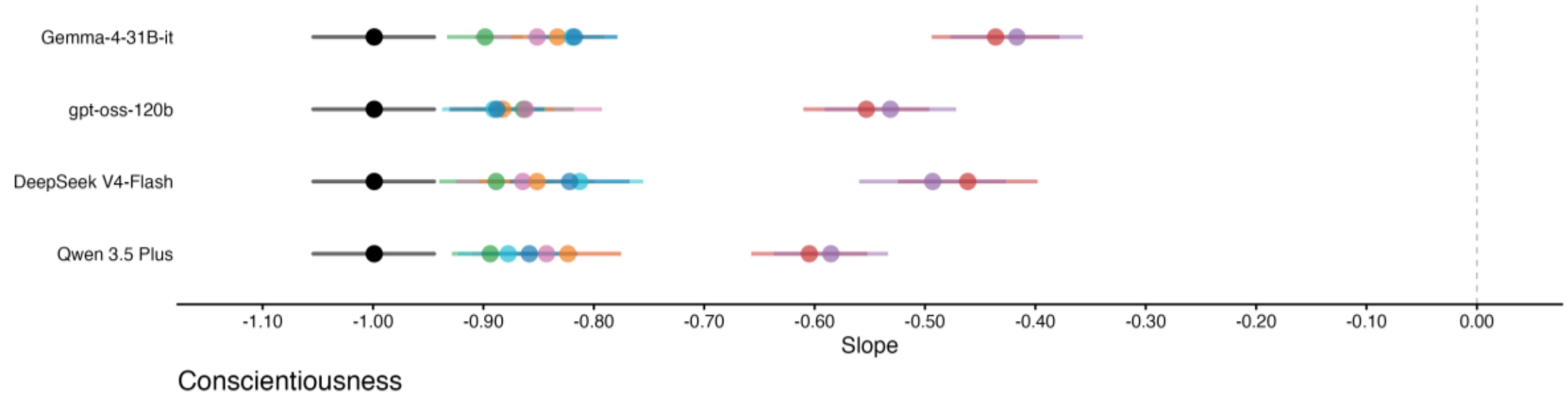

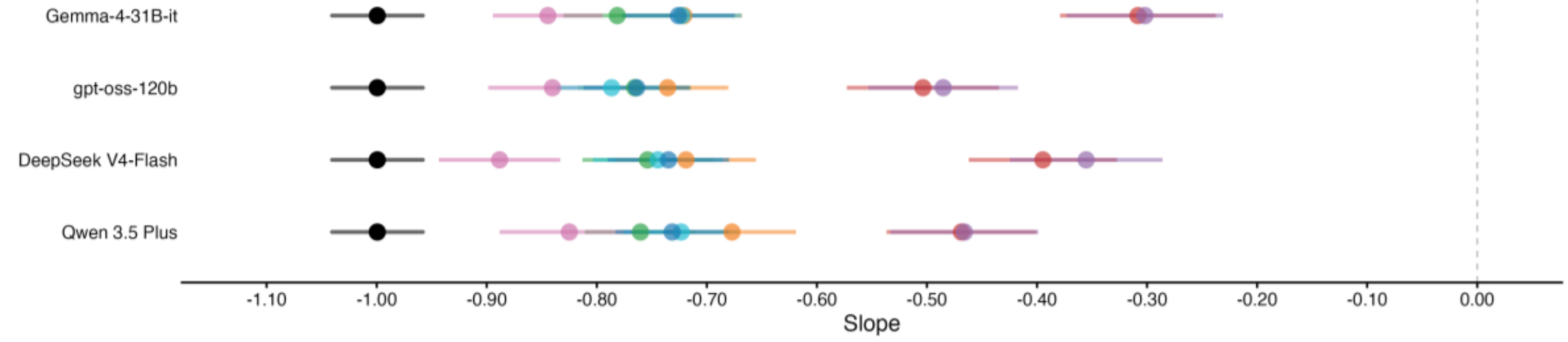

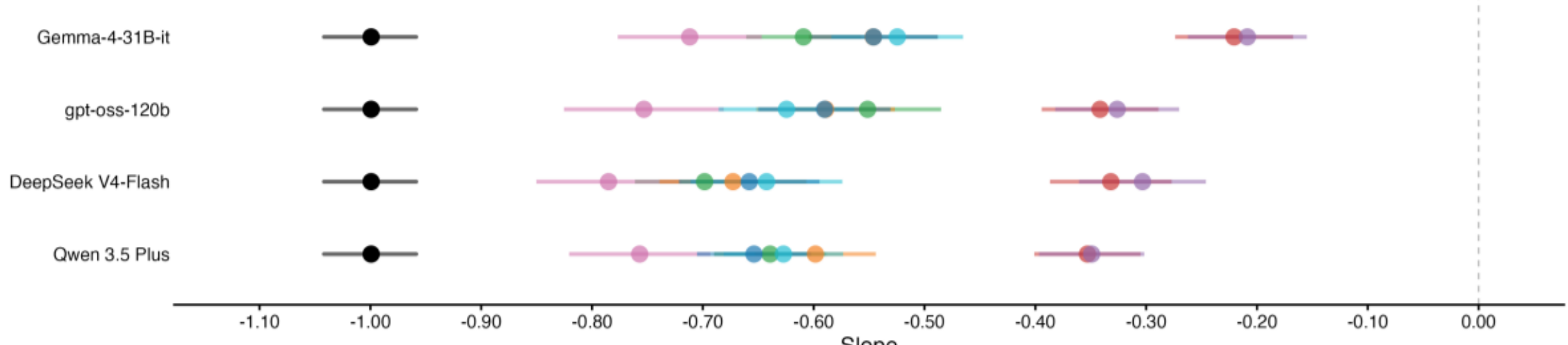

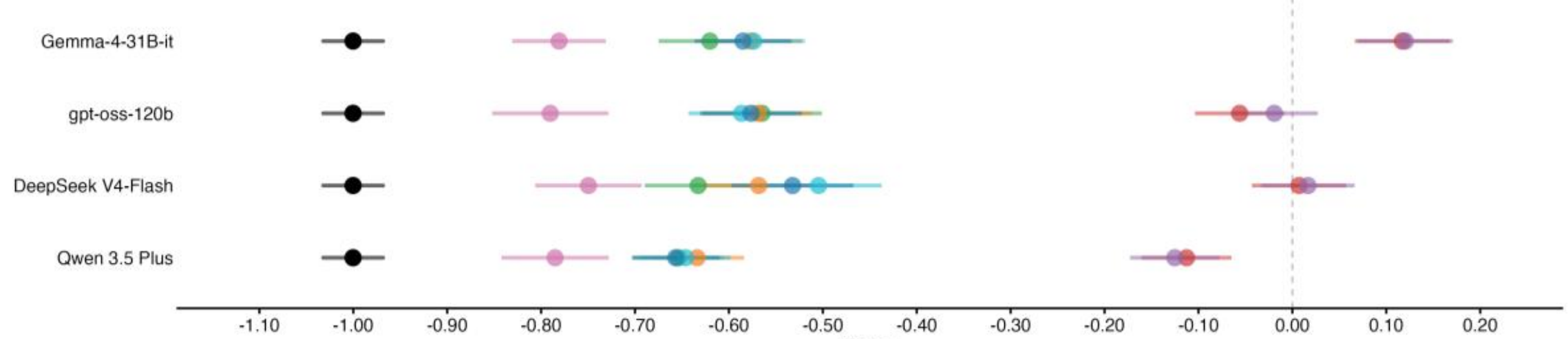

Openness

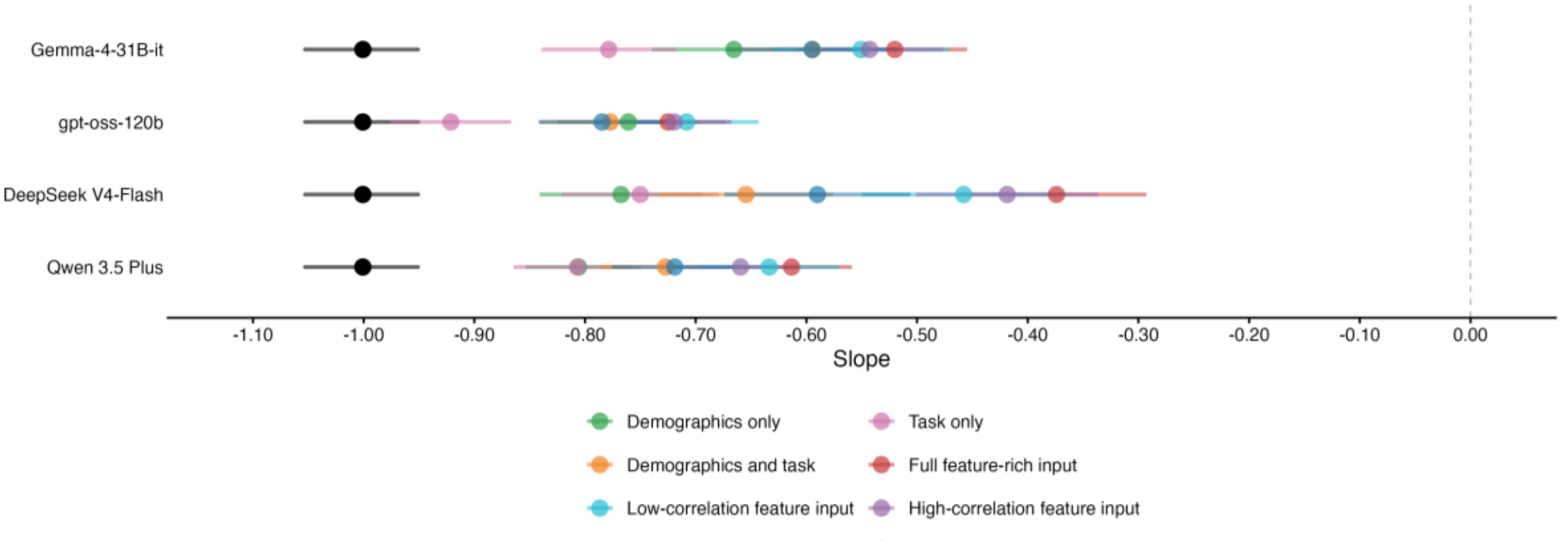

(e)

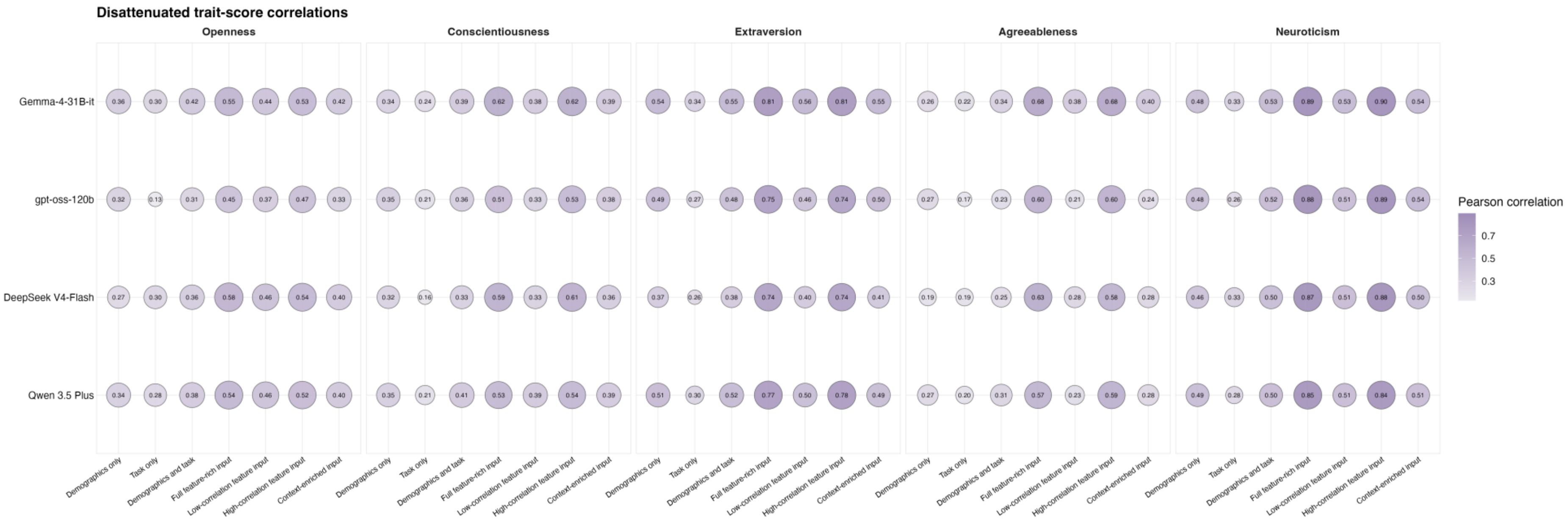

(f)

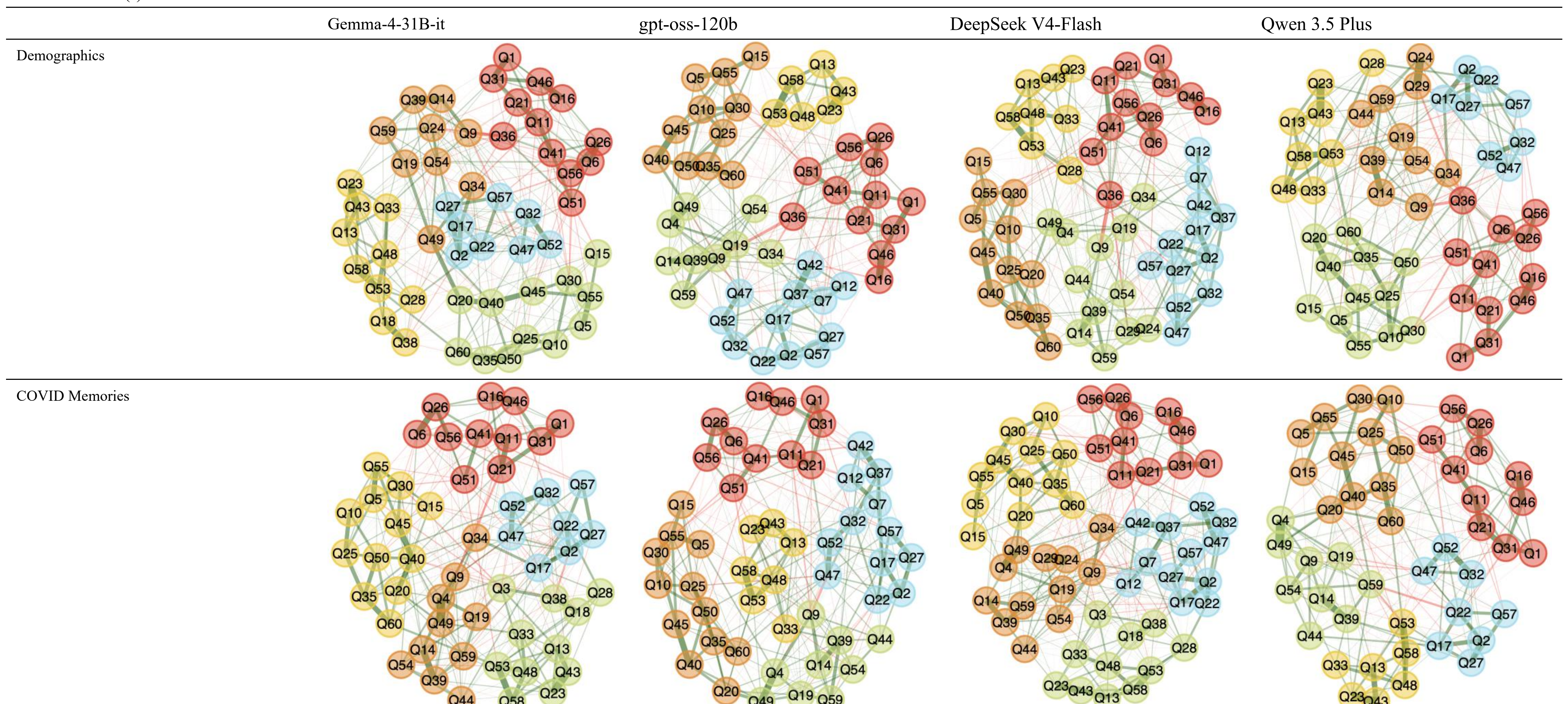

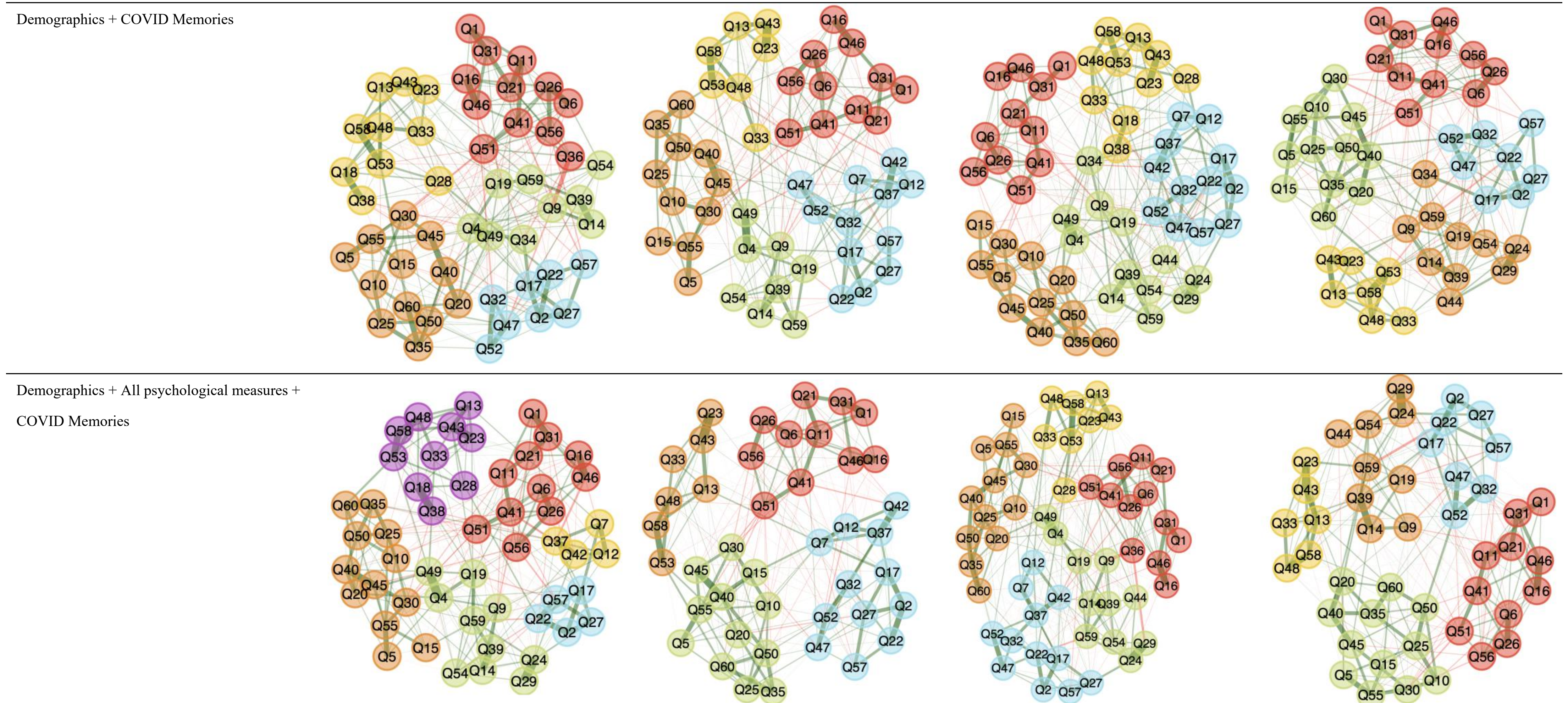

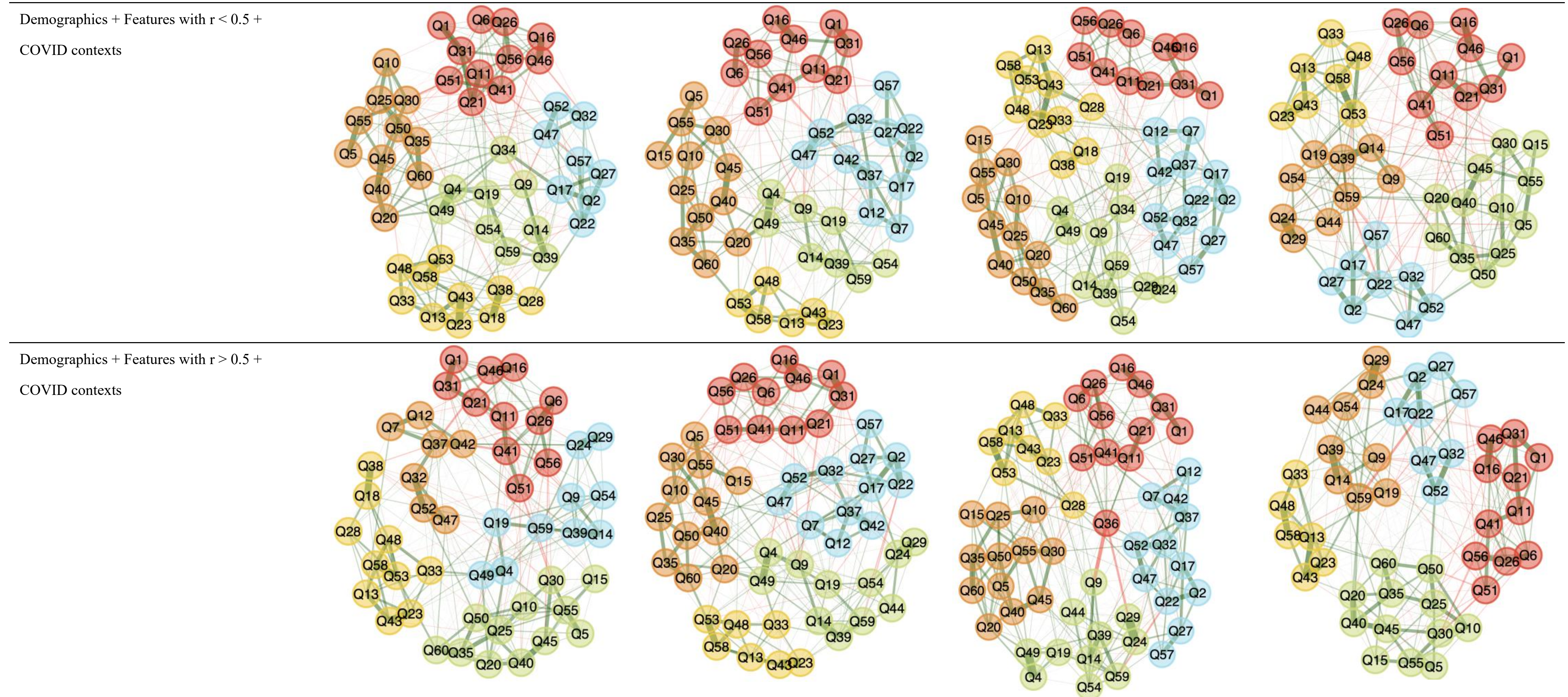

Demographics + COVID Memories + COVID contexts

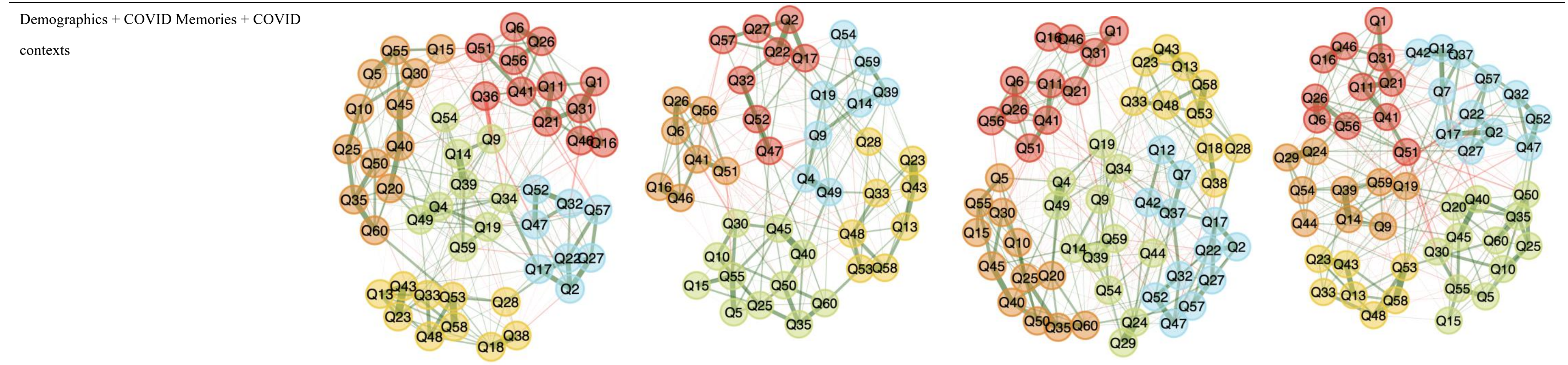

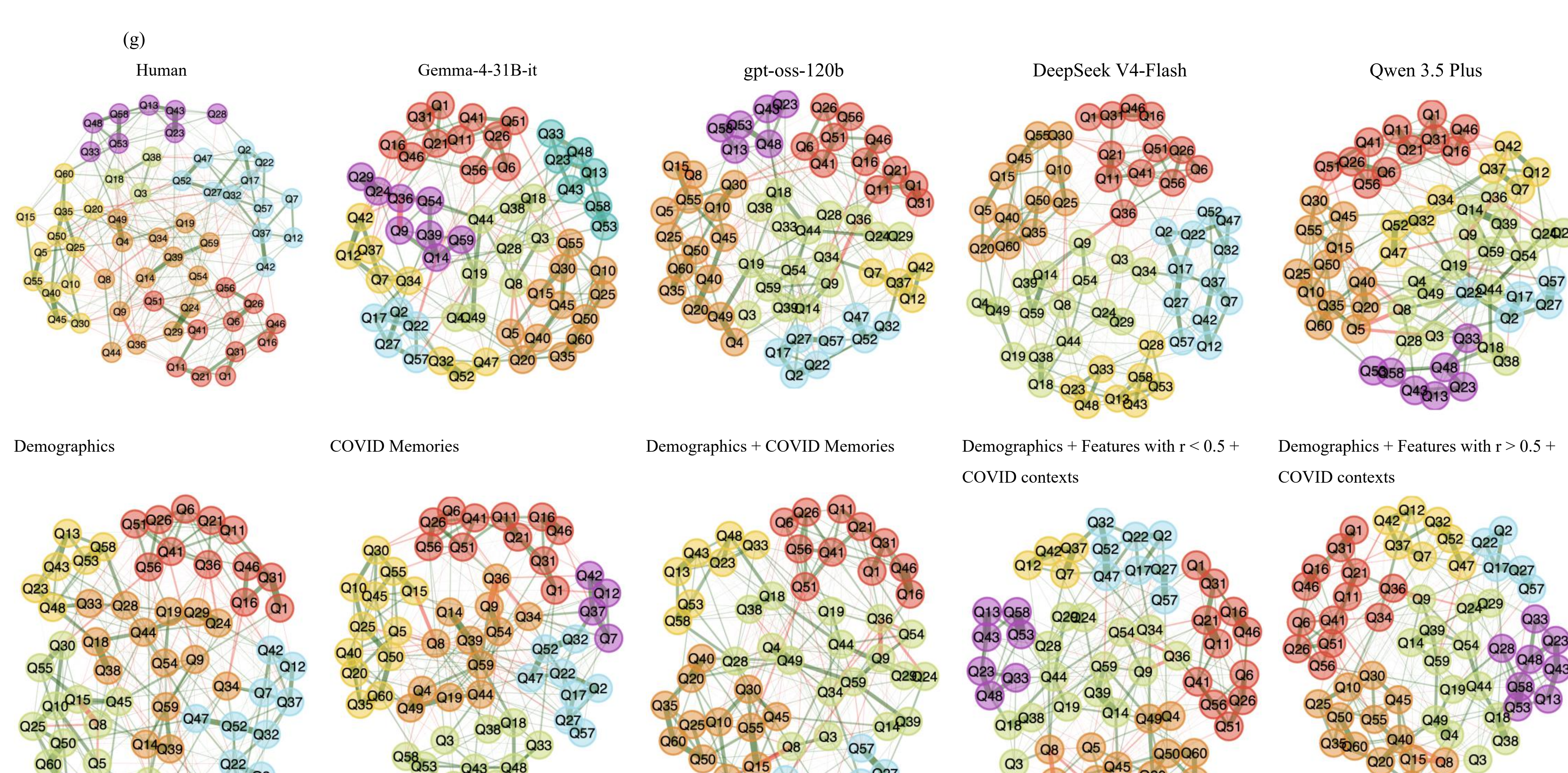

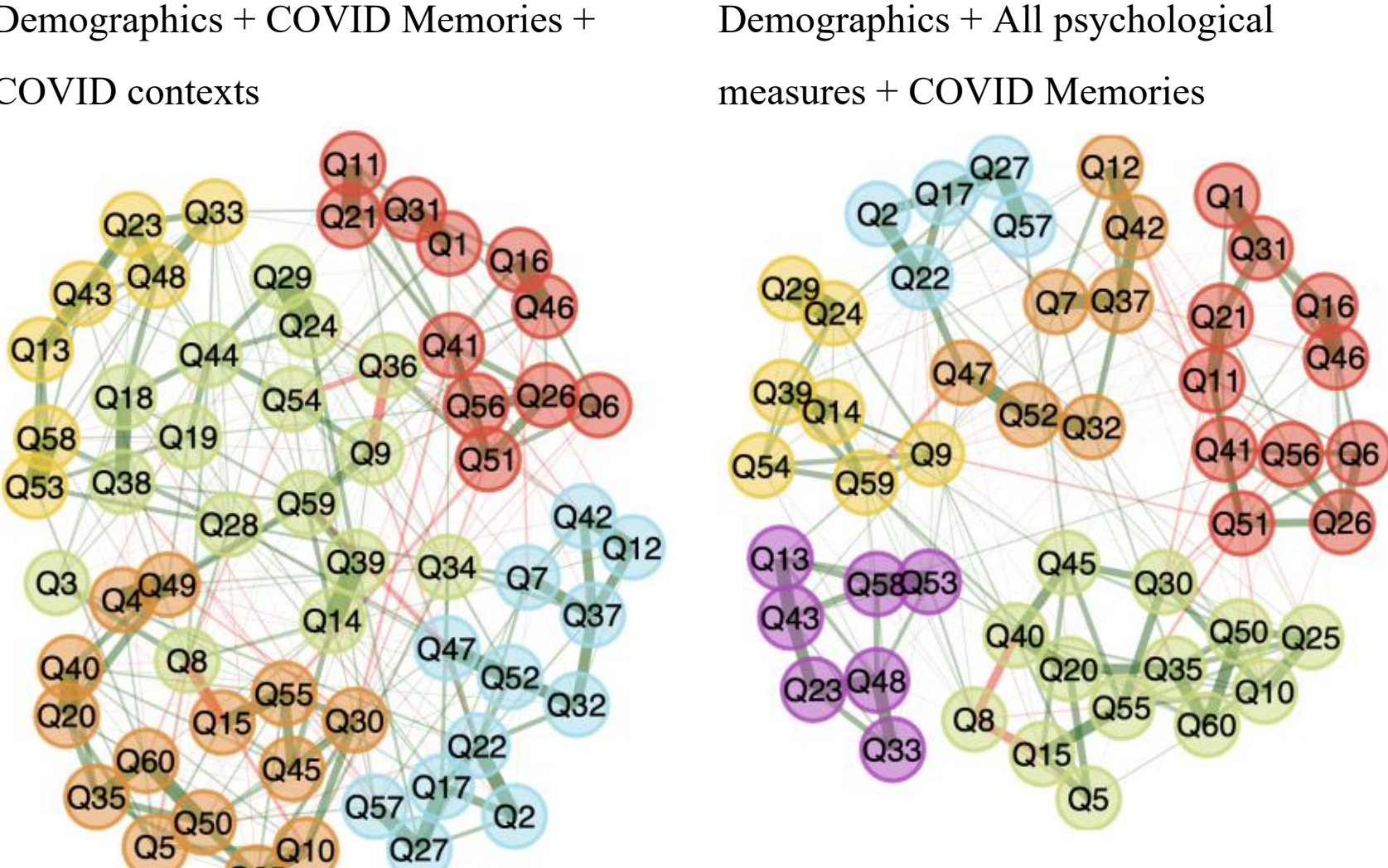

**Supplementary Fig 4. Digital-twin performance against human responses in Study 3b**

(a) Accuracy, (b) item-level correlation, (c) profile correlation, (d) prediction-error slope and (e) disattenuated trait-score correlation for four LLM-based digital twins across seven feature-input conditions, shown separately for the five Big Five domains. Human NEO-FFI-60 responses were treated as the gold standard. Random baseline intervals indicate 95% confidence intervals from 1,000 random simulations. (f–g) Network configural invariance solutions. Nodes represent NEO-FFI-60 items and colors indicate EGA-derived communities. (f) Pairwise comparisons between human responses and each digital-twin model under each input condition. (g) The human EGA solution, within-model comparisons across the seven input conditions, and between-model comparisons among the four models within each input condition.

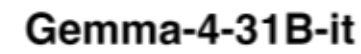

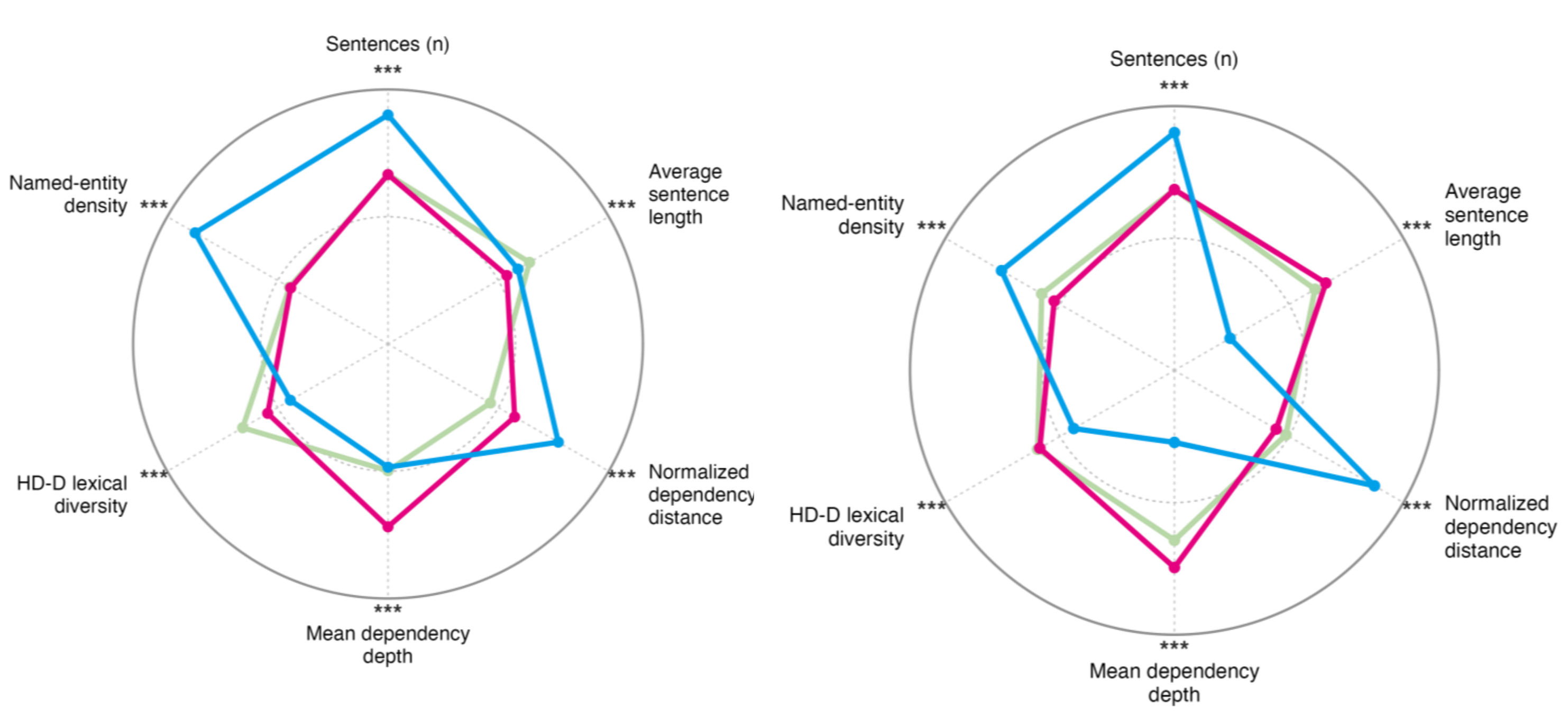

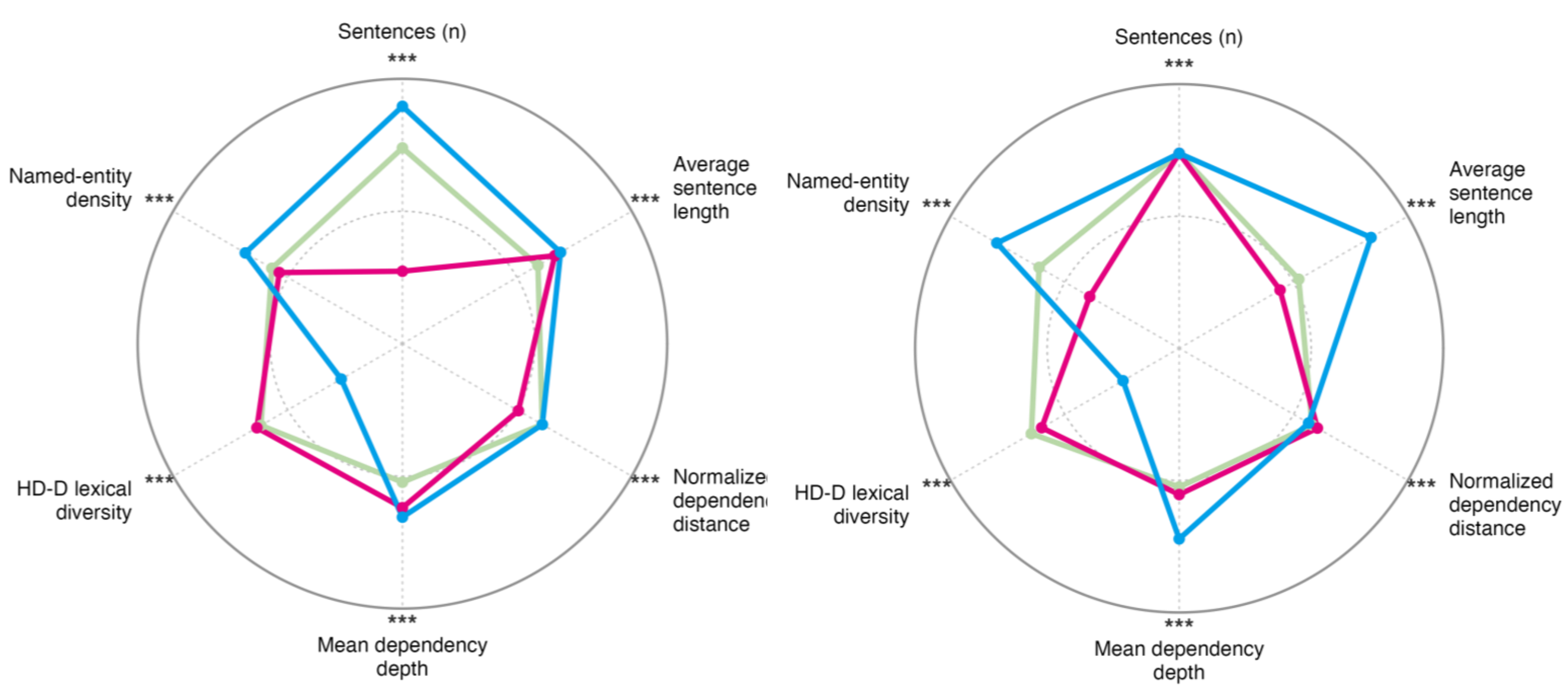

(b)

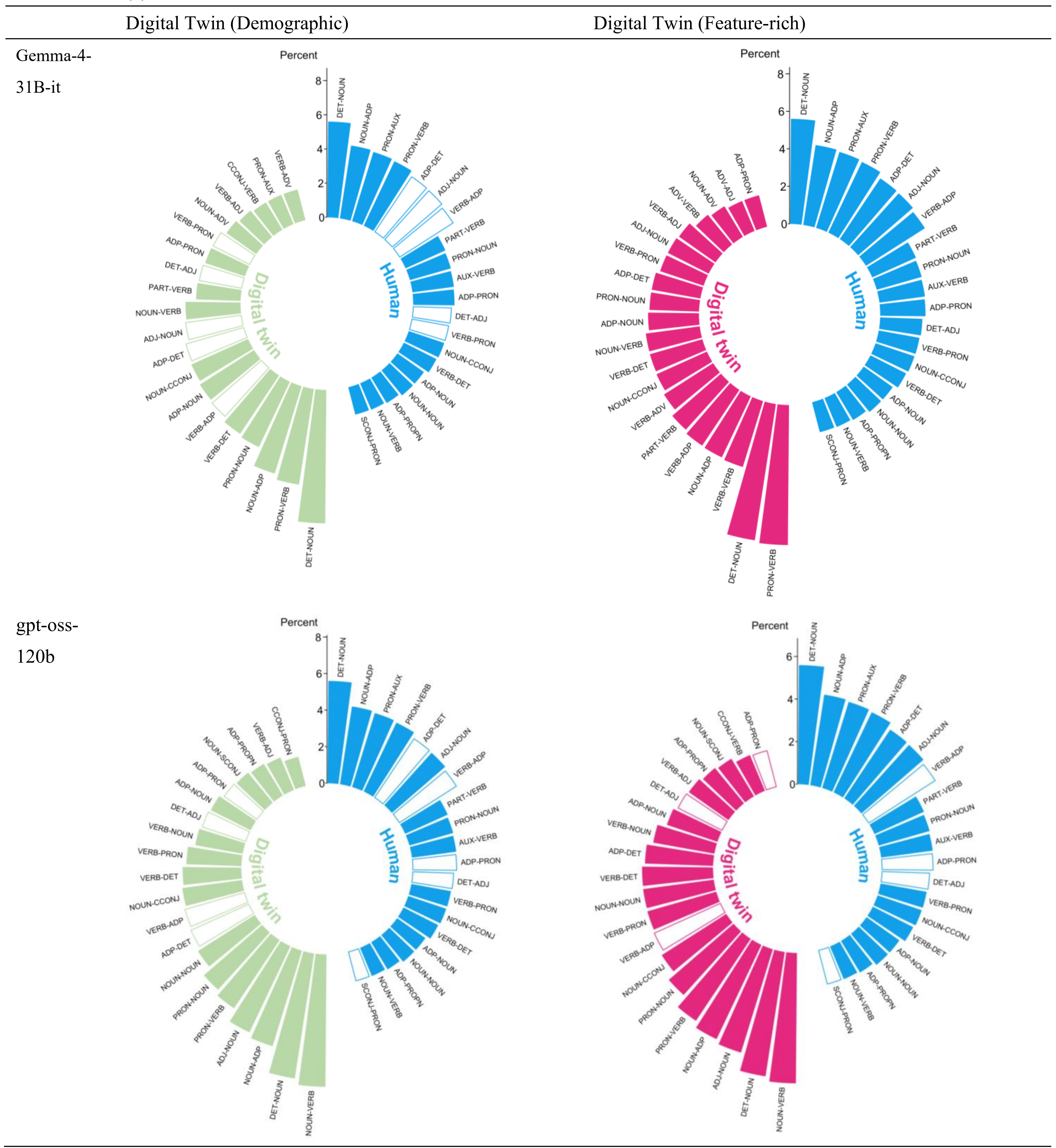

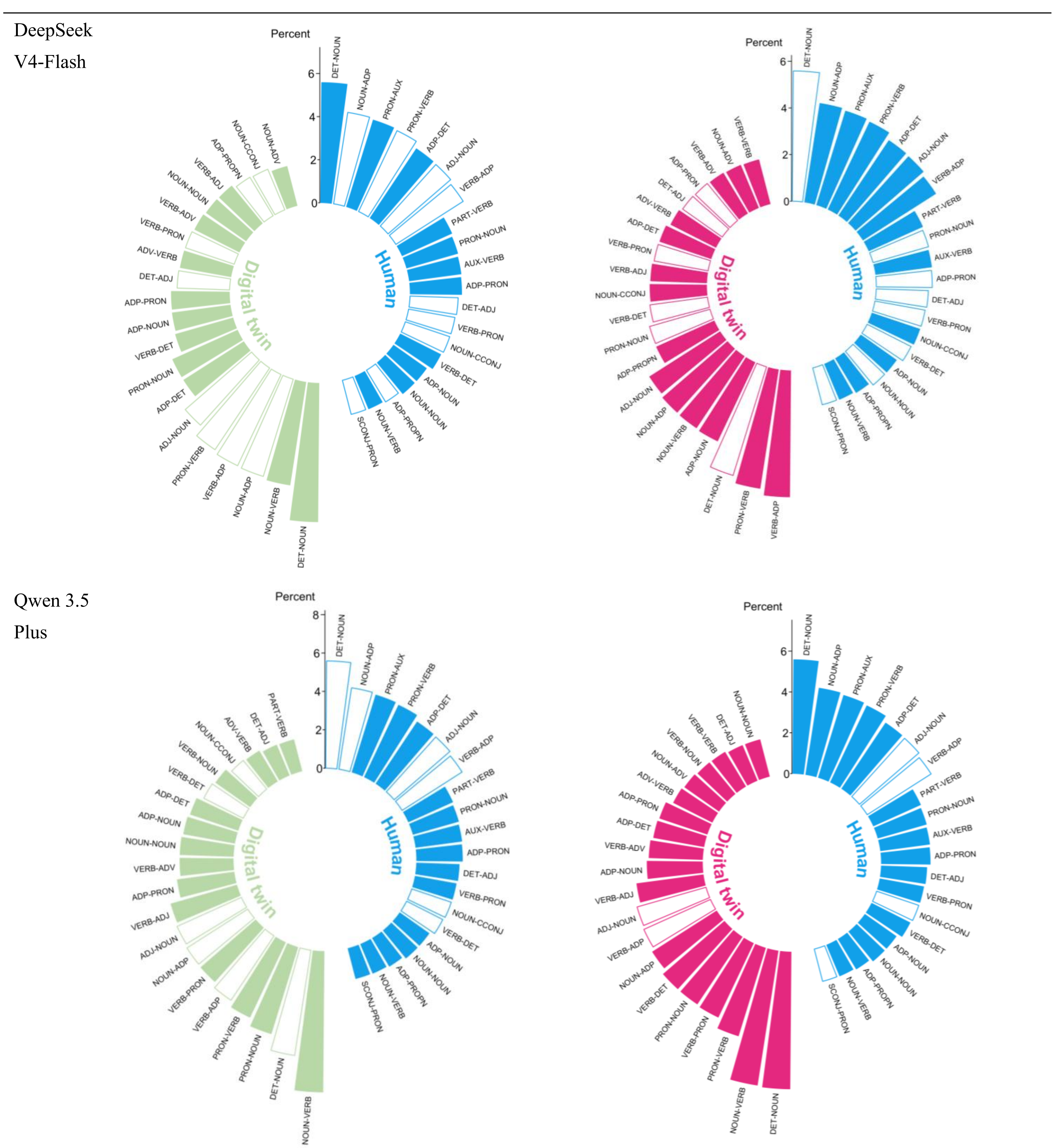

(c)

| | Digital Twin (Demographic) | Digital Twin (Feature-rich) |
| --- | --- | --- |
| Gemma-4-31B-it | | |
| gpt-oss-120b | | |

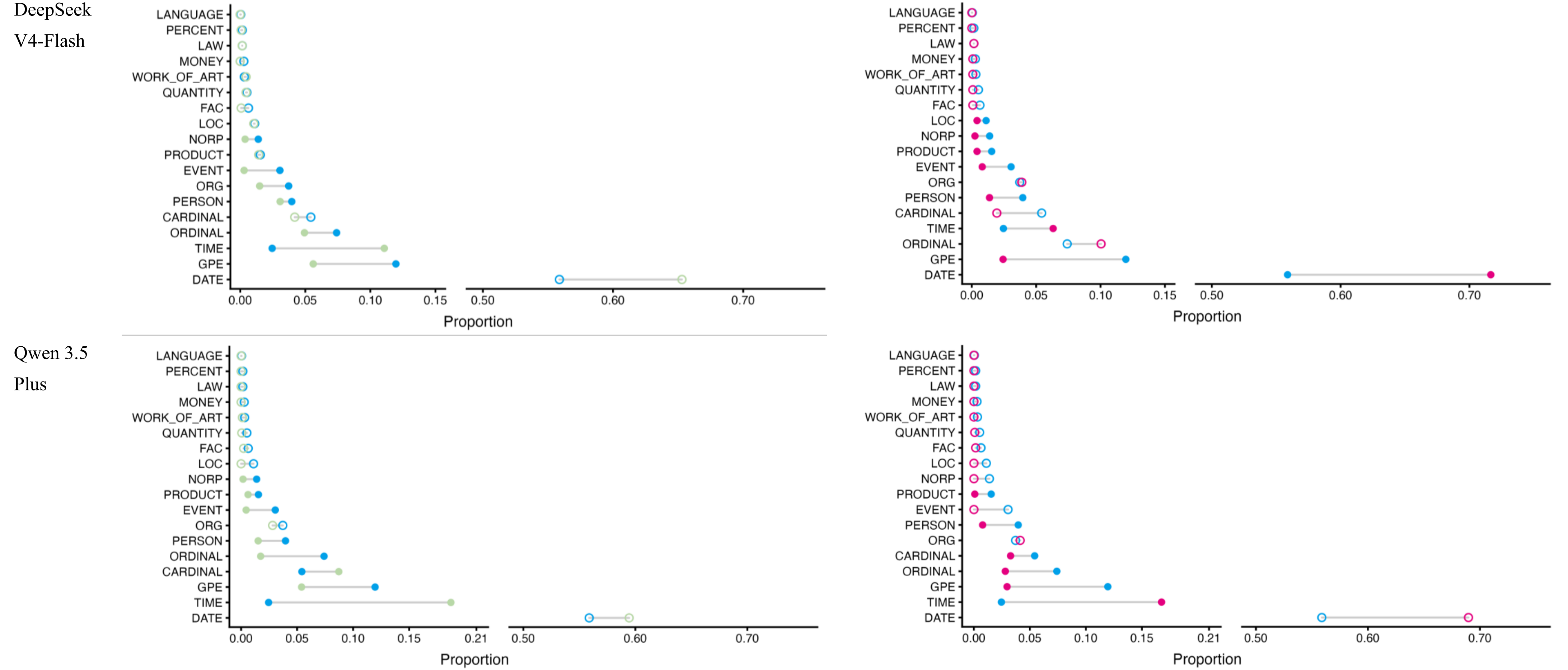

(d)

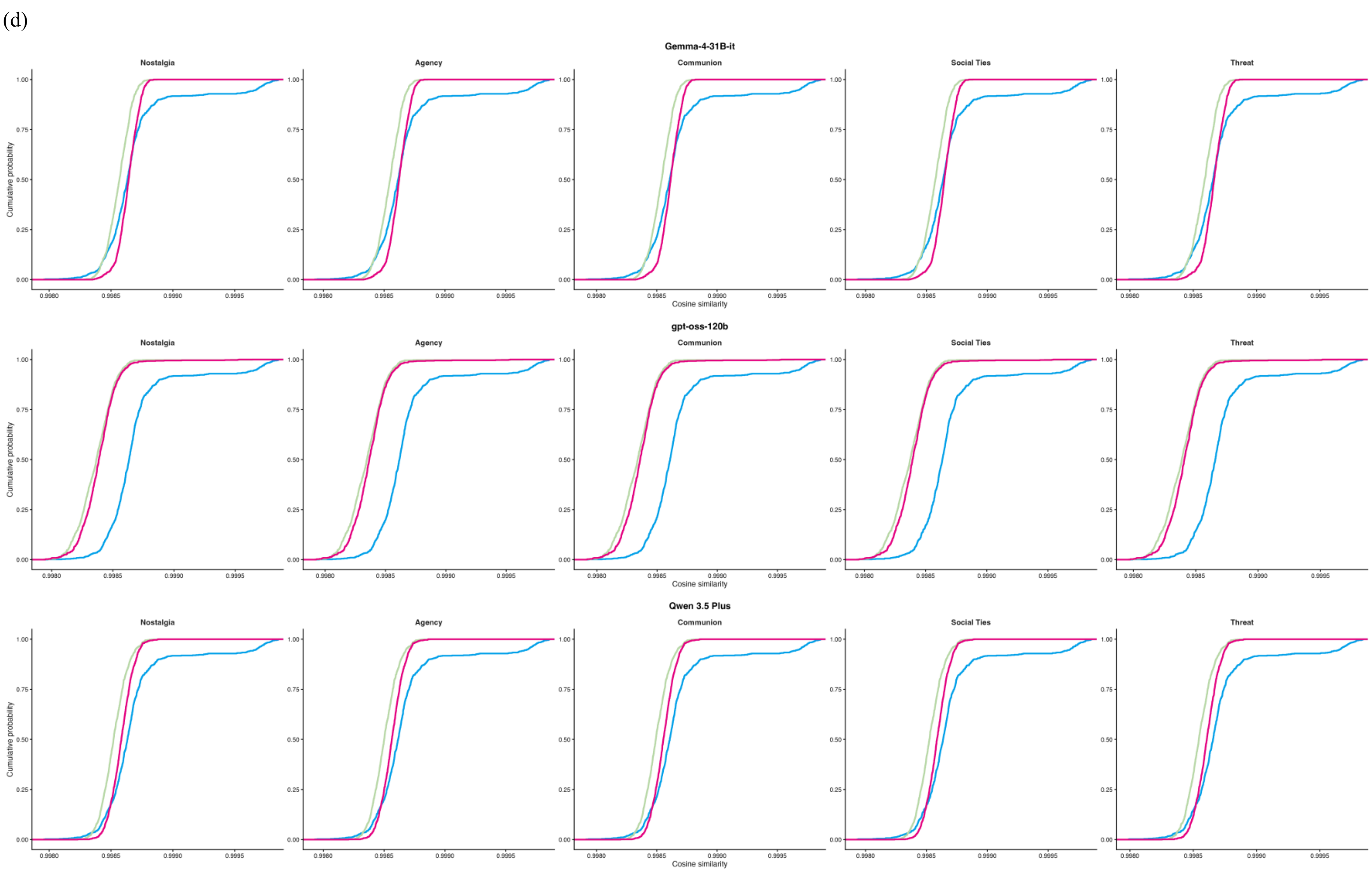

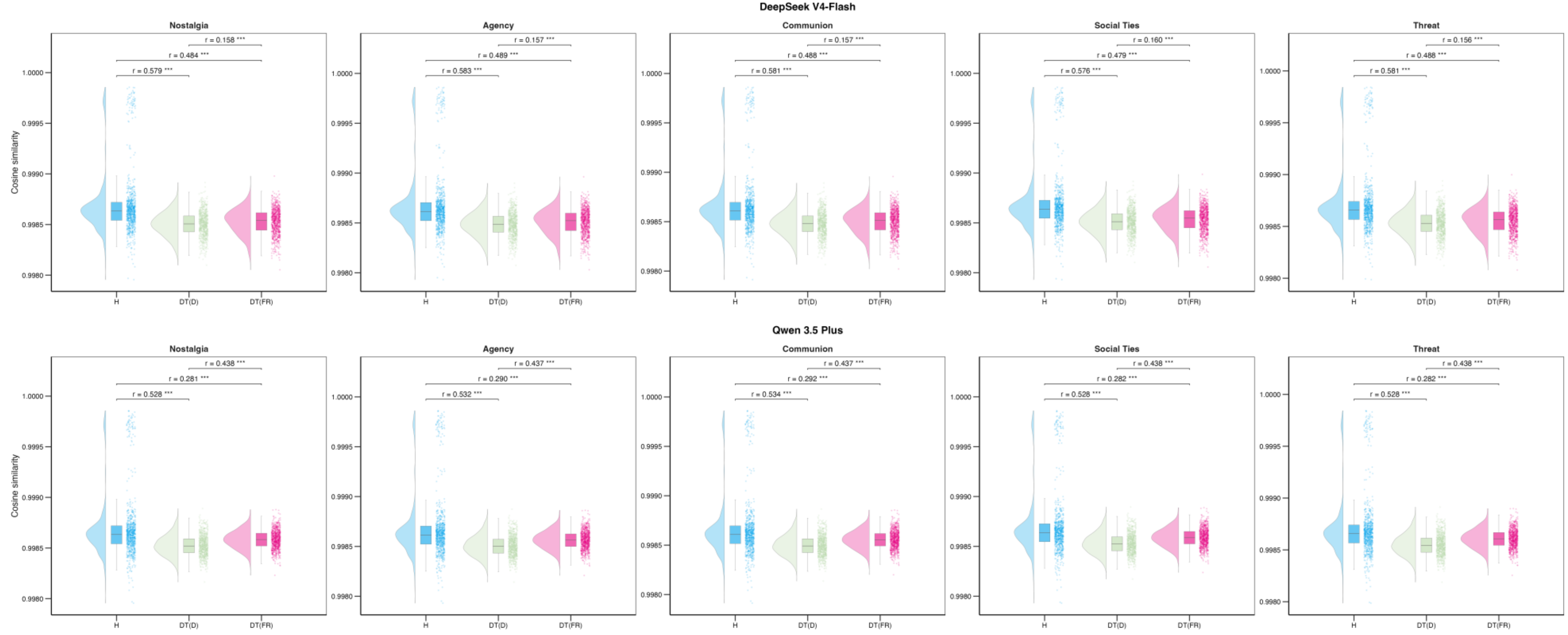

**Supplementary Fig 5. Linguistic evaluation and construct representation of digital twins in Study 4a.**

(a) Percentile ranks of linguistic features (*** $p < .001$, ** $p < .01$). (b) POS-bigram type frequencies for Human vs. Digital twin; filled markers indicate statistical significance at $p < .05$. (c) NER label proportions for Human vs. Digital twin; solid markers indicate statistical significance at $p < .05$. (d) CDFs of cosine similarity index construct representation. (e) Raincloud plots illustrate the distribution of cosine similarity for each dimension. The plots combine a half-violin plot showing the data density, a boxplot showing the median and interquartile range, and jittered raw data points. Three conditions are compared: H (Human), DT(D) (Digital twin - demographic), and DT(FR) (Digital twin - feature-rich). Horizontal bars indicate pairwise comparisons using paired Wilcoxon signed-rank tests, with the effect size reported as Wilcoxon r. *** $p < .001$, ** $p < .01$, * $p < .05$, ns = non-significant. Color: Human (red), Digital twin (feature-rich) (blue), Digital twin (demographic) (green).

(a)

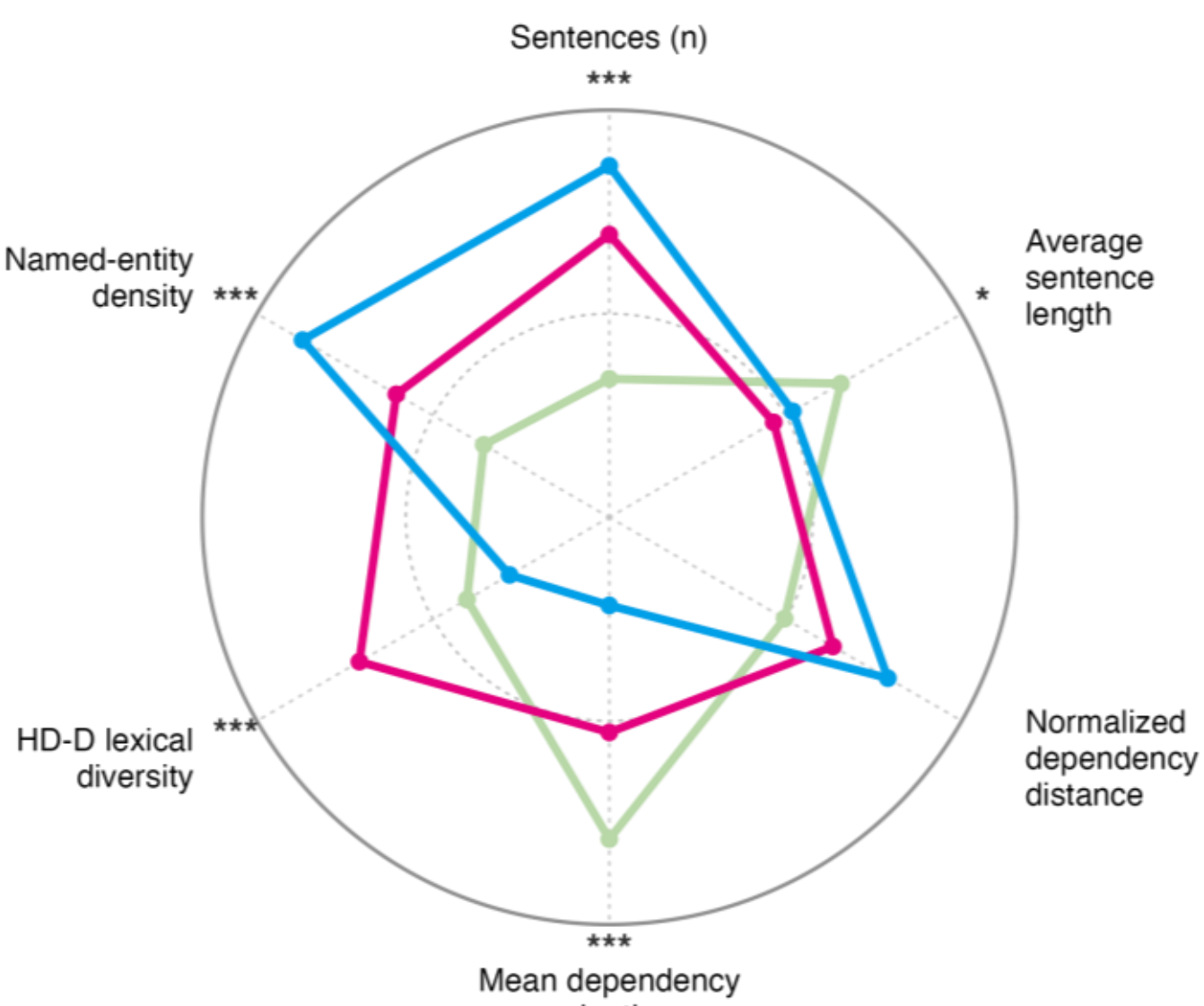

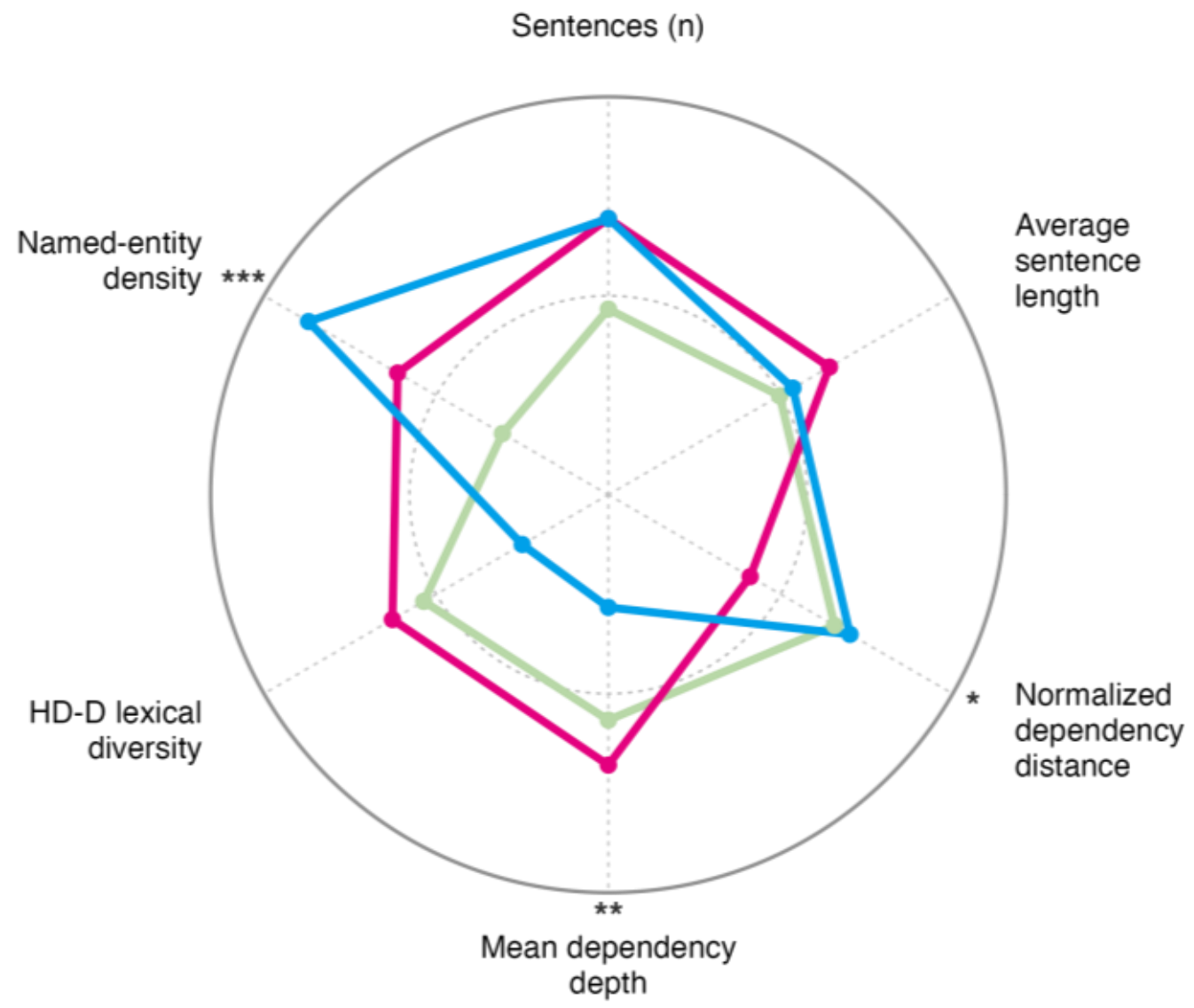

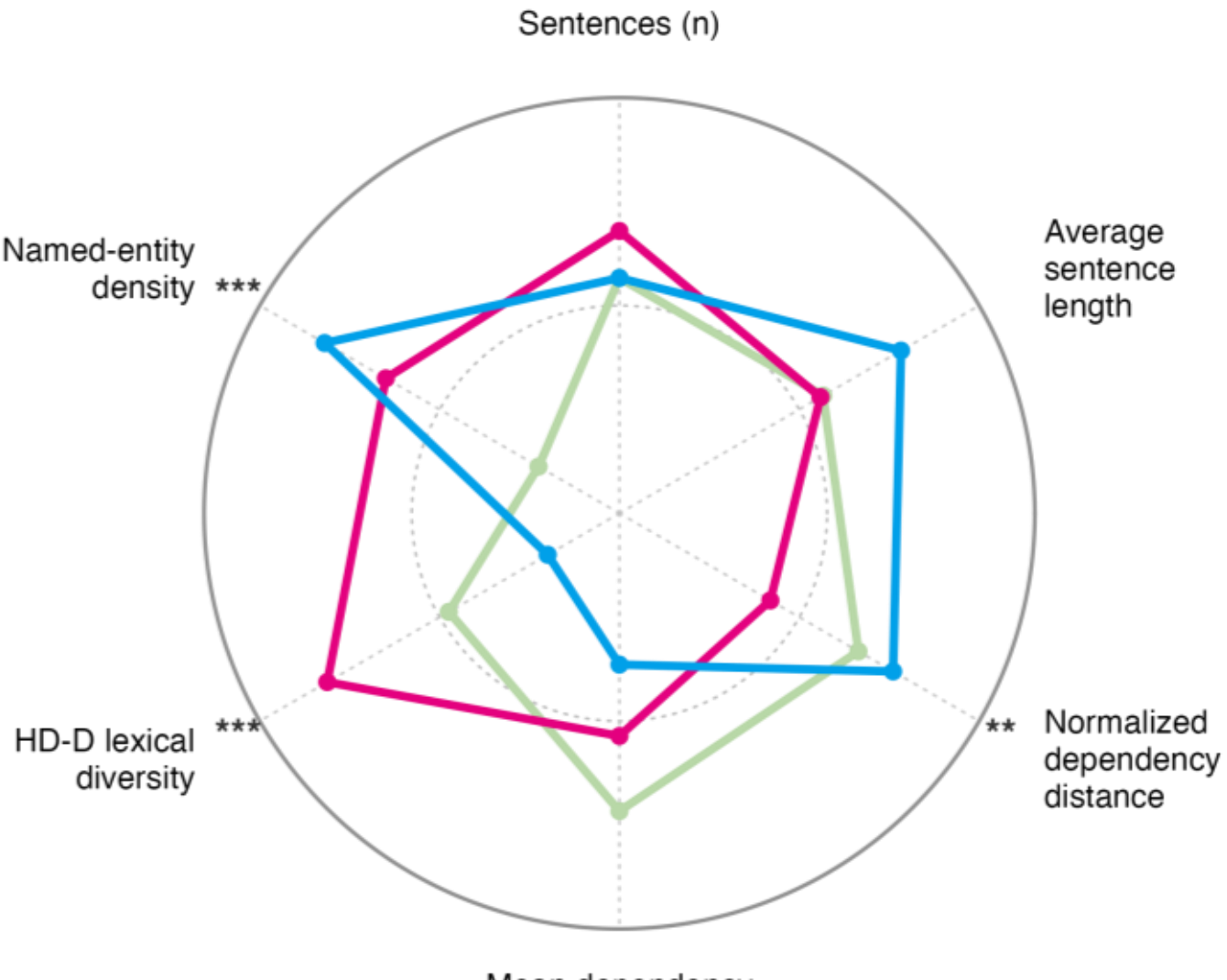

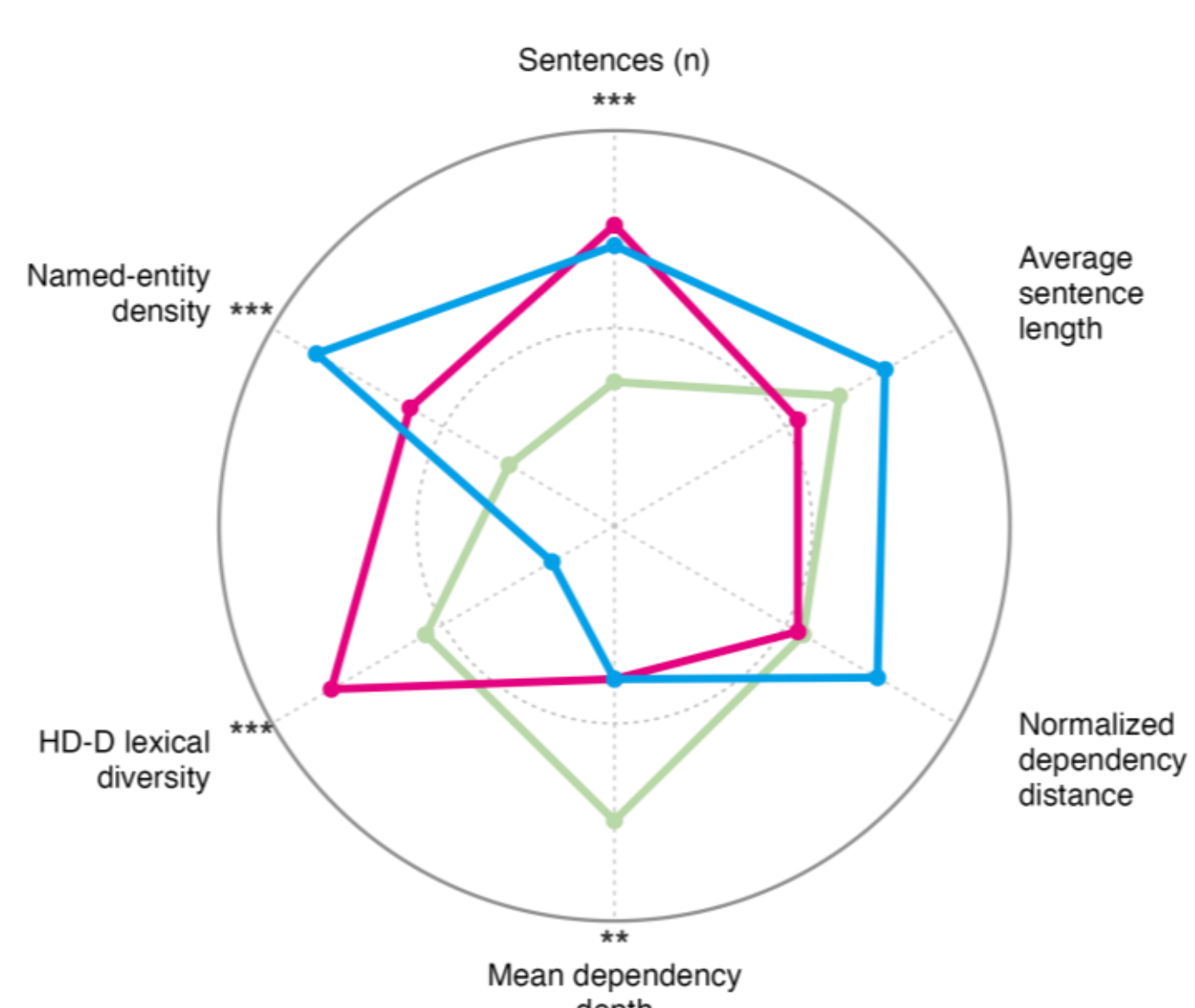

(b)

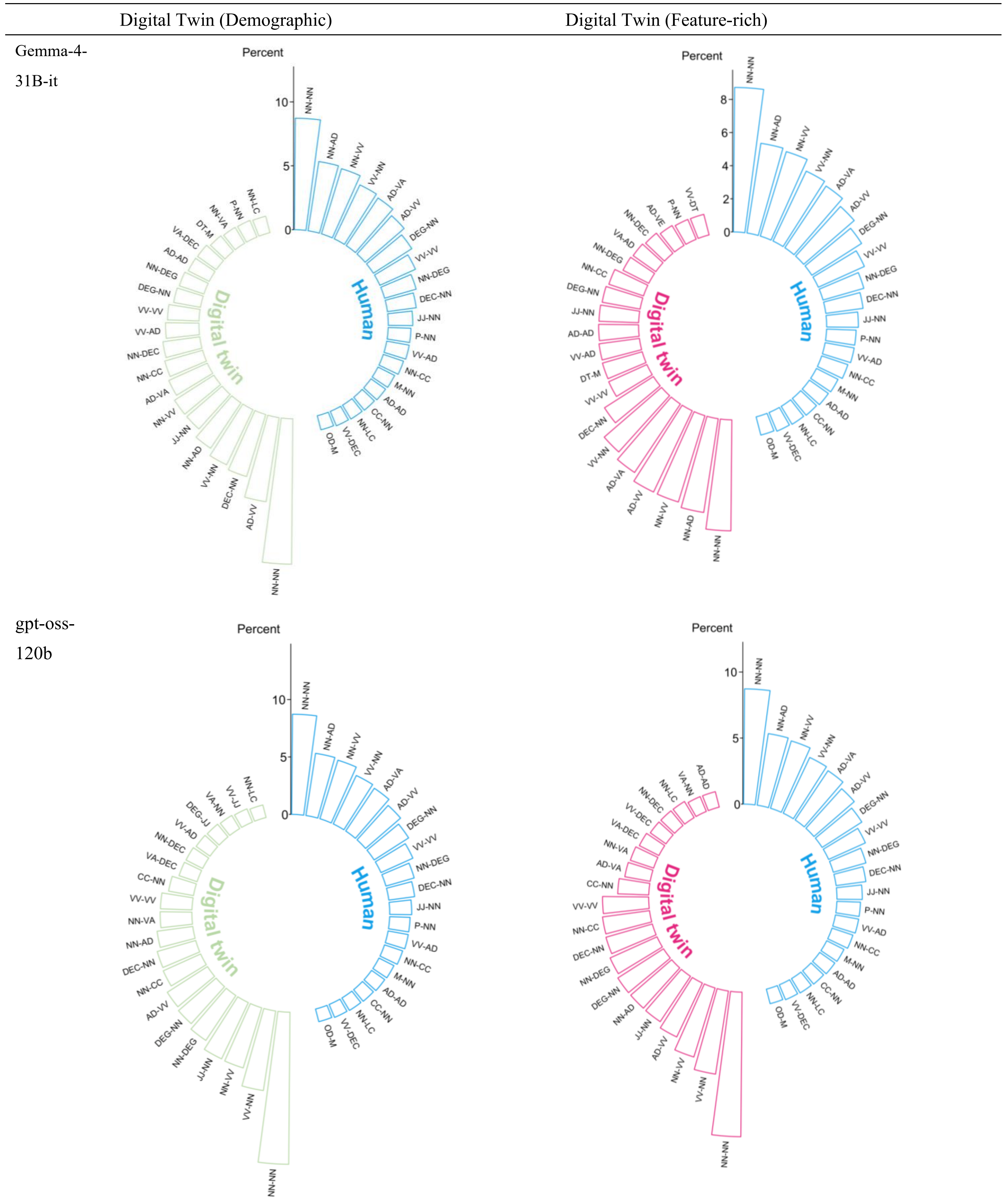

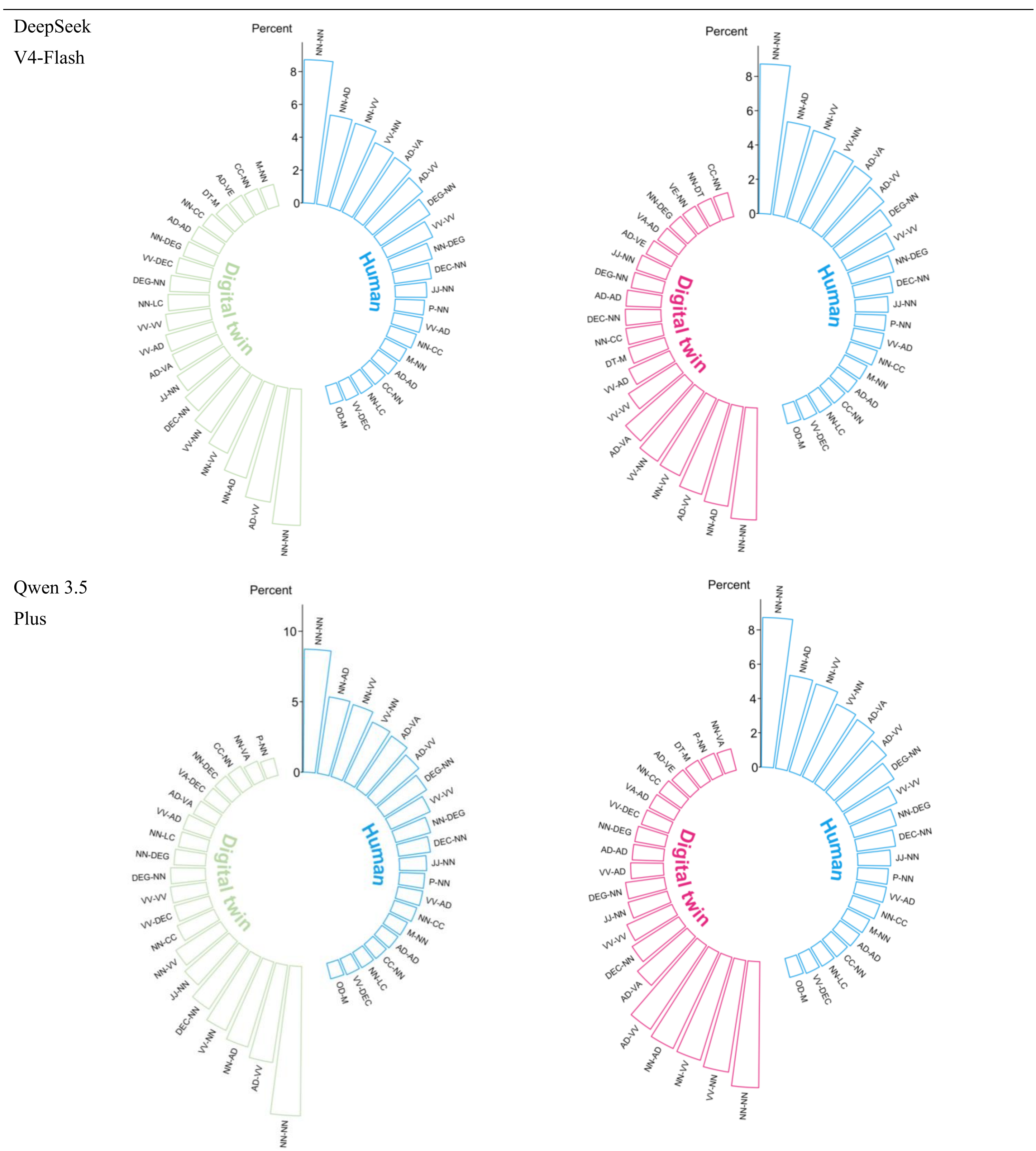

(c)

| | Digital Twin (Demographic) | Digital Twin (Feature-rich) |
|---|---|---|
| Gemma-4-31B-it | LENGTH<br>FRACTION<br>DECIMAL<br>AGE<br>TIME<br>ORGANIZATION<br>INTEGER<br>DATE<br>ORDINAL<br>PERSON<br>LOCATION<br>0.00 0.10 0.20 0.30 0.40 0.50 0.60 0.70<br>Proportion | LENGTH<br>FRACTION<br>DECIMAL<br>AGE<br>TIME<br>INTEGER<br>ORGANIZATION<br>ORDINAL<br>DATE<br>LOCATION<br>PERSON<br>0.00 0.10 0.20 0.30 0.40 0.50 0.60 0.70<br>Proportion |
| gpt-oss-120b | LENGTH<br>DECIMAL<br>AGE<br>TIME<br>ORGANIZATION<br>FRACTION<br>DURATION<br>DATE<br>INTEGER<br>LOCATION<br>ORDINAL<br>PERSON<br>0.00 0.10 0.20 0.30 0.40 0.50 0.60 0.70<br>Proportion | LENGTH<br>FRACTION<br>DECIMAL<br>AGE<br>MONEY<br>TIME<br>INTEGER<br>ORGANIZATION<br>ORDINAL<br>DATE<br>LOCATION<br>PERSON<br>0.00 0.10 0.20 0.30 0.40 0.50 0.60 0.70<br>Proportion |

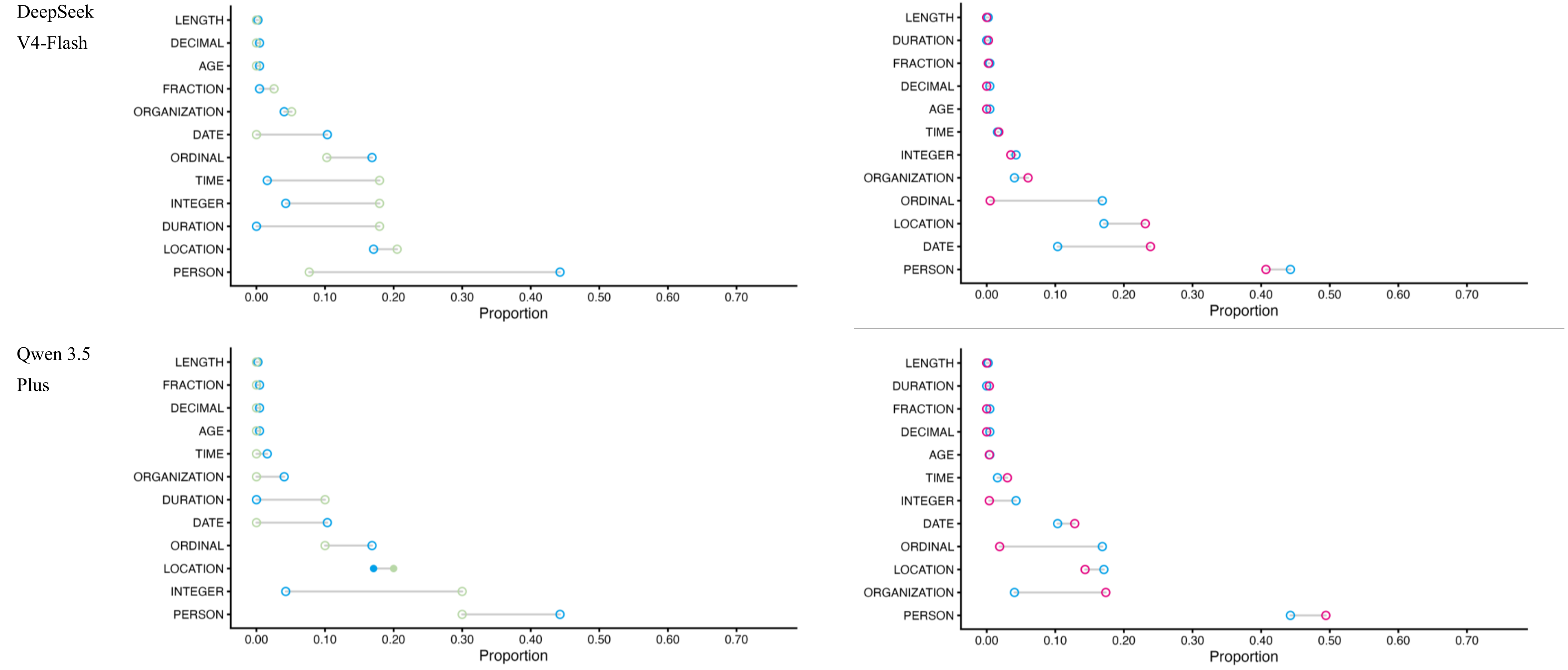

(d)

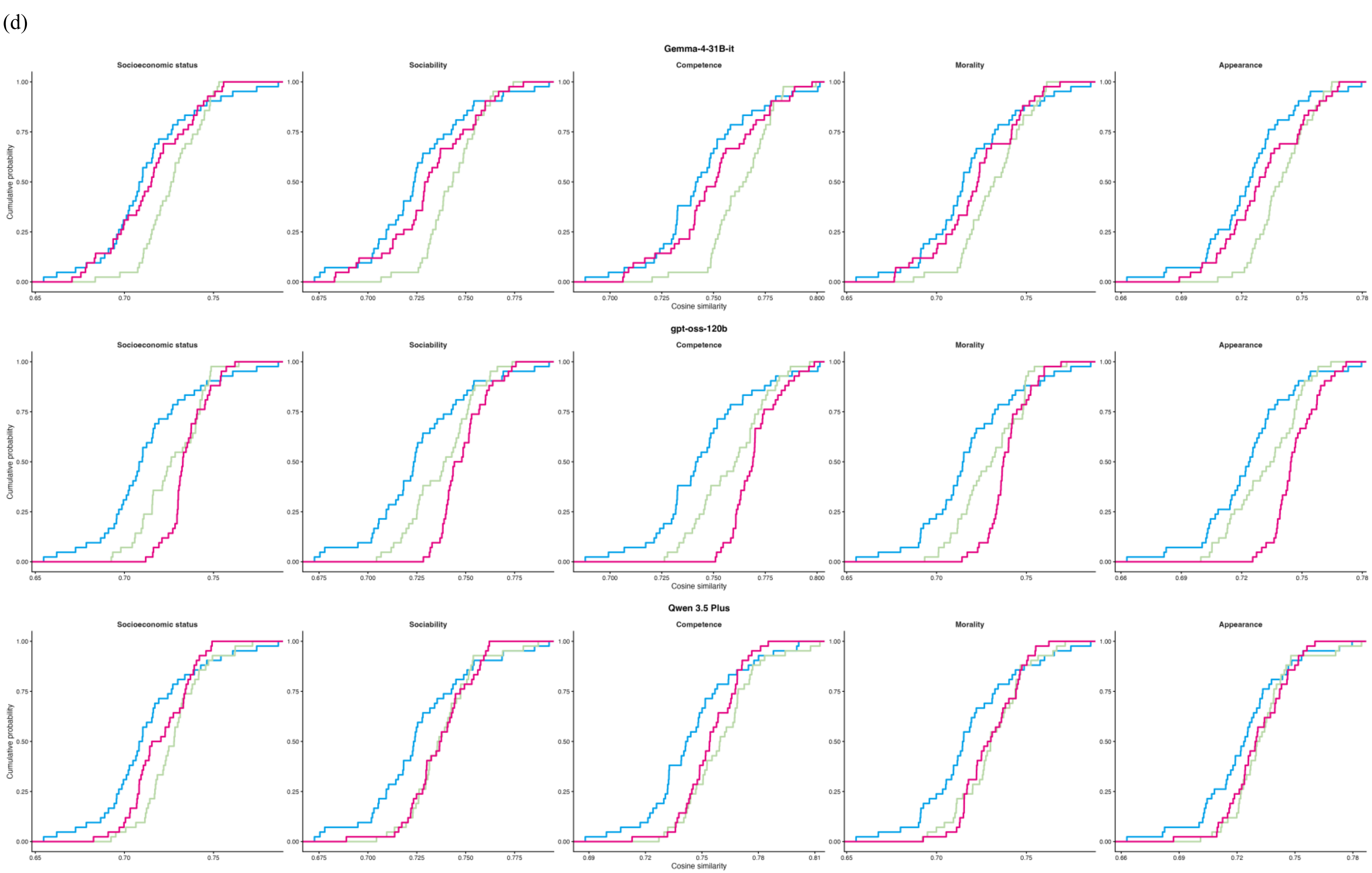

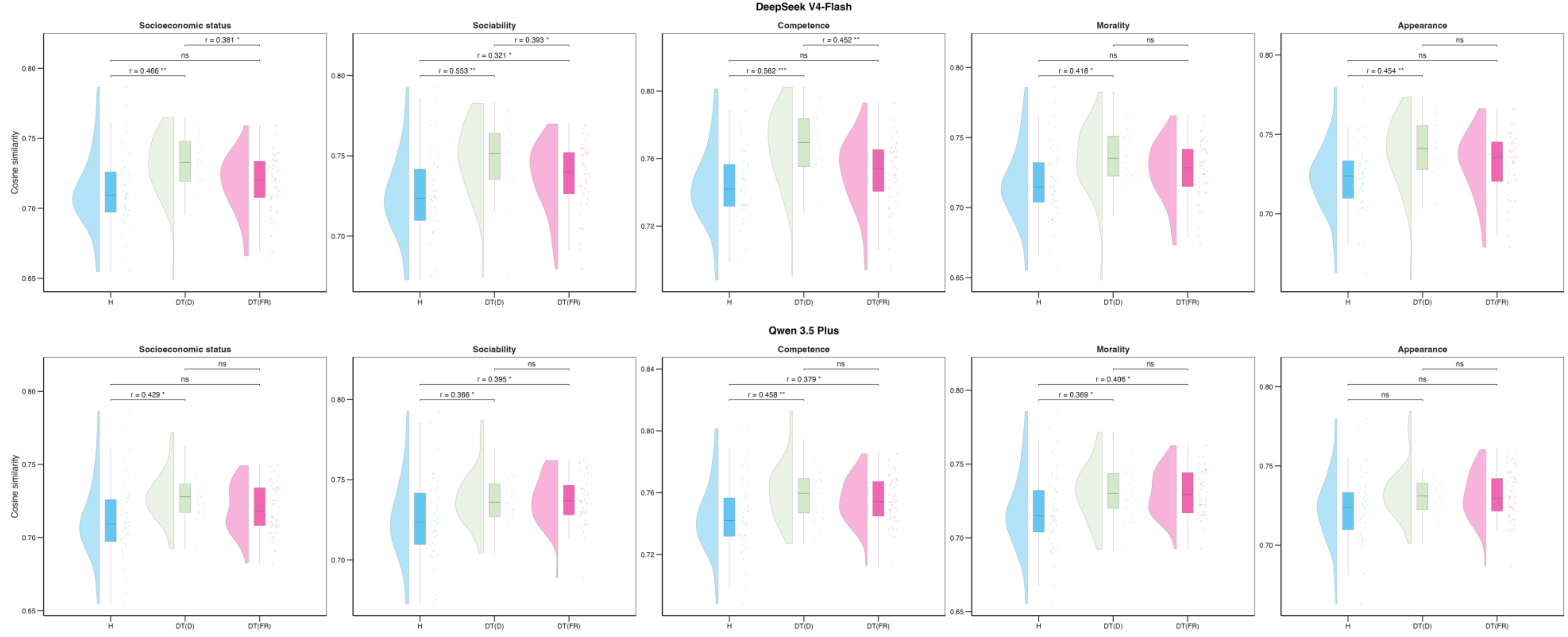

**Supplementary Fig 6. Linguistic evaluation and construct representation of digital twins in Study 4b.**
(a) Percentile ranks of linguistic features (*** $p < .001$, ** $p < .01$). (b) POS-bigram type frequencies for Human vs. Digital twin; filled markers indicate statistical significance at $p < .05$. (c) NER label proportions for Human vs. Digital twin; solid markers indicate statistical significance at $p < .05$. (d) CDFs of cosine similarity index construct representation. (e) Raincloud plots illustrate the distribution of cosine similarity for each dimension. The plots combine a half-violin plot showing the data density, a boxplot showing the median and interquartile range, and jittered raw data points. Three conditions are compared: H (Human), DT(D) (Digital twin - demographic), and DT(FR) (Digital twin - feature-rich). Horizontal bars indicate pairwise comparisons using paired Wilcoxon signed-rank tests, with the effect size reported as Wilcoxon r. *** $p < .001$, ** $p < .01$, * $p < .05$, ns = non-significant. Color: Human (red), Digital twin (feature-rich) (blue), Digital twin (demographic) (green).

**Note regarding abbreviations in Figure 7b**

To enhance visual readability, standard tags were abbreviated as follows:

D: DET (Determiner)

N: NOUN (Noun)

PR: PRON (Pronoun)

PN: PROPN (Proper noun)

PA: PART (Particle)

AU: AUX (Auxiliary)

V: VERB (Verb)

S: SCONJ (Subordinating Conjunction)

C: CCONJ (Coordinating Conjunction)

**SI References**

**References supporting construct-validity recommendations for LLM-based digital-twin studies**